\documentclass[preprints,review,accept,pdftex,moreauthors]{Definitions/mdpi} 

\firstpage{1} 
\pubvolume{1}
\issuenum{1}
\articlenumber{0}
\pubyear{2023}
\copyrightyear{2023}
\datereceived{} 
\dateaccepted{} 
\datepublished{} 
\hreflink{https://doi.org/} % If needed use \linebreak
\usepackage{mathtools}
\usepackage{gensymb, tikz}
\newcommand{\myarrow}{\tikz\draw[thin,black,-latex] (0,2.5ex) -- ++(0,-1.5ex) -- +(2.5ex,0);}

\usepackage[normalem]{ulem}

\Title{Cosmic ray processes in galactic ecosystems}

\TitleCitation{Cosmic ray processes in galactic ecosystems}

\Author{Ellis R. Owen$^{1}$\orcidA{}, Kinwah Wu$^{2}$\orcidB{}, 
Yoshiyuki Inoue$^{1, 3, 4}$\orcidC{}, 
H.-Y. Karen Yang$^{5,6,7}$\orcidD{}, 
Alison M. W. Mitchell$^{8}$\orcidE{}}
\address{%
$^{1}$ \quad Theoretical Astrophysics, Department of Earth and Space Science, Graduate School of Science, Osaka University, Toyonaka,
Osaka 560-0043, Japan\\
$^{2}$ \quad Mullard Space Science Laboratory, University College London, Holmbury St. Mary, Dorking, Surrey, RH5 6NT, United Kingdom \\
$^{3}$ \quad Interdisciplinary Theoretical \& Mathematical Science Program (iTHEMS), RIKEN, 2-1 Hirosawa, Saitama 351-0198, Japan \\
$^{4}$ \quad Kavli Institute for the Physics and Mathematics of the Universe (WPI), UTIAS, The University of Tokyo, Kashiwa, Chiba 277-8583, Japan \\
$^{5}$ \quad Institute of Astronomy, National Tsing Hua University, Hsinchu 30013, Taiwan (ROC)\\
$^{6}$ \quad Center for Informatics and Computation in Astronomy, National Tsing Hua University, Hsinchu 30013, Taiwan (ROC)\\
$^{7}$ \quad Physics Division, National Center for Theoretical Sciences, Taipei 106017, Taiwan (ROC) \\
$^{8}$ \quad Erlangen Centre for Astroparticle Physics, Friedrich-Alexander-Universität Erlangen-Nürnberg, Nikolaus-Fiebiger-Str. 2, 91058 Erlangen, Germany
}

\corres{Correspondence: erowen@astro-osaka.jp}

\abstract{Galaxy evolution is an important topic, and our physical understanding must be complete to establish a correct picture. This includes a thorough treatment of feedback. The effects of thermal-mechanical and radiative feedback have been widely considered, however cosmic rays (CRs) are also powerful energy carriers in galactic ecosystems. Resolving the capability of CRs to operate as a feedback agent is therefore essential to advance our understanding of the processes regulating galaxies. The effects of CRs are yet to be fully understood, and their complex multi-channel feedback mechanisms operating across the hierarchy of galaxy structures pose a significant technical challenge. This review examines the role of CRs in galaxies, 
from the scale of molecular clouds to the circum-galactic medium. An overview of their interaction processes, their implications for galaxy evolution, and their observable signatures is provided and their capability to modify the thermal and hydrodynamic configuration of galactic ecosystems is discussed. We present recent advancements in our understanding of CR processes and interpretation of their signatures, and highlight where technical challenges and unresolved questions persist. We discuss how these may be addressed with upcoming opportunities. 
}

\keyword{cosmic rays; galaxy formation and evolution; interstellar medium; circum-galactic medium; outflows; inflows; star-formation; molecular clouds; feedback} 

\begin{document}

%%%%%%%%%%%%%%%%%%%%%%%%%%%%%%%%%%%%%%%%%%
\section{Introduction}
\label{sec:introduction}

\noindent
Constructing a holistic scenario 
  of the formation and evolution of galaxies 
  requires the integration of information  
  from macroscopic (e.g., dynamical and thermal) and microscopic (e.g., interaction and radiative loss) processes. 
One of the technical hurdles we face in achieving this 
  is to resolve
  the multi-channel feedback processes operating 
  between various internal and external components 
  of galaxies,  
  which themselves reside 
  in a broad range of environments that 
 evolve over cosmic time. 
Despite rapid progress,   
  our current understanding of galactic feedback 
  is still at an infant stage.  
Most existing work to date has been phenomenological. 
  It mainly focuses on either 
 thermo-mechanical and radiative effects. 
An example of the study  
  of thermo-mechanical feedback is 
  in the investigation of 
  energetic of supernova (SN) explosions, 
  where the inputs 
  are based on theoretical and numerical modelling 
  \cite[e.g.][]{Shimizu2019MNRAS, Oku2022ApJS, Ostriker2022ApJ, Orr2022ApJ}, and 
  gauged against observational measurements 
  \citep{Rosado1996AA, SanchezCruces2018MNRAS, SanchezCruces2022MNRAS}. 
An example of the study of radiative feedback 
   is the investigation of star-forming cycles 
   regulated by stellar 
  \cite{Hopkins2020MNRAS} 
  or active galactic nucleus (AGN) activities 
  \cite[see e.g.][]{Qiu2019ApJ, Chen2020ApJ}. 
Radiative feedback played a particularly important role in the early Universe, influencing the modes and efficiency of star-forming episodes and their duty cycles 
  \cite{Yajima2017ApJ}. 

Feedback dynamics in galaxies are intrinsically complex.  
They are extend beyond the effects  
  driven by SN explosions and stellar or AGN activities, 
  which are essentially internal processes  
  operating within or around a galaxy. 
A more complete scenario should encompass effects brought about by all agents 
  that can facilitate the exchange of energy and momentum, including shared processes that operate 
  between galaxies, their neighbours 
  and their surroundings.   
Cosmic rays (CRs) serve this purpose well. 
They deposit energy and momentum 
  as they traverse their parent galaxies,  
  and they also deliver energy and momentum 
  to neighbouring galaxies 
  and their surrounding environments.  

CRs interact with the magnetic and radiation fields of galaxies
  and participate 
  in subatomic hadronic and leptonic processes.     
Their presence in galaxies is evident through observational signatures across the electromagnetic spectrum~\cite{Kornecki2020AA, Kornecki2022AA}. 
These CRs contain a substantial amount of energy, and their contribution to the energy budget in a typical galaxy is comparable to 
that contained by radiation, magnetic fields and thermal gas.
  They are therefore a significant ingredient in a typical galaxy, which can be amplified when CR accelerators are abundant~\cite[e.g. in star-forming galaxies; see][]{Owen2018MNRAS}. Their presence not only modifies outflows from 
  star-forming galaxies~\citep[see e.g.][]{Yu2020MNRAS}
  but also drives global feedback processes 
  within the broader galactic ecosystem 
  \citep[see][]{Krumholz2022arXiv, Owen2019MNRAS}. 

In this \textit{Review}, we provide an overview of the current state of understanding of CR processes in galaxies. We present a cross-section of recent progress in the field, particularly highlighting developments surrounding the feedback impact of CRs. We discuss the origins of CRs, their propagation through different components of galactic ecosystems, and their role in regulating thermal and hydrodynamic configurations of galactic environments. The scope of this paper focuses on scales from molecular clouds to the circum-galactic medium (CGM). We emphasize the implications of various CR process for galaxy evolution and star formation, examining the underlying interaction and propagation microphysics, as well as their observational signatures.  This complements the recent review paper by~\citet{Ruszkowski2023arXiv230603141R}, which provides a detailed introduction to the acceleration and transport physics of CRs, discusses their observable signatures and dynamical impacts, and includes a pedagogical review of the physics of CR feedback from the scales of individual SNe and molecular clouds to the CGM and galaxy clusters. Our focus is primarily on the effects of persistent sources of CRs in galaxies, with detailed discussion of hadronic (pp and p$\gamma$) interactions, including their application to stellar and galactic scale jet sources. We exclude burst-like or transient sources (e.g. neutron star-neutron star mergers, $\gamma$-ray bursts, fast radio bursts or tidal disruption events) and restrict the scope of our discussion to CR sources and processes up to the scale of individual galactic ecosystems. We therefore exclude larger structures such as groups, clusters, and large, strong AGN jets (weak jets in radio-quiet AGNs that terminate at the scale of a kpc are considered in section~\ref{sec:agn_jets_outflows}).
  
  This \textit{Review} is arranged as follows: Section~\ref{sec:cosmic_ray_physics} introduces the relevant CR physics that underlies the subsequent discussions.  Section~\ref{sec:crs_in_ism} discusses the origins and impacts of CRs within the interstellar medium of galaxies. Section~\ref{sec:high_energy_environments} considers CR processes in high-energy environments, with a particular focus on systems associated with relativistic jets on galaxy scales or smaller. Section~\ref{sec:crs_in_gals} discusses CRs in individual galaxies and galactic ecosystems, including the CGM. Finally, Section~\ref{sec:conclusions} presents our concluding remarks and discusses potential future research directions in light of new and up-coming observational and theoretical opportunities and developments. 

%%%%%%%%%%%%%%%%%%%%%%%%%%%%%%%%%%%%%%%%%%
\newpage 
\section{Cosmic ray physics in galaxies} 
\label{sec:cosmic_ray_physics}

\subsection{Particle transport}
\label{sec:particle_transport}

A simple prescription of CR transport 
that retains all essential physical aspects 
can be expressed in terms of the transport equation: 
\begin{align}
 & \bigg\{  \frac{\partial} {\partial t} 
    - {\underbrace{\nabla \cdot \left[ D \nabla  
      - (u + \bar{v}_{\rm A}) \right] }_{\mathclap{\text{Term 1}}}} 
    - {\underbrace{\frac{\partial}{\partial p}\left[ p^2 {\cal D}_{pp}
    \frac{\partial}{\partial p} \frac{1}{p^2} \right] }_{\mathclap{\text{Term 2}}}} 
    + {\underbrace{\frac{\partial}{\partial p}\left[ \dot{p} 
    - \frac{p}{3} \nabla \cdot(u + \bar{v}_{\rm A}) \right] }_{\mathclap{\text{Term 3}}}} \bigg\} \psi_{\rm X}
      \nonumber \\
    & \hspace*{0.5cm} =  \Gamma_{\rm X} - \Lambda_{\rm X} \psi_{\rm X} \ , 
    \label{eq:transport_eq}
\end{align}
where the differential particle density of a CR species ${\rm X}$ per unit of momentum $p$ is written as $\psi_{\rm X}$.  
$\Gamma_{\rm X}$ is a source term, describing the injection of 
CRs of species ${\rm X}$. This may include primary CRs accelerated in a population of sources, or secondary CR species produced by interactions of parent nucleons. 
$\Lambda_{\rm X} \psi_{\rm X}$ is an absorption term. It describes stochastic 
interaction events that 
result in the destruction or absorption of a particle. It can also be used to describe  catastrophic energy losses, where a substantial fraction of the energy of a particle is lost in a single interaction event. 

\textit{Term 1} in equation~\ref{eq:transport_eq}, 
accounts for the spatial transport of the CR distribution. 
This includes diffusion, specified by the spatial diffusion coefficient $D$. 
It also includes advection of CRs in a background bulk flow of velocity $u$. Under advection,  
the total propagation speed of the CRs is be 
$(u + \bar{v}_{\rm A})$. This is the sum of the 
background fluid velocity $u$ that is carrying the CRs, 
and the effective velocity of CRs in their local medium. 
The 
main source of scattering between CRs and their surrounding plasma 
is their interaction with magnetohydrodynamic (MHD) waves. This 
is dominated by CR scattering in 
Alfv\'{e}nic waves, which are driven by the CR gyro-resonant streaming instability. Assuming that the resonant waves are sufficiently energetic, the strong scattering of the CRs practically limits their velocity to the mean Alfv\'{e}n speed $\bar{v}_{\rm A} \approx \langle |B_0| \rangle / \sqrt{4 \pi n_i m_i}$, averaged over the direction of motion of the waves~\cite[see, e.g.][]{Thomas2019MNRAS}. This provides a good 
approximation to their 
 local effective velocity. Here, $n_i$ and $m_i$ are the mass and number density of the background ions, respectively. 
\textit{Term 2} describes the momentum diffusion of the CRs. This is usually dominated by diffusive re-acceleration, characterised by the energy-dependent momentum-space diffusion coefficient ${\cal D}_{pp}$. 
\textit{Term 3} describes the momentum variation of CRs due to gains or losses, including those associated with the movement of the CR fluid and advection (e.g. adiabatic losses). In most galactic scenarios of interest we will discuss in this paper, this term is dominated by particle cooling processes. 

The precise physics that is introduced to each of the terms in equation~\ref{eq:transport_eq} depends on the CR species being considered, ${\rm X}$, and the exact configuration of the local medium. In the context of a galaxy, it is usually sufficient to consider only CR protons and electrons (including any collisional processes that convert one species into another - for example, hadronic interactions). Diffusive re-acceleration processes (term 2) may also often be neglected away from CR sources. Moreover, some parts of the interstellar medium (ISM) are not subject to significant bulk advective flows or strong turbulence~\cite[see][for examples of such simplifications]{Amato2018AdSpR}. These factors allow for the considerable simplification of the transport equation. In many scenarios, analytic or semi-numerical solutions can often be constructed which offer a meaningful description of the dominant physics. In cases where more complex approaches are required to properly capture the subtle details of CR transport, more sophisticated numerical CR propagation and interaction codes are used to solve equation~\ref{eq:transport_eq} without invoking certain assumptions~\cite[e.g., see][]{Strong1998ApJ, Maurin2001ApJ, Evoli2008JCAP, Kissmann2014APh}. Many of the subtle effects handled by these codes fall outside the scope of this review. However, those that are essential to consider for the global evolution of galaxies are discussed in the following sections. 

\subsubsection{The resonant cosmic ray streaming instability and self-confinement}
\label{sec:transport_microphys}

In free space, CRs propagate by streaming at close to the speed of light $c$. However, in 
 magnetized, 
ionized media, they are scattered by MHD waves.  
In the presence of 
CRs streaming at speeds faster than the local Alfv\'{e}n speed, 
MHD waves become unstable to growth. Resonant wave modes corresponding to 
the gyro-radius of passing CRs 
are rapidly amplified. This 
is the \textit{resonant} mode of the CR streaming instability.~\cite{Skilling1975MNRAS}.\footnote{This is different from a \textit{non-resonant} instability (see section~\ref{sec:nr_CRSI}; also called Bell's instability).}  It strengthens CR scattering effects leading to highly non-linear, and locally isotropic CR propagation. CRs are slowed down to the local Alfv\'{e}n speed and 
efficiently confined (the \textit{self-confinement} picture). A dynamical coupling between the CRs and their background plasma is also established, where the medium can be altered by 
 non-thermal pressure gradients associated with the CRs. It can also be heated by CR energy transferred 
 to MHD modes which then undergo damping (see section~\ref{sec:MHD_damping}). 

In the Galactic context, the resonant CR streaming instability has been shown to produce observed breaks in the CR spectrum~\cite{Aloisio2015A&A, Amato2018AdSpR}, while the self-confinement effect can set the CR transport physics within the Galaxy~\cite[for reviews, see][]{Wentzel1974ARA&A, Cesarsky1980ARA&A}. An alternative model of CR transport has also been considered, called the \textit{extrinsic turbulence} picture. In this scenario, CRs also interact resonantly with the turbulence. However, instead of 
scattering off waves that have been amplified by the resonant streaming instability, 
they scatter off pre-existing waves formed by turbulent cascades. In this picture, CR propagation is practically reduced to advection with the background gas, and results in no overall energy transfer between the CR and thermal fluids~\cite{Zweibel2017PhPl}.\footnote{CR energy gains and losses are balanced, as they cancel owing to equal-intensity waves propagating in opposite directions~\cite{Zweibel2017PhPl}}

\subsubsection{Magnetohydrodynamic wave damping and implications for cosmic ray propagation}
\label{sec:MHD_damping}

In the self-confinement picture, the growth of MHD modes is balanced against damping processes. If damping is severe, the growth rate of CR streaming instabilities is moderated. In extreme cases, damping can prevent the growth of MHD modes entirely. More typically, it operates to limit the CR streaming instability and regulates the amplitude of Alfv\'{e}n waves that develop. This determines the strength of CR scattering and confinement, and sets the effective diffusion speed for CR transport~\cite{Plotnikov2021ApJ}. 
The effectiveness of self-confinement and individual damping processes depend on the local conditions and CR energy. In the ISM, the main damping mechanisms often considered are \textit{non-linear} damping, and \textit{ion-neutral} damping.  In non-linear damping, a turbulent cascade develops in wave-wave interactions, leading to dissipation at small scales. Ion-neutral damping instead arises from collisions between neutral particles in a semi-ionized medium, and the ions that are coupled to the MHD waves. Individual collisions transfer kinetic energy from the ions to neutral particles, which damps MHD waves and thermalizes their energy into the background medium. 

Ion-neutral damping is more severe at longer wavelengths and in regions of higher density with a low ionization fraction. It is the dominant means of moderating the growth of MHD modes in most components of the ISM~\cite{Amato2018AdSpR}, with particular suppression expected in molecular clouds (except for waves that are resonant with CRs of very high energies, above a few TeV~\cite{Amato2018AdSpR}). Other damping mechanisms have also been considered. For example, Coulomb collisions can affect MHD waves in ionized or partially-ionized media. In highly ionized ISM phases, simulations have demonstrated they can suppress pressure anisotropies and boost the magnetic field amplification~\cite{Marret2022PhRvL}. Dust grains in the ISM have also been shown to damp Alfv\'{e}n waves. This is more severe in well-ionized parts of the ISM, where CR diffusion can be noticeably enhanced in the presence of dust compared to the standard self-containment picture~\cite{Squire2021MNRAS}.\footnote{It has also been shown that the opposite effect is possible, if dust streams super-Alf\'{e}nically. In this case, CR propagation is suppressed, particularly on scales which are gyro-resonant with the dust~\cite{Squire2021MNRAS}.}

\subsubsection{The non-resonant cosmic ray streaming instability}
\label{sec:nr_CRSI}

The non-resonant streaming instability (NRSI) arises when a flux of CRs passes through a plasma.\footnote{Hadronic CRs are assumed in this discussion. Modified forms of the NRSI are relevant when it is driven by leptonic CRs~\cite{Gupta2021ApJb}.} Their current generates magnetic perturbations, which impart a force into the background field. These drive velocity fluctuations and induce an electric field. The electric field amplifies the magnetic perturbations, creating a feedback loop to cause MHD wave growth. The wavelengths most affected are generally much shorter than the CR gyro-radius. Their growth rate can be very large in some situations, particularly when the CR velocity or abundance is high or the magnetic field is strong. However, wave growth does not occur when the effective CR velocity is below a critical value, $v_{\rm NR} = c U_{\rm B}/U_{\rm CR}$~\cite{Amato2018AdSpR}, where $U_{\rm B}$ and $U_{\rm CR}$ are the energy densities of the magnetic fields and CRs, respectively. The NRSI therefore only operates in certain situations. In typical ISM conditions it is inconsequential. However, it is 
important for shocks propagating in the cold ISM~\cite{Zweibel2010ApJ}. These may arise in an isolated SN remnant (SNR) during its first $\sim 10^4$ yr~\cite{Pelletier2006A&A, Amato2009MNRAS}. For SNR shocks propagating in a hot super-bubble, a modified form of the NRSI may also operate when 
CR densities are high (about 1000 times that of the Galactic ISM), or if the magnetic field in the bubble is below a few $\mu$G~\cite{Amato2011MmSAI}. In this case, the growth of the non-resonant instability may be weaker~\cite{Reville2008IJMPD, Zweibel2010ApJ, Marret2021MNRAS}. 

 Conditions where the NRSI is important are typical of those often associated with CR sources (see also sections~\ref{sec:accelerators_SNRs} and~\ref{sec:pevatrons}). 
 The NRSI has been shown to be essential  
for accelerators to attain the maximum observed CR energies (see~\cite{Blasi2013A&ARv, Amato2014IJMPD} for recent reviews). Specifically, it can account for the fast growth of MHD waves around SNR shocks~\cite{Bell2004MNRAS}, and can produce the inferred strong magnetic fields needed for CR acceleration up to PeV energies (see, e.g.~\cite{Vink2012A&ARv}, or~\cite{Bykov2012SSRv} for a review). 
The wavelengths most strongly amplified by the NRSI are much shorter than the CR gyro-radius. This means 
they do not effectively scatter CRs, and cannot contain them to sustain their acceleration unless the waves amplified by the NRSI undergo an inverse-cascade to larger modes. The mechanism underlying such an inverse cascade is currently unsettled, although some proposals have noted that the dominant wavelength in the non-resonant instability increases with time, in proportion to $(\delta B/B_0)^2$~\cite{Riquelme2009ApJ}. This could boost CR scattering and containment if the inverse-cascade timescale is shorter than advection. 

The NRSI may operate in conjunction with the resonant streaming instability. For example, in the region surrounding CR sources, the excitation of both resonant and non-resonant modes enhances scattering and produces strong CR pressure gradients~\cite{Schroer2021ApJ}. These pressure gradients can drive the formation of CR-dominated bubbles of gas and self-generated magnetic fields. They expand in to the ISM until reaching pressure balance with the surrounding medium, typically at sizes of a few 10s of pc~\cite{Schroer2022MNRAS}. Within the bubbles, 
CR diffusion is suppressed~\cite{Schroer2021ApJ}. If this suppression is sufficiently severe, CRs can become trapped within these structures~\cite{Commercon2019A&A}.

\subsection{Particle interactions}
\label{sec:hadronic_ints}

CRs are inhomogeneous 
  collections of particles consisting of atomic nuclei,  
  bare baryons and mesons, leptons 
  and ultra-relativistic particles 
  such as neutrinos.  
Here we put focus on the CR particles   
  involved in collisional and/or hadronic interactions. This is because they 
   carry the 
  greatest energy density and momentum, and are most 
  strongly engaged with feedback processes in galaxies.  
  Leptonic CRs are primarily composed of 
  electrons and positrons. 
  Since both electrons and positrons undergo similar interactions, we refer to them collectively as "electrons" hereafter for convenience. 
Within galactic environments, electrons experience collisional ionization interactions, free-free cooling, and radiative cooling processes in photon and magnetic fields. Typically, electrons lose their energy rapidly and are less involved in transport throughout galactic ecosystems. Instead, they are more important in mediating energy deposition processes, such as thermalization. 
  
Hadronic CRs are primarily protons, although heavier nuclei are also present. 
The dominant interaction mechanisms for heavy nuclei in galactic ecosystems are essentially the same as those for protons. 
At low energies, below a few hundred MeV, collisional and ionization processes serve as the predominant channels of interaction. However, at GeV energies and above, p$\gamma$ (proton-photon) and pp (proton-proton) processes can become important. These processes result in the rapid production of electrons, neutrinos, and high-energy photons through pion formation. They play a role in transferring energy from a non-thermal hadronic CRs to thermal gas by the formation and thermalization of secondary CR electrons~\cite[e.g.][]{Owen2018MNRAS, Simpson2023MNRAS}.  
Hadronic CRs, from protons to heavy nuclei, are engaged in pion-producing interactions and Bethe-Heitler pair production in radiation fields. Despite the variations in the nuclei of a CR ensemble, their fundamental hadronic interaction processes remain the same. Therefore, without loss of generality, we illustrate these interactions using free protons (p) as a representative example
  of high-energy hadronic CR interactions 
  in matter (baryon)
  and 
  radiation (photon) fields. 

\subsubsection{pp interactions}
\label{sec:pp_interaction}

The pp interaction between CR protons and baryon leads to the following dominant pion production channels: 
\begin{align}\label{eq:pp_interaction}%
{\rm p} + {\rm p} \rightarrow %
	\begin{cases}%
		&{\rm p}  \Delta^{+~\;} \rightarrow\begin{cases}%
				{\rm p} {\rm p} \pi^{0}  \xi_{0}(\pi^{0}) \xi_{\pm}(\pi^{+} \pi^{-}) \\[0.5ex]%
				{\rm p} {\rm p}  \pi^{+}  \pi^{-}  \xi_{0}(\pi^{0}) \xi_{\pm}(\pi^{+} \pi^{-}) \\[0.5ex]%
				{\rm p} {\rm n}  \pi^{+}  \xi_{0}(\pi^{0}) \xi_{\pm}(\pi^{+} \pi^{-})\\[0.5ex]%
			\end{cases} \\%
		&{\rm n} \Delta^{++} \rightarrow\begin{cases}%
				{\rm n} {\rm p} \pi^{+} \xi_{0}(\pi^{0}) \xi_{\pm}(\pi^{+} \pi^{-}) \\[0.5ex]%
				{\rm n} {\rm n} 2\pi^{+} \xi_{0}(\pi^{0}) \xi_{\pm}(\pi^{+} \pi^{-})\\[0.5ex]%
			\end{cases} \\%
	\end{cases} \ .%
\end{align}%
Here, $\xi_{0}$ and $\xi_{\pm}$ are the multiplicities for neutral and charged pions, respectively. 
 The $\Delta^{+}$ and $\Delta^{++}$ baryons are the resonances 
  \citep[see][]{Almeida1968, Skorodko2008EPJA}. This interaction operates above a threshold energy of 
  $E^{\rm{th}}_{\rm p} = (2m_{\pi^{0}}+m^{2}_{\pi^{0}}/2m_{\rm p}) {\rm c}^2 \approx0.28~\rm{GeV}$, which is the proton energy required for the production of a neutral pion through the channel  
   ${\rm p}{\rm p} \rightarrow {\rm p} {\rm p}\pi^{0}$. 
   
The branching ratios across all relevant channels for the production of each pion species $\{\pi^{+}, \pi^{-}, \pi^{0}\}$ is $\{0.6, 0.1, 0.3\}$ at 1 GeV. At higher energies, this levels out to $\{0.3, 0.4, 0.3\}$~\cite{BlattnigPRD2000}. Overall, approximately 30 per-cent of the total interaction energy is transferred to neutral pions, while the remainder goes to charged pion production. 
The neutral pions predominantly decay into two photons through an electromagnetic process: 
\begin{align}%
	\pi^0	&\rightarrow 2\gamma  \ .   % 
\end{align}%	 
 This has a branching ratio of 98.8 per-cent and occurs over a timescale of $8.5 \times 10^{-17}\;\!{\rm s}$~\citep{Patrignani2016ChPh}.   
The charged pions decay into electrons and neutrinos through a weak interaction:
 \begin{align}%	  
 \label{eq:weak_interaction2}
	\pi^+	&\rightarrow \upmu^+ \nu_{\rm \upmu} \rightarrow {\rm e}^+ \nu_{\rm e} \bar{\nu}_{\rm \upmu} \nu_{\rm \upmu}\     \nonumber \\%
	\pi^-	&\rightarrow \upmu^- \bar{\nu}_{\rm \upmu} \rightarrow {\rm e}^- \bar{\nu}_{\rm e} \nu_{\rm \upmu} \bar{\nu}_{\rm \upmu}       % 
 \ .  
\end{align}%
 This channel has a branching ratio of 99.9 per-cent and occurs over a timescale of $2.6\times 10^{-8}\;\!{\rm s}$~\citep{Patrignani2016ChPh}. Around 3/4 of the energy is inherited by neutrinos in this decay process. The remainder is transferred to the secondary electrons.

The total inelastic cross-section 
and formation of secondary products in the pp interaction has been extensively investigated. 
Monte Carlo (MC) event generators have been developed, which simulate high-energy collision events to give a precise description of hadronic interactions in light of new accelerator data, including from the large hadron collider (LHC). 
These generators handle complex calculations and predict many quantities that show good agreement with experimental data~\cite[e.g.][]{Amenomori2019EPJWC}. Several classes of these MC event generator codes exist. General purpose examples include {\tt HERWIG}~\cite{Bahr2008EPJC}, {\tt SHERPA}~\cite{Gleisberg2009JHEP}, {\tt GEANT4}~\cite{Allison2006ITNS, Allison2016NIMPA, Agostinelli2003NIMPA} and {\tt PYTHIA}~\cite{Sjostrand2008CoPhC, Sjostrand2006JHEP, Bierlich2022arXiv220311601B}, which include both Standard Model and beyond Standard Model physics. Other codes include particularly sophisticated high energy hadronic physics models. These include {\tt Phojet}~\cite{Engel1995PhRvD}, {\tt DPMJET}~\cite{Bopp2008PhRvC} and {\tt EPOS}~\cite{Werner2006PhRvC}. Other types of MC codes specialize in air shower simulations, and are particularly concerned with high energy secondary particle production. These include {\tt SIBYLL}~\cite[e.g.][]{Ahn2009PhRvD, Fletcher1994PhRvD, Engel1992PhRvD, Fedynitch2019PhRvD} and {\tt QGSJET}~\cite{Ostapchenko2006NuPhS, Ostapchenko2007AIPC}, and can be widely useful to inform models of secondary particle production in astrophysical applications. 

To reduce computational requirements, parametrizations based on pp interaction MC event generators have also been developed. 
These provide analytical descriptions of key quantities such as the total inelastic cross-section and $\gamma$-ray production spectra~\cite{Kafexhiu2014}. 
Some parametrizations also include inclusive production spectra of secondary electrons and neutrinos~\cite{Kelner2006, Kamae2006ApJ}, making them versatile alternatives to expensive numerical computations. 
Interpolated codes have also been created to provide these parametrizations as user-friendly computational tools~\cite[e.g.][]{Kachelriess2019CoPhC, Koldobskiy2021PhRvD}. While these  
parametrizations are suitable for use in a wide range of astrophysical settings and bypass the need to conduct intensive numerical calculations for pp collisions, they are also inherently limited. In particular, they integrate out intermediate particles in interaction and decay chains, and so do not accurately capture rapidly time-varying systems or situations where conditions lead to noticeable cooling of secondary intermediaries. Therefore, careful consideration should be given to the suitability of these parametrizations in more extreme or highly transient situations. 

\subsubsection{p$\gamma$ interactions}
\label{sec:pgamma_interaction} 
     
p$\gamma$ processes are less straightforward to model. In many astrophysical environments, the contribution of the target photon field to the total interaction energy cannot be neglected. As such, p$\gamma$ interaction calculations depend on both the energy of the CR protons involved in the interaction, and the energy distribution of the target photon field. p$\gamma$ interactions can be categorized into photo-pion production and Bethe-Heitler pair production interactions. 
Photo-pion production, similar to pp interactions, can generate various hadrons, including charged and neutral pions, neutrons, and protons. 
On the other hand, photo-pair production predominantly results in the formation of electron-positron pairs. It can also produce higher-energy lepton pairs such as muons and anti-muons. These higher-mass lepton pairs can subsequently decay into electrons and positrons, with neutrinos as a by-product.\footnote{It has been proposed that such production of muon anti-muon pairs can be part of a purely leptonic mechanism to produce TeV-scale neutrinos in astrophysical environments~\cite{Hooper2023arXiv230506375H}.} 

Photo-pion production occurs when incident protons collide with the photons of a radiation field. The dominant interactions in photo-pion production are resonant single-pion production, direct single-pion production, and multiple-pion production~\citep{Mucke1999PASA}. Although other processes such as diffractive scattering can arise, they are generally less significant. The total cross-section of photo-pion production 
  is therefore the sum of the cross-sections of these three main interactions. 
Resonant single-pion production occurs through the formation of $\Delta^+$ particles. 
 These decay through two major channels that produce charged and neutral pions~\citep[see][]{Berezinsky1993}: 
\begin{align}%
\label{eq:pg_int}%
{\rm p} \gamma \rightarrow \Delta^{+} \rightarrow%
	\begin{cases}%
	    {\rm p} \pi^0 \rightarrow {\rm p}2\gamma				\\[0.5ex]%
		{\rm n} \pi^+ \rightarrow {\rm n} \upmu^+ \nu_{\upmu}		\\%
		\hspace{4.1em} \myarrow {\rm e}^+ \nu_{\rm e} \bar{\nu}_{\upmu}% 
	\end{cases}  \ .
\end{align}%  
The branching ratios for the $\Delta^{+}\rightarrow\pi^0$ and $\Delta^{+}\rightarrow\pi^+$ channels 
   are 2/3 and 1/3, respectively. 
Direct pion production is less efficient. 
While single-pion production dominates at lower energies,  at higher energies  
  multi-pion production becomes important  
  \citep[see][]{Mucke1999PASA}. 

Some neutrons form in process~\ref{eq:pg_int}. These have a half-life of about $880\;\!{\rm s}$~\citep{Nakamura2010}. 
If they do not first collide with other particles or photons, 
  they undergo $\beta^{-}$-decay, 
${\rm n} \rightarrow {\rm p} {\rm e}^- \bar{\nu}_{\rm e}$. 
They may instead undergo further interaction with the radiation field. This leads to additional $\rm p$ and $\pi^-$ production, 
\begin{align}%
{\rm n}\gamma \rightarrow \Delta^0 \rightarrow %
		\begin{cases} %
			{\rm p} \pi^-  \\[0.5ex]%
			{\rm n} \pi^0%
		\end{cases} \ , %
\end{align}%   
with branching ratios of 1/3 for the $\Delta^{0}\rightarrow\pi^-$ channel, and 2/3 for the $\Delta^{0}\rightarrow\pi^0$ channel~\citep{Hummer2010}. When taking the additional charged pions produced in the residual interactions into account
  (including ${\rm p}\gamma \rightarrow \Delta^{++} \pi^{-}, \Delta^{0}\pi^{+}$), 
  each of the pion species are produced in roughly equal proportion 
   \citep[e.g.][]{Dermer2009book}.   

p$\gamma$ interactions in astrophysical settings can be accurately modeled using MC event generator software, such as {\tt SOPHIA}~\cite{Mucke1999PASA, Mucke2000CoPhC}. This provides a comprehensive framework to thoroughly study p$\gamma$ interactions, taking into account detailed cross-section data. Alternatively, to alleviate computational demands, self-consistent models are available that simplify the treatment of interaction kinematics, but adopt the same underlying particle physics as {\tt SOPHIA}~\cite{Hummer2010ApJ}. These models account for the production of the $\Delta$-resonance and higher resonances, multi-pion production, and direct pion production. By separately calculating the contributions of pion intermediaries without integrating out secondaries, they preserve the essential particle physics involved. In situations where the retention of this detail 
is unnecessary and computational efficiency is a priority, analytical parametrizations for p$\gamma$ interactions have also been developed~\cite[e.g.][]{Kelner2008}.

\subsubsection{Comparison between pp and p$\gamma$ channels in galactic environments}
\label{sec:pp_pg_comparison}

\begin{figure}[H]
\begin{adjustwidth}{-\extralength}{0cm}
\centering
\includegraphics[width=17.5cm]{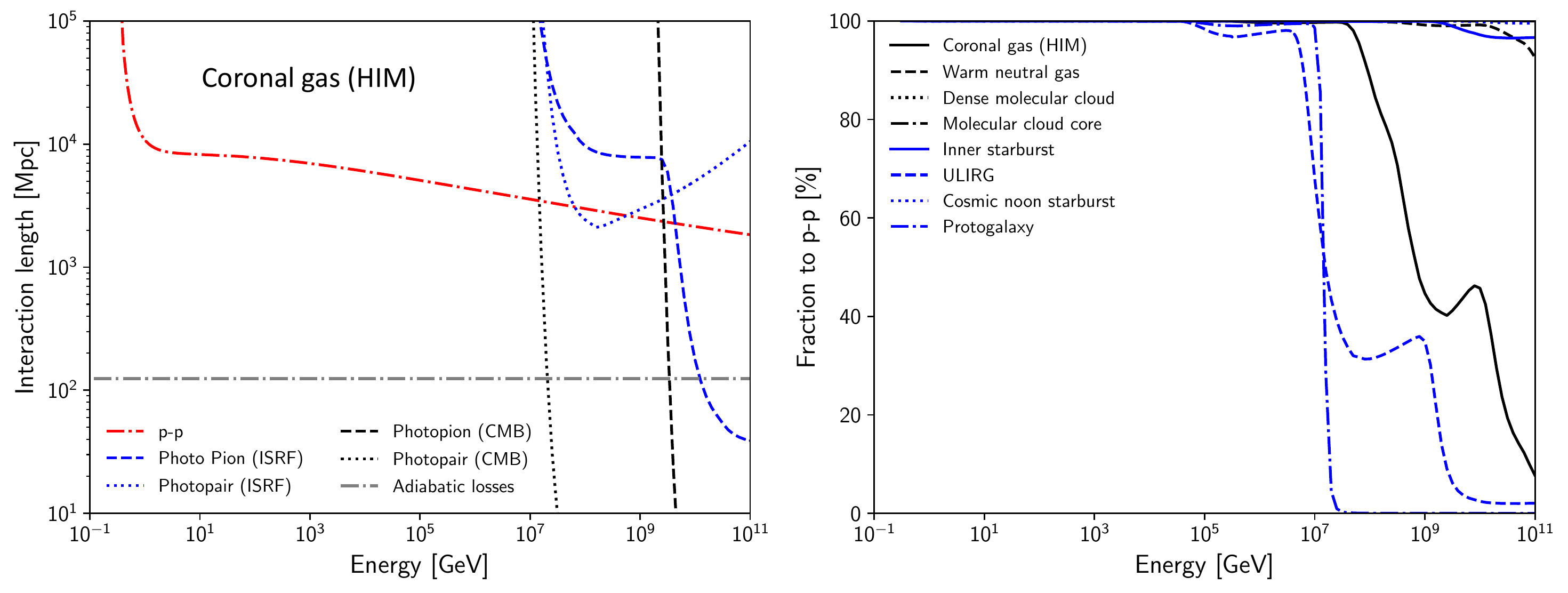}
\end{adjustwidth}
\caption{\textbf{Left:} CR proton losses in terms of effective interaction path lengths for conditions representative of the hot ionized medium (HIM). Losses shown are due to pp interactions, p$\gamma$ (photo-pion and photo-pair) interactions in interstellar radiation fields, p$\gamma$ interactions in the cosmological microwave background (CMB), and adiabatic losses due to the expansion of the Universe. p$\gamma$ processes are inconsequential below energies of $\sim$ 10$^7$ GeV, with pp processes in this hot rarefied gas also being relatively unimportant. The pp interaction rate scales with gas density, and so would become much more severe in warm neutral gas or dense molecular clouds (cf. Table~\ref{tab1:hadr_int_params}). \textbf{Right:} Fractional energy transfer in hadronic collisions to the pp channel. At lower energies, pp losses entirely dominate hadronic processes. At higher energies, hadronic CRs are more likely to engage in p$\gamma$ interactions in environments with strong radiation fields. 
The p$\gamma$ path lengths shown here are computed using analytic approximations rather than by full numerical calculation. This can differ from a thorough treatment by as much as an order of magnitude~\cite{Dermer2009herbbook}, but is not consequential for the conclusions here.}
\label{fig:hadronic_ints}
\end{figure}  

In Fig.~\ref{fig:hadronic_ints}, the relative importance of pp and p$\gamma$ interactions is estimated by comparing the energy loss path lengths of CR protons in different galactic environments. These environments are characterized by parameter values listed in Table~\ref{tab1:hadr_int_params}, and include various components of the ISM of our Galaxy, as well as average conditions expected in starburst galaxies, cosmic noon galaxies, and high-redshift protogalaxies. 
In all of these cases, it can be seen that pp interactions dominate over p$\gamma$ interactions at most energies. The 
p$\gamma$ channel only becomes important at CR energies above $\sim 10^7$ GeV, with the exact transition energy being strongly influenced by the specific physical conditions. The number of CRs at these energies is generally very low. 
It may therefore be considered that the p$\gamma$ interaction only has a negligible involvement in CR feedback in galaxies. Instead, the pp interaction is most involved in regulating the impacts of hadronic CRs in galactic ecosystems. 

\begin{table}[H] 
\caption{Summary of parameters adopted for galaxy and ISM conditions in hadronic pp and p$\gamma$ path length calculations, as shown in Fig.~\ref{fig:hadronic_ints}.}
\begin{adjustwidth}{-\extralength}{0cm}
\newcolumntype{C}{>{\centering\arraybackslash}X}
\begin{tabularx}{\fulllength}{lCCCCCC}
\toprule
\multirow[m]{2}{*}{\textbf{Model}}	& \multicolumn{2}{c}{\textbf{ISRF Energy density}} & \multicolumn{2}{c}{\textbf{ISRF Radiation Temperatures}}& \multirow{2}{*}{\textbf{Gas density}} & \multirow{2}{*}{\textbf{Redshift}} \\
& Stars & Dust & Stars & Dust & & \\
\midrule
Coronal gas/HIM$^{(a)}$		 & 0.66 & 0.31 & 3,000 & 17 & 0.004 & 0 \\
Warm neutral gas$^{(a)}$	 & 0.66 & 0.31 & 3,000 & 17 & 0.6 & 0 \\
Dense molecular cloud$^{(a)}$	 & -- & -- & -- & -- & 10$^{3}$ & 0 \\
Molecular cloud core$^{(a)}$	 & -- & -- & -- & -- & 10$^{6}$ & 0 \\
Inner starburst$^{(b)}$	 & 79.2 $^{(d)}$ & 1.5$\times 10^{6}$ & 18,000$^{(e)}$ & 135 & 6.5$\times 10^{5}$ & 0.018 \\
ULIRG$^{(c)}$	 & 673 & 310 & 18,000$^{(e)}$ & 135 & 0.1$^{(f)}$ & 0.64 \\
Cosmic noon starburst$^{(g)}$ & 2630 & 1230 & 18,000$^{(e)}$ & 57 & 10$^{4}$ & 2.33 \\
Protogalaxy$^{(h)}$ & 2.77 & 1.30 & 18,000$^{(e)}$ & 40 & 0.0003$^{(i)}$ & 9.11 \\
\bottomrule
\end{tabularx}
\noindent{\footnotesize{\textbf{Notes}:\\
$^{(a)}$ Densities for the phases of interstellar gas and Galactic ISRF properties adopted from~\cite{Draine2011piim}. For the dust component of the ISRF, we adopt the properties of the dominant component, although note that around 1/3 of the total dust emission power is thought to be emitted by poly-cyclic aromatic hydrocarbons (PAH) which would not strictly emit radiation of the indicated temperature. \\
$^{(b)}$ Nearby inner starburst parameters are adopted to emulate the inner circum-nuclear disk of Arp 220. Values taken from~\cite{YoastHull2019MNRAS}. For dust emission properties, these values were originally derived from~\cite{Wilson2014ApJ, Scoville2017ApJ}. \\
$^{(c)}$ ULIRG parameters are adopted to emulate the nearby Hyper Luminous IR galaxy (HyLIRG) IRAS F14537+1950, where the ISRF intensity is scaled by bolometric luminosity compared to the Galaxy, from~\cite{RR2000MNRAS}. Dust temperatures are assumed to be the same as the central nuclear disk of Arp 220, as heating by intensive star-formation activity would be similar. \\
$^{(d)}$ Interstellar radiation field (ISRF) stellar energy density is scaled from that of the Milky Way by star-formation rate, assuming a rate of 120 ${\rm M}_{\odot}\;\!{\rm yr}^{-1}$ for Arp 220~\cite[following the estimate by Ref.][]{YoastHull2019MNRAS}. \\
$^{(e)}$ Temperature of starburst galaxy stellar radiation field component adopted from~\cite{Chakraborty2013ApJ}, see also~\cite{Schober2015MNRAS}.\\
$^{(f)}$ Estimated from the Galactic column density towards F1437+1950 from Ref.~\cite{Wilman1998MNRAS}, assuming a kpc sized galaxy. \\
$^{(g)}$ Properties based on the Eyelash galaxy, SMM J2135-0102~\cite{Danielson2011MNRAS}. This is a lensed, dusty cosmic noon submillimeter galaxy where ISRF intensities are scaled against the Milky Way. \\
$^{(h)}$ Based on the properties of the lensed high-redshift galaxy MACS 1149-JD1, with physical parameter choices informed by Ref.~\cite{Hashimoto2018Natur}. The ISRF energy densities are scaled by star-formation rate against Milky Way values. \\
$^{(i)}$ Based on the dust mass obtained by~\cite{Hashimoto2018Natur}, assuming a standard gas-to-dust ratio of 100~\cite{Bohlin1978ApJ}.}}
\label{tab1:hadr_int_params}
\end{adjustwidth}
\end{table}

%%%%%%%%%%%%%%%%%%%%%%%%%%%%%%%%%%%%%%%%%%
\newpage 
\section{Cosmic rays in the interstellar medium}
\label{sec:crs_in_ism}

\subsection{Origins of cosmic rays in galaxies} 
\label{sec:cr_sources}

CRs originate in violent, magnetized astrophysical environments. 
Within galaxies, core-collapse (CC) supernovae (SNe) and their remnants have 
traditionally been regarded as the dominant factories of CRs  
~\cite[see][]{Bykov2018SSRv}. This view 
has been particularly motivated by energy budget arguments, with SNRs among only 
a few candidate source types that can sustain the observed energy flux of CRs in our Galaxy~\cite[e.g.][]{Berezinskii1984acr, Lingenfelter2013AIPC}. 
In recent years, the SNR source paradigm has been increasingly challenged. 
The capability of stellar end products 
to inject 
Galactic CRs with the required spectra and chemical abundances 
to match the observed properties of CRs has become less certain. 
Moreover, the 
lack of clear detections at the highest energies in $\gamma$-rays casts doubt on the capability of SNRs to supply all of the observed CR flux detected on Earth~\cite[for an overview of these challenges, see][]{Cristofari2021Univ}. 
New types of object have therefore been considered, 
including star-forming regions and star clusters. These 
have enjoyed an increasing bank of recent observational support. They have been firmly  established as the origins of at least some CRs in galaxies, and 
have even been proposed as an alternative to SNRs as the dominant source in the Milky Way. 
Our understanding of the origins of CRs in galaxies 
has developed considerably in recent years. 
This section provides an overview of the current status of the field, 
including the main CR source candidates being discussed and developments arising from 
new observational insights at near-PeV energies. 

\subsubsection{Stellar end-products} 
\label{sec:accelerators_SNRs}

SNe in our Galaxy occur at an approximate rate of once per $\sim$35 years~\cite{Li2011MNRAS} and are natural candidates as CR sources. 
CC SNe harbor particularly favourable conditions for particle acceleration. These include 
strong, fast magnetized shocks 
and an abundance of seed particles in the ionized circum-stellar medium (CSM).  
If invoking SNe and their remnants as the only source of CRs in the Galaxy, they must convert around 10 per-cent of their energy into CRs to sustain the observed level of the Galactic CR "sea"~\cite[e.g.][]{Lingenfelter2018AdSpR}. This required efficiency is relatively high. However, it has been shown to be feasible under strong diffusive shock acceleration scenarios~\cite[e.g.][]{Blasi2013A&ARv}.   
CC SNe produce fast shocks that interact with the dense CSM of the progenitor star. Accelerated CRs may be accumulated at these shocks, where they can strengthen magnetic fields through 
streaming instabilities ahead of the shocks. 
 The NRSI (see section~\ref{sec:nr_CRSI}) can be particularly important to amplify magnetic fields in the up-stream region. This mainly happens within a few days of the shock breakout~\cite{Marcowith2018MNRAS, Murase2019ApJ}. 
The strengthened up-stream magnetic fields boosts CR containment and acceleration efficiency~\cite{Marcowith2018MNRAS}. SNe classes hosting  extended shocks, for example type IIb SNe, can experience particularly strong magnetic amplification by this mechanism. 

The maximum energy CR acceleration processes can attain in an expanding SN shell may be estimated from the shock velocity $v_{\rm sh}$, the size of the shell $r_{\rm sh}$, and its mean magnetic field strength, $\langle | B | \rangle$: 
\begin{equation}
    E_{\rm max} \sim 1 \; \left(\frac{r_{\rm sh}}{{\rm pc}} \right) \;\! \left( \frac{v_{\rm sh}}{1,000 {\rm km s}^{-1}} \right) \;\! \left( \frac{\langle | B | \rangle}{\mu{\rm G}} \right) \;\!{\rm TeV} \ , 
    \label{eq:emax_snr}
\end{equation}
~\cite[e.g.][]{Hillas1984ARA&A, Cristofari2021Univ}. 
This indicates that SN shells can 
accelerate CRs to 1-10 TeV energies with typical CSM magnetic field strengths and shock velocities. 
To reach energies in excess of 1 PeV, 
fast ($v_{\rm sh} \sim 10^4$ km s$^{-1}$), 
extended shocks pervading a dense CSM are required. Resonant~\cite{Skilling1975MNRAS, Achterberg1983A&A} and non-resonant~\cite{Bell2004MNRAS, Bell2013MNRAS, Schure2013MNRAS, Schure2014MNRAS} CR streaming instabilities can then build-up the magnetic field to 10s of $\mu$G ahead of the shocks in these circumstances. Such strengths have been 
 observed in young SNR shells (see~\cite{Casanova2022Univ} for a summary). 
These requirements highlight type IIn SNe as viable candidate accelerators capable of operating up to PeV energies~\cite{Marcowith2018MNRAS, Petropoulou2017MNRAS}. In these systems, CR acceleration would start before the shock breakout. A radiatively-dominated shock is initially launched through the hydrostatic core of the progenitor. This propagates outwards to the outer layers of the stellar core, where radiatively-dominated shock is replaced by a collisionless shock~\cite{Waxman2001PhRvL, Chevalier2008ApJ, Ensman1992ApJ, Chevalier1979ApJ}. Diffusive shock acceleration is then possible. However, it has been considered that the formation of a collisionless shock occurs significantly before breakout in some optically thick winds~\cite{Giacinti2015MNRAS} for particles with gyro-radii greater than the shock width~\cite{Bell1978MNRASa, Bell1978MNRASb}.

Remnants of SN explosions have also been regarded as an important source of CRs in galaxies. In SNRs, CRs can be accelerated to a few PeV at the transition between the free expansion phase and the Sedov-Taylor phase, within 100-1000 yr after the explosion~\cite[corresponding to the Sedov time; see][]{Celli2019MNRAS}.\footnote{Prior to the Sedov-Taylor phase, a very small amount of particle escape may also arise during the ejecta-dominated phase~\cite[e.g.][]{Celli2019MNRAS}, when the SNR shock experiences very minor deceleration~\cite{Truelove1999ApJS}.} Particle escape generally arises from the slow-down of the SNR shocks as circum-stellar gas is swept up during the expansion. As the shock becomes slower, it eventually falls below the diffusive propagation speed of the CRs. The CRs then begin to diffuse away upstream ahead of the shock, and the probability of their return from the upstream direction for further acceleration gradually reduces. Eventually the CRs are no longer coupled with the shock~\cite{Drury2011MNRAS}. As particle diffusion is energy-dependent, more energetic CRs diffuse faster and escape from the SNR earlier. Evidence of this can be seen in spectral breaks from SNRs observed in $\gamma$-rays~\cite[e.g.][]{Peron2020ApJ}. 
From equation~\ref{eq:emax_snr}, it can be seen that the maximum energy a SNR can attain will reduce over time as the shocks slow down. 

The accumulated swept-up CSM gas can form a massive shell around the SNR. The dense gas in the shell acts as a target for pp collisions with the accelerated CRs. This can generate $\gamma$-ray production from the decay of neutral pions that form in pp interactions (cf. section~\ref{sec:hadronic_ints}), which has been detected around some SNRs~\cite[e.g.][]{Peron2020ApJ, Feijen2022MNRAS}. 
However SNe with such a dense CSM that can be detected in $\gamma$-rays may not be common~\cite{HESS2019A&A}. The high-energy emission from many SNRs is often instead attributed to the interactions of escaping CRs with nearby molecular clouds~\cite[e.g.][]{Li2023ApJ, Aruga2022ApJ, Zhang2021ApJ, Zhong2023MNRAS, Yeung2023PASJ, Supan2022A&A}. 
Observations of SNR shells at other wavelengths are more informative. In addition to $\gamma$-rays, hadronic pp interactions also generate secondary leptons. These radiate synchrotron emission~\cite{Reynolds2008ARA&A, Murase2014MNRAS}. Together with contributions from primary electrons accelerated by the SNR shocks and/or by turbulent magnetic fields that develop in the vicinity of the shocks~\cite{Bykov2018SSRv}, this emission has been detected at high radio frequencies through to X-rays around young and middle-aged SNRs~\cite{Bykov2018SSRv, Helder2012SSRv, Vink2012A&ARv}. 

\subsubsection{Star-forming regions, star clusters and super-bubbles} 
\label{sec:superbubbles}

Recent studies have put a renewed focus on systems other than SNe and isolated SNRs as the origin of CRs in galaxies. 
This is motivated by differences between the spectra of accelerated particles at SNR shocks and the observed CR spectrum on Earth, as well as difficulties in 
accounting for the CR electron-to-position ratio under a SNR source scenario~\cite[for reviews, see][]{Cristofari2021Univ, Bykov2020SSRv, Tatischeff2018ARNPS}. 
Star-forming regions, including young massive stellar clusters and associated super-bubbles have emerged as particularly promising candidate source populations. 
Star-forming regions hosting massive stars can develop powerful winds. Their massive stellar populations can also rapidly produce SNe. The wind termination shocks, and the shocks associated with the SNe can act as CR accelerators. Star-forming regions have been recently observed in $\gamma$-rays. The Cygnus region has taken centre-stage in many of these observational enquiries, and has perhaps provided the strongest evidence to support 
hadronic CR acceleration in stellar clusters~\cite[e.g.][]{Abeysekara2021NatAs, Cao2021Natur, Ackermann2011Sci, Bartoli2014ApJ}. Particularly notable is the emission associated with Cygnus X. This region includes many young star clusters and OB associations~\cite{Ackermann2011Sci}, and hosts the \textit{Fermi} Cocoon super-bubble which has been detected up to $\sim$ 1.4 PeV~\cite{Abeysekara2021NatAs, Cao2021Natur, Bartoli2014ApJ, Dzhappuev2021ApJ} with the Cygnus OB2 star cluster likely being the origin of the CRs that produce this emission~\cite{Casanova2022Univ}.  Additional studies include 
observations towards W43~\cite{Yang2020A&A}, Westerlund 1~\cite{Aharonian2022A&A, Harer2023A&A}, Westerlund 2~\cite{Mestre2021MNRAS}, the Carina Nebula Complex~\cite{Ge2022MNRAS}, the Aquila Rift and some giant molecular clouds hosting star-formation in the Gould Belt~\cite{Baghmanyan2020ApJ} ($\rho$ Ophiuchus and Cepheus). These studies confirm the role of stellar clusters and star-forming regions as particle accelerators, and demonstrate their capability to operate up to energies well-above a TeV.

The confluence of the shocks and dense winds in tight clusters of star-formation can be particularly important CR factories~\cite{Cesarsky1983SSRv, Voelk1982ApJ}. 
In younger stellar clusters, where a SN event has yet to occur, CR acceleration arises at wind termination shocks. If the cluster is compact, the confluence of winds is decelerated at a spherical termination shock. In looser clusters, this confluence does not generally arise, and each star develops its own wind termination shock.  Older stellar clusters are expected to accelerate CRs through a radically different mechanism. They reside in a super-bubble, which is subjected to the recurrent SNe from the stellar cluster. 
These super-bubbles are regions of hot ($T>10^6\;\!{\rm K}$), low-density ($n<10^{-2}\;\!{\rm cm}^{-3}$) highly turbulent gas carved out by the winds and SN explosions of associations of OB stars, surrounded by a dense communal shell of swept-up ISM gas. The gas internal to the bubble is a combined mixture of chemically enriched SN ejecta and stellar wind material, as well as residual ISM gas~\cite{MacLow1988ApJ}. Within these super-bubbles, particles are accelerated more intermittently, and the process can be more complex than in isolated sources. Acceleration happens 
either in the collective stellar winds, in individual SNe and SNRs~\cite{Reimer2006ApJ, Vieu2022MNRAS}, at the confluence of the shocks and weak reflected shocks from SNRs, in the turbulent medium\footnote{This can also accelerate particles by a second-order Fermi mechanism~\cite[see, e.g.][]{Bykov2001AstL}.} inside the bubble~\cite{Bykov1992MNRAS, Bykov2014A&ARv, Reimer2006ApJ, Parizot2004A&A, Higdon2005ApJ, Bykov2001AstL}, or at the boundary of the bubble. Winds from Wolf-Rayet (WR) stars may also contribute~\cite{Wang2022ApJ, Kamijima2022PhRvD}, with enriched WR wind material being accelerated at stellar wind termination shocks~\cite{Gupta2020MNRAS}. The wind termination shocks in super-bubbles have been shown to be capable of accelerating particles up to PeV energies~\cite{Bykov2014A&ARv, Gupta2018MNRAS, Morlino2021MNRAS}.

Stellar clusters as CR sources provide a more natural explanation of CR chemical composition trends in the ISM~\cite[for brief reviews, see][]{Tatischeff2018ARNPS, Gabici2023arXiv230106505G}.  
Notably, they produce elemental abundances consistent with a mixture of 
primordial and enriched material. This is thought to originate from a seed ISM composed of primordial solar composition material enriched with 
material from stellar outflows and ejecta~\cite{Murphy2016ApJ} and a parent plasma of hot gas~\cite{Tatischeff2021MNRAS}. More subtle indications from CR abundance measurements also lend support to a super-bubble origin. For example, the high abundances of CR refractory elements compared to volatile elements~\cite{Rauch2009ApJ, Murphy2016ApJ, Wiedenbeck2007SSRv} is consistent with dust grains being entrained from the ISM, as would be expected around super-bubbles. These undergo very effective acceleration at shocks due to their typical high charge to mass ratio. 
Dust erosion by sputtering at high velocities releases refractory species into the shock. These have the same velocity of the primary dust grain, and are injected directly into the acceleration process independently of their mass-to-charge ratio~\cite{Meyer1997ApJ, Ellison1997ApJ}. 
A super-bubble origin also provides more consistent predictions for the abundance ratio of B and Be in the observed composition of CRs. 
If their origin is associated with the ISM, these elements are expected to increase in proportion to metalicity (traced by O abundance), following general enrichment trends of the ISM driven by SNe. However their observed increase with the square of O abundance is instead consistent with Be and B being produced by spallation reactions with CRs accelerated out of the ISM, and then interacting with the ISM itself~\cite{Parizot2000A&A, Meneguzzi1971A&A, Reeves1970Natur}.\footnote{O is the main progenitor CR of Be and B in typical spallation reaction chains, so the total production rate becomes proportional to the amount of O released into the ISM by SNe enrichment, and the amount of O as a source of Be and B production by spallation~\cite{Parizot2000A&A}.}

\subsubsection{Searching for \textit{PeVatrons}}
\label{sec:pevatrons}

The identification of Galactic CR origins above a PeV presents a particular challenge. 
Around this energy, the CR flux we receive on Earth transitions from mainly Galactic in origin, to extra-galactic. This is marked by a steepening of the CR spectrum at a few PeV (the so-called \textit{knee}~\cite{kulikov1959size, Parizot2014NuPhS}; for an overview of efforts to resolve this spectral feature, see~\cite{Antoni2005APh}). It is unclear whether this observed steepening is due to the extra-galactic source transition, the effects of different convolved source spectra, or a feature resulting from the competition of different energy-dependent transport effects~\cite[e.g.][]{Adriani2011Sci, Aguilar2015PhRvL, Aguilar2015PhRvL_b}. Moreover, it may be dependent on the CR species, with recent indications that a knee arises for lighter CR species at slightly lower energies of around 0.7 PeV~\cite{Bartoli2015PhRvD}. However, 
the existence of a spectral feature marking the transition to extra-galactic CRs is theoretically expected. This can be 
understood by a comparison between the size of the Galactic halo size or disk thickness (of order a few kpc) with the CR gyro-radius in a $\sim 1 \;\! \mu{\rm G}$ magnetic field, characteristic of the average magnetic field strength in the ISM. At a few 10s of PeV, this gyro radius is $r_{\rm gy} \sim 
1.1 \;\!\left({E_p}/{10 \;\rm{PeV}}\right) \left({B}/{1\;\!\mu {\rm G}}\right)^{-1} \;\! {\rm kpc}$. CRs of lower energies than this would mainly be confined within the Galaxy by its magnetic field. 
This means that the bulk of observed CRs up to the knee must be supplied by sources within the Milky Way. 

A range of plausible candidate source populations can be considered as the potential origins of Galactic PeV CRs (cf. the Hillas criterion~\cite{Hillas1984ARA&A}). 
These meet 
the requirement of being able to accelerate CRs to a PeV. However, this is not sufficient to claim that they are the origin of the bulk of the PeV CRs we observe on Earth. Indeed, nearby well-studied pulsar wind nebulae systems like the Crab nebula have been found to accelerate CR particles to PeV energies~\cite{Lhaaso2021Sci}, yet it remains unsettled whether systems like the Crab could be a dominant source of most of the hadronic Galactic CRs at 10-100 PeV 
(see~\cite{Liu2021ApJ, Lhaaso2021Sci, Peng2022ApJ}). This has motivated the search for CR accelerators operating to PeV energies, called \textit{PeVatrons}, which are capable of sustaining the detected abundance of PeV CRs.

Without an unambiguous determination of the dominant population of Galactic PeVatrons, several possibilities have been considered. Some of these are supported by observational studies, indicating that SNRs~\cite[originally proposed by][]{Baade1934PNAS}, the supermassive black hole in the Galactic Center~\cite[see, e.g.][]{HESS2016Natur}, pulsars~\cite[e.g.][]{Bednarek2002APh}, highly collimated micro-quasar jets~\cite{Escobar2022A&A}, and young massive stellar clusters~\cite[e.g.][]{Parizot2004A&A, Yang2019RLSFN} could all be plausible PeVatrons. 
Interest has particularly increased recently in young star clusters due to the steady increase in the observed associations between star clusters/super-bubbles and high energy $\gamma$-ray sources~\cite{Yang2020A&A, Ackermann2011Sci, Abramowski2012A&A, HESS2015Sci, Cao2021Natur, Yang2018A&A}. 
It has been suggested that these sources may even be a dominant source population for the highest energy Galactic CRs~\cite{Yang2019RLSFN}. Acceleration in fast SNR shocks expanding in the collective wind of a compact cluster may be one mechanism to reach energies well above a PeV~\cite{Vieu2023MNRAS}. However, sufficiently powerful SN explosions to yield the necessary fast SNR shocks in these systems are rare~\cite{Cristofari2020APh} (see also the overview in Ref.~\cite{Gabici2023arXiv230106505G}), and it remains unsettled whether CR energies can be attained that are sufficient to account for the Galactic CR flux above a PeV~\cite{Vieu2022MNRAS, Morlino2021MNRAS}. 
Shocks around isolated SNe have also been considered. However, the requirement for shock velocities of around 40,000 km/s~\cite{Zirakashvili2008ApJ} to reach PeV energies is challenging to achieve. Moreover, it remains unclear whether the non-resonant instability can sufficiently amplify the CSM magnetic field up-stream of most CC SN shocks to sustain PeV CR acceleration.

When discerning the origin of Galactic CRs, the search for hadronic PeVatrons is arguably more pressing than leptonic ones. This is because most of the Galactic CR flux is comprised of hadrons. Moreover, leptons typically experience severe energy losses and generally would not be able to survive substantial propagation through the ISM to sustain a CR sea. This makes their origin a more local question.\footnote{This is with the exception of secondary CRs injected by hadronic primaries, which may be non-negligible~\cite[e.g., see][which shows that the leptonic CR abdunance of galaxies could be a significant or dominant component of the leptonic CR flux in starburst galaxies]{Lacki2013MNRAS}.} Hadronic CR accelerators should yield signatures in $\gamma$-rays persisting to higher energies (from $\pi^0$ decays) and neutrinos (from $\pi^{\pm}$ decays), while leptonic accelerators can only produce electromagnetic emission.  
It is likely that neutrino emission from hadronic PeVatrons arises at levels that are below detection thresholds of current instruments~\cite[for discussion, see, e.g. ][]{Halzen2008PhRvD}. 
Firmly establishing source candidates as hadronic will therefore rely on neutrino detections with up-coming and proposed facilities~\cite{Baikal2018arXiv180810353B, Aartsen2021JPhG, Agostini2020NatAs, Ye2022arXiv220704519Y}. Present observational efforts to discriminate between hadronic and leptonic PeVatrons instead focus on their $\gamma$-ray emission above $\sim$50 TeV. At these energies, Klein-Nishina suppression considerably reduces the efficiency of leptonic $\gamma$-ray emission. 
So far, a number of $\gamma$-ray sources have been detected in this energy range. Tibet-AS${\gamma}$ has reported $\gamma$-rays with energies between 100 TeV and 1 PeV in the Galactic disk~\cite{Amenomori2021PhRvL}, and nine sources have been detected above 56 TeV with the High Altitude Water Cherenkov Observatory (HAWC), including three up to 100 TeV, and 12 above 100 TeV with the Large High Altitude Air Shower Observatory (LHAASO; ~\cite{Cao2021Natur}).\footnote{We note that, at the time of writing, a recent pre-print for the First LHAASO Catalog indicates the number of detections at energies above 100 TeV may 
now have increased to 43 sources with a significance of 4$\sigma$~\cite{Cao2023arXiv230517030C}.} 

The 12 LHAASO sources detected above 100 TeV (which include the three detected above 100\,TeV by HAWC) have been considered as PeVatrons~\cite{Abeysekara2020PhRvL}. The majority of them have plausible associations with energetic pulsars ($\dot{E}\gtrsim 10^{36}$erg/s) and known counterpart very-high-energy (VHE) sources. Pulsar wind nebulae are predominantly leptonic accelerators, for which the Klein-Nishina effect causes an abrupt cut-off at the highest energies \cite{Blumenthal1970}. However, it has been demonstrated that the energy density of high radiation environments can compensate this effect somewhat, such that even leptonic scenarios may account for $\gamma$-ray emission at $\sim100$\,TeV \cite{Breuhaus2021ApJ}. LHAASO\,J2032+4102 associated with the Cygnus region is a promising case as evidence for PeVatron activity by young stellar clusters, laying claim to the highest energy photon measured at $1.42\pm0.13$\,PeV. The vicinity of the Boomerang nebula, LHAASO\,J2226+6057 (also detected by Tibet-AS${\gamma}$ \cite{2021NatAs...5..460TibetG106}) was recently studied in detail with the MAGIC telescopes, exhibiting intriguing energy-dependent morphology. As the location of the emission seems to shift within the SNR G106.3+02.7 with energy, there are indications to support a scenario where high energy particles escape the source and travel further in the surrounding medium \cite{2023A&A...671A..12MAGIC_G106}.

Two sources that do not appear to have a known energetic pulsar counterpart are LHAASO\,J1843-0338, plausibly associated with the SNR G28.6-0.1, and the enigmatic source LHAASO\,J2108+5157; the only source to be first discovered at the highest energies. To date, coincident molecular material as been found at the location of the latter~\cite{2023PASJ...75..546D, 2023arXiv230611921D}, potentially indicating that energetic particles are hadronic in nature, yet no accelerator has as yet been identified in the vicinity \cite{2023A&A...673A..75A_LSTJ2108}.\footnote{However, an old SNR as an accelerator has been proposed as one possible scenario~\cite{2023MNRAS.521L...5D}.} Within a scenario whereby the particles accelerated to the highest energies escape the accelerator, travel through the intervening ISM and subsequently interact with target material such as molecular clouds, one may anticipate a population of such `passive' sources emerging at the highest energies. Future studies from surveys at the highest energies, including LHAASO, HAWC and the forthcoming Southern Wide-Field Gamma-Ray Observatory (SWGO) covering the Southern sky, may be able to shed further light on these phenomena \cite{Albert2019arXiv190208429A}.

Detected $\gamma$-ray sources may not (yet) be able to account for the full budget of Galactic CRs and may not even be hadronic~\cite[e.g.][]{Atoyan1996MNRAS}. Indeed, it has recently been speculated that some hadronic PeVatrons may not yield observable $\gamma$-ray signatures at all. Instead, Ref.~\cite{Sudoh2023PhRvD} discussed how they may be too \textit{thin} in column density for CR hadronic interactions to take place within them at detectable levels. The accelerated CRs then simply escape without local interactions or discernible electromagnetic signatures. Alternatively, they may be too \textit{thick} in matter or radiation density that CRs cannot escape. In this case, the CRs may be attenuated locally by pp or p$\gamma$ interactions (see also section~\ref{sec:hadronic_ints}), or $\gamma$-rays are produced but are attenuated by pair-production with low energy photons in the SN photo-sphere~\cite[e.g.][]{Scannapieco2005ApJ, Cristofari2020MNRAS, Cristofari2022MNRAS}. 
However, it is possible that a larger population of weaker hadronic PeVatrons falls below detection limits for current $\gamma$-ray observatories, and more sources will emerge as integration times increase - particularly with the advances to be made in the 100 TeV domain using the new generation of $\gamma$-ray facilities.

\subsection{Cosmic rays in diffuse interstellar media} 
\label{sec:cr_in_diffuse_ISM}

\subsubsection{Magnetohydrodynamic mode damping and implications for cosmic ray propagation}
\label{sec:damping_ISM}

In conditions typical of the Milky Way's hot ionized medium (HIM), 
the resonant CR streaming instability operates efficiently. 
 However, the growth of MHD waves driven by CRs of energies up to a few GeV 
 is roughly balanced by damping processes~\cite{Amato2018AdSpR}. For these modes, a steady state MHD wave amplitude is established. The CRs driving them will then experience diffusive propagation constrained roughly to the local Alfv\'{e}n speed
 (cf. the self-confinement view; section~\ref{sec:transport_microphys}). 
This self-regulated view of CR transport coupled to 
MHD fluid treatments has been widely adopted in CR propagation simulations. Such studies can recover the characteristics of macroscopic CR transport in a steady-state limit~\cite{Thomas2019MNRAS}, and have been used to inform larger scale simulations of galaxies~\cite[e.g.][]{Thomas2023MNRAS, Armillotta2021ApJ}. Some developments take additional consideration of CR transport variation in different ISM phases~\cite{Armillotta2021ApJ}, where the dominant processes damping MHD wave modes may differ. 
For example, turbulent damping can dominate in the warm ionized medium, but ion-neutral damping is usually operates more quickly in weakly ionized molecular clouds~\cite{Xu2022ApJ}. This can lead to strong suppression of MHD turbulence~\cite{Cesarsky1978AA} and less diffusive~\cite{Kulsrud1969ApJ, Zweibel1982ApJ}, faster CR propagation in these settings. However, recent $\gamma$-ray studies have demonstrated that the situation may be more complex (see also section~\ref{sec:cr_prop_mc}). They have introduced new evidence that diffusive CR propagation persists within molecular clouds, at least down to the sub-pc clump scale below 10 GeV energies~\cite{Yang2023NatAs}. This may suggest damping processes operate less efficiently than expected in ISM molecular complexes, or that MHD wave amplification is more severe. 

\subsubsection{Cosmic ray heating processes in the ionized interstellar medium}
\label{sec:heating_ISM}

The ISM of galaxies is inhomogeneous. 
The largest component is the HIM. 
This fills about half of the ISM volume in the Milky Way~\cite{Draine2011piim}. In other galaxies, it also 
comprises a substantial fraction of the ISM, 
but the exact filling fraction may differ according to local conditions~\cite{Hill2018ApJ}. HIM heating is generally efficient. It is driven mainly by 
starlight and shocks from SN explosions to sustain a temperature above $\sim 10^{5.5}\;\!{\rm K}$ and a high ionization fraction. By contrast, the low gas density means that cooling is inefficient, operating over timescales of Gyrs. The HIM of most galaxies is therefore in a runaway thermal state. 
The HIM of galaxies hosting a rich reservoir of CRs can be subjected to particularly powerful heating. If mediated by hadronic collisions, Coulomb thermalization of CRs can dominate~\cite{Owen2018MNRAS}. By this channel, CRs can have a significant role in 
setting the thermal history of galaxies and their evolution~\cite[e.g.][]{Owen2019AA}.

The other 
major ionized component of the ISM is the warm ionized medium (WIM). In the Milky Way, this occupies around 10 per-cent of the ISM volume. In other galaxies, this fraction can be higher. It depends on local thermal conditions~\cite{Hill2018ApJ}, and can be affected by CR pressure and transport~\cite{Simpson2023MNRAS}. The WIM is comprised of photo-ionized, photo-heated H$~\!$II gas. It is typically held at a temperature of around $\sim 10^{4}\;\!{\rm K}$, with cooling operating relatively slowly via optical line emission and free-free emission. Heating is supplied by photo-electrons and CRs, with CRs believed to be dominant. 
Their role in maintaining the thermal balance of the WIM has been considered in several recent works. This operates through Coulomb collisions of low energy CRs, below $\sim 100$ MeV~\cite{Walker2016ApJ}, Coulomb thermalization of secondary electrons via hadronic collisions~\cite{Owen2018MNRAS, Simpson2023MNRAS}, and the dissipation of MHD waves~\cite{Walker2016ApJ, Minter1997ApJ, Wiener2013ApJ, Wentzel1971ApJ}.

\subsubsection{Cosmic ray effects on the structure of interstellar media}
\label{sec:structure_ISM}

CR pressure gradients and transport effects can modify 
gas structures within the ISM. This can have consequences for galaxy evolution, as the capability of dense 
star-forming clouds to develop can be changed. 
Resolving the relevant CR processes that regulate the phase 
configuration of a galaxy is therefore essential to properly 
establish 
their global feedback impacts. 
CRs have not been found to play a significant role in stellar cluster formation~\cite{Rathjen2021MNRAS}, and their direct impact 
on star-formation in molecular clouds is unsettled, with some studies reporting 
relatively moderate effects~\cite[over 100 Myr timescales][]{Rathjen2021MNRAS}, while others 
have found more severe impacts may be possible in certain  settings~\cite[e.g. in starburst galaxies][]{Owen2021ApJ, Papadopoulos2011MNRAS}. 
Less direct effects on star-formation have been shown to be more consequential. For example, self-generated magnetic fields around CR sources can reduce CR diffusivity in large CR-inflated bubbles~\cite{Schroer2021ApJ}. The accumulation of CRs and strong local pressure gradients prevent the formation of massive gaseous clumps in these regions~\cite{Semenov2021ApJ} and a smoother ISM~\cite{Farcy2022MNRAS}. This frustrates the development of future stellar nurseries.

CR propagation through an established multi-phase ISM can also 
modify the potential for star-formation by disrupting the configuration of existing clouds. 
For example, CRs may propagate less diffusively in dense, relatively neutral structures like clumps and molecular clouds. 
As CRs decouple and experience higher streaming speeds in the semi neutral gas, 
the local CR energy density drops. This creates a CR pressure gradient at the cloud-ISM interface (sometimes referred to as the CR `traffic jam' effect; e.g.~\cite{Bustard2021ApJ}). This can be sufficient to collisionlessly generate waves that are able to overcome local damping mechanisms (e.g. ion-neutral damping) and exert a force on the cloud. 
It can accelerate~\cite{Wiener2017MNRAS, Wiener2019MNRAS, Bruggen2020ApJ} and reshape the cloud~\cite{Bustard2021ApJ},  
and may even boost the gravitational instability to induce earlier collapse and trigger star-formation. 
Cloud compression can also be induced by the Parker instability,\footnote{This instability develops when a perturbation to the magnetic field causes the field lines to bend. Gravity then pulls gas into the valleys of the magnetic field, which sinks and deepens the valleys.} which is more severe and arises more frequently in the presence of CR streaming and self-confinement~\cite{Heintz2020ApJ} (although the effects can be different depending on the CR transport model adopted~\cite{Heintz2018ApJ}). This may increase star-formation. Conversely, the Parker instability is less severe if adopting an extrinsic turbulence model for CR transport~\cite{Heintz2020ApJ}, leading to suppressed star-formation and a more puffed-up ISM overall. These effects are expected to be particularly strong in 
starburst galaxies, where the Parker instability can grow more quickly~\cite{Heintz2020ApJ}.

\subsection{Cosmic rays in molecular clouds}
\label{sec:cr_in_mcs}

\subsubsection{Propagation}
\label{sec:cr_prop_mc}

CR propagation in the ISM can generally be described as a combination of diffusion, ballistic streaming and advection (see section~\ref{sec:particle_transport}). The balance of these processes depends on local conditions. In molecular clouds, advection in bulk flows is usually negligible. However, diffusion and streaming are important. 
In the warm, ionized envelopes around ISM clouds, CR diffusion can be highly variable. Coefficients have been inferred observationally to span at least four orders of magnitude in the Galaxy~\cite[e.g.][]{Protheroe2008MNRAS, Dogiel2015ApJ, Owen2021ApJ}, 
with simulations indicating that no single value is suitable for different locations or for CRs of different energies~\cite{Armillotta2021ApJ}. In molecular clouds, clumps and cores, 
ion-neutral damping can be strong (see section~\ref{sec:damping_ISM}), 
with small-scale resonant MHD turbulence being damped entirely 
in dense cores~\cite[e.g.][]{Ivlev2018ApJ}. 
CR transport should therefore revert increasingly to ballistic streaming in these structures~\cite{Kulsrud1969ApJ, Zweibel1982ApJ}, which increases their effective speed. A local reduction in CR density and pressure then emerges~\cite{Everett2011ApJ, Skilling1976AA, Cesarsky1978AA}, producing a CR deficient zone. 
Effects including modulation by electric fields from charges deposited by CRs~\cite[especially in regions of high electron-dominated ionization; see][]{Silsbee2020ApJ}
and magnetic mirroring~\cite{Cesarsky1978AA, Ko1992AA, Padoan2005ApJ, Chandran2000ApJ, Desch2004ApJ, Owen2021MNRAS}  
can then reduce CR density further, 
especially in multiple-peaked magnetic field structures (where mirroring and focusing effects do not cancel out~\cite{Silsbee2018ApJ}). 

Although CR deficient zones due to fast, less-diffusive propagation can account for observed trends in molecular clouds (e.g. anti-correlated CR ionization rate with density~\cite{Albertsson2018ApJ}), other possibilities have also been considered. Slow CR diffusion in dense star-cluster-forming clumps where stronger magnetic fields sustain turbulent motions~\cite{Li2018MNRAS} would also create a local deficiency of CR density if severe enough. In this scenario, the CR-deficient regions would be more accurately described as exclusion zones, as  
it is harder for CRs to penetrate into them. 
A different possibility is that MHD turbulence on larger clump-scaled structures 
may support more moderate diffusive 
CR propagation. 
Such a scenario is not implausible. Below $\sim 100$ MeV, CR absorption due to ionization processes can be severe, particularly on the scale of dense cores~\cite{Padovani2018A&Ab}.
The CR flux in cloud interiors can therefore be severely reduced, by as much as two orders of magnitude, 
with the CR electron flux being especially affected~\cite{Phan2018MNRAS}. 
The CR flux leaving a clump or core can then be much lower than that entering it. This introduces a flux difference able to sustain MHD turbulence and the generation of Alfv\'{e}n waves via the streaming instability~\cite[e.g.][]{Ivlev2018ApJ, Skilling1976AA}, which can be sufficient to enable some scattering and diffusive propagation. 
 At $>$ GeV energies, attenuation by hadronic interactions is possible. Gas densities associated with clouds are sufficient to cause strong CR attenuation (see section~\ref{sec:hadronic_ints}). This leads to a reduction in CR density overall and is able to account for the $\gamma$-ray `holes' observed with \textit{Fermi}-LAT below 10 GeV~\cite{Yang2023NatAs}. This level of attenuation may also be sufficient to sustain MHD turbulence at a level where some diffusive propagation is possible. 
 A further possibility is that CR propagation may instead be suppressed 
in a small envelope region separating dense clumps from the larger diffuse cloud 
due to self-generated MHD waves~\cite{Morlino2015MNRAS, Dogiel2018ApJ, Ivlev2018ApJ}. 
This would hinder their penetration into the cloud, but boost signatures of their effects in the envelope region. Such an envelope may be challenging to be  disentangle observationally. 

\subsubsection{Cosmic ray interactions in molecular clouds and observable signatures}
\label{sec:CR_MC_ints}

CRs interact with molecular clouds through multiple channels, each of 
which can have distinct physical effects and produce different observable signatures. The dominant interaction process depends on the CR species, its energy and the local physical conditions within a cloud. Collisional ionization 
processes are caused by CRs of all species, and are particularly important at energies between 10-100 MeV~\cite{Padovani2009A&A}. They can modify the physical and chemical conditions deep inside clouds by maintaining a 
low level of ionization through the dense gas. This ionization sustains a coupling between the gas and magnetic field, with implications for the cloud dynamics~\cite{Gabici2022A&ARv}. It also drives rich astrochemical networks, initiated by the production of H$^+$ and H$_2^+$ ions. These lead to the formation a wide selection of molecular ion species that have accessible transition lines at sub-mm wavelengths. 

\begin{figure}[H]
\begin{center}
\includegraphics[width=10.5 cm]{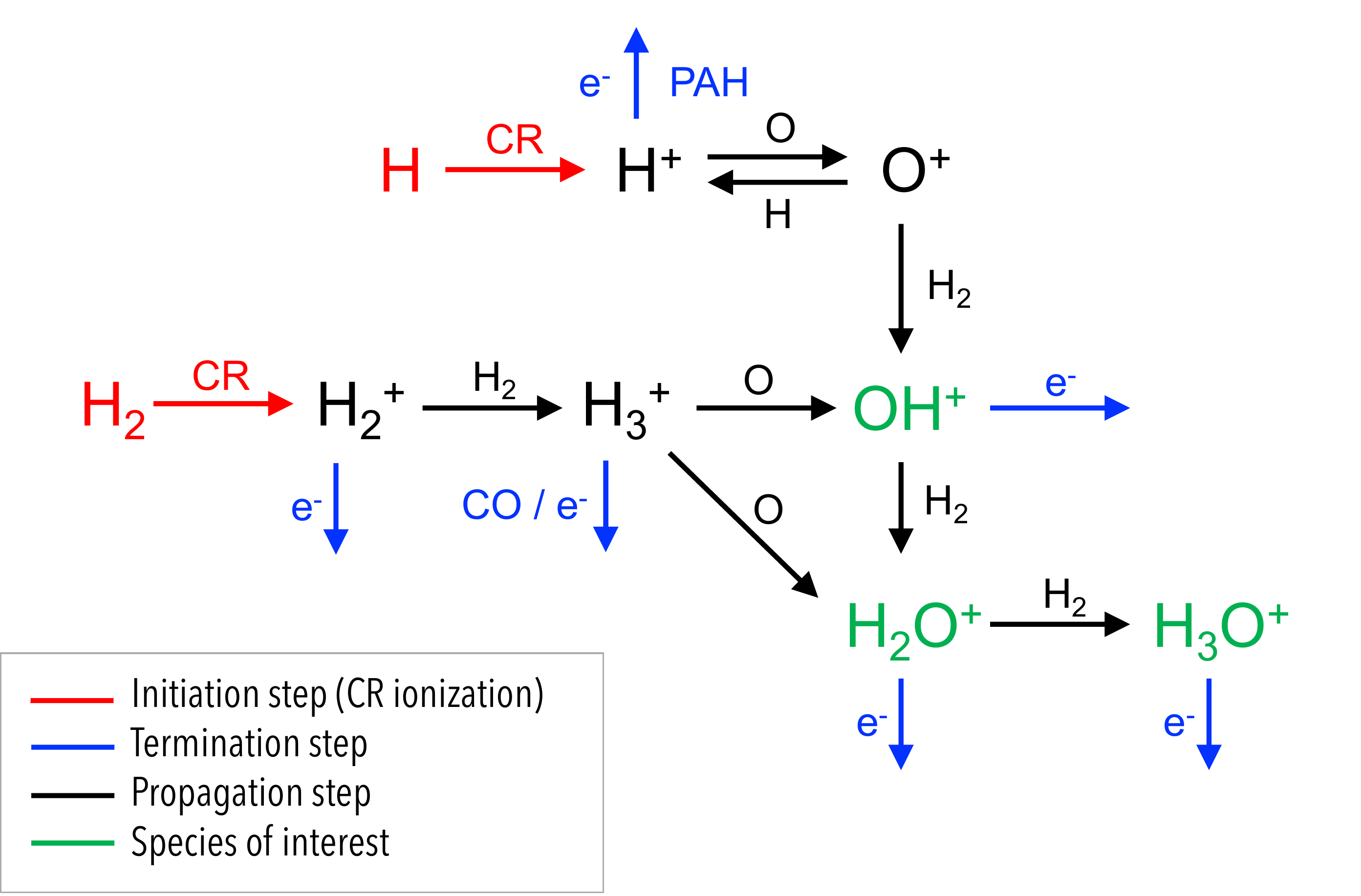} 
\end{center}
\caption{Standard simplified network of chemical processes driven by CR-induced ionisation of H and H$_2$ in a molecular cloud. Pathways leading to the formation of OH$^+$, H$_2$O$^+$ and H$_3$O$^+$ are shown. Termination steps are caused by neutralisation with electrons (which may be ambient free electrons, or even low-energy CRs) within the cloud are indicated in blue. It is also possible for polyaromatic hydrocarbons (PAHs) or carbon monoxide (CO) to facilitate some termination processes, as noted. Adapted from Ref.~\cite{Owen2023A&G}.
}
\label{fig:chemical_network}
\end{figure}

H$_2^+$ rapidly reacts with H$_2$ to produce the H$_3^+$ ion. This can be directly detected to infer the local CR ionization rate along a line of sight~\cite[e.g.][]{Oka2019ApJ,Indriolo2023arXiv, LePetit2016A&A}. 
By using this ion, 
observations with the Atacama Large Millimeter/Sub-Millimeter Array (ALMA) have recently been able to produced intricate maps of CR ionization through molecular cloud complexes~\cite{Sabatini2023arXiv230400329S},  allowing 
new tests of CR propagation models with unprecedented clarity. 
Other molecular ions, such as ArH$^+$~\cite{Schilke2014A&A, Jacob2020A&A, Bialy2019ApJ, Jacob2022ApJ} and HCO$^+$~\cite{Gabici2022A&ARv}, can also be used as probes to measure CR ionization. These are less direct tracers than H$_3^+$, and require additional chemical modeling. This can introduce uncertainties when interpreting data, but provides observational flexibility. For example, in cases where H$_3^+$ emission or absorption lines are not detectable with current instruments, or limitations make detection impractical for a particular target (e.g. a requirement to have a suitable background 
early-type star to obtain an estimate for the column density of the target species~\cite[e.g.][]{Indriolo2012ApJ, Indriolo2023arXiv}), 
other chemically-related species may be accessible instead. Relating observations of indirect tracers to the underlying CR ionization rate 
can often be achieved by simplified gas-phase chemical networks.\footnote{More comprehensive approaches are possible by using sophisticated chemical codes to 
obtain a robust determination of CR ionization (e.g. {\tt UCLCHEM}~\cite{Holdship2017AJ} or {\tt Astrochem}~\cite{Maret2015ascl}) and can 
relax steady-state approximations. Beyond direct studies of CRs themselves, 
other applications require a reliable determination of local CR ionization rate. For example, in the age determinations of molecular cloud cores~\cite{Lin2020A&A}.} Fig.~\ref{fig:chemical_network}  demonstrates an example of an oxygen (O) network which shows how typical observable species like OH$^{+}$ and H$_2$O$^{+}$ can be connected to the CR ionization rate. These reduced networks allow relatively simple analytic relations to be derived that relate the abundance of a target species (e.g. OH$^{+}$) to the CR ionization rate~\cite{Indriolo2010ApJ, Ceccarelli2011ApJ}. 

OH$^{+}$ and H$_2$O$^{+}$ are often used as tracers of CR ionization in diffuse molecular clouds, where their abundances are strongly affected by the ionizing CR flux. Other tracers, such as H$_3$O$^{+}$, DCO$^+$, HCO$^+$, N$_2$H$^+$, N$_2$D$^+$ and their associated isotopologues, are selected depending on the required physical conditions of the target region within a particular cloud complex. 
Observations using these tracers have provided new information about the CR ionization rate at different locations in the Galaxy, with a few hundred diffuse molecular clouds analyzed so far. 
These observations have revealed broad variations in the ionization rate at different locations throughout the Milky Way, ranging from $\sim 10^{-18}\;\!{\rm s}^{-1}$ to $\sim 10^{-14}\;\!{\rm s}^{-1}$~\cite{Caselli1998ApJ, Morales2014A&A, Hezareh2008ApJ, VanDerTak2000A&A, Boisanger1996A&A, Redaelli2021A&A, Indriolo2012ApJ, Porras2014ApJ, Indriolo2007ApJ, Indriolo2015ApJ, Bacalla2019A&A}. 
The ionization rate at a particular cloud column density depends on the propagation model and irradiating CR spectrum adopted~\cite{Padovani2018A&A, Padovani2022AA}. 
Local factors, such as magnetic field strength and structure~\cite[including magnetic mirrors][]{Cesarsky1978AA, Chandran2000ApJ}, also impact the capability for CRs penetrate and ionize interstellar cloud. However, the magnitude of the variation of CR ionization inferred for different molecular clouds 
far exceeds the variation typically attributed to these local factors. Instead, the broad variation of CR ionization rates throughout the Galaxy indicates 
a highly non-uniform irradiating flux of low-energy CRs throughout the ISM. 

Many chemical signatures of CR ionization associated with molecular ions are accessible in the sub-mm band, with instruments such as ALMA. However, their ionizing impacts can also be detected in other parts of the electromagnetic spectrum. In particular, 
MeV CR protons (and heavier species) can collide with neutral to low-ionized Fe atoms in ISM clouds to produce FeI K$\alpha$ X-ray line emission at 6.40 keV~\cite{Tatischeff2012A&A}. Such emission results from the removal of a K-shell electron by a CR ionization event, with the 
vacancy filled rapidly by an electronic transition from the L shell. 
These collisions result in shifts of characteristic X-ray line energies and distinctive spectral structures, making it possible to study the properties and composition of CRs irradiating molecular clouds~\cite{Okon2020PASJ}.

CRs do not only cause ionization when they collide with the gas in molecular clouds. They also excite electronic and ro-vibrational energy levels of H$_2$~\cite{Bialy2020CmPhy}. Primary CRs propagating into a molecular cloud with energy spectra peaking between 10s of keV to a few MeV can produce secondary CR electrons of energies of order 10s of eV~\cite{Padovani2022AA}. These efficiently excite ro-vibrational transitions of H$_2$, particularly to the first vibrational state ($\nu = 1$). The subsequent relaxation leads to the emission of photons that can be detected in infrared (IR) using instruments including the Very Large Telescope (VLT), or the Near InfraRed Spectrograph (NIRSpec) on the \textit{James Webb} Space Telescope (\textit{JWST}).
Four lines in particular, (1-0)O(2), (1-0)Q(2), (1-0)S(0), and (1-0)O(4) at  2.22, 2.41, 2.63 and 3.00 $\mu$m, respectively, have been 
identified as transitions where CR excitation dominates over competing processes like photo-excitation~\cite{Bialy2020CmPhy}. These transitions are therefore viable tracers of CR activity. Recent advancements of this technique have demonstrated that cloud density variations affect the penetrating capability of CRs according to their energy~\cite{Padovani2022AA}. This means that, by measuring ro-vibrational line intensities and H$_2$ column densities towards a molecular complex, the low-energy CR ionization rate, CR spectral index, and spectral normalization can all be deduced. This provides a novel means of studying the variation of CRs of different energies throughout the Milky Way's disk~\cite{Bialy2022AAL, Gaches2022AAb}. In the era of \textit{JWST}, this is a powerful new tool for studying CRs.

High-energy interactions provide an opportunity to study CRs in molecular clouds at energies far above those involved with ionization and ro-vibrational excitations. At $\gtrsim$ GeV, CR protons and electrons are subject to different interaction pathways that generate different signatures. For instance, when the energy of a CR proton exceeds a threshold of about $\sim 0.28$ GeV, it can undergo pion-producing interactions with gas nuclei in clouds (see section~\ref{sec:hadronic_ints}). These interactions release $\gamma$-rays~\cite[e.g.][]{Casanova2010PASJ} and neutrinos~\cite[e.g.][]{Cavasinni2006APh, Banik2021EPJC, Sarmah2023MNRAS, Abbasi2023ApJ} from the decay of neutral and charged pions, respectively. 
These interaction products can be used as a tracer of CR engagement in molecular clouds in violent environments, such as the vicinity of a CR source like a SN  remnant or pulsar wind nebula~\cite{Voisin2019PASA,Banik2021EPJC, HESS2023arXiv}. 
We can gain insight into the properties and composition of the CRs escaping from these sources by studying the emission associated with CRs interacting in these so-called \textit{active} clouds (see also section~\ref{sec:cloud_SNR}). 

High energy CR electrons can drive synchrotron emission when they interact with the magnetic fields of clouds and cores. These magnetic fields can be substantially stronger than the ISM average~\cite[for a review, see][]{Crutcher2012ARAA}. 
 The CR electrons driving this emission can be primary CRs, irradiating a cloud from the ISM. However, given their fast cooling times, they are often secondary CRs, produced from the decay of the charged pions produced in hadronic collisions. 
Detection of synchrotron emission 
from CRs in molecular clouds can be challenging, and is thought to be very low in some clouds, e.g. the Sgr B2 complex~\cite{Protheroe2008MNRAS}. A particular issue is the thermal emission (including that caused by enshrouded stars), which can overwhelm non-thermal emission signatures from the cloud. 
Despite this, 
previous studies have reported non-thermal radio emission from starless infrared dark clouds, such as 
G0.216+0.016, where contamination from thermal radiation is less severe~\cite{Rodriguez2013ApJ}.
This emission is consistent with the radiation produced by secondary CR electrons, with a suppressed diffusion coefficient (by a factor of 0.1-0.01) 
and substantially enhanced local magnetic field of 470$\mu$G~\cite{Jones2014ApJ}.
Other studies have reported correlations between Galactic~\cite[e.g.][]{Zhang2010PASA} and extra-galactic~\cite[e.g.][]{Filho2019MNRAS} CO emission with radio emission, suggesting the radio emission is synchrotron that originates in molecular clouds. 
Future facilities such as the Square Kilometer Array (SKA) may be able to detect synchrotron emission from CRs in a wide range of Galactic molecular complexes~\cite{Strong2014arXiv, Padovani2018A&A}, especially those irradiated by nearby energetic sources such as SNRs~\cite{Gabici2009MNRAS}. Such 
radio observations would provide the opportunity for more spatially-resolved observations than is possible with other signatures.

\subsubsection{Molecular clouds as cosmic ray barometers} 
\label{sec:cr_sea}

\begin{figure}[H]
\begin{center}
\includegraphics[width=9cm]{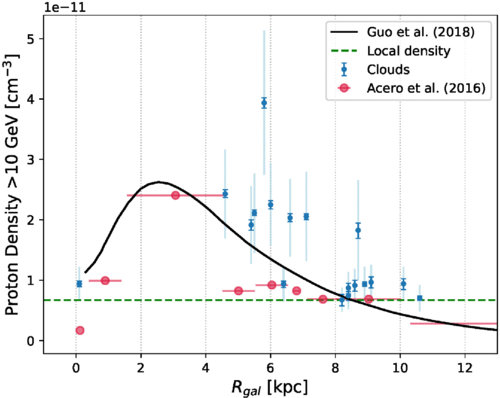} 
\end{center}
\caption{Energy density of CR protons above 10 GeV as a function of distance from the Galactic Centre (GC). Estimates obtained from passive clouds are shown, with statistical (thick lines) and systematic (thin lines) errors shown by the error bars. 
Energy density of protons derived above 10 GeV. For each cloud are indicated the and the  error bars. The data points from Ref.~\cite{Acero2016ApJS} are shown in red. The model adopted from~\cite{Guo2018PhRvD} is shown by the black line, while the local proton density derived from the measurements of~\cite{Aguilar2015PhRvL} is shown by the green dashed line for reference. Figure reproduced from Ref.~\cite{Aharonian2020PhRvD}. 
}
\label{fig:CR_sea}
\end{figure}  

Molecular clouds can serve as CR barometers. 
When CRs interact with them, the resulting products (photons and neutrinos) can be used as observable tracers of the CR distribution throughout the Galaxy. 
\textit{Passive} molecular clouds are irradiated by CRs from the interstellar distribution, or the CR `sea', rather than a nearby source. 
$\gamma$-ray studies have
established the use of these clouds as barometers to measure the variation in the CR `sea level' throughout the ISM~\cite{Aharonian2001SSRv, Casanova2010PASJ, Abrahams2017ApJ, Dogiel2018ApJ, Aharonian2020PhRvD, Peron2022A&A}. They find significant variation in different parts of the Galaxy. This is consistent with the large 
spread of CR ionization rates determined for molecular clouds throughout the Milky Way~\cite[e.g.][]{Caselli1998ApJ, Indriolo2012ApJ, Neufeld2017ApJ}, which indicate that clouds in some parts of the ISM are irradiated by a CR spectrum two orders of magnitude more intense compared to that detected at Earth~\cite{Phan2022arXiv}. This variation may be due in part to the discrete nature of CR sources, which have been shown to produce a distribution of intensities with an expectation value that is not representative of the CR spectrum throughout the Galaxy, but has a median compatible with the CR proton and electron data obtained with \textit{Voyager}~\cite{Phan2021PhRvL}.

Studies of CR ionization rates in clouds have also established a trend with Galactic radius. As seen in Fig.~\ref{fig:CR_sea}, the density of CR protons tends to increase towards smaller Galactic radii, although a significant reduction is observed towards the inner $\sim$ 2 kpc of the Galaxy~\cite{Aharonian2020PhRvD}. 
 Indeed, 
reductions of CR abundance in the central molecular zone (CMZ) have been reported~\cite{Huang2021NatCo}. It has been suggested this may arise due to variations in CR propagation due to e.g. suppression of the diffusion parameter or the presence of a `barrier' formed by the self-excited MHD turbulence at the edges of clouds~\cite{Chernyshov2021Symm} or trapping of CRs in outer cloud layers.

In the GC, the trend of declining CR ionization of clouds appears to reverse. This is likely due to the presence of nearby CR accelerators, rather than a global change of the CR sea level~\cite{Peron2021ApJ}. The GC increase has been studied with 
GeV-TeV $\gamma$-rays, ionization signatures (probed with H$_3^+$, for example) and the X-ray Fe I K$\alpha$ line~\cite{Oka2019ApJ}, targeting cloud complexes such as Sgr B2 (which may vary over time~\cite{Rogers2022ApJ}) and the Arches~\cite{YusefZadeh2002ApJ, Wang2006MNRAS, Kuznetsova2019MNRAS}. The latter of these is located near a cluster of young massive stars near the GC region. It has been found to have time-variable ionization~\cite{Clavel2014MNRAS}, which has been shown to be in agreement with a CR origin~\cite{Krivonos2014ApJ, Chernyshov2018ApJ}. 
These findings complement earlier results obtained with iron K$\alpha$ lines in the Galactic ridge, where the line profile was 
indicative of molecular cloud bombardment by an MeV proton flux~\cite{Nobukawa2019JPhCS1181a2040N}.  

\subsubsection{Thermal balance and star-formation}
\label{sec:MC_heating}

\begin{figure}[H]
\begin{center}
\includegraphics[width=13 cm]{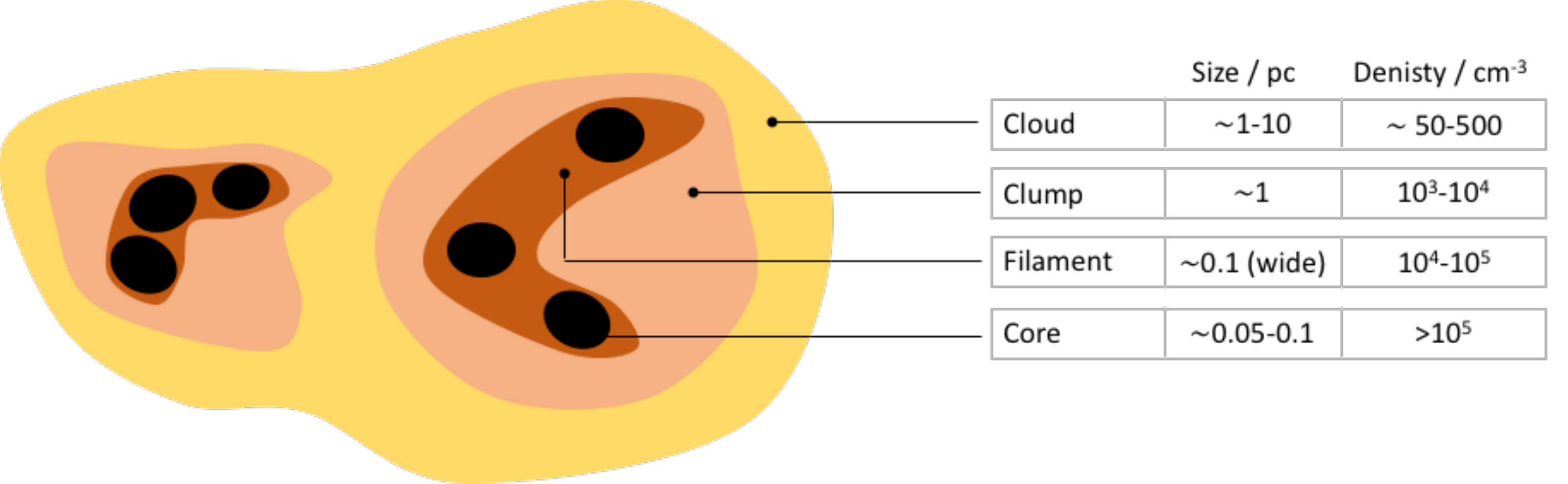} 
\end{center}
\caption{The hierarchical density structure of molecular clouds, ranging from diffuse regions to small dense cores where star-formation can occur~\citep{Bergin2007ARAA}. 
From large to small scales, temperatures vary from a few 10s of K to a few K, and ionization fractions range from nearly $x_i \approx 1$ to $x_i \approx 10^{-7}$~\cite{Draine2011piim}. CR propagation in the clouds is primarily diffusive, becoming less diffusive as higher neutral fractions damp MHD waves~\cite{Kulsrud1969ApJ, Zweibel1982ApJ}. 
The continuous substructure of molecular clouds cannot be fully captured by a simple hierarchy of just a few distinct elements~\cite{Rodriguez2005ASPC}, and alternative distinctions between cloud components are equally valid~\cite[e.g.][]{Myers1995mcsf}
(Figure from Ref.~\cite{Owen2023A&G}).
}
\label{fig:gmc_structure}
\end{figure}

CR heating 
is an important factor in setting the thermal balance of the WIM of galaxies (see section~\ref{sec:heating_ISM}). 
It also plays a role in regulating the thermal structure of molecular clouds. 
The effective power supplied by CR heating is determined 
by the mechanisms responsible for CR thermalization. These depend on the energy distribution of the CRs and the local conditions. 
Molecular clouds have a complex hierarchical structure (as show in Fig.~\ref{fig:gmc_structure}). Different CR heating processes dominate in various regions of the cloud, and the overall efficiency of heating can vary dramatically in each of the components.

In the most dense, neutral regions of clouds, 
collisional ionization is likely to be important in driving CR thermalization~\cite{Spitzer1968ApJ} (see also~\cite{Goldsmith2001ApJ}). This process can be caused by CR protons or electrons, including secondary electrons produced in hadronic collisions. When they collide with gas and cause an ionization event, the electrons released (called `knock-on' electrons) thermalize via electron-neutral collisions and Coulomb scattering. 
The effectiveness of this process depends on the ionization fraction, $x_i$, of the gas. A more direct form of this mechanism can also operate, with lower energy CRs directly undergoing Coulomb and collisional scattering with the cloud medium. This is more important for CRs below 100 MeV, in regions with non-negligible ionization fractions~\cite{Owen2018MNRAS}. 
Another heating process is the excitation and damping of MHD waves, driven by the local CR pressure gradient. This channel is associated with all species of CRs. Strong ion-neutral damping suppresses MHD waves in the densest regions of clouds. However, 
these waves can still be excited by CR pressure gradients 
in dense regions of clouds with strong magnetic fields and/or high ionization fractions. They are damped rapidly so are not usually sufficient to affect CR propagation. Instead, they mediate thermalization from the CRs to the cloud medium with an efficiency that is strongly dependent on the local magnetic field configuration~\cite{Owen2021ApJ}.  

\begin{figure}[h]
\begin{adjustwidth}{-\extralength}{0cm}
\centering
\includegraphics[width=17.5cm]{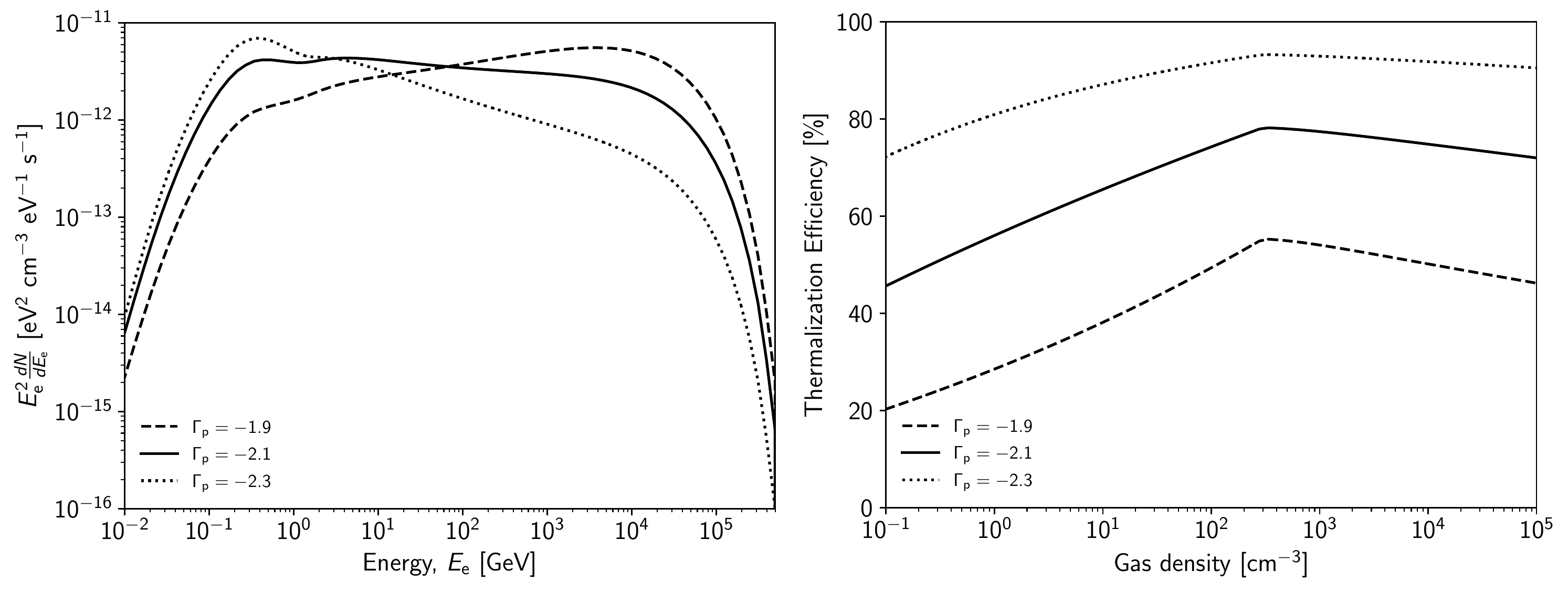}
\end{adjustwidth}
\caption{\textbf{Left panel:}  CR secondary electron spectra produced by CR protons interacting with gas of density $n_{\rm H} = 1.0$ cm$^{-3}$. Proton spectra are defined between 100 MeV and 1 PeV and normalized to an energy density of 1.4 eV cm$^{-3}$. 
Three power-law indices are considered, as labelled. Harder spectra transfer more power to higher energy secondary CRs. \textbf{Right panel:} 
Efficiency of energy transfer from CR secondary electrons to gas heating over a range of densities found in molecular clouds. 
Free-free and radiative synchrotron emission are considered as losses, while heating is driven by ionization and Coulomb interactions. 
The efficiency depends on the local magnetic field structure and is highest at gas densities of $n_{\rm H} \sim 10^3$ cm$^{-3}$, corresponding to clumps in Fig.~\ref{fig:gmc_structure}. 
Softer CR primary proton spectral indices lead to higher heating efficiencies. This is because more electrons are produced at lower energies where thermalization channels are favoured. Heating by thermalization of MHD waves is not considered. It can vary substantially depending on local magnetic field configurations within individual molecular cloud complexes. }
\label{fig:therm_efficiency}
\end{figure}  

The efficiency of CR thermalization is set by the balance between CR interactions that lead to thermalization, and those where energy is radiated away as photons. 
The CRs irradiating ISM clouds are mostly energetic protons~\cite[e.g. as may be inferred from the local CR spectrum,][]{2016arXiv161208562C}. 
They do not thermalize efficiently. Heating instead relies on inelastic hadronic collisions that produce secondary CR electrons. Fig.~\ref{fig:therm_efficiency} shows the typical CR electron spectrum that would be generated by hadronic collisions from an irradiating power-law CR proton flux. These secondary electrons engage more strongly in thermalization. However, 
their production introduces losses to neutrinos and $\gamma$-rays in the pp decay chain. 
This limits the maximum efficiency of CR heating via pp collisions to around 15 per-cent (see, e.g.~\cite{Owen2018MNRAS}). 
The secondary electrons would experience radiative losses, mainly by synchrotron and free-free emission. This further limits their thermalization efficiency. The exact fraction of their energy available for heating the cloud medium is subject to more complicated dependencies on the local environment. This is illustrated by 
Fig.~\ref{fig:therm_efficiency}, which shows how efficiency of thermalization varies through the density structure of a cloud. 
This simplified model adopts an empirically-informed relation between cloud density and magnetic field strength from Ref.~\cite{Crutcher2010ApJ} to determine the synchrotron loss rate, while the ionization fraction of the gas follows from Ref.~\cite{Elmegreen1979ApJ}.
Typically around 50 per-cent of secondary CR energy can go into heating background gas, which can increase to 90 per-cent in certain conditions. The overall thermalization fraction of primary CR protons therefore 
varies between $\sim$5 and 15 per-cent. 

If CR heating is severe, it can elevate the temperature of a cloud.  Observations have found that clouds experiencing significant CR ionization are warmer~\cite{Bisbas2017ApJ, Gaches2018ApJ}. Theoretically, elevated temperatures of 30-50 K have been estimated for molecular clouds with CR ionization rates increased by around $\sim 10^{3}$ times that typical of the Galaxy~\cite{Bisbas2017ApJ}, or $\sim$ 50-100 K for CR energy densities increased by $\sim 10^{3}$ - $10^{4}$~\citep{Papadopoulos2011MNRAS}. 
Warmer clouds are subject to more thermal pressure support against gravitational collapse. This makes star-formation less favourable. The exact effect this has on star formation, quenching~\cite[e.g.][]{Owen2019AA}, clustering~\cite{Owen2021ApJ} and the initial mass function (IMF)~\citep{Papadopoulos2011MNRAS} remains unclear. This is partly because   
the exact temperature increase experienced by a cloud is uncertain. It is strongly dependent on local conditions, model configuration (including the CR transport and interaction micro-physics; see also section~\ref{sec:cr_microphysics_cgm}) and the level of CR bombardment in different regions of a cloud. Some studies have explored these dependencies explicitly, reporting equilibrium temperatures up to $\sim$40 K and varied temperature distributions throughout cloud under different CR irradiation intensities~\cite{Gong2017ApJ}.
More conservative temperature increases ranging from $\sim$6 to $\sim$ 21 K have also been reported~\cite{Pazianotto2021ApJ, Pazianotto2023MNRAS}, with slightly higher values possible in the external layers of clouds~\cite{Pilling2022MNRAS} and strongly heated skins at the boundaries of cold clouds~\cite{Everett2011ApJ}. Other studies did not find an increase in temperature at all in some clouds under Galactic conditions~\cite{Owen2021ApJ}, with CR heating power in diffuse clouds not even being competitive against thermal condition of heat from the hot ambient ISM plasma~\cite{Everett2011ApJ}. 

\subsection{Clouds and diffuse media associated 
with stellar remnants and supernovae} 
\label{sec:cloud_SNR}

Molecular clouds and diffuse ISM structures surrounding stellar remnants and SNe 
provide an opportunity to study CRs in the vicinity of their sources.  
Sometimes referred to as \textit{active} molecular clouds, they are natural laboratories to investigate CR acceleration, propagation, and the feedback impacts CRs can deliver into the dense phases of the ISM.    
$\gamma$-rays provide information about CR interactions in molecular clouds near SNRs at energies above a GeV~\cite{Li2023ApJ, Aruga2022ApJ, Zhang2021ApJ, Zhong2023MNRAS, Yeung2023PASJ, Supan2022A&A} (for early pre-\textit{Fermi} era propositions and reviews, see~\cite{Aharonian1991Ap&SS, Aharonian2001SSRv}).    
The 6.4 keV neutral iron K$\alpha$ line also traces MeV CRs in molecular clouds~\cite{Fujita2021ApJ}, as demonstrated by X-ray
studies of known molecular clouds around SNRs~\cite{Nobukawa2018ApJ, Maxted2018MNRAS, Okon2018PASJ, Saji2018PASJ, Makino2019PASJ, Nobukawa2019PASJ, Shimaguchi2022PASJ}. 
Complementary 
information about the 
engagement of 10-100 MeV CRs can be obtained from molecular ion emission or absorption lines~\cite[e.g.][]{Indriolo2023arXiv}. 
Using multiple tracers sensitive to CRs of different energies 
opens a broad view of the CR spectrum in different parts of the Galaxy, and enables the study of relevant transport and interaction physics involved in CR feedback.  
For example, diffusive propagation of CRs away from their sources may be slower at lower energies. This can enhance the intensity of CR ionization and Iron K$\alpha$ line signatures from the medium near SNRs, but may have less effect on $\gamma$-ray emission from the same region.
To properly resolve the feedback impact CRs have in galaxies, 
a thorough understanding of their 
inhomogeneous distribution at different energies is required. 
Of particular importance is variation of the CR distribution 
with respect to ISM structures like molecular clouds, where CRs deliver feedback. This is set by the CR propagation physics, and the spatial correspondence between clouds and CR source locations.

Within the Milky Way, 
Ref.~\cite{Mitchell2021MNRAS} showed that SNRs and molecular clouds occupy a volume fraction of the ISM of 0.01 per-cent and 0.25-3.3 per-cent, respectively. 
As the volume occupied by SNRs is much less than molecular clouds, the number of clouds in very close proximity to a SNR is low. OH maser measurements suggest only around 15 per-cent of core collapse SNe interact with dense molecular gas directly~\cite{Hewitt2009ApJ}. 
Notable exceptions have been studied. These may be identified from increased cloud temperatures due to shock heating, or high levels of turbulence within a cloud and offset $\gamma$-ray emission away from the cloud direction.  There are several examples of SN shocks interacting directly with molecular clouds \cite[e.g. HESS J1825-137, W49B, and the Boomerang nebula /SNR G106 region; see][]{Voisin2016MNRAS, 2018A&A...612A...1HGPS, 2023A&A...671A..12MAGIC_G106, Sano2021ApJ, 2021NatAs...5..460TibetG106}. However, 
direct interaction of SN shocks, or the associated injection of CRs accelerated by shocks passing through molecular clouds statistically the less common scenario~\cite{Mitchell2021MNRAS}. Instead, hadronic and leptonic CRs accelerated in SN shocks propagate diffusively through the ISM for some distance before encountering a molecular cloud. Non-linear effects, such as the build-up of low-energy ($< 10\; {\rm GeV}$) CRs around sources could lead to self-confinement. In this scenario, the overabundance of CRs generates turbulence that scatters particles. This can lead to the suppression of the diffusion coefficient by up to 2 orders of magnitude~\cite{Jacobs2022JCAP}. $\gamma$-ray observations surrounding a SNR shell~\cite{Feijen2022MNRAS}, and from hadronic $\gamma$-ray emission from molecular clouds in a range of CR-irradiated environments~\cite{Li2023ApJ, Wang2022ApJ} have shown indications of such suppression of CR diffusion. 

\subsection{The interstellar medium of starburst galaxies and implications for cosmic ray feedback}
\label{sec:ISM_starbursts}

The discussion in this section has focused on our local Galactic neighbourhood. However, in starburst galaxies, the configuration of the ISM and its components is different. 
Firstly, the overall energy budget of CR feedback in a starburst galaxy is larger than that of the Milky Way~\cite[e.g.][]{Owen2018MNRAS, Krumholz2022arXiv}. They have more energy to deposit into their host's ISM, where their impacts can be stronger. 
In addition to this, the 
efficiency by which their feedback is  
delivered may 
also be greater. 
This is because of the closer spatial correspondence between sites of CR acceleration and the locations where feedback delivery would affect star-formation. 
For example, massive molecular cloud complexes are common in starburst galaxies~\cite{Fujii2016ApJ}, but not in the Milky Way. These clouds develop into massive stellar clusters, which are important CR accelerators (section~\ref{sec:superbubbles}). 
Consequently, the fraction of CR sources interacting closely with molecular clouds is higher in starburst galaxies.
Moreover, the prevalence of interactions between super-bubbles or SN remnants and clouds is likely to be much higher in intensely star-forming galaxies. This is indicated by the increased volume filling fraction of hot gas, which tracks the SN rate~\cite{Li2015ApJ}. Additionally, the volume filling fraction of SN remnants in the ISM will be increased (e.g. perhaps up to 8-9.5 per-cent in starbursts like M82 and NGC 253~\cite{Krumholz2020MNRAS}), making cloud-SN remnant interactions much more widespread. This is supported by observations, as the total $\gamma$-ray emission of starburst galaxies has a strong dependency on the star-formation rate. By contrast, $\gamma$-ray emission from clouds in the Galaxy shows a strong dependency on mass. This suggests that the $\gamma$-ray emission from star-forming galaxies originates mainly from CRs that are accelerated by local active sources~\cite{Peng2019A&A}. 
These differences imply that CR feedback power is delivered more efficiently in starburst galaxies compared to the Milky Way, as the sources of CRs are generally much closer to the sites of star formation. 

%%%%%%%%%%%%%%%%%%%%%%%%%%%%%%%%%%%%%%%%%
\newpage
\section{High-energy environments}  
\label{sec:high_energy_environments}

Several nearby galaxies exhibit a $\gamma$-ray flux 
  that exceeds the calorimetric limit of their star formation activity associated with SNRs 
  \citep{Eichmann2016ApJ...821...87E, Ajello2020ApJ...894...88A}. 
This section is dedicated to discussing alternative, 
  persistent CR production sites within galaxies. 

\subsection{Active galactic nuclei, jets and outflows}  
\label{sec:agn_jets_outflows}

\begin{figure}[h]
\begin{center}
\includegraphics[width=13.75cm]{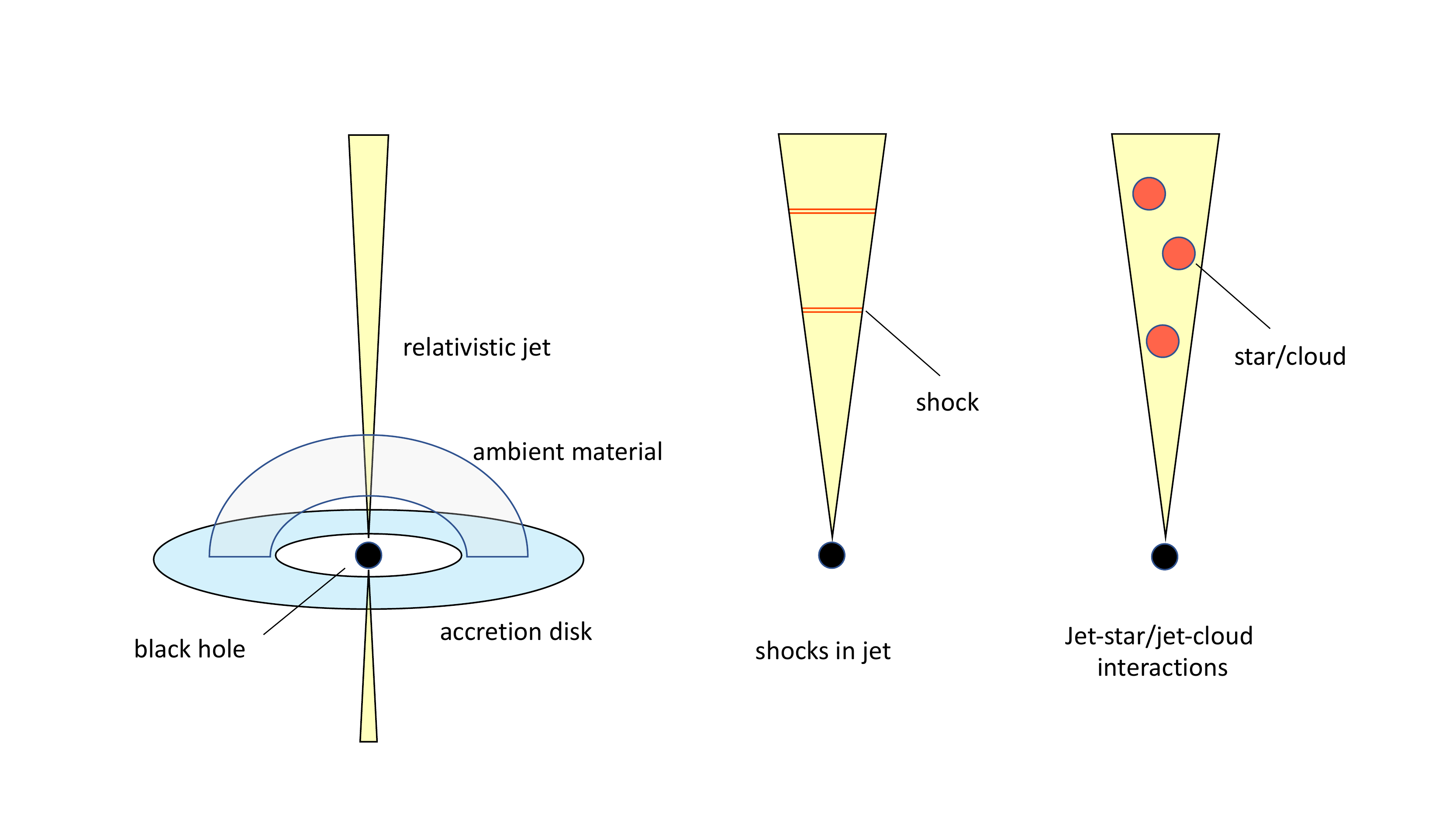} 
\end{center}
\caption{Illustrations of some possible scenarios  
   for hadronic processes in galactic-scale environments 
   associated with jets in a weak AGN (not to scale). 
   \textbf{Left:} 
   The energetic particles in the jets interact 
     with the ambient material, e.g. the ISM, near the central black hole. 
   \textbf{Middle:} Particles are accelerated and/or energised 
     by the shocks formed inside the jet. 
   \textbf{Right:} The AGN jet interacts 
     with stars or clouds trespassing into it. }
 \label{fig:AGN_jet}
\end{figure}

A sizeable fraction of galaxies are found to harbour an 
  AGN or show 
  a certain degree of AGN-like activities 
\citep{Kauffmann2003MNRAS.346.1055K, Schawinski2010ApJ...711..284S}.  
These galaxies, in particular those possessing jets,  
  may have multiple sites 
  of continuous CR production. 
Several scenarios have been proposed.  
  They invoke either p$\gamma$ or pp interactions 
  as the mechanisms 
  shown in Figure~\ref{fig:AGN_jet}, 
  or their combination in some situations.  
As the theme of this \textit{Review} concerns 
  CRs at the galactic scale (i.e. $\lesssim 10~\textrm{kpc}$),  
  we exclude systems 
  with large-scale relativistic jets 
  that extend far beyond the host galaxy.  
The discussions here therefore focus instead  
  on radio-quiet systems, 
  if there is AGN activity in their galactic centre.  
 
%%%%% Weak Jet
Weak jets in radio-quiet AGNs  
  that are CR source candidates  
  would terminate at the scale of $\sim\textrm{kpc}$ 
  \citep[see e.g.,][]{Wilson1987ApJ...319..105W}. 
A recent ALMA observation toward NGC~1068 
  (a radio-quiet Seyfert galaxy) 
   unveiled CR production activity, 
  likely at the `head' of the jet, 
  $\approx670~\textrm{pc}$ away from the supermassive black hole (SMBH) \citep{Michiyama2022ApJ...936L...1M}. 
The CR power in the jet is estimated to exceed 
  that expected from activity directly 
  associated recent star formation episodes in this galaxy.

%%%%% Disk Wind
Wind outflows from the accretion disk 
  are believed to be able to produce CRs. 
Ultra-fast outflows (UFOs) are the 
most powerful among variants of AGN disk winds.
They are also fastest, 
  and their velocities  
  can reach $0.1~c$, i.e. 10 per-cent of the speed of light  \citep{Tombesi2010A&A...521A..57T,Tombesi2012MNRAS.422L...1T}. 
This translates to a large kinetic power, 
  as high as $\sim5$ per-cent of the bolometric luminosity of these systems 
  \citep{Tombesi2012MNRAS.422L...1T, Gofford2015MNRAS.451.4169G}. 
  They are a power bank that 
  feeds the CR production processes. 
Current observations 
  indicate that 
  UFOs are present in roughly 40 per-cent 
  of nearby AGNs~\citep{Tombesi2012MNRAS.422L...1T}.  
However, possible mechanisms to transfer the kinetic energy 
carried in UFOs efficiently into CR energy are yet to be resolved. 
Nonetheless, 
  there is no dispute, in terms of energetics, 
  that UFOs, 
  which share some similarity with relativistic jets, 
  and that it is possible for them to produce high-energy CRs 
  \citep[see][]{Wang2015MNRAS.453..837W, Lamastra2016A&A...596A..68L, Liu2018ApJ...858....9L, Inoue2022arXiv220702097I, Peretti2023arXiv230113689P}.

%%%%% Corona  
The solar corona is known to produce particles of energies 
  many orders of magnitude 
  higher than the photons from the photosphere, through thermal emission process. 
The energetic particles arise as a consequence 
  of a variety of sequences of 
  eruptive magnetic coronal activities, 
  field reconnection, shock formation, flaring emission and plasmoid ejection
  \citep[see][]{Kahler1992ARA&A,Cliver2022LRSP}. 
Such activities could also arise in corona above accretion disk in AGNs. 
  As a consequence, energetic non-thermal particles 
  are produced through macroscopic and microscopic 
  plasma processes or different combinations of both
  in a multi-stage sequence, 
  cf. coronal mass ejection 
  and the associated acceleration of particles.   
Thus, accretion disk corona in AGNs 
  are expected to be CR production sites (e.g., \cite{Kazanas1986ApJ...304..178K, Zdziarski1986ApJ...305...45Z, Sikora1987ApJ...320L..81S, Begelman1990ApJ...362...38B, Stecker1992PhRvL..69.2738S, Inoue2019ApJ...880...40I}; see also \cite{Kimura2015ApJ...806..159K} for low-luminosity AGN cases). 
This scenario is supported by the 
  recent IceCube observations, which identified  
  a neutrino hot-spot (with a significance level of 4.2 $\sigma$) located 
  in the direction of NGC~1068   
  \citep{IceCube2022Sci...378..538I}, 
  which is a known $\gamma$-ray emitter \citep{Lenain2010,3FHL,4FGL}.  
Production of VHE neutrinos 
  in astrophysical systems 
  is thought to arise through pion production channels (see section~\ref{sec:hadronic_ints}). 
There is an equal chance that 
each of the three pion species ($\pi^0$, $\pi^+$ or $\pi^-$) are produced in a hadronic interaction. 
  This implies that neutrino sources 
  would also be intrinsic sources of $\gamma$-rays.  
Moreover, 
  the intrinsic luminosity of the neutrino emission 
  would not significantly exceed 
  the that of the $\gamma$-rays, 
  unless processes are present 
  within the interaction sites 
  that suppress charged pion production 
  compared to neutral pion production. 
 IceCube observations of NGC~1068 
   imply that the neutrino flux 
   is greater than the GeV $\gamma$-ray flux 
    \citep{IceCube2019_NGC1068, IceCube2022Sci...378..538I}. 
This would require a suppression of neutral pion production, 
  which is not easy to achieve. 
  An alternative explanation is the significant attenuation of GeV $\gamma$-rays. 
One possible process to achieve this is the interaction of 
GeV $\gamma$-rays with strong radiation (or matter) fields. 

Studies of X-ray binaries (XRBs)
  have shown them to be X-ray bright with a luminous accretion disk 
  in a high-soft state, and latent jets.  
Taking accretion disks and jets 
  in XRBs as a reference, 
 it can be taken that 
 accretion disks in certain AGN 
  could produce a strong radiation field 
  and that their jets are present but not prominent.   
While weak jets may be radiatively inefficient 
   and are confined near the galactic centre,  
  they would still be capable of accelerating particles.  
  These energetic particles  
  would undergo interactions within the strong radiation 
  from the accretion disk, 
  leading to the development of a particle cascade through the 
  p$\gamma$ interaction. 
In this scenario, $\gamma$-rays would arise from decay of neutral pions formed by CR interactions near 
  the acceleration site, the magnetic disk corona. 
  They would be attenuated 
  when the accretion disk is luminous enough 
  to create a dense radiation field 
\citep[see][]{Inoue2020ApJ...891L..33I,Murase2020PhRvL.125a1101M,Eichmann2022ApJ...939...43E}. 

\begin{figure}
\vspace*{-0.5cm} 
\begin{center}
\includegraphics[width=13cm]{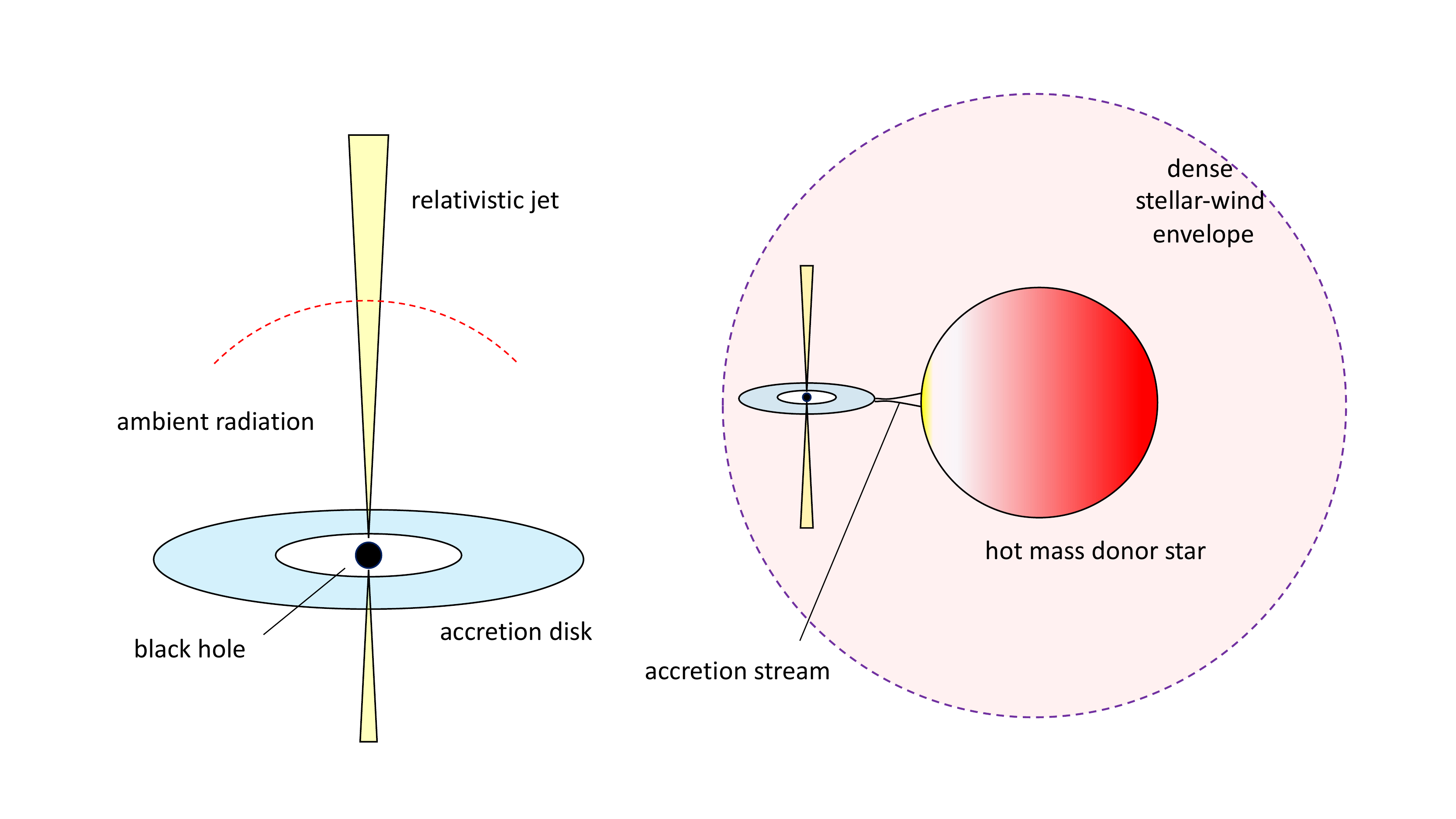} \\ 
\vspace*{-1.0cm}
\includegraphics[width=13cm]{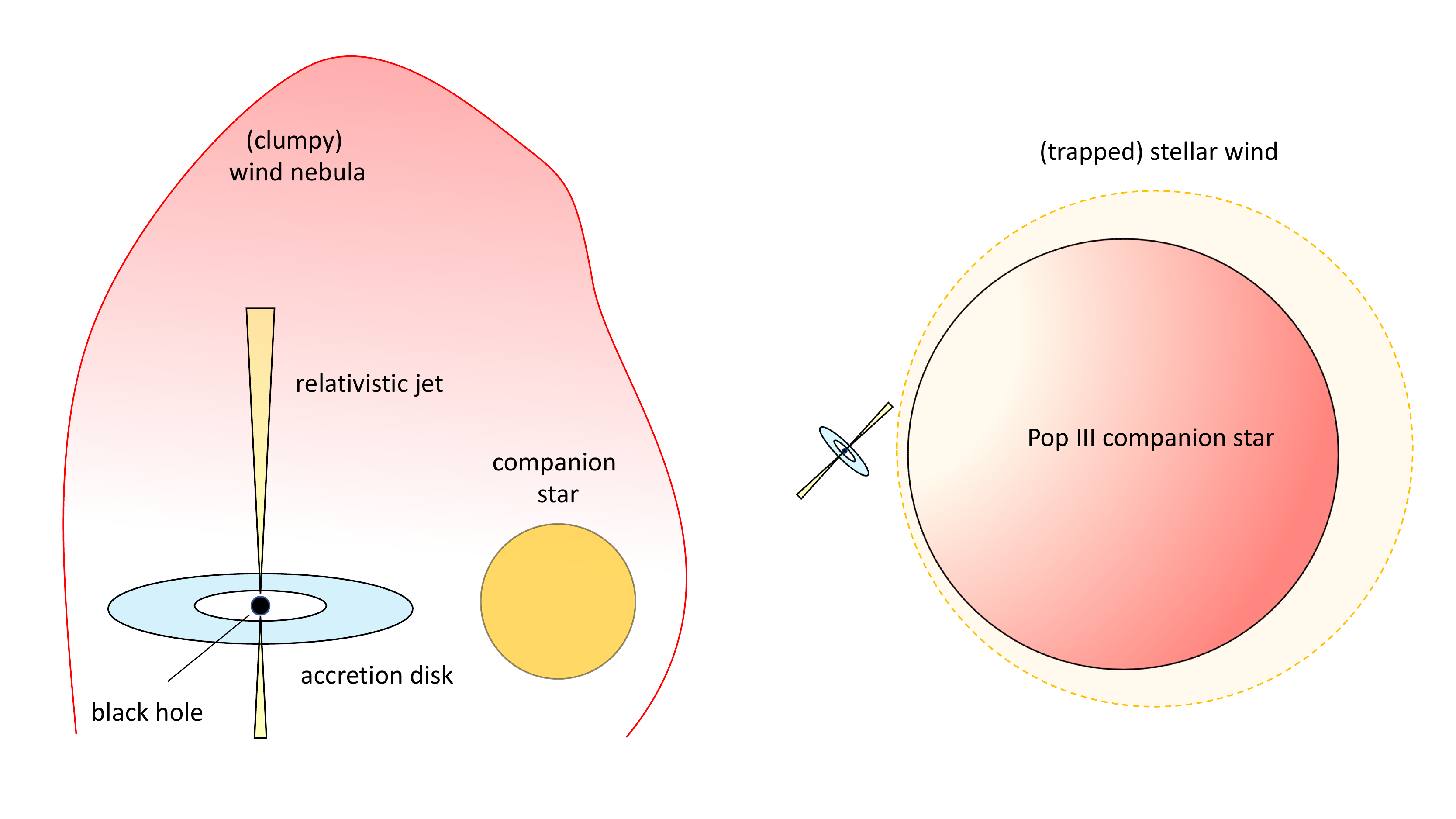}  
\end{center}
\caption{Illustrations
  of example scenarios for hadronic interactions  
  in stellar-scale accretion-powered jet sources (not to scale).  
  \textbf{Top left:} A strong radiation field 
    is present near its central engine, 
    the black hole. 
  The energetic particles in the jets  
    interact with the photons in the ambient radiation field, 
    leading to photo-hadronic processes. 
  \textbf{Top right:} The binary system 
    is embedded within  
    a circum-binary envelope 
    formed from the dense wind 
    of a Wolf-Rayet star.  
    The energetic particles in the jets 
     interact with the baryons in the circum-binary material, through pp processes.    
  \textbf{Bottom left:} The binary system resides 
    inside a wind-blown nebula 
    bounded by neutral or lowly ionised ISM. 
 The nebula could be clumpy, and 
 contains substantial non-thermal particles, 
   which can be hadronic or leptonic. 
 Shocks may arise when 
   the jets propagate through the nebula, 
   which leads to particle acceleration 
   and re-energisation. 
  The energetic particles  
   originally in the jets, and/or 
   accelerated/re-energised in the shocks 
    interact with the nebula material 
   as well as the ISM interfacing 
   with the nebula medium 
   through pp processes. 
\textbf{Bottom right:} A Pop III XRB consists of a compact object accreting material from a very massive companion star. The companion star is expect to drive a wind. This wind is relatively weak compared to line-driven winds from massive stars with high metalicity. The wind material could be somehow trapped near the star. If the jets are not perfectly aligned with the orbital rotational axis, they will be intercepted by the trapped wind envelope or even the companion star itself. 
   }
 \label{fig:mQ-P3XRB}   
\end{figure}  

\subsection{X-ray binaries} 
\label{sec:xrbs_mqs}

Compact objects as individual systems, 
  (e.g. isolated pulsars and magnetars 
  \citep[][]{Michel1984AdSpR,Heyl2010MNRAS,Fang2014PhRvD,Piro2016ApJ})
  or in binaries (e.g. cataclysmic variables \citep[cf.][]{Bowden1992APh,Li2016ApJ,Meintjes2023Galax},  
  XRBs~\citep[][]{Cooper2020MNRAS}
  and millisecond pulsar binaries \citep[][]{Linares2021JCAP,Harding2022ASSL}) 
  could produce photonic  
  and non-photonic particles 
  with energies greatly exceeding 
  tens or hundreds of MeV. 
However, we do not discuss these sources here.  
Instead, we mainly consider microquasars,   
  which are stellar-scale and share similar characteristics to AGN.  
  One of these characteristics 
  is their relativistic jets,  
  which are powered by secular and episodic mass accretion.  
We put focus on SS433 and Cyg X-3,  
  and their probable counterparts  
  in early Universe, i.e. Pop III microquasars. 

\subsubsection{SS433}
\label{sec:ss433}

Highly collimated jets are  
  expected to be efficient particle accelerators,   
  and microquasars (XRBs with jets) 
  are often considered as potential CR sources 
  \citep[][]{Escobar2022A&A}. 
SS433 is one of 
  the most investigated microquasars. It has 
  a pair of well defined jets 
 \cite{Fabrika2004ASPRv..12....1F} and contains 
  a compact object. 
This may be a 
  black hole \citep{Cherepashchuk2021MNRAS},     
  accreting material from  
  a probable super-giant star \citep{Hillwig2004ApJ,Han2020ApJ}. 
  Polarisation observations have indicated 
  that the magnetic field of the jets have helical structures  
  \citep[][]{Roberts2008ApJ,Bowler2018A&A}.  
  They are embedded with the binary in the nebula W50, 
  implying that there would be 
  interaction between the jets 
  and ambient nebula material. 

Observations with the High Altitude Water Cherenkov Observatory (HAWC) 
  have detected of $\gtrsim25$~TeV $\gamma$-rays 
  from the jets of SS433 
  \cite{HAWC2018Natur.562...82A}. 
The emission regions 
  are $\sim30$~pc away 
  from the location of the binary. 
 Both the eastern and western emission lobes 
  are found to coincide with these regions,   
  from which X-ray emission  
  is non-thermal  
  \cite{1983ApJ...273..688W, 1994PASJ...46L.109Y, 1996A&A...312..306B, 1997ApJ...483..868S, 1999ApJ...512..784S, 2022PASJ...74.1143K}. 
These observations 
  suggest that the jet downstream regions 
  are sites of high-energy particle acceleration 
  that produce the TeV $\gamma$-rays 
  and non-thermal X-rays
  \cite{Sudoh2020ApJ...889..146S, Kimura2020ApJ...904..188K}. 

As illustrated in Fig.~\ref{fig:mQ-P3XRB}, 
  extragalactic microquasars, 
similar to SS433,  
  may be associated with expanding bubbles 
  with a velocity 
  that can reach $80-250~ \mathrm{km\ s^{-1}}$ \cite{Pakull2010Natur.466..209P, Cseh2012ApJ...749...17C}. 
These expanding nebulae 
  provide suitable conditions for  
  particle production and CR acceleration, 
  especially through 
  a sequence of processes 
  triggered by an initial hadronic interaction.  
It was shown in \citet{Inoue2017APh....90...14I} 
  that nebulae with fast expansion velocities  
  $\gtrsim120~\mathrm{km\ s^{-1}}$ 
  are capable of accelerating 
  charged particles up to $\sim100$~TeV, 
  making them viable UHE CR sources.

\subsubsection{Cyg X-3 and related systems}
\label{sec:cygx-3}

Cyg~X-3 is an XRB with an orbital period of 4.8~hr 
  \citep{Mason1986ApJ,Stark2003ApJ}.  
It contains a compact object, which is 
  likely to be a black hole, although 
   the possibility of it being a neutron star has not been  ruled out. The compact object is  
  accreting from a companion star,  
  commonly identified as a WR star 
  \citep{vanKerkwijk1996A&A} 
  from the characteristic broad emission line features 
  and lack of Hydrogen lines in the optical spectrum. 
These features make Cyg X-3 unusual among 
 the Galactic XRBs,  
   both in terms 
   of its evolution and orbital dynamics  
  \citep[see e.g.][]{Lommen2005A&A}.   
Cyg X-3 spends most of its time 
  in a hard X-ray spectral state.  
Although its radio emission 
may be relatively latent 
  at that time, 
  it occasionally 
  shows episodic relativistic ejection 
  and giant radio flares \cite[][]{Egron2021ApJ}.  
When it is radio active, 
  Cyg X-3 is the brightest radio source 
  among the Galactic XRBs  
  \citep[see e.g.][]{Koljonen2013MNRAS}.  

The detection of a one-sided jet in Cyg X-3 
  \citep{Mioduszewski2001ApJ,Tudose2010MNRAS} 
  implies that the jets are highly relativistic 
  and that the jet axes are sufficiently 
  aligned with the line-of-sight 
  \citep[with a viewing angle probably $< 15^\circ$, 
  see e.g.][]{Mioduszewski2001ApJ}. 
Cyg X-3 is among the very few XRBs  
  that have been detected in $\gamma$-rays 
  \citep{Ackermann2009Sci,Tavani2009Natur,Dubus2010MNRAS}. 
As such, it has been considered as a potential CR source. 
How particles are accelerated in Cyg X-3 is still unresolved.  
It is also unknown if the material in the Cyg X-3 jets 
  has a substantial amount of baryons.  
With the strong wind from the WR star, 
  Cyg X-3 should be enveloped by a cocoon 
  of dense wind material 
  (see Fig.~\ref{fig:mQ-P3XRB}).  
When the jets of Cyg X-3 encounter dense materials, 
  shocks can be formed. 
  These shocks will in accelerate charged particles, 
  whether they are leptons or baryons. 
These energetic particles are CRs, and 
  their interaction with the WR wind envelope 
  would lead to the production of high-energy photons  
  and particle cascades 
  if there are substantial amounts of baryons in the jet material.  
The subsequent decay of the pions 
  in the particle cascades will  
  give rise to neutrino and $\gamma$-ray emission. 
  
\subsubsection{Pop III X-ray binaries}
\label{sec:p3XRBs}

Pop III stars could have masses in excess of $100~{\rm M}_\odot$ 
 \citep[see][]{Susa2014ApJ,Chantavat2023MNRAS}. 
They are short-lived and evolve rapidly. 
 Some of them would end up forming black holes, 
 through direct collapse 
 or pair-instability SNe 
  \citep[see][]{Haemmerle2018MNRAS,Moriya2019PASJ}. 
A Pop III star paired with a black hole 
  forms a Pop III XRB, 
  when the Pop III star 
  begins to transfer material to the black hole 
   \citep[see][]{Sotomayor_Checa2019}. 
Pop III binaries are often considered 
  as counterparts of Galactic  
  high-mass X-ray binaries (HMXBs)
  in the early Universe. 
However, 
  Pop III stars are metal poor 
  and would not have a strong dense wind 
  driven by large metal line opacity.  
Pop III XRBs are therefore different to  
  present-day HMXBs. 
The mass transfer in Pop III XRBs, 
  at least for X-ray luminous systems, 
  would not be mediated 
  by the accretion of the stellar wind onto the black hole. 
The mass transfer would instead be facilitated 
  by Roche-lobe overflow 
  or focused wind flows near the critical Roche surface 
  and is probably driven by the nuclear evolution 
  of the Pop III mass donor star,  
  instead of orbital shrinkage 
  caused by the loss of orbital angular momentum 
  of the binary 
  \citep[see e.g.][]{Verbunt1981A&A}.  
Thus, 
  Pop III XRBs are semi-detached binaries. 
Some Pop III XRBs would possess 
  a pair of relativistic jets, 
  which are sites of particle acceleration. 
These Pop III XRBs are microquasars, 
  and are potential CR sources. 

The mass ratio ($q = M_2/M_{\rm bh}\;\!$, 
  where $M_2$ is the mass of the Pop III mass donor star) 
  of Pop III XRBs would be around $20- 40$,  
  if considering that the mass of their black holes is  
  $M_{\rm bh} \sim (3- 5)\;\!{\rm M}_\odot$.     
We may obtain 
  the Roche-lobe radius $r_{\rm L}$, 
  in terms of the orbital separation $a$ 
  for semi-detached binaries 
  using the expression given in Ref.~\cite{Eggleton1983ApJ}. 
The ratio $r_{\rm L}/a$ is only weakly-dependent on 
   the mass ratio $q$.  
For the parameters of Pop III XRBs  
   considered here,   
   $r_{\rm L}/a \approx 0.6$,  
   over the range of $q \sim (20-40)$. 
This almost constant value of $r_{\rm L}/a$ 
  implies that a jet/outflow 
  with a $45^\circ$ half-cone opening angle  
  will hit the atmosphere of the Pop III star 
 (see illustration in bottom right panel 
  in Fig.~\ref{fig:mQ-P3XRB}), 
  if the symmetry axis of the jet/outflow  
  has a tilting angle of $> 8^\circ$  
  with respect to the angular momentum vector 
  of the binary. 

While such a jet-tilting angle 
  is not expected 
  to be commonly found in the Galactic microquasars, 
  which are mostly low-mass X-ray binaries (LMXBs)  
  and have evolutionary time scales of order 
  hundreds of Myr, 
  it would not rule out the possibility 
  of such jet orientations to occur 
  in Pop III XRBs. 
However, little is known about the evolutionary trend 
  of the progenitors of the Pop III XRBs.  
It is unlikely that 
  Pop III XRBs would go through 
  the same evolutionary channels 
  that produce Galactic HMXBs and LMXBs 
  \citep[see e.g.][]{Bhattacharya1991PhR,Postnov2014LRR}. 
The short-lived nature of Pop III stars 
  implies that Pop III XRBs are all extremely young systems. 
The black holes in Pop III XRBs 
  would not have sufficient time 
  for their spins to align with 
  the orientation of the angular momentum of the binary orbit. 
  The accretion torque 
   would instead drive the black hole spin axis and the jet axis  
   into precession and nutation. 
If the jets have a substantial amount of protons, 
  the atmosphere of the Pop III star will act as a target 
  for energetic protons in the jet. 
This leads to hadronic pp interactions, 
  which generate cascades of descendent pions and leptons.   
This scenario of CR production does not require 
  the presence of material outside the binary. 

There are more conventional views 
  for CR production 
  where Pop III XRBs are practically direct counterparts  
  of the Galactic microquasars,  
  except that the mass donor star in a Pop III XRB  
  is a massive and ultra-metal-poor Pop III star. 
Thus, 
  the particle acceleration and production scenarios 
  are the same as those proposed for the 
  Galactic XRB jet sources 
  \citep[see e.g.][]{Romero2008A&A,Smponias2021Galax}, 
  using the same prescription developed 
  for CR production in AGN jets 
  \citep[see e.g.][]{Mucke2001APh,Reynoso2011A&A}.  
The energetic particles that 
  result from acceleration in the jets, 
  presumably through shocks, 
  would interact with the ambient  
  radiation field through p$\gamma$ interactions,  
  or pp interactions with baryonic matter in the cocoon/envelope 
  close outside the binary, 
  or in a nebula at a larger distance   
   \citep[see e.g.][]{Carulli2021APh} 
  (see Fig.~\ref{fig:mQ-P3XRB}). 
In this scenario, 
  particle acceleration and hence CR production and interactions 
  in the Pop III XRBs   
  are analogous to those 
  in the microquasars SS433 and Cyg X-3 
  described earlier in this section. 

Compared to low-mass Galactic microquasars,  
  Pop III XRBs might be more likely to reside 
  in crowded young multiple-star environments. 
The evolutionary timescale of Pop III XRBs 
  is expected to be shorter   
  than the dynamical time scale of their host galaxies.  
Thus, Pop III XRBs 
  would not have sufficient 
  time to disperse from their birth places 
  (cf. the distribution of the HMXBs and LMXBs 
  in the Galaxy). 
This also implies that 
  Pop III XRBs 
  might cluster in close proximity to each other. 
When an ISM    
  is bombarded collectively or continually  
  by energetic CR particles and 
  and irradiated by intense radiation 
  from a cluster of Pop III XRBs,  
  it would inevitably be heated.  
This in turn will affect 
 subsequent episodes of star formation 
  and, hence, the next generation of stars 
  and early-Universe XRBs. 

%%%%%%%%%%%%%%%%%%%%%%%%%%%%%%%%%%%%%%
\newpage
\section{Cosmic rays in galaxies and their circum-galactic environments}
\label{sec:crs_in_gals}

\subsection{The Milky Way}
\label{sec:the_mw}

\begin{figure}[H]
\begin{center}
\vspace*{0.5cm}
\includegraphics[width=12 cm]{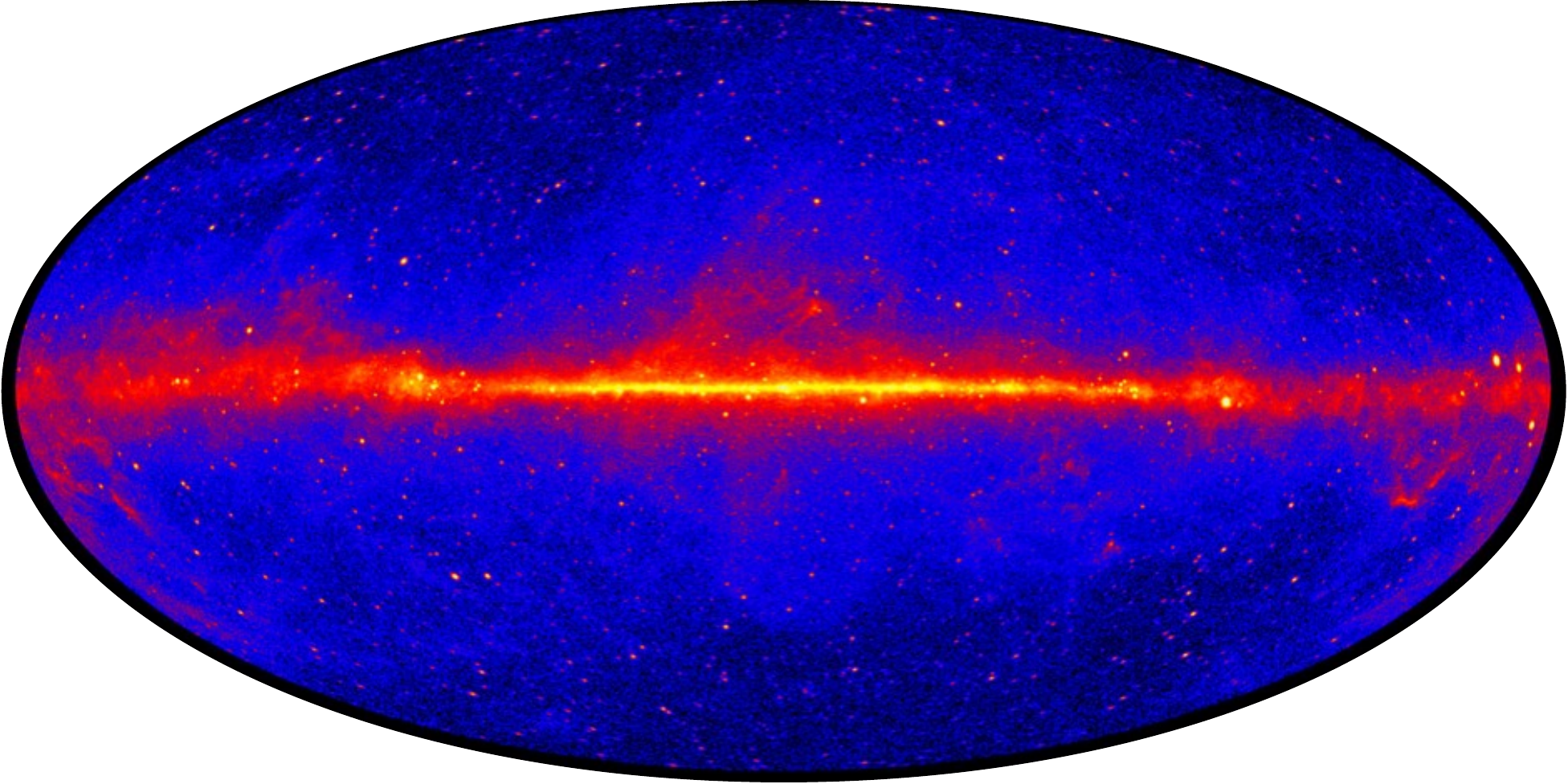}
\end{center}
\caption{Full sky as seen with \textit{Fermi}-LAT at energies above 1 GeV over 12 years of observations (Hammer projection). The Milky Way dominates the emission, and the Galactic Plane appears across the middle of the image as a bright diffuse glow. The Galactic interstellar diffuse $\gamma$-ray emission originates from hadronic CR interactions with gas and leptonic interactions with the interstellar radiation field. It provides a clear signature for the presence of CRs. Image credit: NASA/DOE/\textit{Fermi}-LAT Collaboration. Reproduced in accordance with NASA Media Usage Guidelines.}
 \label{fig:fermi_map}
\end{figure}

The distribution of CRs within and around the Milky Way can be measured using $\gamma$-rays. High quality all-sky data obtained with \textit{Fermi}-LAT now allows a detailed view of CR engagement within the Galaxy to be constructed (see Fig.~\ref{fig:fermi_map}; for a review of interstellar $\gamma$-ray emission studies and its implications for the distribution of CRs in the Galaxy, see~\cite{Tibaldo2021Univ}). 
One of the most striking signatures in the $\gamma$-ray sky above 1 GeV is the Galactic plane itself. Intricate emission structures, which trace the underlying gas distribution, emerge as bright extended diffuse emission. This covers the entire galaxy, while less bright extended emission reaches several degrees above and below the plane. 
Hadronic CRs are contained within the Galactic ISM by the interstellar magnetic field. They interact with the ISM gas by inelastic pp collisions (see section~\ref{sec:hadronic_ints}). The resulting decay of neutral pions, produced in these interactions, generates the observed emission. The intensity of the emission is proportional to the product of the local gas density and the CR density. Thus, the distribution of CRs can be recovered where high-resolution gas maps using an appropriate tracer are available. Some of the more reliable dense gas tracers are 
    CS and dust. 
However, they do not always agree, and they are not guaranteed to provide the same information about the 
precise ISM gas distribution~\citep[e.g.][]{Tsuboi99ApJS, Molinari11ApJL}. This can lead to substantial differences in inferred CR distribution estimations.

$\gamma$-ray emission features in the Galactic plane are particularly strong around the GC. 
The CMZ is located in the middle of the central bright peak. The origin of this $\gamma$-ray emission is unclear. While it is likely both a hadronic and leptonic component would be present, the relative importance of each is unsettled. Strong radio emission has been observed in this region~\cite{YZ2013ApJ}. It has been suggested this originates from predominantly leptonic primary CRs in intense GC magnetic fields. These CRs could also contribute significantly to the $\gamma$-ray emission through inverse Compton scattering~\cite{YZ2013ApJ}. An alternative scenario has also been considered, motivated by the strong magnetic fields (up to 100-250 $\mu$G~\cite{YH2014ApJ, Crocker2010Natur}) and concentrated molecular gas content in the GC. These conditions resemble those typically found in starburst galaxies (see~\cite{YH2014ApJ}, also section~\ref{sec:nearby_starbursts}), leading to the proposition that the CR population could have a significant hadronic composition~\cite{YH2014ApJ}. In this scenario, the CRs experience cooling and advective losses within a single-zone three-phase ISM, including diffuse, hot gas, ionized gas, and dense molecular gas. This can provide a plausible explanation for the TeV emission from the CMZ and account for the radio emission via synchrotron radiation from primary and secondary CR electrons. However, the GeV $\gamma$-ray emission in this scenario is under-predicated and much more difficult to reconcile with observations. This suggests a need for an additional CR component, or a population of unresolved $\gamma$-ray sources. 
The precise details of the CMZ emission require further investigation. However, it is evident that a significant population of CRs exist in that region. They could be supplied by a hidden population of pulsars~\cite[as has recently been proposed by several studies, including, e.g.][]{Calore2016ApJ}, pulsar wind nebulae, processes associated with the Sgr A* supermassive black hole, or SN remnants arising due to the elevated local star-formation rate. 

Above and below the CG, extended diffuse $\gamma$-ray structures can be seen (these can be discerned faintly in  Fig.~\ref{fig:fermi_map}). These so-called 
\textit{Fermi} bubbles extend to heights of 50\degree above and below the GC, with a total 1-100 GeV power of $>10^{37}$ erg s$^{-1}$ and a spectral cut-off at around 110 GeV~\cite{Ackermann2014ApJ} (for a review, see~\cite{Yang2018Galax} and \cite{Yang2023}). The origin of these structures is unsettled, and their non-thermal composition could be hadronic~\cite[e.g.][]{Crocker2012MNRAS, Razzaque2018Galax} or leptonic~\cite[e.g.][]{Yang2012ApJ, Su2010ApJ, Zubovas2011MNRAS}. One possibility is the presence of a galactic outflow driven by feedback from concentrated star-formation activity around the GC~\cite{Crocker2012MNRAS}. This outflow could serve to channel hadronic CRs from the ISM into the Galactic halo and CGM. This may also be achieved by subsonic breezes~\cite{Tourmente2023MNRAS}.
These breezes, although slower, can have a similar role in CR circulation through a combination of advection and diffusion. They 
have been shown to be capable of supplying CRs several kpc into the CGM, where they can sustain CR energy densities as high as a few per-cent as that in the ISM~\cite{Taylor2017PhRvD, Tourmente2023MNRAS}.  Alternatively, the CRs could be supplied by past jet activity of the Sgr A*, where the CR electrons are rapidly transported to high Galactic latitudes before they cool~\citep{Guo2012, Yang2012ApJ, Yang2013}. This leptonic jet model has been shown to reproduce the spatially uniform $\gamma$-ray spectrum~\citep{Yang2017ApJ} as well as the recently discovered X-ray {\it eRosita} bubbles~\citep{Yang2022}.

The presence of CRs in galactic halos has been investigated for some time. This interest stems from observations of 
 kpc-scale synchrotron halos around edge-on galaxies~\cite[e.g.][]{Mulcahy2018A&A}, as well as simulation work indicating a significant population of CRs reside in the CGM of Milky Way mass galaxies~\cite{Butsky2018ApJ, Dashyan2020A&A, Ji2020MNRASd}. 
Recent studies have reported detections 
of extended halo emission 
around M31 
in $\gamma$-rays. It extends up to 200 kpc from the center~\cite{Recchia2021ApJ}. It is possible that some of this extended emission originates from bubbles analogous to the Galactic \textit{Fermi} bubbles~\cite{Pshirkov2016MNRAS}, but most can be attributed to M31's CGM. The exact origin of this emission remains unsettled, however leptonic models with in situ CR acceleration in the acceleration shock due to in-falling matter~\cite{Recchia2021ApJ}, hadronic models involving CR advection~\cite{Zhang2021ApJ} and hadronic models including both CR advection and diffusion have been considered~\cite{Roy2022MNRAS}. 
The presence of CRs in the CGM of the Milky Way has also been observationally confirmed. Studies using $\gamma$-rays emitted from High and Intermediate Velocity Clouds (HVCs and IVCs) at heights ranging from hundreds of pc to a few kpc~\cite{Tibaldo2015ApJ}\footnote{Constraints on their hadronic and leptonic components have been estimated from the isotropic $\gamma$-ray background~\cite{Subrahmanyan2013ApJ}, and the maximum synchrotron flux remaining after subtraction of the Galaxy's contribution  
to the anisotropic radio background~\cite{Jana2020ApJd, Fixsen2011ApJ}.} 
showed their emissivities to exhibit a significant decrease with distance from the Galaxy, up to $\sim$ 2 kpc~\cite{Tibaldo2015ApJ, Joubaud2020A&A}. This is caused by a decreasing CR density with distance. 
It has been suggested that the emission from CRs interacting within the Milky Way's halo gas may contribute to the isotropic $\gamma$-ray background~\cite{Recchia2021ApJ, Feldmann2013ApJ, Blasi2019PhRvL}, and these interactions may also account for some of the diffuse flux of neutrinos observed by IceCube~\cite{Taylor2014PhRvD} (although this contribution is likely sub-dominant~\cite{Kalashev2023JCAP}).

 The existing observations have provided important insights that allow  effective CR transport models in the Galactic halo to be constrained.
However, upcoming instruments like the Cherenkov Telescope Array (CTA), with its increased sensitivity, will enable the detection of more HVCs, IVCs, and halo gas in $\gamma$-rays. With the data to be obtained from these next-generation instruments, it will be possible to conduct more thorough tests of CR transport models within the CGM. 
Efforts have already begun to model CR diffusion in the CGM context more rigorously, in order to relax the approximations associated with effective transport treatments. In particular, consideration has been made of the non-linear interaction of CRs with MHD waves excited through the CR streaming instability~\cite{Recchia2016MNRAS, Holguin2019MNRAS, Dogiel2020ApJ}, and their interplay with fast-mode turbulence~\cite{Kempski2022MNRAS}, or advection from the Galactic disc~\cite{Evoli2018PhRvL}. These models allow the role of CRs in CGM ecosystems, and their influence on the evolution of central galaxies, 
to be explored more thoroughly 
by reducing model uncertainties and establishing more reliable treatments of CR transport.

\subsection{Studies of cosmic ray effects in individual galaxies}
\label{sec:CRs_in_individual_galaxies}

\subsubsection{Cosmic ray containment, calorimetry and galactic magnetic fields}
\label{sec:CR_calorimetry}

CRs emerge as an inevitable consequence of star-formation in galaxies, regardless of the underlying source type or acceleration mechanism 
(see section~\ref{sec:cr_sources}). Starburst galaxies are therefore rich in CRs. These CRs are scattered in galactic magnetic fields (section~\ref{sec:particle_transport}), and typically experience a diffusive propagation regime. They are confined by their host galaxy and leak out gradually through a combination of diffusion and advection, or undergo attenuation and/or complete energy loss through cooling before they can escape. 
A CR confinement scenario can develop quickly in a starburst galaxy. It is controlled by the magnetization of the ISM, which is believed to arise early in its evolution. 
It has been proposed that the growth of magnetic fields in galaxies may be closely connected to their star formation activities. In this scenario, a saturated $\mu$G magnetic field  
  can develop within 
a few Myr after the onset of star formation, if driven by a turbulent dynamo mechanism~\cite{Schober2013A&A}. 
This is consistent with observations, which indicate early magnetic field 
growth in galaxies. In particular, high-redshift galaxies have generally been found to harbor well established ISM magnetic fields with strengths comparable to those in local galaxies~\cite[e.g.][]{Bernet2008Natur, Hammond2012arXiv}. 

The degree of CR confinement a galaxy can practically achieve is set by several factors. In some galaxies, a calorimetric limit can be reached where CRs are completely absorbed, or lose all their energy before they can escape. This results in the conversion of a very high fraction CR energy to ISM heating or non-thermal radiation and neutrinos. The degree of calorimetry varies with CR energy and species, with electron calorimetry generally being more achievable than proton calorimetry. While main sequence or relatively quiescent galaxies like the Milky Way may have low calorimetric fractions, starbursts like NGC 253, Arp 220, or M82 can exhibit very high calorimetry~\cite{Krumholz2020MNRAS}. This is because their strong magnetic fields and dense interiors are very effective in containing, cooling and/or absorbing CRs. 

In highly calorimetric settings, close correlations are expected between CR injection rates (estimated by tracers of a galaxy's star formation rate, e.g. its far-IR luminosity) and signatures for CR activity. Thus, starburst galaxies not dominated by AGN emission generally exhibit strong correlations between their far-IR, $\gamma$-ray, and radio luminosities~\cite[e.g.][]{Kornecki2020AA, Peng2016ApJ, Read2018MNRAS, Ackermann2012ApJ, Sargent2010ApJ, Bourne2011MNRAS, Eichmann2016ApJ}.\footnote{The far-IR radio correlation appears to be valid up to $z\sim$2-3~\cite{Murphy2009ApJ}. It may not hold at higher redshifts due to increased inverse Compton losses experienced by CR electrons interacting with the CMB. Conversely, the far-IR $\gamma$-ray relation should not theoretically see strong variation with redshift, however instrumental constraints and extragalactic background light (EBL) attenuation make all but the nearest starbursts detectable in $\gamma$-rays (see section~\ref{sec:nearby_starbursts}).} 
Electron calorimetry can naturally explain the tightness of the far-infrared (FIR)–radio correlation. However, its validity has been questioned due to conflicts in the observed radio spectral indices for normal galaxies~\cite{Vollmer2010A&A, Vollmer2005A&A} and difficulties to reconcile the inferred CR diffusive escape time in the Milky Way with the typical estimated synchrotron cooling time~\cite{Vollmer2022A&A}. These tensions have motivated the development of modified calorimeter models for normal galaxies that 
include CR escape over comparable timescales to electron cooling~\cite{Lisenfeld1996A&A} and invoke SN remnants as a cause of a flattened radio spectrum~\cite[e.g.][]{Lisenfeld2000A&A}.
Other approaches have proposed to move away from the calorimeter model entirely~\cite[e.g.][]{Helou1993ApJ, Niklas1997A&A}, but these often require some sort of `conspiracy'\footnote{`Conspiracies' typically include efficient cooling of CR electrons in starbursts, and a combination of low effective ulta-violet (UV) dust opacity in lower surface density galaxies. Contributions from secondary CR electrons can also be invoked to counter decreased radio emission from bremsstrahlung, ionization, and inverse Compton cooling in starbursts~\cite{Lacki2010ApJ}. However, these models can still pose a problem where CR 
electron density is  directly proportional to the star formation rate of a galaxy, 
because of the increased radio synchrotron emission associated with the secondary electrons. These issues may instead be resolved by invoking models combining CR escape, cooling and secondary production~\cite{Lacki2010ApJ}.} to maintain the tightness of the FIR–radio correlation, given the enormous dynamic range in physical properties of galaxies that obey it. 

\begin{figure}[h]
\includegraphics[width=13.8 cm]{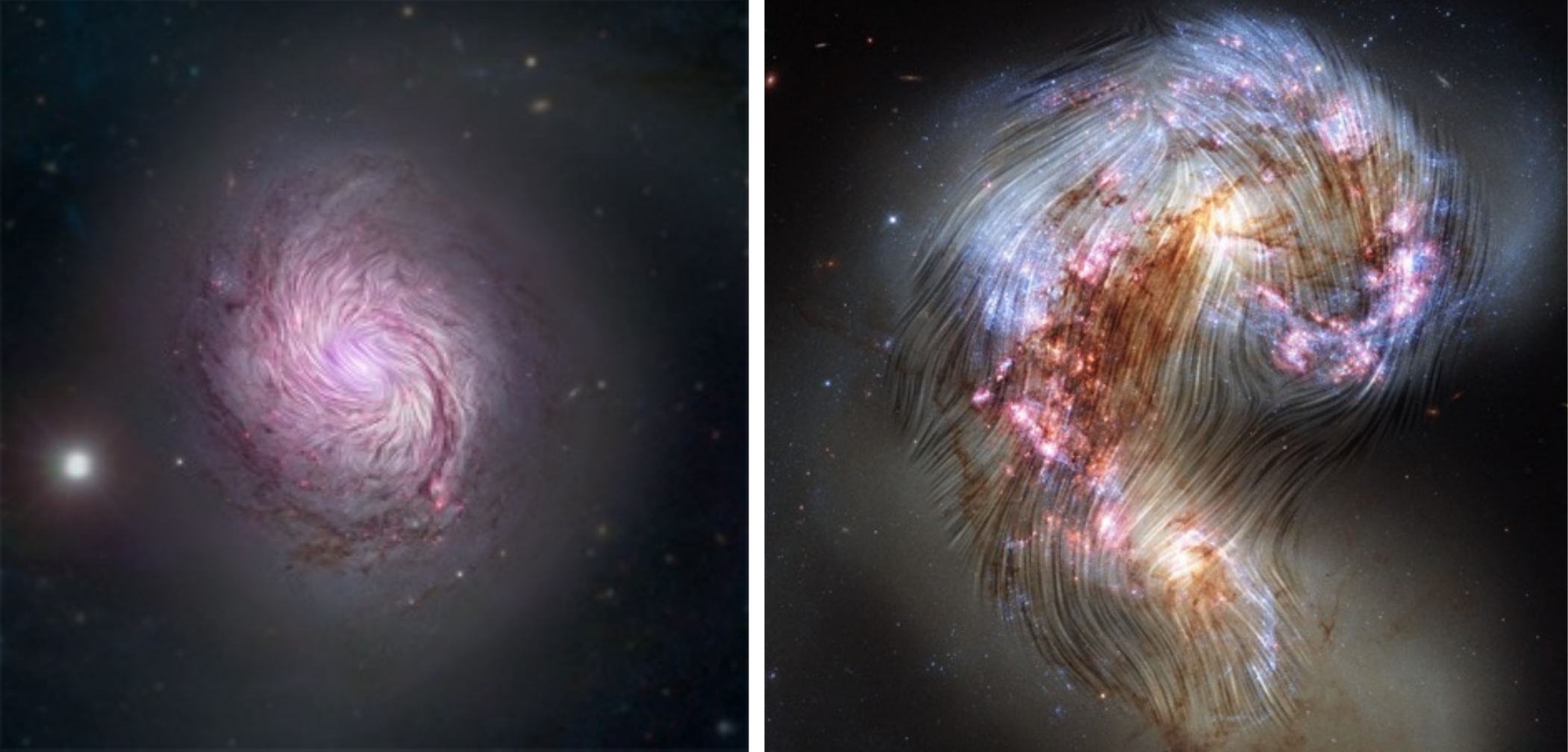}
\caption{\textbf{Left:} Magnetic fields in NGC 1068 shown as streamlines, measured using SOFIA's HAWC+ instrument using polarized far-IR emission to observe dust grains aligned perpendicular to their local magnetic field vector. They are plotted over an X-ray/visible composite image (from \textit{HST}, \textit{NuSTAR} and SDSS). Magnetic fields follow the spiral arms, where elevated star-formation is taking place. The spiral arms in this galaxy will be the main source of CRs, and the magnetic fields aligned with them will preferentially channel CRs. This may introduce  distinctive feedback patterns. Image credit: NASA/SOFIA; NASA/JPL-Caltech/Roma Tre Univ. Reproduced in accordance with NASA Media Usage Guidelines. \textbf{Right:} 
Magnetic fields in Antennae galaxies probed at 154 $\mu$m, overlaid onto an \textit{HST} image (from~\cite{Whitmore2010AJ}). This system is undergoing a collision, with two cores (NGC 4039 below and NGC 4038 above). The magnetic fields roughly align along a tidal tail to the west of the system, and the two cores show inter-connecting magnetic fields. A relic spiral arm persists to the North-East, associated with NGC 4038, which retains some magnetic field alignment. The correspondence between the magnetic field and density/relic spiral structures elsewhere has been completely disrupted. The resulting configuration of the system could channel CRs between the two cores, or release CRs along the tidal tail into the intergalactic medium (IGM). This may reduce CR containment in the individual components of this system. Figure adapted from Lopez-Rodriguez et al.~\cite{LopezRodriguez2023ApJ}.}
 \label{fig:B_field_galaxies}
\end{figure}  

More recent 
studies have 
considered sophisticated analytical models of turbulent clumpy star-forming galactic disks~\cite{Vollmer2022A&A} or
detailed numerical simulations to investigate the physics and development of the 
far-IR-radio and far-IR-$\gamma$-ray relations 
relations in 
various situations. These include 
evolving protogalaxies~\cite{Pfrommer2022MNRAS}, Milky Way-like galaxies~\cite{Pfrommer2017ApJ} and  
isolated galaxies over a broad range of halo masses~\cite{Werhahn2021MNRASb, Werhahn2021MNRAS3}. 
These studies show that the relations can start to break down for low IR luminosity quiescent galaxies~\cite{Wang2018MNRAS}, where CR escape can become more severe~\cite{Zhang2019ApJ}. For example, even though Milky Way-like galaxies are believed to be reasonably efficient electron calorimeters~\cite[e.g.][]{Strong2010ApJ}, they are not likely to be proton calorimeters~\cite{Wang2018MNRAS} and are not guaranteed to closely follow either the far-IR-radio or the far-IR-$\gamma$-ray relation~\cite{Zhang2019ApJ}. 
In very extreme galaxies such as Arp 220 or the Circinus galaxy, calorimetry may exceed 100 per-cent. This could be due to highly efficient CR acceleration, systematically more energetic SN explosions, or contamination from AGN emission or other $\gamma$-ray sources~\cite{Wang2018MNRAS}.

The multi-scale configuration of the magnetic field of a galaxy regulates effective CR propagation parameters (see section~\ref{sec:particle_transport}). This sets the distribution and confinement of CRs, and governs their calorimetry and feedback patterns. 
On macroscopic scales, CR propagation is fundamentally anisotropic~\cite[e.g.][]{Evoli2012PhRvL, Casse2001PhRvD}. Modeling it as an isotropic process on the scale of galaxies can distort our qualitative understanding of the resulting CR feedback effect. For example, invoking isotropic CR diffusion in models can lead to stronger galactic outflows~\cite{Pakmor2016ApJ}, weaken the development of the Parker instability~\cite{Wang2010arXiv}, or significantly affect outflow wind launching and mass loading factors~\cite{Ruszkowski2017ApJ}.  
To properly establish an effective prescription for large-scale anisotropic CR propagation through a galaxy, a thorough understanding of the large-scale magnetic field structure is needed. In recent years, it has become possible to measure this using instruments like the High-Angular Wideband Camera Plus (HAWC+) on the Stratospheric Observatory For Infrared Astronomy (SOFIA) ~\cite{LopezRodriguez2022ApJ} and POL-2 on the James Clerk Maxwell Telescope (JCMT)~\cite{Pattle2021MNRAS}.

From these studies, it has been shown that the large-scale structure of the magnetic field of a galaxy is strongly influenced by its 
dynamical situation. 
For example, barred-spiral galaxies, such as NGC 1068 (Fig.~\ref{fig:B_field_galaxies}, left panel), 
have exhibited an organized magnetic field patterns that align closely with their spiral arms. This alignment 
would direct CR diffusion along the spiral pattern, 
thereby influencing CR feedback patterns. 
Moreover, this alignment may also enhance CR calorimetry on a global scale throughout the galaxy. Anisotropic diffusion of CRs along the arms is favored over inter-arm diffusion, leading to longer propagation distances for CRs and an increased likelihood of cooling or absorption before escaping the galaxy. This effect becomes particularly significant if CRs are predominantly injected into the spiral arms. Indeed, this would be expected in a density-wave origin for the spiral pattern, where the arms represent regions of intensified star-formation and, consequently, higher CR density.

Elevated star-formation levels in galaxies can disrupt the alignment between magnetic fields and spiral arms. This is due to winds and ISM bubbles caused by feedback processes~\cite[e.g.][]{Fletcher2011MNRAS}. 
The disruption becomes even more pronounced in chaotic systems such as the colliding Antennae galaxies (Fig.~\ref{fig:B_field_galaxies}, right panel).
 In this case, the magnetic field is primarily influenced by the interactions between the galaxies themselves, and is stretched between them. 
The strong disruption of magnetic fields during the collision phase of these galaxies would initially release confined CRs into the intergalactic medium.
 However, as interconnected magnetic fields formed between the interacting galaxy cores after the initial collision, CRs would begin to channel between them. CR channeling between the interacting galaxies would become established over a diffusion timescale (a few Myr). This is much shorter than 
 the typical timescale of galaxy interactions, which is on the order of several hundred Myr~\cite{Barnes1992ARA&A}. Consequently, intense episodes of star formation in one galaxy can exert a feedback impact on the other galaxy in this pair. This illustrates the interconnected nature of CR feedback effects in galaxy interactions.

 \subsubsection{Cosmic ray pressure support in individual galaxies and the Eddington limit}
\label{sec:cr_pressure}

CR feedback in galaxies can be invoked as an intrinsic or extrinsic mechanism. Intrinsic processes operate in the ISM, often at the molecular cloud level (see e.g. section~\ref{sec:MC_heating}). They rely on CR containment within a galaxy. 
This containment can also form as part of an extrinsic feedback mechanism. 
The pressure support provided by CRs can help maintain the stability of the ISM. If the density of CRs becomes sufficiently high, it can disrupt hydrostatic equilibrium and give rise to a CR-driven wind~\cite{Socrates2008ApJ}. 
The supply of CRs in a galaxy is linked to its star-formation rate. This connection sets a practical upper limit to the star-formation rate known as the CR Eddington limit, analogous to the case with radiation~\cite{Socrates2008ApJ}. 
Recent studies have re-examined this stability limit and found that galaxies with high gas surface densities, exceeding $10^2 - 10^3$ M$_{\odot}$ pc$^{-2}$, and large star formation rates are unlikely to approach the it. This is because the strong hadronic losses experienced by CRs interacting with the interstellar gas make them less dynamically significant~\cite{Crocker2021MNRAS_a, Crocker2021MNRAS_b}. As a consequence, galaxies in this regime become increasingly calorimetric and present higher $\gamma$-ray emission due to pion production~\cite{Crocker2021MNRAS_a}. For surface densities below $\sim 10^2$ M$_{\odot}$ pc$^{-2}$, CRs can be dynamically important. They are capable of launching winds of cool material from galactic discs, thereby curtailing star-formation through catastrophic losses~\cite{Crocker2021MNRAS_b}. 

Typically, quiescent, low surface density galaxies like the Milky Way and local dwarf galaxies reside within the CR-stable regime. However, many of these systems, particularly Milky Way-like galaxies, are on the cusp of instability. CRs offer significant support to the ISM in these cases~\cite{Huang2022MNRAS, Crocker2021MNRAS_b}. Even slight modifications to the configuration of their ISM or the CR pressure can trigger CR-driven outflows, producing strong limits on their star-formation efficiency~\cite{Crocker2021MNRAS_b}. 
Studies have shown that the wind launching mechanism is relatively robust across different CR transport models~\cite{Huang2022MNRAS}, but the specific model adopted can affect the magnitude of the CR Eddington limit. Advection or diffusion-dominated transport generally leads to sub-Eddington galaxies~\cite{Heintz2022ApJ}, with CR streaming representing the most favorable scenario for galaxies to reach the Eddington limit~\cite{Heintz2022ApJ}. 

\subsubsection{Nearby starbursts}
\label{sec:nearby_starbursts}

High-energy $\gamma$-ray emission has been detected in several nearby star-forming galaxies using \textit{Fermi}-LAT~\cite{Abdo2010ApJ, Peng2016ApJ, Ackermann2012ApJ, Ajello2020ApJ, Xi2020ApJ, Xing2023arXiv230400229X}. Subsequent observations with Imaging Atmospheric Cherenkov Telescopes have confirmed the persistence of this emission up to multi-TeV energies in two starburst galaxies: M82 and NGC 253~\cite{Acero2009Sci, VERITAS2009Natur, HESS2018A&ANGC253}.  Future instruments like CTA, with higher sensitivities, are expected to detect other nearby starburst galaxies at similar energies~\cite{Shimono2021MNRAS}.
High-energy $\gamma$-ray emission serves as an indicator of the presence of CRs within these galaxies. Moreover, the observed correlation between this emission and the IR luminosity of galaxies demonstrates a connection between the injection power of CRs and the level of star formation activity in a galaxy (see also section~\ref{sec:CR_calorimetry}). Traditionally, this correlation has been attributed to CR acceleration in core-collapse SNe and their remnants, as the rate of these events closely tracks the star-formation rate in a galaxy~\cite{Bykov2018SSRv}. However, alternative possibilities, including CR acceleration in young stellar clusters and star-forming regions, have gained increasing consideration in recent years (see section~\ref{sec:cr_sources}). 

M82, NGC 253 and Arp 220 are worthy of dedicated discussion. Their close proximity has allowed extensive study to provide thorough insights into their internal physical conditions, including the effects of CRs. All three galaxies host outflows, which have substantial impacts on CR containment and feedback potential. These outflows are multi-phase~\cite{Zhang2014ApJ, Wu2020MNRAS, Lopez2020ApJ, Lopez2023ApJ, Perna2020A&A} and are 
driven by the confluence of feedback winds from ongoing concentrated bursts of nuclear star-formation. 
M82 maintains a star formation rate of approximately 10 M$_{\odot}\;\!{\rm yr}^{-1}$~\cite{Barker2008A&A} within a central region of diameter 0.3 kpc~\cite{Chevalier1985Natur, Volk1996SSRv}. NGC 253 has a lower star-formation rate, around 5 M$_{\odot}\;\!{\rm yr}^{-1}$. This is a residual burst, initiated by the collision with a gas-rich dwarf galaxy approximately 200 Myr ago~\cite{Bolatto2013Natur}. Although less active than M82, approximately 40 per-cent of NGC 253's star formation is concentrated within the central kpc and is associated with a dense CMZ of size approximately 0.8 kpc~\cite{Leroy2015ApJ}. This concentration of star formation drives an outflow, and may play a driving role in baryon recycling flows throughout the galaxy halo~\cite{Mitsuishi2013PASJ}. In contrast to the tidally-triggered nuclear starbursts in M82 and NGC 253, Arp 220's activity is believed to result from a collision between two spiral galaxies a few hundred Myr ago, leading to a more intense system. The majority of the star-forming activity is concentrated within two nuclei, with rates of 65 M$_{\odot}\;\!{\rm yr}^{-1}$ in an Eastern nucleus and 120 M$_{\odot}\;\!{\rm yr}^{-1}$ in a Western circum-nuclear disk~\cite{YH2015MNRAS}. Both nuclei exhibit fast outflows, with slightly higher velocities observed in the Western nucleus~\cite{Barcos2018ApJ, Perna2020A&A}. 

All three of these starbursts are good electron calorimeters. A regime of strong electron confinement and energy retention is established in their cores. Non-synchrotron losses, such as bremsstrahlung and ionization, play a significant role in establishing electron calorimetry in these galaxies, with bremsstrahlung being particularly important above a GeV~\cite{Lacki2013ApJ}. 
The same level of calorimetry is not achieved for CR protons. None of the three galaxies are believed to be fully calorimetric to protons, although all are more so than the Milky Way. 

NGC 253 has an estimated proton calorimetry fraction of a few tens of per-cent at 1 GeV~\cite{Lacki2013ApJ, YH2013ApJ, Krumholz2020MNRAS}\footnote{Higher values are obtained if the possible advective impacts of outflows on CR containment are excluded~\cite[e.g.][]{Lacki2011ApJ}.} If current estimates are accurate, the timescale for CR protons to undergo hadronic interactions is comparable to the advection escape timescale in the galactic wind. This implies that CRs interact with the ISM in the host galaxy near its mean density, and must be far below the level needed to support the galaxy against gravity. It follows that CRs are not dynamically important in driving the NGC 253 outflow~\cite{Lacki2011ApJ}. Despite this, CR feedback has been estimated to strongly impact the nuclear CMZ region of NGC 253 and potentially dominates the thermal balance of the gas in this area~\cite{Behrens2022ApJ}. 

M82 is a better proton calorimeter compared to NGC 253 at most CR energies~\cite{YH2013ApJ, Lacki2011ApJ, Lacki2013ApJ, Krumholz2020MNRAS}, but a large fraction of protons can still escape, particularly at higher energies~\cite{YH2013ApJ, Buckman2020MNRAS, Lacki2013ApJ}. 
 CR pressure is therefore relatively low, and can sustain only small fraction (around 2 per-cent) of that required for hydrostatic equilibrium if CRs interact with ISM at its mean density~\cite{Lacki2011ApJ}. This suggests that CRs are also not dynamically important in M82’s outflow, and CR pressure gradients are weak compared to gravity~\cite{Buckman2020MNRAS}.

Arp 220 is considerably more abundant in CRs. It is also much denser, with the gas around its nuclear starbursts averaging $\sim 10^4$ cm$^{-3}$~\cite{Downes1998ApJ}, compared to $\sim 10^3\;\!{\rm cm}^{-3}$ in M82's core~\cite{Kennicutt1998ApJ, Buckman2020MNRAS}.  
This high gas density creates excellent conditions for the detection of molecular ions chemically related to low-energy CR ionization. These have been used to confirm the 
 importance of CR heating within this galaxy~\cite{GonzAlf2013A&A}. The high gas density also creates short CR loss times to pp interactions. This boosts proton calorimetry, and a substantial fraction of the proton flux is converted to pions. Of the three examples, Arp 220 is the closest to being a proton calorimeter, even up to CR energies of 100 GeV~\cite{Lacki2013ApJ, YH2015MNRAS, Krumholz2020MNRAS}. At 1 GeV, it may achieve in excess of 99 per-cent calorimetry~\cite{Krumholz2020MNRAS}. While this suggests potential for CRs to play an important dynamical role in Arp 220, their exact importance remains uncertain due to the absence of dedicated work. 

Detailed multi-wavelength studies have been conducted for each of these galaxies to investigate their non-thermal properties. 
These studies include one-zone models that put focus on the role of CRs in multi-wavelength emission~\cite[e.g.,][]{Persic2008A&A, Lacki2013ApJ, Paglione2012ApJ, Domingo2005A&A, YH2014ApJ}. These models are often used to fit or constrain the internal physical configuration of the galaxies~\cite[e.g.,][]{Rephaeli2010MNRAS, Heesen2009A&A_I, Heesen2009A&A_II, Heesen2011A&A_III}. Multi-messenger modelling has also been possible~\cite[e.g.][]{Pozo2009ApJ, Lacki2011ApJ, Eichmann2016ApJ, Ha2021ApJ}, and some of these galaxies have even been considered as potential neutrino source candidates for upcoming instruments such as KM3NeT and IceCube-Gen2, albeit with optimistic model parameter choices~\cite[e.g.,][for NGC 253]{Ha2021ApJ}. More sophisticated two-zone models have also been developed~\cite{YH2015MNRAS, YH2017MNRAS, YoastHull2019MNRAS}. 
 They have revealed that the $\gamma$-ray and radio emission may be inconsistent with a purely starburst origin in certain cases, particularly in Arp 220. In this galaxy, a self-consistent solution may require the presence of an AGN in the western nucleus to account for excess $\gamma$-ray flux~\cite{YH2017MNRAS}. 

\subsubsection{Dusty, and infrared and submillimeter luminous galaxies}
\label{sec:ulirgs}

Luminous infrared galaxies (LIRGs) are a class of galaxy characterized by their high rest-frame IR luminosities, above $10^{11}$ L$_{\odot}$. Ultra and Hyper luminous IR galaxies (ULIRGs and HyLIRGs) are particularly luminous sub-classes, above $10^{12}$ L$_{\odot}$ or $10^{13}$ L$_{\odot}$, respectively. These galaxies are believed to represent an evolutionary phase of merging spirals~\cite{Hung2014ApJ, Larson2016ApJ}. They are powered by intense starbursts, with star formation rates surpassing 100 M$_{\odot}$ yr$^{-1}$~\cite[e.g.,][]{Lonsdale2006asup}. 
Their IR emission mainly originates from the reprocessing by dust of strong interstellar radiation fields associated with the rapidly forming stellar population. Additional contributions from AGN are sub-dominant, typically providing no more than 10 per-cent of total IR galaxy luminosities~\cite{PT2021A&ARv}. 

The vibrant star-formation activity of LIRGs establishes them hosts of large reservoirs of CRs. 
This is seen in Arp 220, the closest example of a (U)LIRG, where the abundant CRs have a role in regulating its physical conditions (see section~\ref{sec:nearby_starbursts}). The interactions of these CRs produce $\gamma$-rays and neutrinos~\cite[e.g.][]{Palladino2019JCAP, He2013PhRvD} which contribute to the extragalactic diffuse backgrounds. These backgrounds have been measured with IceCube~\cite{IceCube2020arXiv200109520I, Abbasi2021PhRvD, Abbasi2022ApJb} and \textit{Fermi}-LAT~\cite{Ackermann2016PhRvL_EGB}. 
LIRGs have recently emerged as a viable 
contributor to these backgrounds following a neutrino detection from the LIRG NGC 1068~\cite{IceCube2022Sci...378..538I} (see also section~\ref{sec:agn_jets_outflows}). 
Combined multi-messenger observations of neutrinos and $\gamma$-rays suggest that some neutrino sources may be opaque to $\gamma$-rays~\cite{Bechtol2017ApJ, Murase2016PhRvL}. If true, it would alleviate the observational constraints on source populations that can generate neutrino backgrounds without violating the observed $\gamma$-ray background~\cite{Ackermann2016PhRvL_EGB}. Dusty LIRGs may experience suppressed $\gamma$-ray emission through pair production interactions within their intense IR radiation fields~\cite[e.g.][]{Owen2021MNRAS} or dense gas~\cite{Vereecken2020arXiv200403435V}. They may constitute an important source class capable of conforming to $\gamma$-ray constraints while still supplying significant flux to the diffuse neutrino background. 

Strong IR continuum emission makes LIRGs excellent natural laboratories for the observation of CR feedback effects. In particular, gas phase species produced from CR ionization in molecular clouds can be observed in absorption against this continuum.  
OH$^{+}$ and H$_2$O$^{+}$ are relatively direct chemical tracers (see section~\ref{sec:CR_MC_ints}) that can be used to probe CR ionization in galaxies~\cite{Hollenbach2012ApJ}. 
These tracers have been detected in the nuclear regions of ULIRGs, revealing that strong CR irradiation is driving substantial ionization. Rates of $10^{-13}$ s$^{-1}$ have been measured in some galaxies~\cite{GonzAlf2013A&A, GonzAlf2018ApJ}. This exceeds the canonical value for the Milky Way by around 4 orders of magnitude. As CR ionization is closely related to CR heating power in ISM gas~\cite[e.g.][]{Owen2021ApJ}, these measurements demonstrate the likely importance of CRs in setting the thermodynamics of IR-luminous galaxies. 

Dusty, bright submillimeter galaxies (SMGs) are even more compelling targets for observational studies of CR ionization and feedback. These are among the brightest and most prolific star-forming galaxies in the Universe, and are believed to form an important stage in the evolution of massive elliptical galaxies seen in the local Universe~\cite{Simpson2014ApJ}. SMGs, being rich in CRs, represent a domain where our understanding of galaxy formation and evolution remains uncertain. Obtaining a clear determination of the impact of CR effects on the development of massive elliptical galaxies is therefore important. Similar to LIRGs, SMGs are very dusty~\cite{Dole2004ApJS, Floc2009ApJ}. The far-IR emission from this dust in SMGs produces a strong continuum at submillimeter wavelengths. This serves as an effective back-light for absorption lines originating from molecular ions associated with CR ionization in their ISM. The shape of the spectral energy distribution (SED) of SMGs overcomes cosmological fading. Their apparent luminosity increases with redshift, resulting in a nearly constant apparent flux density at submm wavelengths between $z = 0.5-7$~\cite{Blain2002PhR}. This so-called `negative k-correction' is a clear observational advantage. When coupled with gravitational lensing 
and the excellent sensitivity of state-of-the-art facilities like ALMA, CR feedback in individual galaxies during the cosmic noon and before can be thoroughly explored. 

Ground-based instruments like ALMA are crucial for high-fidelity spatial observations of CR effects throughout galaxy interiors. Although lines from favoured molecular ion species often used to trace CR ionization usually fall outside accessible frequency ranges in the rest frame, cosmological redshift can bring some of them into observable bands~\cite{Indriolo2018ApJ}. 
Indeed, OH$^{+}$ and H$_2$O$^{+}$ have recently been measured in cosmic noon galaxies with ALMA. CR ionization rates of $\zeta = $ 10$^{-16}$ -- 10$^{-15}$ s$^{-1}$ for SDP 17b and 
10$^{-16}$ for the Eyelash galaxy were found~\cite{Indriolo2018ApJ}, indicative of some CR feedback activity. However, there are uncertainties in these values. 
For example, higher estimates, up to $\zeta = $ 10$^{-13}$ -- 10$^{-11}$ s$^{-1}$, have been obtained for the Eyelash galaxy, but these are more model-dependent, based on the estimated galaxy SN event rate~\cite{Danielson2013MNRAS}. Moreover, the CR ionization rates determined by Ref.~\cite{Indriolo2018ApJ} may be associated with more distant, low-density, extended halo gas surrounding these compact SMGs rather than the galaxy itself.\footnote{This is because CH$^{+}$ generally arises in similar conditions regions of high OH$^{+}$ and H$_2$O$^{+}$ abundances in clouds, and it has been proposed that CH$^{+}$ absorption lines from SMGs originate primarily in halo gas~\cite{Falgarone2017Natur}.}
 Further in-depth studies are therfore required to lift some of these 
 ambiguities and 
 obtain a more definitive measurement of CR ionization within SMGs and their halos.  

\subsubsection{Primordial galaxy evolution}
\label{sec:primordial_galaxies}

CRs have been considered as a potential cause for the quiescent behavior and complex star-formation histories observed in some high-redshift galaxies~\cite{Owen2019MNRAS, Owen2019AA}. A specific focus has been on post-starburst galaxies (PSGs), also known as E+A galaxies, which are found at both low~\cite{French2015ApJ, French2018ApJ, Rowlands2015MNRAS, Alatalo2016ApJ} and high redshifts~\cite{Watson2015Natur, Hashimoto2018Natur, Laporte2022arXiv221205072L}.
PSGs show strong Balmer absorption lines in their SEDs, indicating the presence of young stars. However, they lack the optical emission features typically associated with ongoing star formation. Additionally, the detection of metallic absorption lines (e.g., Ca, H, and K) suggests the existence of older stellar populations within these galaxies. 
These observations imply that PSGs have experienced multiple episodes of star formation, with one having occurred relatively recently. 

The star-formation histories of PSGs is perplexing. Their abundance of interstellar molecular gas would normally sustain star formation~\cite{French2015ApJ}. However, these galaxies evolved through multiple quenched epochs suggesting some kind of transient quenching mechanism arose at different points in their history. 
Conventional quenching mechanisms, such as those associated with AGN feedback, do not seem to play a significant role in shaping the evolution of E+A galaxies~\cite{Lanz2022ApJ}. Additionally, the presence of large molecular reservoirs, dust, and a substantial ISM make the evacuation of gas by outflows an unlikely explanation for their behavior~\cite{Rowlands2015MNRAS, Alatalo2016ApJ, French2018ApJ, Smercina2018ApJ}. 

High-redshift PSGs are challenging to observe in detail. However, lower redshift systems 
 are more accessible and have provided certain insights into the causes of their complex star-formation histories. In particular, they have been found to have a deficiency of dense molecular gas pockets~\cite{French2018ApJ}. This suggests 
 that processes capable of providing additional pressure to their molecular gas reservoirs are at play. These could prevent the fragmentation and collapse of ISM and halo gas. CRs have been proposed as a potential agent able to exert this pressure in both the ISM and CGM. They can act directly on semi-ionized magnetized gas~\cite[e.g.,][]{Farcy2022MNRAS}, or they can operate indirectly by heating gas within~\cite[e.g.,][]{Papadopoulos2010ApJ, Papadopoulos2011MNRAS} or  around~\cite{Yokoyama2023MNRAS} galaxies, or by modifying inflows of gas through the CGM~\cite{Owen2019MNRAS, Owen2019AA}. However, to establish the role of CRs conclusively, a distinctive signature of their effects within these systems is required. 

\begin{figure}[H]
    \centering
    \includegraphics[width=0.85\textwidth]{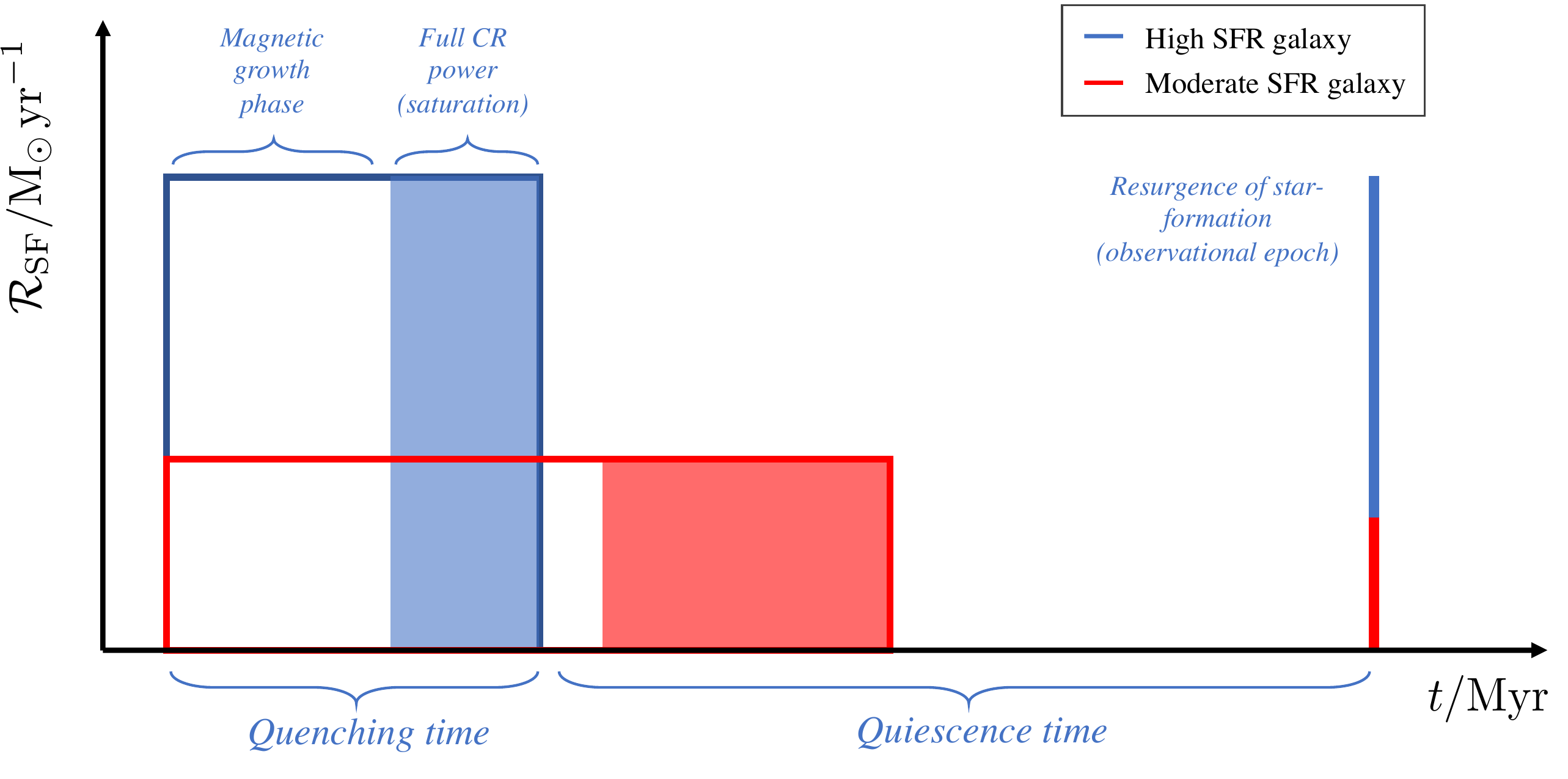}   
        \caption{Illustration of the 
        expected star-formation history of galaxies regulated by CRs. The case with a high star-formation rate shown in blue develops a saturated magnetic field relatively rapidly. Containment of CRs soon brings about the downfall of star-formation in the galaxy, before it later undergoes a resurgence. A galaxy with a lower star-formation rate, shown in red, takes longer to amplify its magnetic field, longer to generate a sufficient abundance of CRs to halt star-formation and sees weaker quenching with a quicker resurgence. Figure reproduced from Ref.~\cite{Owen2022ECRS}.}
\label{fig:burst_sfr}
\end{figure}

Recently, several possibilities have been considered to identify a distinct `smoking gun' signature of CR feedback in PSGs. 
One approach focuses on the progressive and delayed nature of this feedback. If CRs serve as the primary feedback mechanism within a galaxy, their effectiveness relies on the growth and saturation of the galactic magnetic field to levels of a few $\mu$G. This is necessary to contain of CRs and enable a sustained delivery of their feedback power~\cite{Owen2018MNRAS, Owen2019MNRAS}. This produces a progressive feedback effect, characterized by a delayed onset. 
In a population of PSGs, this delayed feedback would be identified from a proportional relationship between the level of star-forming activity during a starburst phase and the duration of a subsequent quenched phase. Additionally, an inverse correlation would be expected between the star-formation rate during the burst phase and the duration of the burst itself (see Fig.\ref{fig:burst_sfr} and Refs.\cite{Owen2019AA, Owen2022ECRS}). By conducting population studies of multiple PSGs experiencing CR regulation, these trends would emerge with minimal scatter to reveal a signature of CR feedback. In contrast, more stochastic feedback mechanisms like energetic hypernova events deliver feedback abruptly and randomly, resulting in widely scattered timescales that exhibit a less pronounced inherent dependency on galaxy properties~\cite{Owen2022ECRS}. 

\subsection{Cosmic ray effects in circum-galactic media}
\label{sec:cr_in_cgm}

\subsubsection{Phase structure of circum-galactic media and cosmic ray effects}
\label{sec:cgm_phase_structure_cr}

\begin{figure}[H] 
\begin{center}
\includegraphics[width=12 cm]{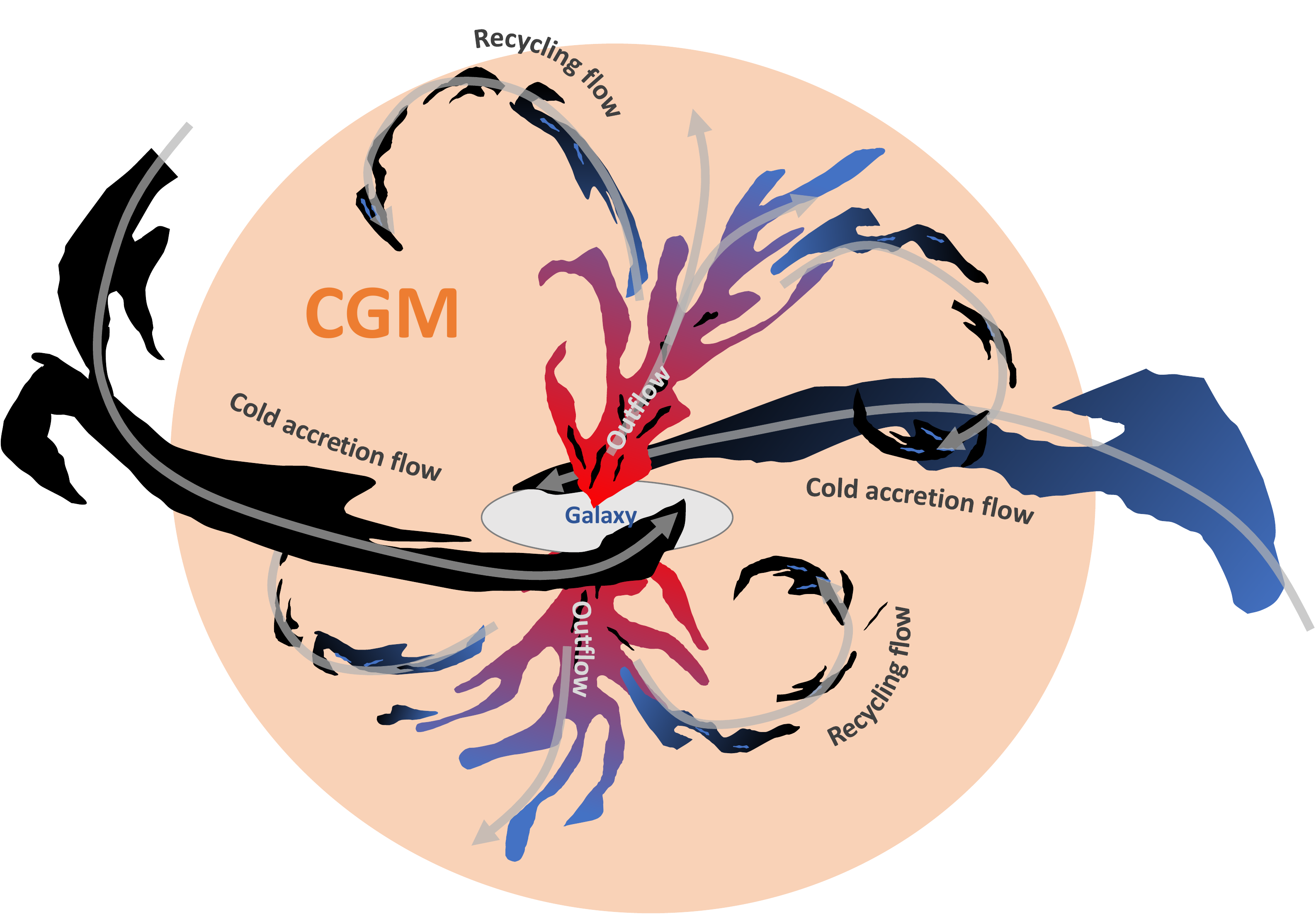} 
\end{center}
\caption{A schematic illustration 
  of a secular galaxy 
  and its circum-galactic environment. 
  The CGM is shown in orange, with flows of gas in red (hot), blue (warm), and black (cold). Typical flow structures circulating baryons, metals, magnetic fields and CRs between the CGM and ISM of a galaxy are indicated. Outflows can be driven by the feedback activity resulting from concentrated nuclear star formation. Cold accretion flows of pristine material from the cosmic web can operate in more primordial settings above $z\sim 2$. These flows may fuel intensive starburst activity. Recycling flows, which consist of multiple phases, facilitate the transfer of gas from the ISM to the CGM, enriching the medium, promoting cooling, and inducing inflow. CRs can modify these processes by providing pressure support, by driving/modifying flows, or by heating CGM gas to raise its thermal support and stability against gravitational collapse.}
 \label{fig:cgm_flows}
\end{figure}   

The CGM is a major component of a galaxy's ecosystem. As a conduit for all gas flows into and out of a galaxy, it can have a strong influence on a galaxy's evolution (for reviews, see~\cite{Tumlinson2017ARAA, Putman2012ARA&A, FaucherGiguere2023arXiv}). In particular, it regulates a galaxy's fuel supply, star formation capacity, and the hydrodynamic state of the gas that ultimately reaches a galaxy's ISM. 
In systems dominated by thermal pressure, the CGM consists mainly of a hot, tenuous gas 
phase with cool gas confined to dense filaments in local pressure equilibrium~\cite{Ji2020MNRASd}.
However, some galaxy-scale simulations including the effects of CRs 
have shown that they may dominate the pressure in the CGM surrounding certain galaxies~\cite{Girichidis2018MNRAS, Ji2020MNRASd}. This opens up the potential for CRs to participate in extrinsic feedback mechanisms which operate on the large-scale processes externally regulating the evolution of a galaxy. 
In cases where CR pressure dominates a galactic halo, CGM baryons primarily exist in a cold phase at approximately $10^4$ K~\cite{Werk2014ApJ, Ji2020MNRASd}. Observations have detected this cold gas in the CGM of galaxies of all types, extending up to around 300 kpc~\cite{Chen2010ApJ, Prochaska2011ApJ, Tumlinson2013ApJ, Keeney2018ApJS}. This is comparable to the full extent of the CGM (traced up to 100-200 kpc using metal-enriched gas~\cite{Fox2017ASSL}).

Some galaxies are observed to sustain massive reservoirs of cold gas in their CGM. This should be able to fuel their star formation. However some of these galaxies appear to be quenched~\cite[e.g.][]{Thom2012ApJ, Berg2019ApJ}. This is indicative of a feedback support mechanism operating to reduce gas in-fall, precipitation, or accretion into the host galaxy to moderate its star formation. 
 This support may be provided by thermal pressure exerted by the hot phase of the CGM (which may be partially heated by CRs through the excitation of short-wavelength Alfv\'{e}n waves~\cite{Wiener2013ApJ}), and/or non-thermal pressure contributed by CRs. The presence of CRs can significantly reduce cold gas supplied to the galaxy. They can also increase the fraction of cold gas mass held in the galactic halo by preventing its precipitation back toward the galaxy~\cite{Butsky2020ApJ}. These effects and their 
 dependence on CR activity offer an opportunity to establish constraints on effective CR transport in the CGM based on parameters such as the total hydrogen column density, average star formation rate, and gas circular velocity~\cite{Butsky2023MNRAS}.

\subsubsection{Cold gas formation from thermal instabilities}
\label{sec:cold_gas_CGM}

The partition of matter between the hot and cold CGM phase components is 
set by the local thermodynamic conditions and the supply of gas. 
The cold phase has multiple possible origins.  One possibility is its direct formation from the hot tenuous phase. In this scenario,  thermal instabilities give rise to runaway cooling and cold gas condensation~\cite{Field1965ApJ, Putman2003ApJ, McCourt2012MNRAS, Sharma2012MNRAS, Voit2015Natur, Wu2020MNRAS}. CRs have been shown to modify this process~\cite{Butsky2020ApJ}. For example, 
CR pressure 
supports more diffuse gas at lower temperatures~\cite{Salem2016MNRAS, Ruszkowski2017ApJ, Butsky2018ApJ} allowing cooling without collapse~\cite{Sharma2010ApJ, Kempski2020MNRAS} and the formation of a cool, low-density CGM component~\cite{Butsky2020ApJ}.
This cold gas component may form as a suspension of small cloudlets, mixed into the warmer phase. Inefficient CR transport can promote the development of larger cloudlets. These can be several orders of magnitude larger in size compared to cloudlets in a purely thermal CGM~\cite{Butsky2020ApJ}. 
When CRs dominate the pressure of a CGM, the halo is comprised of mainly diffuse cool gas ($\sim 10^4$ K), with a large filling factor and a thermal pressure that is insufficient for virial or local pressure balance~\cite{Ji2020MNRASd}. Such conditions have been shown to be more prevalent for larger halo masses and lower redshifts~\cite{Hopkins2020MNRASa}. In these cases, the gas phase configuration and hydrodynamics differ substantially from more multi-phase CGM compositions~\cite{Ji2020MNRASd}. In particular, the additional CR pressure leads to a smoother CGM on small scales, with lower temperatures~\cite{Buck2020MNRAS}.

\subsubsection{Cold gas supply by inflows and the impacts of preventative feedback}
\label{sec:cgm_inflows}

In addition to its formation by thermal instabilities within the CGM, 
cold gas can be supplied to a galactic ecosystem by accretion from the cosmic web (see~\cite{Tumlinson2017ARAA}; also Fig.~\ref{fig:cgm_flows}). At high redshifts, typically above $z\sim 2$, this happens by means of cold accretion flows~\cite{Dekel2006MNRAS, Dekel2009Natur} which pass through the CGM and can directly fuel efficient star formation in the host galaxy~\cite{Keres2005MNRAS, Keres2009MNRAS}.\footnote{Recent studies have found tentative indications that gas inflows may persist around some high-mass galaxies in the nearby Universe~\cite{RobertsBorsani2019MNRAS}.} These flows can fragment as they fall through the CGM towards the host galaxy~\cite{Ceverino2010MNRAS, Emonts2023arXiv230317484E}, or may be reinforced by condensation due to CGM gas cooling~\cite{Mandelker2020MNRAS}. 
Recent observations have found evidence of cold accretion flows with Hydrogen recombination lines~\cite{Martin2015Natur, Daddi2021A&A}. These lines are emitted by warm ($\gtrsim 10^{4}\;\!{\rm K}$) Hydrogen gas cooling as it flows towards the host galaxy~\cite[e.g.][]{Dijkstra2009MNRAS, Rosdahl2012MNRAS}. CI emission has also revealed in-flowing CGM gas around galaxies at much lower temperatures of 10-100 K~\cite{Emonts2023arXiv230317484E}.

Strong feedback mechanisms, such as the confluence of winds from stellar populations~\cite{Pandya2020ApJ}, high-energy processes associated with CRs in the host galaxy~\cite{Owen2019AA}, or outflows (see~\cite{Nelson2015MNRAS}; also section~\ref{sec:outflows_cgm}), can impede the ability of cold accretion flows to penetrate through the CGM. Outflows with large filling factors can introduce particularly severe disruption. These are expected to be especially widespread in CR pressure-dominated systems~\cite{Hopkins2020MNRASa, Ji2020MNRASd}. 
When 
  cold accretion flows are restricted, or even stopped, a 
so-called \textit{preventative} feedback scenario is established. 
 Gas recycling in the CGM and the supply of pristine gas to the host galaxy is strongly curtailed. This preventative scenario can be regulated and enhanced by CRs, leading to considerably extended gas recycling times~\cite{Hopkins2021MNRASb}.\footnote{Similar effects have been reported in other contexts. For example, even modest CR pressures can suppress cooling flows in galaxy clusters~\cite{Su2020MNRAS}.} 
When a preventative feedback scenario has taken hold,  
  gas may instead be supplied to build up the outer CGM. There, it can form a reservoir to 
 fuel star formation at a later time. Stunted cosmic gas supply by preventative feedback has been 
 considered to lead to the emergence of ‘red-and-dead’ massive spheroid galaxies below $z<1$~\cite{Dekel2006MNRAS}. 
 These galaxies undergo a gradual quenching process
 as they deplete their remaining gas reservoir over Gyr timescales~\cite[e.g.][]{Schawinski2014MNRAS}. 
In high redshift settings, transient preventative feedback by CRs has been proposed as a driver for  
more complex star formation histories~\cite{Owen2019MNRAS, Owen2019AA} inferred for candidate $z>7$ galaxies in recent observational studies~\cite[e.g.][]{Hashimoto2018Natur, Laporte2021MNRAS, Laporte2022arXiv221205072L}.

\subsubsection{Cold gas supplied from the interstellar medium}
\label{sec:cold_gas_ISM}

Cold gas within the CGM can also have its origins within the ISM of the host galaxy. 
It is advected into the CGM by multi-phase galactic outflow winds and fountains, which can drive gas far into the CGM.  
A cold phase can develop in 
this advected ISM gas through various mechanisms. These include thermal 
instabilities~\cite{Gronke2018MNRAS, Huang2022ApJ}, 
dynamical instabilities~\cite[e.g.][]{Fujita2009ApJ}
or the entrainment of cold clumps or clouds from the ISM that are accelerated within the outflow~\cite{Cooper2007Ap&SS}. Outflows can be driven by CR pressure, thermal gas pressure or radiation pressure (see Ref.~\cite{Zhang2018Galax} for a review). 
The interaction between these driving mechanisms and entrained cold clouds leads to distinct multi-phase configurations and kinematics within the outflow. 
For example, CR-driven flows tend to be cooler than their thermal-pressure driven counterparts, and 
faster than radiatively-driven winds~\cite[e.g.][]{Yu2020MNRAS}. 
Cold clumps within CR-driven flows are subject to significant CR pressure support. This can favour the development of larger cold clumps~\cite{Gronke2018MNRAS, Butsky2020ApJ, Huang2022ApJ} or a lower density cold phase in the flow~\cite{Wiener2017MNRAS, Wiener2019MNRAS, Huang2022ApJ} than would be expected in a thermally-driven or radiatively-driven system~\cite[see][for a study of cloud properties in thermal and radiation-driven outflows]{Murray2011ApJ}. 

Recent simulation work has set out a picture 
of the acceleration, survival and evolution of cold clouds in galactic outflows, as well as the effects brought about by CRs. In particular, it has been established that 
cold gas clouds can be accelerated by CR streaming
through the CR 
`bottleneck' effect in outflows~\cite{Wiener2019MNRAS, Bruggen2020ApJ, Bustard2021ApJ, Huang2022ApJ} and galactic halos~\cite{Wiener2017MNRAS}. This is
where CRs are forced to slow down when they encounter a cloud, as they cannot stream down their density gradient. 
CR density is enhanced on the side of the cloud facing the CR flux, exerting pressure forces and heating it. This accelerates the cloud, and can affect its structure, e.g. by stretching it~\cite{Wiener2019MNRAS}. 
Moreover, its heating impact on CR-mediated acceleration fronts broadens the cloud-halo interface, 
resulting in detectable changes in ionic abundances~\cite{Wiener2017MNRAS}. 

The CR bottleneck effect preferentially accelerates the side of the cloud experiencing the strongest CR pressure gradient. This is usually facing the CR source~\cite{Huang2022ApJ, Bustard2021ApJ}. The resulting differential acceleration causes the fast-moving gas to detach from the bulk of the cloud to form small clouds. These are less likely to survive the acceleration process~\cite[e.g.][]{Gronke2022MNRAS}.\footnote{Magnetic draping (see~\cite{Lyutikov2006MNRAS}) can can partially mitigate cloud mass loss~\cite{Li2020MNRAS}, and the inclusion of radiative cooling has been shown to enhance the resilience of clouds against CR effects, such as heating~\cite{Wiener2019MNRAS}.}
While this means cloud morphologies will be different in a CR-driven flow compared to a thermally-driven system, this process implies that clumps entrained in a CR-driven wind will also experience a loss of gas mass over time. This is in contrast to clouds entrained in a hot thermally-driven wind, which typically gain mass during their entrainment~\cite{Huang2022ApJ}.

\subsubsection{Cosmic ray impacts on galactic outflow physics}
\label{sec:outflows_cgm}

CRs can play a key role in driving galactic outflows~\cite{Breitschwerdt1991AAP, Uhlig2012MNRAS, Booth2013ApJ, Salem2014MNRAS, Ruszkowski2017ApJ, Wiener2017MNRASb, Chan2019MNRASd, Yu2020MNRAS, Samui2010MNRAS}. Their influence leads to modifications in the flow dynamics and physical properties~\cite{Farber2018ApJ, Holguin2019MNRAS, Yu2020MNRAS, Armillotta2022ApJ, Girichidis2018MNRAS} which 
 can be observed in various ways, including with X-rays~\cite{Yu2021MNRAS} or with Lyman-$\alpha$ spectra~\cite{Gronke2018ApJ}. 
Recent numerical studies have extensively explored CR-driven winds.  These have considered steady-state and time-evolving numerical models, covering a wide range of boundary parameters and physical complexity. 
 Examples include variation of the 
 gas conditions at the base of the flow~\cite[e.g.][]{Mao2018ApJ, Recchia2017MNRAS, Armillotta2022ApJ}, star formation rate in the host galaxy~\cite[e.g.][]{Samui2010MNRAS, Yu2020MNRAS, Chan2019MNRASd, Armillotta2022ApJ}, mass input/loading rates into the flow~\cite{Fujita2018MNRAS, Samui2010MNRAS}, gravitational potential~\cite[e.g.][]{Mao2018ApJ, Yu2020MNRAS, Recchia2017MNRAS, Samui2010MNRAS, Armillotta2022ApJ} or halo mass~\cite{Jacob2018MNRAS}, magnetic and/or gas–to–CR pressure driving ratio~\cite[e.g.][]{Mao2018ApJ, Yu2020MNRAS, Fujita2018MNRAS, Samui2010MNRAS}, and flow angular momentum~\cite[e.g.][]{Mao2018ApJ, Peschken2021MNRAS, Peschken2022arXiv221017328P}. 
 The impacts of the 
 precise CR transport physics has also 
 seen substantial progress, with studies investigating the effects of CR advection and diffusion~\cite{Ruszkowski2017ApJ, Farber2018ApJ, Fujita2018MNRAS, Quataert2022MNRAS, Peschken2021MNRAS, Peschken2022arXiv221017328P, Chan2019MNRASd}, diffusion and streaming~\cite{Wiener2017MNRASb,  Hopkins2021MNRASc, Chan2019MNRASd, Thomas2023MNRAS},\footnote{Note, however that Ref.~\cite{Thomas2023MNRAS} reported CR transport itself cannot reach a steady state and is not well described by either the CR streaming paradigm, the CR diffusion paradigm, or a combination of both.} impacts of driving by CR streaming~\cite{Quataert2022MNRASb, Holguin2019MNRAS, Bai2022ApJ, Recchia2016MNRAS}, streaming suppression due to turbulent and/or wave damping~\cite{Holguin2019MNRAS, Ko2021A&A, Bai2022ApJ, Recchia2016MNRAS}, CR cooling~\cite[e.g.][]{Ko2021A&A}, gas cooling with CR heating~\cite[e.g.][]{Modak2023arXiv230203701M}, and consideration of spectrally resolved treatments of CRs~\cite{Yang2017ApJ, Recchia2016MNRAS, Recchia2017MNRAS, Hopkins2022MNRASa, Girichidis2022MNRASb, Girichidis2023MNRAS_err}. Multi-phase gas flow structure~\cite{Peschken2022arXiv221017328P} and its impact on CR transport~\cite{Farber2018ApJ} and CR acceleration~\cite[e.g.][]{Muller2020MNRAS, Peretti2022MNRAS} have also been investigated. 
 Specific parameter sets to simulate specific classes of galaxies~\cite[e.g.][]{Fujita2018MNRAS} and model fits to individual galaxies (e.g. the M82 superwind) from $\gamma$-ray and radio emission~\cite{YH2013ApJ, Buckman2020MNRAS}\footnote{Constraints on the the underlying wind physics have found that the role of CRs in the case of the M82 superwind is relatively limited~\cite{Buckman2020MNRAS}.} have also been provided. Additionally, 
 simple analytic scaling relations between halo velocity, carrying capacity and mass loading have been obtained~\cite{Mao2018ApJ}. These capture the broad variety of CR-driven and hybrid flow behaviours that are reflective of earlier scaling relations obtained empirically for mass loading factor, thermalization efficiency and flow velocity~\cite[e.g.][]{Chisholm2015ApJ, Heckman2016ApJ, Chisholm2017MNRAS}.

\subsubsection{Cosmic ray effects on circum-galactic baryon recycling flows}
\label{sec:cgm_cr_recycling_flows}

  Outflows play a direct role in the exchange of energy and baryons between a galaxy and its CGM~\cite{Oppenheimer2010MNRAS, Tumlinson2017ARAA}. They propel enriched gas into the CGM and beyond, some of which can return to the galaxy through baryonic recycling flows~\cite{Bertone2007MNRAS, Marinacci2011MNRAS, Zhang2023arXiv230502344Z}. Alternatively, this gas can be completely expelled from the galactic ecosystem, contaminating the surrounding IGM with metals~\cite{Cen2006ApJ, Nelson2018MNRAS} and magnetic fields~\cite{Bertone2006MNRAS, Donnert2009MNRAS, AramburoGarcia2021MNRAS}. Even in cases where gas eventually recycles back to the host galaxy, it can reside in the CGM reservoir for extended periods, possibly exceeding 1 Gyr, especially in the case of very massive galaxies~\cite[e.g.][]{Oppenheimer2008MNRAS, Oppenheimer2010MNRAS, Angles2017MNRAS}.\footnote{Shorter median recycling timescales around 100s Myr have been reported by some studies~\cite[e.g.][]{Angles2017MNRAS}.} 
Despite these long timescales, simulations have demonstrated that a significant fraction, up to half~\cite{Mitchell2021MNRAS} or more~\cite[e.g.][]{Oppenheimer2010MNRAS, Christensen2016ApJ, Angles2017MNRAS}, of the baryons in the CGM within a virial radius pass through the ISM of their host galaxy at least once by $z=0$.  
The influence of CR pressure can modify this recycling process significantly, leading to pronounced changes in outflow morphology~\cite[e.g.][]{Girichidis2016ApJ, Jana2020MNRAS}, and a diminished~\cite{Hopkins2021MNRASb} or even completely suppressed~\cite{Ji2021MNRAS} virial shock. This results in 
substantial alterations to the kinematics of both warm and cool CGM gas~\cite{Butsky2022ApJ, Chan2022MNRAS}. 
CR-driven outflows tend to exhibit cooler temperatures and smoother density structures~\cite{Salem2016MNRAS, Girichidis2018MNRAS, Farcy2022MNRAS}, extending far into the CGM~\cite{Ipavich1975ApJ, Uhlig2012MNRAS, Booth2013ApJ, Pakmor2016ApJ, Girichidis2016ApJ, Simpson2016ApJ, Ruszkowski2017ApJ, Bustard2020ApJ, Jana2020MNRAS, Hopkins2021MNRASa, Chan2022MNRAS}. It is this extended reach which gives them particularly influence over the regulation of 
CGM recycling flows.

 The effect of CRs on CGM recycling can be 
especially consequential in massive halos (above $10^{11}\;\!{\rm M}_{\odot}$) and at low redshifts ($z\leq 1-2$)~\cite{Hopkins2021MNRASa}. CR effects have also been examined in relation to regulating the baryonic content of dwarf galaxies~\cite{Dashyan2020A&A}, where CR diffusion has been found to yield relatively cold, dense winds. 
In the absence of CRs, the high thermal pressure of gas typically traps galactic outflows near the disk of their host galaxy~\cite{Hopkins2021MNRASb}. This forces the development of low-altitude recycling flows and practically mimics a scenario where an outflow is not present at all~\cite{Tumlinson2017ARAA}. 
The presence of CRs leads to 
the formation of a halo with lower thermal pressure, facilitating outflow escape (the CR pressure does not resist outflow expansion due to diffusion).
The continuous flow acceleration provided by CR pressure propels material deep into the CGM, spanning Mpc scales~\cite{Hopkins2020MNRASa}. This reduced confinement of material has been demonstrated to diminish flow recycling in some simulations~\cite[e.g.][]{Hopkins2021MNRASb} with   
significant long-term consequences for the evolution of the host galaxy. Higher CR fluxes counter this effect by transferring more energy to the gas, driving stronger outflows~\cite{Huang2022MNRAS} with 
radically altered morphologies. In strongly CR-dominated halos, outflows become nearly volume-filing and are predominantly comprised of cool gas ($T \sim 10^5$ K)~\cite{Hopkins2020MNRASa, Ji2020MNRASd}. This leads to considerable modification of the dynamical structure of the CGM~\cite{Hopkins2021MNRASb}. 
In multi-phase flows, the CR pressure support within cold clouds reduces their density and alters their kinematics~\cite{Butsky2020ApJ}. The associated change in their buoyancy could have considerable consequences for the enrichment and star formation history of galaxies in CR-dominated ecosystems. 

\subsubsection{Cosmic ray micro-physics on circum-galactic and galactic scales}
\label{sec:cr_microphysics_cgm}

 Many of the large-scale effects of CRs in the CGM 
 are greatly  
influenced by the underlying models used to describe CR propagation and interactions on very small scales, called the \textit{micro-physics}. 
These models are subject to substantial uncertainty and are poorly constrained. This can lead to 
 large variations in predicted galaxy properties, 
 which are even more severe in the CGM~\cite{Butsky2018ApJ, Buck2020MNRAS}. 
Even 
 relatively small differences in CR transport treatments
 or  
 numerical approaches~\cite{Gupta2021MNRAS, Semenov2022ApJS} can produce tremendous variation in simulated galaxies and their CGM~\cite{Butsky2018ApJ, Hopkins2021MNRASc}, and circum-galactic flows~\cite[e.g.][]{Salem2014MNRAS}. 
 Discrepancies also arise when comparing CR propagation models with observed CR scaling relations in the solar-terrestrial environment, such as spectral slopes and the Boron-to-Carbon (B/C) ratio. It is therefore clear that the reliability of most current treatments of CR transport micro-physics is limited~\cite[in particular, see][]{Hopkins2022MNRASb, Kempski2022MNRAS}. 

Observational data can provide some constraints on the CR transport models. However, suitable data for the transport of  relatively low-energy GeV CRs is limited. These CRs typically scatter over gyro-radii of 
$r_{gy}  \approx 0.22 \;\!\left({E_p}/{1\;\!\rm{GeV}}\right) \left({B}/{1\;\!\mu {\rm G}}\right)^{-1}$ au in the magnetic fields associated with galactic ecosystems. This gyro-radius serves as a reference length-scale, indicating where CR transport becomes diffusive (i.e. where the gyro-radius becomes comparable to the local coherence length of magnetic turbulence). Practically, it defines the scale where CR micro-physics must be resolved in order to make reliable macroscopic predictions about their transport. 
Achieving such gyro-scale resolutions in galaxy simulations or observations is 
  extremely challenging. This poses a significant obstacle to the reliable construction and testing of CR transport models derived self-consistently from micro-scale CR processes. The resulting theoretical and observational uncertainties are substantial, and only weak constraints are possible. Even when adopting constraints 
  from Milky Way observations such as diffusion coefficients adjusted to match $\gamma$-ray observations~\cite{Ji2020MNRASd, Chan2019MNRASd}, variations in micro-physical modeling have still 
  demonstrated vast differences in macro-physical predictions~\cite[e.g.][]{Chan2019MNRASd}.

In recent years, significant effort has been devoted to thoroughly exploring the impacts of different model prescriptions and uncertainties on galaxy-scale predictions of CR effects. Work has been conducted to critically examine the impacts of a wide range of standard approximations and physical treatments, including spectrally-resolved CRs~\cite{Yang2017ApJ, Hopkins2022MNRASa, Girichidis2022MNRASb, Girichidis2023MNRAS_err, Girichidis2023arXiv230303417G, Werhahn2023arXiv230104163W}, anisotropic diffusion in CR-driven wind models~\cite{Pakmor2016ApJ, Ruszkowski2017ApJ, Jacob2018MNRAS}, moderately super-Alfv\'{e}nic streaming~\cite{Ruszkowski2017ApJ, Holguin2019MNRAS}\footnote{For a review of Alfv\'{e}n wave damping in MHD turbulence for CR streaming in galactic winds, see Ref.~\cite{Lazarian2022FrP}.}, anisotropic streaming \cite{Ji2020MNRASd}, CR transport with self-confinement or extrinsic turbulence~\cite{Zweibel2013PhPl, Zweibel2017PhPl, Hopkins2021MNRAS}, constant diffusivity, explicitly evolved CR diffusivities, and varying turbulent cascade assumptions~\cite{Hopkins2021MNRAS}. 
Although specific model configurations and parameter choices have been identified where all observational constraints can be reproduced~\cite[in particular, see][]{Hopkins2021MNRAS}, 
these studies suggest that no single approach can be universally adopted for a complete treatment of CR transport while properly capturing all relevant micro-physical effects. 

Despite this, progress has still been possible. Much of this has focused on connecting 
micro-scale CR transport to intermediate-scale effective fluid theories~\cite[e.g.][]{Lazarian2016ApJ, Bai2019ApJ, Holcomb2019ApJ, vanMarle2019MNRAS}. These efforts are providing more robust ways of treating effective CR transport as a function of local plasma properties~\cite{Zweibel2017PhPl, Thomas2019MNRAS}. 
 Moreover, galaxy simulation work has started to reach pc-scale resolution~\cite[e.g.][]{Hopkins2020MNRASa}. 
 This can capture 
 environmental variations through interstellar and circum-galactic media which 
 affect CR transport. 
 It can also reach the deflection length-scales of GeV CRs for observationally-favoured values of the CR scattering rate~\cite{Hopkins2021MNRAS} so CR trajectories can be followed over the structural scales of their medium.
  While a full model of 
 au-scale CR propagation is not yet within reach, 
 current capabilities are now sufficient to construct reliable effective CR transport prescriptions as a function of local physical conditions.  Theoretical studies are now beginning to propose simplified relations that are
physically robust on galaxy or CGM scales and computationally efficient to adopt~\cite{Hopkins2023MNRAS}. 
 These efforts open up the potential for tangible advancements in the development of self-consistent effective transport models on galactic scales.

%%%%%%%%%%%%%%%%%%%%%%%%%%%%%%%%%%%%%%%%%%
\newpage 
\section{Summary and conclusion}
\label{sec:conclusions}

In this \textit{Review}, we have 
provided a cross-section of recent advancements in our understanding of CR processes within galactic environments.    
 By adopting a multi-scale perspective, we have 
explored the significance of CRs in galaxies, with particular focus on their origins, containment, feedback impacts and observable signatures. 
To conclude, we have identified several areas where 
significant challenges exist that will be important to overcome if we are to significantly advance our understanding of CR feedback effects. We also highlight upcoming observational opportunities and emerging theoretical advancements that show promise to support progress in addressing these challenges in the near term. These opportunities and advancements have the capability to elevate our understanding of CR processes in galaxies beyond its current level of maturity, and will likely guide future research directions in the coming years. 

\subsection{Pressing issues}
\label{sec:pressing_issues}

As we seek to consolidate our understanding of the role of CRs in galaxy formation and evolution, certain areas stand-out as focal points requiring particular attention. These represent emerging priorities within the field, where advancements would considerably aid progress in the construction and testing of the next generation of models. 
Addressing these priorities will serve as important milestones in our path forward:

\begin{itemize}
    \item Development of robust connections between multi-wavelength/multi-messenger observables and models of CR propagation and interaction, to support efficient testing of models with the wealth of upcoming new data. 
    \item Establish ways to address the significant numerical challenges involved with  studying the effects of local plasma variations on CR instability growth rates, CR-MHD wave scattering and interaction rates, MHD wave damping, and micro-physical CR transport prescriptions. 
    \item Development of a self-consistent CR transport theory, including self-confinement effects, that aligns with observations. 
    \item Enhancement of CR+MHD numerical simulations to incorporate physically robust CR interaction and transport physics on galactic scales, that correctly account for physics at micro-scales. 
    \item Creation of a comprehensive suite of reliable CR transport theories applicable to galaxies across a wide range of scales and conditions, particularly in the vicinity of CR sources. 
    \item Construction of efficient models of CR interactions and propagation within multi-phase media, including `bottleneck' effects around dense clouds, suitable for integration into MHD simulations. 
    \item Advancement of our understanding of wind-driving effects of CRs, including their interaction with MHD waves undergoing damping, within multi-phase fluid flows, and self-consistent coupling with existing treatments of radiation hydrodynamics. 
    \item Establish a comprehensive understanding of the thermal and dynamical impacts of CR heating and pressure support in the CGM, including their effects on inflows and outflows. 
\end{itemize}

\subsection{Upcoming opportunities}
\label{sec:upcoming_opportunities}

The advancing capabilities of ground-based $\gamma$-ray telescope arrays
will soon allow new science to be conducted at the highest of photon energies with unprecedented angular resolution and sensitivity. 
The principal development in this domain is CTA, 
which is expected to facilitate substantial scientific advancements~\cite[see][]{CTA2019book}. Potential up-coming Southern observatories using the water Cherenkov detector technique, such as SWGO, will provide complementary capabilities such as a first unbiased survey more sensitive to PeVatrons in the Southern hemisphere~\cite{Albert2019arXiv190208429A, Huentemeyer2019BAAS}. 
Up-coming developments in the MeV band are also noteworthy. The Compton Spectrometer and Imager (COSI)~\cite{Tomsick2019BAAS, Tomsick2022icrc} will be of particular importance in the coming decade, as it will open-up the possibility to conduct pioneering studies of $\gamma$-ray polarization, as well as presenting significant improvements in sensitivity, spectral resolution, angular resolution, and sky coverage. COSI will particularly serve as a crucial tool for testing models of CR acceleration in nearby environments. It establishes a more complete picture of CR feedback in galaxies, by offering excellent continuum sensitivity that bridges the gap between the thermal and non-thermal regimes.

The next generation of high-energy instruments will also open-up the multi-messenger domain, principally through advancements in neutrino observations. 
The development of new facilities scheduled to be fully operational in the next decade, including KM3NeT~\citep{Adrian-Martinez16JPG}, Baikal-GVD~\citep{Agostini20NatAst}, IceCube-Gen2~\cite{Aartsen2021JPhG}, and P-ONE~\citep{Baikal19arxiv}, will soon make the exploration of galaxies 
with multi-messengers a reality. 
In addition to this, recently established new facilities form part of an armada of observatories operating across the electromagnetic spectrum. Chief among these are ALMA and \textit{JWST}. These are already allowing us to study external galaxies in unprecedented detail, and have opened-up new ways to pin-down the effects of CRs throughout the hierarchy of structures of galaxy media.

Theoretical and numerical developments form 
another avenue where significant progress is being made, paving the way for new scientific capability. 
 Some especially noteworthy advancements are the emergence of simplified treatments of CR physics that are
physically robust on galaxy or CGM scales and computationally efficient to adopt~\cite[e.g.][]{Hopkins2023MNRAS}, and numerical tools that 
tackle CR transport physics and observational signatures simultaneously~\cite[e.g.][]{Krumholz2022MNRAS}.  
There has also been considerable advancement of suites of simulations that are adopting more sophisticated physical recipes~\cite[for a recent review, see][]{Hanasz2021LRCA}, and moving beyond standard CR+MHD treatments.  
This emerging broader framework encompasses a wider range of physical effects relevant to CRs in galaxies. For instance, spectrally-resolved CRs~\cite[e.g.][]{Yang2017ApJ, Hopkins2022MNRASa, Girichidis2022MNRASb, Girichidis2023MNRAS_err, Girichidis2023arXiv230303417G, Werhahn2023arXiv230104163W}, alternative mechanisms for CR self-confinement and heating of thermal gas (such as pressure anisotropy instability)~\cite{Zweibel2020ApJ},\footnote{This development may be especially important on meso-scales, particularly near sites of CR injection~\cite{Zweibel2020ApJ}.} and numerical simulations of MHD-dust-CR interactions, where the charged dust and CR gyro-radii on au scales are fully resolved~\cite{Ji2022MNRAS}. 

With these ongoing theoretical advancements and the upcoming observational opportunities, we will 
soon be well-placed to confidently investigate the role of CRs in galaxies 
as active agents involved in shaping their evolution. 
This will allow us to refine our treatment of CR feedback dynamics in our understanding of galaxy evolution, bridging the gap from microscopic to macroscopic scales.

%%%%%%%%%%%%%%%%%%%%%%%%%%%%%%%%%%%%%%%%%%
\vspace{6pt} 

%%%%%%%%%%%%%%%%%%%%%%%%%%%%%%%%%%%%%%%%%%
\authorcontributions{E.R.O., K.W. and Y.I. wrote the draft of the paper. H.-Y.K.Y and A.M.W.M. improved the manuscript by contributing text and providing insightful comments.}

\funding{E.R.O. is an overseas researcher under the Postdoctoral Fellowship of Japan Society for the Promotion
of Science (JSPS), supported by JSPS KAKENHI Grant Number JP22F22327. 
 E.R.O. also acknowledges support from the Munich Institute for Astro-, Particle and BioPhysics (MIAPbP), where where some of this work was conducted and 
 where 
 insightful discussions at the "Star-Forming Clumps and Clustered Starbursts across Cosmic Time" program informed the early stages of this work. MIAPbP is funded by the Deutsche Forschungsgemeinschaft (DFG, German Research Foundation) under Germany's Excellence Strategy – EXC-2094 – 390783311. 
K.W. was supported in part by a UK STFC  
 Consolidated Grant awarded to UCL MSSL and 
 by the UCL Cosmoparticle Initiative. 
Y.I. is supported by JSPS KAKENHI Grant Number JP18H05458, JP19K14772, and JP22K18277. This work was supported by World Premier International Research Center Initiative (WPI), MEXT, Japan. 
H.-Y.K.Y. is
supported by the National Science and Technology Council (NSTC) of Taiwan (109-2112-M-007-037-MY3) and the Yushan Scholar Program of the Ministry of Education (MoE) of Taiwan (ROC).  
A.M.W.M. is supported by the DFG - Project Number 452934793.}

\acknowledgments{The authors thank Kentaro Nagamine (Osaka University), Shinsuke Takasao (Osaka University), 
Marcel Strzys (Institute for Cosmic Ray Research, University of Tokyo), Tomonari Michiyama (Shunan University), Sheng-Jun Lin (Academia Sinica Institute for Astronomy and Astrophysics), Ignacio Ferreras (Instituto de Astrof\'{i}sica de Canarias) 
and Anatoli Fedynitch (Academia Sinica Institute of Physics) 
for discussions that helped to inform parts of this work, and two anonymous referees for their constructive comments on this article. Figures 2 and 4 are reproduced from Ref.~\cite{Owen2023A&G}, "The secret agent of galaxy evolution" (Figures 5 and 8), E. R. Owen, \textit{Astronomy \& Geophysics}, Volume 64, Issue 1, February 2023, Pages 1.29–1.35, under Rights and New Business Development - RAS Journals: permissions. Figure 3 is reproduced from Fig. 7 of Ref.~\cite{Aharonian2020PhRvD}, "Probing the sea of galactic cosmic rays with \textit{Fermi}-LAT", \textit{Physical Review D} 101, 083018 (2020), https://doi.org/10.1103/PhysRevD.101.083018, under the Creative Commons Attribution 4.0 International license. Figures 8 and 9 (left panel) are reproduced in accordance with NASA Media Usage Guidelines. Figure 9 (right panel) was adapted from Lopez-Rodriguez et al.~\cite{LopezRodriguez2023ApJ} (their Figure 1, left panel), under the terms of the CC BY 4.0 license. Figure 10 was reproduced from Ref.~\cite{Owen2022ECRS}, under the terms of the CC BY 4.0 license.}

\conflictsofinterest{The authors declare no conflict of interest.} 

\newpage
\abbreviations{Abbreviations}{
The following abbreviations are used in this manuscript:

\noindent 
\begin{tabular}{@{}ll}
AGN & Active galactic nucleus \\
ALMA & Atacama Large Millimeter/Submillimeter Array \\
Baikal-GVD & Baikal–Gigaton Volume Detector \\
CC & Core-collapse \\ 
CGM & Circum-galactic medium \\
CMZ & Central molecular zone \\
CMB & Cosmic microwave background \\
COSI & Compton Spectrometer and Imager \\
CR & Cosmic ray \\ 
CSM & Circum-stellar medium \\
CTA & Cherenkov Telescope Array \\
EBL & Extragalactic background light \\ 
FIR & Far infrared \\
GC & Galactic center \\ 
HAWC & High Altitude Water Cherenkov Observatory \\
HAWC+ & High-Angular Wideband Camera Plus \\ 
HIM & Hot ionized medium \\
HMXB & High-mass X-ray binary \\
HVC & High velocity cloud \\ 
HyLIRG & Hyperluminous infrared galaxy \\
IceCube & IceCube Neutrino Observatory \\
IGM & Intergalactic medium \\ 
IMF & Initial mass function \\ 
IR & Infrared \\
ISRF & Interstellar radiation field \\ 
JCMT & James Clerk Maxwell Telescope \\
JWST & \textit{James Webb} Space Telescope \\ 
KM3NeT & The Cubic Kilometre Neutrino Telescope \\
LAT & Large Area Telescope (on the \textit{Fermi} Gamma-ray Space Telescope) \\
LHC & Large Hadron Collider \\
LHAASO & Large High Altitude Air Shower Observatory \\
LIRG & Luminous infrared galaxy \\
LMXB & Low-mass X-ray binary \\
MC & Monte Carlo \\ 
MHD & Magnetohydrodynamic \\ 
NIRSpec & Near Infra-Red Spectrograph \\ 
NRSI & Non-resonant streaming instability \\
PAH & Poly-cyclic aromatic hydrocarbon \\ 
P-ONE & Pacific Ocean Neutrino Experiment \\
PSG & Post-starburst galaxy \\
SED & Spectral energy distribution \\ 
SKA & Square Kilometer Array \\ 
SMG & Submillimeter galaxy \\ 
SMBH & Supermassive black hole \\
SN & Supernova \\
SNR & Supernova remnant \\
SOFIA & Stratospheric Observatory For Infrared Astronomy \\
SWGO & Southern Wide-Field Gamma-Ray Observatory \\
UFO & Ultra-fast outflow \\
ULIRG & Ultraluminous infrared galaxy \\
UV & Ultra-violet \\
VHE & Very high energy \\ 
VLT & Very large telescope \\ 
WIM & Warm interstellar medium \\
WR & Wolf-Rayet \\ 
XRB & X-ray binary \\
\end{tabular}}

%%%%%%%%%%%%%%%%%%%%%%%%%%%%%%%%%%%%%%%%%%
\begin{adjustwidth}{-\extralength}{0cm}
%\printendnotes[custom] % Un-comment to print a list of endnotes

\reftitle{References}

% Please provide either the correct journal abbreviation (e.g. according to the “List of Title Word Abbreviations” http://www.issn.org/services/online-services/access-to-the-ltwa/) or the full name of the journal.
% Citations and References in Supplementary files are permitted provided that they also appear in the reference list here. 

%=====================================
% References, variant A: external bibliography
%=====================================
\bibliography{references}

\begin{thebibliography}{999}

\bibitem[{Shimizu} \em{et~al.}(2019){Shimizu}, {Todoroki}, {Yajima}, and
  {Nagamine}]{Shimizu2019MNRAS}
{Shimizu}, I.; {Todoroki}, K.; {Yajima}, H.; {Nagamine}, K.
\newblock {Osaka feedback model: isolated disc galaxy simulations}.
\newblock {\em \mnras} {\bf 2019}, {\em 484},~2632--2655,
  \href{http://xxx.lanl.gov/abs/1901.03815}{{\normalfont
  [arXiv:astro-ph.GA/1901.03815]}}.
\newblock
  doi:{\changeurlcolor{black}\href{https://doi.org/10.1093/mnras/stz098}{\detokenize{10.1093/mnras/stz098}}}.

\bibitem[{Oku} \em{et~al.}(2022){Oku}, {Tomida}, {Nagamine}, {Shimizu}, and
  {Cen}]{Oku2022ApJS}
{Oku}, Y.; {Tomida}, K.; {Nagamine}, K.; {Shimizu}, I.; {Cen}, R.
\newblock {Osaka Feedback Model. II. Modeling Supernova Feedback Based on
  High-resolution Simulations}.
\newblock {\em \apjs} {\bf 2022}, {\em 262},~9,
  \href{http://xxx.lanl.gov/abs/2201.00970}{{\normalfont
  [arXiv:astro-ph.GA/2201.00970]}}.
\newblock
  doi:{\changeurlcolor{black}\href{https://doi.org/10.3847/1538-4365/ac77ff}{\detokenize{10.3847/1538-4365/ac77ff}}}.

\bibitem[{Ostriker} and {Kim}(2022)]{Ostriker2022ApJ}
{Ostriker}, E.C.; {Kim}, C.G.
\newblock {Pressure-regulated, Feedback-modulated Star Formation in Disk
  Galaxies}.
\newblock {\em \apj} {\bf 2022}, {\em 936},~137,
  \href{http://xxx.lanl.gov/abs/2206.00681}{{\normalfont
  [arXiv:astro-ph.GA/2206.00681]}}.
\newblock
  doi:{\changeurlcolor{black}\href{https://doi.org/10.3847/1538-4357/ac7de2}{\detokenize{10.3847/1538-4357/ac7de2}}}.

\bibitem[{Orr} \em{et~al.}(2022){Orr}, {Fielding}, {Hayward}, and
  {Burkhart}]{Orr2022ApJ}
{Orr}, M.E.; {Fielding}, D.B.; {Hayward}, C.C.; {Burkhart}, B.
\newblock {Bursting Bubbles: Feedback from Clustered Supernovae and the
  Trade-off Between Turbulence and Outflows}.
\newblock {\em \apj} {\bf 2022}, {\em 932},~88,
  \href{http://xxx.lanl.gov/abs/2109.14656}{{\normalfont
  [arXiv:astro-ph.GA/2109.14656]}}.
\newblock
  doi:{\changeurlcolor{black}\href{https://doi.org/10.3847/1538-4357/ac6c26}{\detokenize{10.3847/1538-4357/ac6c26}}}.

\bibitem[{Rosado} \em{et~al.}(1996){Rosado}, {Ambrocio-Cruz}, {Le Coarer}, and
  {Marcelin}]{Rosado1996AA}
{Rosado}, M.; {Ambrocio-Cruz}, P.; {Le Coarer}, E.; {Marcelin}, M.
\newblock {Kinematics of the galactic supernova remnants RCW 86, MSH 15-56 and
  MSH 11-61A.}
\newblock {\em \aap} {\bf 1996}, {\em 315},~243--252.

\bibitem[{S{\'a}nchez-Cruces} \em{et~al.}(2018){S{\'a}nchez-Cruces}, {Rosado},
  {Fuentes-Carrera}, and {Ambrocio-Cruz}]{SanchezCruces2018MNRAS}
{S{\'a}nchez-Cruces}, M.; {Rosado}, M.; {Fuentes-Carrera}, I.; {Ambrocio-Cruz},
  P.
\newblock {Kinematics of the Galactic Supernova Remnant G109.1-1.0 (CTB 109)}.
\newblock {\em \mnras} {\bf 2018}, {\em 473},~1705--1717,
  \href{http://xxx.lanl.gov/abs/1709.07986}{{\normalfont
  [arXiv:astro-ph.GA/1709.07986]}}.
\newblock
  doi:{\changeurlcolor{black}\href{https://doi.org/10.1093/mnras/stx2460}{\detokenize{10.1093/mnras/stx2460}}}.

\bibitem[{S{\'a}nchez-Cruces} \em{et~al.}(2022){S{\'a}nchez-Cruces},
  {Sardaneta}, {Fuentes-Carrera}, {Rosado}, {C{\'a}rdenas-Mart{\'\i}nez}, and
  {Lara-L{\'o}pez}]{SanchezCruces2022MNRAS}
{S{\'a}nchez-Cruces}, M.; {Sardaneta}, M.M.; {Fuentes-Carrera}, I.; {Rosado},
  M.; {C{\'a}rdenas-Mart{\'\i}nez}, N.; {Lara-L{\'o}pez}, M.A.
\newblock {A kinematical study of the dwarf irregular galaxy NGC 1569 and its
  supernova remnants}.
\newblock {\em \mnras} {\bf 2022}, {\em 513},~1755--1773,
  \href{http://xxx.lanl.gov/abs/2209.06766}{{\normalfont
  [arXiv:astro-ph.GA/2209.06766]}}.
\newblock
  doi:{\changeurlcolor{black}\href{https://doi.org/10.1093/mnras/stac985}{\detokenize{10.1093/mnras/stac985}}}.

\bibitem[{Hopkins} \em{et~al.}(2020){Hopkins}, {Grudi{\'c}}, {Wetzel},
  {Kere{\v{s}}}, {Faucher-Gigu{\`e}re}, {Ma}, {Murray}, and
  {Butcher}]{Hopkins2020MNRAS}
{Hopkins}, P.F.; {Grudi{\'c}}, M.Y.; {Wetzel}, A.; {Kere{\v{s}}}, D.;
  {Faucher-Gigu{\`e}re}, C.A.; {Ma}, X.; {Murray}, N.; {Butcher}, N.
\newblock {Radiative stellar feedback in galaxy formation: Methods and
  physics}.
\newblock {\em \mnras} {\bf 2020}, {\em 491},~3702--3729,
  \href{http://xxx.lanl.gov/abs/1811.12462}{{\normalfont
  [arXiv:astro-ph.GA/1811.12462]}}.
\newblock
  doi:{\changeurlcolor{black}\href{https://doi.org/10.1093/mnras/stz3129}{\detokenize{10.1093/mnras/stz3129}}}.

\bibitem[{Qiu} \em{et~al.}(2019){Qiu}, {Bogdanovi{\'c}}, {Li}, {Park}, and
  {Wise}]{Qiu2019ApJ}
{Qiu}, Y.; {Bogdanovi{\'c}}, T.; {Li}, Y.; {Park}, K.; {Wise}, J.H.
\newblock {The Interplay of Kinetic and Radiative Feedback in Galaxy Clusters}.
\newblock {\em \apj} {\bf 2019}, {\em 877},~47,
  \href{http://xxx.lanl.gov/abs/1810.01857}{{\normalfont
  [arXiv:astro-ph.GA/1810.01857]}}.
\newblock
  doi:{\changeurlcolor{black}\href{https://doi.org/10.3847/1538-4357/ab18fd}{\detokenize{10.3847/1538-4357/ab18fd}}}.

\bibitem[{Chen}(2020)]{Chen2020ApJ}
{Chen}, H.
\newblock {The Role of Quasar Radiative Feedback on Galaxy Formation during
  Cosmic Reionization}.
\newblock {\em \apj} {\bf 2020}, {\em 893},~165,
  \href{http://xxx.lanl.gov/abs/1911.09113}{{\normalfont
  [arXiv:astro-ph.GA/1911.09113]}}.
\newblock
  doi:{\changeurlcolor{black}\href{https://doi.org/10.3847/1538-4357/ab80c6}{\detokenize{10.3847/1538-4357/ab80c6}}}.

\bibitem[{Yajima} \em{et~al.}(2017){Yajima}, {Nagamine}, {Zhu}, {Khochfar}, and
  {Dalla Vecchia}]{Yajima2017ApJ}
{Yajima}, H.; {Nagamine}, K.; {Zhu}, Q.; {Khochfar}, S.; {Dalla Vecchia}, C.
\newblock {Growth of First Galaxies: Impacts of Star Formation and Stellar
  Feedback}.
\newblock {\em \apj} {\bf 2017}, {\em 846},~30,
  \href{http://xxx.lanl.gov/abs/1704.03117}{{\normalfont
  [arXiv:astro-ph.GA/1704.03117]}}.
\newblock
  doi:{\changeurlcolor{black}\href{https://doi.org/10.3847/1538-4357/aa82b5}{\detokenize{10.3847/1538-4357/aa82b5}}}.

\bibitem[{Kornecki} \em{et~al.}(2020){Kornecki}, {Pellizza}, {del Palacio},
  {M{\"u}ller}, {Albacete-Colombo}, and {Romero}]{Kornecki2020AA}
{Kornecki}, P.; {Pellizza}, L.J.; {del Palacio}, S.; {M{\"u}ller}, A.L.;
  {Albacete-Colombo}, J.F.; {Romero}, G.E.
\newblock {{\ensuremath{\gamma}}-ray/infrared luminosity correlation of
  star-forming galaxies}.
\newblock {\em \aap} {\bf 2020}, {\em 641},~A147,
  \href{http://xxx.lanl.gov/abs/2007.07430}{{\normalfont
  [arXiv:astro-ph.HE/2007.07430]}}.
\newblock
  doi:{\changeurlcolor{black}\href{https://doi.org/10.1051/0004-6361/202038428}{\detokenize{10.1051/0004-6361/202038428}}}.

\bibitem[{Kornecki} \em{et~al.}(2022){Kornecki}, {Peretti}, {del Palacio},
  {Benaglia}, and {Pellizza}]{Kornecki2022AA}
{Kornecki}, P.; {Peretti}, E.; {del Palacio}, S.; {Benaglia}, P.; {Pellizza},
  L.J.
\newblock {Exploring the physics behind the non-thermal emission from
  star-forming galaxies detected in {\ensuremath{\gamma}} rays}.
\newblock {\em \aap} {\bf 2022}, {\em 657},~A49,
  \href{http://xxx.lanl.gov/abs/2107.00823}{{\normalfont
  [arXiv:astro-ph.HE/2107.00823]}}.
\newblock
  doi:{\changeurlcolor{black}\href{https://doi.org/10.1051/0004-6361/202141295}{\detokenize{10.1051/0004-6361/202141295}}}.

\bibitem[{Owen} \em{et~al.}(2018){Owen}, {Jacobsen}, {Wu}, and
  {Surajbali}]{Owen2018MNRAS}
{Owen}, E.R.; {Jacobsen}, I.B.; {Wu}, K.; {Surajbali}, P.
\newblock {Interactions between ultra-high-energy particles and protogalactic
  environments}.
\newblock {\em \mnras} {\bf 2018}, {\em 481},~666--687,
  \href{http://xxx.lanl.gov/abs/1808.07837}{{\normalfont
  [arXiv:astro-ph.HE/1808.07837]}}.
\newblock
  doi:{\changeurlcolor{black}\href{https://doi.org/10.1093/mnras/sty2279}{\detokenize{10.1093/mnras/sty2279}}}.

\bibitem[{Yu} \em{et~al.}(2020){Yu}, {Owen}, {Wu}, and {Ferreras}]{Yu2020MNRAS}
{Yu}, B.P.B.; {Owen}, E.R.; {Wu}, K.; {Ferreras}, I.
\newblock {A hydrodynamical study of outflows in starburst galaxies with
  different driving mechanisms}.
\newblock {\em \mnras} {\bf 2020}, {\em 492},~3179--3193,
  \href{http://xxx.lanl.gov/abs/2001.04384}{{\normalfont
  [arXiv:astro-ph.GA/2001.04384]}}.
\newblock
  doi:{\changeurlcolor{black}\href{https://doi.org/10.1093/mnras/staa021}{\detokenize{10.1093/mnras/staa021}}}.

\bibitem[{Krumholz} \em{et~al.}(2023){Krumholz}, {Crocker}, and
  {Offner}]{Krumholz2022arXiv}
{Krumholz}, M.R.; {Crocker}, R.M.; {Offner}, S.S.R.
\newblock {The cosmic ray ionization and {\ensuremath{\gamma}}-ray budgets of
  star-forming galaxies}.
\newblock {\em \mnras} {\bf 2023}, {\em 520},~5126--5143,
  \href{http://xxx.lanl.gov/abs/2211.03488}{{\normalfont
  [arXiv:astro-ph.GA/2211.03488]}}.
\newblock
  doi:{\changeurlcolor{black}\href{https://doi.org/10.1093/mnras/stad459}{\detokenize{10.1093/mnras/stad459}}}.

\bibitem[{Owen} \em{et~al.}(2019){Owen}, {Jin}, {Wu}, and
  {Chan}]{Owen2019MNRAS}
{Owen}, E.R.; {Jin}, X.; {Wu}, K.; {Chan}, S.
\newblock {Hadronic interactions of energetic charged particles in
  protogalactic outflow environments and implications for the early evolution
  of galaxies}.
\newblock {\em \mnras} {\bf 2019}, {\em 484},~1645--1671,
  \href{http://xxx.lanl.gov/abs/1901.01411}{{\normalfont
  [arXiv:astro-ph.GA/1901.01411]}}.
\newblock
  doi:{\changeurlcolor{black}\href{https://doi.org/10.1093/mnras/stz060}{\detokenize{10.1093/mnras/stz060}}}.

\bibitem[{Ruszkowski} and {Pfrommer}(2023)]{Ruszkowski2023arXiv230603141R}
{Ruszkowski}, M.; {Pfrommer}, C.
\newblock {Cosmic ray feedback in galaxies and galaxy clusters -- A pedagogical
  introduction and a topical review of the acceleration, transport,
  observables, and dynamical impact of cosmic rays}.
\newblock {\em arXiv e-prints} {\bf 2023}, p. arXiv:2306.03141,
  \href{http://xxx.lanl.gov/abs/2306.03141}{{\normalfont
  [arXiv:astro-ph.HE/2306.03141]}}.
\newblock
  doi:{\changeurlcolor{black}\href{https://doi.org/10.48550/arXiv.2306.03141}{\detokenize{10.48550/arXiv.2306.03141}}}.

\bibitem[{Thomas} and {Pfrommer}(2019)]{Thomas2019MNRAS}
{Thomas}, T.; {Pfrommer}, C.
\newblock {Cosmic-ray hydrodynamics: Alfv{\'e}n-wave regulated transport of
  cosmic rays}.
\newblock {\em \mnras} {\bf 2019}, {\em 485},~2977--3008,
  \href{http://xxx.lanl.gov/abs/1805.11092}{{\normalfont
  [arXiv:astro-ph.HE/1805.11092]}}.
\newblock
  doi:{\changeurlcolor{black}\href{https://doi.org/10.1093/mnras/stz263}{\detokenize{10.1093/mnras/stz263}}}.

\bibitem[{Amato} and {Blasi}(2018)]{Amato2018AdSpR}
{Amato}, E.; {Blasi}, P.
\newblock {Cosmic ray transport in the Galaxy: A review}.
\newblock {\em Advances in Space Research} {\bf 2018}, {\em 62},~2731--2749,
  \href{http://xxx.lanl.gov/abs/1704.05696}{{\normalfont
  [arXiv:astro-ph.HE/1704.05696]}}.
\newblock
  doi:{\changeurlcolor{black}\href{https://doi.org/10.1016/j.asr.2017.04.019}{\detokenize{10.1016/j.asr.2017.04.019}}}.

\bibitem[{Strong} and {Moskalenko}(1998)]{Strong1998ApJ}
{Strong}, A.W.; {Moskalenko}, I.V.
\newblock {Propagation of Cosmic-Ray Nucleons in the Galaxy}.
\newblock {\em \apj} {\bf 1998}, {\em 509},~212--228,
  \href{http://xxx.lanl.gov/abs/astro-ph/9807150}{{\normalfont
  [arXiv:astro-ph/astro-ph/9807150]}}.
\newblock
  doi:{\changeurlcolor{black}\href{https://doi.org/10.1086/306470}{\detokenize{10.1086/306470}}}.

\bibitem[{Maurin} \em{et~al.}(2001){Maurin}, {Donato}, {Taillet}, and
  {Salati}]{Maurin2001ApJ}
{Maurin}, D.; {Donato}, F.; {Taillet}, R.; {Salati}, P.
\newblock {Cosmic Rays below Z=30 in a Diffusion Model: New Constraints on
  Propagation Parameters}.
\newblock {\em \apj} {\bf 2001}, {\em 555},~585--596,
  \href{http://xxx.lanl.gov/abs/astro-ph/0101231}{{\normalfont
  [arXiv:astro-ph/astro-ph/0101231]}}.
\newblock
  doi:{\changeurlcolor{black}\href{https://doi.org/10.1086/321496}{\detokenize{10.1086/321496}}}.

\bibitem[{Evoli} \em{et~al.}(2008){Evoli}, {Gaggero}, {Grasso}, and
  {Maccione}]{Evoli2008JCAP}
{Evoli}, C.; {Gaggero}, D.; {Grasso}, D.; {Maccione}, L.
\newblock {Cosmic ray nuclei, antiprotons and gamma rays in the galaxy: a new
  diffusion model}.
\newblock {\em \jcap} {\bf 2008}, {\em 2008},~018,
  \href{http://xxx.lanl.gov/abs/0807.4730}{{\normalfont
  [arXiv:astro-ph/0807.4730]}}.
\newblock
  doi:{\changeurlcolor{black}\href{https://doi.org/10.1088/1475-7516/2008/10/018}{\detokenize{10.1088/1475-7516/2008/10/018}}}.

\bibitem[{Kissmann}(2014)]{Kissmann2014APh}
{Kissmann}, R.
\newblock {PICARD: A novel code for the Galactic Cosmic Ray propagation
  problem}.
\newblock {\em Astroparticle Physics} {\bf 2014}, {\em 55},~37--50,
  \href{http://xxx.lanl.gov/abs/1401.4035}{{\normalfont
  [arXiv:astro-ph.HE/1401.4035]}}.
\newblock
  doi:{\changeurlcolor{black}\href{https://doi.org/10.1016/j.astropartphys.2014.02.002}{\detokenize{10.1016/j.astropartphys.2014.02.002}}}.

\bibitem[{Skilling}(1975)]{Skilling1975MNRAS}
{Skilling}, J.
\newblock {Cosmic ray streaming - III. Self-consistent solutions.}
\newblock {\em \mnras} {\bf 1975}, {\em 173},~255--269.
\newblock
  doi:{\changeurlcolor{black}\href{https://doi.org/10.1093/mnras/173.2.255}{\detokenize{10.1093/mnras/173.2.255}}}.

\bibitem[{Aloisio} \em{et~al.}(2015){Aloisio}, {Blasi}, and
  {Serpico}]{Aloisio2015A&A}
{Aloisio}, R.; {Blasi}, P.; {Serpico}, P.D.
\newblock {Nonlinear cosmic ray Galactic transport in the light of AMS-02 and
  Voyager data}.
\newblock {\em \aap} {\bf 2015}, {\em 583},~A95,
  \href{http://xxx.lanl.gov/abs/1507.00594}{{\normalfont
  [arXiv:astro-ph.HE/1507.00594]}}.
\newblock
  doi:{\changeurlcolor{black}\href{https://doi.org/10.1051/0004-6361/201526877}{\detokenize{10.1051/0004-6361/201526877}}}.

\bibitem[{Wentzel}(1974)]{Wentzel1974ARA&A}
{Wentzel}, D.G.
\newblock {Cosmic-ray propagation in the Galaxy: collective effects.}
\newblock {\em \araa} {\bf 1974}, {\em 12},~71--96.
\newblock
  doi:{\changeurlcolor{black}\href{https://doi.org/10.1146/annurev.aa.12.090174.000443}{\detokenize{10.1146/annurev.aa.12.090174.000443}}}.

\bibitem[{Cesarsky}(1980)]{Cesarsky1980ARA&A}
{Cesarsky}, C.J.
\newblock {Cosmic-ray confinement in the galaxy}.
\newblock {\em \araa} {\bf 1980}, {\em 18},~289--319.
\newblock
  doi:{\changeurlcolor{black}\href{https://doi.org/10.1146/annurev.aa.18.090180.001445}{\detokenize{10.1146/annurev.aa.18.090180.001445}}}.

\bibitem[{Zweibel}(2017)]{Zweibel2017PhPl}
{Zweibel}, E.G.
\newblock {The basis for cosmic ray feedback: Written on the wind}.
\newblock {\em Physics of Plasmas} {\bf 2017}, {\em 24},~055402.
\newblock
  doi:{\changeurlcolor{black}\href{https://doi.org/10.1063/1.4984017}{\detokenize{10.1063/1.4984017}}}.

\bibitem[{Plotnikov} \em{et~al.}(2021){Plotnikov}, {Ostriker}, and
  {Bai}]{Plotnikov2021ApJ}
{Plotnikov}, I.; {Ostriker}, E.C.; {Bai}, X.N.
\newblock {Influence of Ion-Neutral Damping on the Cosmic-Ray Streaming
  Instability: Magnetohydrodynamic Particle-in-cell Simulations}.
\newblock {\em \apj} {\bf 2021}, {\em 914},~3,
  \href{http://xxx.lanl.gov/abs/2102.11878}{{\normalfont
  [arXiv:astro-ph.HE/2102.11878]}}.
\newblock
  doi:{\changeurlcolor{black}\href{https://doi.org/10.3847/1538-4357/abf7b3}{\detokenize{10.3847/1538-4357/abf7b3}}}.

\bibitem[{Marret} \em{et~al.}(2022){Marret}, {Ciardi}, {Smets}, {Fuchs}, and
  {Nicolas}]{Marret2022PhRvL}
{Marret}, A.; {Ciardi}, A.; {Smets}, R.; {Fuchs}, J.; {Nicolas}, L.
\newblock {Enhancement of the Nonresonant Streaming Instability by Particle
  Collisions}.
\newblock {\em \prl} {\bf 2022}, {\em 128},~115101,
  \href{http://xxx.lanl.gov/abs/2111.15272}{{\normalfont
  [arXiv:astro-ph.HE/2111.15272]}}.
\newblock
  doi:{\changeurlcolor{black}\href{https://doi.org/10.1103/PhysRevLett.128.115101}{\detokenize{10.1103/PhysRevLett.128.115101}}}.

\bibitem[{Squire} \em{et~al.}(2021){Squire}, {Hopkins}, {Quataert}, and
  {Kempski}]{Squire2021MNRAS}
{Squire}, J.; {Hopkins}, P.F.; {Quataert}, E.; {Kempski}, P.
\newblock {The impact of astrophysical dust grains on the confinement of cosmic
  rays}.
\newblock {\em \mnras} {\bf 2021}, {\em 502},~2630--2644,
  \href{http://xxx.lanl.gov/abs/2011.02497}{{\normalfont
  [arXiv:astro-ph.HE/2011.02497]}}.
\newblock
  doi:{\changeurlcolor{black}\href{https://doi.org/10.1093/mnras/stab179}{\detokenize{10.1093/mnras/stab179}}}.

\bibitem[{Gupta} \em{et~al.}(2021){Gupta}, {Caprioli}, and
  {Haggerty}]{Gupta2021ApJb}
{Gupta}, S.; {Caprioli}, D.; {Haggerty}, C.C.
\newblock {Lepton-driven Nonresonant Streaming Instability}.
\newblock {\em \apj} {\bf 2021}, {\em 923},~208,
  \href{http://xxx.lanl.gov/abs/2106.07672}{{\normalfont
  [arXiv:astro-ph.HE/2106.07672]}}.
\newblock
  doi:{\changeurlcolor{black}\href{https://doi.org/10.3847/1538-4357/ac23cf}{\detokenize{10.3847/1538-4357/ac23cf}}}.

\bibitem[{Zweibel} and {Everett}(2010)]{Zweibel2010ApJ}
{Zweibel}, E.G.; {Everett}, J.E.
\newblock {Environments for Magnetic Field Amplification by Cosmic Rays}.
\newblock {\em \apj} {\bf 2010}, {\em 709},~1412--1419,
  \href{http://xxx.lanl.gov/abs/0912.3511}{{\normalfont
  [arXiv:astro-ph.GA/0912.3511]}}.
\newblock
  doi:{\changeurlcolor{black}\href{https://doi.org/10.1088/0004-637X/709/2/1412}{\detokenize{10.1088/0004-637X/709/2/1412}}}.

\bibitem[{Pelletier} \em{et~al.}(2006){Pelletier}, {Lemoine}, and
  {Marcowith}]{Pelletier2006A&A}
{Pelletier}, G.; {Lemoine}, M.; {Marcowith}, A.
\newblock {Turbulence and particle acceleration in collisionless supernovae
  remnant shocks. I. Anisotropic spectra solutions}.
\newblock {\em \aap} {\bf 2006}, {\em 453},~181--191,
  \href{http://xxx.lanl.gov/abs/astro-ph/0603461}{{\normalfont
  [arXiv:astro-ph/astro-ph/0603461]}}.
\newblock
  doi:{\changeurlcolor{black}\href{https://doi.org/10.1051/0004-6361:20054737}{\detokenize{10.1051/0004-6361:20054737}}}.

\bibitem[{Amato} and {Blasi}(2009)]{Amato2009MNRAS}
{Amato}, E.; {Blasi}, P.
\newblock {A kinetic approach to cosmic-ray-induced streaming instability at
  supernova shocks}.
\newblock {\em \mnras} {\bf 2009}, {\em 392},~1591--1600,
  \href{http://xxx.lanl.gov/abs/0806.1223}{{\normalfont
  [arXiv:astro-ph/0806.1223]}}.
\newblock
  doi:{\changeurlcolor{black}\href{https://doi.org/10.1111/j.1365-2966.2008.14200.x}{\detokenize{10.1111/j.1365-2966.2008.14200.x}}}.

\bibitem[{Amato}(2011)]{Amato2011MmSAI}
{Amato}, E.
\newblock {The streaming instability: a review}.
\newblock {\em \memsai} {\bf 2011}, {\em 82},~806.

\bibitem[{Reville} \em{et~al.}(2008){Reville}, {Kirk}, {Duffy}, and
  {O'Sullivan}]{Reville2008IJMPD}
{Reville}, B.; {Kirk}, J.G.; {Duffy}, P.; {O'Sullivan}, S.
\newblock {Environmental Limits on the Nonresonant Cosmic-Ray Current-Driven
  Instability}.
\newblock {\em International Journal of Modern Physics D} {\bf 2008}, {\em
  17},~1795--1801,  \href{http://xxx.lanl.gov/abs/0802.3322}{{\normalfont
  [arXiv:astro-ph/0802.3322]}}.
\newblock
  doi:{\changeurlcolor{black}\href{https://doi.org/10.1142/S021827180801342X}{\detokenize{10.1142/S021827180801342X}}}.

\bibitem[{Marret} \em{et~al.}(2021){Marret}, {Ciardi}, {Smets}, and
  {Fuchs}]{Marret2021MNRAS}
{Marret}, A.; {Ciardi}, A.; {Smets}, R.; {Fuchs}, J.
\newblock {On the growth of the thermally modified non-resonant streaming
  instability}.
\newblock {\em \mnras} {\bf 2021}, {\em 500},~2302--2315,
  \href{http://xxx.lanl.gov/abs/2010.10237}{{\normalfont
  [arXiv:astro-ph.HE/2010.10237]}}.
\newblock
  doi:{\changeurlcolor{black}\href{https://doi.org/10.1093/mnras/staa3465}{\detokenize{10.1093/mnras/staa3465}}}.

\bibitem[{Blasi}(2013)]{Blasi2013A&ARv}
{Blasi}, P.
\newblock {The origin of galactic cosmic rays}.
\newblock {\em \aapr} {\bf 2013}, {\em 21},~70,
  \href{http://xxx.lanl.gov/abs/1311.7346}{{\normalfont
  [arXiv:astro-ph.HE/1311.7346]}}.
\newblock
  doi:{\changeurlcolor{black}\href{https://doi.org/10.1007/s00159-013-0070-7}{\detokenize{10.1007/s00159-013-0070-7}}}.

\bibitem[{Amato}(2014)]{Amato2014IJMPD}
{Amato}, E.
\newblock {The origin of galactic cosmic rays}.
\newblock {\em International Journal of Modern Physics D} {\bf 2014}, {\em
  23},~1430013,  \href{http://xxx.lanl.gov/abs/1406.7714}{{\normalfont
  [arXiv:astro-ph.HE/1406.7714]}}.
\newblock
  doi:{\changeurlcolor{black}\href{https://doi.org/10.1142/S0218271814300134}{\detokenize{10.1142/S0218271814300134}}}.

\bibitem[{Bell}(2004)]{Bell2004MNRAS}
{Bell}, A.R.
\newblock {Turbulent amplification of magnetic field and diffusive shock
  acceleration of cosmic rays}.
\newblock {\em \mnras} {\bf 2004}, {\em 353},~550--558.
\newblock
  doi:{\changeurlcolor{black}\href{https://doi.org/10.1111/j.1365-2966.2004.08097.x}{\detokenize{10.1111/j.1365-2966.2004.08097.x}}}.

\bibitem[{Vink}(2012)]{Vink2012A&ARv}
{Vink}, J.
\newblock {Supernova remnants: the X-ray perspective}.
\newblock {\em \aapr} {\bf 2012}, {\em 20},~49,
  \href{http://xxx.lanl.gov/abs/1112.0576}{{\normalfont
  [arXiv:astro-ph.HE/1112.0576]}}.
\newblock
  doi:{\changeurlcolor{black}\href{https://doi.org/10.1007/s00159-011-0049-1}{\detokenize{10.1007/s00159-011-0049-1}}}.

\bibitem[{Bykov} \em{et~al.}(2012){Bykov}, {Ellison}, and
  {Renaud}]{Bykov2012SSRv}
{Bykov}, A.M.; {Ellison}, D.C.; {Renaud}, M.
\newblock {Magnetic Fields in Cosmic Particle Acceleration Sources}.
\newblock {\em \ssr} {\bf 2012}, {\em 166},~71--95,
  \href{http://xxx.lanl.gov/abs/1105.0130}{{\normalfont
  [arXiv:astro-ph.HE/1105.0130]}}.
\newblock
  doi:{\changeurlcolor{black}\href{https://doi.org/10.1007/s11214-011-9761-4}{\detokenize{10.1007/s11214-011-9761-4}}}.

\bibitem[{Riquelme} and {Spitkovsky}(2009)]{Riquelme2009ApJ}
{Riquelme}, M.A.; {Spitkovsky}, A.
\newblock {Nonlinear Study of Bell's Cosmic Ray Current-Driven Instability}.
\newblock {\em \apj} {\bf 2009}, {\em 694},~626--642,
  \href{http://xxx.lanl.gov/abs/0810.4565}{{\normalfont
  [arXiv:astro-ph/0810.4565]}}.
\newblock
  doi:{\changeurlcolor{black}\href{https://doi.org/10.1088/0004-637X/694/1/626}{\detokenize{10.1088/0004-637X/694/1/626}}}.

\bibitem[{Schroer} \em{et~al.}(2021){Schroer}, {Pezzi}, {Caprioli}, {Haggerty},
  and {Blasi}]{Schroer2021ApJ}
{Schroer}, B.; {Pezzi}, O.; {Caprioli}, D.; {Haggerty}, C.; {Blasi}, P.
\newblock {Dynamical Effects of Cosmic Rays on the Medium Surrounding Their
  Sources}.
\newblock {\em \apjl} {\bf 2021}, {\em 914},~L13,
  \href{http://xxx.lanl.gov/abs/2011.02238}{{\normalfont
  [arXiv:astro-ph.HE/2011.02238]}}.
\newblock
  doi:{\changeurlcolor{black}\href{https://doi.org/10.3847/2041-8213/ac02cd}{\detokenize{10.3847/2041-8213/ac02cd}}}.

\bibitem[{Schroer} \em{et~al.}(2022){Schroer}, {Pezzi}, {Caprioli}, {Haggerty},
  and {Blasi}]{Schroer2022MNRAS}
{Schroer}, B.; {Pezzi}, O.; {Caprioli}, D.; {Haggerty}, C.C.; {Blasi}, P.
\newblock {Cosmic-ray generated bubbles around their sources}.
\newblock {\em \mnras} {\bf 2022}, {\em 512},~233--244,
  \href{http://xxx.lanl.gov/abs/2202.05814}{{\normalfont
  [arXiv:astro-ph.HE/2202.05814]}}.
\newblock
  doi:{\changeurlcolor{black}\href{https://doi.org/10.1093/mnras/stac466}{\detokenize{10.1093/mnras/stac466}}}.

\bibitem[{Commer{\c{c}}on} \em{et~al.}(2019){Commer{\c{c}}on}, {Marcowith}, and
  {Dubois}]{Commercon2019A&A}
{Commer{\c{c}}on}, B.; {Marcowith}, A.; {Dubois}, Y.
\newblock {Cosmic-ray propagation in the bi-stable interstellar medium. I.
  Conditions for cosmic-ray trapping}.
\newblock {\em \aap} {\bf 2019}, {\em 622},~A143,
  \href{http://xxx.lanl.gov/abs/1811.11509}{{\normalfont
  [arXiv:astro-ph.GA/1811.11509]}}.
\newblock
  doi:{\changeurlcolor{black}\href{https://doi.org/10.1051/0004-6361/201833809}{\detokenize{10.1051/0004-6361/201833809}}}.

\bibitem[{Simpson} \em{et~al.}(2023){Simpson}, {Pakmor}, {Pfrommer}, {Glover},
  and {Smith}]{Simpson2023MNRAS}
{Simpson}, C.M.; {Pakmor}, R.; {Pfrommer}, C.; {Glover}, S.C.O.; {Smith}, R.
\newblock {How cosmic rays mediate the evolution of the interstellar medium}.
\newblock {\em \mnras} {\bf 2023}, {\em 520},~4621--4645.
\newblock
  doi:{\changeurlcolor{black}\href{https://doi.org/10.1093/mnras/stac3601}{\detokenize{10.1093/mnras/stac3601}}}.

\bibitem[Almeida \em{et~al.}(1968)Almeida, Rushbrooke, Scharenguivel, Behrens,
  Blobel, Borecka, Dehne, Dfaz, Knies, Schmitt, Str\"omer, and
  Swanson]{Almeida1968}
Almeida, S.P.; Rushbrooke, J.G.; Scharenguivel, J.H.; Behrens, M.; Blobel, V.;
  Borecka, I.; Dehne, H.C.; Dfaz, J.; Knies, G.; Schmitt, A.;  et~al.
\newblock $\mathrm{pp}$ Interactions at 10 GeV/ \textit{c}.
\newblock {\em Phys. Rev.} {\bf 1968}, {\em 174},~1638.
\newblock
  doi:{\changeurlcolor{black}\href{https://doi.org/10.1103/PhysRev.174.1638}{\detokenize{10.1103/PhysRev.174.1638}}}.

\bibitem[{Skorodko} \em{et~al.}(2008){Skorodko}, {Bashkanov}, {Bogoslawsky},
  {Calen}, {Cappellaro}, {Clement}, {Demiroers}, {Doroshkevich}, {Duniec},
  {Ekstr{\"o}m}, {Franssen}, {Gustafsson}, {H{\"o}istad}, {Ivanov}, {Jacewicz},
  {Jiganov}, {Johansson}, {Khakimova}, {Kaskulov}, {Keleta}, {Koch}, {Kren},
  {Kullander}, {Kup{\'s}{\'c}}, {Kuznetsov}, {Marciniewski}, {Meier},
  {Morosov}, {Pauly}, {Petterson}, {Petukhov}, {Povtorejko}, {Sch{\"o}nning},
  {Scobel}, {Shwartz}, {Sopov}, {Stepeniak}, {Th{\"o}rngren-Engblom},
  {Tikhomirov}, {Wagner}, {Wolke}, {Yamamoto}, {Zabierowski}, and
  {Z{\l}omanczuk}]{Skorodko2008EPJA}
{Skorodko}, T.; {Bashkanov}, M.; {Bogoslawsky}, D.; {Calen}, H.; {Cappellaro},
  F.; {Clement}, H.; {Demiroers}, L.; {Doroshkevich}, E.; {Duniec}, D.;
  {Ekstr{\"o}m}, C.;  et~al.
\newblock {Excitation of the Roper resonance in single- and double-pion
  production in nucleon-nucleon collisions}.
\newblock {\em European Physical Journal A} {\bf 2008}, {\em 35},~317.
\newblock
  doi:{\changeurlcolor{black}\href{https://doi.org/10.1140/epja/i2008-10569-6}{\detokenize{10.1140/epja/i2008-10569-6}}}.

\bibitem[{Blattnig} \em{et~al.}(2000){Blattnig}, {Swaminathan}, {Kruger},
  {Ngom}, and {Norbury}]{BlattnigPRD2000}
{Blattnig}, S.R.; {Swaminathan}, S.R.; {Kruger}, A.T.; {Ngom}, M.; {Norbury},
  J.W.
\newblock {Parametrizations of inclusive cross sections for pion production in
  proton-proton collisions}.
\newblock {\em \prd} {\bf 2000}, {\em 62},~094030,
  \href{http://xxx.lanl.gov/abs/hep-ph/0010170}{{\normalfont
  [hep-ph/0010170]}}.
\newblock
  doi:{\changeurlcolor{black}\href{https://doi.org/10.1103/PhysRevD.62.094030}{\detokenize{10.1103/PhysRevD.62.094030}}}.

\bibitem[Patrignani \em{et~al.}(2016)Patrignani et~al.]{Patrignani2016ChPh}
Patrignani, C.;  et~al.
\newblock {Review of Particle Physics}.
\newblock {\em Chin. Phys.} {\bf 2016}, {\em C40},~100001.
\newblock
  doi:{\changeurlcolor{black}\href{https://doi.org/10.1088/1674-1137/40/10/100001}{\detokenize{10.1088/1674-1137/40/10/100001}}}.

\bibitem[{Amenomori} \em{et~al.}(2019){Amenomori}, {Bi}, {Chen}, {Chen},
  {Chen}, {Cui}, {Danzengluobu}, {Ding}, {Feng}, {Feng}, {Feng}, {Gou}, {Guo},
  {He}, {He}, {Hibino}, {Hotta}, {Hu}, {Hu}, {Huang}, {Jia}, {Jiang}, {Kajino},
  {Kasahara}, {Katayose}, {Kato}, {Kawata}, {Kozai}, {Labaciren}, {Le}, {Li},
  {Li}, {Li}, {Lin}, {Liu}, {Liu}, {Liu}, {Lu}, {Meng}, {Miyazaki}, {Munakata},
  {Nakajima}, {Nakamura}, {Nanjo}, {Nishizawa}, {Niwa}, {Ohnishi}, {Ohta},
  {Ozawa}, {Qian}, {Qu}, {Saito}, {Saito}, {Sakata}, {Sako}, {Shao}, {Shibata},
  {Shiomi}, {Shirai}, {Sugimoto}, {Takita}, {Tan}, {Tateyama}, {Torii},
  {Tsuchiya}, {Udo}, {Wang}, {Wu}, {Xue}, {Yamamoto}, {Yamauchi}, {Yang},
  {Yuan}, {Zhai}, {Zhang}, {Zhang}, {Zhang}, {Zhang}, {Zhang}, {Zhang},
  {Zhaxisangzhu}, and {Zhou}]{Amenomori2019EPJWC}
{Amenomori}, M.; {Bi}, X.J.; {Chen}, D.; {Chen}, T.L.; {Chen}, W.Y.; {Cui},
  S.W.; {Danzengluobu}.; {Ding}, L.K.; {Feng}, C.F.; {Feng}, Z.;  et~al.
\newblock {Test of the hadronic interaction models SIBYLL2.3, EPOS-LHC and
  QGSJETII- 04 with Tibet EAS core data}.
\newblock  European Physical Journal Web of Conferences,  2019, Vol. 208, {\em
  European Physical Journal Web of Conferences}, p. 08013.
\newblock
  doi:{\changeurlcolor{black}\href{https://doi.org/10.1051/epjconf/201920808013}{\detokenize{10.1051/epjconf/201920808013}}}.

\bibitem[{B{\"a}hr} \em{et~al.}(2008){B{\"a}hr}, {Gieseke}, {Gigg},
  {Grellscheid}, {Hamilton}, {Latunde-Dada}, {Pl{\"a}tzer}, {Richardson},
  {Seymour}, {Sherstnev}, and {Webber}]{Bahr2008EPJC}
{B{\"a}hr}, M.; {Gieseke}, S.; {Gigg}, M.A.; {Grellscheid}, D.; {Hamilton}, K.;
  {Latunde-Dada}, O.; {Pl{\"a}tzer}, S.; {Richardson}, P.; {Seymour}, M.H.;
  {Sherstnev}, A.;  et~al.
\newblock {Herwig++ physics and manual}.
\newblock {\em European Physical Journal C} {\bf 2008}, {\em 58},~639--707,
  \href{http://xxx.lanl.gov/abs/0803.0883}{{\normalfont
  [arXiv:hep-ph/0803.0883]}}.
\newblock
  doi:{\changeurlcolor{black}\href{https://doi.org/10.1140/epjc/s10052-008-0798-9}{\detokenize{10.1140/epjc/s10052-008-0798-9}}}.

\bibitem[{Gleisberg} \em{et~al.}(2009){Gleisberg}, {H{\"o}che}, {Krauss},
  {Sch{\"o}nherr}, {Schumann}, {Siegert}, and {Winter}]{Gleisberg2009JHEP}
{Gleisberg}, T.; {H{\"o}che}, S.; {Krauss}, F.; {Sch{\"o}nherr}, M.;
  {Schumann}, S.; {Siegert}, F.; {Winter}, J.
\newblock {Event generation with SHERPA 1.1}.
\newblock {\em Journal of High Energy Physics} {\bf 2009}, {\em 2009},~007,
  \href{http://xxx.lanl.gov/abs/0811.4622}{{\normalfont
  [arXiv:hep-ph/0811.4622]}}.
\newblock
  doi:{\changeurlcolor{black}\href{https://doi.org/10.1088/1126-6708/2009/02/007}{\detokenize{10.1088/1126-6708/2009/02/007}}}.

\bibitem[{Allison} \em{et~al.}(2006){Allison}, {Amako}, {Apostolakis},
  {Araujo}, {Dubois}, {Asai}, {Barrand}, {Capra}, {Chauvie}, {Chytracek},
  {Cirrone}, {Cooperman}, {Cosmo}, {Cuttone}, {Daquino}, {Donszelmann},
  {Dressel}, {Folger}, {Foppiano}, {Generowicz}, {Grichine}, {Guatelli},
  {Gumplinger}, {Heikkinen}, {Hrivnacova}, {Howard}, {Incerti}, {Ivanchenko},
  {Johnson}, {Jones}, {Koi}, {Kokoulin}, {Kossov}, {Kurashige}, {Lara},
  {Larsson}, {Lei}, {Link}, {Longo}, {Maire}, {Mantero}, {Mascialino},
  {McLaren}, {Lorenzo}, {Minamimoto}, {Murakami}, {Nieminen}, {Pandola},
  {Parlati}, {Peralta}, {Perl}, {Pfeiffer}, {Pia}, {Ribon}, {Rodrigues},
  {Russo}, {Sadilov}, {Santin}, {Sasaki}, {Smith}, {Starkov}, {Tanaka},
  {Tcherniaev}, {Tome}, {Trindade}, {Truscott}, {Urban}, {Verderi}, {Walkden},
  {Wellisch}, {Williams}, {Wright}, and {Yoshida}]{Allison2006ITNS}
{Allison}, J.; {Amako}, K.; {Apostolakis}, J.; {Araujo}, H.; {Dubois}, P.A.;
  {Asai}, M.; {Barrand}, G.; {Capra}, R.; {Chauvie}, S.; {Chytracek}, R.;
  et~al.
\newblock {Geant4 developments and applications}.
\newblock {\em IEEE Transactions on Nuclear Science} {\bf 2006}, {\em
  53},~270--278.
\newblock
  doi:{\changeurlcolor{black}\href{https://doi.org/10.1109/TNS.2006.869826}{\detokenize{10.1109/TNS.2006.869826}}}.

\bibitem[{Allison} \em{et~al.}(2016){Allison}, {Amako}, {Apostolakis}, {Arce},
  {Asai}, {Aso}, {Bagli}, {Bagulya}, {Banerjee}, {Barrand}, {Beck}, {Bogdanov},
  {Brandt}, {Brown}, {Burkhardt}, {Canal}, {Cano-Ott}, {Chauvie}, {Cho},
  {Cirrone}, {Cooperman}, {Cort{\'e}s-Giraldo}, {Cosmo}, {Cuttone}, {Depaola},
  {Desorgher}, {Dong}, {Dotti}, {Elvira}, {Folger}, {Francis}, {Galoyan},
  {Garnier}, {Gayer}, {Genser}, {Grichine}, {Guatelli}, {Gu{\`e}ye},
  {Gumplinger}, {Howard}, {H{\v{r}}ivn{\'a}{\v{c}}ov{\'a}}, {Hwang}, {Incerti},
  {Ivanchenko}, {Ivanchenko}, {Jones}, {Jun}, {Kaitaniemi}, {Karakatsanis},
  {Karamitrosi}, {Kelsey}, {Kimura}, {Koi}, {Kurashige}, {Lechner}, {Lee},
  {Longo}, {Maire}, {Mancusi}, {Mantero}, {Mendoza}, {Morgan}, {Murakami},
  {Nikitina}, {Pandola}, {Paprocki}, {Perl}, {Petrovi{\'c}}, {Pia}, {Pokorski},
  {Quesada}, {Raine}, {Reis}, {Ribon}, {Risti{\'c} Fira}, {Romano}, {Russo},
  {Santin}, {Sasaki}, {Sawkey}, {Shin}, {Strakovsky}, {Taborda}, {Tanaka},
  {Tom{\'e}}, {Toshito}, {Tran}, {Truscott}, {Urban}, {Uzhinsky}, {Verbeke},
  {Verderi}, {Wendt}, {Wenzel}, {Wright}, {Wright}, {Yamashita}, {Yarba}, and
  {Yoshida}]{Allison2016NIMPA}
{Allison}, J.; {Amako}, K.; {Apostolakis}, J.; {Arce}, P.; {Asai}, M.; {Aso},
  T.; {Bagli}, E.; {Bagulya}, A.; {Banerjee}, S.; {Barrand}, G.;  et~al.
\newblock {Recent developments in GEANT4}.
\newblock {\em Nuclear Instruments and Methods in Physics Research A} {\bf
  2016}, {\em 835},~186--225.
\newblock
  doi:{\changeurlcolor{black}\href{https://doi.org/10.1016/j.nima.2016.06.125}{\detokenize{10.1016/j.nima.2016.06.125}}}.

\bibitem[{Agostinelli} \em{et~al.}(2003){Agostinelli}, {Allison}, {Amako},
  {Apostolakis}, {Araujo}, {Arce}, {Asai}, {Axen}, {Banerjee}, {Barrand},
  {Behner}, {Bellagamba}, {Boudreau}, {Broglia}, {Brunengo}, {Burkhardt},
  {Chauvie}, {Chuma}, {Chytracek}, {Cooperman}, {Cosmo}, {Degtyarenko},
  {Dell'Acqua}, {Depaola}, {Dietrich}, {Enami}, {Feliciello}, {Ferguson},
  {Fesefeldt}, {Folger}, {Foppiano}, {Forti}, {Garelli}, {Giani},
  {Giannitrapani}, {Gibin}, {G{\'o}mez Cadenas}, {Gonz{\'a}lez}, {Gracia
  Abril}, {Greeniaus}, {Greiner}, {Grichine}, {Grossheim}, {Guatelli},
  {Gumplinger}, {Hamatsu}, {Hashimoto}, {Hasui}, {Heikkinen}, {Howard},
  {Ivanchenko}, {Johnson}, {Jones}, {Kallenbach}, {Kanaya}, {Kawabata},
  {Kawabata}, {Kawaguti}, {Kelner}, {Kent}, {Kimura}, {Kodama}, {Kokoulin},
  {Kossov}, {Kurashige}, {Lamanna}, {Lamp{\'e}n}, {Lara}, {Lefebure}, {Lei},
  {Liendl}, {Lockman}, {Longo}, {Magni}, {Maire}, {Medernach}, {Minamimoto},
  {Mora de Freitas}, {Morita}, {Murakami}, {Nagamatu}, {Nartallo}, {Nieminen},
  {Nishimura}, {Ohtsubo}, {Okamura}, {O'Neale}, {Oohata}, {Paech}, {Perl},
  {Pfeiffer}, {Pia}, {Ranjard}, {Rybin}, {Sadilov}, {Di Salvo}, {Santin},
  {Sasaki}, {Savvas}, {Sawada}, {Scherer}, {Sei}, {Sirotenko}, {Smith},
  {Starkov}, {Stoecker}, {Sulkimo}, {Takahata}, {Tanaka}, {Tcherniaev}, {Safai
  Tehrani}, {Tropeano}, {Truscott}, {Uno}, {Urban}, {Urban}, {Verderi},
  {Walkden}, {Wander}, {Weber}, {Wellisch}, {Wenaus}, {Williams}, {Wright},
  {Yamada}, {Yoshida}, {Zschiesche}, and {G EANT4
  Collaboration}]{Agostinelli2003NIMPA}
{Agostinelli}, S.; {Allison}, J.; {Amako}, K.; {Apostolakis}, J.; {Araujo}, H.;
  {Arce}, P.; {Asai}, M.; {Axen}, D.; {Banerjee}, S.; {Barrand}, G.;  et~al.
\newblock {G EANT4{\textemdash}a simulation toolkit}.
\newblock {\em Nuclear Instruments and Methods in Physics Research A} {\bf
  2003}, {\em 506},~250--303.
\newblock
  doi:{\changeurlcolor{black}\href{https://doi.org/10.1016/S0168-9002(03)01368-8}{\detokenize{10.1016/S0168-9002(03)01368-8}}}.

\bibitem[{Sj{\"o}strand} \em{et~al.}(2008){Sj{\"o}strand}, {Mrenna}, and
  {Skands}]{Sjostrand2008CoPhC}
{Sj{\"o}strand}, T.; {Mrenna}, S.; {Skands}, P.
\newblock {A brief introduction to PYTHIA 8.1}.
\newblock {\em Computer Physics Communications} {\bf 2008}, {\em
  178},~852--867,  \href{http://xxx.lanl.gov/abs/0710.3820}{{\normalfont
  [arXiv:hep-ph/0710.3820]}}.
\newblock
  doi:{\changeurlcolor{black}\href{https://doi.org/10.1016/j.cpc.2008.01.036}{\detokenize{10.1016/j.cpc.2008.01.036}}}.

\bibitem[{Sj{\"o}strand} \em{et~al.}(2006){Sj{\"o}strand}, {Mrenna}, and
  {Skands}]{Sjostrand2006JHEP}
{Sj{\"o}strand}, T.; {Mrenna}, S.; {Skands}, P.
\newblock {PYTHIA 6.4 physics and manual}.
\newblock {\em Journal of High Energy Physics} {\bf 2006}, {\em 2006},~026,
  \href{http://xxx.lanl.gov/abs/hep-ph/0603175}{{\normalfont
  [arXiv:hep-ph/hep-ph/0603175]}}.
\newblock
  doi:{\changeurlcolor{black}\href{https://doi.org/10.1088/1126-6708/2006/05/026}{\detokenize{10.1088/1126-6708/2006/05/026}}}.

\bibitem[{Bierlich} \em{et~al.}(2022){Bierlich}, {Chakraborty}, {Desai},
  {Gellersen}, {Helenius}, {Ilten}, {L{\"o}nnblad}, {Mrenna}, {Prestel},
  {Preuss}, {Sj{\"o}strand}, {Skands}, {Utheim}, and
  {Verheyen}]{Bierlich2022arXiv220311601B}
{Bierlich}, C.; {Chakraborty}, S.; {Desai}, N.; {Gellersen}, L.; {Helenius},
  I.; {Ilten}, P.; {L{\"o}nnblad}, L.; {Mrenna}, S.; {Prestel}, S.; {Preuss},
  C.T.;  et~al.
\newblock {A comprehensive guide to the physics and usage of PYTHIA 8.3}.
\newblock {\em arXiv e-prints} {\bf 2022}, p. arXiv:2203.11601,
  \href{http://xxx.lanl.gov/abs/2203.11601}{{\normalfont
  [arXiv:hep-ph/2203.11601]}}.
\newblock
  doi:{\changeurlcolor{black}\href{https://doi.org/10.48550/arXiv.2203.11601}{\detokenize{10.48550/arXiv.2203.11601}}}.

\bibitem[{Engel} \em{et~al.}(1995){Engel}, {Ranft}, and
  {Roesler}]{Engel1995PhRvD}
{Engel}, R.; {Ranft}, J.; {Roesler}, S.
\newblock {Hard diffraction in hadron-hadron interactions and in
  photoproduction}.
\newblock {\em \prd} {\bf 1995}, {\em 52},~1459--1468,
  \href{http://xxx.lanl.gov/abs/hep-ph/9502319}{{\normalfont
  [arXiv:hep-ph/hep-ph/9502319]}}.
\newblock
  doi:{\changeurlcolor{black}\href{https://doi.org/10.1103/PhysRevD.52.1459}{\detokenize{10.1103/PhysRevD.52.1459}}}.

\bibitem[{Bopp} \em{et~al.}(2008){Bopp}, {Ranft}, {Engel}, and
  {Roesler}]{Bopp2008PhRvC}
{Bopp}, F.W.; {Ranft}, J.; {Engel}, R.; {Roesler}, S.
\newblock {Antiparticle to particle production ratios in hadron-hadron and d-Au
  collisions in the DPMJET-III Monte Carlo model}.
\newblock {\em \prc} {\bf 2008}, {\em 77},~014904,
  \href{http://xxx.lanl.gov/abs/hep-ph/0505035}{{\normalfont
  [arXiv:hep-ph/hep-ph/0505035]}}.
\newblock
  doi:{\changeurlcolor{black}\href{https://doi.org/10.1103/PhysRevC.77.014904}{\detokenize{10.1103/PhysRevC.77.014904}}}.

\bibitem[{Werner} \em{et~al.}(2006){Werner}, {Liu}, and
  {Pierog}]{Werner2006PhRvC}
{Werner}, K.; {Liu}, F.M.; {Pierog}, T.
\newblock {Parton ladder splitting and the rapidity dependence of transverse
  momentum spectra in deuteron-gold collisions at the BNL Relativistic Heavy
  Ion Collider}.
\newblock {\em \prc} {\bf 2006}, {\em 74},~044902,
  \href{http://xxx.lanl.gov/abs/hep-ph/0506232}{{\normalfont
  [arXiv:hep-ph/hep-ph/0506232]}}.
\newblock
  doi:{\changeurlcolor{black}\href{https://doi.org/10.1103/PhysRevC.74.044902}{\detokenize{10.1103/PhysRevC.74.044902}}}.

\bibitem[{Ahn} \em{et~al.}(2009){Ahn}, {Engel}, {Gaisser}, {Lipari}, and
  {Stanev}]{Ahn2009PhRvD}
{Ahn}, E.J.; {Engel}, R.; {Gaisser}, T.K.; {Lipari}, P.; {Stanev}, T.
\newblock {Cosmic ray interaction event generator SIBYLL 2.1}.
\newblock {\em \prd} {\bf 2009}, {\em 80},~094003,
  \href{http://xxx.lanl.gov/abs/0906.4113}{{\normalfont
  [arXiv:hep-ph/0906.4113]}}.
\newblock
  doi:{\changeurlcolor{black}\href{https://doi.org/10.1103/PhysRevD.80.094003}{\detokenize{10.1103/PhysRevD.80.094003}}}.

\bibitem[{Fletcher} \em{et~al.}(1994){Fletcher}, {Gaisser}, {Lipari}, and
  {Stanev}]{Fletcher1994PhRvD}
{Fletcher}, R.S.; {Gaisser}, T.K.; {Lipari}, P.; {Stanev}, T.
\newblock {sibyll: An event generator for simulation of high energy cosmic ray
  cascades}.
\newblock {\em \prd} {\bf 1994}, {\em 50},~5710--5731.
\newblock
  doi:{\changeurlcolor{black}\href{https://doi.org/10.1103/PhysRevD.50.5710}{\detokenize{10.1103/PhysRevD.50.5710}}}.

\bibitem[{Engel} \em{et~al.}(1992){Engel}, {Gaisser}, {Lipari}, and
  {Stanev}]{Engel1992PhRvD}
{Engel}, J.; {Gaisser}, T.K.; {Lipari}, P.; {Stanev}, T.
\newblock {Nucleus-nucleus collisions and interpretation of cosmic-ray
  cascades}.
\newblock {\em \prd} {\bf 1992}, {\em 46},~5013--5025.
\newblock
  doi:{\changeurlcolor{black}\href{https://doi.org/10.1103/PhysRevD.46.5013}{\detokenize{10.1103/PhysRevD.46.5013}}}.

\bibitem[{Fedynitch} \em{et~al.}(2019){Fedynitch}, {Riehn}, {Engel}, {Gaisser},
  and {Stanev}]{Fedynitch2019PhRvD}
{Fedynitch}, A.; {Riehn}, F.; {Engel}, R.; {Gaisser}, T.K.; {Stanev}, T.
\newblock {Hadronic interaction model uc(sibyll) 2.3 c and inclusive lepton
  fluxes}.
\newblock {\em \prd} {\bf 2019}, {\em 100},~103018,
  \href{http://xxx.lanl.gov/abs/1806.04140}{{\normalfont
  [arXiv:hep-ph/1806.04140]}}.
\newblock
  doi:{\changeurlcolor{black}\href{https://doi.org/10.1103/PhysRevD.100.103018}{\detokenize{10.1103/PhysRevD.100.103018}}}.

\bibitem[{Ostapchenko}(2006)]{Ostapchenko2006NuPhS}
{Ostapchenko}, S.
\newblock {QGSJET-II: towards reliable description of very high energy hadronic
  interactions}.
\newblock {\em Nuclear Physics B Proceedings Supplements} {\bf 2006}, {\em
  151},~143--146,  \href{http://xxx.lanl.gov/abs/hep-ph/0412332}{{\normalfont
  [arXiv:hep-ph/hep-ph/0412332]}}.
\newblock
  doi:{\changeurlcolor{black}\href{https://doi.org/10.1016/j.nuclphysbps.2005.07.026}{\detokenize{10.1016/j.nuclphysbps.2005.07.026}}}.

\bibitem[{Ostapchenko}(2007)]{Ostapchenko2007AIPC}
{Ostapchenko}, S.
\newblock {Status of QGSJET}.
\newblock  Collicers to Cosmic Rays; {Tripathi}, M.; {Breedon}, R.E., Eds.,
  2007, Vol. 928, {\em American Institute of Physics Conference Series}, pp.
  118--125,  \href{http://xxx.lanl.gov/abs/0706.3784}{{\normalfont
  [arXiv:hep-ph/0706.3784]}}.
\newblock
  doi:{\changeurlcolor{black}\href{https://doi.org/10.1063/1.2775904}{\detokenize{10.1063/1.2775904}}}.

\bibitem[{Kafexhiu} \em{et~al.}(2014){Kafexhiu}, {Aharonian}, {Taylor}, and
  {Vila}]{Kafexhiu2014}
{Kafexhiu}, E.; {Aharonian}, F.; {Taylor}, A.M.; {Vila}, G.S.
\newblock {Parametrization of gamma-ray production cross sections for p p
  interactions in a broad proton energy range from the kinematic threshold to
  PeV energies}.
\newblock {\em \prd} {\bf 2014}, {\em 90},~123014,
  \href{http://xxx.lanl.gov/abs/1406.7369}{{\normalfont
  [arXiv:astro-ph.HE/1406.7369]}}.
\newblock
  doi:{\changeurlcolor{black}\href{https://doi.org/10.1103/PhysRevD.90.123014}{\detokenize{10.1103/PhysRevD.90.123014}}}.

\bibitem[{Kelner} \em{et~al.}(2006){Kelner}, {Aharonian}, and
  {Bugayov}]{Kelner2006}
{Kelner}, S.R.; {Aharonian}, F.A.; {Bugayov}, V.V.
\newblock {Energy spectra of gamma rays, electrons, and neutrinos produced at
  proton-proton interactions in the very high energy regime}.
\newblock {\em \prd} {\bf 2006}, {\em 74},~034018,
  \href{http://xxx.lanl.gov/abs/astro-ph/0606058}{{\normalfont
  [astro-ph/0606058]}}.
\newblock
  doi:{\changeurlcolor{black}\href{https://doi.org/10.1103/PhysRevD.74.034018}{\detokenize{10.1103/PhysRevD.74.034018}}}.

\bibitem[{Kamae} \em{et~al.}(2006){Kamae}, {Karlsson}, {Mizuno}, {Abe}, and
  {Koi}]{Kamae2006ApJ}
{Kamae}, T.; {Karlsson}, N.; {Mizuno}, T.; {Abe}, T.; {Koi}, T.
\newblock {Parameterization of {\ensuremath{\gamma}}, e$^{+/-}$, and Neutrino
  Spectra Produced by p-p Interaction in Astronomical Environments}.
\newblock {\em \apj} {\bf 2006}, {\em 647},~692--708,
  \href{http://xxx.lanl.gov/abs/astro-ph/0605581}{{\normalfont
  [arXiv:astro-ph/astro-ph/0605581]}}.
\newblock
  doi:{\changeurlcolor{black}\href{https://doi.org/10.1086/505189}{\detokenize{10.1086/505189}}}.

\bibitem[{Kachelrie{\ss}} \em{et~al.}(2019){Kachelrie{\ss}}, {Moskalenko}, and
  {Ostapchenko}]{Kachelriess2019CoPhC}
{Kachelrie{\ss}}, M.; {Moskalenko}, I.V.; {Ostapchenko}, S.
\newblock {AAfrag: Interpolation routines for Monte Carlo results on secondary
  production in proton-proton, proton-nucleus and nucleus-nucleus
  interactions}.
\newblock {\em Computer Physics Communications} {\bf 2019}, {\em 245},~106846,
  \href{http://xxx.lanl.gov/abs/1904.05129}{{\normalfont
  [arXiv:hep-ph/1904.05129]}}.
\newblock
  doi:{\changeurlcolor{black}\href{https://doi.org/10.1016/j.cpc.2019.08.001}{\detokenize{10.1016/j.cpc.2019.08.001}}}.

\bibitem[{Koldobskiy} \em{et~al.}(2021){Koldobskiy}, {Kachelrie{\ss}},
  {Lskavyan}, {Neronov}, {Ostapchenko}, and {Semikoz}]{Koldobskiy2021PhRvD}
{Koldobskiy}, S.; {Kachelrie{\ss}}, M.; {Lskavyan}, A.; {Neronov}, A.;
  {Ostapchenko}, S.; {Semikoz}, D.V.
\newblock {Energy spectra of secondaries in proton-proton interactions}.
\newblock {\em \prd} {\bf 2021}, {\em 104},~123027,
  \href{http://xxx.lanl.gov/abs/2110.00496}{{\normalfont
  [arXiv:astro-ph.HE/2110.00496]}}.
\newblock
  doi:{\changeurlcolor{black}\href{https://doi.org/10.1103/PhysRevD.104.123027}{\detokenize{10.1103/PhysRevD.104.123027}}}.

\bibitem[{Hooper} and {Plant}(2023)]{Hooper2023arXiv230506375H}
{Hooper}, D.; {Plant}, K.
\newblock {A Leptonic Model for Neutrino Emission From Active Galactic Nuclei}.
\newblock {\em arXiv e-prints} {\bf 2023}, p. arXiv:2305.06375,
  \href{http://xxx.lanl.gov/abs/2305.06375}{{\normalfont
  [arXiv:astro-ph.HE/2305.06375]}}.
\newblock
  doi:{\changeurlcolor{black}\href{https://doi.org/10.48550/arXiv.2305.06375}{\detokenize{10.48550/arXiv.2305.06375}}}.

\bibitem[{M\"{u}cke} \em{et~al.}(1999){M\"{u}cke}, {Rachen}, {Engel},
  {Protheroe}, and {Stanev}]{Mucke1999PASA}
{M\"{u}cke}, A.; {Rachen}, J.P.; {Engel}, R.; {Protheroe}, R.J.; {Stanev}, T.
\newblock {Photohadronic Processes in Astrophysical Environments}.
\newblock {\em \pasa} {\bf 1999}, {\em 16},~160,
  \href{http://xxx.lanl.gov/abs/astro-ph/9808279}{{\normalfont
  [astro-ph/9808279]}}.
\newblock
  doi:{\changeurlcolor{black}\href{https://doi.org/10.1071/AS99160}{\detokenize{10.1071/AS99160}}}.

\bibitem[Berezinsky and Gazizov(1993)]{Berezinsky1993}
Berezinsky, V.S.; Gazizov, A.Z.
\newblock Production of high-energy cosmic neutrinos in $p\ensuremath{\gamma}$
  and $n\ensuremath{\gamma}$ scattering. I. Neutrino yields for power-law
  spectra of protons and neutrons.
\newblock {\em Phys. Rev. D} {\bf 1993}, {\em 47},~4206.
\newblock
  doi:{\changeurlcolor{black}\href{https://doi.org/10.1103/PhysRevD.47.4206}{\detokenize{10.1103/PhysRevD.47.4206}}}.

\bibitem[Nakamura(2010)]{Nakamura2010}
Nakamura, K.
\newblock Review of Particle Physics.
\newblock {\em Journal of Physics G: Nuclear and Particle Physics} {\bf 2010},
  {\em 37},~075021.

\bibitem[H\"{u}mmer \em{et~al.}(2010)H\"{u}mmer, R\"{u}ger, Spanier, and
  Winter]{Hummer2010}
H\"{u}mmer, S.; R\"{u}ger, M.; Spanier, F.; Winter, W.
\newblock Simplified Models for Photohadronic Interactions in Cosmic
  Accelerators.
\newblock {\em \apj} {\bf 2010}, {\em 721},~630.
\newblock
  doi:{\changeurlcolor{black}\href{https://doi.org/10.1088/0004-637X/721/1/630}{\detokenize{10.1088/0004-637X/721/1/630}}}.

\bibitem[Dermer and Menon(2009)]{Dermer2009book}
Dermer, C.D.; Menon, G.
\newblock {\em {High energy radiation from black holes: gamma rays, cosmic
  rays, and neutrinos}}; Princeton series in astrophysics, Princeton University
  Press: Princeton, NJ,  2009.

\bibitem[{M{\"u}cke} \em{et~al.}(2000){M{\"u}cke}, {Engel}, {Rachen},
  {Protheroe}, and {Stanev}]{Mucke2000CoPhC}
{M{\"u}cke}, A.; {Engel}, R.; {Rachen}, J.P.; {Protheroe}, R.J.; {Stanev}, T.
\newblock {Monte Carlo simulations of photohadronic processes in astrophysics}.
\newblock {\em Computer Physics Communications} {\bf 2000}, {\em
  124},~290--314,  \href{http://xxx.lanl.gov/abs/astro-ph/9903478}{{\normalfont
  [arXiv:astro-ph/astro-ph/9903478]}}.
\newblock
  doi:{\changeurlcolor{black}\href{https://doi.org/10.1016/S0010-4655(99)00446-4}{\detokenize{10.1016/S0010-4655(99)00446-4}}}.

\bibitem[{H{\"u}mmer} \em{et~al.}(2010){H{\"u}mmer}, {R{\"u}ger}, {Spanier},
  and {Winter}]{Hummer2010ApJ}
{H{\"u}mmer}, S.; {R{\"u}ger}, M.; {Spanier}, F.; {Winter}, W.
\newblock {Simplified Models for Photohadronic Interactions in Cosmic
  Accelerators}.
\newblock {\em \apj} {\bf 2010}, {\em 721},~630--652,
  \href{http://xxx.lanl.gov/abs/1002.1310}{{\normalfont
  [arXiv:astro-ph.HE/1002.1310]}}.
\newblock
  doi:{\changeurlcolor{black}\href{https://doi.org/10.1088/0004-637X/721/1/630}{\detokenize{10.1088/0004-637X/721/1/630}}}.

\bibitem[{Kelner} and {Aharonian}(2008)]{Kelner2008}
{Kelner}, S.R.; {Aharonian}, F.A.
\newblock {Energy spectra of gamma rays, electrons, and neutrinos produced at
  interactions of relativistic protons with low energy radiation}.
\newblock {\em \prd} {\bf 2008}, {\em 78},~034013,
  \href{http://xxx.lanl.gov/abs/0803.0688}{{\normalfont [0803.0688]}}.
\newblock
  doi:{\changeurlcolor{black}\href{https://doi.org/10.1103/PhysRevD.78.034013}{\detokenize{10.1103/PhysRevD.78.034013}}}.

\bibitem[{Dermer} and {Menon}(2009)]{Dermer2009herbbook}
{Dermer}, C.D.; {Menon}, G.
\newblock {\em {High Energy Radiation from Black Holes: Gamma Rays, Cosmic
  Rays, and Neutrinos}};  2009.

\bibitem[{Draine}(2011)]{Draine2011piim}
{Draine}, B.T.
\newblock {\em {Physics of the Interstellar and Intergalactic Medium}};  2011.

\bibitem[{Yoast-Hull} and {Murray}(2019)]{YoastHull2019MNRAS}
{Yoast-Hull}, T.M.; {Murray}, N.
\newblock {Breaking the radio - gamma-ray connection in Arp 220}.
\newblock {\em \mnras} {\bf 2019}, {\em 484},~3665--3680,
  \href{http://xxx.lanl.gov/abs/1901.04564}{{\normalfont
  [arXiv:astro-ph.HE/1901.04564]}}.
\newblock
  doi:{\changeurlcolor{black}\href{https://doi.org/10.1093/mnras/stz223}{\detokenize{10.1093/mnras/stz223}}}.

\bibitem[{Wilson} \em{et~al.}(2014){Wilson}, {Rangwala}, {Glenn}, {Maloney},
  {Spinoglio}, and {Pereira-Santaella}]{Wilson2014ApJ}
{Wilson}, C.D.; {Rangwala}, N.; {Glenn}, J.; {Maloney}, P.R.; {Spinoglio}, L.;
  {Pereira-Santaella}, M.
\newblock {Extreme Dust Disks in Arp 220 as Revealed by ALMA}.
\newblock {\em \apjl} {\bf 2014}, {\em 789},~L36,
  \href{http://xxx.lanl.gov/abs/1406.4530}{{\normalfont
  [arXiv:astro-ph.GA/1406.4530]}}.
\newblock
  doi:{\changeurlcolor{black}\href{https://doi.org/10.1088/2041-8205/789/2/L36}{\detokenize{10.1088/2041-8205/789/2/L36}}}.

\bibitem[{Scoville} \em{et~al.}(2017){Scoville}, {Murchikova}, {Walter},
  {Vlahakis}, {Koda}, {Vanden Bout}, {Barnes}, {Hernquist}, {Sheth}, {Yun},
  {Sanders}, {Armus}, {Cox}, {Thompson}, {Robertson}, {Zschaechner}, {Tacconi},
  {Torrey}, {Hayward}, {Genzel}, {Hopkins}, {van der Werf}, and
  {Decarli}]{Scoville2017ApJ}
{Scoville}, N.; {Murchikova}, L.; {Walter}, F.; {Vlahakis}, C.; {Koda}, J.;
  {Vanden Bout}, P.; {Barnes}, J.; {Hernquist}, L.; {Sheth}, K.; {Yun}, M.;
  et~al.
\newblock {ALMA Resolves the Nuclear Disks of Arp 220}.
\newblock {\em \apj} {\bf 2017}, {\em 836},~66,
  \href{http://xxx.lanl.gov/abs/1605.09381}{{\normalfont
  [arXiv:astro-ph.GA/1605.09381]}}.
\newblock
  doi:{\changeurlcolor{black}\href{https://doi.org/10.3847/1538-4357/836/1/66}{\detokenize{10.3847/1538-4357/836/1/66}}}.

\bibitem[{Rowan-Robinson}(2000)]{RR2000MNRAS}
{Rowan-Robinson}, M.
\newblock {Hyperluminous infrared galaxies}.
\newblock {\em \mnras} {\bf 2000}, {\em 316},~885--900,
  \href{http://xxx.lanl.gov/abs/astro-ph/9912286}{{\normalfont
  [arXiv:astro-ph/astro-ph/9912286]}}.
\newblock
  doi:{\changeurlcolor{black}\href{https://doi.org/10.1046/j.1365-8711.2000.03588.x}{\detokenize{10.1046/j.1365-8711.2000.03588.x}}}.

\bibitem[{Chakraborty} and {Fields}(2013)]{Chakraborty2013ApJ}
{Chakraborty}, N.; {Fields}, B.D.
\newblock {Inverse-Compton Contribution to the Star-forming Extragalactic
  Gamma-Ray Background}.
\newblock {\em \apj} {\bf 2013}, {\em 773},~104,
  \href{http://xxx.lanl.gov/abs/1206.0770}{{\normalfont
  [arXiv:astro-ph.CO/1206.0770]}}.
\newblock
  doi:{\changeurlcolor{black}\href{https://doi.org/10.1088/0004-637X/773/2/104}{\detokenize{10.1088/0004-637X/773/2/104}}}.

\bibitem[{Schober} \em{et~al.}(2015){Schober}, {Schleicher}, and
  {Klessen}]{Schober2015MNRAS}
{Schober}, J.; {Schleicher}, D.R.G.; {Klessen}, R.S.
\newblock {X-ray emission from star-forming galaxies - signatures of cosmic
  rays and magnetic fields}.
\newblock {\em \mnras} {\bf 2015}, {\em 446},~2--17,
  \href{http://xxx.lanl.gov/abs/1404.2578}{{\normalfont
  [arXiv:astro-ph.GA/1404.2578]}}.
\newblock
  doi:{\changeurlcolor{black}\href{https://doi.org/10.1093/mnras/stu1999}{\detokenize{10.1093/mnras/stu1999}}}.

\bibitem[{Wilman} \em{et~al.}(1998){Wilman}, {Fabian}, {Cutri}, {Crawford}, and
  {Brandt}]{Wilman1998MNRAS}
{Wilman}, R.J.; {Fabian}, A.C.; {Cutri}, R.M.; {Crawford}, C.S.; {Brandt}, W.N.
\newblock {Limits on the X-ray emission from several hyperluminous IRAS
  galaxies}.
\newblock {\em \mnras} {\bf 1998}, {\em 300},~L7--L10,
  \href{http://xxx.lanl.gov/abs/astro-ph/9808324}{{\normalfont
  [arXiv:astro-ph/astro-ph/9808324]}}.
\newblock
  doi:{\changeurlcolor{black}\href{https://doi.org/10.1046/j.1365-8711.1998.02047.x}{\detokenize{10.1046/j.1365-8711.1998.02047.x}}}.

\bibitem[{Danielson} \em{et~al.}(2011){Danielson}, {Swinbank}, {Smail}, {Cox},
  {Edge}, {Weiss}, {Harris}, {Baker}, {De Breuck}, {Geach}, {Ivison}, {Krips},
  {Lundgren}, {Longmore}, {Neri}, and {Flaquer}]{Danielson2011MNRAS}
{Danielson}, A.L.R.; {Swinbank}, A.M.; {Smail}, I.; {Cox}, P.; {Edge}, A.C.;
  {Weiss}, A.; {Harris}, A.I.; {Baker}, A.J.; {De Breuck}, C.; {Geach}, J.E.;
  et~al.
\newblock {The properties of the interstellar medium within a star-forming
  galaxy at z= 2.3}.
\newblock {\em \mnras} {\bf 2011}, {\em 410},~1687--1702,
  \href{http://xxx.lanl.gov/abs/1008.3183}{{\normalfont
  [arXiv:astro-ph.CO/1008.3183]}}.
\newblock
  doi:{\changeurlcolor{black}\href{https://doi.org/10.1111/j.1365-2966.2010.17549.x}{\detokenize{10.1111/j.1365-2966.2010.17549.x}}}.

\bibitem[{Hashimoto} \em{et~al.}(2018){Hashimoto}, {Laporte}, {Mawatari},
  {Ellis}, {Inoue}, {Zackrisson}, {Roberts-Borsani}, {Zheng}, {Tamura},
  {Bauer}, {Fletcher}, {Harikane}, {Hatsukade}, {Hayatsu}, {Matsuda}, {Matsuo},
  {Okamoto}, {Ouchi}, {Pell{\'o}}, {Rydberg}, {Shimizu}, {Taniguchi},
  {Umehata}, and {Yoshida}]{Hashimoto2018Natur}
{Hashimoto}, T.; {Laporte}, N.; {Mawatari}, K.; {Ellis}, R.S.; {Inoue}, A.K.;
  {Zackrisson}, E.; {Roberts-Borsani}, G.; {Zheng}, W.; {Tamura}, Y.; {Bauer},
  F.E.;  et~al.
\newblock {The onset of star formation 250 million years after the Big Bang}.
\newblock {\em \nat} {\bf 2018}, {\em 557},~392--395,
  \href{http://xxx.lanl.gov/abs/1805.05966}{{\normalfont
  [arXiv:astro-ph.GA/1805.05966]}}.
\newblock
  doi:{\changeurlcolor{black}\href{https://doi.org/10.1038/s41586-018-0117-z}{\detokenize{10.1038/s41586-018-0117-z}}}.

\bibitem[{Bohlin} \em{et~al.}(1978){Bohlin}, {Savage}, and
  {Drake}]{Bohlin1978ApJ}
{Bohlin}, R.C.; {Savage}, B.D.; {Drake}, J.F.
\newblock {A survey of interstellar H I from Lalpha absorption measurements.
  II.}
\newblock {\em \apj} {\bf 1978}, {\em 224},~132--142.
\newblock
  doi:{\changeurlcolor{black}\href{https://doi.org/10.1086/156357}{\detokenize{10.1086/156357}}}.

\bibitem[{Bykov} \em{et~al.}(2018){Bykov}, {Ellison}, {Marcowith}, and
  {Osipov}]{Bykov2018SSRv}
{Bykov}, A.M.; {Ellison}, D.C.; {Marcowith}, A.; {Osipov}, S.M.
\newblock {Cosmic Ray Production in Supernovae}.
\newblock {\em \ssr} {\bf 2018}, {\em 214},~41,
  \href{http://xxx.lanl.gov/abs/1801.08890}{{\normalfont
  [arXiv:astro-ph.HE/1801.08890]}}.
\newblock
  doi:{\changeurlcolor{black}\href{https://doi.org/10.1007/s11214-018-0479-4}{\detokenize{10.1007/s11214-018-0479-4}}}.

\bibitem[{Berezinskii} \em{et~al.}(1984){Berezinskii}, {Bulanov}, {Ginzburg},
  {Dogel}, and {Ptuskin}]{Berezinskii1984acr}
{Berezinskii}, V.S.; {Bulanov}, S.V.; {Ginzburg}, V.L.; {Dogel}, V.A.;
  {Ptuskin}, V.S.
\newblock {\em {Astrophysics of cosmic rays.}};  1984.

\bibitem[{Lingenfelter}(2013)]{Lingenfelter2013AIPC}
{Lingenfelter}, R.E.
\newblock {Superbubble origin of cosmic rays}.
\newblock  Centenary Symposium 2012: Discovery of Cosmic Rays; {Ormes}, J.F.,
  Ed.,  2013, Vol. 1516, {\em American Institute of Physics Conference Series},
  pp. 162--166,  \href{http://xxx.lanl.gov/abs/1209.5728}{{\normalfont
  [arXiv:astro-ph.HE/1209.5728]}}.
\newblock
  doi:{\changeurlcolor{black}\href{https://doi.org/10.1063/1.4792561}{\detokenize{10.1063/1.4792561}}}.

\bibitem[{Cristofari}(2021)]{Cristofari2021Univ}
{Cristofari}, P.
\newblock {The Hunt for Pevatrons: The Case of Supernova Remnants}.
\newblock {\em Universe} {\bf 2021}, {\em 7},~324,
  \href{http://xxx.lanl.gov/abs/2110.07956}{{\normalfont
  [arXiv:astro-ph.HE/2110.07956]}}.
\newblock
  doi:{\changeurlcolor{black}\href{https://doi.org/10.3390/universe7090324}{\detokenize{10.3390/universe7090324}}}.

\bibitem[{Li} \em{et~al.}(2011){Li}, {Chornock}, {Leaman}, {Filippenko},
  {Poznanski}, {Wang}, {Ganeshalingam}, and {Mannucci}]{Li2011MNRAS}
{Li}, W.; {Chornock}, R.; {Leaman}, J.; {Filippenko}, A.V.; {Poznanski}, D.;
  {Wang}, X.; {Ganeshalingam}, M.; {Mannucci}, F.
\newblock {Nearby supernova rates from the Lick Observatory Supernova Search -
  III. The rate-size relation, and the rates as a function of galaxy Hubble
  type and colour}.
\newblock {\em \mnras} {\bf 2011}, {\em 412},~1473--1507,
  \href{http://xxx.lanl.gov/abs/1006.4613}{{\normalfont
  [arXiv:astro-ph.SR/1006.4613]}}.
\newblock
  doi:{\changeurlcolor{black}\href{https://doi.org/10.1111/j.1365-2966.2011.18162.x}{\detokenize{10.1111/j.1365-2966.2011.18162.x}}}.

\bibitem[{Lingenfelter}(2018)]{Lingenfelter2018AdSpR}
{Lingenfelter}, R.E.
\newblock {Cosmic rays from supernova remnants and superbubbles}.
\newblock {\em Advances in Space Research} {\bf 2018}, {\em 62},~2750--2763,
  \href{http://xxx.lanl.gov/abs/1807.09726}{{\normalfont
  [arXiv:astro-ph.HE/1807.09726]}}.
\newblock
  doi:{\changeurlcolor{black}\href{https://doi.org/10.1016/j.asr.2017.04.006}{\detokenize{10.1016/j.asr.2017.04.006}}}.

\bibitem[{Marcowith} \em{et~al.}(2018){Marcowith}, {Dwarkadas}, {Renaud},
  {Tatischeff}, and {Giacinti}]{Marcowith2018MNRAS}
{Marcowith}, A.; {Dwarkadas}, V.V.; {Renaud}, M.; {Tatischeff}, V.; {Giacinti},
  G.
\newblock {Core-collapse supernovae as cosmic ray sources}.
\newblock {\em \mnras} {\bf 2018}, {\em 479},~4470--4485,
  \href{http://xxx.lanl.gov/abs/1806.09700}{{\normalfont
  [arXiv:astro-ph.HE/1806.09700]}}.
\newblock
  doi:{\changeurlcolor{black}\href{https://doi.org/10.1093/mnras/sty1743}{\detokenize{10.1093/mnras/sty1743}}}.

\bibitem[{Murase} \em{et~al.}(2019){Murase}, {Franckowiak}, {Maeda},
  {Margutti}, and {Beacom}]{Murase2019ApJ}
{Murase}, K.; {Franckowiak}, A.; {Maeda}, K.; {Margutti}, R.; {Beacom}, J.F.
\newblock {High-energy Emission from Interacting Supernovae: New Constraints on
  Cosmic-Ray Acceleration in Dense Circumstellar Environments}.
\newblock {\em \apj} {\bf 2019}, {\em 874},~80,
  \href{http://xxx.lanl.gov/abs/1807.01460}{{\normalfont
  [arXiv:astro-ph.HE/1807.01460]}}.
\newblock
  doi:{\changeurlcolor{black}\href{https://doi.org/10.3847/1538-4357/ab0422}{\detokenize{10.3847/1538-4357/ab0422}}}.

\bibitem[{Hillas}(1984)]{Hillas1984ARA&A}
{Hillas}, A.M.
\newblock {The Origin of Ultra-High-Energy Cosmic Rays}.
\newblock {\em \araa} {\bf 1984}, {\em 22},~425--444.
\newblock
  doi:{\changeurlcolor{black}\href{https://doi.org/10.1146/annurev.aa.22.090184.002233}{\detokenize{10.1146/annurev.aa.22.090184.002233}}}.

\bibitem[{Achterberg}(1983)]{Achterberg1983A&A}
{Achterberg}, A.
\newblock {Modification of scattering waves and its importance for shock
  acceleration}.
\newblock {\em \aap} {\bf 1983}, {\em 119},~274--278.

\bibitem[{Bell} \em{et~al.}(2013){Bell}, {Schure}, {Reville}, and
  {Giacinti}]{Bell2013MNRAS}
{Bell}, A.R.; {Schure}, K.M.; {Reville}, B.; {Giacinti}, G.
\newblock {Cosmic-ray acceleration and escape from supernova remnants}.
\newblock {\em \mnras} {\bf 2013}, {\em 431},~415--429,
  \href{http://xxx.lanl.gov/abs/1301.7264}{{\normalfont
  [arXiv:astro-ph.HE/1301.7264]}}.
\newblock
  doi:{\changeurlcolor{black}\href{https://doi.org/10.1093/mnras/stt179}{\detokenize{10.1093/mnras/stt179}}}.

\bibitem[{Schure} and {Bell}(2013)]{Schure2013MNRAS}
{Schure}, K.M.; {Bell}, A.R.
\newblock {Cosmic ray acceleration in young supernova remnants}.
\newblock {\em \mnras} {\bf 2013}, {\em 435},~1174--1185,
  \href{http://xxx.lanl.gov/abs/1307.6575}{{\normalfont
  [arXiv:astro-ph.HE/1307.6575]}}.
\newblock
  doi:{\changeurlcolor{black}\href{https://doi.org/10.1093/mnras/stt1371}{\detokenize{10.1093/mnras/stt1371}}}.

\bibitem[{Schure} and {Bell}(2014)]{Schure2014MNRAS}
{Schure}, K.M.; {Bell}, A.R.
\newblock {From cosmic ray source to the Galactic pool}.
\newblock {\em \mnras} {\bf 2014}, {\em 437},~2802--2805,
  \href{http://xxx.lanl.gov/abs/1310.7027}{{\normalfont
  [arXiv:astro-ph.HE/1310.7027]}}.
\newblock
  doi:{\changeurlcolor{black}\href{https://doi.org/10.1093/mnras/stt2089}{\detokenize{10.1093/mnras/stt2089}}}.

\bibitem[{Casanova}(2022)]{Casanova2022Univ}
{Casanova}, S.
\newblock {On the Search for the Galactic PeVatrons by Means of Gamma-Ray
  Astronomy}.
\newblock {\em Universe} {\bf 2022}, {\em 8},~505.
\newblock
  doi:{\changeurlcolor{black}\href{https://doi.org/10.3390/universe8100505}{\detokenize{10.3390/universe8100505}}}.

\bibitem[{Petropoulou} \em{et~al.}(2017){Petropoulou}, {Coenders},
  {Vasilopoulos}, {Kamble}, and {Sironi}]{Petropoulou2017MNRAS}
{Petropoulou}, M.; {Coenders}, S.; {Vasilopoulos}, G.; {Kamble}, A.; {Sironi},
  L.
\newblock {Point-source and diffuse high-energy neutrino emission from Type IIn
  supernovae}.
\newblock {\em \mnras} {\bf 2017}, {\em 470},~1881--1893,
  \href{http://xxx.lanl.gov/abs/1705.06752}{{\normalfont
  [arXiv:astro-ph.HE/1705.06752]}}.
\newblock
  doi:{\changeurlcolor{black}\href{https://doi.org/10.1093/mnras/stx1251}{\detokenize{10.1093/mnras/stx1251}}}.

\bibitem[{Waxman} and {Loeb}(2001)]{Waxman2001PhRvL}
{Waxman}, E.; {Loeb}, A.
\newblock {TeV Neutrinos and GeV Photons from Shock Breakout in Supernovae}.
\newblock {\em \prl} {\bf 2001}, {\em 87},~071101,
  \href{http://xxx.lanl.gov/abs/astro-ph/0102317}{{\normalfont
  [arXiv:astro-ph/astro-ph/0102317]}}.
\newblock
  doi:{\changeurlcolor{black}\href{https://doi.org/10.1103/PhysRevLett.87.071101}{\detokenize{10.1103/PhysRevLett.87.071101}}}.

\bibitem[{Chevalier} and {Fransson}(2008)]{Chevalier2008ApJ}
{Chevalier}, R.A.; {Fransson}, C.
\newblock {Shock Breakout Emission from a Type Ib/c Supernova: XRT 080109/SN
  2008D}.
\newblock {\em \apjl} {\bf 2008}, {\em 683},~L135,
  \href{http://xxx.lanl.gov/abs/0806.0371}{{\normalfont
  [arXiv:astro-ph/0806.0371]}}.
\newblock
  doi:{\changeurlcolor{black}\href{https://doi.org/10.1086/591522}{\detokenize{10.1086/591522}}}.

\bibitem[{Ensman} and {Burrows}(1992)]{Ensman1992ApJ}
{Ensman}, L.; {Burrows}, A.
\newblock {Shock Breakout in SN 1987A}.
\newblock {\em \apj} {\bf 1992}, {\em 393},~742.
\newblock
  doi:{\changeurlcolor{black}\href{https://doi.org/10.1086/171542}{\detokenize{10.1086/171542}}}.

\bibitem[{Chevalier} and {Klein}(1979)]{Chevalier1979ApJ}
{Chevalier}, R.A.; {Klein}, R.I.
\newblock {Nonequilibrium processes in the evolution of type II supernovae.}
\newblock {\em \apj} {\bf 1979}, {\em 234},~597--608.
\newblock
  doi:{\changeurlcolor{black}\href{https://doi.org/10.1086/157534}{\detokenize{10.1086/157534}}}.

\bibitem[{Giacinti} and {Bell}(2015)]{Giacinti2015MNRAS}
{Giacinti}, G.; {Bell}, A.R.
\newblock {Collisionless shocks and TeV neutrinos before Supernova shock
  breakout from an optically thick wind}.
\newblock {\em \mnras} {\bf 2015}, {\em 449},~3693--3699,
  \href{http://xxx.lanl.gov/abs/1503.04170}{{\normalfont
  [arXiv:astro-ph.HE/1503.04170]}}.
\newblock
  doi:{\changeurlcolor{black}\href{https://doi.org/10.1093/mnras/stv561}{\detokenize{10.1093/mnras/stv561}}}.

\bibitem[{Bell}(1978{\natexlab{a}})]{Bell1978MNRASa}
{Bell}, A.R.
\newblock {The acceleration of cosmic rays in shock fronts - II.}
\newblock {\em \mnras} {\bf 1978}, {\em 182},~443--455.
\newblock
  doi:{\changeurlcolor{black}\href{https://doi.org/10.1093/mnras/182.3.443}{\detokenize{10.1093/mnras/182.3.443}}}.

\bibitem[{Bell}(1978{\natexlab{b}})]{Bell1978MNRASb}
{Bell}, A.R.
\newblock {The acceleration of cosmic rays in shock fronts - I.}
\newblock {\em \mnras} {\bf 1978}, {\em 182},~147--156.
\newblock
  doi:{\changeurlcolor{black}\href{https://doi.org/10.1093/mnras/182.2.147}{\detokenize{10.1093/mnras/182.2.147}}}.

\bibitem[{Celli} \em{et~al.}(2019){Celli}, {Morlino}, {Gabici}, and
  {Aharonian}]{Celli2019MNRAS}
{Celli}, S.; {Morlino}, G.; {Gabici}, S.; {Aharonian}, F.A.
\newblock {Exploring particle escape in supernova remnants through gamma rays}.
\newblock {\em \mnras} {\bf 2019}, {\em 490},~4317--4333,
  \href{http://xxx.lanl.gov/abs/1906.09454}{{\normalfont
  [arXiv:astro-ph.HE/1906.09454]}}.
\newblock
  doi:{\changeurlcolor{black}\href{https://doi.org/10.1093/mnras/stz2897}{\detokenize{10.1093/mnras/stz2897}}}.

\bibitem[{Truelove} and {McKee}(1999)]{Truelove1999ApJS}
{Truelove}, J.K.; {McKee}, C.F.
\newblock {Evolution of Nonradiative Supernova Remnants}.
\newblock {\em \apjs} {\bf 1999}, {\em 120},~299--326.
\newblock
  doi:{\changeurlcolor{black}\href{https://doi.org/10.1086/313176}{\detokenize{10.1086/313176}}}.

\bibitem[{Drury}(2011)]{Drury2011MNRAS}
{Drury}, L.O.
\newblock {Escaping the accelerator: how, when and in what numbers do cosmic
  rays get out of supernova remnants?}
\newblock {\em \mnras} {\bf 2011}, {\em 415},~1807--1814,
  \href{http://xxx.lanl.gov/abs/1009.4799}{{\normalfont
  [arXiv:astro-ph.GA/1009.4799]}}.
\newblock
  doi:{\changeurlcolor{black}\href{https://doi.org/10.1111/j.1365-2966.2011.18824.x}{\detokenize{10.1111/j.1365-2966.2011.18824.x}}}.

\bibitem[{Peron} \em{et~al.}(2020){Peron}, {Aharonian}, {Casanova}, {Zanin},
  and {Romoli}]{Peron2020ApJ}
{Peron}, G.; {Aharonian}, F.; {Casanova}, S.; {Zanin}, R.; {Romoli}, C.
\newblock {On the Gamma-Ray Emission of W44 and Its Surroundings}.
\newblock {\em \apjl} {\bf 2020}, {\em 896},~L23,
  \href{http://xxx.lanl.gov/abs/2007.04821}{{\normalfont
  [arXiv:astro-ph.HE/2007.04821]}}.
\newblock
  doi:{\changeurlcolor{black}\href{https://doi.org/10.3847/2041-8213/ab93d1}{\detokenize{10.3847/2041-8213/ab93d1}}}.

\bibitem[{Feijen} \em{et~al.}(2022){Feijen}, {Einecke}, {Rowell}, {Braiding},
  {Burton}, and {Wong}]{Feijen2022MNRAS}
{Feijen}, K.; {Einecke}, S.; {Rowell}, G.; {Braiding}, C.; {Burton}, M.G.;
  {Wong}, G.F.
\newblock {Modelling the gamma-ray morphology of HESSJ1804-216 from two
  supernova remnants in a hadronic scenario}.
\newblock {\em \mnras} {\bf 2022}, {\em 511},~5915--5926,
  \href{http://xxx.lanl.gov/abs/2201.11387}{{\normalfont
  [arXiv:astro-ph.HE/2201.11387]}}.
\newblock
  doi:{\changeurlcolor{black}\href{https://doi.org/10.1093/mnras/stac320}{\detokenize{10.1093/mnras/stac320}}}.

\bibitem[{H.~E.~S.~S. Collaboration} \em{et~al.}(2019){H.~E.~S.~S.
  Collaboration}, {Abdalla}, {Aharonian}, {Ait Benkhali}, {Ang{\"u}ner},
  {Arakawa}, {Arcaro}, {Armand}, {Ashkar}, {Backes}, {Barbosa Martins},
  {Barnard}, {Becherini}, {Berge}, {Bernl{\"o}hr}, {Blackwell}, {B{\"o}ttcher},
  {Boisson}, {Bolmont}, {Bonnefoy}, {Bregeon}, {Breuhaus}, {Brun}, {Brun},
  {Bryan}, {B{\"u}chele}, {Bulik}, {Bylund}, {Capasso}, {Caroff}, {Carosi},
  {Casanova}, {Cerruti}, {Chakraborty}, {Chand}, {Chandra}, {Chaves}, {Chen},
  {Colafrancesco}, {Curylo}, {Davids}, {Deil}, {Devin}, {de Wilt}, {Dirson},
  {Djannati-Ata{\"\i}}, {Dmytriiev}, {Donath}, {Doroshenko}, {Drury}, {Dyks},
  {Egberts}, {Emery}, {Ernenwein}, {Eschbach}, {Feijen}, {Fegan}, {Fiasson},
  {Fontaine}, {Funk}, {F{\"u}{\ss}ling}, {Gabici}, {Gallant}, {Gat{\'e}},
  {Giavitto}, {Glawion}, {Glicenstein}, {Gottschall}, {Grondin}, {Hahn},
  {Haupt}, {Heinzelmann}, {Henri}, {Hermann}, {Hinton}, {Hofmann}, {Hoischen},
  {Holch}, {Holler}, {Horns}, {Huber}, {Iwasaki}, {Jamrozy}, {Jankowsky},
  {Jankowsky}, {Jung-Richardt}, {Kastendieck}, {Katarzy{\'n}ski},
  {Katsuragawa}, {Katz}, {Khangulyan}, {Kh{\'e}lifi}, {King}, {Klepser},
  {Klu{\'z}niak}, {Komin}, {Kosack}, {Kostunin}, {Kraus}, {Lamanna}, {Lau},
  {Lemi{\`e}re}, {Lemoine-Goumard}, {Lenain}, {Leser}, {Levy}, {Lohse},
  {L{\'o}pez-Coto}, {Lypova}, {Mackey}, {Majumdar}, {Malyshev}, {Marandon},
  {Marcowith}, {Mares}, {Mariaud}, {Mart{\'\i}-Devesa}, {Marx}, {Maurin},
  {Meintjes}, {Mitchell}, {Moderski}, {Mohamed}, {Mohrmann}, {Muller}, {Moore},
  {Moulin}, {Murach}, {Nakashima}, {de Naurois}, {Ndiyavala}, {Niederwanger},
  {Niemiec}, {Oakes}, {O'Brien}, {Odaka}, {Ohm}, {de Ona Wilhelmi},
  {Ostrowski}, {Oya}, {Panter}, {Parsons}, {Perennes}, {Petrucci}, {Peyaud},
  {Piel}, {Pita}, {Poireau}, {Priyana Noel}, {Prokhorov}, {Prokoph},
  {P{\"u}hlhofer}, {Punch}, {Quirrenbach}, {Raab}, {Rauth}, {Reimer}, {Reimer},
  {Remy}, {Renaud}, {Rieger}, {Rinchiuso}, {Romoli}, {Rowell}, {Rudak},
  {Ruiz-Velasco}, {Sahakian}, {Saito}, {Sanchez}, {Santangelo}, {Sasaki},
  {Schlickeiser}, {Sch{\"u}ssler}, {Schulz}, {Schutte}, {Schwanke},
  {Schwemmer}, {Seglar-Arroyo}, {Senniappan}, {Seyffert}, {Shafi},
  {Shiningayamwe}, {Simoni}, {Sinha}, {Sol}, {Specovius}, {Spir-Jacob},
  {Stawarz}, {Steenkamp}, {Stegmann}, {Steppa}, {Takahashi}, {Tavernier},
  {Taylor}, {Terrier}, {Tiziani}, {Tluczykont}, {Trichard}, {Tsirou}, {Tsuji},
  {Tuffs}, {Uchiyama}, {van der Walt}, {van Eldik}, {van Rensburg}, {van
  Soelen}, {Vasileiadis}, {Veh}, {Venter}, {Vincent}, {Vink}, {Voisin},
  {V{\"o}lk}, {Vuillaume}, {Wadiasingh}, {Wagner}, {White}, {Wierzcholska},
  {Yang}, {Yoneda}, {Zacharias}, {Zanin}, {Zdziarski}, {Zech}, {Ziegler},
  {Zorn}, {{\.Z}ywucka}, and {Maxted}]{HESS2019A&A}
{H.~E.~S.~S. Collaboration}.; {Abdalla}, H.; {Aharonian}, F.; {Ait Benkhali},
  F.; {Ang{\"u}ner}, E.O.; {Arakawa}, M.; {Arcaro}, C.; {Armand}, C.; {Ashkar},
  H.; {Backes}, M.;  et~al.
\newblock {Upper limits on very-high-energy gamma-ray emission from
  core-collapse supernovae observed with H.E.S.S.}
\newblock {\em \aap} {\bf 2019}, {\em 626},~A57,
  \href{http://xxx.lanl.gov/abs/1904.10526}{{\normalfont
  [arXiv:astro-ph.HE/1904.10526]}}.
\newblock
  doi:{\changeurlcolor{black}\href{https://doi.org/10.1051/0004-6361/201935242}{\detokenize{10.1051/0004-6361/201935242}}}.

\bibitem[{Li} \em{et~al.}(2023){Li}, {Xin}, {Liu}, and {He}]{Li2023ApJ}
{Li}, Y.; {Xin}, Y.; {Liu}, S.; {He}, Y.
\newblock {Advanced {\ensuremath{\gamma}}-Ray Emission Studies of G15.4+0.1
  with Fermi-LAT: Evidence of Escaping Cosmic Rays Interacting with Surrounding
  Molecular Clouds}.
\newblock {\em \apj} {\bf 2023}, {\em 945},~21.
\newblock
  doi:{\changeurlcolor{black}\href{https://doi.org/10.3847/1538-4357/acb81d}{\detokenize{10.3847/1538-4357/acb81d}}}.

\bibitem[{Aruga} \em{et~al.}(2022){Aruga}, {Sano}, {Fukui}, {Reynoso},
  {Rowell}, and {Tachihara}]{Aruga2022ApJ}
{Aruga}, M.; {Sano}, H.; {Fukui}, Y.; {Reynoso}, E.M.; {Rowell}, G.;
  {Tachihara}, K.
\newblock {Molecular and Atomic Clouds Associated with the Gamma-Ray Supernova
  Remnant Puppis A}.
\newblock {\em \apj} {\bf 2022}, {\em 938},~94,
  \href{http://xxx.lanl.gov/abs/2206.00211}{{\normalfont
  [arXiv:astro-ph.HE/2206.00211]}}.
\newblock
  doi:{\changeurlcolor{black}\href{https://doi.org/10.3847/1538-4357/ac90c6}{\detokenize{10.3847/1538-4357/ac90c6}}}.

\bibitem[{Zhang} \em{et~al.}(2021){Zhang}, {Liu}, {Li}, {Shao}, {Yan}, and
  {Wang}]{Zhang2021ApJ}
{Zhang}, Y.; {Liu}, R.Y.; {Li}, H.; {Shao}, S.; {Yan}, H.; {Wang}, X.Y.
\newblock {Measuring the Mass of Missing Baryons in the Halo of Andromeda
  Galaxy with Gamma-Ray Observations}.
\newblock {\em \apj} {\bf 2021}, {\em 911},~58,
  \href{http://xxx.lanl.gov/abs/2010.15477}{{\normalfont
  [arXiv:astro-ph.HE/2010.15477]}}.
\newblock
  doi:{\changeurlcolor{black}\href{https://doi.org/10.3847/1538-4357/abe9b6}{\detokenize{10.3847/1538-4357/abe9b6}}}.

\bibitem[{Zhong} \em{et~al.}(2023){Zhong}, {Zhang}, {Chen}, and
  {Zhang}]{Zhong2023MNRAS}
{Zhong}, W.J.; {Zhang}, X.; {Chen}, Y.; {Zhang}, Q.Q.
\newblock {A study of GeV gamma-ray emission towards supernova remnant
  G51.26+0.11 and its molecular environment}.
\newblock {\em \mnras} {\bf 2023}, {\em 521},~1931--1940,
  \href{http://xxx.lanl.gov/abs/2302.12679}{{\normalfont
  [arXiv:astro-ph.HE/2302.12679]}}.
\newblock
  doi:{\changeurlcolor{black}\href{https://doi.org/10.1093/mnras/stad628}{\detokenize{10.1093/mnras/stad628}}}.

\bibitem[{Yeung} \em{et~al.}(2023){Yeung}, {Bamba}, and {Sano}]{Yeung2023PASJ}
{Yeung}, P.K.H.; {Bamba}, A.; {Sano}, H.
\newblock {Multiwavelength studies of G298.6-0.0: An old GeV supernova remnant
  interacting with molecular clouds}.
\newblock {\em \pasj} {\bf 2023}, {\em 75},~384--396,
  \href{http://xxx.lanl.gov/abs/2212.01851}{{\normalfont
  [arXiv:astro-ph.HE/2212.01851]}}.
\newblock
  doi:{\changeurlcolor{black}\href{https://doi.org/10.1093/pasj/psad006}{\detokenize{10.1093/pasj/psad006}}}.

\bibitem[{Supan} \em{et~al.}(2022){Supan}, {Fischetto}, and
  {Castelletti}]{Supan2022A&A}
{Supan}, L.; {Fischetto}, G.; {Castelletti}, G.
\newblock {Supernova remnant G46.8-0.3: A new case of interaction with
  molecular material}.
\newblock {\em \aap} {\bf 2022}, {\em 664},~A89,
  \href{http://xxx.lanl.gov/abs/2203.05537}{{\normalfont
  [arXiv:astro-ph.HE/2203.05537]}}.
\newblock
  doi:{\changeurlcolor{black}\href{https://doi.org/10.1051/0004-6361/202142431}{\detokenize{10.1051/0004-6361/202142431}}}.

\bibitem[{Reynolds}(2008)]{Reynolds2008ARA&A}
{Reynolds}, S.P.
\newblock {Supernova remnants at high energy.}
\newblock {\em \araa} {\bf 2008}, {\em 46},~89--126.
\newblock
  doi:{\changeurlcolor{black}\href{https://doi.org/10.1146/annurev.astro.46.060407.145237}{\detokenize{10.1146/annurev.astro.46.060407.145237}}}.

\bibitem[{Murase} \em{et~al.}(2014){Murase}, {Thompson}, and
  {Ofek}]{Murase2014MNRAS}
{Murase}, K.; {Thompson}, T.A.; {Ofek}, E.O.
\newblock {Probing cosmic ray ion acceleration with radio-submm and gamma-ray
  emission from interaction-powered supernovae}.
\newblock {\em \mnras} {\bf 2014}, {\em 440},~2528--2543,
  \href{http://xxx.lanl.gov/abs/1311.6778}{{\normalfont
  [arXiv:astro-ph.HE/1311.6778]}}.
\newblock
  doi:{\changeurlcolor{black}\href{https://doi.org/10.1093/mnras/stu384}{\detokenize{10.1093/mnras/stu384}}}.

\bibitem[{Helder} \em{et~al.}(2012){Helder}, {Vink}, {Bykov}, {Ohira},
  {Raymond}, and {Terrier}]{Helder2012SSRv}
{Helder}, E.A.; {Vink}, J.; {Bykov}, A.M.; {Ohira}, Y.; {Raymond}, J.C.;
  {Terrier}, R.
\newblock {Observational Signatures of Particle Acceleration in Supernova
  Remnants}.
\newblock {\em \ssr} {\bf 2012}, {\em 173},~369--431,
  \href{http://xxx.lanl.gov/abs/1206.1593}{{\normalfont
  [arXiv:astro-ph.HE/1206.1593]}}.
\newblock
  doi:{\changeurlcolor{black}\href{https://doi.org/10.1007/s11214-012-9919-8}{\detokenize{10.1007/s11214-012-9919-8}}}.

\bibitem[{Bykov} \em{et~al.}(2020){Bykov}, {Marcowith}, {Amato}, {Kalyashova},
  {Kruijssen}, and {Waxman}]{Bykov2020SSRv}
{Bykov}, A.M.; {Marcowith}, A.; {Amato}, E.; {Kalyashova}, M.E.; {Kruijssen},
  J.M.D.; {Waxman}, E.
\newblock {High-Energy Particles and Radiation in Star-Forming Regions}.
\newblock {\em \ssr} {\bf 2020}, {\em 216},~42,
  \href{http://xxx.lanl.gov/abs/2003.11534}{{\normalfont
  [arXiv:astro-ph.HE/2003.11534]}}.
\newblock
  doi:{\changeurlcolor{black}\href{https://doi.org/10.1007/s11214-020-00663-0}{\detokenize{10.1007/s11214-020-00663-0}}}.

\bibitem[{Tatischeff} and {Gabici}(2018)]{Tatischeff2018ARNPS}
{Tatischeff}, V.; {Gabici}, S.
\newblock {Particle Acceleration by Supernova Shocks and Spallogenic
  Nucleosynthesis of Light Elements}.
\newblock {\em Annual Review of Nuclear and Particle Science} {\bf 2018}, {\em
  68},~377--404,  \href{http://xxx.lanl.gov/abs/1803.01794}{{\normalfont
  [arXiv:astro-ph.HE/1803.01794]}}.
\newblock
  doi:{\changeurlcolor{black}\href{https://doi.org/10.1146/annurev-nucl-101917-021151}{\detokenize{10.1146/annurev-nucl-101917-021151}}}.

\bibitem[{Abeysekara} \em{et~al.}(2021){Abeysekara}, {Albert}, {Alfaro},
  {Alvarez}, {Camacho}, {Arteaga-Vel{\'a}zquez}, {Arunbabu}, {Rojas},
  {Solares}, {Baghmanyan}, {Belmont-Moreno}, {BenZvi}, {Blandford}, {Brisbois},
  {Caballero-Mora}, {Capistr{\'a}n}, {Carrami{\~n}ana}, {Casanova}, {Cotti},
  {Le{\'o}n}, {De la Fuente}, {Hernandez}, {Dingus}, {DuVernois}, {Durocher},
  {D{\'\i}az-V{\'e}lez}, {Ellsworth}, {Engel}, {Espinoza}, {Fan}, {Fang},
  {Fleischhack}, {Fraija}, {Galv{\'a}n-G{\'a}mez}, {Garcia},
  {Garc{\'\i}a-Gonz{\'a}lez}, {Garfias}, {Giacinti}, {Gonz{\'a}lez}, {Goodman},
  {Harding}, {Hernandez}, {Hinton}, {Hona}, {Huang}, {Hueyotl-Zahuantitla},
  {H{\"u}ntemeyer}, {Iriarte}, {Jardin-Blicq}, {Joshi}, {Kieda}, {Lara}, {Lee},
  {Vargas}, {Linnemann}, {Longinotti}, {Luis-Raya}, {Lundeen}, {Malone},
  {Martinez}, {Martinez-Castellanos}, {Mart{\'\i}nez-Castro}, {Matthews},
  {Miranda-Romagnoli}, {Morales-Soto}, {Moreno}, {Mostaf{\'a}}, {Nayerhoda},
  {Nellen}, {Newbold}, {Nisa}, {Noriega-Papaqui}, {Olivera-Nieto}, {Omodei},
  {Peisker}, {P{\'e}rez Araujo}, {P{\'e}rez-P{\'e}rez}, {Ren}, {Rho},
  {Rosa-Gonz{\'a}lez}, {Ruiz-Velasco}, {Salazar}, {Greus}, {Sandoval},
  {Schneider}, {Schoorlemmer}, {Serna}, {Smith}, {Springer}, {Surajbali},
  {Tollefson}, {Torres}, {Torres-Escobedo}, {Ure{\~n}a-Mena}, {Weisgarber},
  {Werner}, {Willox}, {Zepeda}, {Zhou}, {De Le{\'o}n}, and
  {{\'A}lvarez}]{Abeysekara2021NatAs}
{Abeysekara}, A.U.; {Albert}, A.; {Alfaro}, R.; {Alvarez}, C.; {Camacho},
  J.R.A.; {Arteaga-Vel{\'a}zquez}, J.C.; {Arunbabu}, K.P.; {Rojas}, D.A.;
  {Solares}, H.A.A.; {Baghmanyan}, V.;  et~al.
\newblock {HAWC observations of the acceleration of very-high-energy cosmic
  rays in the Cygnus Cocoon}.
\newblock {\em Nature Astronomy} {\bf 2021}, {\em 5},~465--471,
  \href{http://xxx.lanl.gov/abs/2103.06820}{{\normalfont
  [arXiv:astro-ph.HE/2103.06820]}}.
\newblock
  doi:{\changeurlcolor{black}\href{https://doi.org/10.1038/s41550-021-01318-y}{\detokenize{10.1038/s41550-021-01318-y}}}.

\bibitem[{Cao} \em{et~al.}(2021){Cao}, {Aharonian}, {An}, {Axikegu}, {Bai},
  {Bao}, {Bastieri}, {Bi}, {Bi}, {Cai}, {Cai}, {Cao}, {Chang}, {Chang},
  {Chang}, {Chen}, {Chen}, {Chen}, {Chen}, {Chen}, {Chen}, {Chen}, {Chen},
  {Chen}, {Chen}, {Chen}, {Chen}, {Chen}, {Cheng}, {Cheng}, {Cui}, {Cui},
  {Cui}, {Dai}, {Dai}, {Dai}, {Danzengluobu}, {della Volpe}, {D'Ettorre
  Piazzoli}, {Dong}, {Fan}, {Fan}, {Fan}, {Fang}, {Fang}, {Feng}, {Feng},
  {Feng}, {Feng}, {Gao}, {Gao}, {Gao}, {Gao}, {Ge}, {Geng}, {Gong}, {Gou},
  {Gu}, {Guo}, {Guo}, {Guo}, {Guo}, {Han}, {He}, {He}, {He}, {He}, {He}, {He},
  {Heller}, {Hor}, {Hou}, {Hou}, {Hu}, {Hu}, {Hu}, {Hu}, {Huang}, {Huang},
  {Huang}, {Huang}, {Huang}, {Ji}, {Ji}, {Jia}, {Jiang}, {Jiang}, {Jin},
  {Kuleshov}, {Levochkin}, {Li}, {Li}, {Li}, {Li}, {Li}, {Li}, {Li}, {Li},
  {Li}, {Li}, {Li}, {Li}, {Li}, {Li}, {Li}, {Li}, {Li}, {Liang}, {Liang},
  {Lin}, {Liu}, {Liu}, {Liu}, {Liu}, {Liu}, {Liu}, {Liu}, {Liu}, {Liu}, {Liu},
  {Liu}, {Liu}, {Liu}, {Liu}, {Liu}, {Long}, {Lu}, {Lv}, {Ma}, {Ma}, {Ma},
  {Mao}, {Masood}, {Mitthumsiri}, {Montaruli}, {Nan}, {Pang},
  {Pattarakijwanich}, {Pei}, {Qi}, {Ruffolo}, {Rulev}, {S{\'a}iz}, {Shao},
  {Shchegolev}, {Sheng}, {Shi}, {Song}, {Stenkin}, {Stepanov}, {Sun}, {Sun},
  {Sun}, {Tam}, {Tang}, {Tian}, {Wang}, {Wang}, {Wang}, {Wang}, {Wang}, {Wang},
  {Wang}, {Wang}, {Wang}, {Wang}, {Wang}, {Wang}, {Wang}, {Wang}, {Wang},
  {Wang}, {Wang}, {Wang}, {Wang}, {Wang}, {Wang}, {Wei}, {Wei}, {Wei}, {Wen},
  {Wu}, {Wu}, {Wu}, {Wu}, {Wu}, {Xi}, {Xia}, {Xia}, {Xiang}, {Xiao}, {Xiao},
  {Xin}, {Xin}, {Xing}, {Xu}, {Xu}, {Xue}, {Yan}, {Yang}, {Yang}, {Yang},
  {Yang}, {Yang}, {Yang}, {Yang}, {Yao}, {Yao}, {Ye}, {Yin}, {Yin}, {You},
  {You}, {Yu}, {Yuan}, {Zeng}, {Zeng}, {Zeng}, {Zeng}, {Zha}, {Zhai}, {Zhang},
  {Zhang}, {Zhang}, {Zhang}, {Zhang}, {Zhang}, {Zhang}, {Zhang}, {Zhang},
  {Zhang}, {Zhang}, {Zhang}, {Zhang}, {Zhang}, {Zhang}, {Zhang}, {Zhang},
  {Zhang}, {Zhang}, {Zhao}, {Zhao}, {Zhao}, {Zhao}, {Zhao}, {Zheng}, {Zheng},
  {Zhou}, {Zhou}, {Zhou}, {Zhou}, {Zhou}, {Zhou}, {Zhu}, {Zhu}, {Zhu}, {Zhu},
  and {Zuo}]{Cao2021Natur}
{Cao}, Z.; {Aharonian}, F.A.; {An}, Q.; {Axikegu}, Bai, L.X.; {Bai}, Y.X.;
  {Bao}, Y.W.; {Bastieri}, D.; {Bi}, X.J.; {Bi}, Y.J.; {Cai}, H.;  et~al.
\newblock {Ultrahigh-energy photons up to 1.4 petaelectronvolts from 12
  {\ensuremath{\gamma}}-ray Galactic sources}.
\newblock {\em \nat} {\bf 2021}, {\em 594},~33--36.
\newblock
  doi:{\changeurlcolor{black}\href{https://doi.org/10.1038/s41586-021-03498-z}{\detokenize{10.1038/s41586-021-03498-z}}}.

\bibitem[{Ackermann} \em{et~al.}(2011){Ackermann}, {Ajello}, {Allafort},
  {Baldini}, {Ballet}, {Barbiellini}, {Bastieri}, {Belfiore}, {Bellazzini},
  {Berenji}, {Blandford}, {Bloom}, {Bonamente}, {Borgland}, {Bottacini},
  {Brigida}, {Bruel}, {Buehler}, {Buson}, {Caliandro}, {Cameron}, {Caraveo},
  {Casandjian}, {Cecchi}, {Chekhtman}, {Cheung}, {Chiang}, {Ciprini}, {Claus},
  {Cohen-Tanugi}, {de Angelis}, {de Palma}, {Dermer}, {do Couto e Silva},
  {Drell}, {Dumora}, {Favuzzi}, {Fegan}, {Focke}, {Fortin}, {Fukazawa},
  {Fusco}, {Gargano}, {Germani}, {Giglietto}, {Giordano}, {Giroletti},
  {Glanzman}, {Godfrey}, {Grenier}, {Guillemot}, {Guiriec}, {Hadasch},
  {Hanabata}, {Harding}, {Hayashida}, {Hayashi}, {Hays}, {J{\'o}hannesson},
  {Johnson}, {Kamae}, {Katagiri}, {Kataoka}, {Kerr}, {Kn{\"o}dlseder}, {Kuss},
  {Lande}, {Latronico}, {Lee}, {Longo}, {Loparco}, {Lott}, {Lovellette},
  {Lubrano}, {Martin}, {Mazziotta}, {McEnery}, {Mehault}, {Michelson},
  {Mitthumsiri}, {Mizuno}, {Monte}, {Monzani}, {Morselli}, {Moskalenko},
  {Murgia}, {Naumann-Godo}, {Nolan}, {Norris}, {Nuss}, {Ohsugi}, {Okumura},
  {Orlando}, {Ormes}, {Ozaki}, {Paneque}, {Parent}, {Pesce-Rollins},
  {Pierbattista}, {Piron}, {Pohl}, {Prokhorov}, {Rain{\`o}}, {Rando},
  {Razzano}, {Reposeur}, {Ritz}, {Parkinson}, {Sgr{\`o}}, {Siskind}, {Smith},
  {Spinelli}, {Strong}, {Takahashi}, {Tanaka}, {Thayer}, {Thayer}, {Thompson},
  {Tibaldo}, {Torres}, {Tosti}, {Tramacere}, {Troja}, {Uchiyama},
  {Vandenbroucke}, {Vasileiou}, {Vianello}, {Vitale}, {Waite}, {Wang}, {Winer},
  {Wood}, {Yang}, {Zimmer}, and {Bontemps}]{Ackermann2011Sci}
{Ackermann}, M.; {Ajello}, M.; {Allafort}, A.; {Baldini}, L.; {Ballet}, J.;
  {Barbiellini}, G.; {Bastieri}, D.; {Belfiore}, A.; {Bellazzini}, R.;
  {Berenji}, B.;  et~al.
\newblock {A Cocoon of Freshly Accelerated Cosmic Rays Detected by Fermi in the
  Cygnus Superbubble}.
\newblock {\em Science} {\bf 2011}, {\em 334},~1103.
\newblock
  doi:{\changeurlcolor{black}\href{https://doi.org/10.1126/science.1210311}{\detokenize{10.1126/science.1210311}}}.

\bibitem[{Bartoli} \em{et~al.}(2014){Bartoli}, {Bernardini}, {Bi}, {Branchini},
  {Budano}, {Camarri}, {Cao}, {Cardarelli}, {Catalanotti}, {Chen}, {Chen},
  {Creti}, {Cui}, {Dai}, {D'Amone}, {Danzengluobu}, {De Mitri}, {D'Ettorre
  Piazzoli}, {Di Girolamo}, {Di Sciascio}, {Feng}, {Feng}, {Feng}, {Gou},
  {Guo}, {He}, {Hu}, {Hu}, {Iacovacci}, {Iuppa}, {Jia}, {Labaciren}, {Li},
  {Liguori}, {Liu}, {Liu}, {Liu}, {Lu}, {Ma}, {Ma}, {Mancarella}, {Mari},
  {Marsella}, {Martello}, {Mastroianni}, {Montini}, {Ning}, {Panareo},
  {Perrone}, {Pistilli}, {Ruggieri}, {Salvini}, {Santonico}, {Shen}, {Sheng},
  {Shi}, {Surdo}, {Tan}, {Vallania}, {Vernetto}, {Vigorito}, {Wang}, {Wu},
  {Wu}, {Xue}, {Yang}, {Yang}, {Yao}, {Yuan}, {Zha}, {Zhang}, {Zhang}, {Zhang},
  {Zhang}, {Zhao}, {Zhaxiciren}, {Zhaxisangzhu}, {Zhou}, {Zhu}, {Zhu}, {Zizzi},
  and {ARGO-YBJ Collaboration}]{Bartoli2014ApJ}
{Bartoli}, B.; {Bernardini}, P.; {Bi}, X.J.; {Branchini}, P.; {Budano}, A.;
  {Camarri}, P.; {Cao}, Z.; {Cardarelli}, R.; {Catalanotti}, S.; {Chen}, S.Z.;
  et~al.
\newblock {Identification of the TeV Gamma-Ray Source ARGO J2031+4157 with the
  Cygnus Cocoon}.
\newblock {\em \apj} {\bf 2014}, {\em 790},~152,
  \href{http://xxx.lanl.gov/abs/1406.6436}{{\normalfont
  [arXiv:astro-ph.HE/1406.6436]}}.
\newblock
  doi:{\changeurlcolor{black}\href{https://doi.org/10.1088/0004-637X/790/2/152}{\detokenize{10.1088/0004-637X/790/2/152}}}.

\bibitem[{Dzhappuev} \em{et~al.}(2021){Dzhappuev}, {Afashokov}, {Dzaparova},
  {Dzhatdoev}, {Gorbacheva}, {Karpikov}, {Khadzhiev}, {Klimenko}, {Kudzhaev},
  {Kurenya}, {Lidvansky}, {Mikhailova}, {Petkov}, {Podlesnyi}, {Romanenko},
  {Rubtsov}, {Troitsky}, {Unatlokov}, {Vaiman}, {Yanin}, {Zhezher},
  {Zhuravleva}, and {Carpet-3 Group}]{Dzhappuev2021ApJ}
{Dzhappuev}, D.D.; {Afashokov}, Y.Z.; {Dzaparova}, I.M.; {Dzhatdoev}, T.A.;
  {Gorbacheva}, E.A.; {Karpikov}, I.S.; {Khadzhiev}, M.M.; {Klimenko}, N.F.;
  {Kudzhaev}, A.U.; {Kurenya}, A.N.;  et~al.
\newblock {Observation of Photons above 300 TeV Associated with a High-energy
  Neutrino from the Cygnus Region}.
\newblock {\em \apjl} {\bf 2021}, {\em 916},~L22,
  \href{http://xxx.lanl.gov/abs/2105.07242}{{\normalfont
  [arXiv:astro-ph.HE/2105.07242]}}.
\newblock
  doi:{\changeurlcolor{black}\href{https://doi.org/10.3847/2041-8213/ac14b2}{\detokenize{10.3847/2041-8213/ac14b2}}}.

\bibitem[{Yang} and {Wang}(2020)]{Yang2020A&A}
{Yang}, R.Z.; {Wang}, Y.
\newblock {The diffuse gamma-ray emission toward the Galactic mini starburst
  W43}.
\newblock {\em \aap} {\bf 2020}, {\em 640},~A60,
  \href{http://xxx.lanl.gov/abs/2007.15295}{{\normalfont
  [arXiv:astro-ph.HE/2007.15295]}}.
\newblock
  doi:{\changeurlcolor{black}\href{https://doi.org/10.1051/0004-6361/202037518}{\detokenize{10.1051/0004-6361/202037518}}}.

\bibitem[{Aharonian} \em{et~al.}(2022){Aharonian}, {Ashkar}, {Backes}, {Barbosa
  Martins}, {Becherini}, {Berge}, {Bi}, {B{\"o}ttcher}, {de Bony de Lavergne},
  {Bradascio}, {Brose}, {Brun}, {Bulik}, {Burger-Scheidlin}, {Cangemi},
  {Caroff}, {Casanova}, {Cerruti}, {Chand}, {Chandra}, {Chen}, {Chibueze},
  {Cristofari}, {Damascene Mbarubucyeye}, {Djannati-Ata{\"\i}}, {Ernenwein},
  {Feijen}, {Fichet de Clairfontaine}, {Fontaine}, {Funk}, {Gabici}, {Gallant},
  {Ghafourizadeh}, {Giavitto}, {Giunti}, {Glawion}, {Glicenstein}, {Goswami},
  {Grondin}, {H{\"a}rer}, {Haupt}, {Hinton}, {H{\"o}rbe}, {Hofmann}, {Holch},
  {Holler}, {Horns}, {Jamrozy}, {Joshi}, {Jung-Richardt}, {Kasai},
  {Katarzy{\'n}ski}, {Katz}, {Kh{\'e}lifi}, {Klu{\'z}niak}, {Komin}, {Kosack},
  {Kostunin}, {Kukec Mezek}, {Lang}, {Le Stum}, {Lemi{\`e}re},
  {Lemoine-Goumard}, {Lenain}, {Leuschner}, {Lohse}, {Luashvili}, {Lypova},
  {Mackey}, {Majumdar}, {Malyshev}, {Marandon}, {Marchegiani}, {Marcowith},
  {Mart{\'\i}-Devesa}, {Marx}, {Maurin}, {Meyer}, {Mitchell}, {Moderski},
  {Mohrmann}, {Montanari}, {Moulin}, {Muller}, {Murach}, {Nakashima}, {de
  Naurois}, {Nayerhoda}, {Niemiec}, {Ohm}, {Olivera-Nieto}, {de Ona Wilhelmi},
  {Ostrowski}, {Panny}, {Panter}, {Parsons}, {Peron}, {Prokhorov},
  {P{\"u}hlhofer}, {Punch}, {Quirrenbach}, {Rauth}, {Reichherzer}, {Reimer},
  {Reimer}, {Renaud}, {Reville}, {Rieger}, {Rowell}, {Rudak}, {Ruiz-Velasco},
  {Sahakian}, {Salzmann}, {Sanchez}, {Santangelo}, {Sasaki}, {Sch{\"u}ssler},
  {Schutte}, {Schwanke}, {Shapopi}, {Specovius}, {Spencer}, {Stawarz},
  {Steenkamp}, {Steinmassl}, {Steppa}, {Sushch}, {Suzuki}, {Takahashi},
  {Tanaka}, {Terrier}, {Thorpe-Morgan}, {Tsirou}, {Tsuji}, {Tuffs}, {Unbehaun},
  {van Eldik}, {van Soelen}, {Vecchi}, {Veh}, {Venter}, {Vink}, {Wagner},
  {White}, {Wierzcholska}, {Wong}, {Zacharias}, {Zargaryan}, {Zdziarski},
  {Zhu}, {Zouari}, {{\.Z}ywucka}, {Blackwell}, {Braiding}, {Burton}, {Cubuk},
  {Filipovi{\'c}}, {Tothill}, and {Wong}]{Aharonian2022A&A}
{Aharonian}, F.; {Ashkar}, H.; {Backes}, M.; {Barbosa Martins}, V.;
  {Becherini}, Y.; {Berge}, D.; {Bi}, B.; {B{\"o}ttcher}, M.; {de Bony de
  Lavergne}, M.; {Bradascio}, F.;  et~al.
\newblock {A deep spectromorphological study of the {\ensuremath{\gamma}}-ray
  emission surrounding the young massive stellar cluster Westerlund 1}.
\newblock {\em \aap} {\bf 2022}, {\em 666},~A124,
  \href{http://xxx.lanl.gov/abs/2207.10921}{{\normalfont
  [arXiv:astro-ph.HE/2207.10921]}}.
\newblock
  doi:{\changeurlcolor{black}\href{https://doi.org/10.1051/0004-6361/202244323}{\detokenize{10.1051/0004-6361/202244323}}}.

\bibitem[{H{\"a}rer} \em{et~al.}(2023){H{\"a}rer}, {Reville}, {Hinton},
  {Mohrmann}, and {Vieu}]{Harer2023A&A}
{H{\"a}rer}, L.K.; {Reville}, B.; {Hinton}, J.; {Mohrmann}, L.; {Vieu}, T.
\newblock {Understanding the TeV {\ensuremath{\gamma}}-ray emission surrounding
  the young massive star cluster Westerlund 1}.
\newblock {\em \aap} {\bf 2023}, {\em 671},~A4,
  \href{http://xxx.lanl.gov/abs/2301.10496}{{\normalfont
  [arXiv:astro-ph.HE/2301.10496]}}.
\newblock
  doi:{\changeurlcolor{black}\href{https://doi.org/10.1051/0004-6361/202245444}{\detokenize{10.1051/0004-6361/202245444}}}.

\bibitem[{Mestre} \em{et~al.}(2021){Mestre}, {de O{\~n}a Wilhelmi}, {Torres},
  {Holch}, {Schwanke}, {Aharonian}, {Parkinson}, {Yang}, and
  {Zanin}]{Mestre2021MNRAS}
{Mestre}, E.; {de O{\~n}a Wilhelmi}, E.; {Torres}, D.F.; {Holch}, T.L.;
  {Schwanke}, U.; {Aharonian}, F.; {Parkinson}, P.S.; {Yang}, R.; {Zanin}, R.
\newblock {Probing the hadronic nature of the gamma-ray emission associated
  with Westerlund 2}.
\newblock {\em \mnras} {\bf 2021}, {\em 505},~2731--2740,
  \href{http://xxx.lanl.gov/abs/2105.09155}{{\normalfont
  [arXiv:astro-ph.HE/2105.09155]}}.
\newblock
  doi:{\changeurlcolor{black}\href{https://doi.org/10.1093/mnras/stab1455}{\detokenize{10.1093/mnras/stab1455}}}.

\bibitem[{Ge} \em{et~al.}(2022){Ge}, {Sun}, {Yang}, {Liang}, and
  {Liang}]{Ge2022MNRAS}
{Ge}, T.T.; {Sun}, X.N.; {Yang}, R.Z.; {Liang}, Y.F.; {Liang}, E.W.
\newblock {Diffuse {\ensuremath{\gamma}}-ray emission around the massive star
  forming region of Carina Nebula Complex}.
\newblock {\em \mnras} {\bf 2022}, {\em 517},~5121--5128,
  \href{http://xxx.lanl.gov/abs/2210.01352}{{\normalfont
  [arXiv:astro-ph.HE/2210.01352]}}.
\newblock
  doi:{\changeurlcolor{black}\href{https://doi.org/10.1093/mnras/stac2885}{\detokenize{10.1093/mnras/stac2885}}}.

\bibitem[{Baghmanyan} \em{et~al.}(2020){Baghmanyan}, {Peron}, {Casanova},
  {Aharonian}, and {Zanin}]{Baghmanyan2020ApJ}
{Baghmanyan}, V.; {Peron}, G.; {Casanova}, S.; {Aharonian}, F.; {Zanin}, R.
\newblock {Evidence of Cosmic-Ray Excess from Local Giant Molecular Clouds}.
\newblock {\em \apjl} {\bf 2020}, {\em 901},~L4,
  \href{http://xxx.lanl.gov/abs/2009.08893}{{\normalfont
  [arXiv:astro-ph.HE/2009.08893]}}.
\newblock
  doi:{\changeurlcolor{black}\href{https://doi.org/10.3847/2041-8213/abb5f8}{\detokenize{10.3847/2041-8213/abb5f8}}}.

\bibitem[{Cesarsky} and {Montmerle}(1983)]{Cesarsky1983SSRv}
{Cesarsky}, C.J.; {Montmerle}, T.
\newblock {Gamma-Rays from Active Regions in the Galaxy - the Possible
  Contribution of Stellar Winds}.
\newblock {\em \ssr} {\bf 1983}, {\em 36},~173--193.
\newblock
  doi:{\changeurlcolor{black}\href{https://doi.org/10.1007/BF00167503}{\detokenize{10.1007/BF00167503}}}.

\bibitem[{Voelk} and {Forman}(1982)]{Voelk1982ApJ}
{Voelk}, H.J.; {Forman}, M.
\newblock {Cosmic rays and gamma-rays from OB stars}.
\newblock {\em \apj} {\bf 1982}, {\em 253},~188--198.
\newblock
  doi:{\changeurlcolor{black}\href{https://doi.org/10.1086/159623}{\detokenize{10.1086/159623}}}.

\bibitem[{Mac Low} and {McCray}(1988)]{MacLow1988ApJ}
{Mac Low}, M.M.; {McCray}, R.
\newblock {Superbubbles in Disk Galaxies}.
\newblock {\em \apj} {\bf 1988}, {\em 324},~776.
\newblock
  doi:{\changeurlcolor{black}\href{https://doi.org/10.1086/165936}{\detokenize{10.1086/165936}}}.

\bibitem[{Reimer} \em{et~al.}(2006){Reimer}, {Pohl}, and
  {Reimer}]{Reimer2006ApJ}
{Reimer}, A.; {Pohl}, M.; {Reimer}, O.
\newblock {Nonthermal High-Energy Emission from Colliding Winds of Massive
  Stars}.
\newblock {\em \apj} {\bf 2006}, {\em 644},~1118--1144,
  \href{http://xxx.lanl.gov/abs/astro-ph/0510701}{{\normalfont
  [arXiv:astro-ph/astro-ph/0510701]}}.
\newblock
  doi:{\changeurlcolor{black}\href{https://doi.org/10.1086/503598}{\detokenize{10.1086/503598}}}.

\bibitem[{Vieu} \em{et~al.}(2022){Vieu}, {Reville}, and
  {Aharonian}]{Vieu2022MNRAS}
{Vieu}, T.; {Reville}, B.; {Aharonian}, F.
\newblock {Can superbubbles accelerate ultrahigh energy protons?}
\newblock {\em \mnras} {\bf 2022}, {\em 515},~2256--2265,
  \href{http://xxx.lanl.gov/abs/2207.01432}{{\normalfont
  [arXiv:astro-ph.HE/2207.01432]}}.
\newblock
  doi:{\changeurlcolor{black}\href{https://doi.org/10.1093/mnras/stac1901}{\detokenize{10.1093/mnras/stac1901}}}.

\bibitem[{Bykov} and {Toptygin}(2001)]{Bykov2001AstL}
{Bykov}, A.M.; {Toptygin}, I.N.
\newblock {A Model of Particle Acceleration to High Energies by Multiple
  Supernova Explosions in OB Associations}.
\newblock {\em Astronomy Letters} {\bf 2001}, {\em 27},~625--633.
\newblock
  doi:{\changeurlcolor{black}\href{https://doi.org/10.1134/1.1404456}{\detokenize{10.1134/1.1404456}}}.

\bibitem[{Bykov} and {Fleishman}(1992)]{Bykov1992MNRAS}
{Bykov}, A.M.; {Fleishman}, G.D.
\newblock {On non-thermal particle generation in superbubbles.}
\newblock {\em \mnras} {\bf 1992}, {\em 255},~269--275.
\newblock
  doi:{\changeurlcolor{black}\href{https://doi.org/10.1093/mnras/255.2.269}{\detokenize{10.1093/mnras/255.2.269}}}.

\bibitem[{Bykov}(2014)]{Bykov2014A&ARv}
{Bykov}, A.M.
\newblock {Nonthermal particles and photons in starburst regions and
  superbubbles}.
\newblock {\em \aapr} {\bf 2014}, {\em 22},~77,
  \href{http://xxx.lanl.gov/abs/1511.04608}{{\normalfont
  [arXiv:astro-ph.HE/1511.04608]}}.
\newblock
  doi:{\changeurlcolor{black}\href{https://doi.org/10.1007/s00159-014-0077-8}{\detokenize{10.1007/s00159-014-0077-8}}}.

\bibitem[{Parizot} \em{et~al.}(2004){Parizot}, {Marcowith}, {van der Swaluw},
  {Bykov}, and {Tatischeff}]{Parizot2004A&A}
{Parizot}, E.; {Marcowith}, A.; {van der Swaluw}, E.; {Bykov}, A.M.;
  {Tatischeff}, V.
\newblock {Superbubbles and energetic particles in the Galaxy. I. Collective
  effects of particle acceleration}.
\newblock {\em \aap} {\bf 2004}, {\em 424},~747--760,
  \href{http://xxx.lanl.gov/abs/astro-ph/0405531}{{\normalfont
  [arXiv:astro-ph/astro-ph/0405531]}}.
\newblock
  doi:{\changeurlcolor{black}\href{https://doi.org/10.1051/0004-6361:20041269}{\detokenize{10.1051/0004-6361:20041269}}}.

\bibitem[{Higdon} and {Lingenfelter}(2005)]{Higdon2005ApJ}
{Higdon}, J.C.; {Lingenfelter}, R.E.
\newblock {OB Associations, Supernova-generated Superbubbles, and the Source of
  Cosmic Rays}.
\newblock {\em \apj} {\bf 2005}, {\em 628},~738--749.
\newblock
  doi:{\changeurlcolor{black}\href{https://doi.org/10.1086/430814}{\detokenize{10.1086/430814}}}.

\bibitem[{Wang} \em{et~al.}(2022){Wang}, {Zhang}, {Liu}, and
  {Wang}]{Wang2022ApJ}
{Wang}, K.; {Zhang}, H.M.; {Liu}, R.Y.; {Wang}, X.Y.
\newblock {Detection of Diffuse {\ensuremath{\gamma}}-Ray Emission toward a
  Massive Star-forming Region Hosting Wolf-Rayet Stars}.
\newblock {\em \apj} {\bf 2022}, {\em 935},~129,
  \href{http://xxx.lanl.gov/abs/2207.06583}{{\normalfont
  [arXiv:astro-ph.HE/2207.06583]}}.
\newblock
  doi:{\changeurlcolor{black}\href{https://doi.org/10.3847/1538-4357/ac815e}{\detokenize{10.3847/1538-4357/ac815e}}}.

\bibitem[{Kamijima} and {Ohira}(2022)]{Kamijima2022PhRvD}
{Kamijima}, S.F.; {Ohira}, Y.
\newblock {Escape of cosmic rays from perpendicular shocks in the circumstellar
  magnetic field}.
\newblock {\em \prd} {\bf 2022}, {\em 106},~123025,
  \href{http://xxx.lanl.gov/abs/2207.13896}{{\normalfont
  [arXiv:astro-ph.HE/2207.13896]}}.
\newblock
  doi:{\changeurlcolor{black}\href{https://doi.org/10.1103/PhysRevD.106.123025}{\detokenize{10.1103/PhysRevD.106.123025}}}.

\bibitem[{Gupta} \em{et~al.}(2020){Gupta}, {Nath}, {Sharma}, and
  {Eichler}]{Gupta2020MNRAS}
{Gupta}, S.; {Nath}, B.B.; {Sharma}, P.; {Eichler}, D.
\newblock {Realistic modelling of wind and supernovae shocks in star clusters:
  addressing $^{22}$Ne/$^{20}$Ne and other problems in Galactic cosmic rays}.
\newblock {\em \mnras} {\bf 2020}, {\em 493},~3159--3177,
  \href{http://xxx.lanl.gov/abs/1910.10168}{{\normalfont
  [arXiv:astro-ph.HE/1910.10168]}}.
\newblock
  doi:{\changeurlcolor{black}\href{https://doi.org/10.1093/mnras/staa286}{\detokenize{10.1093/mnras/staa286}}}.

\bibitem[{Gupta} \em{et~al.}(2018){Gupta}, {Nath}, {Sharma}, and
  {Eichler}]{Gupta2018MNRAS}
{Gupta}, S.; {Nath}, B.B.; {Sharma}, P.; {Eichler}, D.
\newblock {Lack of thermal energy in superbubbles: hint of cosmic rays?}
\newblock {\em \mnras} {\bf 2018}, {\em 473},~1537--1553,
  \href{http://xxx.lanl.gov/abs/1705.10448}{{\normalfont
  [arXiv:astro-ph.GA/1705.10448]}}.
\newblock
  doi:{\changeurlcolor{black}\href{https://doi.org/10.1093/mnras/stx2427}{\detokenize{10.1093/mnras/stx2427}}}.

\bibitem[{Morlino} \em{et~al.}(2021){Morlino}, {Blasi}, {Peretti}, and
  {Cristofari}]{Morlino2021MNRAS}
{Morlino}, G.; {Blasi}, P.; {Peretti}, E.; {Cristofari}, P.
\newblock {Particle acceleration in winds of star clusters}.
\newblock {\em \mnras} {\bf 2021}, {\em 504},~6096--6105,
  \href{http://xxx.lanl.gov/abs/2102.09217}{{\normalfont
  [arXiv:astro-ph.HE/2102.09217]}}.
\newblock
  doi:{\changeurlcolor{black}\href{https://doi.org/10.1093/mnras/stab690}{\detokenize{10.1093/mnras/stab690}}}.

\bibitem[{Gabici}(2023)]{Gabici2023arXiv230106505G}
{Gabici}, S.
\newblock {Star clusters as cosmic ray accelerators}.
\newblock {\em arXiv e-prints} {\bf 2023}, p. arXiv:2301.06505,
  \href{http://xxx.lanl.gov/abs/2301.06505}{{\normalfont
  [arXiv:astro-ph.HE/2301.06505]}}.
\newblock
  doi:{\changeurlcolor{black}\href{https://doi.org/10.48550/arXiv.2301.06505}{\detokenize{10.48550/arXiv.2301.06505}}}.

\bibitem[{Murphy} \em{et~al.}(2016){Murphy}, {Sasaki}, {Binns}, {Brandt},
  {Hams}, {Israel}, {Labrador}, {Link}, {Mewaldt}, {Mitchell}, {Rauch},
  {Sakai}, {Stone}, {Waddington}, {Walsh}, {Ward}, and
  {Wiedenbeck}]{Murphy2016ApJ}
{Murphy}, R.P.; {Sasaki}, M.; {Binns}, W.R.; {Brandt}, T.J.; {Hams}, T.;
  {Israel}, M.H.; {Labrador}, A.W.; {Link}, J.T.; {Mewaldt}, R.A.; {Mitchell},
  J.W.;  et~al.
\newblock {Galactic Cosmic Ray Origins and OB Associations: Evidence from
  SuperTIGER Observations of Elements $_{26}$Fe through $_{40}$Zr}.
\newblock {\em \apj} {\bf 2016}, {\em 831},~148,
  \href{http://xxx.lanl.gov/abs/1608.08183}{{\normalfont
  [arXiv:astro-ph.HE/1608.08183]}}.
\newblock
  doi:{\changeurlcolor{black}\href{https://doi.org/10.3847/0004-637X/831/2/148}{\detokenize{10.3847/0004-637X/831/2/148}}}.

\bibitem[{Tatischeff} \em{et~al.}(2021){Tatischeff}, {Raymond}, {Duprat},
  {Gabici}, and {Recchia}]{Tatischeff2021MNRAS}
{Tatischeff}, V.; {Raymond}, J.C.; {Duprat}, J.; {Gabici}, S.; {Recchia}, S.
\newblock {The origin of Galactic cosmic rays as revealed by their
  composition}.
\newblock {\em \mnras} {\bf 2021}, {\em 508},~1321--1345,
  \href{http://xxx.lanl.gov/abs/2106.15581}{{\normalfont
  [arXiv:astro-ph.HE/2106.15581]}}.
\newblock
  doi:{\changeurlcolor{black}\href{https://doi.org/10.1093/mnras/stab2533}{\detokenize{10.1093/mnras/stab2533}}}.

\bibitem[{Rauch} \em{et~al.}(2009){Rauch}, {Link}, {Lodders}, {Israel},
  {Barbier}, {Binns}, {Christian}, {Cummings}, {de Nolfo}, {Geier}, {Mewaldt},
  {Mitchell}, {Schindler}, {Scott}, {Stone}, {Streitmatter}, {Waddington}, and
  {Wiedenbeck}]{Rauch2009ApJ}
{Rauch}, B.F.; {Link}, J.T.; {Lodders}, K.; {Israel}, M.H.; {Barbier}, L.M.;
  {Binns}, W.R.; {Christian}, E.R.; {Cummings}, J.R.; {de Nolfo}, G.A.;
  {Geier}, S.;  et~al.
\newblock {Cosmic Ray origin in OB Associations and Preferential Acceleration
  of Refractory Elements: Evidence from Abundances of Elements $_{26}$Fe
  through $_{34}$Se}.
\newblock {\em \apj} {\bf 2009}, {\em 697},~2083--2088,
  \href{http://xxx.lanl.gov/abs/0906.2021}{{\normalfont
  [arXiv:astro-ph.HE/0906.2021]}}.
\newblock
  doi:{\changeurlcolor{black}\href{https://doi.org/10.1088/0004-637X/697/2/2083}{\detokenize{10.1088/0004-637X/697/2/2083}}}.

\bibitem[{Wiedenbeck} \em{et~al.}(2007){Wiedenbeck}, {Binns}, {Cummings},
  {Davis}, {de Nolfo}, {Israel}, {Leske}, {Mewaldt}, {Stone}, and {von
  Rosenvinge}]{Wiedenbeck2007SSRv}
{Wiedenbeck}, M.E.; {Binns}, W.R.; {Cummings}, A.C.; {Davis}, A.J.; {de Nolfo},
  G.A.; {Israel}, M.H.; {Leske}, R.A.; {Mewaldt}, R.A.; {Stone}, E.C.; {von
  Rosenvinge}, T.T.
\newblock {An Overview of the Origin of Galactic Cosmic Rays as Inferred from
  Observations of Heavy Ion Composition and Spectra}.
\newblock {\em \ssr} {\bf 2007}, {\em 130},~415--429.
\newblock
  doi:{\changeurlcolor{black}\href{https://doi.org/10.1007/s11214-007-9198-y}{\detokenize{10.1007/s11214-007-9198-y}}}.

\bibitem[{Meyer} \em{et~al.}(1997){Meyer}, {Drury}, and
  {Ellison}]{Meyer1997ApJ}
{Meyer}, J.P.; {Drury}, L.O.; {Ellison}, D.C.
\newblock {Galactic Cosmic Rays from Supernova Remnants. I. A Cosmic-Ray
  Composition Controlled by Volatility and Mass-to-Charge Ratio}.
\newblock {\em \apj} {\bf 1997}, {\em 487},~182--196,
  \href{http://xxx.lanl.gov/abs/astro-ph/9704267}{{\normalfont
  [arXiv:astro-ph/astro-ph/9704267]}}.
\newblock
  doi:{\changeurlcolor{black}\href{https://doi.org/10.1086/304599}{\detokenize{10.1086/304599}}}.

\bibitem[{Ellison} \em{et~al.}(1997){Ellison}, {Drury}, and
  {Meyer}]{Ellison1997ApJ}
{Ellison}, D.C.; {Drury}, L.O.; {Meyer}, J.P.
\newblock {Galactic Cosmic Rays from Supernova Remnants. II. Shock Acceleration
  of Gas and Dust}.
\newblock {\em \apj} {\bf 1997}, {\em 487},~197--217,
  \href{http://xxx.lanl.gov/abs/astro-ph/9704293}{{\normalfont
  [arXiv:astro-ph/astro-ph/9704293]}}.
\newblock
  doi:{\changeurlcolor{black}\href{https://doi.org/10.1086/304580}{\detokenize{10.1086/304580}}}.

\bibitem[{Parizot}(2000)]{Parizot2000A&A}
{Parizot}, E.
\newblock {Superbubbles and the Galactic evolution of Li, Be and B}.
\newblock {\em \aap} {\bf 2000}, {\em 362},~786--798,
  \href{http://xxx.lanl.gov/abs/astro-ph/0006099}{{\normalfont
  [arXiv:astro-ph/astro-ph/0006099]}}.
\newblock
  doi:{\changeurlcolor{black}\href{https://doi.org/10.48550/arXiv.astro-ph/0006099}{\detokenize{10.48550/arXiv.astro-ph/0006099}}}.

\bibitem[{Meneguzzi} \em{et~al.}(1971){Meneguzzi}, {Audouze}, and
  {Reeves}]{Meneguzzi1971A&A}
{Meneguzzi}, M.; {Audouze}, J.; {Reeves}, H.
\newblock {The production of the elements Li, Be, B by galactic cosmic rays in
  space and its relation with stellar observations.}
\newblock {\em \aap} {\bf 1971}, {\em 15},~337.

\bibitem[{Reeves} \em{et~al.}(1970){Reeves}, {Fowler}, and
  {Hoyle}]{Reeves1970Natur}
{Reeves}, H.; {Fowler}, W.A.; {Hoyle}, F.
\newblock {Galactic Cosmic Ray Origin of Li, Be and B in Stars}.
\newblock {\em \nat} {\bf 1970}, {\em 226},~727--729.
\newblock
  doi:{\changeurlcolor{black}\href{https://doi.org/10.1038/226727a0}{\detokenize{10.1038/226727a0}}}.

\bibitem[Kulikov and Khristiansen(1959)]{kulikov1959size}
Kulikov, G.; Khristiansen, G.
\newblock On the size spectrum of extensive air showers.
\newblock {\em Sov. Phys. JETP} {\bf 1959}, {\em 35},~441--444.

\bibitem[{Parizot}(2014)]{Parizot2014NuPhS}
{Parizot}, E.
\newblock {Cosmic Ray Origin: Lessons from Ultra-High-Energy Cosmic Rays and
  the Galactic/Extragalactic Transition}.
\newblock {\em Nuclear Physics B Proceedings Supplements} {\bf 2014}, {\em
  256},~197--212,  \href{http://xxx.lanl.gov/abs/1410.2655}{{\normalfont
  [arXiv:astro-ph.HE/1410.2655]}}.
\newblock
  doi:{\changeurlcolor{black}\href{https://doi.org/10.1016/j.nuclphysbps.2014.10.023}{\detokenize{10.1016/j.nuclphysbps.2014.10.023}}}.

\bibitem[{Antoni} \em{et~al.}(2005){Antoni}, {Apel}, {Badea}, {Bekk},
  {Bercuci}, {Bl{\"u}mer}, {Bozdog}, {Brancus}, {Chilingarian}, {Daumiller},
  {Doll}, {Engel}, {Engler}, {Fe{\ss}ler}, {Gils}, {Glasstetter}, {Haungs},
  {Heck}, {H{\"o}randel}, {Kampert}, {Klages}, {Maier}, {Mathes}, {Mayer},
  {Milke}, {M{\"u}ller}, {Obenland}, {Oehlschl{\"a}ger}, {Ostapchenko},
  {Petcu}, {Rebel}, {Risse}, {Risse}, {Roth}, {Schatz}, {Schieler}, {Scholz},
  {Thouw}, {Ulrich}, {van Buren}, {Vardanyan}, {Weindl}, {Wochele}, and
  {Zabierowski}]{Antoni2005APh}
{Antoni}, T.; {Apel}, W.D.; {Badea}, A.F.; {Bekk}, K.; {Bercuci}, A.;
  {Bl{\"u}mer}, J.; {Bozdog}, H.; {Brancus}, I.M.; {Chilingarian}, A.;
  {Daumiller}, K.;  et~al.
\newblock {KASCADE measurements of energy spectra for elemental groups of
  cosmic rays: Results and open problems}.
\newblock {\em Astroparticle Physics} {\bf 2005}, {\em 24},~1--25,
  \href{http://xxx.lanl.gov/abs/astro-ph/0505413}{{\normalfont
  [arXiv:astro-ph/astro-ph/0505413]}}.
\newblock
  doi:{\changeurlcolor{black}\href{https://doi.org/10.1016/j.astropartphys.2005.04.001}{\detokenize{10.1016/j.astropartphys.2005.04.001}}}.

\bibitem[{Adriani} \em{et~al.}(2011){Adriani}, {Barbarino}, {Bazilevskaya},
  {Bellotti}, {Boezio}, {Bogomolov}, {Bonechi}, {Bongi}, {Bonvicini},
  {Borisov}, {Bottai}, {Bruno}, {Cafagna}, {Campana}, {Carbone}, {Carlson},
  {Casolino}, {Castellini}, {Consiglio}, {De Pascale}, {De Santis}, {De
  Simone}, {Di Felice}, {Galper}, {Gillard}, {Grishantseva}, {Jerse},
  {Karelin}, {Koldashov}, {Krutkov}, {Kvashnin}, {Leonov}, {Malakhov},
  {Malvezzi}, {Marcelli}, {Mayorov}, {Menn}, {Mikhailov}, {Mocchiutti},
  {Monaco}, {Mori}, {Nikonov}, {Osteria}, {Palma}, {Papini}, {Pearce},
  {Picozza}, {Pizzolotto}, {Ricci}, {Ricciarini}, {Rossetto}, {Sarkar},
  {Simon}, {Sparvoli}, {Spillantini}, {Stozhkov}, {Vacchi}, {Vannuccini},
  {Vasilyev}, {Voronov}, {Yurkin}, {Wu}, {Zampa}, {Zampa}, and
  {Zverev}]{Adriani2011Sci}
{Adriani}, O.; {Barbarino}, G.C.; {Bazilevskaya}, G.A.; {Bellotti}, R.;
  {Boezio}, M.; {Bogomolov}, E.A.; {Bonechi}, L.; {Bongi}, M.; {Bonvicini}, V.;
  {Borisov}, S.;  et~al.
\newblock {PAMELA Measurements of Cosmic-Ray Proton and Helium Spectra}.
\newblock {\em Science} {\bf 2011}, {\em 332},~69,
  \href{http://xxx.lanl.gov/abs/1103.4055}{{\normalfont
  [arXiv:astro-ph.HE/1103.4055]}}.
\newblock
  doi:{\changeurlcolor{black}\href{https://doi.org/10.1126/science.1199172}{\detokenize{10.1126/science.1199172}}}.

\bibitem[{Aguilar} \em{et~al.}(2015{\natexlab{a}}){Aguilar}, {Aisa}, {Alpat},
  {Alvino}, {Ambrosi}, {Andeen}, {Arruda}, {Attig}, {Azzarello}, {Bachlechner},
  {Barao}, {Barrau}, {Barrin}, {Bartoloni}, {Basara}, {Battarbee}, {Battiston},
  {Bazo}, {Becker}, {Behlmann}, {Beischer}, {Berdugo}, {Bertucci},
  {Bigongiari}, {Bindi}, {Bizzaglia}, {Bizzarri}, {Boella}, {de Boer},
  {Bollweg}, {Bonnivard}, {Borgia}, {Borsini}, {Boschini}, {Bourquin},
  {Burger}, {Cadoux}, {Cai}, {Capell}, {Caroff}, {Casaus}, {Cascioli},
  {Castellini}, {Cernuda}, {Cerreta}, {Cervelli}, {Chae}, {Chang}, {Chen},
  {Chen}, {Cheng}, {Chen}, {Cheng}, {Chou}, {Choumilov}, {Choutko}, {Chung},
  {Clark}, {Clavero}, {Coignet}, {Consolandi}, {Contin}, {Corti}, {Gil},
  {Coste}, {Creus}, {Crispoltoni}, {Cui}, {Dai}, {Delgado}, {Della Torre},
  {Demirk{\"o}z}, {Derome}, {Di Falco}, {Di Masso}, {Dimiccoli}, {D{\'\i}az},
  {von Doetinchem}, {Donnini}, {Du}, {Duranti}, {D'Urso}, {Eline}, {Eppling},
  {Eronen}, {Fan}, {Farnesini}, {Feng}, {Fiandrini}, {Fiasson}, {Finch},
  {Fisher}, {Galaktionov}, {Gallucci}, {Garc{\'\i}a}, {Garc{\'\i}a-L{\'o}pez},
  {Gargiulo}, {Gast}, {Gebauer}, {Gervasi}, {Ghelfi}, {Gillard}, {Giovacchini},
  {Goglov}, {Gong}, {Goy}, {Grabski}, {Grandi}, {Graziani}, {Guandalini},
  {Guerri}, {Guo}, {Haas}, {Habiby}, {Haino}, {Han}, {He}, {Heil}, {Hoffman},
  {Hsieh}, {Huang}, {Huh}, {Incagli}, {Ionica}, {Jang}, {Jinchi}, {Kanishev},
  {Kim}, {Kim}, {Kirn}, {Kossakowski}, {Kounina}, {Kounine}, {Koutsenko},
  {Krafczyk}, {La Vacca}, {Laudi}, {Laurenti}, {Lazzizzera}, {Lebedev}, {Lee},
  {Lee}, {Leluc}, {Levi}, {Li}, {Li}, {Li}, {Li}, {Li}, {Li}, {Li}, {Li}, {Li},
  {Lim}, {Lin}, {Lipari}, {Lippert}, {Liu}, {Liu}, {Lolli}, {Lomtadze}, {Lu},
  {Lu}, {Lu}, {Luebelsmeyer}, {Luo}, {Lv}, {Majka}, {Ma{\~n}{\'a}},
  {Mar{\'\i}n}, {Martin}, {Mart{\'\i}nez}, {Masi}, {Maurin}, {Menchaca-Rocha},
  {Meng}, {Mo}, {Morescalchi}, {Mott}, {M{\"u}ller}, {Ni}, {Nikonov},
  {Nozzoli}, {Nunes}, {Obermeier}, {Oliva}, {Orcinha}, {Palmonari},
  {Palomares}, {Paniccia}, {Papi}, {Pauluzzi}, {Pedreschi}, {Pensotti},
  {Pereira}, {Picot-Clemente}, {Pilo}, {Piluso}, {Pizzolotto}, {Plyaskin},
  {Pohl}, {Poireau}, {Postaci}, {Putze}, {Quadrani}, {Qi}, {Qin}, {Qu},
  {R{\"a}ih{\"a}}, {Rancoita}, {Rapin}, {Ricol}, {Rodr{\'\i}guez},
  {Rosier-Lees}, {Rozhkov}, {Rozza}, {Sagdeev}, {Sandweiss}, {Saouter},
  {Sbarra}, {Schael}, {Schmidt}, {von Dratzig}, {Schwering}, {Scolieri}, {Seo},
  {Shan}, {Shan}, {Shi}, {Shi}, {Shi}, {Siedenburg}, {Son}, {Spada},
  {Spinella}, {Sun}, {Sun}, {Tacconi}, {Tang}, {Tang}, {Tang}, {Tao},
  {Tescaro}, {Ting}, {Ting}, {Tomassetti}, {Torsti}, {T{\"u}rko{\v{g}}lu},
  {Urban}, {Vagelli}, {Valente}, {Vannini}, {Valtonen}, {Vaurynovich},
  {Vecchi}, {Velasco}, {Vialle}, {Vitale}, {Vitillo}, {Wang}, {Wang}, {Wang},
  {Wang}, {Wang}, {Wang}, {Weng}, {Whitman}, {Wienkenh{\"o}ver}, {Wu}, {Wu},
  {Xia}, {Xie}, {Xie}, {Xiong}, {Xin}, {Xu}, {Xu}, {Yan}, {Yang}, {Yang}, {Ye},
  {Yi}, {Yu}, {Yu}, {Zeissler}, {Zhang}, {Zhang}, {Zhang}, {Zhang}, {Zheng},
  {Zhuang}, {Zhukov}, {Zichichi}, {Zimmermann}, {Zuccon}, {Zurbach}, and {AMS
  Collaboration}]{Aguilar2015PhRvL}
{Aguilar}, M.; {Aisa}, D.; {Alpat}, B.; {Alvino}, A.; {Ambrosi}, G.; {Andeen},
  K.; {Arruda}, L.; {Attig}, N.; {Azzarello}, P.; {Bachlechner}, A.;  et~al.
\newblock {Precision Measurement of the Proton Flux in Primary Cosmic Rays from
  Rigidity 1 GV to 1.8 TV with the Alpha Magnetic Spectrometer on the
  International Space Station}.
\newblock {\em \prl} {\bf 2015}, {\em 114},~171103.
\newblock
  doi:{\changeurlcolor{black}\href{https://doi.org/10.1103/PhysRevLett.114.171103}{\detokenize{10.1103/PhysRevLett.114.171103}}}.

\bibitem[{Aguilar} \em{et~al.}(2015{\natexlab{b}}){Aguilar}, {Aisa}, {Alpat},
  {Alvino}, {Ambrosi}, {Andeen}, {Arruda}, {Attig}, {Azzarello}, {Bachlechner},
  {Barao}, {Barrau}, {Barrin}, {Bartoloni}, {Basara}, {Battarbee}, {Battiston},
  {Bazo}, {Becker}, {Behlmann}, {Beischer}, {Berdugo}, {Bertucci}, {Bindi},
  {Bizzaglia}, {Bizzarri}, {Boella}, {de Boer}, {Bollweg}, {Bonnivard},
  {Borgia}, {Borsini}, {Boschini}, {Bourquin}, {Burger}, {Cadoux}, {Cai},
  {Capell}, {Caroff}, {Casaus}, {Castellini}, {Cernuda}, {Cerreta}, {Cervelli},
  {Chae}, {Chang}, {Chen}, {Chen}, {Chen}, {Chen}, {Cheng}, {Chou},
  {Choumilov}, {Choutko}, {Chung}, {Clark}, {Clavero}, {Coignet}, {Consolandi},
  {Contin}, {Corti}, {Gil}, {Coste}, {Creus}, {Crispoltoni}, {Cui}, {Dai},
  {Delgado}, {Della Torre}, {Demirk{\"o}z}, {Derome}, {Di Falco}, {Di Masso},
  {Dimiccoli}, {D{\'\i}az}, {von Doetinchem}, {Donnini}, {Duranti}, {D'Urso},
  {Egorov}, {Eline}, {Eppling}, {Eronen}, {Fan}, {Farnesini}, {Feng},
  {Fiandrini}, {Fiasson}, {Finch}, {Fisher}, {Formato}, {Galaktionov},
  {Gallucci}, {Garc{\'\i}a}, {Garc{\'\i}a-L{\'o}pez}, {Gargiulo}, {Gast},
  {Gebauer}, {Gervasi}, {Ghelfi}, {Giovacchini}, {Goglov}, {Gong}, {Goy},
  {Grabski}, {Grandi}, {Graziani}, {Guandalini}, {Guerri}, {Guo}, {Haas},
  {Habiby}, {Haino}, {Han}, {He}, {Heil}, {Hoffman}, {Hsieh}, {Huang}, {Huh},
  {Incagli}, {Ionica}, {Jang}, {Jinchi}, {Kanishev}, {Kim}, {Kim}, {Kirn},
  {Korkmaz}, {Kossakowski}, {Kounina}, {Kounine}, {Koutsenko}, {Krafczyk}, {La
  Vacca}, {Laudi}, {Laurenti}, {Lazzizzera}, {Lebedev}, {Lee}, {Lee}, {Leluc},
  {Li}, {Li}, {Li}, {Li}, {Li}, {Li}, {Li}, {Li}, {Li}, {Li}, {Lim}, {Lin},
  {Lipari}, {Lippert}, {Liu}, {Liu}, {Liu}, {Lolli}, {Lomtadze}, {Lu}, {Lu},
  {Lu}, {Luebelsmeyer}, {Luo}, {Luo}, {Lv}, {Majka}, {Ma{\~n}{\'a}},
  {Mar{\'\i}n}, {Martin}, {Mart{\'\i}nez}, {Masi}, {Maurin}, {Menchaca-Rocha},
  {Meng}, {Mo}, {Morescalchi}, {Mott}, {M{\"u}ller}, {Nelson}, {Ni}, {Nikonov},
  {Nozzoli}, {Nunes}, {Obermeier}, {Oliva}, {Orcinha}, {Palmonari},
  {Palomares}, {Paniccia}, {Papi}, {Pauluzzi}, {Pedreschi}, {Pensotti},
  {Pereira}, {Picot-Clemente}, {Pilo}, {Piluso}, {Pizzolotto}, {Plyaskin},
  {Pohl}, {Poireau}, {Putze}, {Quadrani}, {Qi}, {Qin}, {Qu}, {R{\"a}ih{\"a}},
  {Rancoita}, {Rapin}, {Ricol}, {Rodr{\'\i}guez}, {Rosier-Lees}, {Rozhkov},
  {Rozza}, {Sagdeev}, {Sandweiss}, {Saouter}, {Schael}, {Schmidt}, {von
  Dratzig}, {Schwering}, {Scolieri}, {Seo}, {Shan}, {Shan}, {Shi}, {Shi},
  {Shi}, {Siedenburg}, {Son}, {Song}, {Spada}, {Spinella}, {Sun}, {Sun},
  {Tacconi}, {Tang}, {Tang}, {Tang}, {Tao}, {Tescaro}, {Ting}, {Ting},
  {Tomassetti}, {Torsti}, {T{\"u}rko{\v{g}}lu}, {Urban}, {Vagelli}, {Valente},
  {Vannini}, {Valtonen}, {Vaurynovich}, {Vecchi}, {Velasco}, {Vialle},
  {Vitale}, {Vitillo}, {Wang}, {Wang}, {Wang}, {Wang}, {Wang}, {Wang}, {Weng},
  {Whitman}, {Wienkenh{\"o}ver}, {Willenbrock}, {Wu}, {Wu}, {Xia}, {Xie},
  {Xie}, {Xiong}, {Xu}, {Xu}, {Yan}, {Yang}, {Yang}, {Yang}, {Ye}, {Yi}, {Yu},
  {Yu}, {Zeissler}, {Zhang}, {Zhang}, {Zhang}, {Zhang}, {Zhang}, {Zhang},
  {Zhang}, {Zheng}, {Zhuang}, {Zhukov}, {Zichichi}, {Zimmermann}, {Zuccon}, and
  {AMS Collaboration}]{Aguilar2015PhRvL_b}
{Aguilar}, M.; {Aisa}, D.; {Alpat}, B.; {Alvino}, A.; {Ambrosi}, G.; {Andeen},
  K.; {Arruda}, L.; {Attig}, N.; {Azzarello}, P.; {Bachlechner}, A.;  et~al.
\newblock {Precision Measurement of the Helium Flux in Primary Cosmic Rays of
  Rigidities 1.9 GV to 3 TV with the Alpha Magnetic Spectrometer on the
  International Space Station}.
\newblock {\em \prl} {\bf 2015}, {\em 115},~211101.
\newblock
  doi:{\changeurlcolor{black}\href{https://doi.org/10.1103/PhysRevLett.115.211101}{\detokenize{10.1103/PhysRevLett.115.211101}}}.

\bibitem[{Bartoli} \em{et~al.}(2015){Bartoli}, {Bernardini}, {Bi}, {Cao},
  {Catalanotti}, {Chen}, {Chen}, {Cui}, {Dai}, {D'Amone}, {Danzengluobu}, {De
  Mitri}, {D'Ettorre Piazzoli}, {Di Girolamo}, {Di Sciascio}, {Feng}, {Feng},
  {Feng}, {Guo}, {Guo}, {He}, {Hu}, {Hu}, {Iacovacci}, {Iuppa}, {Jia},
  {Labaciren}, {Li}, {Liu}, {Liu}, {Liu}, {Lu}, {Ma}, {Ma}, {Mancarella},
  {Mari}, {Marsella}, {Mastroianni}, {Montini}, {Ning}, {Perrone}, {Pistilli},
  {Salvini}, {Santonico}, {Shen}, {Sheng}, {Shi}, {Surdo}, {Tan}, {Vallania},
  {Vernetto}, {Vigorito}, {Wang}, {Wu}, {Wu}, {Xue}, {Yang}, {Yang}, {Yao},
  {Yuan}, {Zha}, {Zhang}, {Zhang}, {Zhang}, {Zhang}, {Zhao}, {Zhaxiciren},
  {Zhaxisangzhu}, {Zhou}, {Zhu}, {Zhu}, {Bai}, {Chen}, {Feng}, {Gao}, {Gu},
  {Hou}, {Liu}, {Liu}, {Wang}, {Xiao}, {Zhang}, {Zhang}, {Zhou}, {Zuo}, and
  {ARGO-YBJ Collaboration}]{Bartoli2015PhRvD}
{Bartoli}, B.; {Bernardini}, P.; {Bi}, X.J.; {Cao}, Z.; {Catalanotti}, S.;
  {Chen}, S.Z.; {Chen}, T.L.; {Cui}, S.W.; {Dai}, B.Z.; {D'Amone}, A.;  et~al.
\newblock {Knee of the cosmic hydrogen and helium spectrum below 1 PeV measured
  by ARGO-YBJ and a Cherenkov telescope of LHAASO}.
\newblock {\em \prd} {\bf 2015}, {\em 92},~092005,
  \href{http://xxx.lanl.gov/abs/1502.03164}{{\normalfont
  [arXiv:astro-ph.HE/1502.03164]}}.
\newblock
  doi:{\changeurlcolor{black}\href{https://doi.org/10.1103/PhysRevD.92.092005}{\detokenize{10.1103/PhysRevD.92.092005}}}.

\bibitem[{Lhaaso Collaboration} \em{et~al.}(2021){Lhaaso Collaboration}, {Cao},
  {Aharonian}, {An}, {Axikegu}, {Bai}, {Bai}, {Bao}, {Bastieri}, {Bi}, {Bi},
  {Cai}, {Cai}, {Cao}, {Chang}, {Chang}, {Chen}, {Chen}, {Chen}, {Chen},
  {Chen}, {Chen}, {Chen}, {Chen}, {Chen}, {Chen}, {Chen}, {Chen}, {Chen},
  {Chen}, {Cheng}, {Cheng}, {Cui}, {Cui}, {Cui}, {D'Ettorre Piazzoli}, {Dai},
  {Dai}, {Dai}, {Danzengluobu}, {Della Volpe}, {Dong}, {Duan}, {Fan}, {Fan},
  {Fan}, {Fang}, {Fang}, {Feng}, {Feng}, {Feng}, {Feng}, {Gao}, {Gao}, {Gao},
  {Gao}, {Gao}, {Ge}, {Geng}, {Gong}, {Gou}, {Gu}, {Guo}, {Guo}, {Guo}, {Guo},
  {Guo}, {Han}, {He}, {He}, {He}, {He}, {He}, {He}, {Heller}, {Hor}, {Hou},
  {Hou}, {Hu}, {Hu}, {Hu}, {Hu}, {Huang}, {Huang}, {Huang}, {Huang}, {Huang},
  {Huang}, {Ji}, {Ji}, {Jia}, {Jiang}, {Jiang}, {Jin}, {Ke}, {Kuleshov},
  {Levochkin}, {Li}, {Li}, {Li}, {Li}, {Li}, {Li}, {Li}, {Li}, {Li}, {Li},
  {Li}, {Li}, {Li}, {Li}, {Li}, {Li}, {Li}, {Li}, {Liang}, {Liang}, {Lin},
  {Liu}, {Liu}, {Liu}, {Liu}, {Liu}, {Liu}, {Liu}, {Liu}, {Liu}, {Liu}, {Liu},
  {Liu}, {Liu}, {Liu}, {Liu}, {Liu}, {Long}, {Lu}, {Lv}, {Ma}, {Ma}, {Ma},
  {Mao}, {Masood}, {Min}, {Mitthumsiri}, {Montaruli}, {Nan}, {Pang},
  {Pattarakijwanich}, {Pei}, {Qi}, {Qi}, {Qiao}, {Qin}, {Ruffolo}, {Rulev},
  {Saiz}, {Shao}, {Shchegolev}, {Sheng}, {Shi}, {Song}, {Stenkin}, {Stepanov},
  {Su}, {Sun}, {Sun}, {Sun}, {Tam}, {Tang}, {Tian}, {Wang}, {Wang}, {Wang},
  {Wang}, {Wang}, {Wang}, {Wang}, {Wang}, {Wang}, {Wang}, {Wang}, {Wang},
  {Wang}, {Wang}, {Wang}, {Wang}, {Wang}, {Wang}, {Wang}, {Wang}, {Wang},
  {Wang}, {Wei}, {Wei}, {Wei}, {Wen}, {Wu}, {Wu}, {Wu}, {Wu}, {Wu}, {Xi},
  {Xia}, {Xia}, {Xiang}, {Xiao}, {Xiao}, {Xiao}, {Xin}, {Xin}, {Xing}, {Xu},
  {Xu}, {Xue}, {Yan}, {Yan}, {Yang}, {Yang}, {Yang}, {Yang}, {Yang}, {Yang},
  {Yang}, {Yao}, {Yao}, {Ye}, {Yin}, {Yin}, {You}, {You}, {Yu}, {Yuan}, {Zeng},
  {Zeng}, {Zeng}, {Zeng}, {Zha}, {Zhai}, {Zhang}, {Zhang}, {Zhang}, {Zhang},
  {Zhang}, {Zhang}, {Zhang}, {Zhang}, {Zhang}, {Zhang}, {Zhang}, {Zhang},
  {Zhang}, {Zhang}, {Zhang}, {Zhang}, {Zhang}, {Zhang}, {Zhang}, {Zhao},
  {Zhao}, {Zhao}, {Zhao}, {Zhao}, {Zheng}, {Zheng}, {Zhou}, {Zhou}, {Zhou},
  {Zhou}, {Zhou}, {Zhou}, {Zhu}, {Zhu}, {Zhu}, {Zhu}, and {Zuo}]{Lhaaso2021Sci}
{Lhaaso Collaboration}.; {Cao}, Z.; {Aharonian}, F.; {An}, Q.; {Axikegu}.;
  {Bai}, L.X.; {Bai}, Y.X.; {Bao}, Y.W.; {Bastieri}, D.; {Bi}, X.J.;  et~al.
\newblock {Peta-electron volt gamma-ray emission from the Crab Nebula}.
\newblock {\em Science} {\bf 2021}, {\em 373},~425--430,
  \href{http://xxx.lanl.gov/abs/2111.06545}{{\normalfont
  [arXiv:astro-ph.HE/2111.06545]}}.
\newblock
  doi:{\changeurlcolor{black}\href{https://doi.org/10.1126/science.abg5137}{\detokenize{10.1126/science.abg5137}}}.

\bibitem[{Liu} and {Wang}(2021)]{Liu2021ApJ}
{Liu}, R.Y.; {Wang}, X.Y.
\newblock {PeV Emission of the Crab Nebula: Constraints on the Proton Content
  in Pulsar Wind and Implications}.
\newblock {\em \apj} {\bf 2021}, {\em 922},~221,
  \href{http://xxx.lanl.gov/abs/2109.14148}{{\normalfont
  [arXiv:astro-ph.HE/2109.14148]}}.
\newblock
  doi:{\changeurlcolor{black}\href{https://doi.org/10.3847/1538-4357/ac2ba0}{\detokenize{10.3847/1538-4357/ac2ba0}}}.

\bibitem[{Peng} \em{et~al.}(2022){Peng}, {Bao}, {Lu}, and {Zhang}]{Peng2022ApJ}
{Peng}, Q.Y.; {Bao}, B.W.; {Lu}, F.W.; {Zhang}, L.
\newblock {Multiband Emission up to PeV Energy from the Crab Nebula in a
  Spatially Dependent Lepto-hadronic Model}.
\newblock {\em \apj} {\bf 2022}, {\em 926},~7,
  \href{http://xxx.lanl.gov/abs/2112.14939}{{\normalfont
  [arXiv:astro-ph.HE/2112.14939]}}.
\newblock
  doi:{\changeurlcolor{black}\href{https://doi.org/10.3847/1538-4357/ac4161}{\detokenize{10.3847/1538-4357/ac4161}}}.

\bibitem[{Baade} and {Zwicky}(1934)]{Baade1934PNAS}
{Baade}, W.; {Zwicky}, F.
\newblock {Cosmic Rays from Super-novae}.
\newblock {\em Proceedings of the National Academy of Science} {\bf 1934}, {\em
  20},~259--263.
\newblock
  doi:{\changeurlcolor{black}\href{https://doi.org/10.1073/pnas.20.5.259}{\detokenize{10.1073/pnas.20.5.259}}}.

\bibitem[{HESS Collaboration} \em{et~al.}(2016){HESS Collaboration},
  {Abramowski}, {Aharonian}, {Benkhali}, {Akhperjanian}, {Ang{\"u}ner},
  {Backes}, {Balzer}, {Becherini}, {Tjus}, {Berge}, {Bernhard}, {Bernl{\"o}hr},
  {Birsin}, {Blackwell}, {B{\"o}ttcher}, {Boisson}, {Bolmont}, {Bordas},
  {Bregeon}, {Brun}, {Brun}, {Bryan}, {Bulik}, {Carr}, {Casanova},
  {Chakraborty}, {Chalme-Calvet}, {Chaves}, {Chen}, {Chr{\'e}tien},
  {Colafrancesco}, {Cologna}, {Conrad}, {Couturier}, {Cui}, {Davids},
  {Degrange}, {Deil}, {Dewilt}, {Djannati-Ata{\"\i}}, {Domainko}, {Donath},
  {Drury}, {Dubus}, {Dutson}, {Dyks}, {Dyrda}, {Edwards}, {Egberts}, {Eger},
  {Ernenwein}, {Espigat}, {Farnier}, {Fegan}, {Feinstein}, {Fernandes},
  {Fernandez}, {Fiasson}, {Fontaine}, {F{\"o}rster}, {F{\"u}{\ss}ling},
  {Gabici}, {Gajdus}, {Gallant}, {Garrigoux}, {Giavitto}, {Giebels},
  {Glicenstein}, {Gottschall}, {Goyal}, {Grondin}, {Grudzi{\'n}ska}, {Hadasch},
  {H{\"a}ffner}, {Hahn}, {Hawkes}, {Heinzelmann}, {Henri}, {Hermann}, {Hervet},
  {Hillert}, {Hinton}, {Hofmann}, {Hofverberg}, {Hoischen}, {Holler}, {Horns},
  {Ivascenko}, {Jacholkowska}, {Jamrozy}, {Janiak}, {Jankowsky},
  {Jung-Richardt}, {Kastendieck}, {Katarzy{\'n}ski}, {Katz}, {Kerszberg},
  {Kh{\'e}lifi}, {Kieffer}, {Klepser}, {Klochkov}, {Klu{\'z}niak}, {Kolitzus},
  {Komin}, {Kosack}, {Krakau}, {Krayzel}, {Kr{\"u}ger}, {Laffon}, {Lamanna},
  {Lau}, {Lefaucheur}, {Lefranc}, {Lemi{\'e}re}, {Lemoine-Goumard}, {Lenain},
  {Lohse}, {Lopatin}, {Lu}, {Lui}, {Marandon}, {Marcowith}, {Mariaud}, {Marx},
  {Maurin}, {Maxted}, {Mayer}, {Meintjes}, {Menzler}, {Meyer}, {Mitchell},
  {Moderski}, {Mohamed}, {Mor{\r{a}}}, {Moulin}, {Murach}, {de Naurois},
  {Niemiec}, {Oakes}, {Odaka}, {{\"O}ttl}, {Ohm}, {Opitz}, {Ostrowski}, {Oya},
  {Panter}, {Parsons}, {Arribas}, {Pekeur}, {Pelletier}, {Petrucci}, {Peyaud},
  {Pita}, {Poon}, {Prokoph}, {P{\"u}hlhofer}, {Punch}, {Quirrenbach}, {Raab},
  {Reichardt}, {Reimer}, {Reimer}, {Renaud}, {de Los Reyes}, {Rieger},
  {Romoli}, {Rosier-Lees}, {Rowell}, {Rudak}, {Rulten}, {Sahakian}, {Salek},
  {Sanchez}, {Santangelo}, {Sasaki}, {Schlickeiser}, {Sch{\"u}ssler}, {Schulz},
  {Schwanke}, {Schwemmer}, {Seyffert}, {Simoni}, {Sol}, {Spanier}, {Spengler},
  {Spies}, {Stawarz}, {Steenkamp}, {Stegmann}, {Stinzing}, {Stycz}, {Sushch},
  {Tavernet}, {Tavernier}, {Taylor}, {Terrier}, {Tluczykont}, {Trichard},
  {Tuffs}, {Valerius}, {van der Walt}, {van Eldik}, {van Soelen},
  {Vasileiadis}, {Veh}, {Venter}, {Viana}, {Vincent}, {Vink}, {Voisin},
  {V{\"o}lk}, {Vuillaume}, {Wagner}, {Wagner}, {Wagner}, {Weidinger},
  {Weitzel}, {White}, {Wierzcholska}, {Willmann}, {W{\"o}rnlein}, {Wouters},
  {Yang}, {Zabalza}, {Zaborov}, {Zacharias}, {Zdziarski}, {Zech}, {Zefi}, and
  {{\.Z}ywucka}]{HESS2016Natur}
{HESS Collaboration}.; {Abramowski}, A.; {Aharonian}, F.; {Benkhali}, F.A.;
  {Akhperjanian}, A.G.; {Ang{\"u}ner}, E.O.; {Backes}, M.; {Balzer}, A.;
  {Becherini}, Y.; {Tjus}, J.B.;  et~al.
\newblock {Acceleration of petaelectronvolt protons in the Galactic Centre}.
\newblock {\em \nat} {\bf 2016}, {\em 531},~476--479,
  \href{http://xxx.lanl.gov/abs/1603.07730}{{\normalfont
  [arXiv:astro-ph.HE/1603.07730]}}.
\newblock
  doi:{\changeurlcolor{black}\href{https://doi.org/10.1038/nature17147}{\detokenize{10.1038/nature17147}}}.

\bibitem[{Bednarek} and {Protheroe}(2002)]{Bednarek2002APh}
{Bednarek}, W.; {Protheroe}, R.J.
\newblock {Contribution of nuclei accelerated by gamma-ray pulsars to cosmic
  rays in the Galaxy}.
\newblock {\em Astroparticle Physics} {\bf 2002}, {\em 16},~397--409,
  \href{http://xxx.lanl.gov/abs/astro-ph/0103160}{{\normalfont
  [arXiv:astro-ph/astro-ph/0103160]}}.
\newblock
  doi:{\changeurlcolor{black}\href{https://doi.org/10.1016/S0927-6505(01)00124-4}{\detokenize{10.1016/S0927-6505(01)00124-4}}}.

\bibitem[{Escobar} \em{et~al.}(2022){Escobar}, {Pellizza}, and
  {Romero}]{Escobar2022A&A}
{Escobar}, G.J.; {Pellizza}, L.J.; {Romero}, G.E.
\newblock {Highly collimated microquasar jets as efficient cosmic-ray sources}.
\newblock {\em \aap} {\bf 2022}, {\em 665},~A145,
  \href{http://xxx.lanl.gov/abs/2207.08633}{{\normalfont
  [arXiv:astro-ph.HE/2207.08633]}}.
\newblock
  doi:{\changeurlcolor{black}\href{https://doi.org/10.1051/0004-6361/202142753}{\detokenize{10.1051/0004-6361/202142753}}}.

\bibitem[{Yang} \em{et~al.}(2019){Yang}, {Aharonian}, and {de O{\~n}a
  Wilhelmi}]{Yang2019RLSFN}
{Yang}, R.z.; {Aharonian}, F.; {de O{\~n}a Wilhelmi}, E.
\newblock {Massive star clusters as the an alternative source population of
  galactic cosmic rays}.
\newblock {\em Rendiconti Lincei. Scienze Fisiche e Naturali} {\bf 2019}, {\em
  30},~159--164.
\newblock
  doi:{\changeurlcolor{black}\href{https://doi.org/10.1007/s12210-019-00819-3}{\detokenize{10.1007/s12210-019-00819-3}}}.

\bibitem[{Abramowski} \em{et~al.}(2012){Abramowski}, {Acero}, {Aharonian},
  {Akhperjanian}, {Anton}, {Balzer}, {Barnacka}, {Barres de Almeida},
  {Becherini}, {Becker}, {Behera}, {Bernl{\"o}hr}, {Birsin}, {Biteau},
  {Bochow}, {Boisson}, {Bolmont}, {Bordas}, {Brucker}, {Brun}, {Brun}, {Bulik},
  {B{\"u}sching}, {Carrigan}, {Casanova}, {Cerruti}, {Chadwick}, {Charbonnier},
  {Chaves}, {Cheesebrough}, {Chounet}, {Clapson}, {Coignet}, {Cologna},
  {Conrad}, {Dalton}, {Daniel}, {Davids}, {Degrange}, {Deil}, {Dickinson},
  {Djannati-Ata{\"\i}}, {Domainko}, {O'C. Drury}, {Dubois}, {Dubus}, {Dutson},
  {Dyks}, {Dyrda}, {Egberts}, {Eger}, {Espigat}, {Fallon}, {Farnier}, {Fegan},
  {Feinstein}, {Fernandes}, {Fiasson}, {Fontaine}, {F{\"o}rster},
  {F{\"u}{\ss}ling}, {Gallant}, {Gast}, {G{\'e}rard}, {Gerbig}, {Giebels},
  {Glicenstein}, {Gl{\"u}ck}, {Goret}, {G{\"o}ring}, {H{\"a}ffner}, {Hague},
  {Hampf}, {Hauser}, {Heinz}, {Heinzelmann}, {Henri}, {Hermann}, {Hinton},
  {Hoffmann}, {Hofmann}, {Hofverberg}, {Holler}, {Horns}, {Jacholkowska}, {de
  Jager}, {Jahn}, {Jamrozy}, {Jung}, {Kastendieck}, {Katarzy{\'n}ski}, {Katz},
  {Kaufmann}, {Keogh}, {Khangulyan}, {Kh{\'e}lifi}, {Klochkov}, {Klu{\.z}niak},
  {Kneiske}, {Komin}, {Kosack}, {Kossakowski}, {Laffon}, {Lamanna}, {Lennarz},
  {Lohse}, {Lopatin}, {Lu}, {Marandon}, {Marcowith}, {Masbou}, {Maurin},
  {Maxted}, {Mayer}, {McComb}, {Medina}, {M{\'e}hault}, {Moderski}, {Moulin},
  {Naumann}, {Naumann-Godo}, {de Naurois}, {Nedbal}, {Nekrassov}, {Nguyen},
  {Nicholas}, {Niemiec}, {Nolan}, {Ohm}, {de O{\~n}a Wilhelmi}, {Opitz},
  {Ostrowski}, {Oya}, {Panter}, {Paz Arribas}, {Pedaletti}, {Pelletier},
  {Petrucci}, {Pita}, {P{\"u}hlhofer}, {Punch}, {Quirrenbach}, {Raue},
  {Rayner}, {Reimer}, {Reimer}, {Renaud}, {de Los Reyes}, {Rieger}, {Ripken},
  {Rob}, {Rosier-Lees}, {Rowell}, {Rudak}, {Rulten}, {Ruppel}, {Sahakian},
  {Sanchez}, {Santangelo}, {Schlickeiser}, {Sch{\"o}ck}, {Schulz}, {Schwanke},
  {Schwarzburg}, {Schwemmer}, {Sheidaei}, {Sikora}, {Skilton}, {Sol},
  {Spengler}, {Stawarz}, {Steenkamp}, {Stegmann}, {Stinzing}, {Stycz},
  {Sushch}, {Szostek}, {Tavernet}, {Terrier}, {Tluczykont}, {Valerius}, {van
  Eldik}, {Vasileiadis}, {Venter}, {Vialle}, {Viana}, {Vincent}, {V{\"o}lk},
  {Volpe}, {Vorobiov}, {Vorster}, {Wagner}, {Ward}, {White}, {Wierzcholska},
  {Zacharias}, {Zajczyk}, {Zdziarski}, {Zech}, and
  {Zechlin}]{Abramowski2012A&A}
{Abramowski}, A.; {Acero}, F.; {Aharonian}, F.; {Akhperjanian}, A.G.; {Anton},
  G.; {Balzer}, A.; {Barnacka}, A.; {Barres de Almeida}, U.; {Becherini}, Y.;
  {Becker}, J.;  et~al.
\newblock {Discovery of extended VHE {\ensuremath{\gamma}}-ray emission from
  the vicinity of the young massive stellar cluster Westerlund 1}.
\newblock {\em \aap} {\bf 2012}, {\em 537},~A114,
  \href{http://xxx.lanl.gov/abs/1111.2043}{{\normalfont
  [arXiv:astro-ph.HE/1111.2043]}}.
\newblock
  doi:{\changeurlcolor{black}\href{https://doi.org/10.1051/0004-6361/201117928}{\detokenize{10.1051/0004-6361/201117928}}}.

\bibitem[{H.~E.~S.~S. Collaboration} \em{et~al.}(2015){H.~E.~S.~S.
  Collaboration}, {Abramowski}, {Aharonian}, {Ait Benkhali}, {Akhperjanian},
  {Ang{\"u}ner}, {Backes}, {Balenderan}, {Balzer}, {Barnacka}, {Becherini},
  {Becker-Tjus}, {Berge}, {Bernhard}, {Bernl{\"o}hr}, {Birsin}, {Biteau},
  {B{\"o}ttcher}, {Boisson}, {Bolmont}, {Bordas}, {Bregeon}, {Brun}, {Brun},
  {Bryan}, {Bulik}, {Carrigan}, {Casanova}, {Chadwick}, {Chakraborty},
  {Chalme-Calvet}, {Chaves}, {Chr{\'e}tien}, {Colafrancesco}, {Cologna},
  {Conrad}, {Couturier}, {Cui}, {Dalton}, {Davids}, {Degrange}, {Deil}, {de
  Wilt}, {Djannati-Ata{\"\i}}, {Domainko}, {Donath}, {Drury}, {Dubus},
  {Dutson}, {Dyks}, {Dyrda}, {Edwards}, {Egberts}, {Eger}, {Espigat},
  {Farnier}, {Fegan}, {Feinstein}, {Fernandes}, {Fernandez}, {Fiasson},
  {Fontaine}, {F{\"o}rster}, {F{\"u}{\ss}ling}, {Gabici}, {Gajdus}, {Gallant},
  {Garrigoux}, {Giavitto}, {Giebels}, {Glicenstein}, {Gottschall}, {Grondin},
  {Grudzi{\'n}ska}, {Hadasch}, {H{\"a}ffner}, {Hahn}, {Harris}, {Heinzelmann},
  {Henri}, {Hermann}, {Hervet}, {Hillert}, {Hinton}, {Hofmann}, {Hofverberg},
  {Holler}, {Horns}, {Ivascenko}, {Jacholkowska}, {Jahn}, {Jamrozy}, {Janiak},
  {Jankowsky}, {Jung}, {Kastendieck}, {Katarzy{\'n}ski}, {Katz}, {Kaufmann},
  {Kh{\'e}lifi}, {Kieffer}, {Klepser}, {Klochkov}, {Klu{\'z}niak}, {Kolitzus},
  {Komin}, {Kosack}, {Krakau}, {Krayzel}, {Kr{\"u}ger}, {Laffon}, {Lamanna},
  {Lefaucheur}, {Lefranc}, {Lemi{\`e}re}, {Lemoine-Goumard}, {Lenain}, {Lohse},
  {Lopatin}, {Lu}, {Marandon}, {Marcowith}, {Marx}, {Maurin}, {Maxted},
  {Mayer}, {McComb}, {M{\'e}hault}, {Meintjes}, {Menzler}, {Meyer}, {Mitchell},
  {Moderski}, {Mohamed}, {Mor{\r{a}}}, {Moulin}, {Murach}, {de Naurois},
  {Niemiec}, {Nolan}, {Oakes}, {Odaka}, {Ohm}, {Opitz}, {Ostrowski}, {Oya},
  {Panter}, {Parsons}, {Paz Arribas}, {Pekeur}, {Pelletier}, {Perez},
  {Petrucci}, {Peyaud}, {Pita}, {Poon}, {P{\"u}hlhofer}, {Punch},
  {Quirrenbach}, {Raab}, {Reichardt}, {Reimer}, {Reimer}, {Renaud}, {de los
  Reyes}, {Rieger}, {Rob}, {Romoli}, {Rosier-Lees}, {Rowell}, {Rudak},
  {Rulten}, {Sahakian}, {Salek}, {Sanchez}, {Santangelo}, {Schlickeiser},
  {Sch{\"u}ssler}, {Schulz}, {Schwanke}, {Schwarzburg}, {Schwemmer}, {Sol},
  {Spanier}, {Spengler}, {Spies}, {Stawarz}, {Steenkamp}, {Stegmann},
  {Stinzing}, {Stycz}, {Sushch}, {Tavernet}, {Tavernier}, {Taylor}, {Terrier},
  {Tluczykont}, {Trichard}, {Valerius}, {van Eldik}, {van Soelen},
  {Vasileiadis}, {Veh}, {Venter}, {Viana}, {Vincent}, {Vink}, {V{\"o}lk},
  {Volpe}, {Vorster}, {Vuillaume}, {Wagner}, {Wagner}, {Wagner}, {Ward},
  {Weidinger}, {Weitzel}, {White}, {Wierzcholska}, {Willmann}, {W{\"o}rnlein},
  {Wouters}, {Yang}, {Zabalza}, {Zaborov}, {Zacharias}, {Zdziarski}, {Zech},
  and {Zechlin}]{HESS2015Sci}
{H.~E.~S.~S. Collaboration}.; {Abramowski}, A.; {Aharonian}, F.; {Ait
  Benkhali}, F.; {Akhperjanian}, A.G.; {Ang{\"u}ner}, E.O.; {Backes}, M.;
  {Balenderan}, S.; {Balzer}, A.; {Barnacka}, A.;  et~al.
\newblock {The exceptionally powerful TeV {\ensuremath{\gamma}}-ray emitters in
  the Large Magellanic Cloud}.
\newblock {\em Science} {\bf 2015}, {\em 347},~406--412,
  \href{http://xxx.lanl.gov/abs/1501.06578}{{\normalfont
  [arXiv:astro-ph.HE/1501.06578]}}.
\newblock
  doi:{\changeurlcolor{black}\href{https://doi.org/10.1126/science.1261313}{\detokenize{10.1126/science.1261313}}}.

\bibitem[{Yang} \em{et~al.}(2018){Yang}, {de O{\~n}a Wilhelmi}, and
  {Aharonian}]{Yang2018A&A}
{Yang}, R.z.; {de O{\~n}a Wilhelmi}, E.; {Aharonian}, F.
\newblock {Diffuse {\ensuremath{\gamma}}-ray emission in the vicinity of young
  star cluster Westerlund 2}.
\newblock {\em \aap} {\bf 2018}, {\em 611},~A77,
  \href{http://xxx.lanl.gov/abs/1710.02803}{{\normalfont
  [arXiv:astro-ph.HE/1710.02803]}}.
\newblock
  doi:{\changeurlcolor{black}\href{https://doi.org/10.1051/0004-6361/201732045}{\detokenize{10.1051/0004-6361/201732045}}}.

\bibitem[{Vieu} and {Reville}(2023)]{Vieu2023MNRAS}
{Vieu}, T.; {Reville}, B.
\newblock {Massive star cluster origin for the galactic cosmic ray population
  at very-high energies}.
\newblock {\em \mnras} {\bf 2023}, {\em 519},~136--147,
  \href{http://xxx.lanl.gov/abs/2211.11625}{{\normalfont
  [arXiv:astro-ph.HE/2211.11625]}}.
\newblock
  doi:{\changeurlcolor{black}\href{https://doi.org/10.1093/mnras/stac3469}{\detokenize{10.1093/mnras/stac3469}}}.

\bibitem[{Cristofari} \em{et~al.}(2020){Cristofari}, {Blasi}, and
  {Amato}]{Cristofari2020APh}
{Cristofari}, P.; {Blasi}, P.; {Amato}, E.
\newblock {The low rate of Galactic pevatrons}.
\newblock {\em Astroparticle Physics} {\bf 2020}, {\em 123},~102492,
  \href{http://xxx.lanl.gov/abs/2007.04294}{{\normalfont
  [arXiv:astro-ph.HE/2007.04294]}}.
\newblock
  doi:{\changeurlcolor{black}\href{https://doi.org/10.1016/j.astropartphys.2020.102492}{\detokenize{10.1016/j.astropartphys.2020.102492}}}.

\bibitem[{Zirakashvili} and {Ptuskin}(2008)]{Zirakashvili2008ApJ}
{Zirakashvili}, V.N.; {Ptuskin}, V.S.
\newblock {Diffusive Shock Acceleration with Magnetic Amplification by
  Nonresonant Streaming Instability in Supernova Remnants}.
\newblock {\em \apj} {\bf 2008}, {\em 678},~939--949,
  \href{http://xxx.lanl.gov/abs/0801.4488}{{\normalfont
  [arXiv:astro-ph/0801.4488]}}.
\newblock
  doi:{\changeurlcolor{black}\href{https://doi.org/10.1086/529580}{\detokenize{10.1086/529580}}}.

\bibitem[{Lacki} and {Beck}(2013)]{Lacki2013MNRAS}
{Lacki}, B.C.; {Beck}, R.
\newblock {The equipartition magnetic field formula in starburst galaxies:
  accounting for pionic secondaries and strong energy losses}.
\newblock {\em \mnras} {\bf 2013}, {\em 430},~3171--3186,
  \href{http://xxx.lanl.gov/abs/1301.5391}{{\normalfont
  [arXiv:astro-ph.CO/1301.5391]}}.
\newblock
  doi:{\changeurlcolor{black}\href{https://doi.org/10.1093/mnras/stt122}{\detokenize{10.1093/mnras/stt122}}}.

\bibitem[{Halzen} \em{et~al.}(2008){Halzen}, {Kappes}, and {{\'O}
  Murchadha}]{Halzen2008PhRvD}
{Halzen}, F.; {Kappes}, A.; {{\'O} Murchadha}, A.
\newblock {Prospects for identifying the sources of the Galactic cosmic rays
  with IceCube}.
\newblock {\em \prd} {\bf 2008}, {\em 78},~063004,
  \href{http://xxx.lanl.gov/abs/0803.0314}{{\normalfont
  [arXiv:astro-ph/0803.0314]}}.
\newblock
  doi:{\changeurlcolor{black}\href{https://doi.org/10.1103/PhysRevD.78.063004}{\detokenize{10.1103/PhysRevD.78.063004}}}.

\bibitem[{Baikal-GVD Collaboration} \em{et~al.}(2018){Baikal-GVD
  Collaboration}, {:}, {Avrorin}, {Avrorin}, {Aynutdinov}, {Bannash},
  {Belolaptikov}, {Brudanin}, {Budnev}, {Doroshenko}, {Domogatsky},
  {Dvornick{\'y}}, {Dyachok}, {Dzhilkibaev}, {Fajt}, {Fialkovsky}, {Gafarov},
  {Golubkov}, {Gres}, {Honz}, {Kebkal}, {Kebkal}, {Khramov}, {Kolbin},
  {Konischev}, {Korobchenko}, {Koshechkin}, {Kozhin}, {Kulepov}, {Kuleshov},
  {Milenin}, {Mirgazov}, {Osipova}, {Panfilov}, {Pan'kov}, {Petukhov},
  {Pliskovsky}, {Rozanov}, {Rjabov}, {Rushay}, {Safronov}, {Simkovic},
  {Shoibonov}, {Solovjev}, {Sorokovikov}, {Shelepov}, {Suvorova}, {Shtekl},
  {Tabolenko}, {Tarashansky}, {Yakovlev}, {Zagorodnikov}, and
  {Zurbanov}]{Baikal2018arXiv180810353B}
{Baikal-GVD Collaboration}.; {:}.; {Avrorin}, A.D.; {Avrorin}, A.V.;
  {Aynutdinov}, V.M.; {Bannash}, R.; {Belolaptikov}, I.A.; {Brudanin}, V.B.;
  {Budnev}, N.M.; {Doroshenko}, A.A.;  et~al.
\newblock {Baikal-GVD: status and prospects}.
\newblock {\em arXiv e-prints} {\bf 2018}, p. arXiv:1808.10353,
  \href{http://xxx.lanl.gov/abs/1808.10353}{{\normalfont
  [arXiv:astro-ph.IM/1808.10353]}}.
\newblock
  doi:{\changeurlcolor{black}\href{https://doi.org/10.48550/arXiv.1808.10353}{\detokenize{10.48550/arXiv.1808.10353}}}.

\bibitem[{Aartsen} \em{et~al.}(2021){Aartsen}, {Abbasi}, {Ackermann}, {Adams},
  {Aguilar}, {Ahlers}, {Ahrens}, {Alispach}, {Allison}, {Amin}, {Andeen},
  {Anderson}, {Ansseau}, {Anton}, {Arg{\"u}elles}, {Arlen}, {Auffenberg},
  {Axani}, {Bagherpour}, {Bai}, {Balagopal V}, {Barbano}, {Bartos}, {Bastian},
  {Basu}, {Baum}, {Baur}, {Bay}, {Beatty}, {Becker}, {Tjus}, {BenZvi},
  {Berley}, {Bernardini}, {Besson}, {Binder}, {Bindig}, {Blaufuss}, {Blot},
  {Bohm}, {Bohmer}, {B{\"o}ser}, {Botner}, {B{\"o}ttcher}, {Bourbeau},
  {Bourbeau}, {Bradascio}, {Braun}, {Bron}, {Brostean-Kaiser}, {Burgman},
  {Burley}, {Buscher}, {Busse}, {Bustamante}, {Campana}, {Carnie-Bronca},
  {Carver}, {Chen}, {Chen}, {Cheung}, {Chirkin}, {Choi}, {Clark}, {Clark},
  {Classen}, {Coleman}, {Collin}, {Connolly}, {Conrad}, {Coppin}, {Correa},
  {Cowen}, {Cross}, {Dave}, {Deaconu}, {De Clercq}, {DeLaunay}, {De Kockere},
  {Dembinski}, {Deoskar}, {De Ridder}, {Desai}, {Desiati}, {de Vries}, {de
  Wasseige}, {de With}, {DeYoung}, {Dharani}, {Diaz}, {D{\'\i}az-V{\'e}lez},
  {Dujmovic}, {Dunkman}, {DuVernois}, {Dvorak}, {Ehrhardt}, {Eller}, {Engel},
  {Evans}, {Evenson}, {Fahey}, {Farrag}, {Fazely}, {Felde}, {Fienberg},
  {Filimonov}, {Finley}, {Fischer}, {Fox}, {Franckowiak}, {Friedman}, {Fritz},
  {Gaisser}, {Gallagher}, {Ganster}, {Garcia-Fernandez}, {Garrappa}, {Gartner},
  {Gerhard}, {Gernhaeuser}, {Ghadimi}, {Glaser}, {Glauch}, {Gl{\"u}senkamp},
  {Goldschmidt}, {Gonzalez}, {Goswami}, {Grant}, {Gr{\'e}goire}, {Griffith},
  {Griswold}, {G{\"u}nd{\"u}z}, {Haack}, {Hallgren}, {Halliday}, {Halve},
  {Halzen}, {Hanson}, {Hanson}, {Hardin}, {Haugen}, {Haungs}, {Hauser},
  {Hebecker}, {Heinen}, {Heix}, {Helbing}, {Hellauer}, {Henningsen},
  {Hickford}, {Hignight}, {Hill}, {Hill}, {Hoffman}, {Hoffmann}, {Hoffmann},
  {Hoinka}, {Hokanson-Fasig}, {Holzapfel}, {Hoshina}, {Huang}, {Huber},
  {Huber}, {Huege}, {Hughes}, {Hultqvist}, {H{\"u}nnefeld}, {Hussain}, {In},
  {Iovine}, {Ishihara}, {Jansson}, {Japaridze}, {Jeong}, {Jones}, {Jonske},
  {Joppe}, {Kalekin}, {Kang}, {Kang}, {Kang}, {Kappes}, {Kappesser}, {Karg},
  {Karl}, {Karle}, {Katori}, {Katz}, {Kauer}, {Keivani}, {Kellermann},
  {Kelley}, {Kheirandish}, {Kim}, {Kin}, {Kintscher}, {Kiryluk}, {Kittler},
  {Kleifges}, {Klein}, {Koirala}, {Kolanoski}, {K{\"o}pke}, {Kopper}, {Kopper},
  {Koskinen}, {Koundal}, {Kovacevich}, {Kowalski}, {Krauss}, {Krings},
  {Kr{\"u}ckl}, {Kulacz}, {Kurahashi}, {Gualda}, {Lahmann}, {Lanfranchi},
  {Larson}, {Latif}, {Lauber}, {Lazar}, {Leonard}, {Leszczy{\'n}ska}, {Li},
  {Liu}, {Lohfink}, {LoSecco}, {Mariscal}, {Lu}, {Lucarelli}, {Ludwig},
  {L{\"u}nemann}, {Luszczak}, {Lyu}, {Ma}, {Madsen}, {Maggi}, {Mahn}, {Makino},
  {Mallik}, {Mancina}, {Mandalia}, {Mari{\c{s}}}, {Marka}, {Marka}, {Maruyama},
  {Mase}, {Maunu}, {McNally}, {Meagher}, {Medina}, {Meier}, {Meighen-Berger},
  {Merz}, {Meyers}, {Micallef}, {Mockler}, {Moment{\'e}}, {Montaruli}, {Moore},
  {Morse}, {Moulai}, {Muth}, {Naab}, {Nagai}, {Nam}, {Nauman}, {Necker},
  {Neer}, {Nelles}, {Nguyễn}, {Niederhausen}, {Nisa}, {Nowicki}, {Nygren},
  {Oberla}, {Pollmann}, {Oehler}, {Olivas}, {O'Sullivan}, {Pan}, {Pandya},
  {Pankova}, {Papp}, {Park}, {Parker}, {Paudel}, {Peiffer}, {P{\'e}rez de los
  Heros}, {Petersen}, {Philippen}, {Pieloth}, {Pieper}, {Pinfold}, {Pizzuto},
  {Plaisier}, {Plum}, {Popovych}, {Porcelli}, {Rodriguez}, {Price},
  {Przybylski}, {Raab}, {Raissi}, {Rameez}, {Rauch}, {Rawlins}, {Rea},
  {Rehman}, {Reimann}, {Renschler}, {Renzi}, {Resconi}, {Reusch}, {Rhode},
  {Richman}, {Riedel}, {Riegel}, {Roberts}, {Robertson}, {Roellinghoff},
  {Rongen}, {Rott}, {Ruhe}, {Ryckbosch}, {Cantu}, {Safa}, {Herrera},
  {Sandrock}, {Sandroos}, {Sandstrom}, {Santander}, {Sarkar}, {Sarkar},
  {Satalecka}, {Scharf}, {Schaufel}, {Schieler}, {Schlunder}, {Schmidt},
  {Schneider}, {Schneider}, {Schr{\"o}der}, {Schumacher}, {Sclafani}, {Seckel},
  {Seunarine}, {Shaevitz}, {Sharma}, {Shefali}, {Silva}, {Smith}, {Smithers},
  {Snihur}, {Soedingrekso}, {Soldin}, {S{\"o}ldner-Rembold}, {Song},
  {Southall}, {Spiczak}, {Spiering}, {Stachurska}, {Stamatikos}, {Stanev},
  {Stein}, {Stettner}, {Steuer}, {Stezelberger}, {Stokstad}, {Strotjohann},
  {St{\"u}rwald}, {Stuttard}, {Sullivan}, {Taboada}, {Taketa}, {Tanaka},
  {Tenholt}, {Ter-Antonyan}, {Terliuk}, {Tilav}, {Tollefson}, {Tomankova},
  {T{\"o}nnis}, {Torres}, {Toscano}, {Tosi}, {Trettin}, {Tselengidou}, {Tung},
  {Turcati}, {Turcotte}, {Turley}, {Twagirayezu}, {Ty}, {Unger}, {Elorrieta},
  {Vandenbroucke}, {van Eijk}, {van Eijndhoven}, {Vannerom}, {van Santen},
  {Veberic}, {Verpoest}, {Vieregg}, {Vraeghe}, {Walck}, {Watson}, {Weaver},
  {Weindl}, {Weinstock}, {Weiss}, {Weldert}, {Welling}, {Wendt}, {Werthebach},
  {Whitehorn}, {Wiebe}, {Wiebusch}, {Williams}, {Wissel}, {Wolf}, {Wood},
  {Woschnagg}, {Wrede}, {Wren}, {Wulff}, {Xu}, {Xu}, {Yanez}, {Yoshida},
  {Yuan}, {Zhang}, {Zierke}, and {Z{\"o}cklein}]{Aartsen2021JPhG}
{Aartsen}, M.G.; {Abbasi}, R.; {Ackermann}, M.; {Adams}, J.; {Aguilar}, J.A.;
  {Ahlers}, M.; {Ahrens}, M.; {Alispach}, C.; {Allison}, P.; {Amin}, N.M.;
  et~al.
\newblock {IceCube-Gen2: the window to the extreme Universe}.
\newblock {\em Journal of Physics G Nuclear Physics} {\bf 2021}, {\em
  48},~060501,  \href{http://xxx.lanl.gov/abs/2008.04323}{{\normalfont
  [arXiv:astro-ph.HE/2008.04323]}}.
\newblock
  doi:{\changeurlcolor{black}\href{https://doi.org/10.1088/1361-6471/abbd48}{\detokenize{10.1088/1361-6471/abbd48}}}.

\bibitem[{Agostini} \em{et~al.}(2020){Agostini}, {B{\"o}hmer}, {Bosma},
  {Clark}, {Danninger}, {Fruck}, {Gernh{\"a}user}, {G{\"a}rtner}, {Grant},
  {Henningsen}, {Holzapfel}, {Huber}, {Jenkyns}, {Krauss}, {Krings}, {Kopper},
  {Leism{\"u}ller}, {Leys}, {Macoun}, {Meighen-Berger}, {Michel}, {Moore},
  {Morley}, {Padovani}, {Papp}, {Pirenne}, {Qiu}, {Rea}, {Resconi}, {Round},
  {Ruskey}, {Spannfellner}, {Traxler}, {Turcati}, and
  {Yanez}]{Agostini2020NatAs}
{Agostini}, M.; {B{\"o}hmer}, M.; {Bosma}, J.; {Clark}, K.; {Danninger}, M.;
  {Fruck}, C.; {Gernh{\"a}user}, R.; {G{\"a}rtner}, A.; {Grant}, D.;
  {Henningsen}, F.;  et~al.
\newblock {The Pacific Ocean Neutrino Experiment}.
\newblock {\em Nature Astronomy} {\bf 2020}, {\em 4},~913--915,
  \href{http://xxx.lanl.gov/abs/2005.09493}{{\normalfont
  [arXiv:astro-ph.HE/2005.09493]}}.
\newblock
  doi:{\changeurlcolor{black}\href{https://doi.org/10.1038/s41550-020-1182-4}{\detokenize{10.1038/s41550-020-1182-4}}}.

\bibitem[{Ye} \em{et~al.}(2022){Ye}, {Hu}, {Tian}, {Chang}, {Chang}, {Cheng},
  {Gao}, {Ge}, {Gong}, {Guo}, {Guo}, {He}, {Huang}, {Jiang}, {Jiang}, {Jing},
  {Li}, {Li}, {Li}, {Li}, {Li}, {Liao}, {Lin}, {Liu}, {Liu}, {Liu}, {Miao},
  {Mo}, {Morton-Blake}, {Peng}, {Sun}, {Tang}, {Tang}, {Tao}, {Tian}, {Wang},
  {Wang}, {Wang}, {Wei}, {Wei}, {Wu}, {Xian}, {Xiang}, {Xu}, {Xue}, {Yang},
  {Yang}, {Yu}, {Zeng}, {Zhang}, {Zhang}, {Zhang}, {Zhang}, {Zhi}, {Zhong},
  {Zhou}, {Zhu}, and {Zhuang}]{Ye2022arXiv220704519Y}
{Ye}, Z.P.; {Hu}, F.; {Tian}, W.; {Chang}, Q.C.; {Chang}, Y.L.; {Cheng}, Z.S.;
  {Gao}, J.; {Ge}, T.; {Gong}, G.H.; {Guo}, J.;  et~al.
\newblock {Proposal for a neutrino telescope in South China Sea}.
\newblock {\em arXiv e-prints} {\bf 2022}, p. arXiv:2207.04519,
  \href{http://xxx.lanl.gov/abs/2207.04519}{{\normalfont
  [arXiv:astro-ph.HE/2207.04519]}}.
\newblock
  doi:{\changeurlcolor{black}\href{https://doi.org/10.48550/arXiv.2207.04519}{\detokenize{10.48550/arXiv.2207.04519}}}.

\bibitem[{Amenomori} \em{et~al.}(2021){Amenomori}, {Bao}, {Bi}, {Chen}, {Chen},
  {Chen}, {Chen}, {Chen}, {Cirennima}, {Danzengluobu}, {Fang}, {Fang}, {Feng},
  {Feng}, {Feng}, {Gao}, {Gou}, {Guo}, {Guo}, {He}, {He}, {Hibino}, {Hotta},
  {Hu}, {Hu}, {Huang}, {Jia}, {Jiang}, {Jin}, {Kasahara}, {Katayose}, {Kato},
  {Kato}, {Kawata}, {Kihara}, {Ko}, {Kozai}, {Labaciren}, {Li}, {Li}, {Li},
  {Lin}, {Liu}, {Liu}, {Liu}, {Liu}, {Liu}, {Lou}, {Lu}, {Meng}, {Munakata},
  {Nakada}, {Nakamura}, {Nanjo}, {Nishizawa}, {Ohnishi}, {Ohura}, {Ozawa},
  {Qian}, {Qu}, {Saito}, {Sakata}, {Sako}, {Shao}, {Shibata}, {Shiomi},
  {Sugimoto}, {Takano}, {Takita}, {Tan}, {Tateyama}, {Torii}, {Tsuchiya},
  {Udo}, {Wang}, {Wu}, {Xue}, {Yamamoto}, {Yang}, {Yokoe}, {Yuan}, {Zhai},
  {Zhang}, {Zhang}, {Zhang}, {Zhang}, {Zhang}, {Zhang}, {Zhang}, {Zhao},
  {Zhaxisangzhu}, and {Tibet AS<SUB>{\ensuremath{\gamma}}</SUB>
  Collaboration}]{Amenomori2021PhRvL}
{Amenomori}, M.; {Bao}, Y.W.; {Bi}, X.J.; {Chen}, D.; {Chen}, T.L.; {Chen},
  W.Y.; {Chen}, X.; {Chen}, Y.; {Cirennima}, Cui, S.W.; {Danzengluobu}, Ding,
  L.K.;  et~al.
\newblock {First Detection of sub-PeV Diffuse Gamma Rays from the Galactic
  Disk: Evidence for Ubiquitous Galactic Cosmic Rays beyond PeV Energies}.
\newblock {\em \prl} {\bf 2021}, {\em 126},~141101,
  \href{http://xxx.lanl.gov/abs/2104.05181}{{\normalfont
  [arXiv:astro-ph.HE/2104.05181]}}.
\newblock
  doi:{\changeurlcolor{black}\href{https://doi.org/10.1103/PhysRevLett.126.141101}{\detokenize{10.1103/PhysRevLett.126.141101}}}.

\bibitem[{Cao} \em{et~al.}(2023){Cao}, {Aharonian}, {An}, {Axikegu}, {Bai},
  {Bao}, {Bastieri}, {Bi}, {Bi}, {Cai}, {Cao}, {Cao}, {Cao}, {Chang}, {Chang},
  {Chen}, {Chen}, {Chen}, {Chen}, {Chen}, {Chen}, {Chen}, {Chen}, {Chen},
  {Chen}, {Chen}, {Chen}, {Cheng}, {Cheng}, {Cui}, {Cui}, {Cui}, {Cui}, {Dai},
  {Dai}, {Dai}, {Danzengluobu}, {della Volpe}, {Dong}, {Duan}, {Fan}, {Fan},
  {Fang}, {Fang}, {Feng}, {Feng}, {Feng}, {Feng}, {Feng}, {Gabici}, {Gao},
  {Gao}, {Gao}, {Gao}, {Gao}, {Gao}, {Ge}, {Geng}, {Giacinti}, {Gong}, {Gou},
  {Gu}, {Guo}, {Guo}, {Guo}, {Guo}, {Han}, {He}, {He}, {He}, {He}, {He},
  {Heller}, {Hor}, {Hou}, {Hou}, {Hou}, {Hu}, {Hu}, {Hu}, {Huang}, {Huang},
  {Huang}, {Huang}, {Huang}, {Huang}, {Huang}, {Ji}, {Jia}, {Jia}, {Jiang},
  {Jiang}, {Jiang}, {Jin}, {Kang}, {Ke}, {Kuleshov}, {Kurinov}, {Li}, {Li},
  {Li}, {Li}, {Li}, {Li}, {Li}, {Li}, {Li}, {Li}, {Li}, {Li}, {Li}, {Li}, {Li},
  {Li}, {Li}, {Li}, {Li}, {Liang}, {Liang}, {Lin}, {Liu}, {Liu}, {Liu}, {Liu},
  {Liu}, {Liu}, {Liu}, {Liu}, {Liu}, {Liu}, {Liu}, {Liu}, {Liu}, {Liu}, {Lu},
  {Luo}, {Lv}, {Ma}, {Ma}, {Ma}, {Mao}, {Min}, {Mitthumsiri}, {Mu}, {Nan},
  {Neronov}, {Ou}, {Pang}, {Pattarakijwanich}, {Pei}, {Qi}, {Qi}, {Qiao},
  {Qin}, {Ruffolo}, {S{\'a}iz}, {Semikoz}, {Shao}, {Shao}, {Shchegolev},
  {Sheng}, {Shu}, {Song}, {Stenkin}, {Stepanov}, {Su}, {Sun}, {Sun}, {Sun},
  {Tam}, {Tang}, {Tang}, {Tian}, {Wang}, {Wang}, {Wang}, {Wang}, {Wang},
  {Wang}, {Wang}, {Wang}, {Wang}, {Wang}, {Wang}, {Wang}, {Wang}, {Wang},
  {Wang}, {Wang}, {Wang}, {Wang}, {Wang}, {Wang}, {Wang}, {Wei}, {Wei}, {Wei},
  {Wen}, {Wu}, {Wu}, {Wu}, {Wu}, {Wu}, {Xi}, {Xia}, {Xia}, {Xiang}, {Xiao},
  {Xiao}, {Xin}, {Xin}, {Xing}, {Xiong}, {Xu}, {Xu}, {Xu}, {Xu}, {Xue}, {Yan},
  {Yan}, {Yan}, {Yang}, {Yang}, {Yang}, {Yang}, {Yang}, {Yang}, {Yang}, {Yang},
  {Yang}, {Yao}, {Yao}, {Ye}, {Yin}, {Yin}, {You}, {You}, {Yu}, {Yuan}, {Yue},
  {Zeng}, {Zeng}, {Zeng}, {Zha}, {Zhang}, {Zhang}, {Zhang}, {Zhang}, {Zhang},
  {Zhang}, {Zhang}, {Zhang}, {Zhang}, {Zhang}, {Zhang}, {Zhang}, {Zhang},
  {Zhang}, {Zhang}, {Zhang}, {Zhang}, {Zhang}, {Zhao}, {Zhao}, {Zhao}, {Zhao},
  {Zhao}, {Zheng}, {Zhou}, {Zhou}, {Zhou}, {Zhou}, {Zhou}, {Zhou}, {Zhou},
  {Zhu}, {Zhu}, {Zhu}, {Zhu}, and {Zuo.}]{Cao2023arXiv230517030C}
{Cao}, Z.; {Aharonian}, F.; {An}, Q.; {Axikegu}.; {Bai}, Y.X.; {Bao}, Y.W.;
  {Bastieri}, D.; {Bi}, X.J.; {Bi}, Y.J.; {Cai}, J.T.;  et~al.
\newblock {The First LHAASO Catalog of Gamma-Ray Sources}.
\newblock {\em arXiv e-prints} {\bf 2023}, p. arXiv:2305.17030,
  \href{http://xxx.lanl.gov/abs/2305.17030}{{\normalfont
  [arXiv:astro-ph.HE/2305.17030]}}.
\newblock
  doi:{\changeurlcolor{black}\href{https://doi.org/10.48550/arXiv.2305.17030}{\detokenize{10.48550/arXiv.2305.17030}}}.

\bibitem[{Abeysekara} \em{et~al.}(2020){Abeysekara}, {Albert}, {Alfaro},
  {Angeles Camacho}, {Arteaga-Vel{\'a}zquez}, {Arunbabu}, {Avila Rojas}, {Ayala
  Solares}, {Baghmanyan}, {Belmont-Moreno}, {BenZvi}, {Brisbois},
  {Caballero-Mora}, {Capistr{\'a}n}, {Carrami{\~n}ana}, {Casanova}, {Cotti},
  {Cotzomi}, {Couti{\~n}o de Le{\'o}n}, {De la Fuente}, {de Le{\'o}n},
  {Dichiara}, {Dingus}, {DuVernois}, {D{\'\i}az-V{\'e}lez}, {Ellsworth},
  {Engel}, {Espinoza}, {Fleischhack}, {Fraija}, {Galv{\'a}n-G{\'a}mez},
  {Garcia}, {Garc{\'\i}a-Gonz{\'a}lez}, {Garfias}, {Gonz{\'a}lez}, {Goodman},
  {Harding}, {Hernandez}, {Hinton}, {Hona}, {Huang}, {Hueyotl-Zahuantitla},
  {H{\"u}ntemeyer}, {Iriarte}, {Jardin-Blicq}, {Joshi}, {Kaufmann}, {Kieda},
  {Lara}, {Lee}, {Le{\'o}n Vargas}, {Linnemann}, {Longinotti}, {Luis-Raya},
  {Lundeen}, {L{\'o}pez-Coto}, {Malone}, {Marinelli}, {Martinez},
  {Martinez-Castellanos}, {Mart{\'\i}nez-Castro}, {Mart{\'\i}nez-Huerta},
  {Matthews}, {Miranda-Romagnoli}, {Morales-Soto}, {Moreno}, {Mostaf{\'a}},
  {Nayerhoda}, {Nellen}, {Newbold}, {Nisa}, {Noriega-Papaqui}, {Peisker},
  {P{\'e}rez-P{\'e}rez}, {Pretz}, {Ren}, {Rho}, {Rivi{\`e}re},
  {Rosa-Gonz{\'a}lez}, {Rosenberg}, {Ruiz-Velasco}, {Salesa Greus}, {Sandoval},
  {Schneider}, {Schoorlemmer}, {Sinnis}, {Smith}, {Springer}, {Surajbali},
  {Tabachnick}, {Tanner}, {Tibolla}, {Tollefson}, {Torres}, {Torres-Escobedo},
  {Villase{\~n}or}, {Weisgarber}, {Wood}, {Yapici}, {Zhang}, {Zhou}, and {HAWC
  Collaboration}]{Abeysekara2020PhRvL}
{Abeysekara}, A.U.; {Albert}, A.; {Alfaro}, R.; {Angeles Camacho}, J.R.;
  {Arteaga-Vel{\'a}zquez}, J.C.; {Arunbabu}, K.P.; {Avila Rojas}, D.; {Ayala
  Solares}, H.A.; {Baghmanyan}, V.; {Belmont-Moreno}, E.;  et~al.
\newblock {Multiple Galactic Sources with Emission Above 56 TeV Detected by
  HAWC}.
\newblock {\em \prl} {\bf 2020}, {\em 124},~021102,
  \href{http://xxx.lanl.gov/abs/1909.08609}{{\normalfont
  [arXiv:astro-ph.HE/1909.08609]}}.
\newblock
  doi:{\changeurlcolor{black}\href{https://doi.org/10.1103/PhysRevLett.124.021102}{\detokenize{10.1103/PhysRevLett.124.021102}}}.

\bibitem[{Blumenthal} and {Gould}(1970)]{Blumenthal1970}
{Blumenthal}, G.R.; {Gould}, R.J.
\newblock {Bremsstrahlung, Synchrotron Radiation, and Compton Scattering of
  High-Energy Electrons Traversing Dilute Gases}.
\newblock {\em Reviews of Modern Physics} {\bf 1970}, {\em 42},~237--271.
\newblock
  doi:{\changeurlcolor{black}\href{https://doi.org/10.1103/RevModPhys.42.237}{\detokenize{10.1103/RevModPhys.42.237}}}.

\bibitem[{Breuhaus} \em{et~al.}(2021){Breuhaus}, {Hahn}, {Romoli}, {Reville},
  {Giacinti}, {Tuffs}, and {Hinton}]{Breuhaus2021ApJ}
{Breuhaus}, M.; {Hahn}, J.; {Romoli}, C.; {Reville}, B.; {Giacinti}, G.;
  {Tuffs}, R.; {Hinton}, J.A.
\newblock {Ultra-high Energy Inverse Compton Emission from Galactic Electron
  Accelerators}.
\newblock {\em \apjl} {\bf 2021}, {\em 908},~L49,
  \href{http://xxx.lanl.gov/abs/2010.13960}{{\normalfont
  [arXiv:astro-ph.HE/2010.13960]}}.
\newblock
  doi:{\changeurlcolor{black}\href{https://doi.org/10.3847/2041-8213/abe41a}{\detokenize{10.3847/2041-8213/abe41a}}}.

\bibitem[{Tibet AS{\ensuremath{\gamma}} Collaboration} \em{et~al.}(2021){Tibet
  AS{\ensuremath{\gamma}} Collaboration}, {Amenomori}, {Bao}, {Bi}, {Chen},
  {Chen}, {Chen}, {Chen}, {Chen}, {Cirennima}, {Danzengluobu}, {Fang}, {Fang},
  {Feng}, {Feng}, {Feng}, {Gao}, {Gou}, {Guo}, {Guo}, {He}, {He}, {Hibino},
  {Hotta}, {Hu}, {Hu}, {Huang}, {Jia}, {Jiang}, {Jin}, {Kasahara}, {Katayose},
  {Kato}, {Kato}, {Kawata}, {Kihara}, {Ko}, {Kozai}, {Labaciren}, {Li}, {Li},
  {Li}, {Lin}, {Liu}, {Liu}, {Liu}, {Liu}, {Liu}, {Lou}, {Lu}, {Meng},
  {Munakata}, {Nakada}, {Nakamura}, {Nanjo}, {Nishizawa}, {Ohnishi}, {Ohura},
  {Ozawa}, {Qian}, {Qu}, {Saito}, {Sakata}, {Sako}, {Shao}, {Shibata},
  {Shiomi}, {Sugimoto}, {Takano}, {Takita}, {Tan}, {Tateyama}, {Torii},
  {Tsuchiya}, {Udo}, {Wang}, {Wu}, {Xue}, {Yamamoto}, {Yang}, {Yokoe}, {Yuan},
  {Zhai}, {Zhang}, {Zhang}, {Zhang}, {Zhang}, {Zhang}, {Zhang}, {Zhang},
  {Zhao}, and {Zhaxisangzhu}]{2021NatAs...5..460TibetG106}
{Tibet AS{\ensuremath{\gamma}} Collaboration}.; {Amenomori}, M.; {Bao}, Y.W.;
  {Bi}, X.J.; {Chen}, D.; {Chen}, T.L.; {Chen}, W.Y.; {Chen}, X.; {Chen}, Y.;
  {Cirennima}, Cui, S.W.;  et~al.
\newblock {Potential PeVatron supernova remnant G106.3+2.7 seen in the
  highest-energy gamma rays}.
\newblock {\em Nature Astronomy} {\bf 2021}, {\em 5},~460--464,
  \href{http://xxx.lanl.gov/abs/2109.02898}{{\normalfont
  [arXiv:astro-ph.HE/2109.02898]}}.
\newblock
  doi:{\changeurlcolor{black}\href{https://doi.org/10.1038/s41550-020-01294-9}{\detokenize{10.1038/s41550-020-01294-9}}}.

\bibitem[{MAGIC Collaboration} \em{et~al.}(2023){MAGIC Collaboration}, {Abe},
  {Abe}, {Acciari}, {Agudo}, {Aniello}, {Ansoldi}, {Antonelli}, {Arbet Engels},
  {Arcaro}, {Artero}, {Asano}, {Baack}, {Babi{\'c}}, {Baquero}, {Barres de
  Almeida}, {Barrio}, {Batkovi{\'c}}, {Baxter}, {Becerra Gonz{\'a}lez},
  {Bednarek}, {Bernardini}, {Bernardos}, {Berti}, {Besenrieder},
  {Bhattacharyya}, {Bigongiari}, {Biland}, {Blanch}, {Bonnoli},
  {Bo{\v{s}}njak}, {Burelli}, {Busetto}, {Carosi}, {Carretero-Castrillo},
  {Castro-Tirado}, {Ceribella}, {Chai}, {Chilingarian}, {Cikota}, {Colombo},
  {Contreras}, {Cortina}, {Covino}, {D'Amico}, {D'Elia}, {da Vela}, {Dazzi},
  {de Angelis}, {de Lotto}, {Del Popolo}, {Delfino}, {Delgado}, {Delgado
  Mendez}, {Depaoli}, {di Pierro}, {di Venere}, {Do Souto Espi{\~n}eira},
  {Dominis Prester}, {Donini}, {Dorner}, {Doro}, {Elsaesser}, {Emery},
  {Escudero}, {Fallah Ramazani}, {Fari{\~n}a}, {Fattorini}, {Font}, {Fruck},
  {Fukami}, {Fukazawa}, {Garc{\'\i}a L{\'o}pez}, {Garczarczyk}, {Gasparyan},
  {Gaug}, {Giesbrecht Paiva}, {Giglietto}, {Giordano}, {Gliwny},
  {Godinovi{\'c}}, {Grau}, {Green}, {Green}, {Hadasch}, {Hahn}, {Hassan},
  {Heckmann}, {Herrera}, {Hrupec}, {H{\"u}tten}, {Imazawa}, {Inada}, {Iotov},
  {Ishio}, {Jim{\'e}nez Mart{\'\i}nez}, {Jormanainen}, {Kerszberg},
  {Kobayashi}, {Kubo}, {Kushida}, {Lamastra}, {Lelas}, {Leone}, {Lindfors},
  {Linhoff}, {Lombardi}, {Longo}, {L{\'o}pez-Coto}, {L{\'o}pez-Moya},
  {L{\'o}pez-Oramas}, {Loporchio}, {Lorini}, {Lyard}, {Machado de Oliveira
  Fraga}, {Majumdar}, {Makariev}, {Maneva}, {Mang}, {Manganaro}, {Mangano},
  {Mannheim}, {Mariotti}, {Mart{\'\i}nez}, {Mas Aguilar}, {Mazin}, {Menchiari},
  {Mender}, {Mi{\'c}anovi{\'c}}, {Miceli}, {Miener}, {Miranda}, {Mirzoyan},
  {Molina}, {Mondal}, {Moralejo}, {Morcuende}, {Moreno}, {Nakamori}, {Nanci},
  {Nava}, {Neustroev}, {Nievas Rosillo}, {Nigro}, {Nilsson}, {Nishijima}, {Njoh
  Ekoume}, {Noda}, {Nozaki}, {Ohtani}, {Oka}, {Okumura}, {Otero-Santos},
  {Paiano}, {Palatiello}, {Paneque}, {Paoletti}, {Paredes}, {Pavleti{\'c}},
  {Persic}, {Pihet}, {Pirola}, {Podobnik}, {Prada Moroni}, {Prandini},
  {Principe}, {Priyadarshi}, {Rhode}, {Rib{\'o}}, {Rico}, {Righi},
  {Rugliancich}, {Sahakyan}, {Saito}, {Sakurai}, {Satalecka}, {Saturni},
  {Schleicher}, {Schmidt}, {Schmuckermaier}, {Schubert}, {Schweizer},
  {Sitarek}, {Sliusar}, {Sobczynska}, {Spolon}, {Stamerra},
  {Stri{\v{s}}kovi{\'c}}, {Strom}, {Strzys}, {Suda}, {Suri{\'c}}, {Tajima},
  {Takahashi}, {Takeishi}, {Tavecchio}, {Temnikov}, {Terauchi}, {Terzi{\'c}},
  {Teshima}, {Tosti}, {Truzzi}, {Tutone}, {Ubach}, {van Scherpenberg}, {Vazquez
  Acosta}, {Ventura}, {Verguilov}, {Viale}, {Vigorito}, {Vitale}, {Vovk},
  {Walter}, {Will}, {Wunderlich}, {Yamamoto}, and
  {Zari{\'c}}]{2023A&A...671A..12MAGIC_G106}
{MAGIC Collaboration}.; {Abe}, H.; {Abe}, S.; {Acciari}, V.A.; {Agudo}, I.;
  {Aniello}, T.; {Ansoldi}, S.; {Antonelli}, L.A.; {Arbet Engels}, A.;
  {Arcaro}, C.;  et~al.
\newblock {MAGIC observations provide compelling evidence of hadronic multi-TeV
  emission from the putative PeVatron SNR G106.3+2.7}.
\newblock {\em \aap} {\bf 2023}, {\em 671},~A12,
  \href{http://xxx.lanl.gov/abs/2211.15321}{{\normalfont
  [arXiv:astro-ph.HE/2211.15321]}}.
\newblock
  doi:{\changeurlcolor{black}\href{https://doi.org/10.1051/0004-6361/202244931}{\detokenize{10.1051/0004-6361/202244931}}}.

\bibitem[{de la Fuente} \em{et~al.}(2023{\natexlab{a}}){de la Fuente},
  {Toledano-Juarez}, {Kawata}, {Trinidad}, {Tafoya}, {Sano}, {Tokuda},
  {Nishimura}, {Onishi}, {Sako}, {Hona}, {Ohnishi}, and
  {Takita}]{2023PASJ...75..546D}
{de la Fuente}, E.; {Toledano-Juarez}, I.; {Kawata}, K.; {Trinidad}, M.A.;
  {Tafoya}, D.; {Sano}, H.; {Tokuda}, K.; {Nishimura}, A.; {Onishi}, T.;
  {Sako}, T.;  et~al.
\newblock {Detection of a new molecular cloud in the LHAASO J2108+5157 region
  supporting a hadronic PeVatron scenario}.
\newblock {\em \pasj} {\bf 2023}, {\em 75},~546--566,
  \href{http://xxx.lanl.gov/abs/2303.05712}{{\normalfont
  [arXiv:astro-ph.HE/2303.05712]}}.
\newblock
  doi:{\changeurlcolor{black}\href{https://doi.org/10.1093/pasj/psad018}{\detokenize{10.1093/pasj/psad018}}}.

\bibitem[{de la Fuente} \em{et~al.}(2023{\natexlab{b}}){de la Fuente},
  {Toledano-Ju{\'a}rez}, {Kawata}, {Trinidad}, {Yamagishi}, {Takekawa},
  {Tafoya}, {Ohnishi}, {Nishimura}, {Kato}, {Sako}, {Takita}, {Sano}, and
  {Yadav}]{2023arXiv230611921D}
{de la Fuente}, E.; {Toledano-Ju{\'a}rez}, I.; {Kawata}, K.; {Trinidad}, M.A.;
  {Yamagishi}, M.; {Takekawa}, S.; {Tafoya}, D.; {Ohnishi}, M.; {Nishimura},
  A.; {Kato}, S.;  et~al.
\newblock {Evidence for a gamma-ray molecular target in the enigmatic PeVatron
  candidate LHAASO J2108+5157}.
\newblock {\em arXiv e-prints} {\bf 2023}, p. arXiv:2306.11921,
  \href{http://xxx.lanl.gov/abs/2306.11921}{{\normalfont
  [arXiv:astro-ph.HE/2306.11921]}}.
\newblock
  doi:{\changeurlcolor{black}\href{https://doi.org/10.48550/arXiv.2306.11921}{\detokenize{10.48550/arXiv.2306.11921}}}.

\bibitem[{Abe} \em{et~al.}(2023){Abe}, {Aguasca-Cabot}, {Agudo}, {Alvarez
  Crespo}, {Antonelli}, {Aramo}, {Arbet-Engels}, {Artero}, {Asano}, {Aubert},
  {Baktash}, {Bamba}, {Baquero Larriva}, {Baroncelli}, {Barres de Almeida},
  {Barrio}, {Batkovic}, {Baxter}, {Becerra Gonz{\'a}lez}, {Bernardini},
  {Bernardos}, {Bernete Medrano}, {Berti}, {Bhattacharjee}, {Biederbeck},
  {Bigongiari}, {Bissaldi}, {Blanch}, {Bordas}, {Buisson}, {Bulgarelli},
  {Burelli}, {Buscemi}, {Cardillo}, {Caroff}, {Carosi}, {Cassol}, {Cauz},
  {Ceribella}, {Chai}, {Cheng}, {Chiavassa}, {Chikawa}, {Chytka}, {Cifuentes},
  {Contreras}, {Cortina}, {Costantini}, {D'Amico}, {Dalchenko}, {De Angelis},
  {de Bony de Lavergne}, {De Lotto}, {de Menezes}, {Deleglise}, {Delgado},
  {Delgado Mengual}, {della Volpe}, {Dellaiera}, {Di Piano}, {Di Pierro}, {Di
  Tria}, {Di Venere}, {D{\'\i}az}, {Dominik}, {Dominis Prester}, {Donini},
  {Dorner}, {Doro}, {Els{\"a}sser}, {Emery}, {Escudero}, {Fallah Ramazani},
  {Ferrara}, {Fiasson}, {Freixas Coromina}, {Fr{\"o}se}, {Fukami}, {Fukazawa},
  {Garcia}, {Garcia L{\'o}pez}, {Gasparrini}, {Geyer}, {Giesbrecht Paiva},
  {Giglietto}, {Giordano}, {Giro}, {Gliwny}, {Godinovic}, {Grau}, {Green},
  {Green}, {Gunji}, {Hackfeld}, {Hadasch}, {Hahn}, {Hashiyama}, {Hassan},
  {Hayashi}, {Heckmann}, {Heller}, {Herrera Llorente}, {Hirotani}, {Hoffmann},
  {Horns}, {Houles}, {Hrabovsky}, {Hrupec}, {Hui}, {H{\"u}tten}, {Imazawa},
  {Inada}, {Inome}, {Ioka}, {Iori}, {Ishio}, {Iwamura}, {Jacquemont}, {Jimenez
  Martinez}, {Jurysek}, {Kagaya}, {Karas}, {Katagiri}, {Kataoka}, {Kerszberg},
  {Kobayashi}, {Kong}, {Kubo}, {Kushida}, {Lainez}, {Lamanna}, {Lamastra}, {Le
  Flour}, {Linhoff}, {Longo}, {L{\'o}pez-Coto}, {L{\'o}pez-Moya},
  {L{\'o}pez-Oramas}, {Loporchio}, {Lorini}, {Luque-Escamilla}, {Majumdar},
  {Makariev}, {Mandat}, {Manganaro}, {Manic{\`o}}, {Mannheim}, {Mariotti},
  {Marquez}, {Marsella}, {Mart{\'\i}}, {Martinez}, {Mart{\'\i}nez},
  {Mart{\'\i}nez}, {Marusevec}, {Mas-Aguilar}, {Maurin}, {Mazin}, {Mestre
  Guillen}, {Micanovic}, {Miceli}, {Miener}, {Miranda}, {Mirzoyan}, {Mizuno},
  {Molero Gonzalez}, {Molina}, {Montaruli}, {Monteiro}, {Moralejo},
  {Morcuende}, {Morselli}, {Mrakovcic}, {Murase}, {Nagai}, {Nakamori},
  {Nickel}, {Nievas}, {Nishijima}, {Noda}, {Nosek}, {Nozaki}, {Ohishi},
  {Ohtani}, {Okazaki}, {Okumura}, {Orito}, {Otero-Santos}, {Palatiello},
  {Paneque}, {Pantaleo}, {Paoletti}, {Paredes}, {Pavleti{\'c}}, {Pech},
  {Pecimotika}, {Pietropaolo}, {Pirola}, {Podobnik}, {Poireau}, {Polo}, {Pons},
  {Prandini}, {Prast}, {Priyadarshi}, {Prouza}, {Rando}, {Rhode}, {Rib{\'o}},
  {Rizi}, {Rodriguez Fernandez}, {Saito}, {Sakurai}, {Sanchez},
  {{\v{S}}ari{\'c}}, {Saturni}, {Scherpenberg}, {Schleicher}, {Schmuckermaier},
  {Schubert}, {Schussler}, {Schweizer}, {Seglar Arroyo}, {Sitarek}, {Sliusar},
  {Spolon}, {Stri{\v{s}}kovi{\'c}}, {Strzys}, {Suda}, {Sunada}, {Tajima},
  {Takahashi}, {Takahashi}, {Takata}, {Takeishi}, {Tam}, {Tanaka}, {Tateishi},
  {Temnikov}, {Terada}, {Terauchi}, {Terzic}, {Teshima}, {Tluczykont},
  {Tokanai}, {Torres}, {Travnicek}, {Truzzi}, {Tutone}, {Uhlrich}, {Vacula},
  {V{\'a}zquez Acosta}, {Verguilov}, {Viale}, {Vigliano}, {Vigorito}, {Vitale},
  {Voutsinas}, {Vovk}, {Vuillaume}, {Walter}, {Will}, {Yamamoto}, {Yamazaki},
  {Yoshida}, {Yoshikoshi}, {Zywucka (CTA-LST Project)}, {Balbo}, {Eckert}, and
  {Tramacere}]{2023A&A...673A..75A_LSTJ2108}
{Abe}, S.; {Aguasca-Cabot}, A.; {Agudo}, I.; {Alvarez Crespo}, N.; {Antonelli},
  L.A.; {Aramo}, C.; {Arbet-Engels}, A.; {Artero}, M.; {Asano}, K.; {Aubert},
  P.;  et~al.
\newblock {Multiwavelength study of the galactic PeVatron candidate LHAASO
  J2108+5157}.
\newblock {\em \aap} {\bf 2023}, {\em 673},~A75,
  \href{http://xxx.lanl.gov/abs/2210.00775}{{\normalfont
  [arXiv:astro-ph.HE/2210.00775]}}.
\newblock
  doi:{\changeurlcolor{black}\href{https://doi.org/10.1051/0004-6361/202245086}{\detokenize{10.1051/0004-6361/202245086}}}.

\bibitem[{De Sarkar}(2023)]{2023MNRAS.521L...5D}
{De Sarkar}, A.
\newblock {Supernova connection of unidentified ultra-high-energy gamma-ray
  source LHAASO J2108+5157}.
\newblock {\em \mnras} {\bf 2023}, {\em 521},~L5--L10,
  \href{http://xxx.lanl.gov/abs/2301.13451}{{\normalfont
  [arXiv:astro-ph.HE/2301.13451]}}.
\newblock
  doi:{\changeurlcolor{black}\href{https://doi.org/10.1093/mnrasl/slad013}{\detokenize{10.1093/mnrasl/slad013}}}.

\bibitem[{Albert} \em{et~al.}(2019){Albert}, {Alfaro}, {Ashkar}, {Alvarez},
  {{\'A}lvarez}, {Arteaga-Vel{\'a}zquez}, {Ayala Solares}, {Arceo}, {Bellido},
  {BenZvi}, {Bretz}, {Brisbois}, {Brown}, {Brun}, {Caballero-Mora}, {Carosi},
  {Carrami{\~n}ana}, {Casanova}, {Chadwick}, {Cotter}, {Couti{\~n}o De
  Le{\'o}n}, {Cristofari}, {Dasso}, {de la Fuente}, {Dingus}, {Desiati},
  {Salles}, {de Souza}, {Dorner}, {D{\'\i}az-V{\'e}lez},
  {Garc{\'\i}a-Gonz{\'a}lez}, {DuVernois}, {Di Sciascio}, {Engel},
  {Fleischhack}, {Fraija}, {Funk}, {Glicenstein}, {Gonzalez}, {Gonz{\'a}lez},
  {Goodman}, {Harding}, {Haungs}, {Hinton}, {Hona}, {Hoyos}, {Huentemeyer},
  {Iriarte}, {Jardin-Blicq}, {Joshi}, {Kaufmann}, {Kawata}, {Kunwar},
  {Lefaucheur}, {Lenain}, {Link}, {L{\'o}pez-Coto}, {Marandon}, {Mariotti},
  {Mart{\'\i}nez-Castro}, {Mart{\'\i}nez-Huerta}, {Mostaf{\'a}}, {Nayerhoda},
  {Nellen}, {de O{\~n}a Wilhelmi}, {Parsons}, {Patricelli}, {Pichel}, {Piel},
  {Prandini}, {Pueschel}, {Procureur}, {Reisenegger}, {Rivi{\`e}re},
  {Rodriguez}, {Rovero}, {Rowell}, {Ruiz-Velasco}, {Sandoval}, {Santander},
  {Sako}, {Sako}, {Satalecka}, {Schoorlemmer}, {Sch{\"u}ssler},
  {Seglar-Arroyo}, {Smith}, {Spencer}, {Surajbali}, {Tabachnick}, {Taylor},
  {Tibolla}, {Torres}, {Vallage}, {Viana}, {Watson}, {Weisgarber}, {Werner},
  {White}, {Wischnewski}, {Yang}, {Zepeda}, and
  {Zhou}]{Albert2019arXiv190208429A}
{Albert}, A.; {Alfaro}, R.; {Ashkar}, H.; {Alvarez}, C.; {{\'A}lvarez}, J.;
  {Arteaga-Vel{\'a}zquez}, J.C.; {Ayala Solares}, H.A.; {Arceo}, R.; {Bellido},
  J.A.; {BenZvi}, S.;  et~al.
\newblock {Science Case for a Wide Field-of-View Very-High-Energy Gamma-Ray
  Observatory in the Southern Hemisphere}.
\newblock {\em arXiv e-prints} {\bf 2019}, p. arXiv:1902.08429,
  \href{http://xxx.lanl.gov/abs/1902.08429}{{\normalfont
  [arXiv:astro-ph.HE/1902.08429]}}.
\newblock
  doi:{\changeurlcolor{black}\href{https://doi.org/10.48550/arXiv.1902.08429}{\detokenize{10.48550/arXiv.1902.08429}}}.

\bibitem[{Atoyan} and {Aharonian}(1996)]{Atoyan1996MNRAS}
{Atoyan}, A.M.; {Aharonian}, F.A.
\newblock {On the mechanisms of gamma radiation in the Crab Nebula}.
\newblock {\em \mnras} {\bf 1996}, {\em 278},~525--541.
\newblock
  doi:{\changeurlcolor{black}\href{https://doi.org/10.1093/mnras/278.2.525}{\detokenize{10.1093/mnras/278.2.525}}}.

\bibitem[{Sudoh} and {Beacom}(2023)]{Sudoh2023PhRvD}
{Sudoh}, T.; {Beacom}, J.F.
\newblock {Where are Milky Way's hadronic PeVatrons?}
\newblock {\em \prd} {\bf 2023}, {\em 107},~043002,
  \href{http://xxx.lanl.gov/abs/2209.03970}{{\normalfont
  [arXiv:astro-ph.HE/2209.03970]}}.
\newblock
  doi:{\changeurlcolor{black}\href{https://doi.org/10.1103/PhysRevD.107.043002}{\detokenize{10.1103/PhysRevD.107.043002}}}.

\bibitem[{Scannapieco} \em{et~al.}(2005){Scannapieco}, {Madau}, {Woosley},
  {Heger}, and {Ferrara}]{Scannapieco2005ApJ}
{Scannapieco}, E.; {Madau}, P.; {Woosley}, S.; {Heger}, A.; {Ferrara}, A.
\newblock {The Detectability of Pair-Production Supernovae at z
  <\raisebox{-0.5ex}\textasciitilde 6}.
\newblock {\em \apj} {\bf 2005}, {\em 633},~1031--1041,
  \href{http://xxx.lanl.gov/abs/astro-ph/0507182}{{\normalfont
  [arXiv:astro-ph/astro-ph/0507182]}}.
\newblock
  doi:{\changeurlcolor{black}\href{https://doi.org/10.1086/444450}{\detokenize{10.1086/444450}}}.

\bibitem[{Cristofari} \em{et~al.}(2020){Cristofari}, {Renaud}, {Marcowith},
  {Dwarkadas}, and {Tatischeff}]{Cristofari2020MNRAS}
{Cristofari}, P.; {Renaud}, M.; {Marcowith}, A.; {Dwarkadas}, V.V.;
  {Tatischeff}, V.
\newblock {Time-dependent high-energy gamma-ray signal from accelerated
  particles in core-collapse supernovae: the case of SN 1993J}.
\newblock {\em \mnras} {\bf 2020}, {\em 494},~2760--2765,
  \href{http://xxx.lanl.gov/abs/2004.02650}{{\normalfont
  [arXiv:astro-ph.HE/2004.02650]}}.
\newblock
  doi:{\changeurlcolor{black}\href{https://doi.org/10.1093/mnras/staa984}{\detokenize{10.1093/mnras/staa984}}}.

\bibitem[{Cristofari} \em{et~al.}(2022){Cristofari}, {Marcowith}, {Renaud},
  {Dwarkadas}, {Tatischeff}, {Giacinti}, {Peretti}, and
  {Sol}]{Cristofari2022MNRAS}
{Cristofari}, P.; {Marcowith}, A.; {Renaud}, M.; {Dwarkadas}, V.V.;
  {Tatischeff}, V.; {Giacinti}, G.; {Peretti}, E.; {Sol}, H.
\newblock {The first days of Type II-P core collapse supernovae in the
  gamma-ray range}.
\newblock {\em \mnras} {\bf 2022}, {\em 511},~3321--3329,
  \href{http://xxx.lanl.gov/abs/2201.09583}{{\normalfont
  [arXiv:astro-ph.HE/2201.09583]}}.
\newblock
  doi:{\changeurlcolor{black}\href{https://doi.org/10.1093/mnras/stac217}{\detokenize{10.1093/mnras/stac217}}}.

\bibitem[{Thomas} \em{et~al.}(2023){Thomas}, {Pfrommer}, and
  {Pakmor}]{Thomas2023MNRAS}
{Thomas}, T.; {Pfrommer}, C.; {Pakmor}, R.
\newblock {Cosmic-ray-driven galactic winds: transport modes of cosmic rays and
  Alfv{\'e}n-wave dark regions}.
\newblock {\em \mnras} {\bf 2023}, {\em 521},~3023--3042.
\newblock
  doi:{\changeurlcolor{black}\href{https://doi.org/10.1093/mnras/stad472}{\detokenize{10.1093/mnras/stad472}}}.

\bibitem[{Armillotta} \em{et~al.}(2021){Armillotta}, {Ostriker}, and
  {Jiang}]{Armillotta2021ApJ}
{Armillotta}, L.; {Ostriker}, E.C.; {Jiang}, Y.F.
\newblock {Cosmic-Ray Transport in Simulations of Star-forming Galactic Disks}.
\newblock {\em \apj} {\bf 2021}, {\em 922},~11,
  \href{http://xxx.lanl.gov/abs/2108.09356}{{\normalfont
  [arXiv:astro-ph.HE/2108.09356]}}.
\newblock
  doi:{\changeurlcolor{black}\href{https://doi.org/10.3847/1538-4357/ac1db2}{\detokenize{10.3847/1538-4357/ac1db2}}}.

\bibitem[{Xu} and {Lazarian}(2022)]{Xu2022ApJ}
{Xu}, S.; {Lazarian}, A.
\newblock {Cosmic Ray Streaming in the Turbulent Interstellar Medium}.
\newblock {\em \apj} {\bf 2022}, {\em 927},~94,
  \href{http://xxx.lanl.gov/abs/2112.06941}{{\normalfont
  [arXiv:astro-ph.HE/2112.06941]}}.
\newblock
  doi:{\changeurlcolor{black}\href{https://doi.org/10.3847/1538-4357/ac4dfd}{\detokenize{10.3847/1538-4357/ac4dfd}}}.

\bibitem[{Cesarsky} and {Volk}(1978)]{Cesarsky1978AA}
{Cesarsky}, C.J.; {Volk}, H.J.
\newblock {Cosmic Ray Penetration into Molecular Clouds}.
\newblock {\em \aap} {\bf 1978}, {\em 70},~367.

\bibitem[{Kulsrud} and {Pearce}(1969)]{Kulsrud1969ApJ}
{Kulsrud}, R.; {Pearce}, W.P.
\newblock {The Effect of Wave-Particle Interactions on the Propagation of
  Cosmic Rays}.
\newblock {\em \apj} {\bf 1969}, {\em 156},~445.
\newblock
  doi:{\changeurlcolor{black}\href{https://doi.org/10.1086/149981}{\detokenize{10.1086/149981}}}.

\bibitem[{Zweibel} and {Shull}(1982)]{Zweibel1982ApJ}
{Zweibel}, E.G.; {Shull}, J.M.
\newblock {Confinement of cosmic rays in molecular clouds}.
\newblock {\em \apj} {\bf 1982}, {\em 259},~859--868.
\newblock
  doi:{\changeurlcolor{black}\href{https://doi.org/10.1086/160220}{\detokenize{10.1086/160220}}}.

\bibitem[{Yang} \em{et~al.}(2023){Yang}, {Li}, {Wilhelmi}, {Cui}, {Liu}, and
  {Aharonian}]{Yang2023NatAs}
{Yang}, R.z.; {Li}, G.X.; {Wilhelmi}, E.d.O.; {Cui}, Y.D.; {Liu}, B.;
  {Aharonian}, F.
\newblock {Effective shielding of {\ensuremath{\lesssim}}10 GeV cosmic rays
  from dense molecular clumps}.
\newblock {\em Nature Astronomy} {\bf 2023},
  \href{http://xxx.lanl.gov/abs/2301.06716}{{\normalfont
  [arXiv:astro-ph.HE/2301.06716]}}.
\newblock
  doi:{\changeurlcolor{black}\href{https://doi.org/10.1038/s41550-022-01868-9}{\detokenize{10.1038/s41550-022-01868-9}}}.

\bibitem[{Hill} \em{et~al.}(2018){Hill}, {Mac Low}, {Gatto}, and
  {Ib{\'a}{\~n}ez-Mej{\'\i}a}]{Hill2018ApJ}
{Hill}, A.S.; {Mac Low}, M.M.; {Gatto}, A.; {Ib{\'a}{\~n}ez-Mej{\'\i}a}, J.C.
\newblock {Effect of the Heating Rate on the Stability of the Three-phase
  Interstellar Medium}.
\newblock {\em \apj} {\bf 2018}, {\em 862},~55,
  \href{http://xxx.lanl.gov/abs/1806.05571}{{\normalfont
  [arXiv:astro-ph.GA/1806.05571]}}.
\newblock
  doi:{\changeurlcolor{black}\href{https://doi.org/10.3847/1538-4357/aacce2}{\detokenize{10.3847/1538-4357/aacce2}}}.

\bibitem[{Owen} \em{et~al.}(2019){Owen}, {Wu}, {Jin}, {Surajbali}, and
  {Kataoka}]{Owen2019AA}
{Owen}, E.R.; {Wu}, K.; {Jin}, X.; {Surajbali}, P.; {Kataoka}, N.
\newblock {Starburst and post-starburst high-redshift protogalaxies. The
  feedback impact of high energy cosmic rays}.
\newblock {\em \aap} {\bf 2019}, {\em 626},~A85,
  \href{http://xxx.lanl.gov/abs/1905.00338}{{\normalfont
  [arXiv:astro-ph.GA/1905.00338]}}.
\newblock
  doi:{\changeurlcolor{black}\href{https://doi.org/10.1051/0004-6361/201834350}{\detokenize{10.1051/0004-6361/201834350}}}.

\bibitem[{Walker}(2016)]{Walker2016ApJ}
{Walker}, M.A.
\newblock {Heating of the Warm Ionized Medium by Low-energy Cosmic Rays}.
\newblock {\em \apj} {\bf 2016}, {\em 818},~23,
  \href{http://xxx.lanl.gov/abs/1512.07335}{{\normalfont
  [arXiv:astro-ph.GA/1512.07335]}}.
\newblock
  doi:{\changeurlcolor{black}\href{https://doi.org/10.3847/0004-637X/818/1/23}{\detokenize{10.3847/0004-637X/818/1/23}}}.

\bibitem[{Minter} and {Spangler}(1997)]{Minter1997ApJ}
{Minter}, A.H.; {Spangler}, S.R.
\newblock {Heating of the Interstellar Diffuse Ionized Gas via the Dissipation
  of Turbulence}.
\newblock {\em \apj} {\bf 1997}, {\em 485},~182--194.
\newblock
  doi:{\changeurlcolor{black}\href{https://doi.org/10.1086/304396}{\detokenize{10.1086/304396}}}.

\bibitem[{Wiener} \em{et~al.}(2013){Wiener}, {Zweibel}, and
  {Oh}]{Wiener2013ApJ}
{Wiener}, J.; {Zweibel}, E.G.; {Oh}, S.P.
\newblock {Cosmic Ray Heating of the Warm Ionized Medium}.
\newblock {\em \apj} {\bf 2013}, {\em 767},~87,
  \href{http://xxx.lanl.gov/abs/1301.4445}{{\normalfont
  [arXiv:astro-ph.GA/1301.4445]}}.
\newblock
  doi:{\changeurlcolor{black}\href{https://doi.org/10.1088/0004-637X/767/1/87}{\detokenize{10.1088/0004-637X/767/1/87}}}.

\bibitem[{Wentzel}(1971)]{Wentzel1971ApJ}
{Wentzel}, D.G.
\newblock {Acceleration and Heating of Interstellar Gas by Cosmic Rays}.
\newblock {\em \apj} {\bf 1971}, {\em 163},~503.
\newblock
  doi:{\changeurlcolor{black}\href{https://doi.org/10.1086/150794}{\detokenize{10.1086/150794}}}.

\bibitem[{Rathjen} \em{et~al.}(2021){Rathjen}, {Naab}, {Girichidis}, {Walch},
  {W{\"u}nsch}, {Dinnbier}, {Seifried}, {Klessen}, and
  {Glover}]{Rathjen2021MNRAS}
{Rathjen}, T.E.; {Naab}, T.; {Girichidis}, P.; {Walch}, S.; {W{\"u}nsch}, R.;
  {Dinnbier}, F.; {Seifried}, D.; {Klessen}, R.S.; {Glover}, S.C.O.
\newblock {SILCC VI - Multiphase ISM structure, stellar clustering, and
  outflows with supernovae, stellar winds, ionizing radiation, and cosmic
  rays}.
\newblock {\em \mnras} {\bf 2021}, {\em 504},~1039--1061,
  \href{http://xxx.lanl.gov/abs/2103.14128}{{\normalfont
  [arXiv:astro-ph.GA/2103.14128]}}.
\newblock
  doi:{\changeurlcolor{black}\href{https://doi.org/10.1093/mnras/stab900}{\detokenize{10.1093/mnras/stab900}}}.

\bibitem[{Owen} \em{et~al.}(2021){Owen}, {On}, {Lai}, and {Wu}]{Owen2021ApJ}
{Owen}, E.R.; {On}, A.Y.L.; {Lai}, S.P.; {Wu}, K.
\newblock {Observational Signatures of Cosmic-Ray Interactions in Molecular
  Clouds}.
\newblock {\em \apj} {\bf 2021}, {\em 913},~52,
  \href{http://xxx.lanl.gov/abs/2103.06542}{{\normalfont
  [arXiv:astro-ph.GA/2103.06542]}}.
\newblock
  doi:{\changeurlcolor{black}\href{https://doi.org/10.3847/1538-4357/abee1a}{\detokenize{10.3847/1538-4357/abee1a}}}.

\bibitem[{Papadopoulos} \em{et~al.}(2011){Papadopoulos}, {Thi}, {Miniati}, and
  {Viti}]{Papadopoulos2011MNRAS}
{Papadopoulos}, P.P.; {Thi}, W.F.; {Miniati}, F.; {Viti}, S.
\newblock {Extreme cosmic ray dominated regions: a new paradigm for high star
  formation density events in the Universe}.
\newblock {\em \mnras} {\bf 2011}, {\em 414},~1705--1714,
  \href{http://xxx.lanl.gov/abs/1009.2496}{{\normalfont
  [arXiv:astro-ph.CO/1009.2496]}}.
\newblock
  doi:{\changeurlcolor{black}\href{https://doi.org/10.1111/j.1365-2966.2011.18504.x}{\detokenize{10.1111/j.1365-2966.2011.18504.x}}}.

\bibitem[{Semenov} \em{et~al.}(2021){Semenov}, {Kravtsov}, and
  {Caprioli}]{Semenov2021ApJ}
{Semenov}, V.A.; {Kravtsov}, A.V.; {Caprioli}, D.
\newblock {Cosmic-Ray Diffusion Suppression in Star-forming Regions Inhibits
  Clump Formation in Gas-rich Galaxies}.
\newblock {\em \apj} {\bf 2021}, {\em 910},~126,
  \href{http://xxx.lanl.gov/abs/2012.01427}{{\normalfont
  [arXiv:astro-ph.GA/2012.01427]}}.
\newblock
  doi:{\changeurlcolor{black}\href{https://doi.org/10.3847/1538-4357/abe2a6}{\detokenize{10.3847/1538-4357/abe2a6}}}.

\bibitem[{Farcy} \em{et~al.}(2022){Farcy}, {Rosdahl}, {Dubois}, {Blaizot}, and
  {Martin-Alvarez}]{Farcy2022MNRAS}
{Farcy}, M.; {Rosdahl}, J.; {Dubois}, Y.; {Blaizot}, J.; {Martin-Alvarez}, S.
\newblock {Radiation-magnetohydrodynamics simulations of cosmic ray feedback in
  disc galaxies}.
\newblock {\em \mnras} {\bf 2022}, {\em 513},~5000--5019,
  \href{http://xxx.lanl.gov/abs/2202.01245}{{\normalfont
  [arXiv:astro-ph.GA/2202.01245]}}.
\newblock
  doi:{\changeurlcolor{black}\href{https://doi.org/10.1093/mnras/stac1196}{\detokenize{10.1093/mnras/stac1196}}}.

\bibitem[{Bustard} and {Zweibel}(2021)]{Bustard2021ApJ}
{Bustard}, C.; {Zweibel}, E.G.
\newblock {Cosmic-Ray Transport, Energy Loss, and Influence in the Multiphase
  Interstellar Medium}.
\newblock {\em \apj} {\bf 2021}, {\em 913},~106,
  \href{http://xxx.lanl.gov/abs/2012.06585}{{\normalfont
  [arXiv:astro-ph.HE/2012.06585]}}.
\newblock
  doi:{\changeurlcolor{black}\href{https://doi.org/10.3847/1538-4357/abf64c}{\detokenize{10.3847/1538-4357/abf64c}}}.

\bibitem[{Wiener} \em{et~al.}(2017){Wiener}, {Oh}, and
  {Zweibel}]{Wiener2017MNRAS}
{Wiener}, J.; {Oh}, S.P.; {Zweibel}, E.G.
\newblock {Interaction of cosmic rays with cold clouds in galactic haloes}.
\newblock {\em \mnras} {\bf 2017}, {\em 467},~646--660,
  \href{http://xxx.lanl.gov/abs/1610.02041}{{\normalfont
  [arXiv:astro-ph.HE/1610.02041]}}.
\newblock
  doi:{\changeurlcolor{black}\href{https://doi.org/10.1093/mnras/stx109}{\detokenize{10.1093/mnras/stx109}}}.

\bibitem[{Wiener} \em{et~al.}(2019){Wiener}, {Zweibel}, and
  {Ruszkowski}]{Wiener2019MNRAS}
{Wiener}, J.; {Zweibel}, E.G.; {Ruszkowski}, M.
\newblock {Cosmic ray acceleration of cool clouds in the circumgalactic
  medium}.
\newblock {\em \mnras} {\bf 2019}, {\em 489},~205--223,
  \href{http://xxx.lanl.gov/abs/1903.01471}{{\normalfont
  [arXiv:astro-ph.GA/1903.01471]}}.
\newblock
  doi:{\changeurlcolor{black}\href{https://doi.org/10.1093/mnras/stz2007}{\detokenize{10.1093/mnras/stz2007}}}.

\bibitem[{Br{\"u}ggen} and {Scannapieco}(2020)]{Bruggen2020ApJ}
{Br{\"u}ggen}, M.; {Scannapieco}, E.
\newblock {The Launching of Cold Clouds by Galaxy Outflows. IV.
  Cosmic-Ray-driven Acceleration}.
\newblock {\em \apj} {\bf 2020}, {\em 905},~19,
  \href{http://xxx.lanl.gov/abs/2010.07308}{{\normalfont
  [arXiv:astro-ph.GA/2010.07308]}}.
\newblock
  doi:{\changeurlcolor{black}\href{https://doi.org/10.3847/1538-4357/abc00f}{\detokenize{10.3847/1538-4357/abc00f}}}.

\bibitem[{Heintz} \em{et~al.}(2020){Heintz}, {Bustard}, and
  {Zweibel}]{Heintz2020ApJ}
{Heintz}, E.; {Bustard}, C.; {Zweibel}, E.G.
\newblock {The Role of the Parker Instability in Structuring the Interstellar
  Medium}.
\newblock {\em \apj} {\bf 2020}, {\em 891},~157,
  \href{http://xxx.lanl.gov/abs/1910.03588}{{\normalfont
  [arXiv:astro-ph.GA/1910.03588]}}.
\newblock
  doi:{\changeurlcolor{black}\href{https://doi.org/10.3847/1538-4357/ab7453}{\detokenize{10.3847/1538-4357/ab7453}}}.

\bibitem[{Heintz} and {Zweibel}(2018)]{Heintz2018ApJ}
{Heintz}, E.; {Zweibel}, E.G.
\newblock {The Parker Instability with Cosmic-Ray Streaming}.
\newblock {\em \apj} {\bf 2018}, {\em 860},~97,
  \href{http://xxx.lanl.gov/abs/1803.00584}{{\normalfont
  [arXiv:astro-ph.HE/1803.00584]}}.
\newblock
  doi:{\changeurlcolor{black}\href{https://doi.org/10.3847/1538-4357/aac208}{\detokenize{10.3847/1538-4357/aac208}}}.

\bibitem[{Protheroe} \em{et~al.}(2008){Protheroe}, {Ott}, {Ekers}, {Jones}, and
  {Crocker}]{Protheroe2008MNRAS}
{Protheroe}, R.J.; {Ott}, J.; {Ekers}, R.D.; {Jones}, D.I.; {Crocker}, R.M.
\newblock {Interpretation of radio continuum and molecular line observations of
  Sgr B2: free-free and synchrotron emission, and implications for cosmic
  rays}.
\newblock {\em \mnras} {\bf 2008}, {\em 390},~683--692,
  \href{http://xxx.lanl.gov/abs/0807.0127}{{\normalfont
  [arXiv:astro-ph/0807.0127]}}.
\newblock
  doi:{\changeurlcolor{black}\href{https://doi.org/10.1111/j.1365-2966.2008.13752.x}{\detokenize{10.1111/j.1365-2966.2008.13752.x}}}.

\bibitem[{Dogiel} \em{et~al.}(2015){Dogiel}, {Chernyshov}, {Kiselev},
  {Nobukawa}, {Cheng}, {Hui}, {Ko}, {Nobukawa}, and {Tsuru}]{Dogiel2015ApJ}
{Dogiel}, V.A.; {Chernyshov}, D.O.; {Kiselev}, A.M.; {Nobukawa}, M.; {Cheng},
  K.S.; {Hui}, C.Y.; {Ko}, C.M.; {Nobukawa}, K.K.; {Tsuru}, T.G.
\newblock {Spectrum of Relativistic and Subrelativistic Cosmic Rays in the 100
  pc Central Region}.
\newblock {\em \apj} {\bf 2015}, {\em 809},~48,
  \href{http://xxx.lanl.gov/abs/1507.02440}{{\normalfont
  [arXiv:astro-ph.HE/1507.02440]}}.
\newblock
  doi:{\changeurlcolor{black}\href{https://doi.org/10.1088/0004-637X/809/1/48}{\detokenize{10.1088/0004-637X/809/1/48}}}.

\bibitem[{Ivlev} \em{et~al.}(2018){Ivlev}, {Dogiel}, {Chernyshov}, {Caselli},
  {Ko}, and {Cheng}]{Ivlev2018ApJ}
{Ivlev}, A.V.; {Dogiel}, V.A.; {Chernyshov}, D.O.; {Caselli}, P.; {Ko}, C.M.;
  {Cheng}, K.S.
\newblock {Penetration of Cosmic Rays into Dense Molecular Clouds: Role of
  Diffuse Envelopes}.
\newblock {\em \apj} {\bf 2018}, {\em 855},~23,
  \href{http://xxx.lanl.gov/abs/1802.02612}{{\normalfont
  [arXiv:astro-ph.HE/1802.02612]}}.
\newblock
  doi:{\changeurlcolor{black}\href{https://doi.org/10.3847/1538-4357/aaadb9}{\detokenize{10.3847/1538-4357/aaadb9}}}.

\bibitem[{Everett} and {Zweibel}(2011)]{Everett2011ApJ}
{Everett}, J.E.; {Zweibel}, E.G.
\newblock {The Interaction of Cosmic Rays with Diffuse Clouds}.
\newblock {\em \apj} {\bf 2011}, {\em 739},~60,
  \href{http://xxx.lanl.gov/abs/1107.1243}{{\normalfont
  [arXiv:astro-ph.GA/1107.1243]}}.
\newblock
  doi:{\changeurlcolor{black}\href{https://doi.org/10.1088/0004-637X/739/2/60}{\detokenize{10.1088/0004-637X/739/2/60}}}.

\bibitem[{Skilling} and {Strong}(1976)]{Skilling1976AA}
{Skilling}, J.; {Strong}, A.W.
\newblock {Cosmic ray exclusion from dense molecular clouds.}
\newblock {\em \aap} {\bf 1976}, {\em 53},~253--258.

\bibitem[{Silsbee} and {Ivlev}(2020)]{Silsbee2020ApJ}
{Silsbee}, K.; {Ivlev}, A.V.
\newblock {Exclusion of Cosmic Rays from Molecular Clouds by Self-generated
  Electric Fields}.
\newblock {\em \apjl} {\bf 2020}, {\em 902},~L25,
  \href{http://xxx.lanl.gov/abs/2009.14208}{{\normalfont
  [arXiv:astro-ph.HE/2009.14208]}}.
\newblock
  doi:{\changeurlcolor{black}\href{https://doi.org/10.3847/2041-8213/abbc20}{\detokenize{10.3847/2041-8213/abbc20}}}.

\bibitem[{Ko}(1992)]{Ko1992AA}
{Ko}, C.M.
\newblock {A note on the hydrodynamical description of cosmic ray propagation}.
\newblock {\em \aap} {\bf 1992}, {\em 259},~377--381.

\bibitem[{Padoan} and {Scalo}(2005)]{Padoan2005ApJ}
{Padoan}, P.; {Scalo}, J.
\newblock {Confinement-driven Spatial Variations in the Cosmic-Ray Flux}.
\newblock {\em \apjl} {\bf 2005}, {\em 624},~L97--L100,
  \href{http://xxx.lanl.gov/abs/astro-ph/0503585}{{\normalfont
  [arXiv:astro-ph/astro-ph/0503585]}}.
\newblock
  doi:{\changeurlcolor{black}\href{https://doi.org/10.1086/430598}{\detokenize{10.1086/430598}}}.

\bibitem[{Chandran}(2000)]{Chandran2000ApJ}
{Chandran}, B.D.G.
\newblock {Confinement and Isotropization of Galactic Cosmic Rays by
  Molecular-Cloud Magnetic Mirrors When Turbulent Scattering Is Weak}.
\newblock {\em \apj} {\bf 2000}, {\em 529},~513--535.
\newblock
  doi:{\changeurlcolor{black}\href{https://doi.org/10.1086/308232}{\detokenize{10.1086/308232}}}.

\bibitem[{Desch} \em{et~al.}(2004){Desch}, {Connolly}, and
  {Srinivasan}]{Desch2004ApJ}
{Desch}, S.J.; {Connolly}, Harold~C., J.; {Srinivasan}, G.
\newblock {An Interstellar Origin for the Beryllium 10 in Calcium-rich,
  Aluminum-rich Inclusions}.
\newblock {\em \apj} {\bf 2004}, {\em 602},~528--542.
\newblock
  doi:{\changeurlcolor{black}\href{https://doi.org/10.1086/380831}{\detokenize{10.1086/380831}}}.

\bibitem[{Owen} \em{et~al.}(2021){Owen}, {Lee}, and {Kong}]{Owen2021MNRAS}
{Owen}, E.R.; {Lee}, K.G.; {Kong}, A.K.H.
\newblock {Characterizing the signatures of star-forming galaxies in the
  extragalactic {\ensuremath{\gamma}}-ray background}.
\newblock {\em \mnras} {\bf 2021}, {\em 506},~52--72,
  \href{http://xxx.lanl.gov/abs/2106.07308}{{\normalfont
  [arXiv:astro-ph.HE/2106.07308]}}.
\newblock
  doi:{\changeurlcolor{black}\href{https://doi.org/10.1093/mnras/stab1707}{\detokenize{10.1093/mnras/stab1707}}}.

\bibitem[{Silsbee} \em{et~al.}(2018){Silsbee}, {Ivlev}, {Padovani}, and
  {Caselli}]{Silsbee2018ApJ}
{Silsbee}, K.; {Ivlev}, A.V.; {Padovani}, M.; {Caselli}, P.
\newblock {Magnetic Mirroring and Focusing of Cosmic Rays}.
\newblock {\em \apj} {\bf 2018}, {\em 863},~188,
  \href{http://xxx.lanl.gov/abs/1807.05025}{{\normalfont
  [arXiv:astro-ph.HE/1807.05025]}}.
\newblock
  doi:{\changeurlcolor{black}\href{https://doi.org/10.3847/1538-4357/aad3cf}{\detokenize{10.3847/1538-4357/aad3cf}}}.

\bibitem[{Albertsson} \em{et~al.}(2018){Albertsson}, {Kauffmann}, and
  {Menten}]{Albertsson2018ApJ}
{Albertsson}, T.; {Kauffmann}, J.; {Menten}, K.M.
\newblock {Atlas of Cosmic-Ray-induced Astrochemistry}.
\newblock {\em \apj} {\bf 2018}, {\em 868},~40,
  \href{http://xxx.lanl.gov/abs/1811.02862}{{\normalfont
  [arXiv:astro-ph.GA/1811.02862]}}.
\newblock
  doi:{\changeurlcolor{black}\href{https://doi.org/10.3847/1538-4357/aae775}{\detokenize{10.3847/1538-4357/aae775}}}.

\bibitem[{Li} and {Burkert}(2018)]{Li2018MNRAS}
{Li}, G.X.; {Burkert}, A.
\newblock {Quantifying the interplay between gravity and magnetic field in
  molecular clouds - a possible multiscale energy equipartition in NGC 6334}.
\newblock {\em \mnras} {\bf 2018}, {\em 474},~2167--2172,
  \href{http://xxx.lanl.gov/abs/1711.02417}{{\normalfont
  [arXiv:astro-ph.GA/1711.02417]}}.
\newblock
  doi:{\changeurlcolor{black}\href{https://doi.org/10.1093/mnras/stx2827}{\detokenize{10.1093/mnras/stx2827}}}.

\bibitem[{Padovani} \em{et~al.}(2018){Padovani}, {Ivlev}, {Galli}, and
  {Caselli}]{Padovani2018A&Ab}
{Padovani}, M.; {Ivlev}, A.V.; {Galli}, D.; {Caselli}, P.
\newblock {Cosmic-ray ionisation in circumstellar discs}.
\newblock {\em \aap} {\bf 2018}, {\em 614},~A111,
  \href{http://xxx.lanl.gov/abs/1803.09348}{{\normalfont
  [arXiv:astro-ph.HE/1803.09348]}}.
\newblock
  doi:{\changeurlcolor{black}\href{https://doi.org/10.1051/0004-6361/201732202}{\detokenize{10.1051/0004-6361/201732202}}}.

\bibitem[{Phan} \em{et~al.}(2018){Phan}, {Morlino}, and
  {Gabici}]{Phan2018MNRAS}
{Phan}, V.H.M.; {Morlino}, G.; {Gabici}, S.
\newblock {What causes the ionization rates observed in diffuse molecular
  clouds? The role of cosmic ray protons and electrons}.
\newblock {\em \mnras} {\bf 2018}, {\em 480},~5167--5174,
  \href{http://xxx.lanl.gov/abs/1804.10106}{{\normalfont
  [arXiv:astro-ph.HE/1804.10106]}}.
\newblock
  doi:{\changeurlcolor{black}\href{https://doi.org/10.1093/mnras/sty2235}{\detokenize{10.1093/mnras/sty2235}}}.

\bibitem[{Morlino} and {Gabici}(2015)]{Morlino2015MNRAS}
{Morlino}, G.; {Gabici}, S.
\newblock {Cosmic ray penetration in diffuse clouds.}
\newblock {\em \mnras} {\bf 2015}, {\em 451},~L100--L104,
  \href{http://xxx.lanl.gov/abs/1503.02435}{{\normalfont
  [arXiv:astro-ph.HE/1503.02435]}}.
\newblock
  doi:{\changeurlcolor{black}\href{https://doi.org/10.1093/mnrasl/slv074}{\detokenize{10.1093/mnrasl/slv074}}}.

\bibitem[{Dogiel} \em{et~al.}(2018){Dogiel}, {Chernyshov}, {Ivlev}, {Malyshev},
  {Strong}, and {Cheng}]{Dogiel2018ApJ}
{Dogiel}, V.A.; {Chernyshov}, D.O.; {Ivlev}, A.V.; {Malyshev}, D.; {Strong},
  A.W.; {Cheng}, K.S.
\newblock {Gamma-Ray Emission from Molecular Clouds Generated by Penetrating
  Cosmic Rays}.
\newblock {\em \apj} {\bf 2018}, {\em 868},~114,
  \href{http://xxx.lanl.gov/abs/1810.05821}{{\normalfont
  [arXiv:astro-ph.HE/1810.05821]}}.
\newblock
  doi:{\changeurlcolor{black}\href{https://doi.org/10.3847/1538-4357/aae827}{\detokenize{10.3847/1538-4357/aae827}}}.

\bibitem[{Padovani} \em{et~al.}(2009){Padovani}, {Galli}, and
  {Glassgold}]{Padovani2009A&A}
{Padovani}, M.; {Galli}, D.; {Glassgold}, A.E.
\newblock {Cosmic-ray ionization of molecular clouds}.
\newblock {\em \aap} {\bf 2009}, {\em 501},~619--631,
  \href{http://xxx.lanl.gov/abs/0904.4149}{{\normalfont
  [arXiv:astro-ph.SR/0904.4149]}}.
\newblock
  doi:{\changeurlcolor{black}\href{https://doi.org/10.1051/0004-6361/200911794}{\detokenize{10.1051/0004-6361/200911794}}}.

\bibitem[{Gabici}(2022)]{Gabici2022A&ARv}
{Gabici}, S.
\newblock {Low-energy cosmic rays: regulators of the dense interstellar
  medium}.
\newblock {\em \aapr} {\bf 2022}, {\em 30},~4,
  \href{http://xxx.lanl.gov/abs/2203.14620}{{\normalfont
  [arXiv:astro-ph.HE/2203.14620]}}.
\newblock
  doi:{\changeurlcolor{black}\href{https://doi.org/10.1007/s00159-022-00141-2}{\detokenize{10.1007/s00159-022-00141-2}}}.

\bibitem[{Oka} \em{et~al.}(2019){Oka}, {Geballe}, {Goto}, {Usuda}, {Benjamin},
  {McCall}, and {Indriolo}]{Oka2019ApJ}
{Oka}, T.; {Geballe}, T.R.; {Goto}, M.; {Usuda}, T.; {Benjamin}.; {McCall}, J.;
  {Indriolo}, N.
\newblock {The Central 300 pc of the Galaxy Probed by Infrared Spectra of H3+
  and CO. I. Predominance of Warm and Diffuse Gas and High H2 Ionization Rate}.
\newblock {\em \apj} {\bf 2019}, {\em 883},~54,
  \href{http://xxx.lanl.gov/abs/1910.04762}{{\normalfont
  [arXiv:astro-ph.HE/1910.04762]}}.
\newblock
  doi:{\changeurlcolor{black}\href{https://doi.org/10.3847/1538-4357/ab3647}{\detokenize{10.3847/1538-4357/ab3647}}}.

\bibitem[{Indriolo}(2023)]{Indriolo2023arXiv}
{Indriolo}, N.
\newblock {Absorption-line Observations of \{\{H\}\}\_\{3\}(+) and CO in Sight
  Lines Toward the Vela and W28 Supernova Remnants}.
\newblock {\em \apj} {\bf 2023}, {\em 950},~64,
  \href{http://xxx.lanl.gov/abs/2303.13689}{{\normalfont
  [arXiv:astro-ph.HE/2303.13689]}}.
\newblock
  doi:{\changeurlcolor{black}\href{https://doi.org/10.3847/1538-4357/acc6c4}{\detokenize{10.3847/1538-4357/acc6c4}}}.

\bibitem[{Le Petit} \em{et~al.}(2016){Le Petit}, {Ruaud}, {Bron}, {Godard},
  {Roueff}, {Languignon}, and {Le Bourlot}]{LePetit2016A&A}
{Le Petit}, F.; {Ruaud}, M.; {Bron}, E.; {Godard}, B.; {Roueff}, E.;
  {Languignon}, D.; {Le Bourlot}, J.
\newblock {Physical conditions in the central molecular zone inferred by
  H$_{3}$$^{+}$}.
\newblock {\em \aap} {\bf 2016}, {\em 585},~A105,
  \href{http://xxx.lanl.gov/abs/1510.02221}{{\normalfont
  [arXiv:astro-ph.GA/1510.02221]}}.
\newblock
  doi:{\changeurlcolor{black}\href{https://doi.org/10.1051/0004-6361/201526658}{\detokenize{10.1051/0004-6361/201526658}}}.

\bibitem[{Sabatini} \em{et~al.}(2023){Sabatini}, {Bovino}, and
  {Redaelli}]{Sabatini2023arXiv230400329S}
{Sabatini}, G.; {Bovino}, S.; {Redaelli}, E.
\newblock {First ALMA maps of cosmic ray ionisation rate in high-mass
  star-forming regions}.
\newblock {\em arXiv e-prints} {\bf 2023}, p. arXiv:2304.00329,
  \href{http://xxx.lanl.gov/abs/2304.00329}{{\normalfont
  [arXiv:astro-ph.GA/2304.00329]}}.
\newblock
  doi:{\changeurlcolor{black}\href{https://doi.org/10.48550/arXiv.2304.00329}{\detokenize{10.48550/arXiv.2304.00329}}}.

\bibitem[{Schilke} \em{et~al.}(2014){Schilke}, {Neufeld}, {M{\"u}ller},
  {Comito}, {Bergin}, {Lis}, {Gerin}, {Black}, {Wolfire}, {Indriolo},
  {Pearson}, {Menten}, {Winkel}, {S{\'a}nchez-Monge}, {M{\"o}ller}, {Godard},
  and {Falgarone}]{Schilke2014A&A}
{Schilke}, P.; {Neufeld}, D.A.; {M{\"u}ller}, H.S.P.; {Comito}, C.; {Bergin},
  E.A.; {Lis}, D.C.; {Gerin}, M.; {Black}, J.H.; {Wolfire}, M.; {Indriolo}, N.;
   et~al.
\newblock {Ubiquitous argonium (ArH$^{+}$) in the diffuse interstellar medium:
  A molecular tracer of almost purely atomic gas}.
\newblock {\em \aap} {\bf 2014}, {\em 566},~A29,
  \href{http://xxx.lanl.gov/abs/1403.7902}{{\normalfont
  [arXiv:astro-ph.GA/1403.7902]}}.
\newblock
  doi:{\changeurlcolor{black}\href{https://doi.org/10.1051/0004-6361/201423727}{\detokenize{10.1051/0004-6361/201423727}}}.

\bibitem[{Jacob} \em{et~al.}(2020){Jacob}, {Menten}, {Wyrowski}, {Winkel}, and
  {Neufeld}]{Jacob2020A&A}
{Jacob}, A.M.; {Menten}, K.M.; {Wyrowski}, F.; {Winkel}, B.; {Neufeld}, D.A.
\newblock {Extending the view of ArH$^{+}$ chemistry in diffuse clouds}.
\newblock {\em \aap} {\bf 2020}, {\em 643},~A91,
  \href{http://xxx.lanl.gov/abs/2010.02258}{{\normalfont
  [arXiv:astro-ph.GA/2010.02258]}}.
\newblock
  doi:{\changeurlcolor{black}\href{https://doi.org/10.1051/0004-6361/202039197}{\detokenize{10.1051/0004-6361/202039197}}}.

\bibitem[{Bialy} \em{et~al.}(2019){Bialy}, {Neufeld}, {Wolfire}, {Sternberg},
  and {Burkhart}]{Bialy2019ApJ}
{Bialy}, S.; {Neufeld}, D.; {Wolfire}, M.; {Sternberg}, A.; {Burkhart}, B.
\newblock {Chemical Abundances in a Turbulent Medium-H$_{2}$, OH$^{+}$,
  H$_{2}$O$^{+}$, ArH$^{+}$}.
\newblock {\em \apj} {\bf 2019}, {\em 885},~109,
  \href{http://xxx.lanl.gov/abs/1909.12305}{{\normalfont
  [arXiv:astro-ph.GA/1909.12305]}}.
\newblock
  doi:{\changeurlcolor{black}\href{https://doi.org/10.3847/1538-4357/ab487b}{\detokenize{10.3847/1538-4357/ab487b}}}.

\bibitem[{Jacob} \em{et~al.}(2022){Jacob}, {Neufeld}, {Schilke}, {Wiesemeyer},
  {Kim}, {Bialy}, {Busch}, {Elia}, {Falgarone}, {Gerin}, {Godard}, {Higgins},
  {Hennebelle}, {Indriolo}, {Lis}, {Menten}, {Sanchez-Monge}, {M{\"o}ller},
  {Ossenkopf-Okada}, {Rugel}, {Seifried}, {Sonnentrucker}, {Walch}, {Wolfire},
  {Wyrowski}, and {Valdivia}]{Jacob2022ApJ}
{Jacob}, A.M.; {Neufeld}, D.A.; {Schilke}, P.; {Wiesemeyer}, H.; {Kim}, W.J.;
  {Bialy}, S.; {Busch}, M.; {Elia}, D.; {Falgarone}, E.; {Gerin}, M.;  et~al.
\newblock {HyGAL: Characterizing the Galactic Interstellar Medium with
  Observations of Hydrides and Other Small Molecules. I. Survey Description and
  a First Look Toward W3(OH), W3 IRS5, and NGC 7538 IRS1}.
\newblock {\em \apj} {\bf 2022}, {\em 930},~141,
  \href{http://xxx.lanl.gov/abs/2202.05046}{{\normalfont
  [arXiv:astro-ph.GA/2202.05046]}}.
\newblock
  doi:{\changeurlcolor{black}\href{https://doi.org/10.3847/1538-4357/ac5409}{\detokenize{10.3847/1538-4357/ac5409}}}.

\bibitem[{Indriolo} and {McCall}(2012)]{Indriolo2012ApJ}
{Indriolo}, N.; {McCall}, B.J.
\newblock {Investigating the Cosmic-Ray Ionization Rate in the Galactic Diffuse
  Interstellar Medium through Observations of H$^{+}$ $_{3}$}.
\newblock {\em \apj} {\bf 2012}, {\em 745},~91,
  \href{http://xxx.lanl.gov/abs/1111.6936}{{\normalfont
  [arXiv:astro-ph.GA/1111.6936]}}.
\newblock
  doi:{\changeurlcolor{black}\href{https://doi.org/10.1088/0004-637X/745/1/91}{\detokenize{10.1088/0004-637X/745/1/91}}}.

\bibitem[{Holdship} \em{et~al.}(2017){Holdship}, {Viti}, {Jim{\'e}nez-Serra},
  {Makrymallis}, and {Priestley}]{Holdship2017AJ}
{Holdship}, J.; {Viti}, S.; {Jim{\'e}nez-Serra}, I.; {Makrymallis}, A.;
  {Priestley}, F.
\newblock {UCLCHEM: A Gas-grain Chemical Code for Clouds, Cores, and C-Shocks}.
\newblock {\em \aj} {\bf 2017}, {\em 154},~38,
  \href{http://xxx.lanl.gov/abs/1705.10677}{{\normalfont
  [arXiv:astro-ph.GA/1705.10677]}}.
\newblock
  doi:{\changeurlcolor{black}\href{https://doi.org/10.3847/1538-3881/aa773f}{\detokenize{10.3847/1538-3881/aa773f}}}.

\bibitem[{Maret} and {Bergin}(2015)]{Maret2015ascl}
{Maret}, S.; {Bergin}, E.A.
\newblock {Astrochem: Abundances of chemical species in the interstellar
  medium}.
\newblock Astrophysics Source Code Library, record ascl:1507.010,  2015,
  \href{http://xxx.lanl.gov/abs/1507.010}{{\normalfont [1507.010]}}.

\bibitem[{Lin} \em{et~al.}(2020){Lin}, {Pagani}, {Lai}, {Lef{\`e}vre}, and
  {Lique}]{Lin2020A&A}
{Lin}, S.J.; {Pagani}, L.; {Lai}, S.P.; {Lef{\`e}vre}, C.; {Lique}, F.
\newblock {Physical and chemical modeling of the starless core L 1512}.
\newblock {\em \aap} {\bf 2020}, {\em 635},~A188,
  \href{http://xxx.lanl.gov/abs/2002.01346}{{\normalfont
  [arXiv:astro-ph.GA/2002.01346]}}.
\newblock
  doi:{\changeurlcolor{black}\href{https://doi.org/10.1051/0004-6361/201936877}{\detokenize{10.1051/0004-6361/201936877}}}.

\bibitem[{Indriolo} \em{et~al.}(2010){Indriolo}, {Blake}, {Goto}, {Usuda},
  {Oka}, {Geballe}, {Fields}, and {McCall}]{Indriolo2010ApJ}
{Indriolo}, N.; {Blake}, G.A.; {Goto}, M.; {Usuda}, T.; {Oka}, T.; {Geballe},
  T.R.; {Fields}, B.D.; {McCall}, B.J.
\newblock {Investigating the Cosmic-ray Ionization Rate Near the Supernova
  Remnant IC 443 through H$^{+}$ $_{3}$ Observations}.
\newblock {\em \apj} {\bf 2010}, {\em 724},~1357--1365,
  \href{http://xxx.lanl.gov/abs/1010.3252}{{\normalfont
  [arXiv:astro-ph.HE/1010.3252]}}.
\newblock
  doi:{\changeurlcolor{black}\href{https://doi.org/10.1088/0004-637X/724/2/1357}{\detokenize{10.1088/0004-637X/724/2/1357}}}.

\bibitem[{Ceccarelli} \em{et~al.}(2011){Ceccarelli}, {Hily-Blant}, {Montmerle},
  {Dubus}, {Gallant}, and {Fiasson}]{Ceccarelli2011ApJ}
{Ceccarelli}, C.; {Hily-Blant}, P.; {Montmerle}, T.; {Dubus}, G.; {Gallant},
  Y.; {Fiasson}, A.
\newblock {Supernova-enhanced Cosmic-Ray Ionization and Induced Chemistry in a
  Molecular Cloud of W51C}.
\newblock {\em \apjl} {\bf 2011}, {\em 740},~L4,
  \href{http://xxx.lanl.gov/abs/1108.3600}{{\normalfont
  [arXiv:astro-ph.GA/1108.3600]}}.
\newblock
  doi:{\changeurlcolor{black}\href{https://doi.org/10.1088/2041-8205/740/1/L4}{\detokenize{10.1088/2041-8205/740/1/L4}}}.

\bibitem[{Owen}(2023)]{Owen2023A&G}
{Owen}, E.
\newblock {The secret agent of galaxy evolution}.
\newblock {\em Astronomy and Geophysics} {\bf 2023}, {\em 64},~1.29--1.35.
\newblock
  doi:{\changeurlcolor{black}\href{https://doi.org/10.1093/astrogeo/atac090}{\detokenize{10.1093/astrogeo/atac090}}}.

\bibitem[{Caselli} \em{et~al.}(1998){Caselli}, {Walmsley}, {Terzieva}, and
  {Herbst}]{Caselli1998ApJ}
{Caselli}, P.; {Walmsley}, C.M.; {Terzieva}, R.; {Herbst}, E.
\newblock {The Ionization Fraction in Dense Cloud Cores}.
\newblock {\em \apj} {\bf 1998}, {\em 499},~234--249.
\newblock
  doi:{\changeurlcolor{black}\href{https://doi.org/10.1086/305624}{\detokenize{10.1086/305624}}}.

\bibitem[{Morales Ortiz} \em{et~al.}(2014){Morales Ortiz}, {Ceccarelli}, {Lis},
  {Olmi}, {Plume}, and {Schilke}]{Morales2014A&A}
{Morales Ortiz}, J.L.; {Ceccarelli}, C.; {Lis}, D.C.; {Olmi}, L.; {Plume}, R.;
  {Schilke}, P.
\newblock {Ionization toward the high-mass star-forming region NGC 6334 I}.
\newblock {\em \aap} {\bf 2014}, {\em 563},~A127,
  \href{http://xxx.lanl.gov/abs/1306.3012}{{\normalfont
  [arXiv:astro-ph.GA/1306.3012]}}.
\newblock
  doi:{\changeurlcolor{black}\href{https://doi.org/10.1051/0004-6361/201322071}{\detokenize{10.1051/0004-6361/201322071}}}.

\bibitem[{Hezareh} \em{et~al.}(2008){Hezareh}, {Houde}, {McCoey}, {Vastel}, and
  {Peng}]{Hezareh2008ApJ}
{Hezareh}, T.; {Houde}, M.; {McCoey}, C.; {Vastel}, C.; {Peng}, R.
\newblock {Simultaneous Determination of the Cosmic Ray Ionization Rate and
  Fractional Ionization in DR 21(OH)}.
\newblock {\em \apj} {\bf 2008}, {\em 684},~1221--1227,
  \href{http://xxx.lanl.gov/abs/0805.4018}{{\normalfont
  [arXiv:astro-ph/0805.4018]}}.
\newblock
  doi:{\changeurlcolor{black}\href{https://doi.org/10.1086/590365}{\detokenize{10.1086/590365}}}.

\bibitem[{van der Tak} and {van Dishoeck}(2000)]{VanDerTak2000A&A}
{van der Tak}, F.F.S.; {van Dishoeck}, E.F.
\newblock {Limits on the cosmic-ray ionization rate toward massive young
  stars}.
\newblock {\em \aap} {\bf 2000}, {\em 358},~L79--L82,
  \href{http://xxx.lanl.gov/abs/astro-ph/0006246}{{\normalfont
  [arXiv:astro-ph/astro-ph/0006246]}}.
\newblock
  doi:{\changeurlcolor{black}\href{https://doi.org/10.48550/arXiv.astro-ph/0006246}{\detokenize{10.48550/arXiv.astro-ph/0006246}}}.

\bibitem[{de Boisanger} \em{et~al.}(1996){de Boisanger}, {Helmich}, and {van
  Dishoeck}]{Boisanger1996A&A}
{de Boisanger}, C.; {Helmich}, F.P.; {van Dishoeck}, E.F.
\newblock {The ionization fraction in dense clouds.}
\newblock {\em \aap} {\bf 1996}, {\em 310},~315--327,
  \href{http://xxx.lanl.gov/abs/astro-ph/9511091}{{\normalfont
  [arXiv:astro-ph/astro-ph/9511091]}}.
\newblock
  doi:{\changeurlcolor{black}\href{https://doi.org/10.48550/arXiv.astro-ph/9511091}{\detokenize{10.48550/arXiv.astro-ph/9511091}}}.

\bibitem[{Redaelli} \em{et~al.}(2021){Redaelli}, {Sipil{\"a}}, {Padovani},
  {Caselli}, {Galli}, and {Ivlev}]{Redaelli2021A&A}
{Redaelli}, E.; {Sipil{\"a}}, O.; {Padovani}, M.; {Caselli}, P.; {Galli}, D.;
  {Ivlev}, A.V.
\newblock {The cosmic-ray ionisation rate in the pre-stellar core L1544}.
\newblock {\em \aap} {\bf 2021}, {\em 656},~A109,
  \href{http://xxx.lanl.gov/abs/2109.08169}{{\normalfont
  [arXiv:astro-ph.GA/2109.08169]}}.
\newblock
  doi:{\changeurlcolor{black}\href{https://doi.org/10.1051/0004-6361/202141776}{\detokenize{10.1051/0004-6361/202141776}}}.

\bibitem[{Porras} \em{et~al.}(2014){Porras}, {Federman}, {Welty}, and
  {Ritchey}]{Porras2014ApJ}
{Porras}, A.J.; {Federman}, S.R.; {Welty}, D.E.; {Ritchey}, A.M.
\newblock {OH$^{+}$ in Diffuse Molecular Clouds}.
\newblock {\em \apjl} {\bf 2014}, {\em 781},~L8,
  \href{http://xxx.lanl.gov/abs/1312.5239}{{\normalfont
  [arXiv:astro-ph.GA/1312.5239]}}.
\newblock
  doi:{\changeurlcolor{black}\href{https://doi.org/10.1088/2041-8205/781/1/L8}{\detokenize{10.1088/2041-8205/781/1/L8}}}.

\bibitem[{Indriolo} \em{et~al.}(2007){Indriolo}, {Geballe}, {Oka}, and
  {McCall}]{Indriolo2007ApJ}
{Indriolo}, N.; {Geballe}, T.R.; {Oka}, T.; {McCall}, B.J.
\newblock {H$^{+}$$_{3}$ in Diffuse Interstellar Clouds: A Tracer for the
  Cosmic-Ray Ionization Rate}.
\newblock {\em \apj} {\bf 2007}, {\em 671},~1736--1747,
  \href{http://xxx.lanl.gov/abs/0709.1114}{{\normalfont
  [arXiv:astro-ph/0709.1114]}}.
\newblock
  doi:{\changeurlcolor{black}\href{https://doi.org/10.1086/523036}{\detokenize{10.1086/523036}}}.

\bibitem[{Indriolo} \em{et~al.}(2015){Indriolo}, {Neufeld}, {Gerin}, {Schilke},
  {Benz}, {Winkel}, {Menten}, {Chambers}, {Black}, {Bruderer}, {Falgarone},
  {Godard}, {Goicoechea}, {Gupta}, {Lis}, {Ossenkopf}, {Persson},
  {Sonnentrucker}, {van der Tak}, {van Dishoeck}, {Wolfire}, and
  {Wyrowski}]{Indriolo2015ApJ}
{Indriolo}, N.; {Neufeld}, D.A.; {Gerin}, M.; {Schilke}, P.; {Benz}, A.O.;
  {Winkel}, B.; {Menten}, K.M.; {Chambers}, E.T.; {Black}, J.H.; {Bruderer},
  S.;  et~al.
\newblock {Herschel Survey of Galactic OH$^{+}$, H$_{2}$O$^{+}$, and
  H$_{3}$O$^{+}$: Probing the Molecular Hydrogen Fraction and Cosmic-Ray
  Ionization Rate}.
\newblock {\em \apj} {\bf 2015}, {\em 800},~40,
  \href{http://xxx.lanl.gov/abs/1412.1106}{{\normalfont
  [arXiv:astro-ph.GA/1412.1106]}}.
\newblock
  doi:{\changeurlcolor{black}\href{https://doi.org/10.1088/0004-637X/800/1/40}{\detokenize{10.1088/0004-637X/800/1/40}}}.

\bibitem[{Bacalla} \em{et~al.}(2019){Bacalla}, {Linnartz}, {Cox}, {Cami},
  {Roueff}, {Smoker}, {Farhang}, {Bouwman}, and {Zhao}]{Bacalla2019A&A}
{Bacalla}, X.L.; {Linnartz}, H.; {Cox}, N.L.J.; {Cami}, J.; {Roueff}, E.;
  {Smoker}, J.V.; {Farhang}, A.; {Bouwman}, J.; {Zhao}, D.
\newblock {The EDIBLES survey. IV. Cosmic ray ionization rates in diffuse
  clouds from near-ultraviolet observations of interstellar OH$^{+}$}.
\newblock {\em \aap} {\bf 2019}, {\em 622},~A31,
  \href{http://xxx.lanl.gov/abs/1811.08662}{{\normalfont
  [arXiv:astro-ph.GA/1811.08662]}}.
\newblock
  doi:{\changeurlcolor{black}\href{https://doi.org/10.1051/0004-6361/201833039}{\detokenize{10.1051/0004-6361/201833039}}}.

\bibitem[{Padovani} and {Galli}(2018)]{Padovani2018A&A}
{Padovani}, M.; {Galli}, D.
\newblock {Synchrotron emission in molecular cloud cores: the SKA view}.
\newblock {\em \aap} {\bf 2018}, {\em 620},~L4,
  \href{http://xxx.lanl.gov/abs/1811.09698}{{\normalfont
  [arXiv:astro-ph.HE/1811.09698]}}.
\newblock
  doi:{\changeurlcolor{black}\href{https://doi.org/10.1051/0004-6361/201834222}{\detokenize{10.1051/0004-6361/201834222}}}.

\bibitem[{Padovani} \em{et~al.}(2022){Padovani}, {Bialy}, {Galli}, {Ivlev},
  {Grassi}, {Scarlett}, {Rehill}, {Zammit}, {Fursa}, and
  {Bray}]{Padovani2022AA}
{Padovani}, M.; {Bialy}, S.; {Galli}, D.; {Ivlev}, A.V.; {Grassi}, T.;
  {Scarlett}, L.H.; {Rehill}, U.S.; {Zammit}, M.C.; {Fursa}, D.V.; {Bray}, I.
\newblock {Cosmic rays in molecular clouds probed by H$_{2}$ rovibrational
  lines. Perspectives for the James Webb Space Telescope}.
\newblock {\em \aap} {\bf 2022}, {\em 658},~A189,
  \href{http://xxx.lanl.gov/abs/2201.08457}{{\normalfont
  [arXiv:astro-ph.GA/2201.08457]}}.
\newblock
  doi:{\changeurlcolor{black}\href{https://doi.org/10.1051/0004-6361/202142560}{\detokenize{10.1051/0004-6361/202142560}}}.

\bibitem[{Tatischeff} \em{et~al.}(2012){Tatischeff}, {Decourchelle}, and
  {Maurin}]{Tatischeff2012A&A}
{Tatischeff}, V.; {Decourchelle}, A.; {Maurin}, G.
\newblock {Nonthermal X-rays from low-energy cosmic rays: application to the
  6.4 keV line emission from the Arches cluster region}.
\newblock {\em \aap} {\bf 2012}, {\em 546},~A88,
  \href{http://xxx.lanl.gov/abs/1210.2108}{{\normalfont
  [arXiv:astro-ph.HE/1210.2108]}}.
\newblock
  doi:{\changeurlcolor{black}\href{https://doi.org/10.1051/0004-6361/201219016}{\detokenize{10.1051/0004-6361/201219016}}}.

\bibitem[{Okon} \em{et~al.}(2020){Okon}, {Imai}, {Tanaka}, {Uchida}, and
  {Tsuru}]{Okon2020PASJ}
{Okon}, H.; {Imai}, M.; {Tanaka}, T.; {Uchida}, H.; {Tsuru}, T.G.
\newblock {Probing cosmic rays with Fe K{\ensuremath{\alpha}} line structures
  generated by multiple ionization process}.
\newblock {\em \pasj} {\bf 2020}, {\em 72},~L7,
  \href{http://xxx.lanl.gov/abs/2005.08462}{{\normalfont
  [arXiv:astro-ph.HE/2005.08462]}}.
\newblock
  doi:{\changeurlcolor{black}\href{https://doi.org/10.1093/pasj/psaa055}{\detokenize{10.1093/pasj/psaa055}}}.

\bibitem[{Bialy}(2020)]{Bialy2020CmPhy}
{Bialy}, S.
\newblock {Cold clouds as cosmic-ray detectors}.
\newblock {\em Communications Physics} {\bf 2020}, {\em 3},~32,
  \href{http://xxx.lanl.gov/abs/1910.12953}{{\normalfont
  [arXiv:astro-ph.GA/1910.12953]}}.
\newblock
  doi:{\changeurlcolor{black}\href{https://doi.org/10.1038/s42005-020-0293-7}{\detokenize{10.1038/s42005-020-0293-7}}}.

\bibitem[{Bialy} \em{et~al.}(2022){Bialy}, {Belli}, and
  {Padovani}]{Bialy2022AAL}
{Bialy}, S.; {Belli}, S.; {Padovani}, M.
\newblock {Constraining the cosmic-ray ionization rate and spectrum with NIR
  spectroscopy of dense clouds. A testbed for JWST}.
\newblock {\em \aap} {\bf 2022}, {\em 658},~L13,
  \href{http://xxx.lanl.gov/abs/2111.06900}{{\normalfont
  [arXiv:astro-ph.GA/2111.06900]}}.
\newblock
  doi:{\changeurlcolor{black}\href{https://doi.org/10.1051/0004-6361/202142619}{\detokenize{10.1051/0004-6361/202142619}}}.

\bibitem[{Gaches} \em{et~al.}(2022){Gaches}, {Bialy}, {Bisbas}, {Padovani},
  {Seifried}, and {Walch}]{Gaches2022AAb}
{Gaches}, B.A.L.; {Bialy}, S.; {Bisbas}, T.G.; {Padovani}, M.; {Seifried}, D.;
  {Walch}, S.
\newblock {Cosmic-ray-induced H$_{2}$ line emission. Astrochemical modeling and
  implications for JWST observations}.
\newblock {\em \aap} {\bf 2022}, {\em 664},~A150,
  \href{http://xxx.lanl.gov/abs/2206.09780}{{\normalfont
  [arXiv:astro-ph.GA/2206.09780]}}.
\newblock
  doi:{\changeurlcolor{black}\href{https://doi.org/10.1051/0004-6361/202244090}{\detokenize{10.1051/0004-6361/202244090}}}.

\bibitem[{Casanova} \em{et~al.}(2010){Casanova}, {Aharonian}, {Fukui},
  {Gabici}, {Jones}, {Kawamura}, {Onishi}, {Rowell}, {Sano}, {Torii}, and
  {Yamamoto}]{Casanova2010PASJ}
{Casanova}, S.; {Aharonian}, F.A.; {Fukui}, Y.; {Gabici}, S.; {Jones}, D.I.;
  {Kawamura}, A.; {Onishi}, T.; {Rowell}, G.; {Sano}, H.; {Torii}, K.;  et~al.
\newblock {Molecular Clouds as Cosmic-Ray Barometers}.
\newblock {\em \pasj} {\bf 2010}, {\em 62},~769,
  \href{http://xxx.lanl.gov/abs/0904.2887}{{\normalfont
  [arXiv:astro-ph.HE/0904.2887]}}.
\newblock
  doi:{\changeurlcolor{black}\href{https://doi.org/10.1093/pasj/62.3.769}{\detokenize{10.1093/pasj/62.3.769}}}.

\bibitem[{Cavasinni} \em{et~al.}(2006){Cavasinni}, {Grasso}, and
  {Maccione}]{Cavasinni2006APh}
{Cavasinni}, V.; {Grasso}, D.; {Maccione}, L.
\newblock {TeV neutrinos from supernova remnants embedded in giant molecular
  clouds}.
\newblock {\em Astroparticle Physics} {\bf 2006}, {\em 26},~41--49,
  \href{http://xxx.lanl.gov/abs/astro-ph/0604004}{{\normalfont
  [arXiv:astro-ph/astro-ph/0604004]}}.
\newblock
  doi:{\changeurlcolor{black}\href{https://doi.org/10.1016/j.astropartphys.2006.04.009}{\detokenize{10.1016/j.astropartphys.2006.04.009}}}.

\bibitem[{Banik} and {Bhadra}(2021)]{Banik2021EPJC}
{Banik}, P.; {Bhadra}, A.
\newblock {An interacting molecular cloud scenario for production of gamma-rays
  and neutrinos from MAGIC J1835-069, and MAGIC J1837-073}.
\newblock {\em European Physical Journal C} {\bf 2021}, {\em 81},~478,
  \href{http://xxx.lanl.gov/abs/2108.01863}{{\normalfont
  [arXiv:astro-ph.HE/2108.01863]}}.
\newblock
  doi:{\changeurlcolor{black}\href{https://doi.org/10.1140/epjc/s10052-021-09271-w}{\detokenize{10.1140/epjc/s10052-021-09271-w}}}.

\bibitem[{Sarmah} \em{et~al.}(2023){Sarmah}, {Chakraborty}, and
  {Joshi}]{Sarmah2023MNRAS}
{Sarmah}, P.; {Chakraborty}, S.; {Joshi}, J.C.
\newblock {Probing LHAASO galactic PeVatrons through gamma-ray and neutrino
  correspondence}.
\newblock {\em \mnras} {\bf 2023}, {\em 521},~1144--1151,
  \href{http://xxx.lanl.gov/abs/2301.04161}{{\normalfont
  [arXiv:astro-ph.HE/2301.04161]}}.
\newblock
  doi:{\changeurlcolor{black}\href{https://doi.org/10.1093/mnras/stad609}{\detokenize{10.1093/mnras/stad609}}}.

\bibitem[{Abbasi} \em{et~al.}(2023){Abbasi}, {Ackermann}, {Adams}, {Aggarwal},
  {Aguilar}, {Ahlers}, {Alameddine}, {Alves}, {Amin}, {Andeen}, {Anderson},
  {Anton}, {Arg{\"u}elles}, {Ashida}, {Athanasiadou}, {Axani}, {Bai},
  {Balagopal}, {Baricevic}, {Barwick}, {Basu}, {Bay}, {Beatty}, {Becker},
  {Tjus}, {Beise}, {Bellenghi}, {Benda}, {BenZvi}, {Berley}, {Bernardini},
  {Besson}, {Binder}, {Bindig}, {Blaufuss}, {Blot}, {Bontempo}, {Book},
  {Borowka}, {Meneguolo}, {B{\"o}ser}, {Botner}, {B{\"o}ttcher}, {Bourbeau},
  {Braun}, {Brinson}, {Brostean-Kaiser}, {Burley}, {Busse}, {Campana},
  {Carnie-Bronca}, {Chang}, {Chen}, {Chen}, {Chirkin}, {Choi}, {Clark},
  {Classen}, {Coleman}, {Collin}, {Connolly}, {Conrad}, {Coppin}, {Correa},
  {Countryman}, {Cowen}, {Dappen}, {Dave}, {De Clercq}, {DeLaunay},
  {L{\'o}pez}, {Dembinski}, {Deoskar}, {Desai}, {Desiati}, {de Vries}, {de
  Wasseige}, {DeYoung}, {Diaz}, {D{\'\i}az-V{\'e}lez}, {Dittmer}, {Dujmovic},
  {DuVernois}, {Ehrhardt}, {Eller}, {Engel}, {Erpenbeck}, {Evans}, {Evenson},
  {Fan}, {Fazely}, {Fedynitch}, {Feigl}, {Fiedlschuster}, {Fienberg}, {Finley},
  {Fischer}, {Fox}, {Franckowiak}, {Friedman}, {Fritz}, {F{\"u}rst}, {Gaisser},
  {Gallagher}, {Ganster}, {Garcia}, {Garrappa}, {Gerhardt}, {Ghadimi},
  {Glaser}, {Glauch}, {Gl{\"u}senkamp}, {Goehlke}, {Gonzalez}, {Goswami},
  {Grant}, {Gray}, {Gr{\'e}goire}, {Griswold}, {G{\"u}nther}, {Gutjahr},
  {Haack}, {Hallgren}, {Halliday}, {Halve}, {Halzen}, {Hamdaoui}, {Minh},
  {Hanson}, {Hardin}, {Harnisch}, {Hatch}, {Haungs}, {Helbing}, {Hellrung},
  {Henningsen}, {Heuermann}, {Hickford}, {Hidvegi}, {Hill}, {Hill}, {Hoffman},
  {Hoshina}, {Hou}, {Huber}, {Hultqvist}, {H{\"u}nnefeld}, {Hussain}, {Hymon},
  {In}, {Iovine}, {Ishihara}, {Jansson}, {Japaridze}, {Jeong}, {Jin}, {Jones},
  {Kang}, {Kang}, {Kang}, {Kappes}, {Kappesser}, {Kardum}, {Karg}, {Karl},
  {Karle}, {Katz}, {Kauer}, {Kelley}, {Kheirandish}, {Kin}, {Kiryluk}, {Klein},
  {Kochocki}, {Koirala}, {Kolanoski}, {Kontrimas}, {K{\"o}pke}, {Kopper},
  {Koskinen}, {Koundal}, {Kovacevich}, {Kowalski}, {Kozynets}, {Kruiswijk},
  {Krupczak}, {Kun}, {Kurahashi}, {Lad}, {Gualda}, {Lamoureux}, {Larson},
  {Lauber}, {Lazar}, {Lee}, {DeHolton}, {Leszczy{\'n}ska}, {Lincetto}, {Liu},
  {Liubarska}, {Lohfink}, {Love}, {Mariscal}, {Lu}, {Lucarelli}, {Ludwig},
  {Luszczak}, {Lyu}, {Ma}, {Madsen}, {Mahn}, {Makino}, {Mancina}, {Sainte},
  {Mari{\c{s}}}, {Marka}, {Marka}, {Marsee}, {Martinez-Soler}, {Maruyama},
  {Mayhew}, {McElroy}, {McNally}, {Mead}, {Meagher}, {Mechbal}, {Medina},
  {Meier}, {Meighen-Berger}, {Merckx}, {Micallef}, {Mockler}, {Montaruli},
  {Moore}, {Morse}, {Moulai}, {Mukherjee}, {Naab}, {Nagai}, {Naumann},
  {Nayerhoda}, {Necker}, {Neumann}, {Niederhausen}, {Nisa}, {Noell}, {Nowicki},
  {Pollmann}, {Oehler}, {Oeyen}, {Olivas}, {Orsoe}, {Osborn}, {O'Sullivan},
  {Pandya}, {Pankova}, {Park}, {Parker}, {Paudel}, {Paul}, {P{\'e}rez de los
  Heros}, {Peterson}, {Philippen}, {Pieper}, {Pizzuto}, {Plum}, {Popovych},
  {Porcelli}, {Rodriguez}, {Pries}, {Procter-Murphy}, {Przybylski}, {Raab},
  {Rack-Helleis}, {Rameez}, {Rawlins}, {Rechav}, {Rehman}, {Reichherzer},
  {Renzi}, {Resconi}, {Reusch}, {Rhode}, {Richman}, {Riedel}, {Roberts},
  {Robertson}, {Rodan}, {Roellinghoff}, {Rongen}, {Rott}, {Ruhe}, {Ruohan},
  {Ryckbosch}, {Cantu}, {Safa}, {Saffer}, {Salazar-Gallegos}, {Sampathkumar},
  {Sanchez Herrera}, {Sandrock}, {Santander}, {Sarkar}, {Sarkar}, {Savelberg},
  {Schaufel}, {Schieler}, {Schindler}, {Schlueter}, {Schmidt}, {Schneider},
  {Schr{\"o}der}, {Schumacher}, {Schwefer}, {Sclafani}, {Seckel}, {Seunarine},
  {Sharma}, {Shefali}, {Shimizu}, {Silva}, {Skrzypek}, {Smithers}, {Snihur},
  {Soedingrekso}, {S{\o}gaard}, {Soldin}, {Spannfellner}, {Spiczak},
  {Spiering}, {Stamatikos}, {Stanev}, {Stein}, {Stezelberger}, {St{\"u}rwald},
  {Stuttard}, {Sullivan}, {Taboada}, {Ter-Antonyan}, {Thompson}, {Thwaites},
  {Tilav}, {Tollefson}, {T{\"o}nnis}, {Toscano}, {Tosi}, {Trettin}, {Tung},
  {Turcotte}, {Twagirayezu}, {Ty}, {Elorrieta}, {Upshaw}, {Valtonen-Mattila},
  {Vandenbroucke}, {van Eijndhoven}, {Vannerom}, {van Santen}, {Vara},
  {Veitch-Michaelis}, {Verpoest}, {Veske}, {Walck}, {Wang}, {Watson}, {Weaver},
  {Weigel}, {Weindl}, {Weldert}, {Wendt}, {Werthebach}, {Weyrauch},
  {Whitehorn}, {Wiebusch}, {Willey}, {Williams}, {Wolf}, {Wrede}, {Wulff},
  {Xu}, {Xu}, {Yanez}, {Yasutsugu}, {Yildizci}, {Yoshida}, {Yu}, {Yuan},
  {Zhang}, {Zhelnin}, and {IceCube Collaboration}]{Abbasi2023ApJ}
{Abbasi}, R.; {Ackermann}, M.; {Adams}, J.; {Aggarwal}, N.; {Aguilar}, J.A.;
  {Ahlers}, M.; {Alameddine}, J.M.; {Alves}, A.A.; {Amin}, N.M.; {Andeen}, K.;
  et~al.
\newblock {Searches for Neutrinos from Large High Altitude Air Shower
  Observatory Ultra-high-energy {\ensuremath{\gamma}}-Ray Sources Using the
  IceCube Neutrino Observatory}.
\newblock {\em \apjl} {\bf 2023}, {\em 945},~L8,
  \href{http://xxx.lanl.gov/abs/2211.14184}{{\normalfont
  [arXiv:astro-ph.HE/2211.14184]}}.
\newblock
  doi:{\changeurlcolor{black}\href{https://doi.org/10.3847/2041-8213/acb933}{\detokenize{10.3847/2041-8213/acb933}}}.

\bibitem[{Voisin} \em{et~al.}(2019){Voisin}, {Rowell}, {Burton}, {Fukui},
  {Sano}, {Aharonian}, {Maxted}, {Braiding}, {Blackwell}, and
  {Lau}]{Voisin2019PASA}
{Voisin}, F.J.; {Rowell}, G.P.; {Burton}, M.G.; {Fukui}, Y.; {Sano}, H.;
  {Aharonian}, F.; {Maxted}, N.; {Braiding}, C.; {Blackwell}, R.; {Lau}, J.
\newblock {Connecting the ISM to TeV PWNe and PWN candidates}.
\newblock {\em \pasa} {\bf 2019}, {\em 36},~e014,
  \href{http://xxx.lanl.gov/abs/1905.04517}{{\normalfont
  [arXiv:astro-ph.HE/1905.04517]}}.
\newblock
  doi:{\changeurlcolor{black}\href{https://doi.org/10.1017/pasa.2019.7}{\detokenize{10.1017/pasa.2019.7}}}.

\bibitem[{H.~E.~S.~S. Collaboration} \em{et~al.}(2023){H.~E.~S.~S.
  Collaboration}, {:}, {Aharonian}, {Ait Benkhali}, {Aschersleben}, {Ashkar},
  {Backes}, {Barbosa Martins}, {Batzofin}, {Becherini}, {Berge},
  {B{\"o}ttcher}, {Boisson}, {Bolmont}, {Borowska}, {Bouyahiaoui}, {Bradascio},
  {Breuhaus}, {Brose}, {Brun}, {Bruno}, {Bulik}, {Burger-Scheidlin}, {Bylund},
  {Caroff}, {Casanova}, {Celic}, {Cerruti}, {Chambery}, {Chand}, {Chen},
  {Chibueze}, {Chibueze}, {Damascene Mbarubucyeye}, {Djannati-Ata{\"\i}},
  {Dmytriiev}, {Einecke}, {Ernenwein}, {Feijen}, {Filipovic}, {Fontaine},
  {F{\"u}{\ss}ling}, {Funk}, {Gabici}, {Gallant}, {Ghafourizadeh}, {Giavitto},
  {Giunti}, {Glawion}, {Goswami}, {Grolleron}, {Grondin}, {Haerer}, {Hinton},
  {Hofmann}, {Holch}, {Holler}, {Horns}, {Huang}, {Jamrozy}, {Jankowsky},
  {Joshi}, {Jung-Richardt}, {Kasai}, {Katarzy{\'n}ski}, {Kh{\'e}lifi},
  {Klu{\'z}niak}, {Komin}, {Kosack}, {Kostunin}, {Lang}, {Le Stum}, {Leitl},
  {Lemi{\`e}re}, {Lemoine-Goumard}, {Lenain}, {Leuschner}, {Lohse},
  {Luashvili}, {Lypova}, {Mackey}, {Malyshev}, {Malyshev}, {Marandon},
  {Marchegiani}, {Marcowith}, {Marinos}, {Mart{\'\i}-Devesa}, {Marx},
  {Mitchell}, {Moderski}, {Mohrmann}, {Montanari}, {Moulin}, {Muller},
  {Nakashima}, {de Naurois}, {Niemiec}, {Priyana Noel}, {Ohm}, {Olivera-Nieto},
  {de Ona Wilhelmi}, {Ostrowski}, {Panny}, {Panter}, {Parsons}, {Prokhorov},
  {P{\"u}hlhofer}, {Punch}, {Quirrenbach}, {Reichherzer}, {Reimer}, {Reimer},
  {Renaud}, {Reville}, {Rieger}, {Rowell}, {Rudak}, {Sahakian}, {Santangelo},
  {Sasaki}, {Schutte}, {Schwanke}, {Shapopi}, {Sol}, {Specovius}, {Spencer},
  {Stawarz}, {Steenkamp}, {Steinmassl}, {Sushch}, {Suzuki}, {Takahashi},
  {Tanaka}, {Terrier}, {Thorpe-Morgan}, {Tsirou}, {Tsuji}, {Uchiyama}, {van
  Eldik}, {Vecchi}, {Veh}, {Venter}, {Vink}, {Wach}, {Wagner}, {White},
  {Wierzcholska}, {Wun Wong}, {Zacharias}, {Zargaryan}, {Zdziarski}, {Zech},
  {Zouari}, and {{\.Z}ywucka}]{HESS2023arXiv}
{H.~E.~S.~S. Collaboration}.; {:}.; {Aharonian}, F.; {Ait Benkhali}, F.;
  {Aschersleben}, J.; {Ashkar}, H.; {Backes}, M.; {Barbosa Martins}, V.;
  {Batzofin}, R.; {Becherini}, Y.;  et~al.
\newblock {HESS J1809$-$193: a halo of escaped electrons around a pulsar wind
  nebula?}
\newblock {\em arXiv e-prints} {\bf 2023}, p. arXiv:2302.13663,
  \href{http://xxx.lanl.gov/abs/2302.13663}{{\normalfont
  [arXiv:astro-ph.HE/2302.13663]}}.
\newblock
  doi:{\changeurlcolor{black}\href{https://doi.org/10.48550/arXiv.2302.13663}{\detokenize{10.48550/arXiv.2302.13663}}}.

\bibitem[{Crutcher}(2012)]{Crutcher2012ARAA}
{Crutcher}, R.M.
\newblock {Magnetic Fields in Molecular Clouds}.
\newblock {\em \araa} {\bf 2012}, {\em 50},~29--63.
\newblock
  doi:{\changeurlcolor{black}\href{https://doi.org/10.1146/annurev-astro-081811-125514}{\detokenize{10.1146/annurev-astro-081811-125514}}}.

\bibitem[{Rodr{\'\i}guez} and {Zapata}(2013)]{Rodriguez2013ApJ}
{Rodr{\'\i}guez}, L.F.; {Zapata}, L.A.
\newblock {Star Formation in the Massive ``Starless'' Infrared Dark Cloud
  G0.253+0.016}.
\newblock {\em \apjl} {\bf 2013}, {\em 767},~L13,
  \href{http://xxx.lanl.gov/abs/1303.2755}{{\normalfont
  [arXiv:astro-ph.SR/1303.2755]}}.
\newblock
  doi:{\changeurlcolor{black}\href{https://doi.org/10.1088/2041-8205/767/1/L13}{\detokenize{10.1088/2041-8205/767/1/L13}}}.

\bibitem[{Jones}(2014)]{Jones2014ApJ}
{Jones}, D.I.
\newblock {Prospects for Detection of Synchrotron Emission from Secondary
  Electrons and Positrons in Starless Cores: Application to G0.216+0.016}.
\newblock {\em \apjl} {\bf 2014}, {\em 792},~L14,
  \href{http://xxx.lanl.gov/abs/1408.1822}{{\normalfont
  [arXiv:astro-ph.HE/1408.1822]}}.
\newblock
  doi:{\changeurlcolor{black}\href{https://doi.org/10.1088/2041-8205/792/1/L14}{\detokenize{10.1088/2041-8205/792/1/L14}}}.

\bibitem[{Zhang} \em{et~al.}(2010){Zhang}, {Hopkins}, {Barnes}, {Cagnes},
  {Yonekura}, and {Fukui}]{Zhang2010PASA}
{Zhang}, J.; {Hopkins}, A.; {Barnes}, P.J.; {Cagnes}, M.; {Yonekura}, Y.;
  {Fukui}, Y.
\newblock {The Radio-FIR Correlation in the Milky Way}.
\newblock {\em \pasa} {\bf 2010}, {\em 27},~340--346,
  \href{http://xxx.lanl.gov/abs/1006.4282}{{\normalfont
  [arXiv:astro-ph.CO/1006.4282]}}.
\newblock
  doi:{\changeurlcolor{black}\href{https://doi.org/10.1071/AS08072}{\detokenize{10.1071/AS08072}}}.

\bibitem[{Filho} \em{et~al.}(2019){Filho}, {Tabatabaei}, {S{\'a}nchez Almeida},
  {Mu{\~n}oz-Tu{\~n}{\'o}n}, and {Elmegreen}]{Filho2019MNRAS}
{Filho}, M.E.; {Tabatabaei}, F.S.; {S{\'a}nchez Almeida}, J.;
  {Mu{\~n}oz-Tu{\~n}{\'o}n}, C.; {Elmegreen}, B.G.
\newblock {Global correlations between the radio continuum, infrared, and CO
  emissions in dwarf galaxies}.
\newblock {\em \mnras} {\bf 2019}, {\em 484},~543--561,
  \href{http://xxx.lanl.gov/abs/1811.06577}{{\normalfont
  [arXiv:astro-ph.GA/1811.06577]}}.
\newblock
  doi:{\changeurlcolor{black}\href{https://doi.org/10.1093/mnras/sty3199}{\detokenize{10.1093/mnras/sty3199}}}.

\bibitem[{Strong} \em{et~al.}(2014){Strong}, {Dickinson}, and
  {Murphy}]{Strong2014arXiv}
{Strong}, A.W.; {Dickinson}, C.; {Murphy}, E.J.
\newblock {Synchrotron radiation from molecular clouds}.
\newblock {\em arXiv e-prints} {\bf 2014}, p. arXiv:1412.4500,
  \href{http://xxx.lanl.gov/abs/1412.4500}{{\normalfont
  [arXiv:astro-ph.HE/1412.4500]}}.
\newblock
  doi:{\changeurlcolor{black}\href{https://doi.org/10.48550/arXiv.1412.4500}{\detokenize{10.48550/arXiv.1412.4500}}}.

\bibitem[{Gabici} \em{et~al.}(2009){Gabici}, {Aharonian}, and
  {Casanova}]{Gabici2009MNRAS}
{Gabici}, S.; {Aharonian}, F.A.; {Casanova}, S.
\newblock {Broad-band non-thermal emission from molecular clouds illuminated by
  cosmic rays from nearby supernova remnants}.
\newblock {\em \mnras} {\bf 2009}, {\em 396},~1629--1639,
  \href{http://xxx.lanl.gov/abs/0901.4549}{{\normalfont
  [arXiv:astro-ph.HE/0901.4549]}}.
\newblock
  doi:{\changeurlcolor{black}\href{https://doi.org/10.1111/j.1365-2966.2009.14832.x}{\detokenize{10.1111/j.1365-2966.2009.14832.x}}}.

\bibitem[{Acero} \em{et~al.}(2016){Acero}, {Ackermann}, {Ajello}, {Albert},
  {Baldini}, {Ballet}, {Barbiellini}, {Bastieri}, {Bellazzini}, {Bissaldi},
  {Bloom}, {Bonino}, {Bottacini}, {Brandt}, {Bregeon}, {Bruel}, {Buehler},
  {Buson}, {Caliandro}, {Cameron}, {Caragiulo}, {Caraveo}, {Casandjian},
  {Cavazzuti}, {Cecchi}, {Charles}, {Chekhtman}, {Chiang}, {Chiaro}, {Ciprini},
  {Claus}, {Cohen-Tanugi}, {Conrad}, {Cuoco}, {Cutini}, {D'Ammando}, {de
  Angelis}, {de Palma}, {Desiante}, {Digel}, {Di Venere}, {Drell}, {Favuzzi},
  {Fegan}, {Ferrara}, {Focke}, {Franckowiak}, {Funk}, {Fusco}, {Gargano},
  {Gasparrini}, {Giglietto}, {Giordano}, {Giroletti}, {Glanzman}, {Godfrey},
  {Grenier}, {Guiriec}, {Hadasch}, {Harding}, {Hayashi}, {Hays}, {Hewitt},
  {Hill}, {Horan}, {Hou}, {Jogler}, {J{\'o}hannesson}, {Kamae}, {Kuss},
  {Landriu}, {Larsson}, {Latronico}, {Li}, {Li}, {Longo}, {Loparco},
  {Lovellette}, {Lubrano}, {Maldera}, {Malyshev}, {Manfreda}, {Martin},
  {Mayer}, {Mazziotta}, {McEnery}, {Michelson}, {Mirabal}, {Mizuno}, {Monzani},
  {Morselli}, {Nuss}, {Ohsugi}, {Omodei}, {Orienti}, {Orlando}, {Ormes},
  {Paneque}, {Pesce-Rollins}, {Piron}, {Pivato}, {Rain{\`o}}, {Rando},
  {Razzano}, {Razzaque}, {Reimer}, {Reimer}, {Remy}, {Renault},
  {S{\'a}nchez-Conde}, {Schaal}, {Schulz}, {Sgr{\`o}}, {Siskind}, {Spada},
  {Spandre}, {Spinelli}, {Strong}, {Suson}, {Tajima}, {Takahashi}, {Thayer},
  {Thompson}, {Tibaldo}, {Tinivella}, {Torres}, {Tosti}, {Troja}, {Vianello},
  {Werner}, {Wood}, {Wood}, {Zaharijas}, and {Zimmer}]{Acero2016ApJS}
{Acero}, F.; {Ackermann}, M.; {Ajello}, M.; {Albert}, A.; {Baldini}, L.;
  {Ballet}, J.; {Barbiellini}, G.; {Bastieri}, D.; {Bellazzini}, R.;
  {Bissaldi}, E.;  et~al.
\newblock {Development of the Model of Galactic Interstellar Emission for
  Standard Point-source Analysis of Fermi Large Area Telescope Data}.
\newblock {\em \apjs} {\bf 2016}, {\em 223},~26,
  \href{http://xxx.lanl.gov/abs/1602.07246}{{\normalfont
  [arXiv:astro-ph.HE/1602.07246]}}.
\newblock
  doi:{\changeurlcolor{black}\href{https://doi.org/10.3847/0067-0049/223/2/26}{\detokenize{10.3847/0067-0049/223/2/26}}}.

\bibitem[{Guo} and {Yuan}(2018)]{Guo2018PhRvD}
{Guo}, Y.Q.; {Yuan}, Q.
\newblock {Understanding the spectral hardenings and radial distribution of
  Galactic cosmic rays and Fermi diffuse {\ensuremath{\gamma}} rays with
  spatially-dependent propagation}.
\newblock {\em \prd} {\bf 2018}, {\em 97},~063008,
  \href{http://xxx.lanl.gov/abs/1801.05904}{{\normalfont
  [arXiv:astro-ph.HE/1801.05904]}}.
\newblock
  doi:{\changeurlcolor{black}\href{https://doi.org/10.1103/PhysRevD.97.063008}{\detokenize{10.1103/PhysRevD.97.063008}}}.

\bibitem[{Aharonian} \em{et~al.}(2020){Aharonian}, {Peron}, {Yang}, {Casanova},
  and {Zanin}]{Aharonian2020PhRvD}
{Aharonian}, F.; {Peron}, G.; {Yang}, R.; {Casanova}, S.; {Zanin}, R.
\newblock {Probing the sea of galactic cosmic rays with Fermi-LAT}.
\newblock {\em \prd} {\bf 2020}, {\em 101},~083018,
  \href{http://xxx.lanl.gov/abs/1811.12118}{{\normalfont
  [arXiv:astro-ph.HE/1811.12118]}}.
\newblock
  doi:{\changeurlcolor{black}\href{https://doi.org/10.1103/PhysRevD.101.083018}{\detokenize{10.1103/PhysRevD.101.083018}}}.

\bibitem[{Aharonian}(2001)]{Aharonian2001SSRv}
{Aharonian}, F.A.
\newblock {Gamma Rays From Molecular Clouds}.
\newblock {\em \ssr} {\bf 2001}, {\em 99},~187--196,
  \href{http://xxx.lanl.gov/abs/astro-ph/0012290}{{\normalfont
  [arXiv:astro-ph/astro-ph/0012290]}}.
\newblock
  doi:{\changeurlcolor{black}\href{https://doi.org/10.1023/A:1013845015364}{\detokenize{10.1023/A:1013845015364}}}.

\bibitem[{Abrahams} \em{et~al.}(2017){Abrahams}, {Teachey}, and
  {Paglione}]{Abrahams2017ApJ}
{Abrahams}, R.D.; {Teachey}, A.; {Paglione}, T.A.D.
\newblock {Calibrating Column Density Tracers with Gamma-Ray Observations of
  the {\ensuremath{\rho}} Ophiuchi Molecular Cloud}.
\newblock {\em \apj} {\bf 2017}, {\em 834},~91,
  \href{http://xxx.lanl.gov/abs/1611.02265}{{\normalfont
  [arXiv:astro-ph.HE/1611.02265]}}.
\newblock
  doi:{\changeurlcolor{black}\href{https://doi.org/10.3847/1538-4357/834/1/91}{\detokenize{10.3847/1538-4357/834/1/91}}}.

\bibitem[{Peron} and {Aharonian}(2022)]{Peron2022A&A}
{Peron}, G.; {Aharonian}, F.
\newblock {Probing the galactic cosmic-ray density with current and future
  {\ensuremath{\gamma}}-ray instruments}.
\newblock {\em \aap} {\bf 2022}, {\em 659},~A57,
  \href{http://xxx.lanl.gov/abs/2110.08778}{{\normalfont
  [arXiv:astro-ph.HE/2110.08778]}}.
\newblock
  doi:{\changeurlcolor{black}\href{https://doi.org/10.1051/0004-6361/202142416}{\detokenize{10.1051/0004-6361/202142416}}}.

\bibitem[{Neufeld} and {Wolfire}(2017)]{Neufeld2017ApJ}
{Neufeld}, D.A.; {Wolfire}, M.G.
\newblock {The Cosmic-Ray Ionization Rate in the Galactic Disk, as Determined
  from Observations of Molecular Ions}.
\newblock {\em \apj} {\bf 2017}, {\em 845},~163,
  \href{http://xxx.lanl.gov/abs/1704.03877}{{\normalfont
  [arXiv:astro-ph.GA/1704.03877]}}.
\newblock
  doi:{\changeurlcolor{black}\href{https://doi.org/10.3847/1538-4357/aa6d68}{\detokenize{10.3847/1538-4357/aa6d68}}}.

\bibitem[{Phan} \em{et~al.}(2023){Phan}, {Recchia}, {Mertsch}, and
  {Gabici}]{Phan2022arXiv}
{Phan}, V.H.M.; {Recchia}, S.; {Mertsch}, P.; {Gabici}, S.
\newblock {Stochasticity of cosmic rays from supernova remnants and the
  ionization rates in molecular clouds}.
\newblock {\em \prd} {\bf 2023}, {\em 107},~123006,
  \href{http://xxx.lanl.gov/abs/2209.10581}{{\normalfont
  [arXiv:astro-ph.HE/2209.10581]}}.
\newblock
  doi:{\changeurlcolor{black}\href{https://doi.org/10.1103/PhysRevD.107.123006}{\detokenize{10.1103/PhysRevD.107.123006}}}.

\bibitem[{Phan} \em{et~al.}(2021){Phan}, {Schulze}, {Mertsch}, {Recchia}, and
  {Gabici}]{Phan2021PhRvL}
{Phan}, V.H.M.; {Schulze}, F.; {Mertsch}, P.; {Recchia}, S.; {Gabici}, S.
\newblock {Stochastic Fluctuations of Low-Energy Cosmic Rays and the
  Interpretation of Voyager Data}.
\newblock {\em \prl} {\bf 2021}, {\em 127},~141101,
  \href{http://xxx.lanl.gov/abs/2105.00311}{{\normalfont
  [arXiv:astro-ph.HE/2105.00311]}}.
\newblock
  doi:{\changeurlcolor{black}\href{https://doi.org/10.1103/PhysRevLett.127.141101}{\detokenize{10.1103/PhysRevLett.127.141101}}}.

\bibitem[{Huang} \em{et~al.}(2021){Huang}, {Yuan}, and {Fan}]{Huang2021NatCo}
{Huang}, X.; {Yuan}, Q.; {Fan}, Y.Z.
\newblock {A GeV-TeV particle component and the barrier of cosmic-ray sea in
  the Central Molecular Zone}.
\newblock {\em Nature Communications} {\bf 2021}, {\em 12},~6169,
  \href{http://xxx.lanl.gov/abs/2012.05524}{{\normalfont
  [arXiv:astro-ph.HE/2012.05524]}}.
\newblock
  doi:{\changeurlcolor{black}\href{https://doi.org/10.1038/s41467-021-26436-z}{\detokenize{10.1038/s41467-021-26436-z}}}.

\bibitem[{Chernyshov} \em{et~al.}(2021){Chernyshov}, {Egorov}, {Dogiel}, and
  {Ivlev}]{Chernyshov2021Symm}
{Chernyshov}, D.O.; {Egorov}, A.E.; {Dogiel}, V.A.; {Ivlev}, A.V.
\newblock {On a Possible Origin of the Gamma-ray Excess around the Galactic
  Center}.
\newblock {\em Symmetry} {\bf 2021}, {\em 13},~1432,
  \href{http://xxx.lanl.gov/abs/2108.04104}{{\normalfont
  [arXiv:astro-ph.HE/2108.04104]}}.
\newblock
  doi:{\changeurlcolor{black}\href{https://doi.org/10.3390/sym13081432}{\detokenize{10.3390/sym13081432}}}.

\bibitem[{Peron} \em{et~al.}(2021){Peron}, {Aharonian}, {Casanova}, {Yang}, and
  {Zanin}]{Peron2021ApJ}
{Peron}, G.; {Aharonian}, F.; {Casanova}, S.; {Yang}, R.; {Zanin}, R.
\newblock {Probing the Cosmic-Ray Density in the Inner Galaxy}.
\newblock {\em \apjl} {\bf 2021}, {\em 907},~L11,
  \href{http://xxx.lanl.gov/abs/2101.09510}{{\normalfont
  [arXiv:astro-ph.HE/2101.09510]}}.
\newblock
  doi:{\changeurlcolor{black}\href{https://doi.org/10.3847/2041-8213/abcaa9}{\detokenize{10.3847/2041-8213/abcaa9}}}.

\bibitem[{Rogers} \em{et~al.}(2022){Rogers}, {Zhang}, {Perez}, {Clavel}, and
  {Taylor}]{Rogers2022ApJ}
{Rogers}, F.; {Zhang}, S.; {Perez}, K.; {Clavel}, M.; {Taylor}, A.
\newblock {New Constraints on Cosmic Particle Populations at the Galactic
  Center Using X-Ray Observations of the Molecular Cloud Sagittarius B2}.
\newblock {\em \apj} {\bf 2022}, {\em 934},~19.
\newblock
  doi:{\changeurlcolor{black}\href{https://doi.org/10.3847/1538-4357/ac7717}{\detokenize{10.3847/1538-4357/ac7717}}}.

\bibitem[{Yusef-Zadeh} \em{et~al.}(2002){Yusef-Zadeh}, {Law}, {Wardle}, {Wang},
  {Fruscione}, {Lang}, and {Cotera}]{YusefZadeh2002ApJ}
{Yusef-Zadeh}, F.; {Law}, C.; {Wardle}, M.; {Wang}, Q.D.; {Fruscione}, A.;
  {Lang}, C.C.; {Cotera}, A.
\newblock {Detection of X-Ray Emission from the Arches Cluster near the
  Galactic Center}.
\newblock {\em \apj} {\bf 2002}, {\em 570},~665--670,
  \href{http://xxx.lanl.gov/abs/astro-ph/0108174}{{\normalfont
  [arXiv:astro-ph/astro-ph/0108174]}}.
\newblock
  doi:{\changeurlcolor{black}\href{https://doi.org/10.1086/340058}{\detokenize{10.1086/340058}}}.

\bibitem[{Wang} \em{et~al.}(2006){Wang}, {Dong}, and {Lang}]{Wang2006MNRAS}
{Wang}, Q.D.; {Dong}, H.; {Lang}, C.
\newblock {The interplay between star formation and the nuclear environment of
  our Galaxy: deep X-ray observations of the Galactic centre Arches and
  Quintuplet clusters}.
\newblock {\em \mnras} {\bf 2006}, {\em 371},~38--54,
  \href{http://xxx.lanl.gov/abs/astro-ph/0606282}{{\normalfont
  [arXiv:astro-ph/astro-ph/0606282]}}.
\newblock
  doi:{\changeurlcolor{black}\href{https://doi.org/10.1111/j.1365-2966.2006.10656.x}{\detokenize{10.1111/j.1365-2966.2006.10656.x}}}.

\bibitem[{Kuznetsova} \em{et~al.}(2019){Kuznetsova}, {Krivonos}, {Clavel},
  {Lutovinov}, {Chernyshov}, {Hong}, {Mori}, {Ponti}, {Tomsick}, and
  {Zhang}]{Kuznetsova2019MNRAS}
{Kuznetsova}, E.; {Krivonos}, R.; {Clavel}, M.; {Lutovinov}, A.; {Chernyshov},
  D.; {Hong}, J.; {Mori}, K.; {Ponti}, G.; {Tomsick}, J.; {Zhang}, S.
\newblock {Investigating the origin of the faint non-thermal emission of the
  Arches cluster using the 2015-2016 NuSTAR and XMM-Newton X-ray observations}.
\newblock {\em \mnras} {\bf 2019}, {\em 484},~1627--1636,
  \href{http://xxx.lanl.gov/abs/1901.03121}{{\normalfont
  [arXiv:astro-ph.HE/1901.03121]}}.
\newblock
  doi:{\changeurlcolor{black}\href{https://doi.org/10.1093/mnras/stz119}{\detokenize{10.1093/mnras/stz119}}}.

\bibitem[{Clavel} \em{et~al.}(2014){Clavel}, {Soldi}, {Terrier}, {Tatischeff},
  {Maurin}, {Ponti}, {Goldwurm}, and {Decourchelle}]{Clavel2014MNRAS}
{Clavel}, M.; {Soldi}, S.; {Terrier}, R.; {Tatischeff}, V.; {Maurin}, G.;
  {Ponti}, G.; {Goldwurm}, A.; {Decourchelle}, A.
\newblock {Variation of the X-ray non-thermal emission in the Arches cloud.}
\newblock {\em \mnras} {\bf 2014}, {\em 443},~L129--L133,
  \href{http://xxx.lanl.gov/abs/1406.5727}{{\normalfont
  [arXiv:astro-ph.HE/1406.5727]}}.
\newblock
  doi:{\changeurlcolor{black}\href{https://doi.org/10.1093/mnrasl/slu100}{\detokenize{10.1093/mnrasl/slu100}}}.

\bibitem[{Krivonos} \em{et~al.}(2014){Krivonos}, {Tomsick}, {Bauer},
  {Baganoff}, {Barriere}, {Bodaghee}, {Boggs}, {Christensen}, {Craig},
  {Grefenstette}, {Hailey}, {Harrison}, {Hong}, {Madsen}, {Mori}, {Nynka},
  {Stern}, and {Zhang}]{Krivonos2014ApJ}
{Krivonos}, R.A.; {Tomsick}, J.A.; {Bauer}, F.E.; {Baganoff}, F.K.; {Barriere},
  N.M.; {Bodaghee}, A.; {Boggs}, S.E.; {Christensen}, F.E.; {Craig}, W.W.;
  {Grefenstette}, B.W.;  et~al.
\newblock {First Hard X-Ray Detection of the Non-thermal Emission around the
  Arches Cluster: Morphology and Spectral Studies with NuSTAR}.
\newblock {\em \apj} {\bf 2014}, {\em 781},~107,
  \href{http://xxx.lanl.gov/abs/1312.2635}{{\normalfont
  [arXiv:astro-ph.GA/1312.2635]}}.
\newblock
  doi:{\changeurlcolor{black}\href{https://doi.org/10.1088/0004-637X/781/2/107}{\detokenize{10.1088/0004-637X/781/2/107}}}.

\bibitem[{Chernyshov} \em{et~al.}(2018){Chernyshov}, {Ko}, {Krivonos},
  {Dogiel}, and {Cheng}]{Chernyshov2018ApJ}
{Chernyshov}, D.O.; {Ko}, C.M.; {Krivonos}, R.A.; {Dogiel}, V.A.; {Cheng}, K.S.
\newblock {Time Variability of Equivalent Width of 6.4 keV Line from the Arches
  Complex: Reflected X-Rays or Charged Particles?}
\newblock {\em \apj} {\bf 2018}, {\em 863},~85,
  \href{http://xxx.lanl.gov/abs/1807.00526}{{\normalfont
  [arXiv:astro-ph.HE/1807.00526]}}.
\newblock
  doi:{\changeurlcolor{black}\href{https://doi.org/10.3847/1538-4357/aad091}{\detokenize{10.3847/1538-4357/aad091}}}.

\bibitem[{Nobukawa} \em{et~al.}(2019){Nobukawa}, {Saji}, {Hirayama},
  {Nobukawa}, {Yamauchi}, {Matsumoto}, and
  {Koyama}]{Nobukawa2019JPhCS1181a2040N}
{Nobukawa}, K.K.; {Saji}, S.; {Hirayama}, A.; {Nobukawa}, M.; {Yamauchi}, S.;
  {Matsumoto}, H.; {Koyama}, K.
\newblock {Measurement of Low-Energy Cosmic Rays via the Neutral Iron Line}.
\newblock  Journal of Physics Conference Series,  2019, Vol. 1181, {\em Journal
  of Physics Conference Series}, p. 012040.
\newblock
  doi:{\changeurlcolor{black}\href{https://doi.org/10.1088/1742-6596/1181/1/012040}{\detokenize{10.1088/1742-6596/1181/1/012040}}}.

\bibitem[{Bergin} and {Tafalla}(2007)]{Bergin2007ARAA}
{Bergin}, E.A.; {Tafalla}, M.
\newblock {Cold Dark Clouds: The Initial Conditions for Star Formation}.
\newblock {\em \araa} {\bf 2007}, {\em 45},~339--396,
  \href{http://xxx.lanl.gov/abs/0705.3765}{{\normalfont
  [arXiv:astro-ph/0705.3765]}}.
\newblock
  doi:{\changeurlcolor{black}\href{https://doi.org/10.1146/annurev.astro.45.071206.100404}{\detokenize{10.1146/annurev.astro.45.071206.100404}}}.

\bibitem[{Rodr{\'\i}guez}(2005)]{Rodriguez2005ASPC}
{Rodr{\'\i}guez}, L.F.R.
\newblock {Molecular Clouds: Fragmentation, Modeling and Observations}.
\newblock  The Cool Universe: Observing Cosmic Dawn; {Lidman}, C.; {Alloin},
  D., Eds.,  2005, Vol. 344, {\em Astronomical Society of the Pacific
  Conference Series}, p. 146.

\bibitem[{Myers}(1995)]{Myers1995mcsf}
{Myers}, P.C.
\newblock {Star Forming Molecular Clouds}.
\newblock  Molecular Clouds and Star Formation; {Chi}, Y.; {You}, J., Eds.,
  1995, p.~47.

\bibitem[{Spitzer} and {Tomasko}(1968)]{Spitzer1968ApJ}
{Spitzer}, Lyman, J.; {Tomasko}, M.G.
\newblock {Heating of H i Regions by Energetic Particles}.
\newblock {\em \apj} {\bf 1968}, {\em 152},~971.
\newblock
  doi:{\changeurlcolor{black}\href{https://doi.org/10.1086/149610}{\detokenize{10.1086/149610}}}.

\bibitem[{Goldsmith}(2001)]{Goldsmith2001ApJ}
{Goldsmith}, P.F.
\newblock {Molecular Depletion and Thermal Balance in Dark Cloud Cores}.
\newblock {\em \apj} {\bf 2001}, {\em 557},~736--746.
\newblock
  doi:{\changeurlcolor{black}\href{https://doi.org/10.1086/322255}{\detokenize{10.1086/322255}}}.

\bibitem[{Consolandi}(2016)]{2016arXiv161208562C}
{Consolandi}, C.
\newblock {Precision Measurement of the Proton Flux in Primary Cosmic Rays from
  1 GV to 1.8 TV with the Alpha Magnetic Spectrometer on the International
  Space Station}.
\newblock {\em arXiv e-prints} {\bf 2016}, p. arXiv:1612.08562,
  \href{http://xxx.lanl.gov/abs/1612.08562}{{\normalfont
  [arXiv:astro-ph.HE/1612.08562]}}.
\newblock
  doi:{\changeurlcolor{black}\href{https://doi.org/10.48550/arXiv.1612.08562}{\detokenize{10.48550/arXiv.1612.08562}}}.

\bibitem[{Crutcher} \em{et~al.}(2010){Crutcher}, {Wandelt}, {Heiles},
  {Falgarone}, and {Troland}]{Crutcher2010ApJ}
{Crutcher}, R.M.; {Wandelt}, B.; {Heiles}, C.; {Falgarone}, E.; {Troland}, T.H.
\newblock {Magnetic Fields in Interstellar Clouds from Zeeman Observations:
  Inference of Total Field Strengths by Bayesian Analysis}.
\newblock {\em \apj} {\bf 2010}, {\em 725},~466--479.
\newblock
  doi:{\changeurlcolor{black}\href{https://doi.org/10.1088/0004-637X/725/1/466}{\detokenize{10.1088/0004-637X/725/1/466}}}.

\bibitem[{Elmegreen}(1979)]{Elmegreen1979ApJ}
{Elmegreen}, B.G.
\newblock {Magnetic diffusion and ionization fractions in dense molecular
  clouds: the role of charged grains.}
\newblock {\em \apj} {\bf 1979}, {\em 232},~729--739.
\newblock
  doi:{\changeurlcolor{black}\href{https://doi.org/10.1086/157333}{\detokenize{10.1086/157333}}}.

\bibitem[{Bisbas} \em{et~al.}(2017){Bisbas}, {van Dishoeck}, {Papadopoulos},
  {Sz{\H{u}}cs}, {Bialy}, and {Zhang}]{Bisbas2017ApJ}
{Bisbas}, T.G.; {van Dishoeck}, E.F.; {Papadopoulos}, P.P.; {Sz{\H{u}}cs}, L.;
  {Bialy}, S.; {Zhang}, Z.Y.
\newblock {Cosmic-ray Induced Destruction of CO in Star-forming Galaxies}.
\newblock {\em \apj} {\bf 2017}, {\em 839},~90,
  \href{http://xxx.lanl.gov/abs/1703.08598}{{\normalfont
  [arXiv:astro-ph.GA/1703.08598]}}.
\newblock
  doi:{\changeurlcolor{black}\href{https://doi.org/10.3847/1538-4357/aa696d}{\detokenize{10.3847/1538-4357/aa696d}}}.

\bibitem[{Gaches} and {Offner}(2018)]{Gaches2018ApJ}
{Gaches}, B.A.L.; {Offner}, S.S.R.
\newblock {Exploration of Cosmic-ray Acceleration in Protostellar Accretion
  Shocks and a Model for Ionization Rates in Embedded Protoclusters}.
\newblock {\em \apj} {\bf 2018}, {\em 861},~87,
  \href{http://xxx.lanl.gov/abs/1805.03215}{{\normalfont
  [arXiv:astro-ph.GA/1805.03215]}}.
\newblock
  doi:{\changeurlcolor{black}\href{https://doi.org/10.3847/1538-4357/aac94d}{\detokenize{10.3847/1538-4357/aac94d}}}.

\bibitem[{Gong} \em{et~al.}(2017){Gong}, {Ostriker}, and
  {Wolfire}]{Gong2017ApJ}
{Gong}, M.; {Ostriker}, E.C.; {Wolfire}, M.G.
\newblock {A Simple and Accurate Network for Hydrogen and Carbon Chemistry in
  the Interstellar Medium}.
\newblock {\em \apj} {\bf 2017}, {\em 843},~38,
  \href{http://xxx.lanl.gov/abs/1610.09023}{{\normalfont
  [arXiv:astro-ph.GA/1610.09023]}}.
\newblock
  doi:{\changeurlcolor{black}\href{https://doi.org/10.3847/1538-4357/aa7561}{\detokenize{10.3847/1538-4357/aa7561}}}.

\bibitem[{Pazianotto} \em{et~al.}(2021){Pazianotto}, {Pilling}, {Quesada
  Molina}, and {Federico}]{Pazianotto2021ApJ}
{Pazianotto}, M.T.; {Pilling}, S.; {Quesada Molina}, J.M.; {Federico}, C.A.
\newblock {Energy Deposition by Cosmic Rays in the Molecular Cloud Using GEANT4
  Code and Voyager I Data}.
\newblock {\em \apj} {\bf 2021}, {\em 911},~129.
\newblock
  doi:{\changeurlcolor{black}\href{https://doi.org/10.3847/1538-4357/abe7f3}{\detokenize{10.3847/1538-4357/abe7f3}}}.

\bibitem[{Pazianotto} and {Pilling}(2023)]{Pazianotto2023MNRAS}
{Pazianotto}, M.T.; {Pilling}, S.
\newblock {Computational simulation of the bombardment of molecular clump by
  realistic cosmic ray field employing GEANT4 code}.
\newblock {\em \mnras} {\bf 2023}, {\em 518},~1735--1743.
\newblock
  doi:{\changeurlcolor{black}\href{https://doi.org/10.1093/mnras/stac3063}{\detokenize{10.1093/mnras/stac3063}}}.

\bibitem[{Pilling} \em{et~al.}(2022){Pilling}, {Pazianotto}, {de Souza}, and
  {Maciel do Nascimento}]{Pilling2022MNRAS}
{Pilling}, S.; {Pazianotto}, M.T.; {de Souza}, L.A.; {Maciel do Nascimento}, L.
\newblock {Realistic energy deposition and temperature heating in molecular
  clouds due to cosmic rays: a computation simulation with the GEANT4 code
  employing light particles and medium-mass and heavy ions}.
\newblock {\em \mnras} {\bf 2022}, {\em 509},~6169--6178.
\newblock
  doi:{\changeurlcolor{black}\href{https://doi.org/10.1093/mnras/stab3470}{\detokenize{10.1093/mnras/stab3470}}}.

\bibitem[{Aharonian}(1991)]{Aharonian1991Ap&SS}
{Aharonian}, F.A.
\newblock {Vary High and Ultra High Energy Gamma-Rays from Giant Molecular
  Clouds}.
\newblock {\em \apss} {\bf 1991}, {\em 180},~305--320.
\newblock
  doi:{\changeurlcolor{black}\href{https://doi.org/10.1007/BF00648185}{\detokenize{10.1007/BF00648185}}}.

\bibitem[{Fujita} \em{et~al.}(2021){Fujita}, {Nobukawa}, and
  {Sano}]{Fujita2021ApJ}
{Fujita}, Y.; {Nobukawa}, K.K.; {Sano}, H.
\newblock {Intrusion of MeV-TeV Cosmic Rays into Molecular Clouds Studied by
  Ionization, the Neutral Iron Line, and Gamma Rays}.
\newblock {\em \apj} {\bf 2021}, {\em 908},~136,
  \href{http://xxx.lanl.gov/abs/2009.13524}{{\normalfont
  [arXiv:astro-ph.HE/2009.13524]}}.
\newblock
  doi:{\changeurlcolor{black}\href{https://doi.org/10.3847/1538-4357/abce62}{\detokenize{10.3847/1538-4357/abce62}}}.

\bibitem[{Nobukawa} \em{et~al.}(2018){Nobukawa}, {Nobukawa}, {Koyama},
  {Yamauchi}, {Uchiyama}, {Okon}, {Tanaka}, {Uchida}, and
  {Tsuru}]{Nobukawa2018ApJ}
{Nobukawa}, K.K.; {Nobukawa}, M.; {Koyama}, K.; {Yamauchi}, S.; {Uchiyama}, H.;
  {Okon}, H.; {Tanaka}, T.; {Uchida}, H.; {Tsuru}, T.G.
\newblock {Evidence for a Neutral Iron Line Generated by MeV Protons from
  Supernova Remnants Interacting with Molecular Clouds}.
\newblock {\em \apj} {\bf 2018}, {\em 854},~87,
  \href{http://xxx.lanl.gov/abs/1801.07881}{{\normalfont
  [arXiv:astro-ph.HE/1801.07881]}}.
\newblock
  doi:{\changeurlcolor{black}\href{https://doi.org/10.3847/1538-4357/aaa8dc}{\detokenize{10.3847/1538-4357/aaa8dc}}}.

\bibitem[{Maxted} \em{et~al.}(2018){Maxted}, {Braiding}, {Wong}, {Rowell},
  {Burton}, {Filipovi{\'c}}, {Voisin}, {Uro{\v{s}}evi{\'c}}, {Vukoti{\'c}},
  {Pavlovi{\'c}}, {Sano}, and {Fukui}]{Maxted2018MNRAS}
{Maxted}, N.I.; {Braiding}, C.; {Wong}, G.F.; {Rowell}, G.P.; {Burton}, M.G.;
  {Filipovi{\'c}}, M.D.; {Voisin}, F.; {Uro{\v{s}}evi{\'c}}, D.; {Vukoti{\'c}},
  B.; {Pavlovi{\'c}}, M.Z.;  et~al.
\newblock {Searching for an interstellar medium association for HESS J1534 -
  571}.
\newblock {\em \mnras} {\bf 2018}, {\em 480},~134--148,
  \href{http://xxx.lanl.gov/abs/1807.01300}{{\normalfont
  [arXiv:astro-ph.HE/1807.01300]}}.
\newblock
  doi:{\changeurlcolor{black}\href{https://doi.org/10.1093/mnras/sty1797}{\detokenize{10.1093/mnras/sty1797}}}.

\bibitem[{Okon} \em{et~al.}(2018){Okon}, {Uchida}, {Tanaka}, {Matsumura}, and
  {Tsuru}]{Okon2018PASJ}
{Okon}, H.; {Uchida}, H.; {Tanaka}, T.; {Matsumura}, H.; {Tsuru}, T.G.
\newblock {The origin of recombining plasma and the detection of the Fe-K line
  in the supernova remnant W 28}.
\newblock {\em \pasj} {\bf 2018}, {\em 70},~35,
  \href{http://xxx.lanl.gov/abs/1802.02814}{{\normalfont
  [arXiv:astro-ph.HE/1802.02814]}}.
\newblock
  doi:{\changeurlcolor{black}\href{https://doi.org/10.1093/pasj/psy022}{\detokenize{10.1093/pasj/psy022}}}.

\bibitem[{Saji} \em{et~al.}(2018){Saji}, {Matsumoto}, {Nobukawa}, {Nobukawa},
  {Uchiyama}, {Yamauchi}, and {Koyama}]{Saji2018PASJ}
{Saji}, S.; {Matsumoto}, H.; {Nobukawa}, M.; {Nobukawa}, K.K.; {Uchiyama}, H.;
  {Yamauchi}, S.; {Koyama}, K.
\newblock {Discovery of 6.4 keV line and soft X-ray emissions from G323.7-1.0
  with Suzaku}.
\newblock {\em \pasj} {\bf 2018}, {\em 70},~23,
  \href{http://xxx.lanl.gov/abs/1712.09024}{{\normalfont
  [arXiv:astro-ph.HE/1712.09024]}}.
\newblock
  doi:{\changeurlcolor{black}\href{https://doi.org/10.1093/pasj/psx158}{\detokenize{10.1093/pasj/psx158}}}.

\bibitem[{Makino} \em{et~al.}(2019){Makino}, {Fujita}, {Nobukawa}, {Matsumoto},
  and {Ohira}]{Makino2019PASJ}
{Makino}, K.; {Fujita}, Y.; {Nobukawa}, K.K.; {Matsumoto}, H.; {Ohira}, Y.
\newblock {Interaction between molecular clouds and MeV-TeV cosmic-ray protons
  escaped from supernova remnants}.
\newblock {\em \pasj} {\bf 2019}, {\em 71},~78,
  \href{http://xxx.lanl.gov/abs/1901.10477}{{\normalfont
  [arXiv:astro-ph.HE/1901.10477]}}.
\newblock
  doi:{\changeurlcolor{black}\href{https://doi.org/10.1093/pasj/psz058}{\detokenize{10.1093/pasj/psz058}}}.

\bibitem[{Nobukawa} \em{et~al.}(2019){Nobukawa}, {Hirayama}, {Shimaguchi},
  {Fujita}, {Nobukawa}, and {Yamauchi}]{Nobukawa2019PASJ}
{Nobukawa}, K.K.; {Hirayama}, A.; {Shimaguchi}, A.; {Fujita}, Y.; {Nobukawa},
  M.; {Yamauchi}, S.
\newblock {Neutral iron line in the supernova remnant IC 443 and implications
  for MeV cosmic rays}.
\newblock {\em \pasj} {\bf 2019}, {\em 71},~115,
  \href{http://xxx.lanl.gov/abs/1908.05119}{{\normalfont
  [arXiv:astro-ph.HE/1908.05119]}}.
\newblock
  doi:{\changeurlcolor{black}\href{https://doi.org/10.1093/pasj/psz099}{\detokenize{10.1093/pasj/psz099}}}.

\bibitem[{Shimaguchi} \em{et~al.}(2022){Shimaguchi}, {Nobukawa}, {Yamauchi},
  {Nobukawa}, and {Fujita}]{Shimaguchi2022PASJ}
{Shimaguchi}, A.; {Nobukawa}, K.K.; {Yamauchi}, S.; {Nobukawa}, M.; {Fujita},
  Y.
\newblock {Suzaku observations of Fe K-shell lines in the supernova remnant W
  51 C and hard X-ray sources in the proximity}.
\newblock {\em \pasj} {\bf 2022}, {\em 74},~656--663,
  \href{http://xxx.lanl.gov/abs/2203.03136}{{\normalfont
  [arXiv:astro-ph.HE/2203.03136]}}.
\newblock
  doi:{\changeurlcolor{black}\href{https://doi.org/10.1093/pasj/psac026}{\detokenize{10.1093/pasj/psac026}}}.

\bibitem[{Mitchell} \em{et~al.}(2021){Mitchell}, {Rowell}, {Celli}, and
  {Einecke}]{Mitchell2021MNRAS}
{Mitchell}, A.M.W.; {Rowell}, G.P.; {Celli}, S.; {Einecke}, S.
\newblock {Using interstellar clouds to search for Galactic PeVatrons:
  gamma-ray signatures from supernova remnants}.
\newblock {\em \mnras} {\bf 2021}, {\em 503},~3522--3539,
  \href{http://xxx.lanl.gov/abs/2103.01787}{{\normalfont
  [arXiv:astro-ph.HE/2103.01787]}}.
\newblock
  doi:{\changeurlcolor{black}\href{https://doi.org/10.1093/mnras/stab667}{\detokenize{10.1093/mnras/stab667}}}.

\bibitem[{Hewitt} \em{et~al.}(2009){Hewitt}, {Yusef-Zadeh}, and
  {Wardle}]{Hewitt2009ApJ}
{Hewitt}, J.W.; {Yusef-Zadeh}, F.; {Wardle}, M.
\newblock {Correlation of Supernova Remnant Masers and Gamma-Ray Sources}.
\newblock {\em \apjl} {\bf 2009}, {\em 706},~L270--L274,
  \href{http://xxx.lanl.gov/abs/0909.2827}{{\normalfont
  [arXiv:astro-ph.HE/0909.2827]}}.
\newblock
  doi:{\changeurlcolor{black}\href{https://doi.org/10.1088/0004-637X/706/2/L270}{\detokenize{10.1088/0004-637X/706/2/L270}}}.

\bibitem[{Voisin} \em{et~al.}(2016){Voisin}, {Rowell}, {Burton}, {Walsh},
  {Fukui}, and {Aharonian}]{Voisin2016MNRAS}
{Voisin}, F.; {Rowell}, G.; {Burton}, M.G.; {Walsh}, A.; {Fukui}, Y.;
  {Aharonian}, F.
\newblock {ISM gas studies towards the TeV PWN HESS J1825-137 and northern
  region}.
\newblock {\em \mnras} {\bf 2016}, {\em 458},~2813--2835,
  \href{http://xxx.lanl.gov/abs/1604.00090}{{\normalfont
  [arXiv:astro-ph.HE/1604.00090]}}.
\newblock
  doi:{\changeurlcolor{black}\href{https://doi.org/10.1093/mnras/stw473}{\detokenize{10.1093/mnras/stw473}}}.

\bibitem[{H.\,E.\,S.\,S. Collaboration} \em{et~al.}(2018){H.\,E.\,S.\,S.
  Collaboration}, {Abdalla}, {Abramowski}, {Aharonian}, {Ait Benkhali},
  {Ang{\"u}ner}, {Arakawa}, {Arrieta}, {Aubert}, {Backes}, {Balzer}, {Barnard},
  {Becherini}, {Becker Tjus}, {Berge}, {Bernhard}, {Bernl{\"o}hr}, {Blackwell},
  {B{\"o}ttcher}, {Boisson}, {Bolmont}, {Bonnefoy}, {Bordas}, {Bregeon},
  {Brun}, {Brun}, {Bryan}, {B{\"u}chele}, {Bulik}, {Capasso}, {Carrigan},
  {Caroff}, {Carosi}, {Casanova}, {Cerruti}, {Chakraborty}, {Chaves}, {Chen},
  {Chevalier}, {Colafrancesco}, {Condon}, {Conrad}, {Davids}, {Decock}, {Deil},
  {Devin}, {deWilt}, {Dirson}, {Djannati-Ata{\"\i}}, {Domainko}, {Donath},
  {Drury}, {Dutson}, {Dyks}, {Edwards}, {Egberts}, {Eger}, {Emery},
  {Ernenwein}, {Eschbach}, {Farnier}, {Fegan}, {Fernandes}, {Fiasson},
  {Fontaine}, {F{\"o}rster}, {Funk}, {F{\"u}{\ss}ling}, {Gabici}, {Gallant},
  {Garrigoux}, {Gast}, {Gat{\'e}}, {Giavitto}, {Giebels}, {Glawion},
  {Glicenstein}, {Gottschall}, {Grondin}, {Hahn}, {Haupt}, {Hawkes},
  {Heinzelmann}, {Henri}, {Hermann}, {Hinton}, {Hofmann}, {Hoischen}, {Holch},
  {Holler}, {Horns}, {Ivascenko}, {Iwasaki}, {Jacholkowska}, {Jamrozy},
  {Jankowsky}, {Jankowsky}, {Jingo}, {Jouvin}, {Jung-Richardt}, {Kastendieck},
  {Katarzy{\'n}ski}, {Katsuragawa}, {Katz}, {Kerszberg}, {Khangulyan},
  {Kh{\'e}lifi}, {King}, {Klepser}, {Klochkov}, {Klu{\'z}niak}, {Komin},
  {Kosack}, {Krakau}, {Kraus}, {Kr{\"u}ger}, {Laffon}, {Lamanna}, {Lau},
  {Lees}, {Lefaucheur}, {Lemi{\`e}re}, {Lemoine-Goumard}, {Lenain}, {Leser},
  {Lohse}, {Lorentz}, {Liu}, {L{\'o}pez-Coto}, {Lypova}, {Marandon},
  {Malyshev}, {Marcowith}, {Mariaud}, {Marx}, {Maurin}, {Maxted}, {Mayer},
  {Meintjes}, {Meyer}, {Mitchell}, {Moderski}, {Mohamed}, {Mohrmann},
  {Mor{\r{a}}}, {Moulin}, {Murach}, {Nakashima}, {de Naurois}, {Ndiyavala},
  {Niederwanger}, {Niemiec}, {Oakes}, {O'Brien}, {Odaka}, {Ohm}, {Ostrowski},
  {Oya}, {Padovani}, {Panter}, {Parsons}, {Paz Arribas}, {Pekeur}, {Pelletier},
  {Perennes}, {Petrucci}, {Peyaud}, {Piel}, {Pita}, {Poireau}, {Poon},
  {Prokhorov}, {Prokoph}, {P{\"u}hlhofer}, {Punch}, {Quirrenbach}, {Raab},
  {Rauth}, {Reimer}, {Reimer}, {Renaud}, {de los Reyes}, {Rieger}, {Rinchiuso},
  {Romoli}, {Rowell}, {Rudak}, {Rulten}, {Safi-Harb}, {Sahakian}, {Saito},
  {Sanchez}, {Santangelo}, {Sasaki}, {Schandri}, {Schlickeiser},
  {Sch{\"u}ssler}, {Schulz}, {Schwanke}, {Schwemmer}, {Seglar-Arroyo},
  {Settimo}, {Seyffert}, {Shafi}, {Shilon}, {Shiningayamwe}, {Simoni}, {Sol},
  {Spanier}, {Spir-Jacob}, {Stawarz}, {Steenkamp}, {Stegmann}, {Steppa},
  {Sushch}, {Takahashi}, {Tavernet}, {Tavernier}, {Taylor}, {Terrier},
  {Tibaldo}, {Tiziani}, {Tluczykont}, {Trichard}, {Tsirou}, {Tsuji}, {Tuffs},
  {Uchiyama}, {van der Walt}, {van Eldik}, {van Rensburg}, {van Soelen},
  {Vasileiadis}, {Veh}, {Venter}, {Viana}, {Vincent}, {Vink}, {Voisin},
  {V{\"o}lk}, {Vuillaume}, {Wadiasingh}, {Wagner}, {Wagner}, {Wagner}, {White},
  {Wierzcholska}, {Willmann}, {W{\"o}rnlein}, {Wouters}, {Yang}, {Zaborov},
  {Zacharias}, {Zanin}, {Zdziarski}, {Zech}, {Zefi}, {Ziegler}, {Zorn}, and
  {{\.Z}ywucka}]{2018A&A...612A...1HGPS}
{H.\,E.\,S.\,S. Collaboration}.; {Abdalla}, H.; {Abramowski}, A.; {Aharonian},
  F.; {Ait Benkhali}, F.; {Ang{\"u}ner}, E.O.; {Arakawa}, M.; {Arrieta}, M.;
  {Aubert}, P.; {Backes}, M.;  et~al.
\newblock {The H.E.S.S. Galactic plane survey}.
\newblock {\em \aap} {\bf 2018}, {\em 612},~A1,
  \href{http://xxx.lanl.gov/abs/1804.02432}{{\normalfont
  [arXiv:astro-ph.HE/1804.02432]}}.
\newblock
  doi:{\changeurlcolor{black}\href{https://doi.org/10.1051/0004-6361/201732098}{\detokenize{10.1051/0004-6361/201732098}}}.

\bibitem[{Sano} \em{et~al.}(2021){Sano}, {Yoshiike}, {Yamane}, {Hayashi},
  {Enokiya}, {Tokuda}, {Tachihara}, {Rowell}, {Filipovi{\'c}}, and
  {Fukui}]{Sano2021ApJ}
{Sano}, H.; {Yoshiike}, S.; {Yamane}, Y.; {Hayashi}, K.; {Enokiya}, R.;
  {Tokuda}, K.; {Tachihara}, K.; {Rowell}, G.; {Filipovi{\'c}}, M.D.; {Fukui},
  Y.
\newblock {ALMA CO Observations of the Mixed-morphology Supernova Remnant W49B:
  Efficient Production of Recombining Plasma and Hadronic Gamma Rays via
  Shock-Cloud Interactions}.
\newblock {\em \apj} {\bf 2021}, {\em 919},~123,
  \href{http://xxx.lanl.gov/abs/2106.12009}{{\normalfont
  [arXiv:astro-ph.HE/2106.12009]}}.
\newblock
  doi:{\changeurlcolor{black}\href{https://doi.org/10.3847/1538-4357/ac0dba}{\detokenize{10.3847/1538-4357/ac0dba}}}.

\bibitem[{Jacobs} \em{et~al.}(2022){Jacobs}, {Mertsch}, and
  {Phan}]{Jacobs2022JCAP}
{Jacobs}, H.; {Mertsch}, P.; {Phan}, V.H.M.
\newblock {Self-confinement of low-energy cosmic rays around supernova
  remnants}.
\newblock {\em \jcap} {\bf 2022}, {\em 2022},~024,
  \href{http://xxx.lanl.gov/abs/2112.09708}{{\normalfont
  [arXiv:astro-ph.HE/2112.09708]}}.
\newblock
  doi:{\changeurlcolor{black}\href{https://doi.org/10.1088/1475-7516/2022/05/024}{\detokenize{10.1088/1475-7516/2022/05/024}}}.

\bibitem[{Fujii} and {Portegies Zwart}(2016)]{Fujii2016ApJ}
{Fujii}, M.S.; {Portegies Zwart}, S.
\newblock {The Formation and Dynamical Evolution of Young Star Clusters}.
\newblock {\em \apj} {\bf 2016}, {\em 817},~4,
  \href{http://xxx.lanl.gov/abs/1512.00090}{{\normalfont
  [arXiv:astro-ph.GA/1512.00090]}}.
\newblock
  doi:{\changeurlcolor{black}\href{https://doi.org/10.3847/0004-637X/817/1/4}{\detokenize{10.3847/0004-637X/817/1/4}}}.

\bibitem[{Li} \em{et~al.}(2015){Li}, {Ostriker}, {Cen}, {Bryan}, and
  {Naab}]{Li2015ApJ}
{Li}, M.; {Ostriker}, J.P.; {Cen}, R.; {Bryan}, G.L.; {Naab}, T.
\newblock {Supernova Feedback and the Hot Gas Filling Fraction of the
  Interstellar Medium}.
\newblock {\em \apj} {\bf 2015}, {\em 814},~4,
  \href{http://xxx.lanl.gov/abs/1506.07180}{{\normalfont
  [arXiv:astro-ph.GA/1506.07180]}}.
\newblock
  doi:{\changeurlcolor{black}\href{https://doi.org/10.1088/0004-637X/814/1/4}{\detokenize{10.1088/0004-637X/814/1/4}}}.

\bibitem[{Krumholz} \em{et~al.}(2020){Krumholz}, {Crocker}, {Xu}, {Lazarian},
  {Rosevear}, and {Bedwell-Wilson}]{Krumholz2020MNRAS}
{Krumholz}, M.R.; {Crocker}, R.M.; {Xu}, S.; {Lazarian}, A.; {Rosevear}, M.T.;
  {Bedwell-Wilson}, J.
\newblock {Cosmic ray transport in starburst galaxies}.
\newblock {\em \mnras} {\bf 2020}, {\em 493},~2817--2833,
  \href{http://xxx.lanl.gov/abs/1911.09774}{{\normalfont
  [arXiv:astro-ph.HE/1911.09774]}}.
\newblock
  doi:{\changeurlcolor{black}\href{https://doi.org/10.1093/mnras/staa493}{\detokenize{10.1093/mnras/staa493}}}.

\bibitem[{Peng} \em{et~al.}(2019){Peng}, {Xi}, {Wang}, {Zhi}, and
  {Li}]{Peng2019A&A}
{Peng}, F.K.; {Xi}, S.Q.; {Wang}, X.Y.; {Zhi}, Q.J.; {Li}, D.
\newblock {Comparative study of gamma-ray emission from molecular clouds and
  star-forming galaxies}.
\newblock {\em \aap} {\bf 2019}, {\em 621},~A70,
  \href{http://xxx.lanl.gov/abs/1811.07117}{{\normalfont
  [arXiv:astro-ph.HE/1811.07117]}}.
\newblock
  doi:{\changeurlcolor{black}\href{https://doi.org/10.1051/0004-6361/201833859}{\detokenize{10.1051/0004-6361/201833859}}}.

\bibitem[{Eichmann} and {Becker Tjus}(2016)]{Eichmann2016ApJ...821...87E}
{Eichmann}, B.; {Becker Tjus}, J.
\newblock {The Radio-Gamma Correlation in Starburst Galaxies}.
\newblock {\em \apj} {\bf 2016}, {\em 821},~87,
  \href{http://xxx.lanl.gov/abs/1510.03672}{{\normalfont
  [arXiv:astro-ph.HE/1510.03672]}}.
\newblock
  doi:{\changeurlcolor{black}\href{https://doi.org/10.3847/0004-637X/821/2/87}{\detokenize{10.3847/0004-637X/821/2/87}}}.

\bibitem[{Ajello} \em{et~al.}(2020){Ajello}, {Di Mauro}, {Paliya}, and
  {Garrappa}]{Ajello2020ApJ...894...88A}
{Ajello}, M.; {Di Mauro}, M.; {Paliya}, V.S.; {Garrappa}, S.
\newblock {The {\ensuremath{\gamma}}-Ray Emission of Star-forming Galaxies}.
\newblock {\em \apj} {\bf 2020}, {\em 894},~88,
  \href{http://xxx.lanl.gov/abs/2003.05493}{{\normalfont
  [arXiv:astro-ph.GA/2003.05493]}}.
\newblock
  doi:{\changeurlcolor{black}\href{https://doi.org/10.3847/1538-4357/ab86a6}{\detokenize{10.3847/1538-4357/ab86a6}}}.

\bibitem[{Kauffmann} \em{et~al.}(2003){Kauffmann}, {Heckman}, {Tremonti},
  {Brinchmann}, {Charlot}, {White}, {Ridgway}, {Brinkmann}, {Fukugita}, {Hall},
  {Ivezi{\'c}}, {Richards}, and {Schneider}]{Kauffmann2003MNRAS.346.1055K}
{Kauffmann}, G.; {Heckman}, T.M.; {Tremonti}, C.; {Brinchmann}, J.; {Charlot},
  S.; {White}, S.D.M.; {Ridgway}, S.E.; {Brinkmann}, J.; {Fukugita}, M.;
  {Hall}, P.B.;  et~al.
\newblock {The host galaxies of active galactic nuclei}.
\newblock {\em \mnras} {\bf 2003}, {\em 346},~1055--1077,
  \href{http://xxx.lanl.gov/abs/astro-ph/0304239}{{\normalfont
  [arXiv:astro-ph/astro-ph/0304239]}}.
\newblock
  doi:{\changeurlcolor{black}\href{https://doi.org/10.1111/j.1365-2966.2003.07154.x}{\detokenize{10.1111/j.1365-2966.2003.07154.x}}}.

\bibitem[{Schawinski} \em{et~al.}(2010){Schawinski}, {Urry}, {Virani}, {Coppi},
  {Bamford}, {Treister}, {Lintott}, {Sarzi}, {Keel}, {Kaviraj}, {Cardamone},
  {Masters}, {Ross}, {Andreescu}, {Murray}, {Nichol}, {Raddick}, {Slosar},
  {Szalay}, {Thomas}, and {Vandenberg}]{Schawinski2010ApJ...711..284S}
{Schawinski}, K.; {Urry}, C.M.; {Virani}, S.; {Coppi}, P.; {Bamford}, S.P.;
  {Treister}, E.; {Lintott}, C.J.; {Sarzi}, M.; {Keel}, W.C.; {Kaviraj}, S.;
  et~al.
\newblock {Galaxy Zoo: The Fundamentally Different Co-Evolution of Supermassive
  Black Holes and Their Early- and Late-Type Host Galaxies}.
\newblock {\em \apj} {\bf 2010}, {\em 711},~284--302,
  \href{http://xxx.lanl.gov/abs/1001.3141}{{\normalfont
  [arXiv:astro-ph.CO/1001.3141]}}.
\newblock
  doi:{\changeurlcolor{black}\href{https://doi.org/10.1088/0004-637X/711/1/284}{\detokenize{10.1088/0004-637X/711/1/284}}}.

\bibitem[{Wilson} and {Ulvestad}(1987)]{Wilson1987ApJ...319..105W}
{Wilson}, A.S.; {Ulvestad}, J.S.
\newblock {A Radiative Bow Shock Wave (?) Driven by Nuclear Ejecta in a Seyfert
  Galaxy}.
\newblock {\em \apj} {\bf 1987}, {\em 319},~105.
\newblock
  doi:{\changeurlcolor{black}\href{https://doi.org/10.1086/165436}{\detokenize{10.1086/165436}}}.

\bibitem[{Michiyama} \em{et~al.}(2022){Michiyama}, {Inoue}, {Doi}, and
  {Khangulyan}]{Michiyama2022ApJ...936L...1M}
{Michiyama}, T.; {Inoue}, Y.; {Doi}, A.; {Khangulyan}, D.
\newblock {ALMA Detection of Parsec-scale Blobs at the Head of a
  Kiloparsec-scale Jet in the Nearby Seyfert Galaxy NGC 1068}.
\newblock {\em \apjl} {\bf 2022}, {\em 936},~L1,
  \href{http://xxx.lanl.gov/abs/2208.08533}{{\normalfont
  [arXiv:astro-ph.GA/2208.08533]}}.
\newblock
  doi:{\changeurlcolor{black}\href{https://doi.org/10.3847/2041-8213/ac8935}{\detokenize{10.3847/2041-8213/ac8935}}}.

\bibitem[{Tombesi} \em{et~al.}(2010){Tombesi}, {Cappi}, {Reeves}, {Palumbo},
  {Yaqoob}, {Braito}, and {Dadina}]{Tombesi2010A&A...521A..57T}
{Tombesi}, F.; {Cappi}, M.; {Reeves}, J.N.; {Palumbo}, G.G.C.; {Yaqoob}, T.;
  {Braito}, V.; {Dadina}, M.
\newblock {Evidence for ultra-fast outflows in radio-quiet AGNs. I. Detection
  and statistical incidence of Fe K-shell absorption lines}.
\newblock {\em \aap} {\bf 2010}, {\em 521},~A57,
  \href{http://xxx.lanl.gov/abs/1006.2858}{{\normalfont
  [arXiv:astro-ph.HE/1006.2858]}}.
\newblock
  doi:{\changeurlcolor{black}\href{https://doi.org/10.1051/0004-6361/200913440}{\detokenize{10.1051/0004-6361/200913440}}}.

\bibitem[{Tombesi} \em{et~al.}(2012){Tombesi}, {Cappi}, {Reeves}, and
  {Braito}]{Tombesi2012MNRAS.422L...1T}
{Tombesi}, F.; {Cappi}, M.; {Reeves}, J.N.; {Braito}, V.
\newblock {Evidence for ultrafast outflows in radio-quiet AGNs - III. Location
  and energetics}.
\newblock {\em \mnras} {\bf 2012}, {\em 422},~L1--L5,
  \href{http://xxx.lanl.gov/abs/1201.1897}{{\normalfont
  [arXiv:astro-ph.HE/1201.1897]}}.
\newblock
  doi:{\changeurlcolor{black}\href{https://doi.org/10.1111/j.1745-3933.2012.01221.x}{\detokenize{10.1111/j.1745-3933.2012.01221.x}}}.

\bibitem[{Gofford} \em{et~al.}(2015){Gofford}, {Reeves}, {McLaughlin},
  {Braito}, {Turner}, {Tombesi}, and {Cappi}]{Gofford2015MNRAS.451.4169G}
{Gofford}, J.; {Reeves}, J.N.; {McLaughlin}, D.E.; {Braito}, V.; {Turner},
  T.J.; {Tombesi}, F.; {Cappi}, M.
\newblock {The Suzaku view of highly ionized outflows in AGN - II. Location,
  energetics and scalings with bolometric luminosity}.
\newblock {\em \mnras} {\bf 2015}, {\em 451},~4169--4182,
  \href{http://xxx.lanl.gov/abs/1506.00614}{{\normalfont
  [arXiv:astro-ph.HE/1506.00614]}}.
\newblock
  doi:{\changeurlcolor{black}\href{https://doi.org/10.1093/mnras/stv1207}{\detokenize{10.1093/mnras/stv1207}}}.

\bibitem[{Wang} and {Loeb}(2015)]{Wang2015MNRAS.453..837W}
{Wang}, X.; {Loeb}, A.
\newblock {Probing the gaseous halo of galaxies through non-thermal emission
  from AGN-driven outflows}.
\newblock {\em \mnras} {\bf 2015}, {\em 453},~837--848,
  \href{http://xxx.lanl.gov/abs/1506.05470}{{\normalfont
  [arXiv:astro-ph.GA/1506.05470]}}.
\newblock
  doi:{\changeurlcolor{black}\href{https://doi.org/10.1093/mnras/stv1649}{\detokenize{10.1093/mnras/stv1649}}}.

\bibitem[{Lamastra} \em{et~al.}(2016){Lamastra}, {Fiore}, {Guetta},
  {Antonelli}, {Colafrancesco}, {Menci}, {Puccetti}, {Stamerra}, and
  {Zappacosta}]{Lamastra2016A&A...596A..68L}
{Lamastra}, A.; {Fiore}, F.; {Guetta}, D.; {Antonelli}, L.A.; {Colafrancesco},
  S.; {Menci}, N.; {Puccetti}, S.; {Stamerra}, A.; {Zappacosta}, L.
\newblock {Galactic outflow driven by the active nucleus and the origin of the
  gamma-ray emission in NGC 1068}.
\newblock {\em \aap} {\bf 2016}, {\em 596},~A68,
  \href{http://xxx.lanl.gov/abs/1609.09664}{{\normalfont
  [arXiv:astro-ph.HE/1609.09664]}}.
\newblock
  doi:{\changeurlcolor{black}\href{https://doi.org/10.1051/0004-6361/201628667}{\detokenize{10.1051/0004-6361/201628667}}}.

\bibitem[{Liu} \em{et~al.}(2018){Liu}, {Murase}, {Inoue}, {Ge}, and
  {Wang}]{Liu2018ApJ...858....9L}
{Liu}, R.Y.; {Murase}, K.; {Inoue}, S.; {Ge}, C.; {Wang}, X.Y.
\newblock {Can Winds Driven by Active Galactic Nuclei Account for the
  Extragalactic Gamma-Ray and Neutrino Backgrounds?}
\newblock {\em \apj} {\bf 2018}, {\em 858},~9,
  \href{http://xxx.lanl.gov/abs/1712.10168}{{\normalfont
  [arXiv:astro-ph.HE/1712.10168]}}.
\newblock
  doi:{\changeurlcolor{black}\href{https://doi.org/10.3847/1538-4357/aaba74}{\detokenize{10.3847/1538-4357/aaba74}}}.

\bibitem[{Inoue} \em{et~al.}(2022){Inoue}, {Cerruti}, {Murase}, and
  {Liu}]{Inoue2022arXiv220702097I}
{Inoue}, S.; {Cerruti}, M.; {Murase}, K.; {Liu}, R.Y.
\newblock {High-energy neutrinos and gamma rays from winds and tori in active
  galactic nuclei}.
\newblock {\em arXiv e-prints} {\bf 2022}, p. arXiv:2207.02097,
  \href{http://xxx.lanl.gov/abs/2207.02097}{{\normalfont
  [arXiv:astro-ph.HE/2207.02097]}}.
\newblock
  doi:{\changeurlcolor{black}\href{https://doi.org/10.48550/arXiv.2207.02097}{\detokenize{10.48550/arXiv.2207.02097}}}.

\bibitem[{Peretti} \em{et~al.}(2023){Peretti}, {Lamastra}, {Saturni}, {Ahlers},
  {Blasi}, {Morlino}, and {Cristofari}]{Peretti2023arXiv230113689P}
{Peretti}, E.; {Lamastra}, A.; {Saturni}, F.G.; {Ahlers}, M.; {Blasi}, P.;
  {Morlino}, G.; {Cristofari}, P.
\newblock {Diffusive shock acceleration at EeV and associated multimessenger
  flux from ultra-fast outflows driven by Active Galactic Nuclei}.
\newblock {\em arXiv e-prints} {\bf 2023}, p. arXiv:2301.13689,
  \href{http://xxx.lanl.gov/abs/2301.13689}{{\normalfont
  [arXiv:astro-ph.HE/2301.13689]}}.
\newblock
  doi:{\changeurlcolor{black}\href{https://doi.org/10.48550/arXiv.2301.13689}{\detokenize{10.48550/arXiv.2301.13689}}}.

\bibitem[{Kahler}(1992)]{Kahler1992ARA&A}
{Kahler}, S.W.
\newblock {Solar flares and coronal mass ejections.}
\newblock {\em \araa} {\bf 1992}, {\em 30},~113--141.
\newblock
  doi:{\changeurlcolor{black}\href{https://doi.org/10.1146/annurev.aa.30.090192.000553}{\detokenize{10.1146/annurev.aa.30.090192.000553}}}.

\bibitem[{Cliver} \em{et~al.}(2022){Cliver}, {Schrijver}, {Shibata}, and
  {Usoskin}]{Cliver2022LRSP}
{Cliver}, E.W.; {Schrijver}, C.J.; {Shibata}, K.; {Usoskin}, I.G.
\newblock {Extreme solar events}.
\newblock {\em Living Reviews in Solar Physics} {\bf 2022}, {\em 19},~2,
  \href{http://xxx.lanl.gov/abs/2205.09265}{{\normalfont
  [arXiv:astro-ph.SR/2205.09265]}}.
\newblock
  doi:{\changeurlcolor{black}\href{https://doi.org/10.1007/s41116-022-00033-8}{\detokenize{10.1007/s41116-022-00033-8}}}.

\bibitem[{Kazanas} and {Ellison}(1986)]{Kazanas1986ApJ...304..178K}
{Kazanas}, D.; {Ellison}, D.C.
\newblock {The Central Engine of Quasars and Active Galactic Nuclei: Hadronic
  Interactions of Shock-accelerated Relativistic Protons}.
\newblock {\em \apj} {\bf 1986}, {\em 304},~178.
\newblock
  doi:{\changeurlcolor{black}\href{https://doi.org/10.1086/164152}{\detokenize{10.1086/164152}}}.

\bibitem[{Zdziarski}(1986)]{Zdziarski1986ApJ...305...45Z}
{Zdziarski}, A.A.
\newblock {On the Origin of the Infrared and X-Ray Continua of Active Galactic
  Nuclei}.
\newblock {\em \apj} {\bf 1986}, {\em 305},~45.
\newblock
  doi:{\changeurlcolor{black}\href{https://doi.org/10.1086/164227}{\detokenize{10.1086/164227}}}.

\bibitem[{Sikora} \em{et~al.}(1987){Sikora}, {Kirk}, {Begelman}, and
  {Schneider}]{Sikora1987ApJ...320L..81S}
{Sikora}, M.; {Kirk}, J.G.; {Begelman}, M.C.; {Schneider}, P.
\newblock {Electron Injection by Relativistic Protons in Active Galactic
  Nuclei}.
\newblock {\em \apjl} {\bf 1987}, {\em 320},~L81.
\newblock
  doi:{\changeurlcolor{black}\href{https://doi.org/10.1086/184980}{\detokenize{10.1086/184980}}}.

\bibitem[{Begelman} \em{et~al.}(1990){Begelman}, {Rudak}, and
  {Sikora}]{Begelman1990ApJ...362...38B}
{Begelman}, M.C.; {Rudak}, B.; {Sikora}, M.
\newblock {Consequences of Relativistic Proton Injection in Active Galactic
  Nuclei}.
\newblock {\em \apj} {\bf 1990}, {\em 362},~38.
\newblock
  doi:{\changeurlcolor{black}\href{https://doi.org/10.1086/169241}{\detokenize{10.1086/169241}}}.

\bibitem[{Stecker} \em{et~al.}(1992){Stecker}, {Done}, {Salamon}, and
  {Sommers}]{Stecker1992PhRvL..69.2738S}
{Stecker}, F.W.; {Done}, C.; {Salamon}, M.H.; {Sommers}, P.
\newblock {Erratum: ``High-energy neutrinos from active galactic nuclei''
  [Phys. Rev. Lett. 66, 2697 (1991)]}.
\newblock {\em \prl} {\bf 1992}, {\em 69},~2738.
\newblock
  doi:{\changeurlcolor{black}\href{https://doi.org/10.1103/PhysRevLett.69.2738}{\detokenize{10.1103/PhysRevLett.69.2738}}}.

\bibitem[{Inoue} \em{et~al.}(2019){Inoue}, {Khangulyan}, {Inoue}, and
  {Doi}]{Inoue2019ApJ...880...40I}
{Inoue}, Y.; {Khangulyan}, D.; {Inoue}, S.; {Doi}, A.
\newblock {On High-energy Particles in Accretion Disk Coronae of Supermassive
  Black Holes: Implications for MeV Gamma-rays and High-energy Neutrinos from
  AGN Cores}.
\newblock {\em \apj} {\bf 2019}, {\em 880},~40,
  \href{http://xxx.lanl.gov/abs/1904.00554}{{\normalfont
  [arXiv:astro-ph.HE/1904.00554]}}.
\newblock
  doi:{\changeurlcolor{black}\href{https://doi.org/10.3847/1538-4357/ab2715}{\detokenize{10.3847/1538-4357/ab2715}}}.

\bibitem[{Kimura} \em{et~al.}(2015){Kimura}, {Murase}, and
  {Toma}]{Kimura2015ApJ...806..159K}
{Kimura}, S.S.; {Murase}, K.; {Toma}, K.
\newblock {Neutrino and Cosmic-Ray Emission and Cumulative Background from
  Radiatively Inefficient Accretion Flows in Low-luminosity Active Galactic
  Nuclei}.
\newblock {\em \apj} {\bf 2015}, {\em 806},~159,
  \href{http://xxx.lanl.gov/abs/1411.3588}{{\normalfont
  [arXiv:astro-ph.HE/1411.3588]}}.
\newblock
  doi:{\changeurlcolor{black}\href{https://doi.org/10.1088/0004-637X/806/2/159}{\detokenize{10.1088/0004-637X/806/2/159}}}.

\bibitem[{IceCube Collaboration} \em{et~al.}(2022){IceCube Collaboration},
  {Abbasi}, {Ackermann}, {Adams}, {Aguilar}, {Ahlers}, {Ahrens}, {Alameddine},
  {Alispach}, {Alves}, {Amin}, {Andeen}, {Anderson}, {Anton}, {Arg{\"u}elles},
  {Ashida}, {Axani}, {Bai}, {Balagopal}, {Barbano}, {Barwick}, {Bastian},
  {Basu}, {Baur}, {Bay}, {Beatty}, {Becker}, {Becker Tjus}, {Bellenghi},
  {Benzvi}, {Berley}, {Bernardini}, {Besson}, {Binder}, {Bindig}, {Blaufuss},
  {Blot}, {Boddenberg}, {Bontempo}, {Borowka}, {B{\"o}ser}, {Botner},
  {B{\"o}ttcher}, {Bourbeau}, {Bradascio}, {Braun}, {Brinson}, {Bron},
  {Brostean-Kaiser}, {Browne}, {Burgman}, {Burley}, {Busse}, {Campana},
  {Carnie-Bronca}, {Chen}, {Chen}, {Chirkin}, {Choi}, {Clark}, {Clark},
  {Classen}, {Coleman}, {Collin}, {Conrad}, {Coppin}, {Correa}, {Cowen},
  {Cross}, {Dappen}, {Dave}, {de Clercq}, {Delaunay}, {Delgado L{\'o}pez},
  {Dembinski}, {Deoskar}, {Desai}, {Desiati}, {de Vries}, {de Wasseige}, {de
  With}, {Deyoung}, {Diaz}, {D{\'\i}az-V{\'e}lez}, {Dittmer}, {Dujmovic},
  {Dunkman}, {Duvernois}, {Dvorak}, {Ehrhardt}, {Eller}, {Engel}, {Erpenbeck},
  {Evans}, {Evenson}, {Fan}, {Fazely}, {Fedynitch}, {Feigl}, {Fiedlschuster},
  {Fienberg}, {Filimonov}, {Finley}, {Fischer}, {Fox}, {Franckowiak},
  {Friedman}, {Fritz}, {F{\"u}rst}, {Gaisser}, {Gallagher}, {Ganster},
  {Garcia}, {Garrappa}, {Gerhardt}, {Ghadimi}, {Glaser}, {Glauch},
  {Gl{\"u}senkamp}, {Goldschmidt}, {Gonzalez}, {Goswami}, {Grant},
  {Gr{\'e}goire}, {Griswold}, {G{\"u}nther}, {Gutjahr}, {Haack}, {Hallgren},
  {Halliday}, {Halve}, {Halzen}, {Hanson}, {Hardin}, {Harnisch}, {Haungs},
  {Hebecker}, {Helbing}, {Henningsen}, {Hettinger}, {Hickford}, {Hignight},
  {Hill}, {Hill}, {Hoffman}, {Hoffmann}, {Hokanson-Fasig}, {Hoshina}, {Huang},
  {Huber}, {Huber}, {Hultqvist}, {H{\"u}nnefeld}, {Hussain}, {Hymon}, {in},
  {Iovine}, {Ishihara}, {Jansson}, {Japaridze}, {Jeong}, {Jin}, {Jones},
  {Kang}, {Kang}, {Kang}, {Kappes}, {Kappesser}, {Kardum}, {Karg}, {Karl},
  {Karle}, {Katz}, {Kauer}, {Kellermann}, {Kelley}, {Kheirandish}, {Kin},
  {Kintscher}, {Kiryluk}, {Klein}, {Koirala}, {Kolanoski}, {Kontrimas},
  {K{\"o}pke}, {Kopper}, {Kopper}, {Koskinen}, {Koundal}, {Kovacevich},
  {Kowalski}, {Kozynets}, {Kun}, {Kurahashi}, {Lad}, {Lagunas Gualda},
  {Lanfranchi}, {Larson}, {Lauber}, {Lazar}, {Lee}, {Leonard},
  {Leszczy{\'n}ska}, {Li}, {Lincetto}, {Liu}, {Liubarska}, {Lohfink}, {Lozano
  Mariscal}, {Lu}, {Lucarelli}, {Ludwig}, {Luszczak}, {Lyu}, {Ma}, {Madsen},
  {Mahn}, {Makino}, {Mancina}, {Mari{\c{s}}}, {Martinez-Soler}, {Maruyama},
  {Mase}, {McElroy}, {McNally}, {Mead}, {Meagher}, {Mechbal}, {Medina},
  {Meier}, {Meighen-Berger}, {Micallef}, {Mockler}, {Montaruli}, {Moore},
  {Morse}, {Moulai}, {Naab}, {Nagai}, {Nahnhauer}, {Naumann}, {Necker},
  {Nguyen}, {Niederhausen}, {Nisa}, {Nowicki}, {Nygren}, {Obertack},
  {Pollmann}, {Oehler}, {Oeyen}, {Olivas}, {O'Sullivan}, {Pandya}, {Pankova},
  {Park}, {Parker}, {Paudel}, {Paul}, {P{\'e}rez de Los Heros}, {Peters},
  {Peterson}, {Philippen}, {Pieper}, {Pittermann}, {Pizzuto}, {Plum},
  {Popovych}, {Porcelli}, {Prado Rodriguez}, {Price}, {Pries}, {Przybylski},
  {Rack-Helleis}, {Raissi}, {Rameez}, {Rawlins}, {Rea}, {Rehman},
  {Reichherzer}, {Reimann}, {Renzi}, {Resconi}, {Reusch}, {Rhode}, {Richman},
  {Riedel}, {Roberts}, {Robertson}, {Roellinghoff}, {Rongen}, {Rott}, {Ruhe},
  {Ryckbosch}, {Rysewyk Cantu}, {Safa}, {Saffer}, {Sanchez Herrera},
  {Sandrock}, {Sandroos}, {Santander}, {Sarkar}, {Sarkar}, {Satalecka},
  {Schaufel}, {Schieler}, {Schindler}, {Schmidt}, {Schneider}, {Schneider},
  {Schr{\"o}der}, {Schumacher}, {Schwefer}, {Sclafani}, {Seckel}, {Seunarine},
  {Sharma}, {Shefali}, {Silva}, {Skrzypek}, {Smithers}, {Snihur},
  {Soedingrekso}, {Soldin}, {Spannfellner}, {Spiczak}, {Spiering},
  {Stachurska}, {Stamatikos}, {Stanev}, {Stein}, {Stettner}, {Steuer},
  {Stezelberger}, {Stokstad}, {St{\"u}rwald}, {Stuttard}, {Sullivan},
  {Taboada}, {Ter-Antonyan}, {Tilav}, {Tischbein}, {Tollefson}, {T{\"o}nnis},
  {Toscano}, {Tosi}, {Trettin}, {Tselengidou}, {Tung}, {Turcati}, {Turcotte},
  {Turley}, {Twagirayezu}, {Ty}, {Unland Elorrieta}, {Valtonen-Mattila},
  {Vandenbroucke}, {van Eijndhoven}, {Vannerom}, {van Santen}, {Verpoest},
  {Walck}, {Watson}, {Weaver}, {Weigel}, {Weindl}, {Weiss}, {Weldert}, {Wendt},
  {Werthebach}, {Weyrauch}, {Whitehorn}, {Wiebusch}, {Williams}, {Wolf},
  {Woschnagg}, {Wrede}, {Wulff}, {Xu}, {Yanez}, {Yoshida}, {Yu}, {Yuan},
  {Zhangan}, and {Zhelnin}]{IceCube2022Sci...378..538I}
{IceCube Collaboration}.; {Abbasi}, R.; {Ackermann}, M.; {Adams}, J.;
  {Aguilar}, J.A.; {Ahlers}, M.; {Ahrens}, M.; {Alameddine}, J.M.; {Alispach},
  C.; {Alves}, A.~A., J.;  et~al.
\newblock {Evidence for neutrino emission from the nearby active galaxy NGC
  1068}.
\newblock {\em Science} {\bf 2022}, {\em 378},~538--543,
  \href{http://xxx.lanl.gov/abs/2211.09972}{{\normalfont
  [arXiv:astro-ph.HE/2211.09972]}}.
\newblock
  doi:{\changeurlcolor{black}\href{https://doi.org/10.1126/science.abg3395}{\detokenize{10.1126/science.abg3395}}}.

\bibitem[{Lenain} \em{et~al.}(2010){Lenain}, {Ricci}, {T{\"u}rler}, {Dorner},
  and {Walter}]{Lenain2010}
{Lenain}, J.P.; {Ricci}, C.; {T{\"u}rler}, M.; {Dorner}, D.; {Walter}, R.
\newblock {Seyfert 2 galaxies in the GeV band: jets and starburst}.
\newblock {\em \aap} {\bf 2010}, {\em 524},~A72,
  \href{http://xxx.lanl.gov/abs/1008.5164}{{\normalfont
  [arXiv:astro-ph.CO/1008.5164]}}.
\newblock
  doi:{\changeurlcolor{black}\href{https://doi.org/10.1051/0004-6361/201015644}{\detokenize{10.1051/0004-6361/201015644}}}.

\bibitem[{Ajello} \em{et~al.}(2017){Ajello}, {Atwood}, {Baldini}, {Ballet},
  {Barbiellini}, {Bastieri}, {Bellazzini}, {Bissaldi}, {Blandford}, {Bloom},
  {Bonino}, {Bregeon}, {Britto}, {Bruel}, {Buehler}, {Buson}, {Cameron},
  {Caputo}, {Caragiulo}, {Caraveo}, {Cavazzuti}, {Cecchi}, {Charles},
  {Chekhtman}, {Cheung}, {Chiaro}, {Ciprini}, {Cohen}, {Costantin}, {Costanza},
  {Cuoco}, {Cutini}, {D'Ammando}, {de Palma}, {Desiante}, {Digel}, {Di Lalla},
  {Di Mauro}, {Di Venere}, {Dom{\'\i}nguez}, {Drell}, {Dumora}, {Favuzzi},
  {Fegan}, {Ferrara}, {Fortin}, {Franckowiak}, {Fukazawa}, {Funk}, {Fusco},
  {Gargano}, {Gasparrini}, {Giglietto}, {Giommi}, {Giordano}, {Giroletti},
  {Glanzman}, {Green}, {Grenier}, {Grondin}, {Grove}, {Guillemot}, {Guiriec},
  {Harding}, {Hays}, {Hewitt}, {Horan}, {J{\'o}hannesson}, {Kensei}, {Kuss},
  {La Mura}, {Larsson}, {Latronico}, {Lemoine-Goumard}, {Li}, {Longo},
  {Loparco}, {Lott}, {Lubrano}, {Magill}, {Maldera}, {Manfreda}, {Mazziotta},
  {McEnery}, {Meyer}, {Michelson}, {Mirabal}, {Mitthumsiri}, {Mizuno},
  {Moiseev}, {Monzani}, {Morselli}, {Moskalenko}, {Negro}, {Nuss}, {Ohsugi},
  {Omodei}, {Orienti}, {Orlando}, {Palatiello}, {Paliya}, {Paneque}, {Perkins},
  {Persic}, {Pesce-Rollins}, {Piron}, {Porter}, {Principe}, {Rain{\`o}},
  {Rando}, {Razzano}, {Razzaque}, {Reimer}, {Reimer}, {Reposeur}, {Saz
  Parkinson}, {Sgr{\`o}}, {Simone}, {Siskind}, {Spada}, {Spandre}, {Spinelli},
  {Stawarz}, {Suson}, {Takahashi}, {Tak}, {Thayer}, {Thayer}, {Thompson},
  {Torres}, {Torresi}, {Troja}, {Vianello}, {Wood}, and {Wood}]{3FHL}
{Ajello}, M.; {Atwood}, W.B.; {Baldini}, L.; {Ballet}, J.; {Barbiellini}, G.;
  {Bastieri}, D.; {Bellazzini}, R.; {Bissaldi}, E.; {Blandford}, R.D.; {Bloom},
  E.D.;  et~al.
\newblock {3FHL: The Third Catalog of Hard Fermi-LAT Sources}.
\newblock {\em \apjs} {\bf 2017}, {\em 232},~18,
  \href{http://xxx.lanl.gov/abs/1702.00664}{{\normalfont
  [arXiv:astro-ph.HE/1702.00664]}}.
\newblock
  doi:{\changeurlcolor{black}\href{https://doi.org/10.3847/1538-4365/aa8221}{\detokenize{10.3847/1538-4365/aa8221}}}.

\bibitem[{The Fermi-LAT collaboration}(2019)]{4FGL}
{The Fermi-LAT collaboration}.
\newblock {Fermi Large Area Telescope Fourth Source Catalog}.
\newblock {\em arXiv e-prints} {\bf 2019}, p. arXiv:1902.10045,
  \href{http://xxx.lanl.gov/abs/1902.10045}{{\normalfont
  [arXiv:astro-ph.HE/1902.10045]}}.

\bibitem[{IceCube Collaboration} \em{et~al.}(2019){IceCube Collaboration},
  {Aartsen}, {Ackermann}, {Adams}, {Aguilar}, {Ahlers}, {Ahrens}, {Alispach},
  {Andeen}, {Anderson}, {Ansseau}, {Anton}, {Arg{\"u}elles}, {Auffenberg},
  {Axani}, {Backes}, {Bagherpour}, {Bai}, {Balagopal V.}, {Barbano}, {Barwick},
  {Bastian}, {Baum}, {Baur}, {Bay}, {Beatty}, {Becker}, {Becker Tjus},
  {BenZvi}, {Berley}, {Bernardini}, {Besson}, {Binder}, {Bindig}, {Blaufuss},
  {Blot}, {Bohm}, {B{\"o}rner}, {B{\"o}ser}, {Botner}, {B{\"o}ttcher},
  {Bourbeau}, {Bourbeau}, {Bradascio}, {Braun}, {Bron}, {Brostean-Kaiser},
  {Burgman}, {Buscher}, {Busse}, {Carver}, {Chen}, {Cheung}, {Chirkin}, {Choi},
  {Clark}, {Classen}, {Coleman}, {Collin}, {Conrad}, {Coppin}, {Correa},
  {Cowen}, {Cross}, {Dave}, {De Clercq}, {DeLaunay}, {Dembinski}, {Deoskar},
  {De Ridder}, {Desiati}, {de Vries}, {de Wasseige}, {de With}, {DeYoung},
  {Diaz}, {D{\'\i}az-V{\'e}lez}, {Dujmovic}, {Dunkman}, {Dvorak}, {Eberhardt},
  {Ehrhardt}, {Eller}, {Engel}, {Evenson}, {Fahey}, {Fazely}, {Felde},
  {Filimonov}, {Finley}, {Fox}, {Franckowiak}, {Friedman}, {Fritz}, {Gaisser},
  {Gallagher}, {Ganster}, {Garrappa}, {Gerhardt}, {Ghorbani}, {Glauch},
  {Gl{\"u}senkamp}, {Goldschmidt}, {Gonzalez}, {Grant}, {Griffith}, {Griswold},
  {G{\"u}nder}, {G{\"u}nd{\"u}z}, {Haack}, {Hallgren}, {Halliday}, {Halve},
  {Halzen}, {Hanson}, {Haungs}, {Hebecker}, {Heereman}, {Heix}, {Helbing},
  {Hellauer}, {Henningsen}, {Hickford}, {Hignight}, {Hill}, {Hoffman},
  {Hoffmann}, {Hoinka}, {Hokanson-Fasig}, {Hoshina}, {Huang}, {Huber}, {Huber},
  {Hultqvist}, {H{\"u}nnefeld}, {Hussain}, {In}, {Iovine}, {Ishihara},
  {Japaridze}, {Jeong}, {Jero}, {Jones}, {Jonske}, {Joppe}, {Kang}, {Kang},
  {Kappes}, {Kappesser}, {Karg}, {Karl}, {Karle}, {Katz}, {Kauer}, {Kelley},
  {Kheirandish}, {Kim}, {Kintscher}, {Kiryluk}, {Kittler}, {Klein}, {Koirala},
  {Kolanoski}, {K{\"o}pke}, {Kopper}, {Kopper}, {Koskinen}, {Kowalski},
  {Krings}, {Kr{\"u}ckl}, {Kulacz}, {Kurahashi}, {Kyriacou}, {Lanfranchi},
  {Larson}, {Lauber}, {Lazar}, {Leonard}, {Leszczy{\'n}ska}, {Leuermann},
  {Liu}, {Lohfink}, {Lozano Mariscal}, {Lu}, {Lucarelli}, {L{\"u}nemann},
  {Luszczak}, {Lyu}, {Ma}, {Madsen}, {Maggi}, {Mahn}, {Makino}, {Mallik},
  {Mallot}, {Mancina}, {Mari\{{\textcommabelow s}\}}, {Maruyama}, {Mase},
  {Maunu}, {McNally}, {Meagher}, {Medici}, {Medina}, {Meier}, {Meighen-Berger},
  {Menne}, {Merino}, {Meures}, {Micallef}, {Mockler}, {Moment{\'e}},
  {Montaruli}, {Moore}, {Morse}, {Moulai}, {Muth}, {Nagai}, {Naumann}, {Neer},
  {Niederhausen}, {Nisa}, {Nowicki}, {Nygren}, {Obertacke Pollmann}, {Oehler},
  {Olivas}, {O'Murchadha}, {O'Sullivan}, {Palczewski}, {Pandya}, {Pankova},
  {Park}, {Peiffer}, {P{\'e}rez de los Heros}, {Philippen}, {Pieloth}, {Pinat},
  {Pizzuto}, {Plum}, {Porcelli}, {Price}, {Przybylski}, {Raab}, {Raissi},
  {Rameez}, {Rauch}, {Rawlins}, {Rea}, {Reimann}, {Relethford}, {Renschler},
  {Renzi}, {Resconi}, {Rhode}, {Richman}, {Robertson}, {Rongen}, {Rott},
  {Ruhe}, {Ryckbosch}, {Rysewyk}, {Safa}, {Sanchez Herrera}, {Sandrock},
  {Sandroos}, {Santander}, {Sarkar}, {Sarkar}, {Satalecka}, {Schaufel},
  {Schieler}, {Schlunder}, {Schmidt}, {Schneider}, {Schneider}, {Schr{\"o}der},
  {Schumacher}, {Sclafani}, {Seckel}, {Seunarine}, {Shefali}, {Silva},
  {Snihur}, {Soedingrekso}, {Soldin}, {Song}, {Spiczak}, {Spiering},
  {Stachurska}, {Stamatikos}, {Stanev}, {Stein}, {Steinm{\"u}ller}, {Stettner},
  {Steuer}, {Stezelberger}, {Stokstad}, {St{\"o}{\ss}l}, {Strotjohann},
  {St{\"u}rwald}, {Stuttard}, {Sullivan}, {Taboada}, {Tenholt}, {Ter-Antonyan},
  {Terliuk}, {Tilav}, {Tollefson}, {Tomankova}, {T{\"o}nnis}, {Toscano},
  {Tosi}, {Trettin}, {Tselengidou}, {Tung}, {Turcati}, {Turcotte}, {Turley},
  {Ty}, {Unger}, {Unland Elorrieta}, {Usner}, {Vandenbroucke}, {Van Driessche},
  {van Eijk}, {van Eijndhoven}, {van Santen}, {Verpoest}, {Vraeghe}, {Walck},
  {Wallace}, {Wallraff}, {Wandkowsky}, {Watson}, {Weaver}, {Weindl}, {Weiss},
  {Weldert}, {Wendt}, {Werthebach}, {Whelan}, {Whitehorn}, {Wiebe}, {Wiebusch},
  {Wille}, {Williams}, {Wills}, {Wolf}, {Wood}, {Wood}, {Woschnagg}, {Wrede},
  {Xu}, {Xu}, {Xu}, {Yanez}, {Yodh}, {Yoshida}, {Yuan}, and
  {Z{\"o}cklein}]{IceCube2019_NGC1068}
{IceCube Collaboration}.; {Aartsen}, M.G.; {Ackermann}, M.; {Adams}, J.;
  {Aguilar}, J.A.; {Ahlers}, M.; {Ahrens}, M.; {Alispach}, C.; {Andeen}, K.;
  {Anderson}, T.;  et~al.
\newblock {Time-integrated Neutrino Source Searches with 10 years of IceCube
  Data}.
\newblock {\em arXiv e-prints} {\bf 2019}, p. arXiv:1910.08488,
  \href{http://xxx.lanl.gov/abs/1910.08488}{{\normalfont
  [arXiv:astro-ph.HE/1910.08488]}}.

\bibitem[{Inoue} \em{et~al.}(2020){Inoue}, {Khangulyan}, and
  {Doi}]{Inoue2020ApJ...891L..33I}
{Inoue}, Y.; {Khangulyan}, D.; {Doi}, A.
\newblock {On the Origin of High-energy Neutrinos from NGC 1068: The Role of
  Nonthermal Coronal Activity}.
\newblock {\em \apjl} {\bf 2020}, {\em 891},~L33,
  \href{http://xxx.lanl.gov/abs/1909.02239}{{\normalfont
  [arXiv:astro-ph.HE/1909.02239]}}.
\newblock
  doi:{\changeurlcolor{black}\href{https://doi.org/10.3847/2041-8213/ab7661}{\detokenize{10.3847/2041-8213/ab7661}}}.

\bibitem[{Murase} \em{et~al.}(2020){Murase}, {Kimura}, and
  {M{\'e}sz{\'a}ros}]{Murase2020PhRvL.125a1101M}
{Murase}, K.; {Kimura}, S.S.; {M{\'e}sz{\'a}ros}, P.
\newblock {Hidden Cores of Active Galactic Nuclei as the Origin of
  Medium-Energy Neutrinos: Critical Tests with the MeV Gamma-Ray Connection}.
\newblock {\em \prl} {\bf 2020}, {\em 125},~011101,
  \href{http://xxx.lanl.gov/abs/1904.04226}{{\normalfont
  [arXiv:astro-ph.HE/1904.04226]}}.
\newblock
  doi:{\changeurlcolor{black}\href{https://doi.org/10.1103/PhysRevLett.125.011101}{\detokenize{10.1103/PhysRevLett.125.011101}}}.

\bibitem[{Eichmann} \em{et~al.}(2022){Eichmann}, {Oikonomou}, {Salvatore},
  {Dettmar}, and {Tjus}]{Eichmann2022ApJ...939...43E}
{Eichmann}, B.; {Oikonomou}, F.; {Salvatore}, S.; {Dettmar}, R.J.; {Tjus}, J.B.
\newblock {Solving the Multimessenger Puzzle of the AGN-starburst Composite
  Galaxy NGC 1068}.
\newblock {\em \apj} {\bf 2022}, {\em 939},~43,
  \href{http://xxx.lanl.gov/abs/2207.00102}{{\normalfont
  [arXiv:astro-ph.HE/2207.00102]}}.
\newblock
  doi:{\changeurlcolor{black}\href{https://doi.org/10.3847/1538-4357/ac9588}{\detokenize{10.3847/1538-4357/ac9588}}}.

\bibitem[{Michel}(1984)]{Michel1984AdSpR}
{Michel}, F.C.
\newblock {Cosmic-ray acceleration by pulsars}.
\newblock {\em Advances in Space Research} {\bf 1984}, {\em 4},~387--391.
\newblock
  doi:{\changeurlcolor{black}\href{https://doi.org/10.1016/0273-1177(84)90336-3}{\detokenize{10.1016/0273-1177(84)90336-3}}}.

\bibitem[{Heyl} \em{et~al.}(2010){Heyl}, {Gill}, and
  {Hernquist}]{Heyl2010MNRAS}
{Heyl}, J.S.; {Gill}, R.; {Hernquist}, L.
\newblock {Cosmic rays from pulsars and magnetars}.
\newblock {\em \mnras} {\bf 2010}, {\em 406},~L25--L29,
  \href{http://xxx.lanl.gov/abs/1005.1003}{{\normalfont
  [arXiv:astro-ph.HE/1005.1003]}}.
\newblock
  doi:{\changeurlcolor{black}\href{https://doi.org/10.1111/j.1745-3933.2010.00874.x}{\detokenize{10.1111/j.1745-3933.2010.00874.x}}}.

\bibitem[{Fang} \em{et~al.}(2014){Fang}, {Kotera}, {Murase}, and
  {Olinto}]{Fang2014PhRvD}
{Fang}, K.; {Kotera}, K.; {Murase}, K.; {Olinto}, A.V.
\newblock {Testing the newborn pulsar origin of ultrahigh energy cosmic rays
  with EeV neutrinos}.
\newblock {\em \prd} {\bf 2014}, {\em 90},~103005,
  \href{http://xxx.lanl.gov/abs/1311.2044}{{\normalfont
  [arXiv:astro-ph.HE/1311.2044]}}.
\newblock
  doi:{\changeurlcolor{black}\href{https://doi.org/10.1103/PhysRevD.90.103005}{\detokenize{10.1103/PhysRevD.90.103005}}}.

\bibitem[{Piro} and {Kollmeier}(2016)]{Piro2016ApJ}
{Piro}, A.L.; {Kollmeier}, J.A.
\newblock {Ultrahigh-energy Cosmic Rays from the ``En Caul'' Birth of
  Magnetars}.
\newblock {\em \apj} {\bf 2016}, {\em 826},~97,
  \href{http://xxx.lanl.gov/abs/1601.02625}{{\normalfont
  [arXiv:astro-ph.HE/1601.02625]}}.
\newblock
  doi:{\changeurlcolor{black}\href{https://doi.org/10.3847/0004-637X/826/1/97}{\detokenize{10.3847/0004-637X/826/1/97}}}.

\bibitem[{Bowden} \em{et~al.}(1992){Bowden}, {Bradbury}, {Chadwick},
  {Dickinson}, {Dipper}, {Edwards}, {Lincoln}, {McComb}, {Orford}, {Rayner},
  and {Turver}]{Bowden1992APh}
{Bowden}, C.C.G.; {Bradbury}, S.M.; {Chadwick}, P.M.; {Dickinson}, J.E.;
  {Dipper}, N.A.; {Edwards}, P.J.; {Lincoln}, E.W.; {McComb}, T.J.L.; {Orford},
  K.J.; {Rayner}, S.M.;  et~al.
\newblock {350 GeV gamma rays from AE Aqr}.
\newblock {\em Astroparticle Physics} {\bf 1992}, {\em 1},~47--59.
\newblock
  doi:{\changeurlcolor{black}\href{https://doi.org/10.1016/0927-6505(92)90008-N}{\detokenize{10.1016/0927-6505(92)90008-N}}}.

\bibitem[{Li} \em{et~al.}(2016){Li}, {Torres}, {Rea}, {de O{\~n}a Wilhelmi},
  {Papitto}, {Hou}, and {Mauche}]{Li2016ApJ}
{Li}, J.; {Torres}, D.F.; {Rea}, N.; {de O{\~n}a Wilhelmi}, E.; {Papitto}, A.;
  {Hou}, X.; {Mauche}, C.W.
\newblock {Search for Gamma-Ray Emission from AE Aquarii with Seven Years of
  Fermi-LAT Observations}.
\newblock {\em \apj} {\bf 2016}, {\em 832},~35,
  \href{http://xxx.lanl.gov/abs/1608.06662}{{\normalfont
  [arXiv:astro-ph.HE/1608.06662]}}.
\newblock
  doi:{\changeurlcolor{black}\href{https://doi.org/10.3847/0004-637X/832/1/35}{\detokenize{10.3847/0004-637X/832/1/35}}}.

\bibitem[{Meintjes} \em{et~al.}(2023){Meintjes}, {Madzime}, {Kaplan}, and {van
  Heerden}]{Meintjes2023Galax}
{Meintjes}, P.J.; {Madzime}, S.T.; {Kaplan}, Q.; {van Heerden}, H.J.
\newblock {Spun-Up Rotation-Powered Magnetized White Dwarfs in Close Binaries
  as Possible Gamma-ray Sources: Signatures of Pulsed Modulation from AE
  Aquarii and AR Scorpii in Fermi-LAT Data}.
\newblock {\em Galaxies} {\bf 2023}, {\em 11},~14.
\newblock
  doi:{\changeurlcolor{black}\href{https://doi.org/10.3390/galaxies11010014}{\detokenize{10.3390/galaxies11010014}}}.

\bibitem[{Cooper} \em{et~al.}(2020){Cooper}, {Gaggero}, {Markoff}, and
  {Zhang}]{Cooper2020MNRAS}
{Cooper}, A.J.; {Gaggero}, D.; {Markoff}, S.; {Zhang}, S.
\newblock {High-energy cosmic ray production in X-ray binary jets}.
\newblock {\em \mnras} {\bf 2020}, {\em 493},~3212--3222,
  \href{http://xxx.lanl.gov/abs/2002.01477}{{\normalfont
  [arXiv:astro-ph.HE/2002.01477]}}.
\newblock
  doi:{\changeurlcolor{black}\href{https://doi.org/10.1093/mnras/staa373}{\detokenize{10.1093/mnras/staa373}}}.

\bibitem[{Linares} and {Kachelrie{\ss}}(2021)]{Linares2021JCAP}
{Linares}, M.; {Kachelrie{\ss}}, M.
\newblock {Cosmic ray positrons from compact binary millisecond pulsars}.
\newblock {\em \jcap} {\bf 2021}, {\em 2021},~030,
  \href{http://xxx.lanl.gov/abs/2010.02844}{{\normalfont
  [arXiv:astro-ph.HE/2010.02844]}}.
\newblock
  doi:{\changeurlcolor{black}\href{https://doi.org/10.1088/1475-7516/2021/02/030}{\detokenize{10.1088/1475-7516/2021/02/030}}}.

\bibitem[{Harding}(2022)]{Harding2022ASSL}
{Harding}, A.K.
\newblock {The Emission Physics of Millisecond Pulsars}.
\newblock  Astrophysics and Space Science Library; {Bhattacharyya}, S.;
  {Papitto}, A.; {Bhattacharya}, D., Eds.,  2022, Vol. 465, {\em Astrophysics
  and Space Science Library}, pp. 57--85,
  \href{http://xxx.lanl.gov/abs/2101.05751}{{\normalfont
  [arXiv:astro-ph.HE/2101.05751]}}.
\newblock
  doi:{\changeurlcolor{black}\href{https://doi.org/10.1007/978-3-030-85198-9_3}{\detokenize{10.1007/978-3-030-85198-9_3}}}.

\bibitem[{Fabrika}(2004)]{Fabrika2004ASPRv..12....1F}
{Fabrika}, S.
\newblock {The jets and supercritical accretion disk in SS433}.
\newblock {\em \apspr} {\bf 2004}, {\em 12},~1--152,
  \href{http://xxx.lanl.gov/abs/astro-ph/0603390}{{\normalfont
  [arXiv:astro-ph/astro-ph/0603390]}}.
\newblock
  doi:{\changeurlcolor{black}\href{https://doi.org/10.48550/arXiv.astro-ph/0603390}{\detokenize{10.48550/arXiv.astro-ph/0603390}}}.

\bibitem[{Cherepashchuk} \em{et~al.}(2021){Cherepashchuk}, {Belinski}, {Dodin},
  and {Postnov}]{Cherepashchuk2021MNRAS}
{Cherepashchuk}, A.M.; {Belinski}, A.A.; {Dodin}, A.V.; {Postnov}, K.A.
\newblock {Discovery of orbital eccentricity and evidence for orbital period
  increase of SS433}.
\newblock {\em \mnras} {\bf 2021}, {\em 507},~L19--L23,
  \href{http://xxx.lanl.gov/abs/2107.09005}{{\normalfont
  [arXiv:astro-ph.SR/2107.09005]}}.
\newblock
  doi:{\changeurlcolor{black}\href{https://doi.org/10.1093/mnrasl/slab083}{\detokenize{10.1093/mnrasl/slab083}}}.

\bibitem[{Hillwig} \em{et~al.}(2004){Hillwig}, {Gies}, {Huang}, {McSwain},
  {Stark}, {van der Meer}, and {Kaper}]{Hillwig2004ApJ}
{Hillwig}, T.C.; {Gies}, D.R.; {Huang}, W.; {McSwain}, M.V.; {Stark}, M.A.;
  {van der Meer}, A.; {Kaper}, L.
\newblock {Identification of the Mass Donor Star's Spectrum in SS 433}.
\newblock {\em \apj} {\bf 2004}, {\em 615},~422--431,
  \href{http://xxx.lanl.gov/abs/astro-ph/0403634}{{\normalfont
  [arXiv:astro-ph/astro-ph/0403634]}}.
\newblock
  doi:{\changeurlcolor{black}\href{https://doi.org/10.1086/423927}{\detokenize{10.1086/423927}}}.

\bibitem[{} and {Li}(2020)]{Han2020ApJ}
{}, Q.; {Li}, X.D.
\newblock {On the Formation of SS433}.
\newblock {\em \apj} {\bf 2020}, {\em 896},~34,
  \href{http://xxx.lanl.gov/abs/2004.12547}{{\normalfont
  [arXiv:astro-ph.HE/2004.12547]}}.
\newblock
  doi:{\changeurlcolor{black}\href{https://doi.org/10.3847/1538-4357/ab8d3d}{\detokenize{10.3847/1538-4357/ab8d3d}}}.

\bibitem[{Roberts} \em{et~al.}(2008){Roberts}, {Wardle}, {Lipnick},
  {Selesnick}, and {Slutsky}]{Roberts2008ApJ}
{Roberts}, D.H.; {Wardle}, J.F.C.; {Lipnick}, S.L.; {Selesnick}, P.L.;
  {Slutsky}, S.
\newblock {Structure and Magnetic Fields in the Precessing Jet System SS 433.
  I. Multifrequency Imaging from 1998}.
\newblock {\em \apj} {\bf 2008}, {\em 676},~584--593,
  \href{http://xxx.lanl.gov/abs/0712.2005}{{\normalfont
  [arXiv:astro-ph/0712.2005]}}.
\newblock
  doi:{\changeurlcolor{black}\href{https://doi.org/10.1086/527544}{\detokenize{10.1086/527544}}}.

\bibitem[{Bowler} and {Keppens}(2018)]{Bowler2018A&A}
{Bowler}, M.G.; {Keppens}, R.
\newblock {W 50 and SS 433}.
\newblock {\em \aap} {\bf 2018}, {\em 617},~A29,
  \href{http://xxx.lanl.gov/abs/1805.10094}{{\normalfont
  [arXiv:astro-ph.GA/1805.10094]}}.
\newblock
  doi:{\changeurlcolor{black}\href{https://doi.org/10.1051/0004-6361/201732488}{\detokenize{10.1051/0004-6361/201732488}}}.

\bibitem[{Abeysekara} \em{et~al.}(2018){Abeysekara}, {Albert}, {Alfaro},
  {Alvarez}, {{\'A}lvarez}, {Arceo}, {Arteaga-Vel{\'a}zquez}, {Avila Rojas},
  {Ayala Solares}, {Belmont-Moreno}, {BenZvi}, {Brisbois}, {Caballero-Mora},
  {Capistr{\'a}n}, {Carrami{\~n}ana}, {Casanova}, {Castillo}, {Cotti},
  {Cotzomi}, {Couti{\~n}o de Le{\'o}n}, {De Le{\'o}n}, {De la Fuente},
  {D{\'\i}az-V{\'e}lez}, {Dichiara}, {Dingus}, {DuVernois}, {Ellsworth},
  {Engel}, {Espinoza}, {Fang}, {Fleischhack}, {Fraija}, {Galv{\'a}n-G{\'a}mez},
  {Garc{\'\i}a-Gonz{\'a}lez}, {Garfias}, {Gonz{\'a}lez-Mu{\~n}oz},
  {Gonz{\'a}lez}, {Goodman}, {Hampel-Arias}, {Harding}, {Hernandez}, {Hinton},
  {Hona}, {Hueyotl-Zahuantitla}, {Hui}, {H{\"u}ntemeyer}, {Iriarte},
  {Jardin-Blicq}, {Joshi}, {Kaufmann}, {Kar}, {Kunde}, {Lauer}, {Lee},
  {Le{\'o}n Vargas}, {Li}, {Linnemann}, {Longinotti}, {Luis-Raya},
  {L{\'o}pez-Coto}, {Malone}, {Marinelli}, {Martinez}, {Martinez-Castellanos},
  {Mart{\'\i}nez-Castro}, {Matthews}, {Miranda-Romagnoli}, {Moreno},
  {Mostaf{\'a}}, {Nayerhoda}, {Nellen}, {Newbold}, {Nisa}, {Noriega-Papaqui},
  {Pretz}, {P{\'e}rez-P{\'e}rez}, {Ren}, {Rho}, {Rivi{\`e}re},
  {Rosa-Gonz{\'a}lez}, {Rosenberg}, {Ruiz-Velasco}, {Salesa Greus}, {Sandoval},
  {Schneider}, {Schoorlemmer}, {Seglar Arroyo}, {Sinnis}, {Smith}, {Springer},
  {Surajbali}, {Taboada}, {Tibolla}, {Tollefson}, {Torres}, {Vianello},
  {Villase{\~n}or}, {Weisgarber}, {Werner}, {Westerhoff}, {Wood}, {Yapici},
  {Yodh}, {Zepeda}, {Zhang}, and {Zhou}]{HAWC2018Natur.562...82A}
{Abeysekara}, A.U.; {Albert}, A.; {Alfaro}, R.; {Alvarez}, C.; {{\'A}lvarez},
  J.D.; {Arceo}, R.; {Arteaga-Vel{\'a}zquez}, J.C.; {Avila Rojas}, D.; {Ayala
  Solares}, H.A.; {Belmont-Moreno}, E.;  et~al.
\newblock {Very-high-energy particle acceleration powered by the jets of the
  microquasar SS 433}.
\newblock {\em \nat} {\bf 2018}, {\em 562},~82--85.
\newblock
  doi:{\changeurlcolor{black}\href{https://doi.org/10.1038/s41586-018-0565-5}{\detokenize{10.1038/s41586-018-0565-5}}}.

\bibitem[{Watson} \em{et~al.}(1983){Watson}, {Willingale}, {Grindlay}, and
  {Seward}]{1983ApJ...273..688W}
{Watson}, M.G.; {Willingale}, R.; {Grindlay}, J.E.; {Seward}, F.D.
\newblock {The X-ray lobes of SS 433.}
\newblock {\em \apj} {\bf 1983}, {\em 273},~688--696.
\newblock
  doi:{\changeurlcolor{black}\href{https://doi.org/10.1086/161403}{\detokenize{10.1086/161403}}}.

\bibitem[{Yamauchi} \em{et~al.}(1994){Yamauchi}, {Kawai}, and
  {Aoki}]{1994PASJ...46L.109Y}
{Yamauchi}, S.; {Kawai}, N.; {Aoki}, T.
\newblock {A Non-Thermal X-Ray Spectrum from the Supernova Remnant W 50}.
\newblock {\em \pasj} {\bf 1994}, {\em 46},~L109--L113.

\bibitem[{Brinkmann} \em{et~al.}(1996){Brinkmann}, {Aschenbach}, and
  {Kawai}]{1996A&A...312..306B}
{Brinkmann}, W.; {Aschenbach}, B.; {Kawai}, N.
\newblock {ROSAT observations of the W 50/SS 433 system.}
\newblock {\em \aap} {\bf 1996}, {\em 312},~306--316.

\bibitem[{Safi-Harb} and {{\"O}gelman}(1997)]{1997ApJ...483..868S}
{Safi-Harb}, S.; {{\"O}gelman}, H.
\newblock {ROSAT and ASCA Observations of W50 Associated with the Peculiar
  Source SS 433}.
\newblock {\em \apj} {\bf 1997}, {\em 483},~868--881.
\newblock
  doi:{\changeurlcolor{black}\href{https://doi.org/10.1086/304274}{\detokenize{10.1086/304274}}}.

\bibitem[{Safi-Harb} and {Petre}(1999)]{1999ApJ...512..784S}
{Safi-Harb}, S.; {Petre}, R.
\newblock {Rossi X-Ray Timing Explorer Observations of the Eastern Lobe of W50
  Associated with SS 433}.
\newblock {\em \apj} {\bf 1999}, {\em 512},~784--792.
\newblock
  doi:{\changeurlcolor{black}\href{https://doi.org/10.1086/306803}{\detokenize{10.1086/306803}}}.

\bibitem[{Kayama} \em{et~al.}(2022){Kayama}, {Tanaka}, {Uchida}, {Tsuru},
  {Sudoh}, {Inoue}, {Khangulyan}, {Tsuji}, and {Yamamoto}]{2022PASJ...74.1143K}
{Kayama}, K.; {Tanaka}, T.; {Uchida}, H.; {Tsuru}, T.G.; {Sudoh}, T.; {Inoue},
  Y.; {Khangulyan}, D.; {Tsuji}, N.; {Yamamoto}, H.
\newblock {Spatially resolved study of the SS 433/W 50 west region with
  Chandra: X-ray structure and spectral variation of non-thermal emission}.
\newblock {\em \pasj} {\bf 2022}, {\em 74},~1143--1156,
  \href{http://xxx.lanl.gov/abs/2207.05924}{{\normalfont
  [arXiv:astro-ph.HE/2207.05924]}}.
\newblock
  doi:{\changeurlcolor{black}\href{https://doi.org/10.1093/pasj/psac060}{\detokenize{10.1093/pasj/psac060}}}.

\bibitem[{Sudoh} \em{et~al.}(2020){Sudoh}, {Inoue}, and
  {Khangulyan}]{Sudoh2020ApJ...889..146S}
{Sudoh}, T.; {Inoue}, Y.; {Khangulyan}, D.
\newblock {Multiwavelength Emission from Galactic Jets: The Case of the
  Microquasar SS433}.
\newblock {\em \apj} {\bf 2020}, {\em 889},~146,
  \href{http://xxx.lanl.gov/abs/1911.00013}{{\normalfont
  [arXiv:astro-ph.HE/1911.00013]}}.
\newblock
  doi:{\changeurlcolor{black}\href{https://doi.org/10.3847/1538-4357/ab6442}{\detokenize{10.3847/1538-4357/ab6442}}}.

\bibitem[{Kimura} \em{et~al.}(2020){Kimura}, {Murase}, and
  {M{\'e}sz{\'a}ros}]{Kimura2020ApJ...904..188K}
{Kimura}, S.S.; {Murase}, K.; {M{\'e}sz{\'a}ros}, P.
\newblock {Deciphering the Origin of the GeV-TeV Gamma-Ray Emission from SS
  433}.
\newblock {\em \apj} {\bf 2020}, {\em 904},~188,
  \href{http://xxx.lanl.gov/abs/2008.04515}{{\normalfont
  [arXiv:astro-ph.HE/2008.04515]}}.
\newblock
  doi:{\changeurlcolor{black}\href{https://doi.org/10.3847/1538-4357/abbe00}{\detokenize{10.3847/1538-4357/abbe00}}}.

\bibitem[{Pakull} \em{et~al.}(2010){Pakull}, {Soria}, and
  {Motch}]{Pakull2010Natur.466..209P}
{Pakull}, M.W.; {Soria}, R.; {Motch}, C.
\newblock {A 300-parsec-long jet-inflated bubble around a powerful microquasar
  in the galaxy NGC 7793}.
\newblock {\em \nat} {\bf 2010}, {\em 466},~209--212.
\newblock
  doi:{\changeurlcolor{black}\href{https://doi.org/10.1038/nature09168}{\detokenize{10.1038/nature09168}}}.

\bibitem[{Cseh} \em{et~al.}(2012){Cseh}, {Corbel}, {Kaaret}, {Lang},
  {Gris{\'e}}, {Paragi}, {Tzioumis}, {Tudose}, and
  {Feng}]{Cseh2012ApJ...749...17C}
{Cseh}, D.; {Corbel}, S.; {Kaaret}, P.; {Lang}, C.; {Gris{\'e}}, F.; {Paragi},
  Z.; {Tzioumis}, A.; {Tudose}, V.; {Feng}, H.
\newblock {Black Hole Powered Nebulae and a Case Study of the Ultraluminous
  X-Ray Source IC 342 X-1}.
\newblock {\em \apj} {\bf 2012}, {\em 749},~17,
  \href{http://xxx.lanl.gov/abs/1201.4473}{{\normalfont
  [arXiv:astro-ph.HE/1201.4473]}}.
\newblock
  doi:{\changeurlcolor{black}\href{https://doi.org/10.1088/0004-637X/749/1/17}{\detokenize{10.1088/0004-637X/749/1/17}}}.

\bibitem[{Inoue} \em{et~al.}(2017){Inoue}, {Lee}, {Tanaka}, and
  {Kobayashi}]{Inoue2017APh....90...14I}
{Inoue}, Y.; {Lee}, S.H.; {Tanaka}, Y.T.; {Kobayashi}, S.B.
\newblock {High energy gamma rays from nebulae associated with extragalactic
  microquasars and ultra-luminous X-ray sources}.
\newblock {\em Astroparticle Physics} {\bf 2017}, {\em 90},~14--19,
  \href{http://xxx.lanl.gov/abs/1701.08882}{{\normalfont
  [arXiv:astro-ph.HE/1701.08882]}}.
\newblock
  doi:{\changeurlcolor{black}\href{https://doi.org/10.1016/j.astropartphys.2017.01.012}{\detokenize{10.1016/j.astropartphys.2017.01.012}}}.

\bibitem[{Mason} \em{et~al.}(1986){Mason}, {Cordova}, and
  {White}]{Mason1986ApJ}
{Mason}, K.O.; {Cordova}, F.A.; {White}, N.E.
\newblock {Simultaneous X-Ray and Infrared Observations of Cygnus X-3}.
\newblock {\em \apj} {\bf 1986}, {\em 309},~700.
\newblock
  doi:{\changeurlcolor{black}\href{https://doi.org/10.1086/164638}{\detokenize{10.1086/164638}}}.

\bibitem[{Stark} and {Saia}(2003)]{Stark2003ApJ}
{Stark}, M.J.; {Saia}, M.
\newblock {Doppler Modulation of X-Ray Lines in Cygnus X-3}.
\newblock {\em \apjl} {\bf 2003}, {\em 587},~L101--L104,
  \href{http://xxx.lanl.gov/abs/astro-ph/0301554}{{\normalfont
  [arXiv:astro-ph/astro-ph/0301554]}}.
\newblock
  doi:{\changeurlcolor{black}\href{https://doi.org/10.1086/375287}{\detokenize{10.1086/375287}}}.

\bibitem[{van Kerkwijk} \em{et~al.}(1996){van Kerkwijk}, {Geballe}, {King},
  {van der Klis}, and {van Paradijs}]{vanKerkwijk1996A&A}
{van Kerkwijk}, M.H.; {Geballe}, T.R.; {King}, D.L.; {van der Klis}, M.; {van
  Paradijs}, J.
\newblock {The Wolf-Rayet counterpart of Cygnus X-3.}
\newblock {\em \aap} {\bf 1996}, {\em 314},~521--540,
  \href{http://xxx.lanl.gov/abs/astro-ph/9604100}{{\normalfont
  [arXiv:astro-ph/astro-ph/9604100]}}.
\newblock
  doi:{\changeurlcolor{black}\href{https://doi.org/10.48550/arXiv.astro-ph/9604100}{\detokenize{10.48550/arXiv.astro-ph/9604100}}}.

\bibitem[{Lommen} \em{et~al.}(2005){Lommen}, {Yungelson}, {van den Heuvel},
  {Nelemans}, and {Portegies Zwart}]{Lommen2005A&A}
{Lommen}, D.; {Yungelson}, L.; {van den Heuvel}, E.; {Nelemans}, G.; {Portegies
  Zwart}, S.
\newblock {Cygnus X-3 and the problem of the missing Wolf-Rayet X-ray
  binaries}.
\newblock {\em \aap} {\bf 2005}, {\em 443},~231--241,
  \href{http://xxx.lanl.gov/abs/astro-ph/0507304}{{\normalfont
  [arXiv:astro-ph/astro-ph/0507304]}}.
\newblock
  doi:{\changeurlcolor{black}\href{https://doi.org/10.1051/0004-6361:20052824}{\detokenize{10.1051/0004-6361:20052824}}}.

\bibitem[{Egron} \em{et~al.}(2021){Egron}, {Pellizzoni}, {Righini},
  {Giroletti}, {Koljonen}, {Pottschmidt}, {Trushkin}, {Lobina}, {Pilia},
  {Wilms}, {Corbel}, {Grinberg}, {Loru}, {Trois}, {Rodriguez},
  {L{\"a}hteenm{\"a}ki}, {Tornikoski}, {Enestam}, and
  {J{\"a}rvel{\"a}}]{Egron2021ApJ}
{Egron}, E.; {Pellizzoni}, A.; {Righini}, S.; {Giroletti}, M.; {Koljonen}, K.;
  {Pottschmidt}, K.; {Trushkin}, S.; {Lobina}, J.; {Pilia}, M.; {Wilms}, J.;
  et~al.
\newblock {Investigating the Mini and Giant Radio Flare Episodes of Cygnus
  X-3}.
\newblock {\em \apj} {\bf 2021}, {\em 906},~10,
  \href{http://xxx.lanl.gov/abs/2010.15002}{{\normalfont
  [arXiv:astro-ph.HE/2010.15002]}}.
\newblock
  doi:{\changeurlcolor{black}\href{https://doi.org/10.3847/1538-4357/abc5b1}{\detokenize{10.3847/1538-4357/abc5b1}}}.

\bibitem[{Koljonen} \em{et~al.}(2013){Koljonen}, {McCollough}, {Hannikainen},
  and {Droulans}]{Koljonen2013MNRAS}
{Koljonen}, K.I.I.; {McCollough}, M.L.; {Hannikainen}, D.C.; {Droulans}, R.
\newblock {2006 May-July major radio flare episodes in Cygnus X-3:
  spectrotiming analysis of the X-ray data}.
\newblock {\em \mnras} {\bf 2013}, {\em 429},~1173--1188,
  \href{http://xxx.lanl.gov/abs/1211.3300}{{\normalfont
  [arXiv:astro-ph.HE/1211.3300]}}.
\newblock
  doi:{\changeurlcolor{black}\href{https://doi.org/10.1093/mnras/sts404}{\detokenize{10.1093/mnras/sts404}}}.

\bibitem[{Mioduszewski} \em{et~al.}(2001){Mioduszewski}, {Rupen}, {Hjellming},
  {Pooley}, and {Waltman}]{Mioduszewski2001ApJ}
{Mioduszewski}, A.J.; {Rupen}, M.P.; {Hjellming}, R.M.; {Pooley}, G.G.;
  {Waltman}, E.B.
\newblock {A One-sided Highly Relativistic Jet from Cygnus X-3}.
\newblock {\em \apj} {\bf 2001}, {\em 553},~766--775,
  \href{http://xxx.lanl.gov/abs/astro-ph/0102018}{{\normalfont
  [arXiv:astro-ph/astro-ph/0102018]}}.
\newblock
  doi:{\changeurlcolor{black}\href{https://doi.org/10.1086/320965}{\detokenize{10.1086/320965}}}.

\bibitem[{Tudose} \em{et~al.}(2010){Tudose}, {Miller-Jones}, {Fender},
  {Paragi}, {Sakari}, {Szostek}, {Garrett}, {Dhawan}, {Rushton}, {Spencer}, and
  {van der Klis}]{Tudose2010MNRAS}
{Tudose}, V.; {Miller-Jones}, J.C.A.; {Fender}, R.P.; {Paragi}, Z.; {Sakari},
  C.; {Szostek}, A.; {Garrett}, M.A.; {Dhawan}, V.; {Rushton}, A.; {Spencer},
  R.E.;  et~al.
\newblock {Probing the behaviour of the X-ray binary Cygnus X-3 with very long
  baseline radio interferometry}.
\newblock {\em \mnras} {\bf 2010}, {\em 401},~890--900,
  \href{http://xxx.lanl.gov/abs/0909.2790}{{\normalfont
  [arXiv:astro-ph.HE/0909.2790]}}.
\newblock
  doi:{\changeurlcolor{black}\href{https://doi.org/10.1111/j.1365-2966.2009.15719.x}{\detokenize{10.1111/j.1365-2966.2009.15719.x}}}.

\bibitem[{Fermi LAT Collaboration} \em{et~al.}(2009){Fermi LAT Collaboration},
  {Abdo}, {Ackermann}, {Ajello}, {Axelsson}, {Baldini}, {Ballet},
  {Barbiellini}, {Bastieri}, {Baughman}, {Bechtol}, {Bellazzini}, {Berenji},
  {Blandford}, {Bloom}, {Bonamente}, {Borgland}, {Brez}, {Brigida}, {Bruel},
  {Burnett}, {Buson}, {Caliandro}, {Cameron}, {Caraveo}, {Casandjian},
  {Cecchi}, {{\c{C}}elik}, {Chaty}, {Cheung}, {Chiang}, {Ciprini}, {Claus},
  {Cohen-Tanugi}, {Cominsky}, {Conrad}, {Corbel}, {Corbet}, {Dermer}, {de
  Palma}, {Digel}, {do Couto e Silva}, {Drell}, {Dubois}, {Dubus}, {Dumora},
  {Farnier}, {Favuzzi}, {Fegan}, {Focke}, {Fortin}, {Frailis}, {Fusco},
  {Gargano}, {Gehrels}, {Germani}, {Giavitto}, {Giebels}, {Giglietto},
  {Giordano}, {Glanzman}, {Godfrey}, {Grenier}, {Grondin}, {Grove},
  {Guillemot}, {Guiriec}, {Hanabata}, {Harding}, {Hayashida}, {Hays}, {Hill},
  {Hjalmarsdotter}, {Horan}, {Hughes}, {Jackson}, {J{\'o}hannesson}, {Johnson},
  {Johnson}, {Johnson}, {Kamae}, {Katagiri}, {Kawai}, {Kerr}, {Kn{\"o}dlseder},
  {Kocian}, {Koerding}, {Kuss}, {Lande}, {Latronico}, {Lemoine-Goumard},
  {Longo}, {Loparco}, {Lott}, {Lovellette}, {Lubrano}, {Madejski}, {Makeev},
  {Marchand}, {Marelli}, {Max-Moerbeck}, {Mazziotta}, {McColl}, {McEnery},
  {Meurer}, {Michelson}, {Migliari}, {Mitthumsiri}, {Mizuno}, {Monte},
  {Monzani}, {Morselli}, {Moskalenko}, {Murgia}, {Nolan}, {Norris}, {Nuss},
  {Ohsugi}, {Omodei}, {Ong}, {Ormes}, {Paneque}, {Parent}, {Pelassa}, {Pepe},
  {Pesce-Rollins}, {Piron}, {Pooley}, {Porter}, {Pottschmidt}, {Rain{\`o}},
  {Rando}, {Ray}, {Razzano}, {Rea}, {Readhead}, {Reimer}, {Reimer}, {Richards},
  {Rochester}, {Rodriguez}, {Rodriguez}, {Romani}, {Ryde}, {Sadrozinski},
  {Sander}, {Saz Parkinson}, {Sgr{\`o}}, {Siskind}, {Smith}, {Smith},
  {Spinelli}, {Starck}, {Stevenson}, {Strickman}, {Suson}, {Takahashi},
  {Tanaka}, {Thayer}, {Thompson}, {Tibaldo}, {Tomsick}, {Torres}, {Tosti},
  {Tramacere}, {Uchiyama}, {Usher}, {Vasileiou}, {Vilchez}, {Vitale}, {Waite},
  {Wang}, {Wilms}, {Winer}, {Wood}, {Ylinen}, and {Ziegler}]{Ackermann2009Sci}
{Fermi LAT Collaboration}.; {Abdo}, A.A.; {Ackermann}, M.; {Ajello}, M.;
  {Axelsson}, M.; {Baldini}, L.; {Ballet}, J.; {Barbiellini}, G.; {Bastieri},
  D.; {Baughman}, B.M.;  et~al.
\newblock {Modulated High-Energy Gamma-Ray Emission from the Microquasar Cygnus
  X-3}.
\newblock {\em Science} {\bf 2009}, {\em 326},~1512.
\newblock
  doi:{\changeurlcolor{black}\href{https://doi.org/10.1126/science.1182174}{\detokenize{10.1126/science.1182174}}}.

\bibitem[{Tavani} \em{et~al.}(2009){Tavani}, {Bulgarelli}, {Piano}, {Sabatini},
  {Striani}, {Evangelista}, {Trois}, {Pooley}, {Trushkin}, {Nizhelskij},
  {McCollough}, {Koljonen}, {Pucella}, {Giuliani}, {Chen}, {Costa},
  {Vittorini}, {Trifoglio}, {Gianotti}, {Argan}, {Barbiellini}, {Caraveo},
  {Cattaneo}, {Cocco}, {Contessi}, {D'Ammando}, {Del Monte}, {de Paris}, {Di
  Cocco}, {di Persio}, {Donnarumma}, {Feroci}, {Ferrari}, {Fuschino}, {Galli},
  {Labanti}, {Lapshov}, {Lazzarotto}, {Lipari}, {Longo}, {Mattaini},
  {Marisaldi}, {Mastropietro}, {Mauri}, {Mereghetti}, {Morelli}, {Morselli},
  {Pacciani}, {Pellizzoni}, {Perotti}, {Picozza}, {Pilia}, {Prest},
  {Rapisarda}, {Rappoldi}, {Rossi}, {Rubini}, {Scalise}, {Soffitta},
  {Vallazza}, {Vercellone}, {Zambra}, {Zanello}, {Pittori}, {Verrecchia},
  {Giommi}, {Colafrancesco}, {Santolamazza}, {Antonelli}, and
  {Salotti}]{Tavani2009Natur}
{Tavani}, M.; {Bulgarelli}, A.; {Piano}, G.; {Sabatini}, S.; {Striani}, E.;
  {Evangelista}, Y.; {Trois}, A.; {Pooley}, G.; {Trushkin}, S.; {Nizhelskij},
  N.A.;  et~al.
\newblock {Extreme particle acceleration in the microquasar CygnusX-3}.
\newblock {\em \nat} {\bf 2009}, {\em 462},~620--623,
  \href{http://xxx.lanl.gov/abs/0910.5344}{{\normalfont
  [arXiv:astro-ph.HE/0910.5344]}}.
\newblock
  doi:{\changeurlcolor{black}\href{https://doi.org/10.1038/nature08578}{\detokenize{10.1038/nature08578}}}.

\bibitem[{Dubus} \em{et~al.}(2010){Dubus}, {Cerutti}, and
  {Henri}]{Dubus2010MNRAS}
{Dubus}, G.; {Cerutti}, B.; {Henri}, G.
\newblock {The relativistic jet of Cygnus X-3 in gamma-rays}.
\newblock {\em \mnras} {\bf 2010}, {\em 404},~L55--L59,
  \href{http://xxx.lanl.gov/abs/1002.3888}{{\normalfont
  [arXiv:astro-ph.HE/1002.3888]}}.
\newblock
  doi:{\changeurlcolor{black}\href{https://doi.org/10.1111/j.1745-3933.2010.00834.x}{\detokenize{10.1111/j.1745-3933.2010.00834.x}}}.

\bibitem[{Susa} \em{et~al.}(2014){Susa}, {Hasegawa}, and
  {Tominaga}]{Susa2014ApJ}
{Susa}, H.; {Hasegawa}, K.; {Tominaga}, N.
\newblock {The Mass Spectrum of the First Stars}.
\newblock {\em \apj} {\bf 2014}, {\em 792},~32,
  \href{http://xxx.lanl.gov/abs/1407.1374}{{\normalfont
  [arXiv:astro-ph.GA/1407.1374]}}.
\newblock
  doi:{\changeurlcolor{black}\href{https://doi.org/10.1088/0004-637X/792/1/32}{\detokenize{10.1088/0004-637X/792/1/32}}}.

\bibitem[{Chantavat} \em{et~al.}(2023){Chantavat}, {Chongchitnan}, and
  {Silk}]{Chantavat2023MNRAS}
{Chantavat}, T.; {Chongchitnan}, S.; {Silk}, J.
\newblock {The most massive Population III stars}.
\newblock {\em \mnras} {\bf 2023}, {\em 522},~3256--3262,
  \href{http://xxx.lanl.gov/abs/2302.09763}{{\normalfont
  [arXiv:astro-ph.SR/2302.09763]}}.
\newblock
  doi:{\changeurlcolor{black}\href{https://doi.org/10.1093/mnras/stad1196}{\detokenize{10.1093/mnras/stad1196}}}.

\bibitem[{Haemmerl{\'e}} \em{et~al.}(2018){Haemmerl{\'e}}, {Woods}, {Klessen},
  {Heger}, and {Whalen}]{Haemmerle2018MNRAS}
{Haemmerl{\'e}}, L.; {Woods}, T.E.; {Klessen}, R.S.; {Heger}, A.; {Whalen},
  D.J.
\newblock {The evolution of supermassive Population III stars}.
\newblock {\em \mnras} {\bf 2018}, {\em 474},~2757--2773,
  \href{http://xxx.lanl.gov/abs/1705.09301}{{\normalfont
  [arXiv:astro-ph.SR/1705.09301]}}.
\newblock
  doi:{\changeurlcolor{black}\href{https://doi.org/10.1093/mnras/stx2919}{\detokenize{10.1093/mnras/stx2919}}}.

\bibitem[{Moriya} \em{et~al.}(2019){Moriya}, {Wong}, {Koyama}, {Tanaka},
  {Oguri}, {Hilbert}, and {Nomoto}]{Moriya2019PASJ}
{Moriya}, T.J.; {Wong}, K.C.; {Koyama}, Y.; {Tanaka}, M.; {Oguri}, M.;
  {Hilbert}, S.; {Nomoto}, K.
\newblock {Searches for Population III pair-instability supernovae: Predictions
  for ULTIMATE-Subaru and WFIRST}.
\newblock {\em \pasj} {\bf 2019}, {\em 71},~59,
  \href{http://xxx.lanl.gov/abs/1903.01613}{{\normalfont
  [arXiv:astro-ph.HE/1903.01613]}}.
\newblock
  doi:{\changeurlcolor{black}\href{https://doi.org/10.1093/pasj/psz035}{\detokenize{10.1093/pasj/psz035}}}.

\bibitem[{Sotomayor Checa} and {Romero}(2019)]{Sotomayor_Checa2019}
{Sotomayor Checa}, P.; {Romero}, G.E.
\newblock {Model for Population III microquasars}.
\newblock {\em \aap} {\bf 2019}, {\em 629},~A76,
  \href{http://xxx.lanl.gov/abs/1906.05184}{{\normalfont
  [arXiv:astro-ph.HE/1906.05184]}}.
\newblock
  doi:{\changeurlcolor{black}\href{https://doi.org/10.1051/0004-6361/201834191}{\detokenize{10.1051/0004-6361/201834191}}}.

\bibitem[{Verbunt} and {Zwaan}(1981)]{Verbunt1981A&A}
{Verbunt}, F.; {Zwaan}, C.
\newblock {Magnetic braking in low-mass X-ray binaries.}
\newblock {\em \aap} {\bf 1981}, {\em 100},~L7--L9.

\bibitem[{Eggleton}(1983)]{Eggleton1983ApJ}
{Eggleton}, P.P.
\newblock {Aproximations to the radii of Roche lobes.}
\newblock {\em \apj} {\bf 1983}, {\em 268},~368--369.
\newblock
  doi:{\changeurlcolor{black}\href{https://doi.org/10.1086/160960}{\detokenize{10.1086/160960}}}.

\bibitem[{Bhattacharya} and {van den Heuvel}(1991)]{Bhattacharya1991PhR}
{Bhattacharya}, D.; {van den Heuvel}, E.P.J.
\newblock {Formation and evolution of binary and millisecond radio pulsars}.
\newblock {\em \physrep} {\bf 1991}, {\em 203},~1--124.
\newblock
  doi:{\changeurlcolor{black}\href{https://doi.org/10.1016/0370-1573(91)90064-S}{\detokenize{10.1016/0370-1573(91)90064-S}}}.

\bibitem[{Postnov} and {Yungelson}(2014)]{Postnov2014LRR}
{Postnov}, K.A.; {Yungelson}, L.R.
\newblock {The Evolution of Compact Binary Star Systems}.
\newblock {\em Living Reviews in Relativity} {\bf 2014}, {\em 17},~3,
  \href{http://xxx.lanl.gov/abs/1403.4754}{{\normalfont
  [arXiv:astro-ph.HE/1403.4754]}}.
\newblock
  doi:{\changeurlcolor{black}\href{https://doi.org/10.12942/lrr-2014-3}{\detokenize{10.12942/lrr-2014-3}}}.

\bibitem[{Romero} and {Vila}(2008)]{Romero2008A&A}
{Romero}, G.E.; {Vila}, G.S.
\newblock {The proton low-mass microquasar: high-energy emission}.
\newblock {\em \aap} {\bf 2008}, {\em 485},~623--631,
  \href{http://xxx.lanl.gov/abs/0804.4606}{{\normalfont
  [arXiv:astro-ph/0804.4606]}}.
\newblock
  doi:{\changeurlcolor{black}\href{https://doi.org/10.1051/0004-6361:200809563}{\detokenize{10.1051/0004-6361:200809563}}}.

\bibitem[{Smponias}(2021)]{Smponias2021Galax}
{Smponias}, T.
\newblock {Synthetic Neutrino Imaging of a Microquasar}.
\newblock {\em Galaxies} {\bf 2021}, {\em 9},~80.
\newblock
  doi:{\changeurlcolor{black}\href{https://doi.org/10.3390/galaxies9040080}{\detokenize{10.3390/galaxies9040080}}}.

\bibitem[{M{\"u}cke} and {Protheroe}(2001)]{Mucke2001APh}
{M{\"u}cke}, A.; {Protheroe}, R.J.
\newblock {A proton synchrotron blazar model for flaring in Markarian 501}.
\newblock {\em Astroparticle Physics} {\bf 2001}, {\em 15},~121--136,
  \href{http://xxx.lanl.gov/abs/astro-ph/0004052}{{\normalfont
  [arXiv:astro-ph/astro-ph/0004052]}}.
\newblock
  doi:{\changeurlcolor{black}\href{https://doi.org/10.1016/S0927-6505(00)00141-9}{\detokenize{10.1016/S0927-6505(00)00141-9}}}.

\bibitem[{Reynoso} \em{et~al.}(2011){Reynoso}, {Medina}, and
  {Romero}]{Reynoso2011A&A}
{Reynoso}, M.M.; {Medina}, M.C.; {Romero}, G.E.
\newblock {A lepto-hadronic model for high-energy emission from FR I
  radiogalaxies}.
\newblock {\em \aap} {\bf 2011}, {\em 531},~A30,
  \href{http://xxx.lanl.gov/abs/1005.3025}{{\normalfont
  [arXiv:astro-ph.HE/1005.3025]}}.
\newblock
  doi:{\changeurlcolor{black}\href{https://doi.org/10.1051/0004-6361/201014998}{\detokenize{10.1051/0004-6361/201014998}}}.

\bibitem[{Carulli} \em{et~al.}(2021){Carulli}, {Reynoso}, and
  {Romero}]{Carulli2021APh}
{Carulli}, A.M.; {Reynoso}, M.M.; {Romero}, G.E.
\newblock {Neutrino production in Population III microquasars}.
\newblock {\em Astroparticle Physics} {\bf 2021}, {\em 128},~102557,
  \href{http://xxx.lanl.gov/abs/2101.02999}{{\normalfont
  [arXiv:astro-ph.HE/2101.02999]}}.
\newblock
  doi:{\changeurlcolor{black}\href{https://doi.org/10.1016/j.astropartphys.2021.102557}{\detokenize{10.1016/j.astropartphys.2021.102557}}}.

\bibitem[{Tibaldo} \em{et~al.}(2021){Tibaldo}, {Gaggero}, and
  {Martin}]{Tibaldo2021Univ}
{Tibaldo}, L.; {Gaggero}, D.; {Martin}, P.
\newblock {Gamma Rays as Probes of Cosmic-Ray Propagation and Interactions in
  Galaxies}.
\newblock {\em Universe} {\bf 2021}, {\em 7},~141,
  \href{http://xxx.lanl.gov/abs/2103.16423}{{\normalfont
  [arXiv:astro-ph.HE/2103.16423]}}.
\newblock
  doi:{\changeurlcolor{black}\href{https://doi.org/10.3390/universe7050141}{\detokenize{10.3390/universe7050141}}}.

\bibitem[{Tsuboi} \em{et~al.}(1999){Tsuboi}, {Handa}, and
  {Ukita}]{Tsuboi99ApJS}
{Tsuboi}, M.; {Handa}, T.; {Ukita}, N.
\newblock {Dense Molecular Clouds in the Galactic Center Region. I.
  Observations and Data}.
\newblock {\em \apjs} {\bf 1999}, {\em 120},~1--39.
\newblock
  doi:{\changeurlcolor{black}\href{https://doi.org/10.1086/313165}{\detokenize{10.1086/313165}}}.

\bibitem[{Molinari} \em{et~al.}(2011){Molinari}, {Bally}, {Noriega-Crespo},
  {Compi{\`e}gne}, {Bernard}, {Paradis}, {Martin}, {Testi}, {Barlow}, {Moore},
  {Plume}, {Swinyard}, {Zavagno}, {Calzoletti}, {Di Giorgio}, {Elia},
  {Faustini}, {Natoli}, {Pestalozzi}, {Pezzuto}, {Piacentini}, {Polenta},
  {Polychroni}, {Schisano}, {Traficante}, {Veneziani}, {Battersby}, {Burton},
  {Carey}, {Fukui}, {Li}, {Lord}, {Morgan}, {Motte}, {Schuller},
  {Stringfellow}, {Tan}, {Thompson}, {Ward-Thompson}, {White}, and
  {Umana}]{Molinari11ApJL}
{Molinari}, S.; {Bally}, J.; {Noriega-Crespo}, A.; {Compi{\`e}gne}, M.;
  {Bernard}, J.P.; {Paradis}, D.; {Martin}, P.; {Testi}, L.; {Barlow}, M.;
  {Moore}, T.;  et~al.
\newblock {A 100 pc Elliptical and Twisted Ring of Cold and Dense Molecular
  Clouds Revealed by Herschel Around the Galactic Center}.
\newblock {\em \apjl} {\bf 2011}, {\em 735},~L33,
  \href{http://xxx.lanl.gov/abs/1105.5486}{{\normalfont
  [arXiv:astro-ph.GA/1105.5486]}}.
\newblock
  doi:{\changeurlcolor{black}\href{https://doi.org/10.1088/2041-8205/735/2/L33}{\detokenize{10.1088/2041-8205/735/2/L33}}}.

\bibitem[{Yusef-Zadeh} \em{et~al.}(2013){Yusef-Zadeh}, {Hewitt}, {Wardle},
  {Tatischeff}, {Roberts}, {Cotton}, {Uchiyama}, {Nobukawa}, {Tsuru}, {Heinke},
  and {Royster}]{YZ2013ApJ}
{Yusef-Zadeh}, F.; {Hewitt}, J.W.; {Wardle}, M.; {Tatischeff}, V.; {Roberts},
  D.A.; {Cotton}, W.; {Uchiyama}, H.; {Nobukawa}, M.; {Tsuru}, T.G.; {Heinke},
  C.;  et~al.
\newblock {Interacting Cosmic Rays with Molecular Clouds: A Bremsstrahlung
  Origin of Diffuse High-energy Emission from the Inner
  2{\textdegree}{\texttimes}1{\textdegree} of the Galactic Center}.
\newblock {\em \apj} {\bf 2013}, {\em 762},~33,
  \href{http://xxx.lanl.gov/abs/1206.6882}{{\normalfont
  [arXiv:astro-ph.HE/1206.6882]}}.
\newblock
  doi:{\changeurlcolor{black}\href{https://doi.org/10.1088/0004-637X/762/1/33}{\detokenize{10.1088/0004-637X/762/1/33}}}.

\bibitem[{Yoast-Hull} \em{et~al.}(2014){Yoast-Hull}, {Gallagher}, and
  {Zweibel}]{YH2014ApJ}
{Yoast-Hull}, T.M.; {Gallagher}, J.~S., I.; {Zweibel}, E.G.
\newblock {The Cosmic-Ray Population of the Galactic Central Molecular Zone}.
\newblock {\em \apj} {\bf 2014}, {\em 790},~86,
  \href{http://xxx.lanl.gov/abs/1405.7059}{{\normalfont
  [arXiv:astro-ph.HE/1405.7059]}}.
\newblock
  doi:{\changeurlcolor{black}\href{https://doi.org/10.1088/0004-637X/790/2/86}{\detokenize{10.1088/0004-637X/790/2/86}}}.

\bibitem[{Crocker} \em{et~al.}(2010){Crocker}, {Jones}, {Melia}, {Ott}, and
  {Protheroe}]{Crocker2010Natur}
{Crocker}, R.M.; {Jones}, D.I.; {Melia}, F.; {Ott}, J.; {Protheroe}, R.J.
\newblock {A lower limit of 50 microgauss for the magnetic field near the
  Galactic Centre}.
\newblock {\em \nat} {\bf 2010}, {\em 463},~65--67,
  \href{http://xxx.lanl.gov/abs/1001.1275}{{\normalfont
  [arXiv:astro-ph.GA/1001.1275]}}.
\newblock
  doi:{\changeurlcolor{black}\href{https://doi.org/10.1038/nature08635}{\detokenize{10.1038/nature08635}}}.

\bibitem[{Calore} \em{et~al.}(2016){Calore}, {Di Mauro}, {Donato}, {Hessels},
  and {Weniger}]{Calore2016ApJ}
{Calore}, F.; {Di Mauro}, M.; {Donato}, F.; {Hessels}, J.W.T.; {Weniger}, C.
\newblock {Radio Detection Prospects for a Bulge Population of Millisecond
  Pulsars as Suggested by Fermi-LAT Observations of the Inner Galaxy}.
\newblock {\em \apj} {\bf 2016}, {\em 827},~143,
  \href{http://xxx.lanl.gov/abs/1512.06825}{{\normalfont
  [arXiv:astro-ph.HE/1512.06825]}}.
\newblock
  doi:{\changeurlcolor{black}\href{https://doi.org/10.3847/0004-637X/827/2/143}{\detokenize{10.3847/0004-637X/827/2/143}}}.

\bibitem[{Ackermann} \em{et~al.}(2014){Ackermann}, {Albert}, {Atwood},
  {Baldini}, {Ballet}, {Barbiellini}, {Bastieri}, {Bellazzini}, {Bissaldi},
  {Blandford}, {Bloom}, {Bottacini}, {Brandt}, {Bregeon}, {Bruel}, {Buehler},
  {Buson}, {Caliandro}, {Cameron}, {Caragiulo}, {Caraveo}, {Cavazzuti},
  {Cecchi}, {Charles}, {Chekhtman}, {Chiang}, {Chiaro}, {Ciprini}, {Claus},
  {Cohen-Tanugi}, {Conrad}, {Cutini}, {D'Ammando}, {de Angelis}, {de Palma},
  {Dermer}, {Digel}, {Di Venere}, {Silva}, {Drell}, {Favuzzi}, {Ferrara},
  {Focke}, {Franckowiak}, {Fukazawa}, {Funk}, {Fusco}, {Gargano}, {Gasparrini},
  {Germani}, {Giglietto}, {Giordano}, {Giroletti}, {Godfrey}, {Gomez-Vargas},
  {Grenier}, {Guiriec}, {Hadasch}, {Harding}, {Hays}, {Hewitt}, {Hou},
  {Jogler}, {J{\'o}hannesson}, {Johnson}, {Johnson}, {Kamae}, {Kataoka},
  {Kn{\"o}dlseder}, {Kocevski}, {Kuss}, {Larsson}, {Latronico}, {Longo},
  {Loparco}, {Lovellette}, {Lubrano}, {Malyshev}, {Manfreda}, {Massaro},
  {Mayer}, {Mazziotta}, {McEnery}, {Michelson}, {Mitthumsiri}, {Mizuno},
  {Monzani}, {Morselli}, {Moskalenko}, {Murgia}, {Nemmen}, {Nuss}, {Ohsugi},
  {Omodei}, {Orienti}, {Orlando}, {Ormes}, {Paneque}, {Panetta}, {Perkins},
  {Pesce-Rollins}, {Petrosian}, {Piron}, {Pivato}, {Rain{\`o}}, {Rando},
  {Razzano}, {Razzaque}, {Reimer}, {Reimer}, {S{\'a}nchez-Conde}, {Schaal},
  {Schulz}, {Sgr{\`o}}, {Siskind}, {Spandre}, {Spinelli}, {Stawarz}, {Strong},
  {Suson}, {Tahara}, {Takahashi}, {Thayer}, {Tibaldo}, {Tinivella}, {Torres},
  {Tosti}, {Troja}, {Uchiyama}, {Vianello}, {Werner}, {Winer}, {Wood}, {Wood},
  and {Zaharijas}]{Ackermann2014ApJ}
{Ackermann}, M.; {Albert}, A.; {Atwood}, W.B.; {Baldini}, L.; {Ballet}, J.;
  {Barbiellini}, G.; {Bastieri}, D.; {Bellazzini}, R.; {Bissaldi}, E.;
  {Blandford}, R.D.;  et~al.
\newblock {The Spectrum and Morphology of the Fermi Bubbles}.
\newblock {\em \apj} {\bf 2014}, {\em 793},~64,
  \href{http://xxx.lanl.gov/abs/1407.7905}{{\normalfont
  [arXiv:astro-ph.HE/1407.7905]}}.
\newblock
  doi:{\changeurlcolor{black}\href{https://doi.org/10.1088/0004-637X/793/1/64}{\detokenize{10.1088/0004-637X/793/1/64}}}.

\bibitem[{Yang} \em{et~al.}(2018){Yang}, {Ruszkowski}, and
  {Zweibel}]{Yang2018Galax}
{Yang}, H.Y.; {Ruszkowski}, M.; {Zweibel}, E.
\newblock {Unveiling the Origin of the Fermi Bubbles}.
\newblock {\em Galaxies} {\bf 2018}, {\em 6},~29,
  \href{http://xxx.lanl.gov/abs/1802.03890}{{\normalfont
  [arXiv:astro-ph.HE/1802.03890]}}.
\newblock
  doi:{\changeurlcolor{black}\href{https://doi.org/10.3390/galaxies6010029}{\detokenize{10.3390/galaxies6010029}}}.

\bibitem[Yang \em{et~al.}(2023)Yang, Ruszkowski, and Zweibel]{Yang2023}
Yang, H.Y.K.; Ruszkowski, M.; Zweibel, E.
\newblock {Unveiling the Origin of the Fermi/eRosita Bubbles}.
\newblock {\em PoS} {\bf 2023}, {\em ECRS},~023.
\newblock
  doi:{\changeurlcolor{black}\href{https://doi.org/10.22323/1.423.0023}{\detokenize{10.22323/1.423.0023}}}.

\bibitem[{Crocker}(2012)]{Crocker2012MNRAS}
{Crocker}, R.M.
\newblock {Non-thermal insights on mass and energy flows through the Galactic
  Centre and into the Fermi bubbles}.
\newblock {\em \mnras} {\bf 2012}, {\em 423},~3512--3539,
  \href{http://xxx.lanl.gov/abs/1112.6247}{{\normalfont
  [arXiv:astro-ph.GA/1112.6247]}}.
\newblock
  doi:{\changeurlcolor{black}\href{https://doi.org/10.1111/j.1365-2966.2012.21149.x}{\detokenize{10.1111/j.1365-2966.2012.21149.x}}}.

\bibitem[{Razzaque} and {Yang}(2018)]{Razzaque2018Galax}
{Razzaque}, S.; {Yang}, L.
\newblock {Hadronic Models of the Fermi Bubbles: Future Perspectives}.
\newblock {\em Galaxies} {\bf 2018}, {\em 6},~47,
  \href{http://xxx.lanl.gov/abs/1802.05636}{{\normalfont
  [arXiv:astro-ph.HE/1802.05636]}}.
\newblock
  doi:{\changeurlcolor{black}\href{https://doi.org/10.3390/galaxies6020047}{\detokenize{10.3390/galaxies6020047}}}.

\bibitem[{Yang} \em{et~al.}(2012){Yang}, {Ruszkowski}, {Ricker}, {Zweibel}, and
  {Lee}]{Yang2012ApJ}
{Yang}, H.Y.K.; {Ruszkowski}, M.; {Ricker}, P.M.; {Zweibel}, E.; {Lee}, D.
\newblock {The Fermi Bubbles: Supersonic Active Galactic Nucleus Jets with
  Anisotropic Cosmic-Ray Diffusion}.
\newblock {\em \apj} {\bf 2012}, {\em 761},~185,
  \href{http://xxx.lanl.gov/abs/1207.4185}{{\normalfont
  [arXiv:astro-ph.GA/1207.4185]}}.
\newblock
  doi:{\changeurlcolor{black}\href{https://doi.org/10.1088/0004-637X/761/2/185}{\detokenize{10.1088/0004-637X/761/2/185}}}.

\bibitem[{Su} \em{et~al.}(2010){Su}, {Slatyer}, and {Finkbeiner}]{Su2010ApJ}
{Su}, M.; {Slatyer}, T.R.; {Finkbeiner}, D.P.
\newblock {Giant Gamma-ray Bubbles from Fermi-LAT: Active Galactic Nucleus
  Activity or Bipolar Galactic Wind?}
\newblock {\em \apj} {\bf 2010}, {\em 724},~1044--1082,
  \href{http://xxx.lanl.gov/abs/1005.5480}{{\normalfont
  [arXiv:astro-ph.HE/1005.5480]}}.
\newblock
  doi:{\changeurlcolor{black}\href{https://doi.org/10.1088/0004-637X/724/2/1044}{\detokenize{10.1088/0004-637X/724/2/1044}}}.

\bibitem[{Zubovas} \em{et~al.}(2011){Zubovas}, {King}, and
  {Nayakshin}]{Zubovas2011MNRAS}
{Zubovas}, K.; {King}, A.R.; {Nayakshin}, S.
\newblock {The Milky Way's Fermi bubbles: echoes of the last quasar outburst?}
\newblock {\em \mnras} {\bf 2011}, {\em 415},~L21--L25,
  \href{http://xxx.lanl.gov/abs/1104.5443}{{\normalfont
  [arXiv:astro-ph.GA/1104.5443]}}.
\newblock
  doi:{\changeurlcolor{black}\href{https://doi.org/10.1111/j.1745-3933.2011.01070.x}{\detokenize{10.1111/j.1745-3933.2011.01070.x}}}.

\bibitem[{Tourmente} \em{et~al.}(2023){Tourmente}, {Rodgers-Lee}, and
  {Taylor}]{Tourmente2023MNRAS}
{Tourmente}, O.; {Rodgers-Lee}, D.; {Taylor}, A.M.
\newblock {A galactic breeze origin for the Fermi bubbles emission}.
\newblock {\em \mnras} {\bf 2023}, {\em 518},~6083--6091,
  \href{http://xxx.lanl.gov/abs/2207.09189}{{\normalfont
  [arXiv:astro-ph.HE/2207.09189]}}.
\newblock
  doi:{\changeurlcolor{black}\href{https://doi.org/10.1093/mnras/stac3517}{\detokenize{10.1093/mnras/stac3517}}}.

\bibitem[{Taylor} and {Giacinti}(2017)]{Taylor2017PhRvD}
{Taylor}, A.M.; {Giacinti}, G.
\newblock {Cosmic rays in a galactic breeze}.
\newblock {\em \prd} {\bf 2017}, {\em 95},~023001,
  \href{http://xxx.lanl.gov/abs/1607.08862}{{\normalfont
  [arXiv:astro-ph.HE/1607.08862]}}.
\newblock
  doi:{\changeurlcolor{black}\href{https://doi.org/10.1103/PhysRevD.95.023001}{\detokenize{10.1103/PhysRevD.95.023001}}}.

\bibitem[{Guo} and {Mathews}(2012)]{Guo2012}
{Guo}, F.; {Mathews}, W.G.
\newblock {The Fermi Bubbles. I. Possible Evidence for Recent AGN Jet Activity
  in the Galaxy}.
\newblock {\em \apj} {\bf 2012}, {\em 756},~181,
  \href{http://xxx.lanl.gov/abs/1103.0055}{{\normalfont
  [arXiv:astro-ph.HE/1103.0055]}}.
\newblock
  doi:{\changeurlcolor{black}\href{https://doi.org/10.1088/0004-637X/756/2/181}{\detokenize{10.1088/0004-637X/756/2/181}}}.

\bibitem[{Yang} \em{et~al.}(2013){Yang}, {Ruszkowski}, and {Zweibel}]{Yang2013}
{Yang}, H.Y.K.; {Ruszkowski}, M.; {Zweibel}, E.
\newblock {The Fermi bubbles: gamma-ray, microwave and polarization signatures
  of leptonic AGN jets}.
\newblock {\em \mnras} {\bf 2013}, {\em 436},~2734--2746,
  \href{http://xxx.lanl.gov/abs/1307.3551}{{\normalfont
  [arXiv:astro-ph.GA/1307.3551]}}.
\newblock
  doi:{\changeurlcolor{black}\href{https://doi.org/10.1093/mnras/stt1772}{\detokenize{10.1093/mnras/stt1772}}}.

\bibitem[{Yang} and {Ruszkowski}(2017)]{Yang2017ApJ}
{Yang}, H.Y.K.; {Ruszkowski}, M.
\newblock {The Spatially Uniform Spectrum of the Fermi Bubbles: The Leptonic
  Active Galactic Nucleus Jet Scenario}.
\newblock {\em \apj} {\bf 2017}, {\em 850},~2,
  \href{http://xxx.lanl.gov/abs/1706.05025}{{\normalfont
  [arXiv:astro-ph.HE/1706.05025]}}.
\newblock
  doi:{\changeurlcolor{black}\href{https://doi.org/10.3847/1538-4357/aa9434}{\detokenize{10.3847/1538-4357/aa9434}}}.

\bibitem[{Yang} \em{et~al.}(2022){Yang}, {Ruszkowski}, and {Zweibel}]{Yang2022}
{Yang}, H.Y.K.; {Ruszkowski}, M.; {Zweibel}, E.G.
\newblock {Fermi and eROSITA bubbles as relics of the past activity of the
  Galaxy's central black hole}.
\newblock {\em Nature Astronomy} {\bf 2022}, {\em 6},~584--591,
  \href{http://xxx.lanl.gov/abs/2203.02526}{{\normalfont
  [arXiv:astro-ph.HE/2203.02526]}}.
\newblock
  doi:{\changeurlcolor{black}\href{https://doi.org/10.1038/s41550-022-01618-x}{\detokenize{10.1038/s41550-022-01618-x}}}.

\bibitem[{Mulcahy} \em{et~al.}(2018){Mulcahy}, {Horneffer}, {Beck}, {Krause},
  {Schmidt}, {Basu}, {Chy{\.z}y}, {Dettmar}, {Haverkorn}, {Heald}, {Heesen},
  {Horellou}, {Iacobelli}, {Nikiel-Wroczy{\'n}ski}, {Paladino}, {Scaife},
  {Sridhar}, {Strom}, {Tabatabaei}, {Cantwell}, {Carey}, {Grainge}, {Hickish},
  {Perrot}, {Razavi-Ghods}, {Scott}, and {Titterington}]{Mulcahy2018A&A}
{Mulcahy}, D.D.; {Horneffer}, A.; {Beck}, R.; {Krause}, M.; {Schmidt}, P.;
  {Basu}, A.; {Chy{\.z}y}, K.T.; {Dettmar}, R.J.; {Haverkorn}, M.; {Heald}, G.;
   et~al.
\newblock {Investigation of the cosmic ray population and magnetic field
  strength in the halo of NGC 891}.
\newblock {\em \aap} {\bf 2018}, {\em 615},~A98,
  \href{http://xxx.lanl.gov/abs/1804.00752}{{\normalfont
  [arXiv:astro-ph.GA/1804.00752]}}.
\newblock
  doi:{\changeurlcolor{black}\href{https://doi.org/10.1051/0004-6361/201832837}{\detokenize{10.1051/0004-6361/201832837}}}.

\bibitem[{Butsky} and {Quinn}(2018)]{Butsky2018ApJ}
{Butsky}, I.S.; {Quinn}, T.R.
\newblock {The Role of Cosmic-ray Transport in Shaping the Simulated
  Circumgalactic Medium}.
\newblock {\em \apj} {\bf 2018}, {\em 868},~108,
  \href{http://xxx.lanl.gov/abs/1803.06345}{{\normalfont
  [arXiv:astro-ph.GA/1803.06345]}}.
\newblock
  doi:{\changeurlcolor{black}\href{https://doi.org/10.3847/1538-4357/aaeac2}{\detokenize{10.3847/1538-4357/aaeac2}}}.

\bibitem[{Dashyan} and {Dubois}(2020)]{Dashyan2020A&A}
{Dashyan}, G.; {Dubois}, Y.
\newblock {Cosmic ray feedback from supernovae in dwarf galaxies}.
\newblock {\em \aap} {\bf 2020}, {\em 638},~A123,
  \href{http://xxx.lanl.gov/abs/2003.09900}{{\normalfont
  [arXiv:astro-ph.GA/2003.09900]}}.
\newblock
  doi:{\changeurlcolor{black}\href{https://doi.org/10.1051/0004-6361/201936339}{\detokenize{10.1051/0004-6361/201936339}}}.

\bibitem[{Ji} \em{et~al.}(2020){Ji}, {Chan}, {Hummels}, {Hopkins}, {Stern},
  {Kere{\v{s}}}, {Quataert}, {Faucher-Gigu{\`e}re}, and {Murray}]{Ji2020MNRASd}
{Ji}, S.; {Chan}, T.K.; {Hummels}, C.B.; {Hopkins}, P.F.; {Stern}, J.;
  {Kere{\v{s}}}, D.; {Quataert}, E.; {Faucher-Gigu{\`e}re}, C.A.; {Murray}, N.
\newblock {Properties of the circumgalactic medium in cosmic ray-dominated
  galaxy haloes}.
\newblock {\em \mnras} {\bf 2020}, {\em 496},~4221--4238,
  \href{http://xxx.lanl.gov/abs/1909.00003}{{\normalfont
  [arXiv:astro-ph.GA/1909.00003]}}.
\newblock
  doi:{\changeurlcolor{black}\href{https://doi.org/10.1093/mnras/staa1849}{\detokenize{10.1093/mnras/staa1849}}}.

\bibitem[{Recchia} \em{et~al.}(2021){Recchia}, {Gabici}, {Aharonian}, and
  {Niro}]{Recchia2021ApJ}
{Recchia}, S.; {Gabici}, S.; {Aharonian}, F.A.; {Niro}, V.
\newblock {Giant Cosmic-Ray Halos around M31 and the Milky Way}.
\newblock {\em \apj} {\bf 2021}, {\em 914},~135,
  \href{http://xxx.lanl.gov/abs/2101.05016}{{\normalfont
  [arXiv:astro-ph.HE/2101.05016]}}.
\newblock
  doi:{\changeurlcolor{black}\href{https://doi.org/10.3847/1538-4357/abfda4}{\detokenize{10.3847/1538-4357/abfda4}}}.

\bibitem[{Pshirkov} \em{et~al.}(2016){Pshirkov}, {Vasiliev}, and
  {Postnov}]{Pshirkov2016MNRAS}
{Pshirkov}, M.S.; {Vasiliev}, V.V.; {Postnov}, K.A.
\newblock {Evidence of Fermi bubbles around M31}.
\newblock {\em \mnras} {\bf 2016}, {\em 459},~L76--L80,
  \href{http://xxx.lanl.gov/abs/1603.07245}{{\normalfont
  [arXiv:astro-ph.HE/1603.07245]}}.
\newblock
  doi:{\changeurlcolor{black}\href{https://doi.org/10.1093/mnrasl/slw045}{\detokenize{10.1093/mnrasl/slw045}}}.

\bibitem[{Roy} and {Nath}(2022)]{Roy2022MNRAS}
{Roy}, M.; {Nath}, B.B.
\newblock {Gamma-rays from the circumgalactic medium of M31}.
\newblock {\em \mnras} {\bf 2022}, {\em 514},~1412--1421,
  \href{http://xxx.lanl.gov/abs/2205.12291}{{\normalfont
  [arXiv:astro-ph.GA/2205.12291]}}.
\newblock
  doi:{\changeurlcolor{black}\href{https://doi.org/10.1093/mnras/stac1465}{\detokenize{10.1093/mnras/stac1465}}}.

\bibitem[{Tibaldo} \em{et~al.}(2015){Tibaldo}, {Digel}, {Casandjian},
  {Franckowiak}, {Grenier}, {J{\'o}hannesson}, {Marshall}, {Moskalenko},
  {Negro}, {Orlando}, {Porter}, {Reimer}, and {Strong}]{Tibaldo2015ApJ}
{Tibaldo}, L.; {Digel}, S.W.; {Casandjian}, J.M.; {Franckowiak}, A.; {Grenier},
  I.A.; {J{\'o}hannesson}, G.; {Marshall}, D.J.; {Moskalenko}, I.V.; {Negro},
  M.; {Orlando}, E.;  et~al.
\newblock {Fermi-LAT Observations of High- and Intermediate-velocity Clouds:
  Tracing Cosmic Rays in the Halo of the Milky Way}.
\newblock {\em \apj} {\bf 2015}, {\em 807},~161,
  \href{http://xxx.lanl.gov/abs/1505.04223}{{\normalfont
  [arXiv:astro-ph.HE/1505.04223]}}.
\newblock
  doi:{\changeurlcolor{black}\href{https://doi.org/10.1088/0004-637X/807/2/161}{\detokenize{10.1088/0004-637X/807/2/161}}}.

\bibitem[{Subrahmanyan} and {Cowsik}(2013)]{Subrahmanyan2013ApJ}
{Subrahmanyan}, R.; {Cowsik}, R.
\newblock {Is there an Unaccounted for Excess in the Extragalactic Cosmic Radio
  Background?}
\newblock {\em \apj} {\bf 2013}, {\em 776},~42,
  \href{http://xxx.lanl.gov/abs/1305.7060}{{\normalfont
  [arXiv:astro-ph.CO/1305.7060]}}.
\newblock
  doi:{\changeurlcolor{black}\href{https://doi.org/10.1088/0004-637X/776/1/42}{\detokenize{10.1088/0004-637X/776/1/42}}}.

\bibitem[{Jana} \em{et~al.}(2020){Jana}, {Roy}, and {Nath}]{Jana2020ApJd}
{Jana}, R.; {Roy}, M.; {Nath}, B.B.
\newblock {Gamma-Ray and Radio Background Constraints on Cosmic Rays in Milky
  Way Circumgalactic Medium}.
\newblock {\em \apjl} {\bf 2020}, {\em 903},~L9,
  \href{http://xxx.lanl.gov/abs/2007.11015}{{\normalfont
  [arXiv:astro-ph.GA/2007.11015]}}.
\newblock
  doi:{\changeurlcolor{black}\href{https://doi.org/10.3847/2041-8213/abbee4}{\detokenize{10.3847/2041-8213/abbee4}}}.

\bibitem[{Fixsen} \em{et~al.}(2011){Fixsen}, {Kogut}, {Levin}, {Limon},
  {Lubin}, {Mirel}, {Seiffert}, {Singal}, {Wollack}, {Villela}, and
  {Wuensche}]{Fixsen2011ApJ}
{Fixsen}, D.J.; {Kogut}, A.; {Levin}, S.; {Limon}, M.; {Lubin}, P.; {Mirel},
  P.; {Seiffert}, M.; {Singal}, J.; {Wollack}, E.; {Villela}, T.;  et~al.
\newblock {ARCADE 2 Measurement of the Absolute Sky Brightness at 3-90 GHz}.
\newblock {\em \apj} {\bf 2011}, {\em 734},~5,
  \href{http://xxx.lanl.gov/abs/0901.0555}{{\normalfont
  [arXiv:astro-ph.CO/0901.0555]}}.
\newblock
  doi:{\changeurlcolor{black}\href{https://doi.org/10.1088/0004-637X/734/1/5}{\detokenize{10.1088/0004-637X/734/1/5}}}.

\bibitem[{Joubaud} \em{et~al.}(2020){Joubaud}, {Grenier}, {Casandjian},
  {Tolksdorf}, and {Schlickeiser}]{Joubaud2020A&A}
{Joubaud}, T.; {Grenier}, I.A.; {Casandjian}, J.M.; {Tolksdorf}, T.;
  {Schlickeiser}, R.
\newblock {The cosmic-ray content of the Orion-Eridanus superbubble}.
\newblock {\em \aap} {\bf 2020}, {\em 635},~A96,
  \href{http://xxx.lanl.gov/abs/2001.10139}{{\normalfont
  [arXiv:astro-ph.HE/2001.10139]}}.
\newblock
  doi:{\changeurlcolor{black}\href{https://doi.org/10.1051/0004-6361/201937205}{\detokenize{10.1051/0004-6361/201937205}}}.

\bibitem[{Feldmann} \em{et~al.}(2013){Feldmann}, {Hooper}, and
  {Gnedin}]{Feldmann2013ApJ}
{Feldmann}, R.; {Hooper}, D.; {Gnedin}, N.Y.
\newblock {Circum-galactic Gas and the Isotropic Gamma-Ray Background}.
\newblock {\em \apj} {\bf 2013}, {\em 763},~21,
  \href{http://xxx.lanl.gov/abs/1205.0249}{{\normalfont
  [arXiv:astro-ph.HE/1205.0249]}}.
\newblock
  doi:{\changeurlcolor{black}\href{https://doi.org/10.1088/0004-637X/763/1/21}{\detokenize{10.1088/0004-637X/763/1/21}}}.

\bibitem[{Blasi} and {Amato}(2019)]{Blasi2019PhRvL}
{Blasi}, P.; {Amato}, E.
\newblock {Escape of Cosmic Rays from the Galaxy and Effects on the
  Circumgalactic Medium}.
\newblock {\em \prl} {\bf 2019}, {\em 122},~051101,
  \href{http://xxx.lanl.gov/abs/1901.03609}{{\normalfont
  [arXiv:astro-ph.HE/1901.03609]}}.
\newblock
  doi:{\changeurlcolor{black}\href{https://doi.org/10.1103/PhysRevLett.122.051101}{\detokenize{10.1103/PhysRevLett.122.051101}}}.

\bibitem[{Taylor} \em{et~al.}(2014){Taylor}, {Gabici}, and
  {Aharonian}]{Taylor2014PhRvD}
{Taylor}, A.M.; {Gabici}, S.; {Aharonian}, F.
\newblock {Galactic halo origin of the neutrinos detected by IceCube}.
\newblock {\em \prd} {\bf 2014}, {\em 89},~103003,
  \href{http://xxx.lanl.gov/abs/1403.3206}{{\normalfont
  [arXiv:astro-ph.HE/1403.3206]}}.
\newblock
  doi:{\changeurlcolor{black}\href{https://doi.org/10.1103/PhysRevD.89.103003}{\detokenize{10.1103/PhysRevD.89.103003}}}.

\bibitem[{Kalashev} \em{et~al.}(2023){Kalashev}, {Martynenko}, and
  {Troitsky}]{Kalashev2023JCAP}
{Kalashev}, O.; {Martynenko}, N.; {Troitsky}, S.
\newblock {On the contribution of cosmic-ray interactions in the circumgalactic
  gas to the observed high-energy neutrino flux}.
\newblock {\em \jcap} {\bf 2023}, {\em 2023},~053,
  \href{http://xxx.lanl.gov/abs/2207.12458}{{\normalfont
  [arXiv:astro-ph.HE/2207.12458]}}.
\newblock
  doi:{\changeurlcolor{black}\href{https://doi.org/10.1088/1475-7516/2023/03/053}{\detokenize{10.1088/1475-7516/2023/03/053}}}.

\bibitem[{Recchia} \em{et~al.}(2016){Recchia}, {Blasi}, and
  {Morlino}]{Recchia2016MNRAS}
{Recchia}, S.; {Blasi}, P.; {Morlino}, G.
\newblock {Cosmic ray driven Galactic winds}.
\newblock {\em \mnras} {\bf 2016}, {\em 462},~4227--4239,
  \href{http://xxx.lanl.gov/abs/1603.06746}{{\normalfont
  [arXiv:astro-ph.HE/1603.06746]}}.
\newblock
  doi:{\changeurlcolor{black}\href{https://doi.org/10.1093/mnras/stw1966}{\detokenize{10.1093/mnras/stw1966}}}.

\bibitem[{Holguin} \em{et~al.}(2019){Holguin}, {Ruszkowski}, {Lazarian},
  {Farber}, and {Yang}]{Holguin2019MNRAS}
{Holguin}, F.; {Ruszkowski}, M.; {Lazarian}, A.; {Farber}, R.; {Yang}, H.Y.K.
\newblock {Role of cosmic-ray streaming and turbulent damping in driving
  galactic winds}.
\newblock {\em \mnras} {\bf 2019}, {\em 490},~1271--1282,
  \href{http://xxx.lanl.gov/abs/1807.05494}{{\normalfont
  [arXiv:astro-ph.HE/1807.05494]}}.
\newblock
  doi:{\changeurlcolor{black}\href{https://doi.org/10.1093/mnras/stz2568}{\detokenize{10.1093/mnras/stz2568}}}.

\bibitem[{Dogiel} \em{et~al.}(2020){Dogiel}, {Ivlev}, {Chernyshov}, and
  {Ko}]{Dogiel2020ApJ}
{Dogiel}, V.A.; {Ivlev}, A.V.; {Chernyshov}, D.O.; {Ko}, C.M.
\newblock {Formation of the Cosmic-Ray Halo: Galactic Spectrum of Primary
  Cosmic Rays}.
\newblock {\em \apj} {\bf 2020}, {\em 903},~135,
  \href{http://xxx.lanl.gov/abs/2009.08799}{{\normalfont
  [arXiv:astro-ph.HE/2009.08799]}}.
\newblock
  doi:{\changeurlcolor{black}\href{https://doi.org/10.3847/1538-4357/abba31}{\detokenize{10.3847/1538-4357/abba31}}}.

\bibitem[{Kempski} and {Quataert}(2022)]{Kempski2022MNRAS}
{Kempski}, P.; {Quataert}, E.
\newblock {Reconciling cosmic ray transport theory with phenomenological models
  motivated by Milky-Way data}.
\newblock {\em \mnras} {\bf 2022}, {\em 514},~657--674,
  \href{http://xxx.lanl.gov/abs/2109.10977}{{\normalfont
  [arXiv:astro-ph.HE/2109.10977]}}.
\newblock
  doi:{\changeurlcolor{black}\href{https://doi.org/10.1093/mnras/stac1240}{\detokenize{10.1093/mnras/stac1240}}}.

\bibitem[{Evoli} \em{et~al.}(2018){Evoli}, {Blasi}, {Morlino}, and
  {Aloisio}]{Evoli2018PhRvL}
{Evoli}, C.; {Blasi}, P.; {Morlino}, G.; {Aloisio}, R.
\newblock {Origin of the Cosmic Ray Galactic Halo Driven by Advected Turbulence
  and Self-Generated Waves}.
\newblock {\em \prl} {\bf 2018}, {\em 121},~021102,
  \href{http://xxx.lanl.gov/abs/1806.04153}{{\normalfont
  [arXiv:astro-ph.HE/1806.04153]}}.
\newblock
  doi:{\changeurlcolor{black}\href{https://doi.org/10.1103/PhysRevLett.121.021102}{\detokenize{10.1103/PhysRevLett.121.021102}}}.

\bibitem[{Schober} \em{et~al.}(2013){Schober}, {Schleicher}, and
  {Klessen}]{Schober2013A&A}
{Schober}, J.; {Schleicher}, D.R.G.; {Klessen}, R.S.
\newblock {Magnetic field amplification in young galaxies}.
\newblock {\em \aap} {\bf 2013}, {\em 560},~A87,
  \href{http://xxx.lanl.gov/abs/1310.0853}{{\normalfont
  [arXiv:astro-ph.GA/1310.0853]}}.
\newblock
  doi:{\changeurlcolor{black}\href{https://doi.org/10.1051/0004-6361/201322185}{\detokenize{10.1051/0004-6361/201322185}}}.

\bibitem[{Bernet} \em{et~al.}(2008){Bernet}, {Miniati}, {Lilly}, {Kronberg},
  and {Dessauges-Zavadsky}]{Bernet2008Natur}
{Bernet}, M.L.; {Miniati}, F.; {Lilly}, S.J.; {Kronberg}, P.P.;
  {Dessauges-Zavadsky}, M.
\newblock {Strong magnetic fields in normal galaxies at high redshift}.
\newblock {\em \nat} {\bf 2008}, {\em 454},~302--304,
  \href{http://xxx.lanl.gov/abs/0807.3347}{{\normalfont
  [arXiv:astro-ph/0807.3347]}}.
\newblock
  doi:{\changeurlcolor{black}\href{https://doi.org/10.1038/nature07105}{\detokenize{10.1038/nature07105}}}.

\bibitem[{Hammond} \em{et~al.}(2012){Hammond}, {Robishaw}, and
  {Gaensler}]{Hammond2012arXiv}
{Hammond}, A.M.; {Robishaw}, T.; {Gaensler}, B.M.
\newblock {A New Catalog of Faraday Rotation Measures and Redshifts for
  Extragalactic Radio Sources}.
\newblock {\em arXiv e-prints} {\bf 2012}, p. arXiv:1209.1438,
  \href{http://xxx.lanl.gov/abs/1209.1438}{{\normalfont
  [arXiv:astro-ph.CO/1209.1438]}}.
\newblock
  doi:{\changeurlcolor{black}\href{https://doi.org/10.48550/arXiv.1209.1438}{\detokenize{10.48550/arXiv.1209.1438}}}.

\bibitem[{Peng} \em{et~al.}(2016){Peng}, {Wang}, {Liu}, {Tang}, and
  {Wang}]{Peng2016ApJ}
{Peng}, F.K.; {Wang}, X.Y.; {Liu}, R.Y.; {Tang}, Q.W.; {Wang}, J.F.
\newblock {First Detection of GeV Emission from an Ultraluminous Infrared
  Galaxy: Arp 220 as Seen with the Fermi Large Area Telescope}.
\newblock {\em \apjl} {\bf 2016}, {\em 821},~L20,
  \href{http://xxx.lanl.gov/abs/1603.06355}{{\normalfont
  [arXiv:astro-ph.HE/1603.06355]}}.
\newblock
  doi:{\changeurlcolor{black}\href{https://doi.org/10.3847/2041-8205/821/2/L20}{\detokenize{10.3847/2041-8205/821/2/L20}}}.

\bibitem[{Read} \em{et~al.}(2018){Read}, {Smith}, {G{\"u}rkan}, {Hardcastle},
  {Williams}, {Best}, {Brinks}, {Calistro-Rivera}, {Chy{\.Z}y}, {Duncan},
  {Dunne}, {Jarvis}, {Morabito}, {Prandoni}, {R{\"o}ttgering}, {Sabater}, and
  {Viaene}]{Read2018MNRAS}
{Read}, S.C.; {Smith}, D.J.B.; {G{\"u}rkan}, G.; {Hardcastle}, M.J.;
  {Williams}, W.L.; {Best}, P.N.; {Brinks}, E.; {Calistro-Rivera}, G.;
  {Chy{\.Z}y}, K.T.; {Duncan}, K.;  et~al.
\newblock {The Far-Infrared Radio Correlation at low radio frequency with
  LOFAR/H-ATLAS}.
\newblock {\em \mnras} {\bf 2018}, {\em 480},~5625--5644,
  \href{http://xxx.lanl.gov/abs/1808.10452}{{\normalfont
  [arXiv:astro-ph.GA/1808.10452]}}.
\newblock
  doi:{\changeurlcolor{black}\href{https://doi.org/10.1093/mnras/sty2198}{\detokenize{10.1093/mnras/sty2198}}}.

\bibitem[{Ackermann} \em{et~al.}(2012){Ackermann}, {Ajello}, {Allafort},
  {Baldini}, {Ballet}, {Bastieri}, {Bechtol}, {Bellazzini}, {Berenji}, {Bloom},
  {Bonamente}, {Borgland}, {Bouvier}, {Bregeon}, {Brigida}, {Bruel}, {Buehler},
  {Buson}, {Caliandro}, {Cameron}, {Caraveo}, {Casandjian}, {Cecchi},
  {Charles}, {Chekhtman}, {Cheung}, {Chiang}, {Cillis}, {Ciprini}, {Claus},
  {Cohen-Tanugi}, {Conrad}, {Cutini}, {de Palma}, {Dermer}, {Digel}, {Silva},
  {Drell}, {Drlica-Wagner}, {Favuzzi}, {Fegan}, {Fortin}, {Fukazawa}, {Funk},
  {Fusco}, {Gargano}, {Gasparrini}, {Germani}, {Giglietto}, {Giordano},
  {Glanzman}, {Godfrey}, {Grenier}, {Guiriec}, {Gustafsson}, {Hadasch},
  {Hayashida}, {Hays}, {Hughes}, {J{\'o}hannesson}, {Johnson}, {Kamae},
  {Katagiri}, {Kataoka}, {Kn{\"o}dlseder}, {Kuss}, {Lande}, {Longo}, {Loparco},
  {Lott}, {Lovellette}, {Lubrano}, {Madejski}, {Martin}, {Mazziotta},
  {McEnery}, {Michelson}, {Mizuno}, {Monte}, {Monzani}, {Morselli},
  {Moskalenko}, {Murgia}, {Nishino}, {Norris}, {Nuss}, {Ohno}, {Ohsugi},
  {Okumura}, {Omodei}, {Orlando}, {Ozaki}, {Parent}, {Persic}, {Pesce-Rollins},
  {Petrosian}, {Pierbattista}, {Piron}, {Pivato}, {Porter}, {Rain{\`o}},
  {Rando}, {Razzano}, {Reimer}, {Reimer}, {Ritz}, {Roth}, {Sbarra}, {Sgr{\`o}},
  {Siskind}, {Spandre}, {Spinelli}, {Stawarz}, {Strong}, {Takahashi}, {Tanaka},
  {Thayer}, {Tibaldo}, {Tinivella}, {Torres}, {Tosti}, {Troja}, {Uchiyama},
  {Vandenbroucke}, {Vianello}, {Vitale}, {Waite}, {Wood}, and
  {Yang}]{Ackermann2012ApJ}
{Ackermann}, M.; {Ajello}, M.; {Allafort}, A.; {Baldini}, L.; {Ballet}, J.;
  {Bastieri}, D.; {Bechtol}, K.; {Bellazzini}, R.; {Berenji}, B.; {Bloom},
  E.D.;  et~al.
\newblock {GeV Observations of Star-forming Galaxies with the Fermi Large Area
  Telescope}.
\newblock {\em \apj} {\bf 2012}, {\em 755},~164,
  \href{http://xxx.lanl.gov/abs/1206.1346}{{\normalfont
  [arXiv:astro-ph.HE/1206.1346]}}.
\newblock
  doi:{\changeurlcolor{black}\href{https://doi.org/10.1088/0004-637X/755/2/164}{\detokenize{10.1088/0004-637X/755/2/164}}}.

\bibitem[{Sargent} \em{et~al.}(2010){Sargent}, {Schinnerer}, {Murphy},
  {Carilli}, {Helou}, {Aussel}, {Le Floc'h}, {Frayer}, {Ilbert}, {Oesch},
  {Salvato}, {Smol{\v{c}}i{\'c}}, {Kartaltepe}, and {Sanders}]{Sargent2010ApJ}
{Sargent}, M.T.; {Schinnerer}, E.; {Murphy}, E.; {Carilli}, C.L.; {Helou}, G.;
  {Aussel}, H.; {Le Floc'h}, E.; {Frayer}, D.T.; {Ilbert}, O.; {Oesch}, P.;
  et~al.
\newblock {No Evolution in the IR-Radio Relation for IR-luminous Galaxies at z
  < 2 in the COSMOS Field}.
\newblock {\em \apjl} {\bf 2010}, {\em 714},~L190--L195,
  \href{http://xxx.lanl.gov/abs/1003.4271}{{\normalfont
  [arXiv:astro-ph.CO/1003.4271]}}.
\newblock
  doi:{\changeurlcolor{black}\href{https://doi.org/10.1088/2041-8205/714/2/L190}{\detokenize{10.1088/2041-8205/714/2/L190}}}.

\bibitem[{Bourne} \em{et~al.}(2011){Bourne}, {Dunne}, {Ivison}, {Maddox},
  {Dickinson}, and {Frayer}]{Bourne2011MNRAS}
{Bourne}, N.; {Dunne}, L.; {Ivison}, R.J.; {Maddox}, S.J.; {Dickinson}, M.;
  {Frayer}, D.T.
\newblock {Evolution of the far-infrared-radio correlation and infrared
  spectral energy distributions of massive galaxies over z= 0-2}.
\newblock {\em \mnras} {\bf 2011}, {\em 410},~1155--1173,
  \href{http://xxx.lanl.gov/abs/1005.3155}{{\normalfont
  [arXiv:astro-ph.CO/1005.3155]}}.
\newblock
  doi:{\changeurlcolor{black}\href{https://doi.org/10.1111/j.1365-2966.2010.17517.x}{\detokenize{10.1111/j.1365-2966.2010.17517.x}}}.

\bibitem[{Eichmann} and {Becker Tjus}(2016)]{Eichmann2016ApJ}
{Eichmann}, B.; {Becker Tjus}, J.
\newblock {The Radio-Gamma Correlation in Starburst Galaxies}.
\newblock {\em \apj} {\bf 2016}, {\em 821},~87,
  \href{http://xxx.lanl.gov/abs/1510.03672}{{\normalfont
  [arXiv:astro-ph.HE/1510.03672]}}.
\newblock
  doi:{\changeurlcolor{black}\href{https://doi.org/10.3847/0004-637X/821/2/87}{\detokenize{10.3847/0004-637X/821/2/87}}}.

\bibitem[{Murphy}(2009)]{Murphy2009ApJ}
{Murphy}, E.J.
\newblock {The Far-Infrared-Radio Correlation at High Redshifts: Physical
  Considerations and Prospects for the Square Kilometer Array}.
\newblock {\em \apj} {\bf 2009}, {\em 706},~482--496,
  \href{http://xxx.lanl.gov/abs/0910.0011}{{\normalfont
  [arXiv:astro-ph.CO/0910.0011]}}.
\newblock
  doi:{\changeurlcolor{black}\href{https://doi.org/10.1088/0004-637X/706/1/482}{\detokenize{10.1088/0004-637X/706/1/482}}}.

\bibitem[{Vollmer} \em{et~al.}(2010){Vollmer}, {Gassmann}, {Derri{\`e}re},
  {Boch}, {Louys}, {Bonnarel}, {Dubois}, {Genova}, and
  {Ochsenbein}]{Vollmer2010A&A}
{Vollmer}, B.; {Gassmann}, B.; {Derri{\`e}re}, S.; {Boch}, T.; {Louys}, M.;
  {Bonnarel}, F.; {Dubois}, P.; {Genova}, F.; {Ochsenbein}, F.
\newblock {The SPECFIND V2.0 catalogue of radio cross-identifications and
  spectra. SPECFIND meets the Virtual Observatory}.
\newblock {\em \aap} {\bf 2010}, {\em 511},~A53,
  \href{http://xxx.lanl.gov/abs/0912.4174}{{\normalfont
  [arXiv:astro-ph.CO/0912.4174]}}.
\newblock
  doi:{\changeurlcolor{black}\href{https://doi.org/10.1051/0004-6361/200913460}{\detokenize{10.1051/0004-6361/200913460}}}.

\bibitem[{Vollmer} \em{et~al.}(2005){Vollmer}, {Davoust}, {Dubois}, {Genova},
  {Ochsenbein}, and {van Driel}]{Vollmer2005A&A}
{Vollmer}, B.; {Davoust}, E.; {Dubois}, P.; {Genova}, F.; {Ochsenbein}, F.;
  {van Driel}, W.
\newblock {A method for determining radio continuum spectra and its application
  to large surveys}.
\newblock {\em \aap} {\bf 2005}, {\em 431},~1177--1187,
  \href{http://xxx.lanl.gov/abs/astro-ph/0410538}{{\normalfont
  [arXiv:astro-ph/astro-ph/0410538]}}.
\newblock
  doi:{\changeurlcolor{black}\href{https://doi.org/10.1051/0004-6361:20040562}{\detokenize{10.1051/0004-6361:20040562}}}.

\bibitem[{Vollmer} \em{et~al.}(2022){Vollmer}, {Soida}, and
  {Dallant}]{Vollmer2022A&A}
{Vollmer}, B.; {Soida}, M.; {Dallant}, J.
\newblock {Deciphering the radio-star formation correlation on kpc scales. II.
  The integrated infrared-radio continuum and star formation-radio continuum
  correlations}.
\newblock {\em \aap} {\bf 2022}, {\em 667},~A30,
  \href{http://xxx.lanl.gov/abs/2207.06173}{{\normalfont
  [arXiv:astro-ph.GA/2207.06173]}}.
\newblock
  doi:{\changeurlcolor{black}\href{https://doi.org/10.1051/0004-6361/202142877}{\detokenize{10.1051/0004-6361/202142877}}}.

\bibitem[{Lisenfeld} \em{et~al.}(1996){Lisenfeld}, {Voelk}, and
  {Xu}]{Lisenfeld1996A&A}
{Lisenfeld}, U.; {Voelk}, H.J.; {Xu}, C.
\newblock {A quantitative model of the FIR/radio correlation for normal
  late-type galaxies.}
\newblock {\em \aap} {\bf 1996}, {\em 306},~677,
  \href{http://xxx.lanl.gov/abs/astro-ph/9603130}{{\normalfont
  [arXiv:astro-ph/astro-ph/9603130]}}.
\newblock
  doi:{\changeurlcolor{black}\href{https://doi.org/10.48550/arXiv.astro-ph/9603130}{\detokenize{10.48550/arXiv.astro-ph/9603130}}}.

\bibitem[{Lisenfeld} and {V{\"o}lk}(2000)]{Lisenfeld2000A&A}
{Lisenfeld}, U.; {V{\"o}lk}, H.J.
\newblock {On the radio spectral index of galaxies}.
\newblock {\em \aap} {\bf 2000}, {\em 354},~423--430,
  \href{http://xxx.lanl.gov/abs/astro-ph/9912232}{{\normalfont
  [arXiv:astro-ph/astro-ph/9912232]}}.
\newblock
  doi:{\changeurlcolor{black}\href{https://doi.org/10.48550/arXiv.astro-ph/9912232}{\detokenize{10.48550/arXiv.astro-ph/9912232}}}.

\bibitem[{Helou} and {Bicay}(1993)]{Helou1993ApJ}
{Helou}, G.; {Bicay}, M.D.
\newblock {A Physical Model of the Infrared-to-Radio Correlation in Galaxies}.
\newblock {\em \apj} {\bf 1993}, {\em 415},~93.
\newblock
  doi:{\changeurlcolor{black}\href{https://doi.org/10.1086/173146}{\detokenize{10.1086/173146}}}.

\bibitem[{Niklas} and {Beck}(1997)]{Niklas1997A&A}
{Niklas}, S.; {Beck}, R.
\newblock {A new approach to the radio-far infrared correlation for
  non-calorimeter galaxies.}
\newblock {\em \aap} {\bf 1997}, {\em 320},~54--64.

\bibitem[{Lacki} \em{et~al.}(2010){Lacki}, {Thompson}, and
  {Quataert}]{Lacki2010ApJ}
{Lacki}, B.C.; {Thompson}, T.A.; {Quataert}, E.
\newblock {The Physics of the Far-infrared-Radio Correlation. I. Calorimetry,
  Conspiracy, and Implications}.
\newblock {\em \apj} {\bf 2010}, {\em 717},~1--28,
  \href{http://xxx.lanl.gov/abs/0907.4161}{{\normalfont
  [arXiv:astro-ph.CO/0907.4161]}}.
\newblock
  doi:{\changeurlcolor{black}\href{https://doi.org/10.1088/0004-637X/717/1/1}{\detokenize{10.1088/0004-637X/717/1/1}}}.

\bibitem[{Whitmore} \em{et~al.}(2010){Whitmore}, {Chandar}, {Schweizer},
  {Rothberg}, {Leitherer}, {Rieke}, {Rieke}, {Blair}, {Mengel}, and
  {Alonso-Herrero}]{Whitmore2010AJ}
{Whitmore}, B.C.; {Chandar}, R.; {Schweizer}, F.; {Rothberg}, B.; {Leitherer},
  C.; {Rieke}, M.; {Rieke}, G.; {Blair}, W.P.; {Mengel}, S.; {Alonso-Herrero},
  A.
\newblock {The Antennae Galaxies (NGC 4038/4039) Revisited: Advanced Camera for
  Surveys and NICMOS Observations of a Prototypical Merger}.
\newblock {\em \aj} {\bf 2010}, {\em 140},~75--109,
  \href{http://xxx.lanl.gov/abs/1005.0629}{{\normalfont
  [arXiv:astro-ph.EP/1005.0629]}}.
\newblock
  doi:{\changeurlcolor{black}\href{https://doi.org/10.1088/0004-6256/140/1/75}{\detokenize{10.1088/0004-6256/140/1/75}}}.

\bibitem[{Lopez-Rodriguez} \em{et~al.}(2023){Lopez-Rodriguez}, {Borlaff},
  {Beck}, {Reach}, {Mao}, {Ntormousi}, {Tassis}, {Martin-Alvarez}, {Clark},
  {Dale}, and {del Moral-Castro}]{LopezRodriguez2023ApJ}
{Lopez-Rodriguez}, E.; {Borlaff}, A.S.; {Beck}, R.; {Reach}, W.T.; {Mao}, S.A.;
  {Ntormousi}, E.; {Tassis}, K.; {Martin-Alvarez}, S.; {Clark}, S.E.; {Dale},
  D.A.;  et~al.
\newblock {Extragalactic Magnetism with SOFIA (SALSA Legacy Program): The
  Magnetic Fields in the Multiphase Interstellar Medium of the Antennae
  Galaxies}.
\newblock {\em \apjl} {\bf 2023}, {\em 942},~L13,
  \href{http://xxx.lanl.gov/abs/2211.00012}{{\normalfont
  [arXiv:astro-ph.GA/2211.00012]}}.
\newblock
  doi:{\changeurlcolor{black}\href{https://doi.org/10.3847/2041-8213/acaaa2}{\detokenize{10.3847/2041-8213/acaaa2}}}.

\bibitem[{Pfrommer} \em{et~al.}(2022){Pfrommer}, {Werhahn}, {Pakmor},
  {Girichidis}, and {Simpson}]{Pfrommer2022MNRAS}
{Pfrommer}, C.; {Werhahn}, M.; {Pakmor}, R.; {Girichidis}, P.; {Simpson}, C.M.
\newblock {Simulating radio synchrotron emission in star-forming galaxies:
  small-scale magnetic dynamo and the origin of the far-infrared-radio
  correlation}.
\newblock {\em \mnras} {\bf 2022}, {\em 515},~4229--4264,
  \href{http://xxx.lanl.gov/abs/2105.12132}{{\normalfont
  [arXiv:astro-ph.GA/2105.12132]}}.
\newblock
  doi:{\changeurlcolor{black}\href{https://doi.org/10.1093/mnras/stac1808}{\detokenize{10.1093/mnras/stac1808}}}.

\bibitem[{Pfrommer} \em{et~al.}(2017){Pfrommer}, {Pakmor}, {Simpson}, and
  {Springel}]{Pfrommer2017ApJ}
{Pfrommer}, C.; {Pakmor}, R.; {Simpson}, C.M.; {Springel}, V.
\newblock {Simulating Gamma-Ray Emission in Star-forming Galaxies}.
\newblock {\em \apjl} {\bf 2017}, {\em 847},~L13,
  \href{http://xxx.lanl.gov/abs/1709.05343}{{\normalfont
  [arXiv:astro-ph.GA/1709.05343]}}.
\newblock
  doi:{\changeurlcolor{black}\href{https://doi.org/10.3847/2041-8213/aa8bb1}{\detokenize{10.3847/2041-8213/aa8bb1}}}.

\bibitem[{Werhahn} \em{et~al.}(2021{\natexlab{a}}){Werhahn}, {Pfrommer},
  {Girichidis}, and {Winner}]{Werhahn2021MNRASb}
{Werhahn}, M.; {Pfrommer}, C.; {Girichidis}, P.; {Winner}, G.
\newblock {Cosmic rays and non-thermal emission in simulated galaxies - II.
  {\ensuremath{\gamma}}-ray maps, spectra, and the
  far-infrared-{\ensuremath{\gamma}}-ray relation}.
\newblock {\em \mnras} {\bf 2021}, {\em 505},~3295--3313,
  \href{http://xxx.lanl.gov/abs/2105.11463}{{\normalfont
  [arXiv:astro-ph.HE/2105.11463]}}.
\newblock
  doi:{\changeurlcolor{black}\href{https://doi.org/10.1093/mnras/stab1325}{\detokenize{10.1093/mnras/stab1325}}}.

\bibitem[{Werhahn} \em{et~al.}(2021{\natexlab{b}}){Werhahn}, {Pfrommer}, and
  {Girichidis}]{Werhahn2021MNRAS3}
{Werhahn}, M.; {Pfrommer}, C.; {Girichidis}, P.
\newblock {Cosmic rays and non-thermal emission in simulated galaxies - III.
  Probing cosmic-ray calorimetry with radio spectra and the FIR-radio
  correlation}.
\newblock {\em \mnras} {\bf 2021}, {\em 508},~4072--4095,
  \href{http://xxx.lanl.gov/abs/2105.12134}{{\normalfont
  [arXiv:astro-ph.GA/2105.12134]}}.
\newblock
  doi:{\changeurlcolor{black}\href{https://doi.org/10.1093/mnras/stab2535}{\detokenize{10.1093/mnras/stab2535}}}.

\bibitem[{Wang} and {Fields}(2018)]{Wang2018MNRAS}
{Wang}, X.; {Fields}, B.D.
\newblock {Are starburst galaxies proton calorimeters?}
\newblock {\em \mnras} {\bf 2018}, {\em 474},~4073--4088,
  \href{http://xxx.lanl.gov/abs/1612.07290}{{\normalfont
  [arXiv:astro-ph.HE/1612.07290]}}.
\newblock
  doi:{\changeurlcolor{black}\href{https://doi.org/10.1093/mnras/stx2917}{\detokenize{10.1093/mnras/stx2917}}}.

\bibitem[{Zhang} \em{et~al.}(2019){Zhang}, {Peng}, and {Wang}]{Zhang2019ApJ}
{Zhang}, Y.; {Peng}, F.K.; {Wang}, X.Y.
\newblock {Interpreting the Relation between the Gamma-Ray and Infrared
  Luminosities of Star-forming Galaxies}.
\newblock {\em \apj} {\bf 2019}, {\em 874},~173,
  \href{http://xxx.lanl.gov/abs/1902.09654}{{\normalfont
  [arXiv:astro-ph.HE/1902.09654]}}.
\newblock
  doi:{\changeurlcolor{black}\href{https://doi.org/10.3847/1538-4357/ab0ae2}{\detokenize{10.3847/1538-4357/ab0ae2}}}.

\bibitem[{Strong} \em{et~al.}(2010){Strong}, {Porter}, {Digel},
  {J{\'o}hannesson}, {Martin}, {Moskalenko}, {Murphy}, and
  {Orlando}]{Strong2010ApJ}
{Strong}, A.W.; {Porter}, T.A.; {Digel}, S.W.; {J{\'o}hannesson}, G.; {Martin},
  P.; {Moskalenko}, I.V.; {Murphy}, E.J.; {Orlando}, E.
\newblock {Global Cosmic-ray-related Luminosity and Energy Budget of the Milky
  Way}.
\newblock {\em \apjl} {\bf 2010}, {\em 722},~L58--L63,
  \href{http://xxx.lanl.gov/abs/1008.4330}{{\normalfont
  [arXiv:astro-ph.HE/1008.4330]}}.
\newblock
  doi:{\changeurlcolor{black}\href{https://doi.org/10.1088/2041-8205/722/1/L58}{\detokenize{10.1088/2041-8205/722/1/L58}}}.

\bibitem[{Evoli} \em{et~al.}(2012){Evoli}, {Gaggero}, {Grasso}, and
  {Maccione}]{Evoli2012PhRvL}
{Evoli}, C.; {Gaggero}, D.; {Grasso}, D.; {Maccione}, L.
\newblock {Common Solution to the Cosmic Ray Anisotropy and Gradient Problems}.
\newblock {\em \prl} {\bf 2012}, {\em 108},~211102,
  \href{http://xxx.lanl.gov/abs/1203.0570}{{\normalfont
  [arXiv:astro-ph.HE/1203.0570]}}.
\newblock
  doi:{\changeurlcolor{black}\href{https://doi.org/10.1103/PhysRevLett.108.211102}{\detokenize{10.1103/PhysRevLett.108.211102}}}.

\bibitem[{Casse} \em{et~al.}(2001){Casse}, {Lemoine}, and
  {Pelletier}]{Casse2001PhRvD}
{Casse}, F.; {Lemoine}, M.; {Pelletier}, G.
\newblock {Transport of cosmic rays in chaotic magnetic fields}.
\newblock {\em \prd} {\bf 2001}, {\em 65},~023002,
  \href{http://xxx.lanl.gov/abs/astro-ph/0109223}{{\normalfont
  [arXiv:astro-ph/astro-ph/0109223]}}.
\newblock
  doi:{\changeurlcolor{black}\href{https://doi.org/10.1103/PhysRevD.65.023002}{\detokenize{10.1103/PhysRevD.65.023002}}}.

\bibitem[{Pakmor} \em{et~al.}(2016){Pakmor}, {Pfrommer}, {Simpson}, and
  {Springel}]{Pakmor2016ApJ}
{Pakmor}, R.; {Pfrommer}, C.; {Simpson}, C.M.; {Springel}, V.
\newblock {Galactic Winds Driven by Isotropic and Anisotropic Cosmic-Ray
  Diffusion in Disk Galaxies}.
\newblock {\em \apjl} {\bf 2016}, {\em 824},~L30,
  \href{http://xxx.lanl.gov/abs/1605.00643}{{\normalfont
  [arXiv:astro-ph.GA/1605.00643]}}.
\newblock
  doi:{\changeurlcolor{black}\href{https://doi.org/10.3847/2041-8205/824/2/L30}{\detokenize{10.3847/2041-8205/824/2/L30}}}.

\bibitem[{Wang} \em{et~al.}(2010){Wang}, {Lo}, and {Ko}]{Wang2010arXiv}
{Wang}, C.Y.; {Lo}, Y.Y.; {Ko}, C.M.
\newblock {MHD Simulations of Parker Instability Undergoing Cosmic-Ray
  Diffusion}.
\newblock {\em arXiv e-prints} {\bf 2010}, p. arXiv:1011.0162,
  \href{http://xxx.lanl.gov/abs/1011.0162}{{\normalfont
  [arXiv:astro-ph.HE/1011.0162]}}.
\newblock
  doi:{\changeurlcolor{black}\href{https://doi.org/10.48550/arXiv.1011.0162}{\detokenize{10.48550/arXiv.1011.0162}}}.

\bibitem[{Ruszkowski} \em{et~al.}(2017){Ruszkowski}, {Yang}, and
  {Zweibel}]{Ruszkowski2017ApJ}
{Ruszkowski}, M.; {Yang}, H.Y.K.; {Zweibel}, E.
\newblock {Global Simulations of Galactic Winds Including Cosmic-ray
  Streaming}.
\newblock {\em \apj} {\bf 2017}, {\em 834},~208,
  \href{http://xxx.lanl.gov/abs/1602.04856}{{\normalfont
  [arXiv:astro-ph.GA/1602.04856]}}.
\newblock
  doi:{\changeurlcolor{black}\href{https://doi.org/10.3847/1538-4357/834/2/208}{\detokenize{10.3847/1538-4357/834/2/208}}}.

\bibitem[{Lopez-Rodriguez} \em{et~al.}(2022){Lopez-Rodriguez}, {Mao}, {Beck},
  {Borlaff}, {Ntormousi}, {Tassis}, {Dale}, {Roman-Duval}, {Subramanian},
  {Martin-Alvarez}, {Marcum}, {Clark}, {Reach}, {Harper}, and
  {Zweibel}]{LopezRodriguez2022ApJ}
{Lopez-Rodriguez}, E.; {Mao}, S.A.; {Beck}, R.; {Borlaff}, A.S.; {Ntormousi},
  E.; {Tassis}, K.; {Dale}, D.A.; {Roman-Duval}, J.; {Subramanian}, K.;
  {Martin-Alvarez}, S.;  et~al.
\newblock {Extragalactic Magnetism with SOFIA (SALSA Legacy Program). IV.
  Program Overview and First Results on the Polarization Fraction}.
\newblock {\em \apj} {\bf 2022}, {\em 936},~92,
  \href{http://xxx.lanl.gov/abs/2205.01105}{{\normalfont
  [arXiv:astro-ph.GA/2205.01105]}}.
\newblock
  doi:{\changeurlcolor{black}\href{https://doi.org/10.3847/1538-4357/ac7f9d}{\detokenize{10.3847/1538-4357/ac7f9d}}}.

\bibitem[{Pattle} \em{et~al.}(2021){Pattle}, {Gear}, {Redman}, {Smith}, and
  {Greaves}]{Pattle2021MNRAS}
{Pattle}, K.; {Gear}, W.; {Redman}, M.; {Smith}, M.W.L.; {Greaves}, J.
\newblock {Submillimetre observations of the two-component magnetic field in
  M82}.
\newblock {\em \mnras} {\bf 2021}, {\em 505},~684--688,
  \href{http://xxx.lanl.gov/abs/2105.01989}{{\normalfont
  [arXiv:astro-ph.GA/2105.01989]}}.
\newblock
  doi:{\changeurlcolor{black}\href{https://doi.org/10.1093/mnras/stab1300}{\detokenize{10.1093/mnras/stab1300}}}.

\bibitem[{Fletcher} \em{et~al.}(2011){Fletcher}, {Beck}, {Shukurov},
  {Berkhuijsen}, and {Horellou}]{Fletcher2011MNRAS}
{Fletcher}, A.; {Beck}, R.; {Shukurov}, A.; {Berkhuijsen}, E.M.; {Horellou}, C.
\newblock {Magnetic fields and spiral arms in the galaxy M51}.
\newblock {\em \mnras} {\bf 2011}, {\em 412},~2396--2416,
  \href{http://xxx.lanl.gov/abs/1001.5230}{{\normalfont
  [arXiv:astro-ph.CO/1001.5230]}}.
\newblock
  doi:{\changeurlcolor{black}\href{https://doi.org/10.1111/j.1365-2966.2010.18065.x}{\detokenize{10.1111/j.1365-2966.2010.18065.x}}}.

\bibitem[{Barnes} and {Hernquist}(1992)]{Barnes1992ARA&A}
{Barnes}, J.E.; {Hernquist}, L.
\newblock {Dynamics of interacting galaxies.}
\newblock {\em \araa} {\bf 1992}, {\em 30},~705--742.
\newblock
  doi:{\changeurlcolor{black}\href{https://doi.org/10.1146/annurev.aa.30.090192.003421}{\detokenize{10.1146/annurev.aa.30.090192.003421}}}.

\bibitem[{Socrates} \em{et~al.}(2008){Socrates}, {Davis}, and
  {Ramirez-Ruiz}]{Socrates2008ApJ}
{Socrates}, A.; {Davis}, S.W.; {Ramirez-Ruiz}, E.
\newblock {The Eddington Limit in Cosmic Rays: An Explanation for the Observed
  Faintness of Starbursting Galaxies}.
\newblock {\em \apj} {\bf 2008}, {\em 687},~202--215,
  \href{http://xxx.lanl.gov/abs/astro-ph/0609796}{{\normalfont
  [arXiv:astro-ph/astro-ph/0609796]}}.
\newblock
  doi:{\changeurlcolor{black}\href{https://doi.org/10.1086/590046}{\detokenize{10.1086/590046}}}.

\bibitem[{Crocker} \em{et~al.}(2021{\natexlab{a}}){Crocker}, {Krumholz}, and
  {Thompson}]{Crocker2021MNRAS_a}
{Crocker}, R.M.; {Krumholz}, M.R.; {Thompson}, T.A.
\newblock {Cosmic rays across the star-forming galaxy sequence - I. Cosmic ray
  pressures and calorimetry}.
\newblock {\em \mnras} {\bf 2021}, {\em 502},~1312--1333,
  \href{http://xxx.lanl.gov/abs/2006.15819}{{\normalfont
  [arXiv:astro-ph.GA/2006.15819]}}.
\newblock
  doi:{\changeurlcolor{black}\href{https://doi.org/10.1093/mnras/stab148}{\detokenize{10.1093/mnras/stab148}}}.

\bibitem[{Crocker} \em{et~al.}(2021{\natexlab{b}}){Crocker}, {Krumholz}, and
  {Thompson}]{Crocker2021MNRAS_b}
{Crocker}, R.M.; {Krumholz}, M.R.; {Thompson}, T.A.
\newblock {Cosmic rays across the star-forming galaxy sequence - II. Stability
  limits and the onset of cosmic ray-driven outflows}.
\newblock {\em \mnras} {\bf 2021}, {\em 503},~2651--2664,
  \href{http://xxx.lanl.gov/abs/2006.15821}{{\normalfont
  [arXiv:astro-ph.GA/2006.15821]}}.
\newblock
  doi:{\changeurlcolor{black}\href{https://doi.org/10.1093/mnras/stab502}{\detokenize{10.1093/mnras/stab502}}}.

\bibitem[{Huang} and {Davis}(2022)]{Huang2022MNRAS}
{Huang}, X.; {Davis}, S.W.
\newblock {The launching of cosmic ray-driven outflows}.
\newblock {\em \mnras} {\bf 2022}, {\em 511},~5125--5141,
  \href{http://xxx.lanl.gov/abs/2105.11506}{{\normalfont
  [arXiv:astro-ph.GA/2105.11506]}}.
\newblock
  doi:{\changeurlcolor{black}\href{https://doi.org/10.1093/mnras/stac059}{\detokenize{10.1093/mnras/stac059}}}.

\bibitem[{Heintz} and {Zweibel}(2022)]{Heintz2022ApJ}
{Heintz}, E.; {Zweibel}, E.G.
\newblock {Galaxies at a Cosmic Ray Eddington Limit}.
\newblock {\em \apj} {\bf 2022}, {\em 941},~78,
  \href{http://xxx.lanl.gov/abs/2206.04082}{{\normalfont
  [arXiv:astro-ph.GA/2206.04082]}}.
\newblock
  doi:{\changeurlcolor{black}\href{https://doi.org/10.3847/1538-4357/ac9e9e}{\detokenize{10.3847/1538-4357/ac9e9e}}}.

\bibitem[{Abdo} \em{et~al.}(2010){Abdo}, {Ackermann}, {Ajello}, {Atwood},
  {Axelsson}, {Baldini}, {Ballet}, {Barbiellini}, {Bastieri}, {Bechtol},
  {Bellazzini}, {Berenji}, {Bloom}, {Bonamente}, {Borgland}, {Bregeon}, {Brez},
  {Brigida}, {Bruel}, {Burnett}, {Caliandro}, {Cameron}, {Caraveo},
  {Casandjian}, {Cavazzuti}, {Cecchi}, {{\c{C}}elik}, {Charles}, {Chekhtman},
  {Cheung}, {Chiang}, {Ciprini}, {Claus}, {Cohen-Tanugi}, {Conrad}, {Dermer},
  {de Angelis}, {de Palma}, {Digel}, {Silva}, {Drell}, {Drlica-Wagner},
  {Dubois}, {Dumora}, {Farnier}, {Favuzzi}, {Fegan}, {Focke}, {Foschini},
  {Frailis}, {Fukazawa}, {Funk}, {Fusco}, {Gargano}, {Gasparrini}, {Gehrels},
  {Germani}, {Giebels}, {Giglietto}, {Giordano}, {Glanzman}, {Godfrey},
  {Grenier}, {Grondin}, {Grove}, {Guillemot}, {Guiriec}, {Hanabata}, {Harding},
  {Hayashida}, {Hays}, {Hughes}, {J{\'o}hannesson}, {Johnson}, {Johnson},
  {Johnson}, {Kamae}, {Katagiri}, {Kataoka}, {Kawai}, {Kerr}, {Kn{\"o}dlseder},
  {Kocian}, {Kuss}, {Lande}, {Latronico}, {Lemoine-Goumard}, {Longo},
  {Loparco}, {Lott}, {Lovellette}, {Lubrano}, {Madejski}, {Makeev},
  {Mazziotta}, {McConville}, {McEnery}, {Meurer}, {Michelson}, {Mitthumsiri},
  {Mizuno}, {Moiseev}, {Monte}, {Monzani}, {Morselli}, {Moskalenko}, {Murgia},
  {Nakamori}, {Nolan}, {Norris}, {Nuss}, {Ohsugi}, {Omodei}, {Orlando},
  {Ormes}, {Ozaki}, {Paneque}, {Panetta}, {Parent}, {Pelassa}, {Pepe},
  {Pesce-Rollins}, {Piron}, {Porter}, {Rain{\`o}}, {Rando}, {Razzano},
  {Reimer}, {Reimer}, {Reposeur}, {Ritz}, {Rodriguez}, {Romani}, {Roth},
  {Ryde}, {Sadrozinski}, {Sander}, {Saz Parkinson}, {Scargle}, {Sellerholm},
  {Sgr{\`o}}, {Shaw}, {Smith}, {Smith}, {Spandre}, {Spinelli}, {Strickman},
  {Strong}, {Suson}, {Takahashi}, {Tanaka}, {Thayer}, {Thayer}, {Thompson},
  {Tibaldo}, {Tibolla}, {Torres}, {Tosti}, {Tramacere}, {Uchiyama}, {Usher},
  {Vasileiou}, {Vilchez}, {Vitale}, {Waite}, {Wang}, {Winer}, {Wood}, {Ylinen},
  {Ziegler}, and {Fermi LAT Collaboration}]{Abdo2010ApJ}
{Abdo}, A.A.; {Ackermann}, M.; {Ajello}, M.; {Atwood}, W.B.; {Axelsson}, M.;
  {Baldini}, L.; {Ballet}, J.; {Barbiellini}, G.; {Bastieri}, D.; {Bechtol},
  K.;  et~al.
\newblock {Detection of Gamma-Ray Emission from the Starburst Galaxies M82 and
  NGC 253 with the Large Area Telescope on Fermi}.
\newblock {\em \apjl} {\bf 2010}, {\em 709},~L152--L157,
  \href{http://xxx.lanl.gov/abs/0911.5327}{{\normalfont
  [arXiv:astro-ph.HE/0911.5327]}}.
\newblock
  doi:{\changeurlcolor{black}\href{https://doi.org/10.1088/2041-8205/709/2/L152}{\detokenize{10.1088/2041-8205/709/2/L152}}}.

\bibitem[{Ajello} \em{et~al.}(2020){Ajello}, {Di Mauro}, {Paliya}, and
  {Garrappa}]{Ajello2020ApJ}
{Ajello}, M.; {Di Mauro}, M.; {Paliya}, V.S.; {Garrappa}, S.
\newblock {The {\ensuremath{\gamma}}-Ray Emission of Star-forming Galaxies}.
\newblock {\em \apj} {\bf 2020}, {\em 894},~88,
  \href{http://xxx.lanl.gov/abs/2003.05493}{{\normalfont
  [arXiv:astro-ph.GA/2003.05493]}}.
\newblock
  doi:{\changeurlcolor{black}\href{https://doi.org/10.3847/1538-4357/ab86a6}{\detokenize{10.3847/1538-4357/ab86a6}}}.

\bibitem[{Xi} \em{et~al.}(2020){Xi}, {Zhang}, {Liu}, and {Wang}]{Xi2020ApJ}
{Xi}, S.Q.; {Zhang}, H.M.; {Liu}, R.Y.; {Wang}, X.Y.
\newblock {GeV {\ensuremath{\gamma}}-Ray Emission from M33 and Arp 299}.
\newblock {\em \apj} {\bf 2020}, {\em 901},~158,
  \href{http://xxx.lanl.gov/abs/2003.07830}{{\normalfont
  [arXiv:astro-ph.HE/2003.07830]}}.
\newblock
  doi:{\changeurlcolor{black}\href{https://doi.org/10.3847/1538-4357/aba043}{\detokenize{10.3847/1538-4357/aba043}}}.

\bibitem[{Xing} and {Wang}(2023)]{Xing2023arXiv230400229X}
{Xing}, Y.; {Wang}, Z.
\newblock {Identifying the Gamma-ray Emission of the Nearby Galaxy M83}.
\newblock {\em arXiv e-prints} {\bf 2023}, p. arXiv:2304.00229,
  \href{http://xxx.lanl.gov/abs/2304.00229}{{\normalfont
  [arXiv:astro-ph.HE/2304.00229]}}.
\newblock
  doi:{\changeurlcolor{black}\href{https://doi.org/10.48550/arXiv.2304.00229}{\detokenize{10.48550/arXiv.2304.00229}}}.

\bibitem[{Acero} \em{et~al.}(2009){Acero}, {Aharonian}, {Akhperjanian},
  {Anton}, {Barres de Almeida}, {Bazer-Bachi}, {Becherini}, {Behera},
  {Bernl{\"o}hr}, {Bochow}, {Boisson}, {Bolmont}, {Borrel}, {Brucker}, {Brun},
  {Brun}, {B{\"u}hler}, {Bulik}, {B{\"u}sching}, {Boutelier}, {Chadwick},
  {Charbonnier}, {Chaves}, {Cheesebrough}, {Chounet}, {Clapson}, {Coignet},
  {Dalton}, {Daniel}, {Davids}, {Degrange}, {Deil}, {Dickinson},
  {Djannati-Ata{\"\i}}, {Domainko}, {Drury}, {Dubois}, {Dubus}, {Dyks},
  {Dyrda}, {Egberts}, {Emmanoulopoulos}, {Espigat}, {Farnier}, {Fegan},
  {Feinstein}, {Fiasson}, {F{\"o}rster}, {Fontaine}, {F{\"u}{\ss}ling},
  {Gabici}, {Gallant}, {G{\'e}rard}, {Gerbig}, {Giebels}, {Glicenstein},
  {Gl{\"u}ck}, {Goret}, {G{\"o}ring}, {Hauser}, {Hauser}, {Heinz},
  {Heinzelmann}, {Henri}, {Hermann}, {Hinton}, {Hoffmann}, {Hofmann},
  {Hofverberg}, {Hoppe}, {Horns}, {Jacholkowska}, {de Jager}, {Jahn}, {Jung},
  {Katarzy{\'n}ski}, {Katz}, {Kaufmann}, {Kerschhaggl}, {Khangulyan},
  {Kh{\'e}lifi}, {Keogh}, {Klochkov}, {Klu{\'z}niak}, {Kneiske}, {Komin},
  {Kosack}, {Kossakowski}, {Lamanna}, {Lenain}, {Lohse}, {Marandon},
  {Martineau-Huynh}, {Marcowith}, {Masbou}, {Maurin}, {McComb}, {Medina},
  {M{\'e}hault}, {Moderski}, {Moulin}, {Naumann-Godo}, {de Naurois}, {Nedbal},
  {Nekrassov}, {Nicholas}, {Niemiec}, {Nolan}, {Ohm}, {Olive}, {Wilhelmi},
  {Orford}, {Ostrowski}, {Panter}, {Arribas}, {Pedaletti}, {Pelletier},
  {Petrucci}, {Pita}, {P{\"u}hlhofer}, {Punch}, {Quirrenbach}, {Raubenheimer},
  {Raue}, {Rayner}, {Reimer}, {Renaud}, {Rieger}, {Ripken}, {Rob},
  {Rosier-Lees}, {Rowell}, {Rudak}, {Rulten}, {Ruppel}, {Sahakian},
  {Santangelo}, {Schlickeiser}, {Sch{\"o}ck}, {Schwanke}, {Schwarzburg},
  {Schwemmer}, {Shalchi}, {Sikora}, {Skilton}, {Sol}, {Stawarz}, {Steenkamp},
  {Stegmann}, {Stinzing}, {Superina}, {Szostek}, {Tam}, {Tavernet}, {Terrier},
  {Tibolla}, {Tluczykont}, {van Eldik}, {Vasileiadis}, {Venter}, {Venter},
  {Vialle}, {Vincent}, {Vivier}, {V{\"o}lk}, {Volpe}, {Wagner}, {Ward},
  {Zdziarski}, and {Zech}]{Acero2009Sci}
{Acero}, F.; {Aharonian}, F.; {Akhperjanian}, A.G.; {Anton}, G.; {Barres de
  Almeida}, U.; {Bazer-Bachi}, A.R.; {Becherini}, Y.; {Behera}, B.;
  {Bernl{\"o}hr}, K.; {Bochow}, A.;  et~al.
\newblock {Detection of Gamma Rays from a Starburst Galaxy}.
\newblock {\em Science} {\bf 2009}, {\em 326},~1080,
  \href{http://xxx.lanl.gov/abs/0909.4651}{{\normalfont
  [arXiv:astro-ph.HE/0909.4651]}}.
\newblock
  doi:{\changeurlcolor{black}\href{https://doi.org/10.1126/science.1178826}{\detokenize{10.1126/science.1178826}}}.

\bibitem[{VERITAS Collaboration} \em{et~al.}(2009){VERITAS Collaboration},
  {Acciari}, {Aliu}, {Arlen}, {Aune}, {Bautista}, {Beilicke}, {Benbow},
  {Boltuch}, {Bradbury}, {Buckley}, {Bugaev}, {Byrum}, {Cannon}, {Celik},
  {Cesarini}, {Chow}, {Ciupik}, {Cogan}, {Colin}, {Cui}, {Dickherber}, {Duke},
  {Fegan}, {Finley}, {Finnegan}, {Fortin}, {Fortson}, {Furniss}, {Galante},
  {Gall}, {Gibbs}, {Gillanders}, {Godambe}, {Grube}, {Guenette}, {Gyuk},
  {Hanna}, {Holder}, {Horan}, {Hui}, {Humensky}, {Imran}, {Kaaret}, {Karlsson},
  {Kertzman}, {Kieda}, {Kildea}, {Konopelko}, {Krawczynski}, {Krennrich},
  {Lang}, {Lebohec}, {Maier}, {McArthur}, {McCann}, {McCutcheon}, {Millis},
  {Moriarty}, {Mukherjee}, {Nagai}, {Ong}, {Otte}, {Pandel}, {Perkins},
  {Pizlo}, {Pohl}, {Quinn}, {Ragan}, {Reyes}, {Reynolds}, {Roache}, {Rose},
  {Schroedter}, {Sembroski}, {Smith}, {Steele}, {Swordy}, {Theiling},
  {Thibadeau}, {Varlotta}, {Vassiliev}, {Vincent}, {Wagner}, {Wakely}, {Ward},
  {Weekes}, {Weinstein}, {Weisgarber}, {Williams}, {Wissel}, {Wood}, and
  {Zitzer}]{VERITAS2009Natur}
{VERITAS Collaboration}.; {Acciari}, V.A.; {Aliu}, E.; {Arlen}, T.; {Aune}, T.;
  {Bautista}, M.; {Beilicke}, M.; {Benbow}, W.; {Boltuch}, D.; {Bradbury},
  S.M.;  et~al.
\newblock {A connection between star formation activity and cosmic rays in the
  starburst galaxy M82}.
\newblock {\em \nat} {\bf 2009}, {\em 462},~770--772,
  \href{http://xxx.lanl.gov/abs/0911.0873}{{\normalfont
  [arXiv:astro-ph.CO/0911.0873]}}.
\newblock
  doi:{\changeurlcolor{black}\href{https://doi.org/10.1038/nature08557}{\detokenize{10.1038/nature08557}}}.

\bibitem[{H.~E.~S.~S. Collaboration} \em{et~al.}(2018){H.~E.~S.~S.
  Collaboration}, {Abdalla}, {Aharonian}, {Ait Benkhali}, {Ang{\"u}ner},
  {Arakawa}, {Arcaro}, {Armand}, {Arrieta}, {Backes}, {Barnard}, {Becherini},
  {Becker Tjus}, {Berge}, {Bernhard}, {Bernl{\"o}hr}, {Blackwell},
  {B{\"o}ttcher}, {Boisson}, {Bolmont}, {Bonnefoy}, {Bordas}, {Bregeon},
  {Brun}, {Brun}, {Bryan}, {B{\"u}chele}, {Bulik}, {Bylund}, {Capasso},
  {Caroff}, {Carosi}, {Casanova}, {Cerruti}, {Chakraborty}, {Chandra},
  {Chaves}, {Chen}, {Colafrancesco}, {Condon}, {Davids}, {Deil}, {Devin},
  {deWilt}, {Dirson}, {Djannati-Ata{\"\i}}, {Dmytriiev}, {Donath}, {Drury},
  {Dyks}, {Egberts}, {Emery}, {Ernenwein}, {Eschbach}, {Fegan}, {Fiasson},
  {Fontaine}, {Funk}, {F{\"u}{\ss}ling}, {Gabici}, {Gallant}, {Garrigoux},
  {Gat{\'e}}, {Giavitto}, {Glawion}, {Glicenstein}, {Gottschall}, {Grondin},
  {Hahn}, {Haupt}, {Heinzelmann}, {Henri}, {Hermann}, {Hinton}, {Hofmann},
  {Hoischen}, {Holch}, {Holler}, {Horns}, {Huber}, {Iwasaki}, {Jacholkowska},
  {Jamrozy}, {Jankowsky}, {Jankowsky}, {Jouvin}, {Jung-Richardt},
  {Kastendieck}, {Kat{\textasciiacute}nski}, {Katsuragawa}, {Katz},
  {Kerszberg}, {Khangulyan}, {Kh{\'e}lifi}, {King}, {Klepser},
  {K{\textasciiacute}zniak}, {Komin}, {Kosack}, {Krakau}, {Kraus},
  {Kr{\"u}ger}, {Lamanna}, {Lau}, {Lefaucheur}, {Lemi{\`e}re},
  {Lemoine-Goumard}, {Lenain}, {Leser}, {Lohse}, {Lorentz}, {L{\'o}pez-Coto},
  {Lypova}, {Malyshev}, {Marandon}, {Marcowith}, {Mariaud},
  {Mart{\'\i}-Devesa}, {Marx}, {Maurin}, {Meintjes}, {Mitchell}, {Moderski},
  {Mohamed}, {Mohrmann}, {Moulin}, {Murach}, {Nakashima}, {de Naurois},
  {Ndiyavala}, {Niederwanger}, {Niemiec}, {Oakes}, {O'Brien}, {Odaka}, {Ohm},
  {Ostrowski}, {Oya}, {Padovani}, {Panter}, {Parsons}, {Perennes}, {Petrucci},
  {Peyaud}, {Piel}, {Pita}, {Poireau}, {Priyana Noel}, {Prokhorov}, {Prokoph},
  {P{\"u}hlhofer}, {Punch}, {Quirrenbach}, {Raab}, {Rauth}, {Reimer}, {Reimer},
  {Renaud}, {Rieger}, {Rinchiuso}, {Romoli}, {Rowell}, {Rudak}, {Ruiz-Velasco},
  {Sahakian}, {Saito}, {Sanchez}, {Santangelo}, {Sasaki}, {Schlickeiser},
  {Sch{\"u}ssler}, {Schulz}, {Schwanke}, {Schwemmer}, {Seglar-Arroyo},
  {Senniappan}, {Seyffert}, {Shafi}, {Shilon}, {Shiningayamwe}, {Simoni},
  {Sinha}, {Sol}, {Spanier}, {Specovius}, {Spir-Jacob}, {Stawarz}, {Steenkamp},
  {Stegmann}, {Steppa}, {Sushch}, {Takahashi}, {Tavernet}, {Tavernier},
  {Taylor}, {Terrier}, {Tibaldo}, {Tiziani}, {Tluczykont}, {Trichard},
  {Tsirou}, {Tsuji}, {Tuffs}, {Uchiyama}, {van der Walt}, {van Eldik}, {van
  Rensburg}, {van Soelen}, {Vasileiadis}, {Veh}, {Venter}, {Viana}, {Vincent},
  {Vink}, {Voisin}, {V{\"o}lk}, {Vuillaume}, {Wadiasingh}, {Wagner}, {Wagner},
  {Wagner}, {White}, {Wierzcholska}, {W{\"o}rnlein}, {Yang}, {Zaborov},
  {Zacharias}, {Zanin}, {Zdziarski}, {Zech}, {Zefi}, {Ziegler}, {Zorn}, and
  {{\.Z}ywucka}]{HESS2018A&ANGC253}
{H.~E.~S.~S. Collaboration}.; {Abdalla}, H.; {Aharonian}, F.; {Ait Benkhali},
  F.; {Ang{\"u}ner}, E.O.; {Arakawa}, M.; {Arcaro}, C.; {Armand}, C.;
  {Arrieta}, M.; {Backes}, M.;  et~al.
\newblock {The starburst galaxy NGC 253 revisited by H.E.S.S. and Fermi-LAT}.
\newblock {\em \aap} {\bf 2018}, {\em 617},~A73,
  \href{http://xxx.lanl.gov/abs/1806.03866}{{\normalfont
  [arXiv:astro-ph.HE/1806.03866]}}.
\newblock
  doi:{\changeurlcolor{black}\href{https://doi.org/10.1051/0004-6361/201833202}{\detokenize{10.1051/0004-6361/201833202}}}.

\bibitem[{Shimono} \em{et~al.}(2021){Shimono}, {Totani}, and
  {Sudoh}]{Shimono2021MNRAS}
{Shimono}, N.; {Totani}, T.; {Sudoh}, T.
\newblock {Prospects of newly detecting nearby star-forming galaxies by the
  Cherenkov Telescope Array}.
\newblock {\em \mnras} {\bf 2021}, {\em 506},~6212--6222,
  \href{http://xxx.lanl.gov/abs/2103.08287}{{\normalfont
  [arXiv:astro-ph.HE/2103.08287]}}.
\newblock
  doi:{\changeurlcolor{black}\href{https://doi.org/10.1093/mnras/stab2118}{\detokenize{10.1093/mnras/stab2118}}}.

\bibitem[{Zhang} \em{et~al.}(2014){Zhang}, {Wang}, {Ji}, {Smith}, {Foster}, and
  {Zhou}]{Zhang2014ApJ}
{Zhang}, S.; {Wang}, Q.D.; {Ji}, L.; {Smith}, R.K.; {Foster}, A.R.; {Zhou}, X.
\newblock {Spectral Modeling of the Charge-exchange X-Ray Emission from M82}.
\newblock {\em \apj} {\bf 2014}, {\em 794},~61,
  \href{http://xxx.lanl.gov/abs/1408.3207}{{\normalfont
  [arXiv:astro-ph.GA/1408.3207]}}.
\newblock
  doi:{\changeurlcolor{black}\href{https://doi.org/10.1088/0004-637X/794/1/61}{\detokenize{10.1088/0004-637X/794/1/61}}}.

\bibitem[{Wu} \em{et~al.}(2020){Wu}, {Li}, {Owen}, {Ji}, {Zhang}, and
  {Branduardi-Raymont}]{Wu2020MNRAS}
{Wu}, K.; {Li}, K.J.; {Owen}, E.R.; {Ji}, L.; {Zhang}, S.;
  {Branduardi-Raymont}, G.
\newblock {Charge-exchange emission and cold clumps in multiphase galactic
  outflows}.
\newblock {\em \mnras} {\bf 2020}, {\em 491},~5621--5635,
  \href{http://xxx.lanl.gov/abs/1911.11860}{{\normalfont
  [arXiv:astro-ph.GA/1911.11860]}}.
\newblock
  doi:{\changeurlcolor{black}\href{https://doi.org/10.1093/mnras/stz3301}{\detokenize{10.1093/mnras/stz3301}}}.

\bibitem[{Lopez} \em{et~al.}(2020){Lopez}, {Mathur}, {Nguyen}, {Thompson}, and
  {Olivier}]{Lopez2020ApJ}
{Lopez}, L.A.; {Mathur}, S.; {Nguyen}, D.D.; {Thompson}, T.A.; {Olivier}, G.M.
\newblock {Temperature and Metallicity Gradients in the Hot Gas Outflows of
  M82}.
\newblock {\em \apj} {\bf 2020}, {\em 904},~152,
  \href{http://xxx.lanl.gov/abs/2006.08623}{{\normalfont
  [arXiv:astro-ph.HE/2006.08623]}}.
\newblock
  doi:{\changeurlcolor{black}\href{https://doi.org/10.3847/1538-4357/abc010}{\detokenize{10.3847/1538-4357/abc010}}}.

\bibitem[{Lopez} \em{et~al.}(2023){Lopez}, {Lopez}, {Nguyen}, {Thompson},
  {Mathur}, {Bolatto}, {Vulic}, and {Sardone}]{Lopez2023ApJ}
{Lopez}, S.; {Lopez}, L.A.; {Nguyen}, D.D.; {Thompson}, T.A.; {Mathur}, S.;
  {Bolatto}, A.D.; {Vulic}, N.; {Sardone}, A.
\newblock {X-Ray Properties of NGC 253's Starburst-driven Outflow}.
\newblock {\em \apj} {\bf 2023}, {\em 942},~108,
  \href{http://xxx.lanl.gov/abs/2209.09260}{{\normalfont
  [arXiv:astro-ph.HE/2209.09260]}}.
\newblock
  doi:{\changeurlcolor{black}\href{https://doi.org/10.3847/1538-4357/aca65e}{\detokenize{10.3847/1538-4357/aca65e}}}.

\bibitem[{Perna} \em{et~al.}(2020){Perna}, {Arribas}, {Catal{\'a}n-Torrecilla},
  {Colina}, {Bellocchi}, {Fluetsch}, {Maiolino}, {Cazzoli}, {Hern{\'a}n
  Caballero}, {Pereira Santaella}, {Piqueras L{\'o}pez}, and {Rodr{\'\i}guez
  del Pino}]{Perna2020A&A}
{Perna}, M.; {Arribas}, S.; {Catal{\'a}n-Torrecilla}, C.; {Colina}, L.;
  {Bellocchi}, E.; {Fluetsch}, A.; {Maiolino}, R.; {Cazzoli}, S.; {Hern{\'a}n
  Caballero}, A.; {Pereira Santaella}, M.;  et~al.
\newblock {MUSE view of Arp220: Kpc-scale multi-phase outflow and evidence for
  positive feedback}.
\newblock {\em \aap} {\bf 2020}, {\em 643},~A139,
  \href{http://xxx.lanl.gov/abs/2009.03353}{{\normalfont
  [arXiv:astro-ph.GA/2009.03353]}}.
\newblock
  doi:{\changeurlcolor{black}\href{https://doi.org/10.1051/0004-6361/202038328}{\detokenize{10.1051/0004-6361/202038328}}}.

\bibitem[{Barker} \em{et~al.}(2008){Barker}, {de Grijs}, and
  {Cervi{\~n}o}]{Barker2008A&A}
{Barker}, S.; {de Grijs}, R.; {Cervi{\~n}o}, M.
\newblock {Star cluster versus field star formation in the nucleus of the
  prototype starburst galaxy M 82}.
\newblock {\em \aap} {\bf 2008}, {\em 484},~711--720,
  \href{http://xxx.lanl.gov/abs/0804.1913}{{\normalfont
  [arXiv:astro-ph/0804.1913]}}.
\newblock
  doi:{\changeurlcolor{black}\href{https://doi.org/10.1051/0004-6361:200809653}{\detokenize{10.1051/0004-6361:200809653}}}.

\bibitem[{Chevalier} and {Clegg}(1985)]{Chevalier1985Natur}
{Chevalier}, R.A.; {Clegg}, A.W.
\newblock {Wind from a starburst galaxy nucleus}.
\newblock {\em \nat} {\bf 1985}, {\em 317},~44--45.
\newblock
  doi:{\changeurlcolor{black}\href{https://doi.org/10.1038/317044a0}{\detokenize{10.1038/317044a0}}}.

\bibitem[{V{\"o}lk} \em{et~al.}(1996){V{\"o}lk}, {Aharonian}, and
  {Breitschwerdt}]{Volk1996SSRv}
{V{\"o}lk}, H.J.; {Aharonian}, F.A.; {Breitschwerdt}, D.
\newblock {The Nonthermal Energy Content and Gamma-Ray Emission of Starburst
  Galaxies and Clusters of Galaxies}.
\newblock {\em \ssr} {\bf 1996}, {\em 75},~279--297.
\newblock
  doi:{\changeurlcolor{black}\href{https://doi.org/10.1007/BF00195040}{\detokenize{10.1007/BF00195040}}}.

\bibitem[{Bolatto} \em{et~al.}(2013){Bolatto}, {Warren}, {Leroy}, {Walter},
  {Veilleux}, {Ostriker}, {Ott}, {Zwaan}, {Fisher}, {Weiss}, {Rosolowsky}, and
  {Hodge}]{Bolatto2013Natur}
{Bolatto}, A.D.; {Warren}, S.R.; {Leroy}, A.K.; {Walter}, F.; {Veilleux}, S.;
  {Ostriker}, E.C.; {Ott}, J.; {Zwaan}, M.; {Fisher}, D.B.; {Weiss}, A.;
  et~al.
\newblock {Suppression of star formation in the galaxy NGC 253 by a
  starburst-driven molecular wind}.
\newblock {\em \nat} {\bf 2013}, {\em 499},~450--453,
  \href{http://xxx.lanl.gov/abs/1307.6259}{{\normalfont
  [arXiv:astro-ph.CO/1307.6259]}}.
\newblock
  doi:{\changeurlcolor{black}\href{https://doi.org/10.1038/nature12351}{\detokenize{10.1038/nature12351}}}.

\bibitem[{Leroy} \em{et~al.}(2015){Leroy}, {Bolatto}, {Ostriker}, {Rosolowsky},
  {Walter}, {Warren}, {Donovan Meyer}, {Hodge}, {Meier}, {Ott}, {Sandstrom},
  {Schruba}, {Veilleux}, and {Zwaan}]{Leroy2015ApJ}
{Leroy}, A.K.; {Bolatto}, A.D.; {Ostriker}, E.C.; {Rosolowsky}, E.; {Walter},
  F.; {Warren}, S.R.; {Donovan Meyer}, J.; {Hodge}, J.; {Meier}, D.S.; {Ott},
  J.;  et~al.
\newblock {ALMA Reveals the Molecular Medium Fueling the Nearest Nuclear
  Starburst}.
\newblock {\em \apj} {\bf 2015}, {\em 801},~25,
  \href{http://xxx.lanl.gov/abs/1411.2836}{{\normalfont
  [arXiv:astro-ph.GA/1411.2836]}}.
\newblock
  doi:{\changeurlcolor{black}\href{https://doi.org/10.1088/0004-637X/801/1/25}{\detokenize{10.1088/0004-637X/801/1/25}}}.

\bibitem[{Mitsuishi} \em{et~al.}(2013){Mitsuishi}, {Yamasaki}, and
  {Takei}]{Mitsuishi2013PASJ}
{Mitsuishi}, I.; {Yamasaki}, N.Y.; {Takei}, Y.
\newblock {An X-Ray Study of the Galactic-Scale Starburst-Driven Outflow in NGC
  253}.
\newblock {\em \pasj} {\bf 2013}, {\em 65},~44,
  \href{http://xxx.lanl.gov/abs/1212.1904}{{\normalfont
  [arXiv:astro-ph.CO/1212.1904]}}.
\newblock
  doi:{\changeurlcolor{black}\href{https://doi.org/10.1093/pasj/65.2.44}{\detokenize{10.1093/pasj/65.2.44}}}.

\bibitem[{Yoast-Hull} \em{et~al.}(2015){Yoast-Hull}, {Gallagher}, and
  {Zweibel}]{YH2015MNRAS}
{Yoast-Hull}, T.M.; {Gallagher}, J.S.; {Zweibel}, E.G.
\newblock {Cosmic rays, {\ensuremath{\gamma}}-rays, and neutrinos in the
  starburst nuclei of Arp 220}.
\newblock {\em \mnras} {\bf 2015}, {\em 453},~222--228,
  \href{http://xxx.lanl.gov/abs/1506.05133}{{\normalfont
  [arXiv:astro-ph.HE/1506.05133]}}.
\newblock
  doi:{\changeurlcolor{black}\href{https://doi.org/10.1093/mnras/stv1525}{\detokenize{10.1093/mnras/stv1525}}}.

\bibitem[{Barcos-Mu{\~n}oz} \em{et~al.}(2018){Barcos-Mu{\~n}oz}, {Aalto},
  {Thompson}, {Sakamoto}, {Mart{\'\i}n}, {Leroy}, {Privon}, {Evans}, and
  {Kepley}]{Barcos2018ApJ}
{Barcos-Mu{\~n}oz}, L.; {Aalto}, S.; {Thompson}, T.A.; {Sakamoto}, K.;
  {Mart{\'\i}n}, S.; {Leroy}, A.K.; {Privon}, G.C.; {Evans}, A.S.; {Kepley}, A.
\newblock {Fast, Collimated Outflow in the Western Nucleus of Arp 220}.
\newblock {\em \apjl} {\bf 2018}, {\em 853},~L28,
  \href{http://xxx.lanl.gov/abs/1712.06381}{{\normalfont
  [arXiv:astro-ph.GA/1712.06381]}}.
\newblock
  doi:{\changeurlcolor{black}\href{https://doi.org/10.3847/2041-8213/aaa28d}{\detokenize{10.3847/2041-8213/aaa28d}}}.

\bibitem[{Lacki} and {Thompson}(2013)]{Lacki2013ApJ}
{Lacki}, B.C.; {Thompson}, T.A.
\newblock {Diffuse Hard X-Ray Emission in Starburst Galaxies as Synchrotron
  from Very High Energy Electrons}.
\newblock {\em \apj} {\bf 2013}, {\em 762},~29,
  \href{http://xxx.lanl.gov/abs/1010.3030}{{\normalfont
  [arXiv:astro-ph.HE/1010.3030]}}.
\newblock
  doi:{\changeurlcolor{black}\href{https://doi.org/10.1088/0004-637X/762/1/29}{\detokenize{10.1088/0004-637X/762/1/29}}}.

\bibitem[{Yoast-Hull} \em{et~al.}(2013){Yoast-Hull}, {Everett}, {Gallagher},
  and {Zweibel}]{YH2013ApJ}
{Yoast-Hull}, T.M.; {Everett}, J.E.; {Gallagher}, J.~S., I.; {Zweibel}, E.G.
\newblock {Winds, Clumps, and Interacting Cosmic Rays in M82}.
\newblock {\em \apj} {\bf 2013}, {\em 768},~53,
  \href{http://xxx.lanl.gov/abs/1303.4305}{{\normalfont
  [arXiv:astro-ph.HE/1303.4305]}}.
\newblock
  doi:{\changeurlcolor{black}\href{https://doi.org/10.1088/0004-637X/768/1/53}{\detokenize{10.1088/0004-637X/768/1/53}}}.

\bibitem[{Lacki} \em{et~al.}(2011){Lacki}, {Thompson}, {Quataert}, {Loeb}, and
  {Waxman}]{Lacki2011ApJ}
{Lacki}, B.C.; {Thompson}, T.A.; {Quataert}, E.; {Loeb}, A.; {Waxman}, E.
\newblock {On the GeV and TeV Detections of the Starburst Galaxies M82 and NGC
  253}.
\newblock {\em \apj} {\bf 2011}, {\em 734},~107,
  \href{http://xxx.lanl.gov/abs/1003.3257}{{\normalfont
  [arXiv:astro-ph.HE/1003.3257]}}.
\newblock
  doi:{\changeurlcolor{black}\href{https://doi.org/10.1088/0004-637X/734/2/107}{\detokenize{10.1088/0004-637X/734/2/107}}}.

\bibitem[{Behrens} \em{et~al.}(2022){Behrens}, {Mangum}, {Holdship}, {Viti},
  {Harada}, {Mart{\'\i}n}, {Sakamoto}, {Muller}, {Tanaka}, {Nakanishi},
  {Herrero-Illana}, {Yoshimura}, {Aladro}, {Colzi}, {Emig}, {Henkel}, {Huang},
  {Humire}, {Meier}, {Rivilla}, {van der Werf}, and {Alma Comprehensive
  High-Resolution Extragalactic Molecular Inventory (Alchemi)
  Collaboration}]{Behrens2022ApJ}
{Behrens}, E.; {Mangum}, J.G.; {Holdship}, J.; {Viti}, S.; {Harada}, N.;
  {Mart{\'\i}n}, S.; {Sakamoto}, K.; {Muller}, S.; {Tanaka}, K.; {Nakanishi},
  K.;  et~al.
\newblock {Tracing Interstellar Heating: An ALCHEMI Measurement of the HCN
  Isomers in NGC 253}.
\newblock {\em \apj} {\bf 2022}, {\em 939},~119,
  \href{http://xxx.lanl.gov/abs/2209.06244}{{\normalfont
  [arXiv:astro-ph.GA/2209.06244]}}.
\newblock
  doi:{\changeurlcolor{black}\href{https://doi.org/10.3847/1538-4357/ac91ce}{\detokenize{10.3847/1538-4357/ac91ce}}}.

\bibitem[{Buckman} \em{et~al.}(2020){Buckman}, {Linden}, and
  {Thompson}]{Buckman2020MNRAS}
{Buckman}, B.J.; {Linden}, T.; {Thompson}, T.A.
\newblock {Cosmic rays and magnetic fields in the core and halo of the
  starburst M82: implications for galactic wind physics}.
\newblock {\em \mnras} {\bf 2020}, {\em 494},~2679--2705,
  \href{http://xxx.lanl.gov/abs/1908.09824}{{\normalfont
  [arXiv:astro-ph.GA/1908.09824]}}.
\newblock
  doi:{\changeurlcolor{black}\href{https://doi.org/10.1093/mnras/staa875}{\detokenize{10.1093/mnras/staa875}}}.

\bibitem[{Downes} and {Solomon}(1998)]{Downes1998ApJ}
{Downes}, D.; {Solomon}, P.M.
\newblock {Rotating Nuclear Rings and Extreme Starbursts in Ultraluminous
  Galaxies}.
\newblock {\em \apj} {\bf 1998}, {\em 507},~615--654,
  \href{http://xxx.lanl.gov/abs/astro-ph/9806377}{{\normalfont
  [arXiv:astro-ph/astro-ph/9806377]}}.
\newblock
  doi:{\changeurlcolor{black}\href{https://doi.org/10.1086/306339}{\detokenize{10.1086/306339}}}.

\bibitem[{Kennicutt}(1998)]{Kennicutt1998ApJ}
{Kennicutt}, Robert~C., J.
\newblock {The Global Schmidt Law in Star-forming Galaxies}.
\newblock {\em \apj} {\bf 1998}, {\em 498},~541--552,
  \href{http://xxx.lanl.gov/abs/astro-ph/9712213}{{\normalfont
  [arXiv:astro-ph/astro-ph/9712213]}}.
\newblock
  doi:{\changeurlcolor{black}\href{https://doi.org/10.1086/305588}{\detokenize{10.1086/305588}}}.

\bibitem[{Gonz{\'a}lez-Alfonso} \em{et~al.}(2013){Gonz{\'a}lez-Alfonso},
  {Fischer}, {Bruderer}, {M{\"u}ller}, {Graci{\'a}-Carpio}, {Sturm}, {Lutz},
  {Poglitsch}, {Feuchtgruber}, {Veilleux}, {Contursi}, {Sternberg},
  {Hailey-Dunsheath}, {Verma}, {Christopher}, {Davies}, {Genzel}, and
  {Tacconi}]{GonzAlf2013A&A}
{Gonz{\'a}lez-Alfonso}, E.; {Fischer}, J.; {Bruderer}, S.; {M{\"u}ller},
  H.S.P.; {Graci{\'a}-Carpio}, J.; {Sturm}, E.; {Lutz}, D.; {Poglitsch}, A.;
  {Feuchtgruber}, H.; {Veilleux}, S.;  et~al.
\newblock {Excited OH$^{+}$, H$_{2}$O$^{+}$, and H$_{3}$O$^{+}$ in NGC 4418 and
  Arp 220}.
\newblock {\em \aap} {\bf 2013}, {\em 550},~A25,
  \href{http://xxx.lanl.gov/abs/1211.5064}{{\normalfont
  [arXiv:astro-ph.GA/1211.5064]}}.
\newblock
  doi:{\changeurlcolor{black}\href{https://doi.org/10.1051/0004-6361/201220466}{\detokenize{10.1051/0004-6361/201220466}}}.

\bibitem[{Persic} \em{et~al.}(2008){Persic}, {Rephaeli}, and
  {Arieli}]{Persic2008A&A}
{Persic}, M.; {Rephaeli}, Y.; {Arieli}, Y.
\newblock {Very-high-energy emission from M 82}.
\newblock {\em \aap} {\bf 2008}, {\em 486},~143--149.
\newblock
  doi:{\changeurlcolor{black}\href{https://doi.org/10.1051/0004-6361:200809525}{\detokenize{10.1051/0004-6361:200809525}}}.

\bibitem[{Paglione} and {Abrahams}(2012)]{Paglione2012ApJ}
{Paglione}, T.A.D.; {Abrahams}, R.D.
\newblock {Properties of nearby Starburst Galaxies Based on their Diffuse
  Gamma-Ray Emission}.
\newblock {\em \apj} {\bf 2012}, {\em 755},~106,
  \href{http://xxx.lanl.gov/abs/1206.3530}{{\normalfont
  [arXiv:astro-ph.HE/1206.3530]}}.
\newblock
  doi:{\changeurlcolor{black}\href{https://doi.org/10.1088/0004-637X/755/2/106}{\detokenize{10.1088/0004-637X/755/2/106}}}.

\bibitem[{Domingo-Santamar{\'\i}a} and {Torres}(2005)]{Domingo2005A&A}
{Domingo-Santamar{\'\i}a}, E.; {Torres}, D.F.
\newblock {High energy {\ensuremath{\gamma}}-ray emission from the starburst
  nucleus of NGC 253}.
\newblock {\em \aap} {\bf 2005}, {\em 444},~403--415,
  \href{http://xxx.lanl.gov/abs/astro-ph/0506240}{{\normalfont
  [arXiv:astro-ph/astro-ph/0506240]}}.
\newblock
  doi:{\changeurlcolor{black}\href{https://doi.org/10.1051/0004-6361:20053613}{\detokenize{10.1051/0004-6361:20053613}}}.

\bibitem[{Rephaeli} \em{et~al.}(2010){Rephaeli}, {Arieli}, and
  {Persic}]{Rephaeli2010MNRAS}
{Rephaeli}, Y.; {Arieli}, Y.; {Persic}, M.
\newblock {High-energy emission from the starburst galaxy NGC 253}.
\newblock {\em \mnras} {\bf 2010}, {\em 401},~473--478,
  \href{http://xxx.lanl.gov/abs/0906.1921}{{\normalfont
  [arXiv:astro-ph.HE/0906.1921]}}.
\newblock
  doi:{\changeurlcolor{black}\href{https://doi.org/10.1111/j.1365-2966.2009.15661.x}{\detokenize{10.1111/j.1365-2966.2009.15661.x}}}.

\bibitem[{Heesen} \em{et~al.}(2009{\natexlab{a}}){Heesen}, {Beck}, {Krause},
  and {Dettmar}]{Heesen2009A&A_I}
{Heesen}, V.; {Beck}, R.; {Krause}, M.; {Dettmar}, R.J.
\newblock {Cosmic rays and the magnetic field in the nearby starburst galaxy
  NGC 253. I. The distribution and transport of cosmic rays}.
\newblock {\em \aap} {\bf 2009}, {\em 494},~563--577,
  \href{http://xxx.lanl.gov/abs/0812.0346}{{\normalfont
  [arXiv:astro-ph/0812.0346]}}.
\newblock
  doi:{\changeurlcolor{black}\href{https://doi.org/10.1051/0004-6361:200810543}{\detokenize{10.1051/0004-6361:200810543}}}.

\bibitem[{Heesen} \em{et~al.}(2009{\natexlab{b}}){Heesen}, {Krause}, {Beck},
  and {Dettmar}]{Heesen2009A&A_II}
{Heesen}, V.; {Krause}, M.; {Beck}, R.; {Dettmar}, R.J.
\newblock {Cosmic rays and the magnetic field in the nearby starburst galaxy
  NGC 253. II. The magnetic field structure}.
\newblock {\em \aap} {\bf 2009}, {\em 506},~1123--1135,
  \href{http://xxx.lanl.gov/abs/0908.2985}{{\normalfont
  [arXiv:astro-ph.CO/0908.2985]}}.
\newblock
  doi:{\changeurlcolor{black}\href{https://doi.org/10.1051/0004-6361/200911698}{\detokenize{10.1051/0004-6361/200911698}}}.

\bibitem[{Heesen} \em{et~al.}(2011){Heesen}, {Beck}, {Krause}, and
  {Dettmar}]{Heesen2011A&A_III}
{Heesen}, V.; {Beck}, R.; {Krause}, M.; {Dettmar}, R.J.
\newblock {Cosmic rays and the magnetic field in the nearby starburst galaxy
  NGC 253 III. Helical magnetic fields in the nuclear outflow}.
\newblock {\em \aap} {\bf 2011}, {\em 535},~A79,
  \href{http://xxx.lanl.gov/abs/1109.0255}{{\normalfont
  [arXiv:astro-ph.CO/1109.0255]}}.
\newblock
  doi:{\changeurlcolor{black}\href{https://doi.org/10.1051/0004-6361/201117618}{\detokenize{10.1051/0004-6361/201117618}}}.

\bibitem[{de Cea del Pozo} \em{et~al.}(2009){de Cea del Pozo}, {Torres}, and
  {Rodriguez Marrero}]{Pozo2009ApJ}
{de Cea del Pozo}, E.; {Torres}, D.F.; {Rodriguez Marrero}, A.Y.
\newblock {Multimessenger Model for the Starburst Galaxy M82}.
\newblock {\em \apj} {\bf 2009}, {\em 698},~1054--1060,
  \href{http://xxx.lanl.gov/abs/0901.2688}{{\normalfont
  [arXiv:astro-ph.GA/0901.2688]}}.
\newblock
  doi:{\changeurlcolor{black}\href{https://doi.org/10.1088/0004-637X/698/2/1054}{\detokenize{10.1088/0004-637X/698/2/1054}}}.

\bibitem[{Ha} \em{et~al.}(2021){Ha}, {Ryu}, and {Kang}]{Ha2021ApJ}
{Ha}, J.H.; {Ryu}, D.; {Kang}, H.
\newblock {Modeling of Cosmic-Ray Production and Transport and Estimation of
  Gamma-Ray and Neutrino Emissions in Starburst Galaxies}.
\newblock {\em \apj} {\bf 2021}, {\em 907},~26,
  \href{http://xxx.lanl.gov/abs/2008.06650}{{\normalfont
  [arXiv:astro-ph.HE/2008.06650]}}.
\newblock
  doi:{\changeurlcolor{black}\href{https://doi.org/10.3847/1538-4357/abd247}{\detokenize{10.3847/1538-4357/abd247}}}.

\bibitem[{Yoast-Hull} \em{et~al.}(2017){Yoast-Hull}, {Gallagher}, {Aalto}, and
  {Varenius}]{YH2017MNRAS}
{Yoast-Hull}, T.M.; {Gallagher}, John~S., I.; {Aalto}, S.; {Varenius}, E.
\newblock {{\ensuremath{\gamma}}-Ray emission from Arp 220: indications of an
  active galactic nucleus}.
\newblock {\em \mnras} {\bf 2017}, {\em 469},~L89--L93,
  \href{http://xxx.lanl.gov/abs/1704.03791}{{\normalfont
  [arXiv:astro-ph.HE/1704.03791]}}.
\newblock
  doi:{\changeurlcolor{black}\href{https://doi.org/10.1093/mnrasl/slx054}{\detokenize{10.1093/mnrasl/slx054}}}.

\bibitem[{Hung} \em{et~al.}(2014){Hung}, {Sanders}, {Casey}, {Koss}, {Larson},
  {Lee}, {Li}, {Lockhart}, {Shih}, {Barnes}, {Kartaltepe}, and
  {Smith}]{Hung2014ApJ}
{Hung}, C.L.; {Sanders}, D.B.; {Casey}, C.M.; {Koss}, M.; {Larson}, K.L.;
  {Lee}, N.; {Li}, Y.; {Lockhart}, K.; {Shih}, H.Y.; {Barnes}, J.E.;  et~al.
\newblock {A Comparison of the Morphological Properties between Local and z
  \raisebox{-0.5ex}\textasciitilde 1 Infrared Luminous Galaxies: Are Local and
  High-z (U)LIRGs Different?}
\newblock {\em \apj} {\bf 2014}, {\em 791},~63,
  \href{http://xxx.lanl.gov/abs/1406.4509}{{\normalfont
  [arXiv:astro-ph.GA/1406.4509]}}.
\newblock
  doi:{\changeurlcolor{black}\href{https://doi.org/10.1088/0004-637X/791/1/63}{\detokenize{10.1088/0004-637X/791/1/63}}}.

\bibitem[{Larson} \em{et~al.}(2016){Larson}, {Sanders}, {Barnes}, {Ishida},
  {Evans}, {U}, {Mazzarella}, {Kim}, {Privon}, {Mirabel}, and
  {Flewelling}]{Larson2016ApJ}
{Larson}, K.L.; {Sanders}, D.B.; {Barnes}, J.E.; {Ishida}, C.M.; {Evans}, A.S.;
  {U}, V.; {Mazzarella}, J.M.; {Kim}, D.C.; {Privon}, G.C.; {Mirabel}, I.F.;
  et~al.
\newblock {Morphology and Molecular Gas Fractions of Local Luminous Infrared
  Galaxies as a Function of Infrared Luminosity and Merger Stage}.
\newblock {\em \apj} {\bf 2016}, {\em 825},~128,
  \href{http://xxx.lanl.gov/abs/1605.05417}{{\normalfont
  [arXiv:astro-ph.GA/1605.05417]}}.
\newblock
  doi:{\changeurlcolor{black}\href{https://doi.org/10.3847/0004-637X/825/2/128}{\detokenize{10.3847/0004-637X/825/2/128}}}.

\bibitem[{Lonsdale} \em{et~al.}(2006){Lonsdale}, {Farrah}, and
  {Smith}]{Lonsdale2006asup}
{Lonsdale}, C.J.; {Farrah}, D.; {Smith}, H.E.
\newblock {Ultraluminous Infrared Galaxies}. In {\em Astrophysics Update 2};
  {Mason}, J.W., Ed.;  2006; p. 285.
\newblock
  doi:{\changeurlcolor{black}\href{https://doi.org/10.1007/3-540-30313-8_9}{\detokenize{10.1007/3-540-30313-8_9}}}.

\bibitem[{P{\'e}rez-Torres} \em{et~al.}(2021){P{\'e}rez-Torres}, {Mattila},
  {Alonso-Herrero}, {Aalto}, and {Efstathiou}]{PT2021A&ARv}
{P{\'e}rez-Torres}, M.; {Mattila}, S.; {Alonso-Herrero}, A.; {Aalto}, S.;
  {Efstathiou}, A.
\newblock {Star formation and nuclear activity in luminous infrared galaxies:
  an infrared through radio review}.
\newblock {\em \aapr} {\bf 2021}, {\em 29},~2,
  \href{http://xxx.lanl.gov/abs/2010.05072}{{\normalfont
  [arXiv:astro-ph.GA/2010.05072]}}.
\newblock
  doi:{\changeurlcolor{black}\href{https://doi.org/10.1007/s00159-020-00128-x}{\detokenize{10.1007/s00159-020-00128-x}}}.

\bibitem[{Palladino} \em{et~al.}(2019){Palladino}, {Fedynitch}, {Rasmussen},
  and {Taylor}]{Palladino2019JCAP}
{Palladino}, A.; {Fedynitch}, A.; {Rasmussen}, R.W.; {Taylor}, A.M.
\newblock {IceCube neutrinos from hadronically powered gamma-ray galaxies}.
\newblock {\em \jcap} {\bf 2019}, {\em 2019},~004,
  \href{http://xxx.lanl.gov/abs/1812.04685}{{\normalfont
  [arXiv:astro-ph.HE/1812.04685]}}.
\newblock
  doi:{\changeurlcolor{black}\href{https://doi.org/10.1088/1475-7516/2019/09/004}{\detokenize{10.1088/1475-7516/2019/09/004}}}.

\bibitem[{He} \em{et~al.}(2013){He}, {Wang}, {Fan}, {Liu}, and
  {Wei}]{He2013PhRvD}
{He}, H.N.; {Wang}, T.; {Fan}, Y.Z.; {Liu}, S.M.; {Wei}, D.M.
\newblock {Diffuse PeV neutrino emission from ultraluminous infrared galaxies}.
\newblock {\em \prd} {\bf 2013}, {\em 87},~063011,
  \href{http://xxx.lanl.gov/abs/1303.1253}{{\normalfont
  [arXiv:astro-ph.HE/1303.1253]}}.
\newblock
  doi:{\changeurlcolor{black}\href{https://doi.org/10.1103/PhysRevD.87.063011}{\detokenize{10.1103/PhysRevD.87.063011}}}.

\bibitem[{IceCube Collaboration} \em{et~al.}(2020){IceCube Collaboration},
  {Aartsen}, {Ackermann}, {Adams}, {Aguilar}, {Ahlers}, {Ahrens}, {Alispach},
  {Andeen}, {Anderson}, {Ansseau}, {Anton}, {Arg{\"u}elles}, {Auffenberg},
  {Axani}, {Backes}, {Bagherpour}, {Bai}, {Balagopal V.}, {Barbano}, {Barwick},
  {Bastian}, {Baum}, {Baur}, {Bay}, {Beatty}, {Becker}, {Becker Tjus},
  {BenZvi}, {Berley}, {Bernardini}, {Besson}, {Binder}, {Bindig}, {Blaufuss},
  {Blot}, {Bohm}, {B{\"o}ser}, {Botner}, {B{\"o}ttcher}, {Bourbeau},
  {Bourbeau}, {Bradascio}, {Braun}, {Bron}, {Brostean-Kaiser}, {Burgman},
  {Buscher}, {Busse}, {Carver}, {Chen}, {Cheung}, {Chirkin}, {Choi}, {Clark},
  {Classen}, {Coleman}, {Collin}, {Conrad}, {Coppin}, {Correa}, {Cowen},
  {Cross}, {Dave}, {De Clercq}, {DeLaunay}, {Dembinski}, {Deoskar}, {De
  Ridder}, {Desiati}, {de Vries}, {de Wasseige}, {de With}, {DeYoung}, {Diaz},
  {D{\'\i}az-V{\'e}lez}, {Dujmovic}, {Dunkman}, {Dvorak}, {Eberhardt},
  {Ehrhardt}, {Eller}, {Engel}, {Evenson}, {Fahey}, {Fazely}, {Felde},
  {Filimonov}, {Finley}, {Fox}, {Franckowiak}, {Friedman}, {Fritz}, {Gaisser},
  {Gallagher}, {Ganster}, {Garrappa}, {Gerhardt}, {Ghorbani}, {Glauch},
  {Gl{\"u}senkamp}, {Goldschmidt}, {Gonzalez}, {Grant}, {Gr{\'e}goire},
  {Griffith}, {Griswold}, {G{\"u}nder}, {G{\"u}nd{\"u}z}, {Haack}, {Hallgren},
  {Halliday}, {Halve}, {Halzen}, {Hanson}, {Haungs}, {Hebecker}, {Heereman},
  {Heix}, {Helbing}, {Hellauer}, {Henningsen}, {Hickford}, {Hignight}, {Hill},
  {Hoffman}, {Hoffmann}, {Hoinka}, {Hokanson-Fasig}, {Hoshina}, {Huang},
  {Huber}, {Huber}, {Hultqvist}, {H{\"u}nnefeld}, {Hussain}, {In}, {Iovine},
  {Ishihara}, {Jansson}, {Japaridze}, {Jeong}, {Jero}, {Jones}, {Jonske},
  {Joppe}, {Kang}, {Kang}, {Kappes}, {Kappesser}, {Karg}, {Karl}, {Karle},
  {Katz}, {Kauer}, {Kelley}, {Kheirandish}, {Kim}, {Kintscher}, {Kiryluk},
  {Kittler}, {Klein}, {Koirala}, {Kolanoski}, {K{\"o}pke}, {Kopper}, {Kopper},
  {Koskinen}, {Kowalski}, {Krings}, {Kr{\"u}ckl}, {Kulacz}, {Kurahashi},
  {Kyriacou}, {Lanfranchi}, {Larson}, {Lauber}, {Lazar}, {Leonard},
  {Lesiak-Bzdak}, {Leszczy{\'n}ska}, {Leuermann}, {Liu}, {Lohfink}, {Lozano
  Mariscal}, {Lu}, {Lucarelli}, {L{\"u}nemann}, {Luszczak}, {Lyu}, {Ma},
  {Madsen}, {Maggi}, {Mahn}, {Makino}, {Mallik}, {Mallot}, {Mancina},
  {Mari\{{\c{s}}\}}, {Maruyama}, {Mase}, {Maunu}, {McNally}, {Meagher},
  {Medici}, {Medina}, {Meier}, {Meighen-Berger}, {Merino}, {Meures},
  {Micallef}, {Mockler}, {Moment{\'e}}, {Montaruli}, {Moore}, {Morse},
  {Moulai}, {Muth}, {Nagai}, {Naumann}, {Neer}, {Niederhausen}, {Nisa},
  {Nowicki}, {Nygren}, {Obertacke Pollmann}, {Oehler}, {Olivas}, {O'Murchadha},
  {O'Sullivan}, {Palczewski}, {Pandya}, {Pankova}, {Park}, {Peiffer},
  {P{\'e}rez de los Heros}, {Philippen}, {Pieloth}, {Pinat}, {Pizzuto}, {Plum},
  {Porcelli}, {Price}, {Przybylski}, {Raab}, {Raissi}, {Rameez}, {Rauch},
  {Rawlins}, {Rea}, {Rehman}, {Reimann}, {Relethford}, {Renschler}, {Renzi},
  {Resconi}, {Rhode}, {Richman}, {Robertson}, {Rongen}, {Rott}, {Ruhe},
  {Ryckbosch}, {Rysewyk}, {Safa}, {Sanchez Herrera}, {Sandrock}, {Sandroos},
  {Santander}, {Sarkar}, {Sarkar}, {Satalecka}, {Schaufel}, {Schieler},
  {Schlunder}, {Schmidt}, {Schneider}, {Schneider}, {Schr{\"o}der},
  {Schumacher}, {Sclafani}, {Seckel}, {Seunarine}, {Shefali}, {Silva},
  {Snihur}, {Soedingrekso}, {Soldin}, {Song}, {Spiczak}, {Spiering},
  {Stachurska}, {Stamatikos}, {Stanev}, {Stein}, {Stettner}, {Steuer},
  {Stezelberger}, {Stokstad}, {St{\"o}{\ss}l}, {Strotjohann}, {St{\"u}rwald},
  {Stuttard}, {Sullivan}, {Taboada}, {Tenholt}, {Ter-Antonyan}, {Terliuk},
  {Tilav}, {Tollefson}, {Tomankova}, {T{\"o}nnis}, {Toscano}, {Tosi},
  {Trettin}, {Tselengidou}, {Tung}, {Turcati}, {Turcotte}, {Turley}, {Ty},
  {Unger}, {Unland Elorrieta}, {Usner}, {Vandenbroucke}, {Van Driessche}, {van
  Eijk}, {van Eijndhoven}, {van Santen}, {Verpoest}, {Vraeghe}, {Walck},
  {Wallace}, {Wallraff}, {Wandkowsky}, {Watson}, {Weaver}, {Weindl}, {Weiss},
  {Weldert}, {Wendt}, {Werthebach}, {Whelan}, {Whitehorn}, {Wiebe}, {Wiebusch},
  {Wille}, {Williams}, {Wills}, {Wolf}, {Wood}, {Wood}, {Woschnagg}, {Wrede},
  {Xu}, {Xu}, {Xu}, {Yanez}, {Yodh}, {Yoshida}, {Yuan}, and
  {Z{\"o}cklein}]{IceCube2020arXiv200109520I}
{IceCube Collaboration}.; {Aartsen}, M.G.; {Ackermann}, M.; {Adams}, J.;
  {Aguilar}, J.A.; {Ahlers}, M.; {Ahrens}, M.; {Alispach}, C.; {Andeen}, K.;
  {Anderson}, T.;  et~al.
\newblock {Characteristics of the diffuse astrophysical electron and tau
  neutrino flux with six years of IceCube high energy cascade data}.
\newblock {\em arXiv e-prints} {\bf 2020}, p. arXiv:2001.09520,
  \href{http://xxx.lanl.gov/abs/2001.09520}{{\normalfont
  [arXiv:astro-ph.HE/2001.09520]}}.
\newblock
  doi:{\changeurlcolor{black}\href{https://doi.org/10.48550/arXiv.2001.09520}{\detokenize{10.48550/arXiv.2001.09520}}}.

\bibitem[{Abbasi} \em{et~al.}(2021){Abbasi}, {Ackermann}, {Adams}, {Aguilar},
  {Ahlers}, {Ahrens}, {Alispach}, {Alves}, {Amin}, {Andeen}, {Anderson},
  {Ansseau}, {Anton}, {Arg{\"u}elles}, {Axani}, {Bai}, {Balagopal V.},
  {Barbano}, {Barwick}, {Bastian}, {Basu}, {Baum}, {Baur}, {Bay}, {Beatty},
  {Becker}, {Becker Tjus}, {Bellenghi}, {BenZvi}, {Berley}, {Bernardini},
  {Besson}, {Binder}, {Bindig}, {Blaufuss}, {Blot}, {B{\"o}ser}, {Botner},
  {B{\"o}ttcher}, {Bourbeau}, {Bourbeau}, {Bradascio}, {Braun}, {Bron},
  {Brostean-Kaiser}, {Burgman}, {Busse}, {Campana}, {Chen}, {Chirkin}, {Choi},
  {Clark}, {Clark}, {Classen}, {Coleman}, {Collin}, {Conrad}, {Coppin},
  {Correa}, {Cowen}, {Cross}, {Dave}, {De Clercq}, {DeLaunay}, {Dembinski},
  {Deoskar}, {De Ridder}, {Desai}, {Desiati}, {de Vries}, {de Wasseige}, {de
  With}, {DeYoung}, {Dharani}, {Diaz}, {D{\'\i}az-V{\'e}lez}, {Dujmovic},
  {Dunkman}, {DuVernois}, {Dvorak}, {Ehrhardt}, {Eller}, {Engel}, {Evans},
  {Evenson}, {Fahey}, {Fazely}, {Fiedlschuster}, {Fienberg}, {Filimonov},
  {Finley}, {Fischer}, {Fox}, {Franckowiak}, {Friedman}, {Fritz}, {F{\"u}rst},
  {Gaisser}, {Gallagher}, {Ganster}, {Garrappa}, {Gerhardt}, {Ghadimi},
  {Glauch}, {Gl{\"u}senkamp}, {Goldschmidt}, {Gonzalez}, {Goswami}, {Grant},
  {Gr{\'e}goire}, {Griffith}, {Griswold}, {G{\"u}nd{\"u}z}, {Haack},
  {Hallgren}, {Halliday}, {Halve}, {Halzen}, {Ha Minh}, {Hanson}, {Hardin},
  {Haungs}, {Hauser}, {Hebecker}, {Helbing}, {Henningsen}, {Hickford},
  {Hignight}, {Hill}, {Hill}, {Hoffman}, {Hoffmann}, {Hoinka},
  {Hokanson-Fasig}, {Hoshina}, {Huang}, {Huber}, {Huber}, {Hultqvist},
  {H{\"u}nnefeld}, {Hussain}, {In}, {Iovine}, {Ishihara}, {Jansson},
  {Japaridze}, {Jeong}, {Jones}, {Joppe}, {Kang}, {Kang}, {Kang}, {Kappes},
  {Kappesser}, {Karg}, {Karl}, {Karle}, {Katori}, {Katz}, {Kauer},
  {Kellermann}, {Kelley}, {Kheirandish}, {Kim}, {Kin}, {Kintscher}, {Kiryluk},
  {Klein}, {Koirala}, {Kolanoski}, {K{\"o}pke}, {Kopper}, {Kopper}, {Koskinen},
  {Koundal}, {Kovacevich}, {Kowalski}, {Krings}, {Kr{\"u}ckl}, {Kulacz},
  {Kurahashi}, {Kyriacou}, {Lagunas Gualda}, {Lanfranchi}, {Larson}, {Lauber},
  {Lazar}, {Leonard}, {Leszczy{\'n}ska}, {Li}, {Liu}, {Lohfink}, {Lozano
  Mariscal}, {Lu}, {Lucarelli}, {Ludwig}, {Luszczak}, {Lyu}, {Ma}, {Madsen},
  {Mahn}, {Makino}, {Mallik}, {Mancina}, {Mandalia}, {Mari{\c{s}}}, {Maruyama},
  {Mase}, {McNally}, {Meagher}, {Medina}, {Meier}, {Meighen-Berger}, {Merz},
  {Micallef}, {Mockler}, {Moment{\'e}}, {Montaruli}, {Moore}, {Morse},
  {Moulai}, {Naab}, {Nagai}, {Naumann}, {Necker}, {Neer},
  {Nguy{\'a}{\guillemotright} n}, {Niederhausen}, {Nisa}, {Nowicki}, {Nygren},
  {Obertacke Pollmann}, {Oehler}, {Olivas}, {O'Sullivan}, {Pandya}, {Pankova},
  {Park}, {Parker}, {Paudel}, {Peiffer}, {P{\'e}rez de los Heros}, {Philippen},
  {Pieloth}, {Pieper}, {Pizzuto}, {Plum}, {Popovych}, {Porcelli}, {Prado
  Rodriguez}, {Price}, {Przybylski}, {Raab}, {Raissi}, {Rameez}, {Rawlins},
  {Rea}, {Rehman}, {Reimann}, {Renschler}, {Renzi}, {Resconi}, {Reusch},
  {Rhode}, {Richman}, {Riedel}, {Robertson}, {Roellinghoff}, {Rongen}, {Rott},
  {Ruhe}, {Ryckbosch}, {Rysewyk Cantu}, {Safa}, {Sanchez Herrera}, {Sandrock},
  {Sandroos}, {Santander}, {Sarkar}, {Sarkar}, {Satalecka}, {Scharf},
  {Schaufel}, {Schieler}, {Schlunder}, {Schmidt}, {Schneider}, {Schneider},
  {Schr{\"o}der}, {Schumacher}, {Sclafani}, {Seckel}, {Seunarine}, {Shefali},
  {Silva}, {Smithers}, {Snihur}, {Soedingrekso}, {Soldin}, {Spiczak},
  {Spiering}, {Stachurska}, {Stamatikos}, {Stanev}, {Stein}, {Stettner},
  {Steuer}, {Stezelberger}, {Stokstad}, {Strotjohann}, {Stuttard}, {Sullivan},
  {Taboada}, {Tenholt}, {Ter-Antonyan}, {Tilav}, {Tischbein}, {Tollefson},
  {Tomankova}, {T{\"o}nnis}, {Toscano}, {Tosi}, {Trettin}, {Tselengidou},
  {Tung}, {Turcati}, {Turcotte}, {Turley}, {Twagirayezu}, {Ty}, {Unger},
  {Unland Elorrieta}, {Vandenbroucke}, {van Eijk}, {van Eijndhoven},
  {Vannerom}, {van Santen}, {Verpoest}, {Vraeghe}, {Walck}, {Wallace},
  {Wandkowsky}, {Watson}, {Weaver}, {Weindl}, {Weiss}, {Weldert}, {Wendt},
  {Werthebach}, {Weyrauch}, {Whelan}, {Whitehorn}, {Wiebe}, {Wiebusch},
  {Williams}, {Wolf}, {Wood}, {Woschnagg}, {Wrede}, {Wulff}, {Xu}, {Xu},
  {Yanez}, {Yoshida}, {Yuan}, {Zhang}, and {IceCube
  Collaboration}]{Abbasi2021PhRvD}
{Abbasi}, R.; {Ackermann}, M.; {Adams}, J.; {Aguilar}, J.A.; {Ahlers}, M.;
  {Ahrens}, M.; {Alispach}, C.; {Alves}, A.A.; {Amin}, N.M.; {Andeen}, K.;
  et~al.
\newblock {IceCube high-energy starting event sample: Description and flux
  characterization with 7.5 years of data}.
\newblock {\em \prd} {\bf 2021}, {\em 104},~022002,
  \href{http://xxx.lanl.gov/abs/2011.03545}{{\normalfont
  [arXiv:astro-ph.HE/2011.03545]}}.
\newblock
  doi:{\changeurlcolor{black}\href{https://doi.org/10.1103/PhysRevD.104.022002}{\detokenize{10.1103/PhysRevD.104.022002}}}.

\bibitem[{Abbasi} \em{et~al.}(2022){Abbasi}, {Ackermann}, {Adams}, {Aguilar},
  {Ahlers}, {Ahrens}, {Alameddine}, {Alispach}, {Alves}, {Amin}, {Andeen},
  {Anderson}, {Anton}, {Arg{\"u}elles}, {Ashida}, {Axani}, {Bai}, {Balagopal
  V.}, {Barbano}, {Barwick}, {Bastian}, {Basu}, {Baur}, {Bay}, {Beatty},
  {Becker}, {Tjus}, {Bellenghi}, {BenZvi}, {Berley}, {Bernardini}, {Besson},
  {Binder}, {Bindig}, {Blaufuss}, {Blot}, {Boddenberg}, {Bontempo}, {Borowka},
  {B{\"o}ser}, {Botner}, {B{\"o}ttcher}, {Bourbeau}, {Bradascio}, {Braun},
  {Brinson}, {Bron}, {Brostean-Kaiser}, {Browne}, {Burgman}, {Burley}, {Busse},
  {Campana}, {Carnie-Bronca}, {Chen}, {Chen}, {Chirkin}, {Choi}, {Clark},
  {Clark}, {Classen}, {Coleman}, {Collin}, {Conrad}, {Coppin}, {Correa},
  {Cowen}, {Cross}, {Dappen}, {Dave}, {De Clercq}, {DeLaunay}, {L{\'o}pez},
  {Dembinski}, {Deoskar}, {Desai}, {Desiati}, {de Vries}, {de Wasseige}, {de
  With}, {DeYoung}, {Diaz}, {D{\'\i}az-V{\'e}lez}, {Dittmer}, {Dujmovic},
  {Dunkman}, {DuVernois}, {Dvorak}, {Ehrhardt}, {Eller}, {Engel}, {Erpenbeck},
  {Evans}, {Evenson}, {Fan}, {Fazely}, {Feigl}, {Fiedlschuster}, {Fienberg},
  {Filimonov}, {Finley}, {Fischer}, {Fox}, {Franckowiak}, {Friedman}, {Fritz},
  {F{\"u}rst}, {Gaisser}, {Gallagher}, {Ganster}, {Garcia}, {Garrappa},
  {Gerhardt}, {Ghadimi}, {Glaser}, {Glauch}, {Gl{\"u}senkamp}, {Gonzalez},
  {Goswami}, {Grant}, {Gr{\'e}goire}, {Griswold}, {G{\"u}nther}, {Gutjahr},
  {Haack}, {Hallgren}, {Halliday}, {Halve}, {Halzen}, {Minh}, {Hanson},
  {Hardin}, {Harnisch}, {Haungs}, {Hebecker}, {Helbing}, {Henningsen},
  {Hettinger}, {Hickford}, {Hignight}, {Hill}, {Hill}, {Hoffman}, {Hoffmann},
  {Hokanson-Fasig}, {Hoshina}, {Huang}, {Huber}, {Huber}, {Hultqvist},
  {H{\"u}nnefeld}, {Hussain}, {Hymon}, {In}, {Iovine}, {Ishihara}, {Jansson},
  {Japaridze}, {Jeong}, {Jin}, {Jones}, {Kang}, {Kang}, {Kang}, {Kappes},
  {Kappesser}, {Kardum}, {Karg}, {Karl}, {Karle}, {Katz}, {Kauer},
  {Kellermann}, {Kelley}, {Kheirandish}, {Kin}, {Kintscher}, {Kiryluk},
  {Klein}, {Koirala}, {Kolanoski}, {Kontrimas}, {K{\"o}pke}, {Kopper},
  {Kopper}, {Koskinen}, {Koundal}, {Kovacevich}, {Kowalski}, {Kozynets}, {Kun},
  {Kurahashi}, {Lad}, {Gualda}, {Lanfranchi}, {Larson}, {Lauber}, {Lazar},
  {Lee}, {Leonard}, {Leszczy{\'n}ska}, {Li}, {Lincetto}, {Liu}, {Liubarska},
  {Lohfink}, {Mariscal}, {Lu}, {Lucarelli}, {Ludwig}, {Luszczak}, {Lyu}, {Ma},
  {Madsen}, {Mahn}, {Makino}, {Mancina}, {Mari{\c{s}}}, {Martinez-Soler},
  {Maruyama}, {Mase}, {McElroy}, {McNally}, {Mead}, {Meagher}, {Mechbal},
  {Medina}, {Meier}, {Meighen-Berger}, {Micallef}, {Mockler}, {Montaruli},
  {Moore}, {Morse}, {Moulai}, {Naab}, {Nagai}, {Naumann}, {Necker}, {Nguyễn},
  {Niederhausen}, {Nisa}, {Nowicki}, {Pollmann}, {Oehler}, {Oeyen}, {Olivas},
  {O'Sullivan}, {Pandya}, {Pankova}, {Park}, {Parker}, {Paudel}, {Paul}, {de
  los Heros}, {Peters}, {Peterson}, {Philippen}, {Pieper}, {Pittermann},
  {Pizzuto}, {Plum}, {Popovych}, {Porcelli}, {Rodriguez}, {Price}, {Pries},
  {Przybylski}, {Raab}, {Raissi}, {Rameez}, {Rawlins}, {Rea}, {Rehman},
  {Reichherzer}, {Reimann}, {Renzi}, {Resconi}, {Reusch}, {Rhode}, {Richman},
  {Riedel}, {Roberts}, {Robertson}, {Roellinghoff}, {Rongen}, {Rott}, {Ruhe},
  {Ryckbosch}, {Cantu}, {Safa}, {Saffer}, {Herrera}, {Sandrock}, {Sandroos},
  {Santander}, {Sarkar}, {Sarkar}, {Satalecka}, {Schaufel}, {Schieler},
  {Schindler}, {Schmidt}, {Schneider}, {Schneider}, {Schr{\"o}der},
  {Schumacher}, {Schwefer}, {Sclafani}, {Seckel}, {Seunarine}, {Sharma},
  {Shefali}, {Silva}, {Skrzypek}, {Smithers}, {Snihur}, {Soedingrekso},
  {Soldin}, {Spannfellner}, {Spiczak}, {Spiering}, {Stachurska}, {Stamatikos},
  {Stanev}, {Stein}, {Stettner}, {Steuer}, {Stezelberger}, {St{\"u}rwald},
  {Stuttard}, {Sullivan}, {Taboada}, {Ter-Antonyan}, {Tilav}, {Tischbein},
  {Tollefson}, {T{\"o}nnis}, {Toscano}, {Tosi}, {Trettin}, {Tselengidou},
  {Tung}, {Turcati}, {Turcotte}, {Turley}, {Twagirayezu}, {Ty}, {Elorrieta},
  {Valtonen-Mattila}, {Vandenbroucke}, {van Eijndhoven}, {Vannerom}, {van
  Santen}, {Verpoest}, {Walck}, {Watson}, {Weaver}, {Weigel}, {Weindl},
  {Weiss}, {Weldert}, {Wendt}, {Werthebach}, {Weyrauch}, {Whitehorn},
  {Wiebusch}, {Williams}, {Wolf}, {Woschnagg}, {Wrede}, {Wulff}, {Xu}, {Yanez},
  {Yoshida}, {Yu}, {Yuan}, {Zhang}, {Zhelnin}, and {IceCube
  Collaboration}]{Abbasi2022ApJb}
{Abbasi}, R.; {Ackermann}, M.; {Adams}, J.; {Aguilar}, J.A.; {Ahlers}, M.;
  {Ahrens}, M.; {Alameddine}, J.M.; {Alispach}, C.; {Alves}, A.~A., J.; {Amin},
  N.M.;  et~al.
\newblock {Improved Characterization of the Astrophysical Muon-neutrino Flux
  with 9.5 Years of IceCube Data}.
\newblock {\em \apj} {\bf 2022}, {\em 928},~50,
  \href{http://xxx.lanl.gov/abs/2111.10299}{{\normalfont
  [arXiv:astro-ph.HE/2111.10299]}}.
\newblock
  doi:{\changeurlcolor{black}\href{https://doi.org/10.3847/1538-4357/ac4d29}{\detokenize{10.3847/1538-4357/ac4d29}}}.

\bibitem[{Ackermann} \em{et~al.}(2016){Ackermann}, {Ajello}, {Albert},
  {Atwood}, {Baldini}, {Ballet}, {Barbiellini}, {Bastieri}, {Bechtol},
  {Bellazzini}, {Bissaldi}, {Blandford}, {Bloom}, {Bonino}, {Bregeon},
  {Britto}, {Bruel}, {Buehler}, {Caliandro}, {Cameron}, {Caragiulo}, {Caraveo},
  {Cavazzuti}, {Cecchi}, {Charles}, {Chekhtman}, {Chiang}, {Chiaro}, {Ciprini},
  {Cohen-Tanugi}, {Cominsky}, {Costanza}, {Cutini}, {D'Ammando}, {de Angelis},
  {de Palma}, {Desiante}, {Digel}, {Di Mauro}, {Di Venere}, {Dom{\'\i}nguez},
  {Drell}, {Favuzzi}, {Fegan}, {Ferrara}, {Franckowiak}, {Fukazawa}, {Funk},
  {Fusco}, {Gargano}, {Gasparrini}, {Giglietto}, {Giommi}, {Giordano},
  {Giroletti}, {Godfrey}, {Green}, {Grenier}, {Guiriec}, {Hays}, {Horan},
  {Iafrate}, {Jogler}, {J{\'o}hannesson}, {Kuss}, {La Mura}, {Larsson},
  {Latronico}, {Li}, {Li}, {Longo}, {Loparco}, {Lott}, {Lovellette}, {Lubrano},
  {Madejski}, {Magill}, {Maldera}, {Manfreda}, {Mayer}, {Mazziotta},
  {Michelson}, {Mitthumsiri}, {Mizuno}, {Moiseev}, {Monzani}, {Morselli},
  {Moskalenko}, {Murgia}, {Negro}, {Nuss}, {Ohsugi}, {Okada}, {Omodei},
  {Orlando}, {Ormes}, {Paneque}, {Perkins}, {Pesce-Rollins}, {Petrosian},
  {Piron}, {Pivato}, {Porter}, {Rain{\`o}}, {Rando}, {Razzano}, {Razzaque},
  {Reimer}, {Reimer}, {Reposeur}, {Romani}, {S{\'a}nchez-Conde}, {Schmid},
  {Schulz}, {Sgr{\`o}}, {Simone}, {Siskind}, {Spada}, {Spandre}, {Spinelli},
  {Suson}, {Takahashi}, {Thayer}, {Tibaldo}, {Torres}, {Troja}, {Vianello},
  {Yassine}, and {Zimmer}]{Ackermann2016PhRvL_EGB}
{Ackermann}, M.; {Ajello}, M.; {Albert}, A.; {Atwood}, W.B.; {Baldini}, L.;
  {Ballet}, J.; {Barbiellini}, G.; {Bastieri}, D.; {Bechtol}, K.; {Bellazzini},
  R.;  et~al.
\newblock {Resolving the Extragalactic {\ensuremath{\gamma}} -Ray Background
  above 50 GeV with the Fermi Large Area Telescope}.
\newblock {\em \prl} {\bf 2016}, {\em 116},~151105,
  \href{http://xxx.lanl.gov/abs/1511.00693}{{\normalfont
  [arXiv:astro-ph.CO/1511.00693]}}.
\newblock
  doi:{\changeurlcolor{black}\href{https://doi.org/10.1103/PhysRevLett.116.151105}{\detokenize{10.1103/PhysRevLett.116.151105}}}.

\bibitem[{Bechtol} \em{et~al.}(2017){Bechtol}, {Ahlers}, {Di Mauro}, {Ajello},
  and {Vandenbroucke}]{Bechtol2017ApJ}
{Bechtol}, K.; {Ahlers}, M.; {Di Mauro}, M.; {Ajello}, M.; {Vandenbroucke}, J.
\newblock {Evidence against Star-forming Galaxies as the Dominant Source of
  Icecube Neutrinos}.
\newblock {\em \apj} {\bf 2017}, {\em 836},~47,
  \href{http://xxx.lanl.gov/abs/1511.00688}{{\normalfont
  [arXiv:astro-ph.HE/1511.00688]}}.
\newblock
  doi:{\changeurlcolor{black}\href{https://doi.org/10.3847/1538-4357/836/1/47}{\detokenize{10.3847/1538-4357/836/1/47}}}.

\bibitem[{Murase} \em{et~al.}(2016){Murase}, {Guetta}, and
  {Ahlers}]{Murase2016PhRvL}
{Murase}, K.; {Guetta}, D.; {Ahlers}, M.
\newblock {Hidden Cosmic-Ray Accelerators as an Origin of TeV-PeV Cosmic
  Neutrinos}.
\newblock {\em \prl} {\bf 2016}, {\em 116},~071101,
  \href{http://xxx.lanl.gov/abs/1509.00805}{{\normalfont
  [arXiv:astro-ph.HE/1509.00805]}}.
\newblock
  doi:{\changeurlcolor{black}\href{https://doi.org/10.1103/PhysRevLett.116.071101}{\detokenize{10.1103/PhysRevLett.116.071101}}}.

\bibitem[{Vereecken} and {de Vries}(2020)]{Vereecken2020arXiv200403435V}
{Vereecken}, M.; {de Vries}, K.D.
\newblock {Obscured $pp$-channel neutrino sources}.
\newblock {\em arXiv e-prints} {\bf 2020}, p. arXiv:2004.03435,
  \href{http://xxx.lanl.gov/abs/2004.03435}{{\normalfont
  [arXiv:astro-ph.HE/2004.03435]}}.
\newblock
  doi:{\changeurlcolor{black}\href{https://doi.org/10.48550/arXiv.2004.03435}{\detokenize{10.48550/arXiv.2004.03435}}}.

\bibitem[{Hollenbach} \em{et~al.}(2012){Hollenbach}, {Kaufman}, {Neufeld},
  {Wolfire}, and {Goicoechea}]{Hollenbach2012ApJ}
{Hollenbach}, D.; {Kaufman}, M.J.; {Neufeld}, D.; {Wolfire}, M.; {Goicoechea},
  J.R.
\newblock {The Chemistry of Interstellar OH$^{+}$, H$_{2}$O$^{+}$, and
  H$_{3}$O$^{+}$: Inferring the Cosmic-Ray Ionization Rates from Observations
  of Molecular Ions}.
\newblock {\em \apj} {\bf 2012}, {\em 754},~105,
  \href{http://xxx.lanl.gov/abs/1205.6446}{{\normalfont
  [arXiv:astro-ph.GA/1205.6446]}}.
\newblock
  doi:{\changeurlcolor{black}\href{https://doi.org/10.1088/0004-637X/754/2/105}{\detokenize{10.1088/0004-637X/754/2/105}}}.

\bibitem[{Gonz{\'a}lez-Alfonso} \em{et~al.}(2018){Gonz{\'a}lez-Alfonso},
  {Fischer}, {Bruderer}, {Ashby}, {Smith}, {Veilleux}, {M{\"u}ller}, {Stewart},
  and {Sturm}]{GonzAlf2018ApJ}
{Gonz{\'a}lez-Alfonso}, E.; {Fischer}, J.; {Bruderer}, S.; {Ashby}, M.L.N.;
  {Smith}, H.A.; {Veilleux}, S.; {M{\"u}ller}, H.S.P.; {Stewart}, K.P.;
  {Sturm}, E.
\newblock {Outflowing OH$^{+}$ in Markarian 231: The Ionization Rate of the
  Molecular Gas}.
\newblock {\em \apj} {\bf 2018}, {\em 857},~66,
  \href{http://xxx.lanl.gov/abs/1803.04690}{{\normalfont
  [arXiv:astro-ph.GA/1803.04690]}}.
\newblock
  doi:{\changeurlcolor{black}\href{https://doi.org/10.3847/1538-4357/aab6b8}{\detokenize{10.3847/1538-4357/aab6b8}}}.

\bibitem[{Simpson} \em{et~al.}(2014){Simpson}, {Swinbank}, {Smail},
  {Alexander}, {Brandt}, {Bertoldi}, {de Breuck}, {Chapman}, {Coppin}, {da
  Cunha}, {Danielson}, {Dannerbauer}, {Greve}, {Hodge}, {Ivison}, {Karim},
  {Knudsen}, {Poggianti}, {Schinnerer}, {Thomson}, {Walter}, {Wardlow},
  {Wei{\ss}}, and {van der Werf}]{Simpson2014ApJ}
{Simpson}, J.M.; {Swinbank}, A.M.; {Smail}, I.; {Alexander}, D.M.; {Brandt},
  W.N.; {Bertoldi}, F.; {de Breuck}, C.; {Chapman}, S.C.; {Coppin}, K.E.K.; {da
  Cunha}, E.;  et~al.
\newblock {An ALMA Survey of Submillimeter Galaxies in the Extended Chandra
  Deep Field South: The Redshift Distribution and Evolution of Submillimeter
  Galaxies}.
\newblock {\em \apj} {\bf 2014}, {\em 788},~125,
  \href{http://xxx.lanl.gov/abs/1310.6363}{{\normalfont
  [arXiv:astro-ph.CO/1310.6363]}}.
\newblock
  doi:{\changeurlcolor{black}\href{https://doi.org/10.1088/0004-637X/788/2/125}{\detokenize{10.1088/0004-637X/788/2/125}}}.

\bibitem[{Dole} \em{et~al.}(2004){Dole}, {Le Floc'h}, {P{\'e}rez-Gonz{\'a}lez},
  {Papovich}, {Egami}, {Lagache}, {Alonso-Herrero}, {Engelbracht}, {Gordon},
  {Hines}, {Krause}, {Misselt}, {Morrison}, {Rieke}, {Rieke}, {Rigby}, {Young},
  {Bai}, {Blaylock}, {Neugebauer}, {Beichman}, {Frayer}, {Mould}, and
  {Richards}]{Dole2004ApJS}
{Dole}, H.; {Le Floc'h}, E.; {P{\'e}rez-Gonz{\'a}lez}, P.G.; {Papovich}, C.;
  {Egami}, E.; {Lagache}, G.; {Alonso-Herrero}, A.; {Engelbracht}, C.W.;
  {Gordon}, K.D.; {Hines}, D.C.;  et~al.
\newblock {Far-infrared Source Counts at 70 and 160 Microns in Spitzer Deep
  Surveys}.
\newblock {\em \apjs} {\bf 2004}, {\em 154},~87--92,
  \href{http://xxx.lanl.gov/abs/astro-ph/0406021}{{\normalfont
  [arXiv:astro-ph/astro-ph/0406021]}}.
\newblock
  doi:{\changeurlcolor{black}\href{https://doi.org/10.1086/422472}{\detokenize{10.1086/422472}}}.

\bibitem[{Le Floc'h} \em{et~al.}(2009){Le Floc'h}, {Aussel}, {Ilbert},
  {Riguccini}, {Frayer}, {Salvato}, {Arnouts}, {Surace}, {Feruglio},
  {Rodighiero}, {Capak}, {Kartaltepe}, {Heinis}, {Sheth}, {Yan}, {McCracken},
  {Thompson}, {Sanders}, {Scoville}, and {Koekemoer}]{Floc2009ApJ}
{Le Floc'h}, E.; {Aussel}, H.; {Ilbert}, O.; {Riguccini}, L.; {Frayer}, D.T.;
  {Salvato}, M.; {Arnouts}, S.; {Surace}, J.; {Feruglio}, C.; {Rodighiero}, G.;
   et~al.
\newblock {Deep Spitzer 24 {\ensuremath{\mu}}m COSMOS Imaging. I. The Evolution
  of Luminous Dusty Galaxies{\textemdash}Confronting the Models}.
\newblock {\em \apj} {\bf 2009}, {\em 703},~222--239,
  \href{http://xxx.lanl.gov/abs/0909.4303}{{\normalfont
  [arXiv:astro-ph.CO/0909.4303]}}.
\newblock
  doi:{\changeurlcolor{black}\href{https://doi.org/10.1088/0004-637X/703/1/222}{\detokenize{10.1088/0004-637X/703/1/222}}}.

\bibitem[{Blain} \em{et~al.}(2002){Blain}, {Smail}, {Ivison}, {Kneib}, and
  {Frayer}]{Blain2002PhR}
{Blain}, A.W.; {Smail}, I.; {Ivison}, R.J.; {Kneib}, J.P.; {Frayer}, D.T.
\newblock {Submillimeter galaxies}.
\newblock {\em \physrep} {\bf 2002}, {\em 369},~111--176,
  \href{http://xxx.lanl.gov/abs/astro-ph/0202228}{{\normalfont
  [arXiv:astro-ph/astro-ph/0202228]}}.
\newblock
  doi:{\changeurlcolor{black}\href{https://doi.org/10.1016/S0370-1573(02)00134-5}{\detokenize{10.1016/S0370-1573(02)00134-5}}}.

\bibitem[{Indriolo} \em{et~al.}(2018){Indriolo}, {Bergin}, {Falgarone},
  {Godard}, {Zwaan}, {Neufeld}, and {Wolfire}]{Indriolo2018ApJ}
{Indriolo}, N.; {Bergin}, E.A.; {Falgarone}, E.; {Godard}, B.; {Zwaan}, M.A.;
  {Neufeld}, D.A.; {Wolfire}, M.G.
\newblock {Constraints on the Cosmic-Ray Ionization Rate in the z
  {\ensuremath{\sim}} 2.3 Lensed Galaxies SMM J2135-0102 and SDP 17b from
  Observations of OH$^{+}$ and H$_{2}$O$^{+}$}.
\newblock {\em \apj} {\bf 2018}, {\em 865},~127,
  \href{http://xxx.lanl.gov/abs/1808.04852}{{\normalfont
  [arXiv:astro-ph.GA/1808.04852]}}.
\newblock
  doi:{\changeurlcolor{black}\href{https://doi.org/10.3847/1538-4357/aad7b3}{\detokenize{10.3847/1538-4357/aad7b3}}}.

\bibitem[{Danielson} \em{et~al.}(2013){Danielson}, {Swinbank}, {Smail},
  {Bayet}, {van der Werf}, {Cox}, {Edge}, {Henkel}, and
  {Ivison}]{Danielson2013MNRAS}
{Danielson}, A.L.R.; {Swinbank}, A.M.; {Smail}, I.; {Bayet}, E.; {van der
  Werf}, P.P.; {Cox}, P.; {Edge}, A.C.; {Henkel}, C.; {Ivison}, R.J.
\newblock {$^{13}$CO and C$^{18}$O emission from a dense gas disc at z = 2.3:
  abundance variations, cosmic rays and the initial conditions for star
  formation}.
\newblock {\em \mnras} {\bf 2013}, {\em 436},~2793--2809,
  \href{http://xxx.lanl.gov/abs/1309.5952}{{\normalfont
  [arXiv:astro-ph.CO/1309.5952]}}.
\newblock
  doi:{\changeurlcolor{black}\href{https://doi.org/10.1093/mnras/stt1775}{\detokenize{10.1093/mnras/stt1775}}}.

\bibitem[{Falgarone} \em{et~al.}(2017){Falgarone}, {Zwaan}, {Godard}, {Bergin},
  {Ivison}, {Andreani}, {Bournaud}, {Bussmann}, {Elbaz}, {Omont}, {Oteo}, and
  {Walter}]{Falgarone2017Natur}
{Falgarone}, E.; {Zwaan}, M.A.; {Godard}, B.; {Bergin}, E.; {Ivison}, R.J.;
  {Andreani}, P.M.; {Bournaud}, F.; {Bussmann}, R.S.; {Elbaz}, D.; {Omont}, A.;
   et~al.
\newblock {Large turbulent reservoirs of cold molecular gas around
  high-redshift starburst galaxies}.
\newblock {\em \nat} {\bf 2017}, {\em 548},~430--433,
  \href{http://xxx.lanl.gov/abs/1708.08851}{{\normalfont
  [arXiv:astro-ph.GA/1708.08851]}}.
\newblock
  doi:{\changeurlcolor{black}\href{https://doi.org/10.1038/nature23298}{\detokenize{10.1038/nature23298}}}.

\bibitem[{French} \em{et~al.}(2015){French}, {Yang}, {Zabludoff}, {Narayanan},
  {Shirley}, {Walter}, {Smith}, and {Tremonti}]{French2015ApJ}
{French}, K.D.; {Yang}, Y.; {Zabludoff}, A.; {Narayanan}, D.; {Shirley}, Y.;
  {Walter}, F.; {Smith}, J.D.; {Tremonti}, C.A.
\newblock {Discovery of Large Molecular Gas Reservoirs in Post-starburst
  Galaxies}.
\newblock {\em \apj} {\bf 2015}, {\em 801},~1,
  \href{http://xxx.lanl.gov/abs/1501.00983}{{\normalfont
  [arXiv:astro-ph.GA/1501.00983]}}.
\newblock
  doi:{\changeurlcolor{black}\href{https://doi.org/10.1088/0004-637X/801/1/1}{\detokenize{10.1088/0004-637X/801/1/1}}}.

\bibitem[{French} \em{et~al.}(2018){French}, {Zabludoff}, {Yoon}, {Shirley},
  {Yang}, {Smercina}, {Smith}, and {Narayanan}]{French2018ApJ}
{French}, K.D.; {Zabludoff}, A.I.; {Yoon}, I.; {Shirley}, Y.; {Yang}, Y.;
  {Smercina}, A.; {Smith}, J.D.; {Narayanan}, D.
\newblock {Why Post-starburst Galaxies Are Now Quiescent}.
\newblock {\em \apj} {\bf 2018}, {\em 861},~123,
  \href{http://xxx.lanl.gov/abs/1805.12132}{{\normalfont
  [arXiv:astro-ph.GA/1805.12132]}}.
\newblock
  doi:{\changeurlcolor{black}\href{https://doi.org/10.3847/1538-4357/aac8de}{\detokenize{10.3847/1538-4357/aac8de}}}.

\bibitem[{Rowlands} \em{et~al.}(2015){Rowlands}, {Wild}, {Nesvadba},
  {Sibthorpe}, {Mortier}, {Lehnert}, and {da Cunha}]{Rowlands2015MNRAS}
{Rowlands}, K.; {Wild}, V.; {Nesvadba}, N.; {Sibthorpe}, B.; {Mortier}, A.;
  {Lehnert}, M.; {da Cunha}, E.
\newblock {The evolution of the cold interstellar medium in galaxies following
  a starburst}.
\newblock {\em \mnras} {\bf 2015}, {\em 448},~258--279,
  \href{http://xxx.lanl.gov/abs/1412.6090}{{\normalfont
  [arXiv:astro-ph.GA/1412.6090]}}.
\newblock
  doi:{\changeurlcolor{black}\href{https://doi.org/10.1093/mnras/stu2714}{\detokenize{10.1093/mnras/stu2714}}}.

\bibitem[{Alatalo} \em{et~al.}(2016){Alatalo}, {Lisenfeld}, {Lanz}, {Appleton},
  {Ardila}, {Cales}, {Kewley}, {Lacy}, {Medling}, {Nyland}, {Rich}, and
  {Urry}]{Alatalo2016ApJ}
{Alatalo}, K.; {Lisenfeld}, U.; {Lanz}, L.; {Appleton}, P.N.; {Ardila}, F.;
  {Cales}, S.L.; {Kewley}, L.J.; {Lacy}, M.; {Medling}, A.M.; {Nyland}, K.;
  et~al.
\newblock {Shocked POststarburst Galaxy Survey. II. The Molecular Gas Content
  and Properties of a Subset of SPOGs}.
\newblock {\em \apj} {\bf 2016}, {\em 827},~106,
  \href{http://xxx.lanl.gov/abs/1604.01122}{{\normalfont
  [arXiv:astro-ph.GA/1604.01122]}}.
\newblock
  doi:{\changeurlcolor{black}\href{https://doi.org/10.3847/0004-637X/827/2/106}{\detokenize{10.3847/0004-637X/827/2/106}}}.

\bibitem[{Watson} \em{et~al.}(2015){Watson}, {Christensen}, {Knudsen},
  {Richard}, {Gallazzi}, and {Micha{\l}owski}]{Watson2015Natur}
{Watson}, D.; {Christensen}, L.; {Knudsen}, K.K.; {Richard}, J.; {Gallazzi},
  A.; {Micha{\l}owski}, M.J.
\newblock {A dusty, normal galaxy in the epoch of reionization}.
\newblock {\em \nat} {\bf 2015}, {\em 519},~327--330,
  \href{http://xxx.lanl.gov/abs/1503.00002}{{\normalfont
  [arXiv:astro-ph.GA/1503.00002]}}.
\newblock
  doi:{\changeurlcolor{black}\href{https://doi.org/10.1038/nature14164}{\detokenize{10.1038/nature14164}}}.

\bibitem[{Laporte} \em{et~al.}(2023){Laporte}, {Ellis}, {Witten}, and
  {Roberts-Borsani}]{Laporte2022arXiv221205072L}
{Laporte}, N.; {Ellis}, R.S.; {Witten}, C.E.C.; {Roberts-Borsani}, G.
\newblock {Resolving ambiguities in the inferred star formation histories of
  intense [O III] emitters in the reionization Era}.
\newblock {\em \mnras} {\bf 2023}, {\em 523},~3018--3024,
  \href{http://xxx.lanl.gov/abs/2212.05072}{{\normalfont
  [arXiv:astro-ph.GA/2212.05072]}}.
\newblock
  doi:{\changeurlcolor{black}\href{https://doi.org/10.1093/mnras/stad1597}{\detokenize{10.1093/mnras/stad1597}}}.

\bibitem[{Lanz} \em{et~al.}(2022){Lanz}, {Stepanoff}, {Hickox}, {Alatalo},
  {French}, {Rowlands}, {Nyland}, {Appleton}, {Lacy}, {Medling}, {Mulchaey},
  {Sazonova}, and {Urry}]{Lanz2022ApJ}
{Lanz}, L.; {Stepanoff}, S.; {Hickox}, R.C.; {Alatalo}, K.; {French}, K.D.;
  {Rowlands}, K.; {Nyland}, K.; {Appleton}, P.N.; {Lacy}, M.; {Medling}, A.;
  et~al.
\newblock {Are Active Galactic Nuclei in Post-starburst Galaxies Driving the
  Change or Along for the Ride?}
\newblock {\em \apj} {\bf 2022}, {\em 935},~29,
  \href{http://xxx.lanl.gov/abs/2207.00607}{{\normalfont
  [arXiv:astro-ph.GA/2207.00607]}}.
\newblock
  doi:{\changeurlcolor{black}\href{https://doi.org/10.3847/1538-4357/ac7d56}{\detokenize{10.3847/1538-4357/ac7d56}}}.

\bibitem[{Smercina} \em{et~al.}(2018){Smercina}, {Smith}, {Dale}, {French},
  {Croxall}, {Zhukovska}, {Togi}, {Bell}, {Crocker}, {Draine}, {Jarrett},
  {Tremonti}, {Yang}, and {Zabludoff}]{Smercina2018ApJ}
{Smercina}, A.; {Smith}, J.D.T.; {Dale}, D.A.; {French}, K.D.; {Croxall}, K.V.;
  {Zhukovska}, S.; {Togi}, A.; {Bell}, E.F.; {Crocker}, A.F.; {Draine}, B.T.;
  et~al.
\newblock {After the Fall: The Dust and Gas in E+A Post-starburst Galaxies}.
\newblock {\em \apj} {\bf 2018}, {\em 855},~51,
  \href{http://xxx.lanl.gov/abs/1802.04798}{{\normalfont
  [arXiv:astro-ph.GA/1802.04798]}}.
\newblock
  doi:{\changeurlcolor{black}\href{https://doi.org/10.3847/1538-4357/aaafcd}{\detokenize{10.3847/1538-4357/aaafcd}}}.

\bibitem[{Papadopoulos}(2010)]{Papadopoulos2010ApJ}
{Papadopoulos}, P.P.
\newblock {A Cosmic-ray-dominated Interstellar Medium in Ultra Luminous
  Infrared Galaxies: New Initial Conditions for Star Formation}.
\newblock {\em \apj} {\bf 2010}, {\em 720},~226--232,
  \href{http://xxx.lanl.gov/abs/1009.1134}{{\normalfont
  [arXiv:astro-ph.CO/1009.1134]}}.
\newblock
  doi:{\changeurlcolor{black}\href{https://doi.org/10.1088/0004-637X/720/1/226}{\detokenize{10.1088/0004-637X/720/1/226}}}.

\bibitem[{Yokoyama} and {Ohira}(2023)]{Yokoyama2023MNRAS}
{Yokoyama}, S.L.; {Ohira}, Y.
\newblock {Resistive heating induced by streaming cosmic rays around a galaxy
  in the early Universe}.
\newblock {\em \mnras} {\bf 2023}, {\em 523},~3671--3677,
  \href{http://xxx.lanl.gov/abs/2302.07028}{{\normalfont
  [arXiv:astro-ph.HE/2302.07028]}}.
\newblock
  doi:{\changeurlcolor{black}\href{https://doi.org/10.1093/mnras/stad1596}{\detokenize{10.1093/mnras/stad1596}}}.

\bibitem[{Owen}(2022)]{Owen2022ECRS}
{Owen}, E.R.
\newblock {Cosmic rays as a feedback agent in primordial galactic ecosystems}.
\newblock {\em arXiv e-prints} {\bf 2022}, p. arXiv:2212.06469,
  \href{http://xxx.lanl.gov/abs/2212.06469}{{\normalfont
  [arXiv:astro-ph.GA/2212.06469]}}.
\newblock
  doi:{\changeurlcolor{black}\href{https://doi.org/10.48550/arXiv.2212.06469}{\detokenize{10.48550/arXiv.2212.06469}}}.

\bibitem[{Tumlinson} \em{et~al.}(2017){Tumlinson}, {Peeples}, and
  {Werk}]{Tumlinson2017ARAA}
{Tumlinson}, J.; {Peeples}, M.S.; {Werk}, J.K.
\newblock {The Circumgalactic Medium}.
\newblock {\em \araa} {\bf 2017}, {\em 55},~389--432,
  \href{http://xxx.lanl.gov/abs/1709.09180}{{\normalfont
  [arXiv:astro-ph.GA/1709.09180]}}.
\newblock
  doi:{\changeurlcolor{black}\href{https://doi.org/10.1146/annurev-astro-091916-055240}{\detokenize{10.1146/annurev-astro-091916-055240}}}.

\bibitem[{Putman} \em{et~al.}(2012){Putman}, {Peek}, and
  {Joung}]{Putman2012ARA&A}
{Putman}, M.E.; {Peek}, J.E.G.; {Joung}, M.R.
\newblock {Gaseous Galaxy Halos}.
\newblock {\em \araa} {\bf 2012}, {\em 50},~491--529,
  \href{http://xxx.lanl.gov/abs/1207.4837}{{\normalfont
  [arXiv:astro-ph.GA/1207.4837]}}.
\newblock
  doi:{\changeurlcolor{black}\href{https://doi.org/10.1146/annurev-astro-081811-125612}{\detokenize{10.1146/annurev-astro-081811-125612}}}.

\bibitem[{Faucher-Giguere} and {Oh}(2023)]{FaucherGiguere2023arXiv}
{Faucher-Giguere}, C.A.; {Oh}, S.P.
\newblock {Key Physical Processes in the Circumgalactic Medium}.
\newblock {\em arXiv e-prints} {\bf 2023}, p. arXiv:2301.10253,
  \href{http://xxx.lanl.gov/abs/2301.10253}{{\normalfont
  [arXiv:astro-ph.GA/2301.10253]}}.
\newblock
  doi:{\changeurlcolor{black}\href{https://doi.org/10.48550/arXiv.2301.10253}{\detokenize{10.48550/arXiv.2301.10253}}}.

\bibitem[{Girichidis} \em{et~al.}(2018){Girichidis}, {Naab}, {Hanasz}, and
  {Walch}]{Girichidis2018MNRAS}
{Girichidis}, P.; {Naab}, T.; {Hanasz}, M.; {Walch}, S.
\newblock {Cooler and smoother - the impact of cosmic rays on the phase
  structure of galactic outflows}.
\newblock {\em \mnras} {\bf 2018}, {\em 479},~3042--3067,
  \href{http://xxx.lanl.gov/abs/1805.09333}{{\normalfont
  [arXiv:astro-ph.GA/1805.09333]}}.
\newblock
  doi:{\changeurlcolor{black}\href{https://doi.org/10.1093/mnras/sty1653}{\detokenize{10.1093/mnras/sty1653}}}.

\bibitem[{Werk} \em{et~al.}(2014){Werk}, {Prochaska}, {Tumlinson}, {Peeples},
  {Tripp}, {Fox}, {Lehner}, {Thom}, {O'Meara}, {Ford}, {Bordoloi}, {Katz},
  {Tejos}, {Oppenheimer}, {Dav{\'e}}, and {Weinberg}]{Werk2014ApJ}
{Werk}, J.K.; {Prochaska}, J.X.; {Tumlinson}, J.; {Peeples}, M.S.; {Tripp},
  T.M.; {Fox}, A.J.; {Lehner}, N.; {Thom}, C.; {O'Meara}, J.M.; {Ford}, A.B.;
  et~al.
\newblock {The COS-Halos Survey: Physical Conditions and Baryonic Mass in the
  Low-redshift Circumgalactic Medium}.
\newblock {\em \apj} {\bf 2014}, {\em 792},~8,
  \href{http://xxx.lanl.gov/abs/1403.0947}{{\normalfont
  [arXiv:astro-ph.CO/1403.0947]}}.
\newblock
  doi:{\changeurlcolor{black}\href{https://doi.org/10.1088/0004-637X/792/1/8}{\detokenize{10.1088/0004-637X/792/1/8}}}.

\bibitem[{Chen} \em{et~al.}(2010){Chen}, {Helsby}, {Gauthier}, {Shectman},
  {Thompson}, and {Tinker}]{Chen2010ApJ}
{Chen}, H.W.; {Helsby}, J.E.; {Gauthier}, J.R.; {Shectman}, S.A.; {Thompson},
  I.B.; {Tinker}, J.L.
\newblock {An Empirical Characterization of Extended Cool Gas Around Galaxies
  Using Mg II Absorption Features}.
\newblock {\em \apj} {\bf 2010}, {\em 714},~1521--1541,
  \href{http://xxx.lanl.gov/abs/1004.0705}{{\normalfont
  [arXiv:astro-ph.CO/1004.0705]}}.
\newblock
  doi:{\changeurlcolor{black}\href{https://doi.org/10.1088/0004-637X/714/2/1521}{\detokenize{10.1088/0004-637X/714/2/1521}}}.

\bibitem[{Prochaska} \em{et~al.}(2011){Prochaska}, {Weiner}, {Chen},
  {Mulchaey}, and {Cooksey}]{Prochaska2011ApJ}
{Prochaska}, J.X.; {Weiner}, B.; {Chen}, H.W.; {Mulchaey}, J.; {Cooksey}, K.
\newblock {Probing the Intergalactic Medium/Galaxy Connection. V. On the Origin
  of Ly{\ensuremath{\alpha}} and O VI Absorption at z < 0.2}.
\newblock {\em \apj} {\bf 2011}, {\em 740},~91,
  \href{http://xxx.lanl.gov/abs/1103.1891}{{\normalfont
  [arXiv:astro-ph.CO/1103.1891]}}.
\newblock
  doi:{\changeurlcolor{black}\href{https://doi.org/10.1088/0004-637X/740/2/91}{\detokenize{10.1088/0004-637X/740/2/91}}}.

\bibitem[{Tumlinson} \em{et~al.}(2013){Tumlinson}, {Thom}, {Werk}, {Prochaska},
  {Tripp}, {Katz}, {Dav{\'e}}, {Oppenheimer}, {Meiring}, {Ford}, {O'Meara},
  {Peeples}, {Sembach}, and {Weinberg}]{Tumlinson2013ApJ}
{Tumlinson}, J.; {Thom}, C.; {Werk}, J.K.; {Prochaska}, J.X.; {Tripp}, T.M.;
  {Katz}, N.; {Dav{\'e}}, R.; {Oppenheimer}, B.D.; {Meiring}, J.D.; {Ford},
  A.B.;  et~al.
\newblock {The COS-Halos Survey: Rationale, Design, and a Census of
  Circumgalactic Neutral Hydrogen}.
\newblock {\em \apj} {\bf 2013}, {\em 777},~59,
  \href{http://xxx.lanl.gov/abs/1309.6317}{{\normalfont
  [arXiv:astro-ph.CO/1309.6317]}}.
\newblock
  doi:{\changeurlcolor{black}\href{https://doi.org/10.1088/0004-637X/777/1/59}{\detokenize{10.1088/0004-637X/777/1/59}}}.

\bibitem[{Keeney} \em{et~al.}(2018){Keeney}, {Stocke}, {Pratt}, {Davis},
  {Syphers}, {Danforth}, {Shull}, {Froning}, {Green}, {Penton}, and
  {Savage}]{Keeney2018ApJS}
{Keeney}, B.A.; {Stocke}, J.T.; {Pratt}, C.T.; {Davis}, J.D.; {Syphers}, D.;
  {Danforth}, C.W.; {Shull}, J.M.; {Froning}, C.S.; {Green}, J.C.; {Penton},
  S.V.;  et~al.
\newblock {A Galaxy Redshift Survey Near HST/COS AGN Sight Lines}.
\newblock {\em \apjs} {\bf 2018}, {\em 237},~11,
  \href{http://xxx.lanl.gov/abs/1805.08767}{{\normalfont
  [arXiv:astro-ph.GA/1805.08767]}}.
\newblock
  doi:{\changeurlcolor{black}\href{https://doi.org/10.3847/1538-4365/aac727}{\detokenize{10.3847/1538-4365/aac727}}}.

\bibitem[Fox(2017)]{Fox2017ASSL}
{\em {Gas Accretion onto Galaxies}}; Vol. 430, {\em Astrophysics and Space
  Science Library},  2017.
\newblock
  doi:{\changeurlcolor{black}\href{https://doi.org/10.1007/978-3-319-52512-9}{\detokenize{10.1007/978-3-319-52512-9}}}.

\bibitem[{Thom} \em{et~al.}(2012){Thom}, {Tumlinson}, {Werk}, {Prochaska},
  {Oppenheimer}, {Peeples}, {Tripp}, {Katz}, {O'Meara}, {Ford}, {Dav{\'e}},
  {Sembach}, and {Weinberg}]{Thom2012ApJ}
{Thom}, C.; {Tumlinson}, J.; {Werk}, J.K.; {Prochaska}, J.X.; {Oppenheimer},
  B.D.; {Peeples}, M.S.; {Tripp}, T.M.; {Katz}, N.S.; {O'Meara}, J.M.; {Ford},
  A.B.;  et~al.
\newblock {Not Dead Yet: Cool Circumgalactic Gas in the Halos of Early-type
  Galaxies}.
\newblock {\em \apjl} {\bf 2012}, {\em 758},~L41,
  \href{http://xxx.lanl.gov/abs/1209.5442}{{\normalfont
  [arXiv:astro-ph.CO/1209.5442]}}.
\newblock
  doi:{\changeurlcolor{black}\href{https://doi.org/10.1088/2041-8205/758/2/L41}{\detokenize{10.1088/2041-8205/758/2/L41}}}.

\bibitem[{Berg} \em{et~al.}(2019){Berg}, {Howk}, {Lehner}, {Wotta}, {O'Meara},
  {Bowen}, {Burchett}, {Peeples}, and {Tejos}]{Berg2019ApJ}
{Berg}, M.A.; {Howk}, J.C.; {Lehner}, N.; {Wotta}, C.B.; {O'Meara}, J.M.;
  {Bowen}, D.V.; {Burchett}, J.N.; {Peeples}, M.S.; {Tejos}, N.
\newblock {The Red Dead Redemption Survey of Circumgalactic Gas about Massive
  Galaxies. I. Mass and Metallicity of the Cool Phase}.
\newblock {\em \apj} {\bf 2019}, {\em 883},~5,
  \href{http://xxx.lanl.gov/abs/1811.10717}{{\normalfont
  [arXiv:astro-ph.GA/1811.10717]}}.
\newblock
  doi:{\changeurlcolor{black}\href{https://doi.org/10.3847/1538-4357/ab378e}{\detokenize{10.3847/1538-4357/ab378e}}}.

\bibitem[{Butsky} \em{et~al.}(2020){Butsky}, {Fielding}, {Hayward}, {Hummels},
  {Quinn}, and {Werk}]{Butsky2020ApJ}
{Butsky}, I.S.; {Fielding}, D.B.; {Hayward}, C.C.; {Hummels}, C.B.; {Quinn},
  T.R.; {Werk}, J.K.
\newblock {The Impact of Cosmic Rays on Thermal Instability in the
  Circumgalactic Medium}.
\newblock {\em \apj} {\bf 2020}, {\em 903},~77,
  \href{http://xxx.lanl.gov/abs/2008.04915}{{\normalfont
  [arXiv:astro-ph.GA/2008.04915]}}.
\newblock
  doi:{\changeurlcolor{black}\href{https://doi.org/10.3847/1538-4357/abbad2}{\detokenize{10.3847/1538-4357/abbad2}}}.

\bibitem[{Butsky} \em{et~al.}(2023){Butsky}, {Nakum}, {Ponnada}, {Hummels},
  {Ji}, and {Hopkins}]{Butsky2023MNRAS}
{Butsky}, I.S.; {Nakum}, S.; {Ponnada}, S.B.; {Hummels}, C.B.; {Ji}, S.;
  {Hopkins}, P.F.
\newblock {Constraining cosmic ray transport with observations of the
  circumgalactic medium}.
\newblock {\em \mnras} {\bf 2023}, {\em 521},~2477--2483,
  \href{http://xxx.lanl.gov/abs/2210.14232}{{\normalfont
  [arXiv:astro-ph.GA/2210.14232]}}.
\newblock
  doi:{\changeurlcolor{black}\href{https://doi.org/10.1093/mnras/stad671}{\detokenize{10.1093/mnras/stad671}}}.

\bibitem[{Field}(1965)]{Field1965ApJ}
{Field}, G.B.
\newblock {Thermal Instability.}
\newblock {\em \apj} {\bf 1965}, {\em 142},~531.
\newblock
  doi:{\changeurlcolor{black}\href{https://doi.org/10.1086/148317}{\detokenize{10.1086/148317}}}.

\bibitem[{Putman} \em{et~al.}(2003){Putman}, {Staveley-Smith}, {Freeman},
  {Gibson}, and {Barnes}]{Putman2003ApJ}
{Putman}, M.E.; {Staveley-Smith}, L.; {Freeman}, K.C.; {Gibson}, B.K.;
  {Barnes}, D.G.
\newblock {The Magellanic Stream, High-Velocity Clouds, and the Sculptor
  Group}.
\newblock {\em \apj} {\bf 2003}, {\em 586},~170--194,
  \href{http://xxx.lanl.gov/abs/astro-ph/0209127}{{\normalfont
  [arXiv:astro-ph/astro-ph/0209127]}}.
\newblock
  doi:{\changeurlcolor{black}\href{https://doi.org/10.1086/344477}{\detokenize{10.1086/344477}}}.

\bibitem[{McCourt} \em{et~al.}(2012){McCourt}, {Sharma}, {Quataert}, and
  {Parrish}]{McCourt2012MNRAS}
{McCourt}, M.; {Sharma}, P.; {Quataert}, E.; {Parrish}, I.J.
\newblock {Thermal instability in gravitationally stratified plasmas:
  implications for multiphase structure in clusters and galaxy haloes}.
\newblock {\em \mnras} {\bf 2012}, {\em 419},~3319--3337,
  \href{http://xxx.lanl.gov/abs/1105.2563}{{\normalfont
  [arXiv:astro-ph.CO/1105.2563]}}.
\newblock
  doi:{\changeurlcolor{black}\href{https://doi.org/10.1111/j.1365-2966.2011.19972.x}{\detokenize{10.1111/j.1365-2966.2011.19972.x}}}.

\bibitem[{Sharma} \em{et~al.}(2012){Sharma}, {McCourt}, {Quataert}, and
  {Parrish}]{Sharma2012MNRAS}
{Sharma}, P.; {McCourt}, M.; {Quataert}, E.; {Parrish}, I.J.
\newblock {Thermal instability and the feedback regulation of hot haloes in
  clusters, groups and galaxies}.
\newblock {\em \mnras} {\bf 2012}, {\em 420},~3174--3194,
  \href{http://xxx.lanl.gov/abs/1106.4816}{{\normalfont
  [arXiv:astro-ph.CO/1106.4816]}}.
\newblock
  doi:{\changeurlcolor{black}\href{https://doi.org/10.1111/j.1365-2966.2011.20246.x}{\detokenize{10.1111/j.1365-2966.2011.20246.x}}}.

\bibitem[{Voit} \em{et~al.}(2015){Voit}, {Donahue}, {Bryan}, and
  {McDonald}]{Voit2015Natur}
{Voit}, G.M.; {Donahue}, M.; {Bryan}, G.L.; {McDonald}, M.
\newblock {Regulation of star formation in giant galaxies by precipitation,
  feedback and conduction}.
\newblock {\em \nat} {\bf 2015}, {\em 519},~203--206,
  \href{http://xxx.lanl.gov/abs/1409.1598}{{\normalfont
  [arXiv:astro-ph.GA/1409.1598]}}.
\newblock
  doi:{\changeurlcolor{black}\href{https://doi.org/10.1038/nature14167}{\detokenize{10.1038/nature14167}}}.

\bibitem[{Salem} \em{et~al.}(2016){Salem}, {Bryan}, and
  {Corlies}]{Salem2016MNRAS}
{Salem}, M.; {Bryan}, G.L.; {Corlies}, L.
\newblock {Role of cosmic rays in the circumgalactic medium}.
\newblock {\em \mnras} {\bf 2016}, {\em 456},~582--601,
  \href{http://xxx.lanl.gov/abs/1511.05144}{{\normalfont
  [arXiv:astro-ph.GA/1511.05144]}}.
\newblock
  doi:{\changeurlcolor{black}\href{https://doi.org/10.1093/mnras/stv2641}{\detokenize{10.1093/mnras/stv2641}}}.

\bibitem[{Sharma} \em{et~al.}(2010){Sharma}, {Parrish}, and
  {Quataert}]{Sharma2010ApJ}
{Sharma}, P.; {Parrish}, I.J.; {Quataert}, E.
\newblock {Thermal Instability with Anisotropic Thermal Conduction and
  Adiabatic Cosmic Rays: Implications for Cold Filaments in Galaxy Clusters}.
\newblock {\em \apj} {\bf 2010}, {\em 720},~652--665,
  \href{http://xxx.lanl.gov/abs/1003.5546}{{\normalfont
  [arXiv:astro-ph.GA/1003.5546]}}.
\newblock
  doi:{\changeurlcolor{black}\href{https://doi.org/10.1088/0004-637X/720/1/652}{\detokenize{10.1088/0004-637X/720/1/652}}}.

\bibitem[{Kempski} and {Quataert}(2020)]{Kempski2020MNRAS}
{Kempski}, P.; {Quataert}, E.
\newblock {Thermal instability of halo gas heated by streaming cosmic rays}.
\newblock {\em \mnras} {\bf 2020}, {\em 493},~1801--1817,
  \href{http://xxx.lanl.gov/abs/1908.10367}{{\normalfont
  [arXiv:astro-ph.GA/1908.10367]}}.
\newblock
  doi:{\changeurlcolor{black}\href{https://doi.org/10.1093/mnras/staa385}{\detokenize{10.1093/mnras/staa385}}}.

\bibitem[{Hopkins} \em{et~al.}(2020){Hopkins}, {Chan}, {Garrison-Kimmel}, {Ji},
  {Su}, {Hummels}, {Kere{\v{s}}}, {Quataert}, and
  {Faucher-Gigu{\`e}re}]{Hopkins2020MNRASa}
{Hopkins}, P.F.; {Chan}, T.K.; {Garrison-Kimmel}, S.; {Ji}, S.; {Su}, K.Y.;
  {Hummels}, C.B.; {Kere{\v{s}}}, D.; {Quataert}, E.; {Faucher-Gigu{\`e}re},
  C.A.
\newblock {But what about...: cosmic rays, magnetic fields, conduction, and
  viscosity in galaxy formation}.
\newblock {\em \mnras} {\bf 2020}, {\em 492},~3465--3498,
  \href{http://xxx.lanl.gov/abs/1905.04321}{{\normalfont
  [arXiv:astro-ph.GA/1905.04321]}}.
\newblock
  doi:{\changeurlcolor{black}\href{https://doi.org/10.1093/mnras/stz3321}{\detokenize{10.1093/mnras/stz3321}}}.

\bibitem[{Buck} \em{et~al.}(2020){Buck}, {Pfrommer}, {Pakmor}, {Grand}, and
  {Springel}]{Buck2020MNRAS}
{Buck}, T.; {Pfrommer}, C.; {Pakmor}, R.; {Grand}, R.J.J.; {Springel}, V.
\newblock {The effects of cosmic rays on the formation of Milky Way-mass
  galaxies in a cosmological context}.
\newblock {\em \mnras} {\bf 2020}, {\em 497},~1712--1737,
  \href{http://xxx.lanl.gov/abs/1911.00019}{{\normalfont
  [arXiv:astro-ph.GA/1911.00019]}}.
\newblock
  doi:{\changeurlcolor{black}\href{https://doi.org/10.1093/mnras/staa1960}{\detokenize{10.1093/mnras/staa1960}}}.

\bibitem[{Dekel} and {Birnboim}(2006)]{Dekel2006MNRAS}
{Dekel}, A.; {Birnboim}, Y.
\newblock {Galaxy bimodality due to cold flows and shock heating}.
\newblock {\em \mnras} {\bf 2006}, {\em 368},~2--20,
  \href{http://xxx.lanl.gov/abs/astro-ph/0412300}{{\normalfont
  [arXiv:astro-ph/astro-ph/0412300]}}.
\newblock
  doi:{\changeurlcolor{black}\href{https://doi.org/10.1111/j.1365-2966.2006.10145.x}{\detokenize{10.1111/j.1365-2966.2006.10145.x}}}.

\bibitem[{Dekel} \em{et~al.}(2009){Dekel}, {Birnboim}, {Engel}, {Freundlich},
  {Goerdt}, {Mumcuoglu}, {Neistein}, {Pichon}, {Teyssier}, and
  {Zinger}]{Dekel2009Natur}
{Dekel}, A.; {Birnboim}, Y.; {Engel}, G.; {Freundlich}, J.; {Goerdt}, T.;
  {Mumcuoglu}, M.; {Neistein}, E.; {Pichon}, C.; {Teyssier}, R.; {Zinger}, E.
\newblock {Cold streams in early massive hot haloes as the main mode of galaxy
  formation}.
\newblock {\em \nat} {\bf 2009}, {\em 457},~451--454,
  \href{http://xxx.lanl.gov/abs/0808.0553}{{\normalfont
  [arXiv:astro-ph/0808.0553]}}.
\newblock
  doi:{\changeurlcolor{black}\href{https://doi.org/10.1038/nature07648}{\detokenize{10.1038/nature07648}}}.

\bibitem[{Kere{\v{s}}} \em{et~al.}(2005){Kere{\v{s}}}, {Katz}, {Weinberg}, and
  {Dav{\'e}}]{Keres2005MNRAS}
{Kere{\v{s}}}, D.; {Katz}, N.; {Weinberg}, D.H.; {Dav{\'e}}, R.
\newblock {How do galaxies get their gas?}
\newblock {\em \mnras} {\bf 2005}, {\em 363},~2--28,
  \href{http://xxx.lanl.gov/abs/astro-ph/0407095}{{\normalfont
  [arXiv:astro-ph/astro-ph/0407095]}}.
\newblock
  doi:{\changeurlcolor{black}\href{https://doi.org/10.1111/j.1365-2966.2005.09451.x}{\detokenize{10.1111/j.1365-2966.2005.09451.x}}}.

\bibitem[{Kere{\v{s}}} \em{et~al.}(2009){Kere{\v{s}}}, {Katz}, {Fardal},
  {Dav{\'e}}, and {Weinberg}]{Keres2009MNRAS}
{Kere{\v{s}}}, D.; {Katz}, N.; {Fardal}, M.; {Dav{\'e}}, R.; {Weinberg}, D.H.
\newblock {Galaxies in a simulated {\ensuremath{\Lambda}}CDM Universe - I. Cold
  mode and hot cores}.
\newblock {\em \mnras} {\bf 2009}, {\em 395},~160--179,
  \href{http://xxx.lanl.gov/abs/0809.1430}{{\normalfont
  [arXiv:astro-ph/0809.1430]}}.
\newblock
  doi:{\changeurlcolor{black}\href{https://doi.org/10.1111/j.1365-2966.2009.14541.x}{\detokenize{10.1111/j.1365-2966.2009.14541.x}}}.

\bibitem[{Roberts-Borsani} and {Saintonge}(2019)]{RobertsBorsani2019MNRAS}
{Roberts-Borsani}, G.W.; {Saintonge}, A.
\newblock {The prevalence and properties of cold gas inflows and outflows
  around galaxies in the local Universe}.
\newblock {\em \mnras} {\bf 2019}, {\em 482},~4111--4145,
  \href{http://xxx.lanl.gov/abs/1807.07575}{{\normalfont
  [arXiv:astro-ph.GA/1807.07575]}}.
\newblock
  doi:{\changeurlcolor{black}\href{https://doi.org/10.1093/mnras/sty2824}{\detokenize{10.1093/mnras/sty2824}}}.

\bibitem[{Ceverino} \em{et~al.}(2010){Ceverino}, {Dekel}, and
  {Bournaud}]{Ceverino2010MNRAS}
{Ceverino}, D.; {Dekel}, A.; {Bournaud}, F.
\newblock {High-redshift clumpy discs and bulges in cosmological simulations}.
\newblock {\em \mnras} {\bf 2010}, {\em 404},~2151--2169,
  \href{http://xxx.lanl.gov/abs/0907.3271}{{\normalfont
  [arXiv:astro-ph.CO/0907.3271]}}.
\newblock
  doi:{\changeurlcolor{black}\href{https://doi.org/10.1111/j.1365-2966.2010.16433.x}{\detokenize{10.1111/j.1365-2966.2010.16433.x}}}.

\bibitem[{Emonts} \em{et~al.}(2023){Emonts}, {Lehnert}, {Yoon}, {Mandelker},
  {Villar-Mart{\'\i}n}, {Miley}, {De Breuck}, {P{\'e}rez-Torres}, {Hatch}, and
  {Guillard}]{Emonts2023arXiv230317484E}
{Emonts}, B.H.C.; {Lehnert}, M.D.; {Yoon}, I.; {Mandelker}, N.;
  {Villar-Mart{\'\i}n}, M.; {Miley}, G.K.; {De Breuck}, C.; {P{\'e}rez-Torres},
  M.A.; {Hatch}, N.A.; {Guillard}, P.
\newblock {A cosmic stream of atomic carbon gas connected to a massive radio
  galaxy at redshift 3.8}.
\newblock {\em Science} {\bf 2023}, {\em 379},~1323--1326,
  \href{http://xxx.lanl.gov/abs/2303.17484}{{\normalfont
  [arXiv:astro-ph.GA/2303.17484]}}.
\newblock
  doi:{\changeurlcolor{black}\href{https://doi.org/10.1126/science.abh2150}{\detokenize{10.1126/science.abh2150}}}.

\bibitem[{Mandelker} \em{et~al.}(2020){Mandelker}, {Nagai}, {Aung}, {Dekel},
  {Birnboim}, and {van den Bosch}]{Mandelker2020MNRAS}
{Mandelker}, N.; {Nagai}, D.; {Aung}, H.; {Dekel}, A.; {Birnboim}, Y.; {van den
  Bosch}, F.C.
\newblock {Instability of supersonic cold streams feeding galaxies - IV.
  Survival of radiatively cooling streams}.
\newblock {\em \mnras} {\bf 2020}, {\em 494},~2641--2663,
  \href{http://xxx.lanl.gov/abs/1910.05344}{{\normalfont
  [arXiv:astro-ph.GA/1910.05344]}}.
\newblock
  doi:{\changeurlcolor{black}\href{https://doi.org/10.1093/mnras/staa812}{\detokenize{10.1093/mnras/staa812}}}.

\bibitem[{Martin} \em{et~al.}(2015){Martin}, {Matuszewski}, {Morrissey},
  {Neill}, {Moore}, {Cantalupo}, {Prochaska}, and {Chang}]{Martin2015Natur}
{Martin}, D.C.; {Matuszewski}, M.; {Morrissey}, P.; {Neill}, J.D.; {Moore}, A.;
  {Cantalupo}, S.; {Prochaska}, J.X.; {Chang}, D.
\newblock {A giant protogalactic disk linked to the cosmic web}.
\newblock {\em \nat} {\bf 2015}, {\em 524},~192--195.
\newblock
  doi:{\changeurlcolor{black}\href{https://doi.org/10.1038/nature14616}{\detokenize{10.1038/nature14616}}}.

\bibitem[{Daddi} \em{et~al.}(2021){Daddi}, {Valentino}, {Rich}, {Neill},
  {Gronke}, {O'Sullivan}, {Elbaz}, {Bournaud}, {Finoguenov}, {Marchal},
  {Delvecchio}, {Jin}, {Liu}, {Strazzullo}, {Calabro}, {Coogan}, {D'Eugenio},
  {Gobat}, {Kalita}, {Laursen}, {Martin}, {Puglisi}, {Schinnerer}, and
  {Wang}]{Daddi2021A&A}
{Daddi}, E.; {Valentino}, F.; {Rich}, R.M.; {Neill}, J.D.; {Gronke}, M.;
  {O'Sullivan}, D.; {Elbaz}, D.; {Bournaud}, F.; {Finoguenov}, A.; {Marchal},
  A.;  et~al.
\newblock {Three Lyman-{\ensuremath{\alpha}}-emitting filaments converging to a
  massive galaxy group at z = 2.91: discussing the case for cold gas infall}.
\newblock {\em \aap} {\bf 2021}, {\em 649},~A78,
  \href{http://xxx.lanl.gov/abs/2006.11089}{{\normalfont
  [arXiv:astro-ph.GA/2006.11089]}}.
\newblock
  doi:{\changeurlcolor{black}\href{https://doi.org/10.1051/0004-6361/202038700}{\detokenize{10.1051/0004-6361/202038700}}}.

\bibitem[{Dijkstra} and {Loeb}(2009)]{Dijkstra2009MNRAS}
{Dijkstra}, M.; {Loeb}, A.
\newblock {Ly{\ensuremath{\alpha}} blobs as an observational signature of cold
  accretion streams into galaxies}.
\newblock {\em \mnras} {\bf 2009}, {\em 400},~1109--1120,
  \href{http://xxx.lanl.gov/abs/0902.2999}{{\normalfont
  [arXiv:astro-ph.CO/0902.2999]}}.
\newblock
  doi:{\changeurlcolor{black}\href{https://doi.org/10.1111/j.1365-2966.2009.15533.x}{\detokenize{10.1111/j.1365-2966.2009.15533.x}}}.

\bibitem[{Rosdahl} and {Blaizot}(2012)]{Rosdahl2012MNRAS}
{Rosdahl}, J.; {Blaizot}, J.
\newblock {Extended Ly{\ensuremath{\alpha}} emission from cold accretion
  streams}.
\newblock {\em \mnras} {\bf 2012}, {\em 423},~344--366,
  \href{http://xxx.lanl.gov/abs/1112.4408}{{\normalfont
  [arXiv:astro-ph.CO/1112.4408]}}.
\newblock
  doi:{\changeurlcolor{black}\href{https://doi.org/10.1111/j.1365-2966.2012.20883.x}{\detokenize{10.1111/j.1365-2966.2012.20883.x}}}.

\bibitem[{Pandya} \em{et~al.}(2020){Pandya}, {Somerville},
  {Angl{\'e}s-Alc{\'a}zar}, {Hayward}, {Bryan}, {Fielding}, {Forbes},
  {Burkhart}, {Genel}, {Hernquist}, {Kim}, {Tonnesen}, and
  {Starkenburg}]{Pandya2020ApJ}
{Pandya}, V.; {Somerville}, R.S.; {Angl{\'e}s-Alc{\'a}zar}, D.; {Hayward},
  C.C.; {Bryan}, G.L.; {Fielding}, D.B.; {Forbes}, J.C.; {Burkhart}, B.;
  {Genel}, S.; {Hernquist}, L.;  et~al.
\newblock {First Results from SMAUG: The Need for Preventative Stellar Feedback
  and Improved Baryon Cycling in Semianalytic Models of Galaxy Formation}.
\newblock {\em \apj} {\bf 2020}, {\em 905},~4,
  \href{http://xxx.lanl.gov/abs/2006.16317}{{\normalfont
  [arXiv:astro-ph.GA/2006.16317]}}.
\newblock
  doi:{\changeurlcolor{black}\href{https://doi.org/10.3847/1538-4357/abc3c1}{\detokenize{10.3847/1538-4357/abc3c1}}}.

\bibitem[{Nelson} \em{et~al.}(2015){Nelson}, {Genel}, {Vogelsberger},
  {Springel}, {Sijacki}, {Torrey}, and {Hernquist}]{Nelson2015MNRAS}
{Nelson}, D.; {Genel}, S.; {Vogelsberger}, M.; {Springel}, V.; {Sijacki}, D.;
  {Torrey}, P.; {Hernquist}, L.
\newblock {The impact of feedback on cosmological gas accretion}.
\newblock {\em \mnras} {\bf 2015}, {\em 448},~59--74,
  \href{http://xxx.lanl.gov/abs/1410.5425}{{\normalfont
  [arXiv:astro-ph.CO/1410.5425]}}.
\newblock
  doi:{\changeurlcolor{black}\href{https://doi.org/10.1093/mnras/stv017}{\detokenize{10.1093/mnras/stv017}}}.

\bibitem[{Hopkins} \em{et~al.}(2021){Hopkins}, {Chan}, {Ji}, {Hummels},
  {Kere{\v{s}}}, {Quataert}, and {Faucher-Gigu{\`e}re}]{Hopkins2021MNRASb}
{Hopkins}, P.F.; {Chan}, T.K.; {Ji}, S.; {Hummels}, C.B.; {Kere{\v{s}}}, D.;
  {Quataert}, E.; {Faucher-Gigu{\`e}re}, C.A.
\newblock {Cosmic ray driven outflows to Mpc scales from L$_{*}$ galaxies}.
\newblock {\em \mnras} {\bf 2021}, {\em 501},~3640--3662,
  \href{http://xxx.lanl.gov/abs/2002.02462}{{\normalfont
  [arXiv:astro-ph.GA/2002.02462]}}.
\newblock
  doi:{\changeurlcolor{black}\href{https://doi.org/10.1093/mnras/staa3690}{\detokenize{10.1093/mnras/staa3690}}}.

\bibitem[{Su} \em{et~al.}(2020){Su}, {Hopkins}, {Hayward},
  {Faucher-Gigu{\`e}re}, {Kere{\v{s}}}, {Ma}, {Orr}, {Chan}, and
  {Robles}]{Su2020MNRAS}
{Su}, K.Y.; {Hopkins}, P.F.; {Hayward}, C.C.; {Faucher-Gigu{\`e}re}, C.A.;
  {Kere{\v{s}}}, D.; {Ma}, X.; {Orr}, M.E.; {Chan}, T.K.; {Robles}, V.H.
\newblock {Cosmic rays or turbulence can suppress cooling flows (where thermal
  heating or momentum injection fail)}.
\newblock {\em \mnras} {\bf 2020}, {\em 491},~1190--1212,
  \href{http://xxx.lanl.gov/abs/1812.03997}{{\normalfont
  [arXiv:astro-ph.GA/1812.03997]}}.
\newblock
  doi:{\changeurlcolor{black}\href{https://doi.org/10.1093/mnras/stz3011}{\detokenize{10.1093/mnras/stz3011}}}.

\bibitem[{Schawinski} \em{et~al.}(2014){Schawinski}, {Urry}, {Simmons},
  {Fortson}, {Kaviraj}, {Keel}, {Lintott}, {Masters}, {Nichol}, {Sarzi},
  {Skibba}, {Treister}, {Willett}, {Wong}, and {Yi}]{Schawinski2014MNRAS}
{Schawinski}, K.; {Urry}, C.M.; {Simmons}, B.D.; {Fortson}, L.; {Kaviraj}, S.;
  {Keel}, W.C.; {Lintott}, C.J.; {Masters}, K.L.; {Nichol}, R.C.; {Sarzi}, M.;
  et~al.
\newblock {The green valley is a red herring: Galaxy Zoo reveals two
  evolutionary pathways towards quenching of star formation in early- and
  late-type galaxies}.
\newblock {\em \mnras} {\bf 2014}, {\em 440},~889--907,
  \href{http://xxx.lanl.gov/abs/1402.4814}{{\normalfont
  [arXiv:astro-ph.GA/1402.4814]}}.
\newblock
  doi:{\changeurlcolor{black}\href{https://doi.org/10.1093/mnras/stu327}{\detokenize{10.1093/mnras/stu327}}}.

\bibitem[{Laporte} \em{et~al.}(2021){Laporte}, {Meyer}, {Ellis}, {Robertson},
  {Chisholm}, and {Roberts-Borsani}]{Laporte2021MNRAS}
{Laporte}, N.; {Meyer}, R.A.; {Ellis}, R.S.; {Robertson}, B.E.; {Chisholm}, J.;
  {Roberts-Borsani}, G.W.
\newblock {Probing cosmic dawn: Ages and star formation histories of candidate
  z {\ensuremath{\geq}} 9 galaxies}.
\newblock {\em \mnras} {\bf 2021}, {\em 505},~3336--3346,
  \href{http://xxx.lanl.gov/abs/2104.08168}{{\normalfont
  [arXiv:astro-ph.GA/2104.08168]}}.
\newblock
  doi:{\changeurlcolor{black}\href{https://doi.org/10.1093/mnras/stab1239}{\detokenize{10.1093/mnras/stab1239}}}.

\bibitem[{Gronke} and {Oh}(2018)]{Gronke2018MNRAS}
{Gronke}, M.; {Oh}, S.P.
\newblock {The growth and entrainment of cold gas in a hot wind}.
\newblock {\em \mnras} {\bf 2018}, {\em 480},~L111--L115,
  \href{http://xxx.lanl.gov/abs/1806.02728}{{\normalfont
  [arXiv:astro-ph.GA/1806.02728]}}.
\newblock
  doi:{\changeurlcolor{black}\href{https://doi.org/10.1093/mnrasl/sly131}{\detokenize{10.1093/mnrasl/sly131}}}.

\bibitem[{Huang} \em{et~al.}(2022){Huang}, {Jiang}, and {Davis}]{Huang2022ApJ}
{Huang}, X.; {Jiang}, Y.f.; {Davis}, S.W.
\newblock {Cosmic-Ray-driven Multiphase Gas Formed via Thermal Instability}.
\newblock {\em \apj} {\bf 2022}, {\em 931},~140,
  \href{http://xxx.lanl.gov/abs/2204.08543}{{\normalfont
  [arXiv:astro-ph.GA/2204.08543]}}.
\newblock
  doi:{\changeurlcolor{black}\href{https://doi.org/10.3847/1538-4357/ac69dc}{\detokenize{10.3847/1538-4357/ac69dc}}}.

\bibitem[{Fujita} \em{et~al.}(2009){Fujita}, {Martin}, {Mac Low}, {New}, and
  {Weaver}]{Fujita2009ApJ}
{Fujita}, A.; {Martin}, C.L.; {Mac Low}, M.M.; {New}, K.C.B.; {Weaver}, R.
\newblock {The Origin and Kinematics of Cold Gas in Galactic Winds: Insight
  from Numerical Simulations}.
\newblock {\em \apj} {\bf 2009}, {\em 698},~693--714,
  \href{http://xxx.lanl.gov/abs/0803.2892}{{\normalfont
  [arXiv:astro-ph/0803.2892]}}.
\newblock
  doi:{\changeurlcolor{black}\href{https://doi.org/10.1088/0004-637X/698/1/693}{\detokenize{10.1088/0004-637X/698/1/693}}}.

\bibitem[{Cooper} \em{et~al.}(2007){Cooper}, {Bicknell}, {Sutherland}, and
  {Bland-Hawthorn}]{Cooper2007Ap&SS}
{Cooper}, J.L.; {Bicknell}, G.V.; {Sutherland}, R.S.; {Bland-Hawthorn}, J.
\newblock {Three-dimensional simulations of a starburst wind}.
\newblock {\em \apss} {\bf 2007}, {\em 311},~99--103,
  \href{http://xxx.lanl.gov/abs/0710.5437}{{\normalfont
  [arXiv:astro-ph/0710.5437]}}.
\newblock
  doi:{\changeurlcolor{black}\href{https://doi.org/10.1007/s10509-007-9526-4}{\detokenize{10.1007/s10509-007-9526-4}}}.

\bibitem[{Zhang}(2018)]{Zhang2018Galax}
{Zhang}, D.
\newblock {A Review of the Theory of Galactic Winds Driven by Stellar
  Feedback}.
\newblock {\em Galaxies} {\bf 2018}, {\em 6},~114,
  \href{http://xxx.lanl.gov/abs/1811.00558}{{\normalfont
  [arXiv:astro-ph.GA/1811.00558]}}.
\newblock
  doi:{\changeurlcolor{black}\href{https://doi.org/10.3390/galaxies6040114}{\detokenize{10.3390/galaxies6040114}}}.

\bibitem[{Murray} \em{et~al.}(2011){Murray}, {M{\'e}nard}, and
  {Thompson}]{Murray2011ApJ}
{Murray}, N.; {M{\'e}nard}, B.; {Thompson}, T.A.
\newblock {Radiation Pressure from Massive Star Clusters as a Launching
  Mechanism for Super-galactic Winds}.
\newblock {\em \apj} {\bf 2011}, {\em 735},~66,
  \href{http://xxx.lanl.gov/abs/1005.4419}{{\normalfont
  [arXiv:astro-ph.CO/1005.4419]}}.
\newblock
  doi:{\changeurlcolor{black}\href{https://doi.org/10.1088/0004-637X/735/1/66}{\detokenize{10.1088/0004-637X/735/1/66}}}.

\bibitem[{Gronke} \em{et~al.}(2022){Gronke}, {Oh}, {Ji}, and
  {Norman}]{Gronke2022MNRAS}
{Gronke}, M.; {Oh}, S.P.; {Ji}, S.; {Norman}, C.
\newblock {Survival and mass growth of cold gas in a turbulent, multiphase
  medium}.
\newblock {\em \mnras} {\bf 2022}, {\em 511},~859--876,
  \href{http://xxx.lanl.gov/abs/2107.13012}{{\normalfont
  [arXiv:astro-ph.GA/2107.13012]}}.
\newblock
  doi:{\changeurlcolor{black}\href{https://doi.org/10.1093/mnras/stab3351}{\detokenize{10.1093/mnras/stab3351}}}.

\bibitem[{Lyutikov}(2006)]{Lyutikov2006MNRAS}
{Lyutikov}, M.
\newblock {Magnetic draping of merging cores and radio bubbles in clusters of
  galaxies}.
\newblock {\em \mnras} {\bf 2006}, {\em 373},~73--78,
  \href{http://xxx.lanl.gov/abs/astro-ph/0604178}{{\normalfont
  [arXiv:astro-ph/astro-ph/0604178]}}.
\newblock
  doi:{\changeurlcolor{black}\href{https://doi.org/10.1111/j.1365-2966.2006.10835.x}{\detokenize{10.1111/j.1365-2966.2006.10835.x}}}.

\bibitem[{Li} \em{et~al.}(2020){Li}, {Hopkins}, {Squire}, and
  {Hummels}]{Li2020MNRAS}
{Li}, Z.; {Hopkins}, P.F.; {Squire}, J.; {Hummels}, C.
\newblock {On the survival of cool clouds in the circumgalactic medium}.
\newblock {\em \mnras} {\bf 2020}, {\em 492},~1841--1854,
  \href{http://xxx.lanl.gov/abs/1909.02632}{{\normalfont
  [arXiv:astro-ph.GA/1909.02632]}}.
\newblock
  doi:{\changeurlcolor{black}\href{https://doi.org/10.1093/mnras/stz3567}{\detokenize{10.1093/mnras/stz3567}}}.

\bibitem[{Breitschwerdt} \em{et~al.}(1991){Breitschwerdt}, {McKenzie}, and
  {Voelk}]{Breitschwerdt1991AAP}
{Breitschwerdt}, D.; {McKenzie}, J.F.; {Voelk}, H.J.
\newblock {Galactic winds. I - Cosmic ray and wave-driven winds from the
  Galaxy}.
\newblock {\em \aap} {\bf 1991}, {\em 245},~79--98.

\bibitem[{Uhlig} \em{et~al.}(2012){Uhlig}, {Pfrommer}, {Sharma}, {Nath},
  {En{\ss}lin}, and {Springel}]{Uhlig2012MNRAS}
{Uhlig}, M.; {Pfrommer}, C.; {Sharma}, M.; {Nath}, B.B.; {En{\ss}lin}, T.A.;
  {Springel}, V.
\newblock {Galactic winds driven by cosmic ray streaming}.
\newblock {\em \mnras} {\bf 2012}, {\em 423},~2374--2396,
  \href{http://xxx.lanl.gov/abs/1203.1038}{{\normalfont
  [arXiv:astro-ph.CO/1203.1038]}}.
\newblock
  doi:{\changeurlcolor{black}\href{https://doi.org/10.1111/j.1365-2966.2012.21045.x}{\detokenize{10.1111/j.1365-2966.2012.21045.x}}}.

\bibitem[{Booth} \em{et~al.}(2013){Booth}, {Agertz}, {Kravtsov}, and
  {Gnedin}]{Booth2013ApJ}
{Booth}, C.M.; {Agertz}, O.; {Kravtsov}, A.V.; {Gnedin}, N.Y.
\newblock {Simulations of Disk Galaxies with Cosmic Ray Driven Galactic Winds}.
\newblock {\em \apjl} {\bf 2013}, {\em 777},~L16,
  \href{http://xxx.lanl.gov/abs/1308.4974}{{\normalfont
  [arXiv:astro-ph.GA/1308.4974]}}.
\newblock
  doi:{\changeurlcolor{black}\href{https://doi.org/10.1088/2041-8205/777/1/L16}{\detokenize{10.1088/2041-8205/777/1/L16}}}.

\bibitem[{Salem} and {Bryan}(2014)]{Salem2014MNRAS}
{Salem}, M.; {Bryan}, G.L.
\newblock {Cosmic ray driven outflows in global galaxy disc models}.
\newblock {\em \mnras} {\bf 2014}, {\em 437},~3312--3330,
  \href{http://xxx.lanl.gov/abs/1307.6215}{{\normalfont
  [arXiv:astro-ph.CO/1307.6215]}}.
\newblock
  doi:{\changeurlcolor{black}\href{https://doi.org/10.1093/mnras/stt2121}{\detokenize{10.1093/mnras/stt2121}}}.

\bibitem[{Wiener} \em{et~al.}(2017){Wiener}, {Pfrommer}, and
  {Oh}]{Wiener2017MNRASb}
{Wiener}, J.; {Pfrommer}, C.; {Oh}, S.P.
\newblock {Cosmic ray-driven galactic winds: streaming or diffusion?}
\newblock {\em \mnras} {\bf 2017}, {\em 467},~906--921,
  \href{http://xxx.lanl.gov/abs/1608.02585}{{\normalfont
  [arXiv:astro-ph.GA/1608.02585]}}.
\newblock
  doi:{\changeurlcolor{black}\href{https://doi.org/10.1093/mnras/stx127}{\detokenize{10.1093/mnras/stx127}}}.

\bibitem[{Chan} \em{et~al.}(2019){Chan}, {Kere{\v{s}}}, {Hopkins}, {Quataert},
  {Su}, {Hayward}, and {Faucher-Gigu{\`e}re}]{Chan2019MNRASd}
{Chan}, T.K.; {Kere{\v{s}}}, D.; {Hopkins}, P.F.; {Quataert}, E.; {Su}, K.Y.;
  {Hayward}, C.C.; {Faucher-Gigu{\`e}re}, C.A.
\newblock {Cosmic ray feedback in the FIRE simulations: constraining cosmic ray
  propagation with GeV {\ensuremath{\gamma}}-ray emission}.
\newblock {\em \mnras} {\bf 2019}, {\em 488},~3716--3744,
  \href{http://xxx.lanl.gov/abs/1812.10496}{{\normalfont
  [arXiv:astro-ph.GA/1812.10496]}}.
\newblock
  doi:{\changeurlcolor{black}\href{https://doi.org/10.1093/mnras/stz1895}{\detokenize{10.1093/mnras/stz1895}}}.

\bibitem[{Samui} \em{et~al.}(2010){Samui}, {Subramanian}, and
  {Srianand}]{Samui2010MNRAS}
{Samui}, S.; {Subramanian}, K.; {Srianand}, R.
\newblock {Cosmic ray driven outflows from high-redshift galaxies}.
\newblock {\em \mnras} {\bf 2010}, {\em 402},~2778--2791,
  \href{http://xxx.lanl.gov/abs/0909.3854}{{\normalfont
  [arXiv:astro-ph.CO/0909.3854]}}.
\newblock
  doi:{\changeurlcolor{black}\href{https://doi.org/10.1111/j.1365-2966.2009.16099.x}{\detokenize{10.1111/j.1365-2966.2009.16099.x}}}.

\bibitem[{Farber} \em{et~al.}(2018){Farber}, {Ruszkowski}, {Yang}, and
  {Zweibel}]{Farber2018ApJ}
{Farber}, R.; {Ruszkowski}, M.; {Yang}, H.Y.K.; {Zweibel}, E.G.
\newblock {Impact of Cosmic-Ray Transport on Galactic Winds}.
\newblock {\em \apj} {\bf 2018}, {\em 856},~112,
  \href{http://xxx.lanl.gov/abs/1707.04579}{{\normalfont
  [arXiv:astro-ph.HE/1707.04579]}}.
\newblock
  doi:{\changeurlcolor{black}\href{https://doi.org/10.3847/1538-4357/aab26d}{\detokenize{10.3847/1538-4357/aab26d}}}.

\bibitem[{Armillotta} \em{et~al.}(2022){Armillotta}, {Ostriker}, and
  {Jiang}]{Armillotta2022ApJ}
{Armillotta}, L.; {Ostriker}, E.C.; {Jiang}, Y.F.
\newblock {Cosmic-Ray Transport in Varying Galactic Environments}.
\newblock {\em \apj} {\bf 2022}, {\em 929},~170,
  \href{http://xxx.lanl.gov/abs/2203.11949}{{\normalfont
  [arXiv:astro-ph.GA/2203.11949]}}.
\newblock
  doi:{\changeurlcolor{black}\href{https://doi.org/10.3847/1538-4357/ac5fa9}{\detokenize{10.3847/1538-4357/ac5fa9}}}.

\bibitem[{Yu} \em{et~al.}(2021){Yu}, {Owen}, {Pan}, {Wu}, and
  {Ferreras}]{Yu2021MNRAS}
{Yu}, B.P.B.; {Owen}, E.R.; {Pan}, K.C.; {Wu}, K.; {Ferreras}, I.
\newblock {Outflows from starburst galaxies with various driving mechanisms and
  their X-ray properties}.
\newblock {\em \mnras} {\bf 2021}, {\em 508},~5092--5113,
  \href{http://xxx.lanl.gov/abs/2109.09764}{{\normalfont
  [arXiv:astro-ph.GA/2109.09764]}}.
\newblock
  doi:{\changeurlcolor{black}\href{https://doi.org/10.1093/mnras/stab2738}{\detokenize{10.1093/mnras/stab2738}}}.

\bibitem[{Gronke} \em{et~al.}(2018){Gronke}, {Girichidis}, {Naab}, and
  {Walch}]{Gronke2018ApJ}
{Gronke}, M.; {Girichidis}, P.; {Naab}, T.; {Walch}, S.
\newblock {The Imprint of Cosmic Ray Driven Outflows on
  Lyman-{\ensuremath{\alpha}} Spectra}.
\newblock {\em \apjl} {\bf 2018}, {\em 862},~L7,
  \href{http://xxx.lanl.gov/abs/1805.12251}{{\normalfont
  [arXiv:astro-ph.GA/1805.12251]}}.
\newblock
  doi:{\changeurlcolor{black}\href{https://doi.org/10.3847/2041-8213/aad286}{\detokenize{10.3847/2041-8213/aad286}}}.

\bibitem[{Mao} and {Ostriker}(2018)]{Mao2018ApJ}
{Mao}, S.A.; {Ostriker}, E.C.
\newblock {Galactic Disk Winds Driven by Cosmic Ray Pressure}.
\newblock {\em \apj} {\bf 2018}, {\em 854},~89,
  \href{http://xxx.lanl.gov/abs/1801.06544}{{\normalfont
  [arXiv:astro-ph.GA/1801.06544]}}.
\newblock
  doi:{\changeurlcolor{black}\href{https://doi.org/10.3847/1538-4357/aaa88e}{\detokenize{10.3847/1538-4357/aaa88e}}}.

\bibitem[{Recchia} \em{et~al.}(2017){Recchia}, {Blasi}, and
  {Morlino}]{Recchia2017MNRAS}
{Recchia}, S.; {Blasi}, P.; {Morlino}, G.
\newblock {Cosmic ray-driven winds in the Galactic environment and the cosmic
  ray spectrum}.
\newblock {\em \mnras} {\bf 2017}, {\em 470},~865--881,
  \href{http://xxx.lanl.gov/abs/1703.04490}{{\normalfont
  [arXiv:astro-ph.HE/1703.04490]}}.
\newblock
  doi:{\changeurlcolor{black}\href{https://doi.org/10.1093/mnras/stx1214}{\detokenize{10.1093/mnras/stx1214}}}.

\bibitem[{Fujita} and {Mac Low}(2018)]{Fujita2018MNRAS}
{Fujita}, A.; {Mac Low}, M.M.
\newblock {Cosmic ray driven outflows in an ultraluminous galaxy}.
\newblock {\em \mnras} {\bf 2018}, {\em 477},~531--538,
  \href{http://xxx.lanl.gov/abs/1804.10302}{{\normalfont
  [arXiv:astro-ph.GA/1804.10302]}}.
\newblock
  doi:{\changeurlcolor{black}\href{https://doi.org/10.1093/mnras/sty715}{\detokenize{10.1093/mnras/sty715}}}.

\bibitem[{Jacob} \em{et~al.}(2018){Jacob}, {Pakmor}, {Simpson}, {Springel}, and
  {Pfrommer}]{Jacob2018MNRAS}
{Jacob}, S.; {Pakmor}, R.; {Simpson}, C.M.; {Springel}, V.; {Pfrommer}, C.
\newblock {The dependence of cosmic ray-driven galactic winds on halo mass}.
\newblock {\em \mnras} {\bf 2018}, {\em 475},~570--584,
  \href{http://xxx.lanl.gov/abs/1712.04947}{{\normalfont
  [arXiv:astro-ph.GA/1712.04947]}}.
\newblock
  doi:{\changeurlcolor{black}\href{https://doi.org/10.1093/mnras/stx3221}{\detokenize{10.1093/mnras/stx3221}}}.

\bibitem[{Peschken} \em{et~al.}(2021){Peschken}, {Hanasz}, {Naab},
  {W{\'o}lta{\'n}ski}, and {Gawryszczak}]{Peschken2021MNRAS}
{Peschken}, N.; {Hanasz}, M.; {Naab}, T.; {W{\'o}lta{\'n}ski}, D.;
  {Gawryszczak}, A.
\newblock {The angular momentum structure of CR-driven galactic outflows
  triggered by stream accretion}.
\newblock {\em \mnras} {\bf 2021}, {\em 508},~4269--4281,
  \href{http://xxx.lanl.gov/abs/2109.12360}{{\normalfont
  [arXiv:astro-ph.GA/2109.12360]}}.
\newblock
  doi:{\changeurlcolor{black}\href{https://doi.org/10.1093/mnras/stab2784}{\detokenize{10.1093/mnras/stab2784}}}.

\bibitem[{Peschken} \em{et~al.}(2023){Peschken}, {Hanasz}, {Naab},
  {W{\'o}lta{\'n}ski}, and {Gawryszczak}]{Peschken2022arXiv221017328P}
{Peschken}, N.; {Hanasz}, M.; {Naab}, T.; {W{\'o}lta{\'n}ski}, D.;
  {Gawryszczak}, A.
\newblock {The phase structure of cosmic ray driven outflows in stream fed disc
  galaxies}.
\newblock {\em \mnras} {\bf 2023}, {\em 522},~5529--5545,
  \href{http://xxx.lanl.gov/abs/2210.17328}{{\normalfont
  [arXiv:astro-ph.GA/2210.17328]}}.
\newblock
  doi:{\changeurlcolor{black}\href{https://doi.org/10.1093/mnras/stad1358}{\detokenize{10.1093/mnras/stad1358}}}.

\bibitem[{Quataert} \em{et~al.}(2022){Quataert}, {Thompson}, and
  {Jiang}]{Quataert2022MNRAS}
{Quataert}, E.; {Thompson}, T.A.; {Jiang}, Y.F.
\newblock {The physics of galactic winds driven by cosmic rays I: Diffusion}.
\newblock {\em \mnras} {\bf 2022}, {\em 510},~1184--1203,
  \href{http://xxx.lanl.gov/abs/2102.05696}{{\normalfont
  [arXiv:astro-ph.GA/2102.05696]}}.
\newblock
  doi:{\changeurlcolor{black}\href{https://doi.org/10.1093/mnras/stab3273}{\detokenize{10.1093/mnras/stab3273}}}.

\bibitem[{Hopkins} \em{et~al.}(2021){Hopkins}, {Chan}, {Squire}, {Quataert},
  {Ji}, {Kere{\v{s}}}, and {Faucher-Gigu{\`e}re}]{Hopkins2021MNRASc}
{Hopkins}, P.F.; {Chan}, T.K.; {Squire}, J.; {Quataert}, E.; {Ji}, S.;
  {Kere{\v{s}}}, D.; {Faucher-Gigu{\`e}re}, C.A.
\newblock {Effects of different cosmic ray transport models on galaxy
  formation}.
\newblock {\em \mnras} {\bf 2021}, {\em 501},~3663--3669,
  \href{http://xxx.lanl.gov/abs/2004.02897}{{\normalfont
  [arXiv:astro-ph.GA/2004.02897]}}.
\newblock
  doi:{\changeurlcolor{black}\href{https://doi.org/10.1093/mnras/staa3692}{\detokenize{10.1093/mnras/staa3692}}}.

\bibitem[{Quataert} \em{et~al.}(2022){Quataert}, {Jiang}, and
  {Thompson}]{Quataert2022MNRASb}
{Quataert}, E.; {Jiang}, Y.F.; {Thompson}, T.A.
\newblock {The physics of galactic winds driven by cosmic rays - II. Isothermal
  streaming solutions}.
\newblock {\em \mnras} {\bf 2022}, {\em 510},~920--945,
  \href{http://xxx.lanl.gov/abs/2106.08404}{{\normalfont
  [arXiv:astro-ph.GA/2106.08404]}}.
\newblock
  doi:{\changeurlcolor{black}\href{https://doi.org/10.1093/mnras/stab3274}{\detokenize{10.1093/mnras/stab3274}}}.

\bibitem[{Bai}(2022)]{Bai2022ApJ}
{Bai}, X.N.
\newblock {Toward First-principles Characterization of Cosmic-Ray Transport
  Coefficients from Multiscale Kinetic Simulations}.
\newblock {\em \apj} {\bf 2022}, {\em 928},~112,
  \href{http://xxx.lanl.gov/abs/2112.14782}{{\normalfont
  [arXiv:astro-ph.HE/2112.14782]}}.
\newblock
  doi:{\changeurlcolor{black}\href{https://doi.org/10.3847/1538-4357/ac56e1}{\detokenize{10.3847/1538-4357/ac56e1}}}.

\bibitem[{Ko} \em{et~al.}(2021){Ko}, {Ramzan}, and {Chernyshov}]{Ko2021A&A}
{Ko}, C.M.; {Ramzan}, B.; {Chernyshov}, D.O.
\newblock {Outflows in the presence of cosmic rays and waves with cooling}.
\newblock {\em \aap} {\bf 2021}, {\em 654},~A63,
  \href{http://xxx.lanl.gov/abs/2110.06170}{{\normalfont
  [arXiv:astro-ph.HE/2110.06170]}}.
\newblock
  doi:{\changeurlcolor{black}\href{https://doi.org/10.1051/0004-6361/202141047}{\detokenize{10.1051/0004-6361/202141047}}}.

\bibitem[{Modak} \em{et~al.}(2023){Modak}, {Quataert}, {Jiang}, and
  {Thompson}]{Modak2023arXiv230203701M}
{Modak}, S.; {Quataert}, E.; {Jiang}, Y.F.; {Thompson}, T.A.
\newblock {Cosmic-Ray Driven Galactic Winds from the Warm Interstellar Medium}.
\newblock {\em arXiv e-prints} {\bf 2023}, p. arXiv:2302.03701,
  \href{http://xxx.lanl.gov/abs/2302.03701}{{\normalfont
  [arXiv:astro-ph.GA/2302.03701]}}.
\newblock
  doi:{\changeurlcolor{black}\href{https://doi.org/10.48550/arXiv.2302.03701}{\detokenize{10.48550/arXiv.2302.03701}}}.

\bibitem[{Hopkins} \em{et~al.}(2022){Hopkins}, {Butsky}, {Panopoulou}, {Ji},
  {Quataert}, {Faucher-Gigu{\`e}re}, and {Kere{\v{s}}}]{Hopkins2022MNRASa}
{Hopkins}, P.F.; {Butsky}, I.S.; {Panopoulou}, G.V.; {Ji}, S.; {Quataert}, E.;
  {Faucher-Gigu{\`e}re}, C.A.; {Kere{\v{s}}}, D.
\newblock {First predicted cosmic ray spectra, primary-to-secondary ratios, and
  ionization rates from MHD galaxy formation simulations}.
\newblock {\em \mnras} {\bf 2022}, {\em 516},~3470--3514,
  \href{http://xxx.lanl.gov/abs/2109.09762}{{\normalfont
  [arXiv:astro-ph.HE/2109.09762]}}.
\newblock
  doi:{\changeurlcolor{black}\href{https://doi.org/10.1093/mnras/stac1791}{\detokenize{10.1093/mnras/stac1791}}}.

\bibitem[{Girichidis} \em{et~al.}(2022){Girichidis}, {Pfrommer}, {Pakmor}, and
  {Springel}]{Girichidis2022MNRASb}
{Girichidis}, P.; {Pfrommer}, C.; {Pakmor}, R.; {Springel}, V.
\newblock {Spectrally resolved cosmic rays - II. Momentum-dependent cosmic ray
  diffusion drives powerful galactic winds}.
\newblock {\em \mnras} {\bf 2022}, {\em 510},~3917--3938,
  \href{http://xxx.lanl.gov/abs/2109.13250}{{\normalfont
  [arXiv:astro-ph.GA/2109.13250]}}.
\newblock
  doi:{\changeurlcolor{black}\href{https://doi.org/10.1093/mnras/stab3462}{\detokenize{10.1093/mnras/stab3462}}}.

\bibitem[{Girichidis} \em{et~al.}(2023){Girichidis}, {Pfrommer}, {Pakmor}, and
  {Springel}]{Girichidis2023MNRAS_err}
{Girichidis}, P.; {Pfrommer}, C.; {Pakmor}, R.; {Springel}, V.
\newblock {Correction to: Spectrally resolved cosmic rays - II.
  Momentum-dependent cosmic ray diffusion drives powerful galactic winds}.
\newblock {\em \mnras} {\bf 2023}, {\em 521},~5410--5417.
\newblock
  doi:{\changeurlcolor{black}\href{https://doi.org/10.1093/mnras/stad810}{\detokenize{10.1093/mnras/stad810}}}.

\bibitem[{M{\"u}ller} \em{et~al.}(2020){M{\"u}ller}, {Romero}, and
  {Roth}]{Muller2020MNRAS}
{M{\"u}ller}, A.L.; {Romero}, G.E.; {Roth}, M.
\newblock {High-energy processes in starburst-driven winds}.
\newblock {\em \mnras} {\bf 2020}, {\em 496},~2474--2481,
  \href{http://xxx.lanl.gov/abs/2006.12259}{{\normalfont
  [arXiv:astro-ph.HE/2006.12259]}}.
\newblock
  doi:{\changeurlcolor{black}\href{https://doi.org/10.1093/mnras/staa1720}{\detokenize{10.1093/mnras/staa1720}}}.

\bibitem[{Peretti} \em{et~al.}(2022){Peretti}, {Morlino}, {Blasi}, and
  {Cristofari}]{Peretti2022MNRAS}
{Peretti}, E.; {Morlino}, G.; {Blasi}, P.; {Cristofari}, P.
\newblock {Particle acceleration and multimessenger emission from
  starburst-driven galactic winds}.
\newblock {\em \mnras} {\bf 2022}, {\em 511},~1336--1348,
  \href{http://xxx.lanl.gov/abs/2104.10978}{{\normalfont
  [arXiv:astro-ph.HE/2104.10978]}}.
\newblock
  doi:{\changeurlcolor{black}\href{https://doi.org/10.1093/mnras/stac084}{\detokenize{10.1093/mnras/stac084}}}.

\bibitem[{Chisholm} \em{et~al.}(2015){Chisholm}, {Tremonti}, {Leitherer},
  {Chen}, {Wofford}, and {Lundgren}]{Chisholm2015ApJ}
{Chisholm}, J.; {Tremonti}, C.A.; {Leitherer}, C.; {Chen}, Y.; {Wofford}, A.;
  {Lundgren}, B.
\newblock {Scaling Relations Between Warm Galactic Outflows and Their Host
  Galaxies}.
\newblock {\em \apj} {\bf 2015}, {\em 811},~149,
  \href{http://xxx.lanl.gov/abs/1412.2139}{{\normalfont
  [arXiv:astro-ph.GA/1412.2139]}}.
\newblock
  doi:{\changeurlcolor{black}\href{https://doi.org/10.1088/0004-637X/811/2/149}{\detokenize{10.1088/0004-637X/811/2/149}}}.

\bibitem[{Heckman} and {Borthakur}(2016)]{Heckman2016ApJ}
{Heckman}, T.M.; {Borthakur}, S.
\newblock {The Implications of Extreme Outflows from Extreme Starbursts}.
\newblock {\em \apj} {\bf 2016}, {\em 822},~9,
  \href{http://xxx.lanl.gov/abs/1603.03036}{{\normalfont
  [arXiv:astro-ph.GA/1603.03036]}}.
\newblock
  doi:{\changeurlcolor{black}\href{https://doi.org/10.3847/0004-637X/822/1/9}{\detokenize{10.3847/0004-637X/822/1/9}}}.

\bibitem[{Chisholm} \em{et~al.}(2017){Chisholm}, {Tremonti}, {Leitherer}, and
  {Chen}]{Chisholm2017MNRAS}
{Chisholm}, J.; {Tremonti}, C.A.; {Leitherer}, C.; {Chen}, Y.
\newblock {The mass and momentum outflow rates of photoionized galactic
  outflows}.
\newblock {\em \mnras} {\bf 2017}, {\em 469},~4831--4849,
  \href{http://xxx.lanl.gov/abs/1702.07351}{{\normalfont
  [arXiv:astro-ph.GA/1702.07351]}}.
\newblock
  doi:{\changeurlcolor{black}\href{https://doi.org/10.1093/mnras/stx1164}{\detokenize{10.1093/mnras/stx1164}}}.

\bibitem[{Oppenheimer} \em{et~al.}(2010){Oppenheimer}, {Dav{\'e}},
  {Kere{\v{s}}}, {Fardal}, {Katz}, {Kollmeier}, and
  {Weinberg}]{Oppenheimer2010MNRAS}
{Oppenheimer}, B.D.; {Dav{\'e}}, R.; {Kere{\v{s}}}, D.; {Fardal}, M.; {Katz},
  N.; {Kollmeier}, J.A.; {Weinberg}, D.H.
\newblock {Feedback and recycled wind accretion: assembling the z = 0 galaxy
  mass function}.
\newblock {\em \mnras} {\bf 2010}, {\em 406},~2325--2338,
  \href{http://xxx.lanl.gov/abs/0912.0519}{{\normalfont
  [arXiv:astro-ph.CO/0912.0519]}}.
\newblock
  doi:{\changeurlcolor{black}\href{https://doi.org/10.1111/j.1365-2966.2010.16872.x}{\detokenize{10.1111/j.1365-2966.2010.16872.x}}}.

\bibitem[{Bertone} \em{et~al.}(2007){Bertone}, {De Lucia}, and
  {Thomas}]{Bertone2007MNRAS}
{Bertone}, S.; {De Lucia}, G.; {Thomas}, P.A.
\newblock {The recycling of gas and metals in galaxy formation: predictions of
  a dynamical feedback model}.
\newblock {\em \mnras} {\bf 2007}, {\em 379},~1143--1154,
  \href{http://xxx.lanl.gov/abs/astro-ph/0701407}{{\normalfont
  [arXiv:astro-ph/astro-ph/0701407]}}.
\newblock
  doi:{\changeurlcolor{black}\href{https://doi.org/10.1111/j.1365-2966.2007.11997.x}{\detokenize{10.1111/j.1365-2966.2007.11997.x}}}.

\bibitem[{Marinacci} \em{et~al.}(2011){Marinacci}, {Fraternali}, {Nipoti},
  {Binney}, {Ciotti}, and {Londrillo}]{Marinacci2011MNRAS}
{Marinacci}, F.; {Fraternali}, F.; {Nipoti}, C.; {Binney}, J.; {Ciotti}, L.;
  {Londrillo}, P.
\newblock {Galactic fountains and the rotation of disc-galaxy coronae}.
\newblock {\em \mnras} {\bf 2011}, {\em 415},~1534--1542,
  \href{http://xxx.lanl.gov/abs/1103.5358}{{\normalfont
  [arXiv:astro-ph.GA/1103.5358]}}.
\newblock
  doi:{\changeurlcolor{black}\href{https://doi.org/10.1111/j.1365-2966.2011.18810.x}{\detokenize{10.1111/j.1365-2966.2011.18810.x}}}.

\bibitem[{Zhang} \em{et~al.}(2023){Zhang}, {Cai}, {Xu}, {Shimakawa}, {Arrigoni
  Battaia}, {Prochaska}, {Cen}, {Zheng}, {Wu}, {Li}, {Dou}, {Wu}, {Zabludoff},
  {Fan}, {Ai}, {Golden-Marx}, {Li}, {Lu}, {Ma}, {Wang}, {Wang}, and
  {Yuan}]{Zhang2023arXiv230502344Z}
{Zhang}, S.; {Cai}, Z.; {Xu}, D.; {Shimakawa}, R.; {Arrigoni Battaia}, F.;
  {Prochaska}, J.X.; {Cen}, R.; {Zheng}, Z.; {Wu}, Y.; {Li}, Q.;  et~al.
\newblock {Inspiraling streams of enriched gas observed around a massive galaxy
  11 billion years ago}.
\newblock {\em Science} {\bf 2023}, {\em 380},~494--498,
  \href{http://xxx.lanl.gov/abs/2305.02344}{{\normalfont
  [arXiv:astro-ph.GA/2305.02344]}}.
\newblock
  doi:{\changeurlcolor{black}\href{https://doi.org/10.1126/science.abj9192}{\detokenize{10.1126/science.abj9192}}}.

\bibitem[{Cen} and {Ostriker}(2006)]{Cen2006ApJ}
{Cen}, R.; {Ostriker}, J.P.
\newblock {Where Are the Baryons? II. Feedback Effects}.
\newblock {\em \apj} {\bf 2006}, {\em 650},~560--572,
  \href{http://xxx.lanl.gov/abs/astro-ph/0601008}{{\normalfont
  [arXiv:astro-ph/astro-ph/0601008]}}.
\newblock
  doi:{\changeurlcolor{black}\href{https://doi.org/10.1086/506505}{\detokenize{10.1086/506505}}}.

\bibitem[{Nelson} \em{et~al.}(2018){Nelson}, {Kauffmann}, {Pillepich}, {Genel},
  {Springel}, {Pakmor}, {Hernquist}, {Weinberger}, {Torrey}, {Vogelsberger},
  and {Marinacci}]{Nelson2018MNRAS}
{Nelson}, D.; {Kauffmann}, G.; {Pillepich}, A.; {Genel}, S.; {Springel}, V.;
  {Pakmor}, R.; {Hernquist}, L.; {Weinberger}, R.; {Torrey}, P.;
  {Vogelsberger}, M.;  et~al.
\newblock {The abundance, distribution, and physical nature of highly ionized
  oxygen O VI, O VII, and O VIII in IllustrisTNG}.
\newblock {\em \mnras} {\bf 2018}, {\em 477},~450--479,
  \href{http://xxx.lanl.gov/abs/1712.00016}{{\normalfont
  [arXiv:astro-ph.GA/1712.00016]}}.
\newblock
  doi:{\changeurlcolor{black}\href{https://doi.org/10.1093/mnras/sty656}{\detokenize{10.1093/mnras/sty656}}}.

\bibitem[{Bertone} \em{et~al.}(2006){Bertone}, {Vogt}, and
  {En{\ss}lin}]{Bertone2006MNRAS}
{Bertone}, S.; {Vogt}, C.; {En{\ss}lin}, T.
\newblock {Magnetic field seeding by galactic winds}.
\newblock {\em \mnras} {\bf 2006}, {\em 370},~319--330,
  \href{http://xxx.lanl.gov/abs/astro-ph/0604462}{{\normalfont
  [arXiv:astro-ph/astro-ph/0604462]}}.
\newblock
  doi:{\changeurlcolor{black}\href{https://doi.org/10.1111/j.1365-2966.2006.10474.x}{\detokenize{10.1111/j.1365-2966.2006.10474.x}}}.

\bibitem[{Donnert} \em{et~al.}(2009){Donnert}, {Dolag}, {Lesch}, and
  {M{\"u}ller}]{Donnert2009MNRAS}
{Donnert}, J.; {Dolag}, K.; {Lesch}, H.; {M{\"u}ller}, E.
\newblock {Cluster magnetic fields from galactic outflows}.
\newblock {\em \mnras} {\bf 2009}, {\em 392},~1008--1021,
  \href{http://xxx.lanl.gov/abs/0808.0919}{{\normalfont
  [arXiv:astro-ph/0808.0919]}}.
\newblock
  doi:{\changeurlcolor{black}\href{https://doi.org/10.1111/j.1365-2966.2008.14132.x}{\detokenize{10.1111/j.1365-2966.2008.14132.x}}}.

\bibitem[{Ar{\'a}mburo-Garc{\'\i}a}
  \em{et~al.}(2021){Ar{\'a}mburo-Garc{\'\i}a}, {Bondarenko}, {Boyarsky},
  {Nelson}, {Pillepich}, and {Sokolenko}]{AramburoGarcia2021MNRAS}
{Ar{\'a}mburo-Garc{\'\i}a}, A.; {Bondarenko}, K.; {Boyarsky}, A.; {Nelson}, D.;
  {Pillepich}, A.; {Sokolenko}, A.
\newblock {Magnetization of the intergalactic medium in the IllustrisTNG
  simulations: the importance of extended, outflow-driven bubbles}.
\newblock {\em \mnras} {\bf 2021}, {\em 505},~5038--5057,
  \href{http://xxx.lanl.gov/abs/2011.11581}{{\normalfont
  [arXiv:astro-ph.CO/2011.11581]}}.
\newblock
  doi:{\changeurlcolor{black}\href{https://doi.org/10.1093/mnras/stab1632}{\detokenize{10.1093/mnras/stab1632}}}.

\bibitem[{Oppenheimer} and {Dav{\'e}}(2008)]{Oppenheimer2008MNRAS}
{Oppenheimer}, B.D.; {Dav{\'e}}, R.
\newblock {Mass, metal, and energy feedback in cosmological simulations}.
\newblock {\em \mnras} {\bf 2008}, {\em 387},~577--600,
  \href{http://xxx.lanl.gov/abs/0712.1827}{{\normalfont
  [arXiv:astro-ph/0712.1827]}}.
\newblock
  doi:{\changeurlcolor{black}\href{https://doi.org/10.1111/j.1365-2966.2008.13280.x}{\detokenize{10.1111/j.1365-2966.2008.13280.x}}}.

\bibitem[{Angl{\'e}s-Alc{\'a}zar} \em{et~al.}(2017){Angl{\'e}s-Alc{\'a}zar},
  {Faucher-Gigu{\`e}re}, {Kere{\v{s}}}, {Hopkins}, {Quataert}, and
  {Murray}]{Angles2017MNRAS}
{Angl{\'e}s-Alc{\'a}zar}, D.; {Faucher-Gigu{\`e}re}, C.A.; {Kere{\v{s}}}, D.;
  {Hopkins}, P.F.; {Quataert}, E.; {Murray}, N.
\newblock {The cosmic baryon cycle and galaxy mass assembly in the FIRE
  simulations}.
\newblock {\em \mnras} {\bf 2017}, {\em 470},~4698--4719,
  \href{http://xxx.lanl.gov/abs/1610.08523}{{\normalfont
  [arXiv:astro-ph.GA/1610.08523]}}.
\newblock
  doi:{\changeurlcolor{black}\href{https://doi.org/10.1093/mnras/stx1517}{\detokenize{10.1093/mnras/stx1517}}}.

\bibitem[{Christensen} \em{et~al.}(2016){Christensen}, {Dav{\'e}}, {Governato},
  {Pontzen}, {Brooks}, {Munshi}, {Quinn}, and {Wadsley}]{Christensen2016ApJ}
{Christensen}, C.R.; {Dav{\'e}}, R.; {Governato}, F.; {Pontzen}, A.; {Brooks},
  A.; {Munshi}, F.; {Quinn}, T.; {Wadsley}, J.
\newblock {In-N-Out: The Gas Cycle from Dwarfs to Spiral Galaxies}.
\newblock {\em \apj} {\bf 2016}, {\em 824},~57,
  \href{http://xxx.lanl.gov/abs/1508.00007}{{\normalfont
  [arXiv:astro-ph.GA/1508.00007]}}.
\newblock
  doi:{\changeurlcolor{black}\href{https://doi.org/10.3847/0004-637X/824/1/57}{\detokenize{10.3847/0004-637X/824/1/57}}}.

\bibitem[{Girichidis} \em{et~al.}(2016){Girichidis}, {Naab}, {Walch}, {Hanasz},
  {Mac Low}, {Ostriker}, {Gatto}, {Peters}, {W{\"u}nsch}, {Glover}, {Klessen},
  {Clark}, and {Baczynski}]{Girichidis2016ApJ}
{Girichidis}, P.; {Naab}, T.; {Walch}, S.; {Hanasz}, M.; {Mac Low}, M.M.;
  {Ostriker}, J.P.; {Gatto}, A.; {Peters}, T.; {W{\"u}nsch}, R.; {Glover},
  S.C.O.;  et~al.
\newblock {Launching Cosmic-Ray-driven Outflows from the Magnetized
  Interstellar Medium}.
\newblock {\em \apjl} {\bf 2016}, {\em 816},~L19,
  \href{http://xxx.lanl.gov/abs/1509.07247}{{\normalfont
  [arXiv:astro-ph.GA/1509.07247]}}.
\newblock
  doi:{\changeurlcolor{black}\href{https://doi.org/10.3847/2041-8205/816/2/L19}{\detokenize{10.3847/2041-8205/816/2/L19}}}.

\bibitem[{Jana} \em{et~al.}(2020){Jana}, {Gupta}, and {Nath}]{Jana2020MNRAS}
{Jana}, R.; {Gupta}, S.; {Nath}, B.B.
\newblock {Role of cosmic rays in the early stages of galactic outflows}.
\newblock {\em \mnras} {\bf 2020}, {\em 497},~2623--2640,
  \href{http://xxx.lanl.gov/abs/2007.00696}{{\normalfont
  [arXiv:astro-ph.GA/2007.00696]}}.
\newblock
  doi:{\changeurlcolor{black}\href{https://doi.org/10.1093/mnras/staa2025}{\detokenize{10.1093/mnras/staa2025}}}.

\bibitem[{Ji} \em{et~al.}(2021){Ji}, {Kere{\v{s}}}, {Chan}, {Stern}, {Hummels},
  {Hopkins}, {Quataert}, and {Faucher-Gigu{\`e}re}]{Ji2021MNRAS}
{Ji}, S.; {Kere{\v{s}}}, D.; {Chan}, T.K.; {Stern}, J.; {Hummels}, C.B.;
  {Hopkins}, P.F.; {Quataert}, E.; {Faucher-Gigu{\`e}re}, C.A.
\newblock {Virial shocks are suppressed in cosmic ray-dominated galaxy haloes}.
\newblock {\em \mnras} {\bf 2021}, {\em 505},~259--273,
  \href{http://xxx.lanl.gov/abs/2011.04706}{{\normalfont
  [arXiv:astro-ph.GA/2011.04706]}}.
\newblock
  doi:{\changeurlcolor{black}\href{https://doi.org/10.1093/mnras/stab1264}{\detokenize{10.1093/mnras/stab1264}}}.

\bibitem[{Butsky} \em{et~al.}(2022){Butsky}, {Werk}, {Tchernyshyov},
  {Fielding}, {Breneman}, {Piacitelli}, {Quinn}, {Sanchez}, {Cruz}, {Hummels},
  {Burchett}, and {Tremmel}]{Butsky2022ApJ}
{Butsky}, I.S.; {Werk}, J.K.; {Tchernyshyov}, K.; {Fielding}, D.B.; {Breneman},
  J.; {Piacitelli}, D.R.; {Quinn}, T.R.; {Sanchez}, N.N.; {Cruz}, A.;
  {Hummels}, C.B.;  et~al.
\newblock {The Impact of Cosmic Rays on the Kinematics of the Circumgalactic
  Medium}.
\newblock {\em \apj} {\bf 2022}, {\em 935},~69,
  \href{http://xxx.lanl.gov/abs/2106.14889}{{\normalfont
  [arXiv:astro-ph.GA/2106.14889]}}.
\newblock
  doi:{\changeurlcolor{black}\href{https://doi.org/10.3847/1538-4357/ac7ebd}{\detokenize{10.3847/1538-4357/ac7ebd}}}.

\bibitem[{Chan} \em{et~al.}(2022){Chan}, {Kere{\v{s}}}, {Gurvich}, {Hopkins},
  {Trapp}, {Ji}, and {Faucher-Gigu{\`e}re}]{Chan2022MNRAS}
{Chan}, T.K.; {Kere{\v{s}}}, D.; {Gurvich}, A.B.; {Hopkins}, P.F.; {Trapp}, C.;
  {Ji}, S.; {Faucher-Gigu{\`e}re}, C.A.
\newblock {The impact of cosmic rays on dynamical balance and disc-halo
  interaction in L{\ensuremath{\star}} disc galaxies}.
\newblock {\em \mnras} {\bf 2022}, {\em 517},~597--615,
  \href{http://xxx.lanl.gov/abs/2110.06231}{{\normalfont
  [arXiv:astro-ph.GA/2110.06231]}}.
\newblock
  doi:{\changeurlcolor{black}\href{https://doi.org/10.1093/mnras/stac2236}{\detokenize{10.1093/mnras/stac2236}}}.

\bibitem[{Ipavich}(1975)]{Ipavich1975ApJ}
{Ipavich}, F.M.
\newblock {Galactic winds driven by cosmic rays.}
\newblock {\em \apj} {\bf 1975}, {\em 196},~107--120.
\newblock
  doi:{\changeurlcolor{black}\href{https://doi.org/10.1086/153397}{\detokenize{10.1086/153397}}}.

\bibitem[{Simpson} \em{et~al.}(2016){Simpson}, {Pakmor}, {Marinacci},
  {Pfrommer}, {Springel}, {Glover}, {Clark}, and {Smith}]{Simpson2016ApJ}
{Simpson}, C.M.; {Pakmor}, R.; {Marinacci}, F.; {Pfrommer}, C.; {Springel}, V.;
  {Glover}, S.C.O.; {Clark}, P.C.; {Smith}, R.J.
\newblock {The Role of Cosmic-Ray Pressure in Accelerating Galactic Outflows}.
\newblock {\em \apjl} {\bf 2016}, {\em 827},~L29,
  \href{http://xxx.lanl.gov/abs/1606.02324}{{\normalfont
  [arXiv:astro-ph.GA/1606.02324]}}.
\newblock
  doi:{\changeurlcolor{black}\href{https://doi.org/10.3847/2041-8205/827/2/L29}{\detokenize{10.3847/2041-8205/827/2/L29}}}.

\bibitem[{Bustard} \em{et~al.}(2020){Bustard}, {Zweibel}, {D'Onghia},
  {Gallagher}, and {Farber}]{Bustard2020ApJ}
{Bustard}, C.; {Zweibel}, E.G.; {D'Onghia}, E.; {Gallagher}, J.~S., I.;
  {Farber}, R.
\newblock {Cosmic-Ray-driven Outflows from the Large Magellanic Cloud:
  Contributions to the LMC Filament}.
\newblock {\em \apj} {\bf 2020}, {\em 893},~29,
  \href{http://xxx.lanl.gov/abs/1911.02021}{{\normalfont
  [arXiv:astro-ph.GA/1911.02021]}}.
\newblock
  doi:{\changeurlcolor{black}\href{https://doi.org/10.3847/1538-4357/ab7fa3}{\detokenize{10.3847/1538-4357/ab7fa3}}}.

\bibitem[{Hopkins} \em{et~al.}(2021){Hopkins}, {Chan}, {Ji}, {Hummels},
  {Kere{\v{s}}}, {Quataert}, and {Faucher-Gigu{\`e}re}]{Hopkins2021MNRASa}
{Hopkins}, P.F.; {Chan}, T.K.; {Ji}, S.; {Hummels}, C.B.; {Kere{\v{s}}}, D.;
  {Quataert}, E.; {Faucher-Gigu{\`e}re}, C.A.
\newblock {Cosmic ray driven outflows to Mpc scales from L$_{*}$ galaxies}.
\newblock {\em \mnras} {\bf 2021}, {\em 501},~3640--3662,
  \href{http://xxx.lanl.gov/abs/2002.02462}{{\normalfont
  [arXiv:astro-ph.GA/2002.02462]}}.
\newblock
  doi:{\changeurlcolor{black}\href{https://doi.org/10.1093/mnras/staa3690}{\detokenize{10.1093/mnras/staa3690}}}.

\bibitem[{Gupta} \em{et~al.}(2021){Gupta}, {Sharma}, and
  {Mignone}]{Gupta2021MNRAS}
{Gupta}, S.; {Sharma}, P.; {Mignone}, A.
\newblock {A numerical approach to the non-uniqueness problem of cosmic ray
  two-fluid equations at shocks}.
\newblock {\em \mnras} {\bf 2021}, {\em 502},~2733--2749,
  \href{http://xxx.lanl.gov/abs/1906.07200}{{\normalfont
  [arXiv:astro-ph.HE/1906.07200]}}.
\newblock
  doi:{\changeurlcolor{black}\href{https://doi.org/10.1093/mnras/stab142}{\detokenize{10.1093/mnras/stab142}}}.

\bibitem[{Semenov} \em{et~al.}(2022){Semenov}, {Kravtsov}, and
  {Diemer}]{Semenov2022ApJS}
{Semenov}, V.A.; {Kravtsov}, A.V.; {Diemer}, B.
\newblock {Entropy-conserving Scheme for Modeling Nonthermal Energies in Fluid
  Dynamics Simulations}.
\newblock {\em \apjs} {\bf 2022}, {\em 261},~16,
  \href{http://xxx.lanl.gov/abs/2107.14240}{{\normalfont
  [arXiv:astro-ph.GA/2107.14240]}}.
\newblock
  doi:{\changeurlcolor{black}\href{https://doi.org/10.3847/1538-4365/ac69e1}{\detokenize{10.3847/1538-4365/ac69e1}}}.

\bibitem[{Hopkins} \em{et~al.}(2022){Hopkins}, {Squire}, {Butsky}, and
  {Ji}]{Hopkins2022MNRASb}
{Hopkins}, P.F.; {Squire}, J.; {Butsky}, I.S.; {Ji}, S.
\newblock {Standard self-confinement and extrinsic turbulence models for cosmic
  ray transport are fundamentally incompatible with observations}.
\newblock {\em \mnras} {\bf 2022}, {\em 517},~5413--5448,
  \href{http://xxx.lanl.gov/abs/2112.02153}{{\normalfont
  [arXiv:astro-ph.HE/2112.02153]}}.
\newblock
  doi:{\changeurlcolor{black}\href{https://doi.org/10.1093/mnras/stac2909}{\detokenize{10.1093/mnras/stac2909}}}.

\bibitem[{Girichidis} \em{et~al.}(2023){Girichidis}, {Werhahn}, {Pfrommer},
  {Pakmor}, and {Springel}]{Girichidis2023arXiv230303417G}
{Girichidis}, P.; {Werhahn}, M.; {Pfrommer}, C.; {Pakmor}, R.; {Springel}, V.
\newblock {Spectrally resolved cosmic rays -- III. Dynamical impact and
  properties of the circumgalactic medium}.
\newblock {\em arXiv e-prints} {\bf 2023}, p. arXiv:2303.03417,
  \href{http://xxx.lanl.gov/abs/2303.03417}{{\normalfont
  [arXiv:astro-ph.GA/2303.03417]}}.
\newblock
  doi:{\changeurlcolor{black}\href{https://doi.org/10.48550/arXiv.2303.03417}{\detokenize{10.48550/arXiv.2303.03417}}}.

\bibitem[{Werhahn} \em{et~al.}(2023){Werhahn}, {Girichidis}, {Pfrommer}, and
  {Whittingham}]{Werhahn2023arXiv230104163W}
{Werhahn}, M.; {Girichidis}, P.; {Pfrommer}, C.; {Whittingham}, J.
\newblock {Gamma-ray emission from spectrally resolved cosmic rays in
  galaxies}.
\newblock {\em arXiv e-prints} {\bf 2023}, p. arXiv:2301.04163,
  \href{http://xxx.lanl.gov/abs/2301.04163}{{\normalfont
  [arXiv:astro-ph.HE/2301.04163]}}.
\newblock
  doi:{\changeurlcolor{black}\href{https://doi.org/10.48550/arXiv.2301.04163}{\detokenize{10.48550/arXiv.2301.04163}}}.

\bibitem[{Lazarian} and {Xu}(2022)]{Lazarian2022FrP}
{Lazarian}, A.; {Xu}, S.
\newblock {Damping of Alfv{\'e}n Waves in MHD Turbulence and Implications for
  Cosmic Ray Streaming Instability and Galactic Winds}.
\newblock {\em Frontiers in Physics} {\bf 2022}, {\em 10},~702799,
  \href{http://xxx.lanl.gov/abs/2201.05168}{{\normalfont
  [arXiv:astro-ph.GA/2201.05168]}}.
\newblock
  doi:{\changeurlcolor{black}\href{https://doi.org/10.3389/fphy.2022.702799}{\detokenize{10.3389/fphy.2022.702799}}}.

\bibitem[{Zweibel}(2013)]{Zweibel2013PhPl}
{Zweibel}, E.G.
\newblock {The microphysics and macrophysics of cosmic rays}.
\newblock {\em Physics of Plasmas} {\bf 2013}, {\em 20},~055501.
\newblock
  doi:{\changeurlcolor{black}\href{https://doi.org/10.1063/1.4807033}{\detokenize{10.1063/1.4807033}}}.

\bibitem[{Hopkins} \em{et~al.}(2021){Hopkins}, {Squire}, {Chan}, {Quataert},
  {Ji}, {Kere{\v{s}}}, and {Faucher-Gigu{\`e}re}]{Hopkins2021MNRAS}
{Hopkins}, P.F.; {Squire}, J.; {Chan}, T.K.; {Quataert}, E.; {Ji}, S.;
  {Kere{\v{s}}}, D.; {Faucher-Gigu{\`e}re}, C.A.
\newblock {Testing physical models for cosmic ray transport coefficients on
  galactic scales: self-confinement and extrinsic turbulence at
  {\ensuremath{\sim}}GeV energies}.
\newblock {\em \mnras} {\bf 2021}, {\em 501},~4184--4213,
  \href{http://xxx.lanl.gov/abs/2002.06211}{{\normalfont
  [arXiv:astro-ph.HE/2002.06211]}}.
\newblock
  doi:{\changeurlcolor{black}\href{https://doi.org/10.1093/mnras/staa3691}{\detokenize{10.1093/mnras/staa3691}}}.

\bibitem[{Lazarian}(2016)]{Lazarian2016ApJ}
{Lazarian}, A.
\newblock {Damping of Alfv{\'e}n Waves by Turbulence and Its Consequences: From
  Cosmic-ray Streaming to Launching Winds}.
\newblock {\em \apj} {\bf 2016}, {\em 833},~131,
  \href{http://xxx.lanl.gov/abs/1607.02042}{{\normalfont
  [arXiv:astro-ph.HE/1607.02042]}}.
\newblock
  doi:{\changeurlcolor{black}\href{https://doi.org/10.3847/1538-4357/833/2/131}{\detokenize{10.3847/1538-4357/833/2/131}}}.

\bibitem[{Bai} \em{et~al.}(2019){Bai}, {Ostriker}, {Plotnikov}, and
  {Stone}]{Bai2019ApJ}
{Bai}, X.N.; {Ostriker}, E.C.; {Plotnikov}, I.; {Stone}, J.M.
\newblock {Magnetohydrodynamic Particle-in-cell Simulations of the Cosmic-Ray
  Streaming Instability: Linear Growth and Quasi-linear Evolution}.
\newblock {\em \apj} {\bf 2019}, {\em 876},~60,
  \href{http://xxx.lanl.gov/abs/1902.10219}{{\normalfont
  [arXiv:astro-ph.HE/1902.10219]}}.
\newblock
  doi:{\changeurlcolor{black}\href{https://doi.org/10.3847/1538-4357/ab1648}{\detokenize{10.3847/1538-4357/ab1648}}}.

\bibitem[{Holcomb} and {Spitkovsky}(2019)]{Holcomb2019ApJ}
{Holcomb}, C.; {Spitkovsky}, A.
\newblock {On the Growth and Saturation of the Gyroresonant Streaming
  Instabilities}.
\newblock {\em \apj} {\bf 2019}, {\em 882},~3,
  \href{http://xxx.lanl.gov/abs/1811.01951}{{\normalfont
  [arXiv:astro-ph.HE/1811.01951]}}.
\newblock
  doi:{\changeurlcolor{black}\href{https://doi.org/10.3847/1538-4357/ab328a}{\detokenize{10.3847/1538-4357/ab328a}}}.

\bibitem[{van Marle} \em{et~al.}(2019){van Marle}, {Casse}, and
  {Marcowith}]{vanMarle2019MNRAS}
{van Marle}, A.J.; {Casse}, F.; {Marcowith}, A.
\newblock {Three-dimensional simulations of non-resonant streaming instability
  and particle acceleration near non-relativistic astrophysical shocks}.
\newblock {\em \mnras} {\bf 2019}, {\em 490},~1156--1165,
  \href{http://xxx.lanl.gov/abs/1909.06931}{{\normalfont
  [arXiv:astro-ph.HE/1909.06931]}}.
\newblock
  doi:{\changeurlcolor{black}\href{https://doi.org/10.1093/mnras/stz2624}{\detokenize{10.1093/mnras/stz2624}}}.

\bibitem[{Hopkins} \em{et~al.}(2023){Hopkins}, {Butsky}, {Ji}, and
  {Kere{\v{s}}}]{Hopkins2023MNRAS}
{Hopkins}, P.F.; {Butsky}, I.S.; {Ji}, S.; {Kere{\v{s}}}, D.
\newblock {A simple sub-grid model for cosmic ray effects on galactic scales}.
\newblock {\em \mnras} {\bf 2023},
  \href{http://xxx.lanl.gov/abs/2211.05811}{{\normalfont
  [arXiv:astro-ph.GA/2211.05811]}}.
\newblock
  doi:{\changeurlcolor{black}\href{https://doi.org/10.1093/mnras/stad976}{\detokenize{10.1093/mnras/stad976}}}.

\bibitem[{Cherenkov Telescope Array Consortium} \em{et~al.}(2019){Cherenkov
  Telescope Array Consortium}, {Acharya}, {Agudo}, {Al Samarai}, {Alfaro},
  {Alfaro}, {Alispach}, {Alves Batista}, {Amans}, {Amato}, {Ambrosi},
  {Antolini}, {Antonelli}, {Aramo}, {Araya}, {Armstrong}, {Arqueros},
  {Arrabito}, {Asano}, {Ashley}, {Backes}, {Balazs}, {Balbo}, {Ballester},
  {Ballet}, {Bamba}, {Barkov}, {Barres de Almeida}, {Barrio}, {Bastieri},
  {Becherini}, {Belfiore}, {Benbow}, {Berge}, {Bernardini}, {Bernardini},
  {Bernardos}, {Bernl{\"o}hr}, {Bertucci}, {Biasuzzi}, {Bigongiari}, {Biland},
  {Bissaldi}, {Biteau}, {Blanch}, {Blazek}, {Boisson}, {Bolmont}, {Bonanno},
  {Bonardi}, {Bonavolont{\`a}}, {Bonnoli}, {Bosnjak}, {B{\"o}ttcher},
  {Braiding}, {Bregeon}, {Brill}, {Brown}, {Brun}, {Brunetti}, {Buanes},
  {Buckley}, {Bugaev}, {B{\"u}hler}, {Bulgarelli}, {Bulik}, {Burton},
  {Burtovoi}, {Busetto}, {Canestrari}, {Capalbi}, {Capitanio}, {Caproni},
  {Caraveo}, {C{\'a}rdenas}, {Carlile}, {Carosi}, {Carqu{\'\i}n}, {Carr},
  {Casanova}, {Cascone}, {Catalani}, {Catalano}, {Cauz}, {Cerruti}, {Chadwick},
  {Chaty}, {Chaves}, {Chen}, {Chen}, {Chernyakova}, {Chikawa}, {Christov},
  {Chudoba}, {Cie{\'s}lar}, {Coco}, {Colafrancesco}, {Colin}, {Conforti},
  {Connaughton}, {Conrad}, {Contreras}, {Cortina}, {Costa}, {Costantini},
  {Cotter}, {Covino}, {Crocker}, {Cuadra}, {Cuevas}, {Cumani}, {D'A{\`\i}},
  {D'Ammando}, {D'Avanzo}, {D'Urso}, {Daniel}, {Davids}, {Dawson}, {Dazzi}, {De
  Angelis}, {de C{\'a}ssia dos Anjos}, {De Cesare}, {De Franco}, {de Gouveia
  Dal Pino}, {de la Calle}, {de los Reyes Lopez}, {De Lotto}, {De Luca}, {De
  Lucia}, {de Naurois}, {de O{\~n}a Wilhelmi}, {De Palma}, {De Persio}, {de
  Souza}, {Deil}, {Del Santo}, {Delgado}, {della Volpe}, {Di Girolamo}, {Di
  Pierro}, {Di Venere}, {D{\'\i}az}, {Dib}, {Diebold}, {Djannati-Ata{\"\i}},
  {Dom{\'\i}nguez}, {Dominis Prester}, {Dorner}, {Doro}, {Drass}, {Dravins},
  {Dubus}, {Dwarkadas}, {Ebr}, {Eckner}, {Egberts}, {Einecke}, {Ekoume},
  {Els{\"a}sser}, {Ernenwein}, {Espinoza}, {Evoli}, {Fairbairn},
  {Falceta-Goncalves}, {Falcone}, {Farnier}, {Fasola}, {Fedorova}, {Fegan},
  {Fernandez-Alonso}, {Fern{\'a}ndez-Barral}, {Ferrand}, {Fesquet},
  {Filipovic}, {Fioretti}, {Fontaine}, {Fornasa}, {Fortson}, {Freixas
  Coromina}, {Fruck}, {Fujita}, {Fukazawa}, {Funk}, {F{\"u}{\ss}ling},
  {Gabici}, {Gadola}, {Gallant}, {Garcia}, {Garcia L{\'o}pez}, {Garczarczyk},
  {Gaskins}, {Gasparetto}, {Gaug}, {Gerard}, {Giavitto}, {Giglietto}, {Giommi},
  {Giordano}, {Giro}, {Giroletti}, {Giuliani}, {Glicenstein}, {Gnatyk},
  {Godinovic}, {Goldoni}, {G{\'o}mez-Vargas}, {Gonz{\'a}lez}, {Gonz{\'a}lez},
  {G{\"o}tz}, {Graham}, {Grandi}, {Granot}, {Green}, {Greenshaw}, {Griffiths},
  {Gunji}, {Hadasch}, {Hara}, {Hardcastle}, {Hassan}, {Hayashi}, {Hayashida},
  {Heller}, {Helo}, {Hermann}, {Hinton}, {Hnatyk}, {Hofmann}, {Holder},
  {Horan}, {H{\"o}randel}, {Horns}, {Horvath}, {Hovatta}, {Hrabovsky},
  {Hrupec}, {Humensky}, {H{\"u}tten}, {Iarlori}, {Inada}, {Inome}, {Inoue},
  {Inoue}, {Inoue}, {Iocco}, {Ioka}, {Iori}, {Ishio}, {Iwamura}, {Jamrozy},
  {Janecek}, {Jankowsky}, {Jean}, {Jung-Richardt}, {Jurysek}, {Kaaret},
  {Karkar}, {Katagiri}, {Katz}, {Kawanaka}, {Kazanas}, {Kh{\'e}lifi}, {Kieda},
  {Kimeswenger}, {Kimura}, {Kisaka}, {Knapp}, {Kn{\"o}dlseder}, {Koch},
  {Kohri}, {Komin}, {Kosack}, {Kraus}, {Krause}, {Krau{\ss}}, {Kubo}, {Kukec
  Mezek}, {Kuroda}, {Kushida}, {La Palombara}, {Lamanna}, {Lang}, {Lapington},
  {Le Blanc}, {Leach}, {Lees}, {Lefaucheur}, {Leigui de Oliveira}, {Lenain},
  {Lico}, {Limon}, {Lindfors}, {Lohse}, {Lombardi}, {Longo}, {L{\'o}pez},
  {L{\'o}pez-Coto}, {Lu}, {Lucarelli}, {Luque-Escamilla}, {Lyard}, {Maccarone},
  {Maier}, {Majumdar}, {Malaguti}, {Mandat}, {Maneva}, {Manganaro}, {Mangano},
  {Marcowith}, {Mar{\'\i}n}, {Markoff}, {Mart{\'\i}}, {Martin},
  {Mart{\'\i}nez}, {Mart{\'\i}nez}, {Masetti}, {Masuda}, {Maurin}, {Maxted},
  {Mazin}, {Medina}, {Melandri}, {Mereghetti}, {Meyer}, {Minaya}, {Mirabal},
  {Mirzoyan}, {Mitchell}, {Mizuno}, {Moderski}, {Mohammed}, {Mohrmann},
  {Montaruli}, {Moralejo}, {Morcuende-Parrilla}, {Mori}, {Morlino}, {Morris},
  {Morselli}, {Moulin}, {Mukherjee}, {Mundell}, {Murach}, {Muraishi}, {Murase},
  {Nagai}, {Nagataki}, {Nagayoshi}, {Naito}, {Nakamori}, {Nakamura}, {Niemiec},
  {Nieto}, {Niko{\l}ajuk}, {Nishijima}, {Noda}, {Nosek}, {Novosyadlyj},
  {Nozaki}, {O'Brien}, {Oakes}, {Ohira}, {Ohishi}, {Ohm}, {Okazaki}, {Okumura},
  {Ong}, {Orienti}, {Orito}, {Osborne}, {Ostrowski}, {Otte}, {Oya}, {Padovani},
  {Paizis}, {Palatiello}, {Palatka}, {Paoletti}, {Paredes}, {Pareschi},
  {Parsons}, {Pe'er}, {Pech}, {Pedaletti}, {Perri}, {Persic}, {Petrashyk},
  {Petrucci}, {Petruk}, {Peyaud}, {Pfeifer}, {Piano}, {Pisarski}, {Pita},
  {Pohl}, {Polo}, {Pozo}, {Prandini}, {Prast}, {Principe}, {Prokhorov},
  {Prokoph}, {Prouza}, {P{\"u}hlhofer}, {Punch}, {P{\"u}rckhauer}, {Queiroz},
  {Quirrenbach}, {Rain{\`o}}, {Razzaque}, {Reimer}, {Reimer}, {Reisenegger},
  {Renaud}, {Rezaeian}, {Rhode}, {Ribeiro}, {Rib{\'o}}, {Richtler}, {Rico},
  {Rieger}, {Riquelme}, {Rivoire}, {Rizi}, {Rodriguez}, {Rodriguez Fernandez},
  {Rodr{\'\i}guez V{\'a}zquez}, {Rojas}, {Romano}, {Romeo}, {Rosado}, {Rovero},
  {Rowell}, {Rudak}, {Rugliancich}, {Rulten}, {Sadeh}, {Safi-Harb}, {Saito},
  {Sakaki}, {Sakurai}, {Salina}, {S{\'a}nchez-Conde}, {Sandaker}, {Sandoval},
  {Sangiorgi}, {Sanguillon}, {Sano}, {Santander}, {Sarkar}, {Satalecka},
  {Saturni}, {Schioppa}, {Schlenstedt}, {Schneider}, {Schoorlemmer},
  {Schovanek}, {Schulz}, {Schussler}, {Schwanke}, {Sciacca}, {Scuderi},
  {Seitenzahl}, {Semikoz}, {Sergijenko}, {Servillat}, {Shalchi}, {Shellard},
  {Sidoli}, {Siejkowski}, {Sillanp{\"a}{\"a}}, {Sironi}, {Sitarek}, {Sliusar},
  {Slowikowska}, {Sol}, {Stamerra}, {Stani{\v{c}}}, {Starling}, {Stawarz},
  {Stefanik}, {Stephan}, {Stolarczyk}, {Stratta}, {Straumann}, {Suomijarvi},
  {Supanitsky}, {Tagliaferri}, {Tajima}, {Tavani}, {Tavecchio}, {Tavernet},
  {Tayabaly}, {Tejedor}, {Temnikov}, {Terada}, {Terrier}, {Terzic}, {Teshima},
  {Testa}, {Thoudam}, {Tian}, {Tibaldo}, {Tluczykont}, {Todero Peixoto},
  {Tokanai}, {Tomastik}, {Tonev}, {Tornikoski}, {Torres}, {Torresi}, {Tosti},
  {Tothill}, {Tovmassian}, {Travnicek}, {Trichard}, {Trifoglio}, {Troyano
  Pujadas}, {Tsujimoto}, {Umana}, {Vagelli}, {Vagnetti}, {Valentino},
  {Vallania}, {Valore}, {van Eldik}, {Vandenbroucke}, {Varner}, {Vasileiadis},
  {Vassiliev}, {V{\'a}zquez Acosta}, {Vecchi}, {Vega}, {Vercellone}, {Veres},
  {Vergani}, {Verzi}, {Vettolani}, {Viana}, {Vigorito}, {Villanueva}, {Voelk},
  {Vollhardt}, {Vorobiov}, {Vrastil}, {Vuillaume}, {Wagner}, {Wagner},
  {Walter}, {Ward}, {Warren}, {Watson}, {Werner}, {White}, {White},
  {Wierzcholska}, {Wilcox}, {Will}, {Williams}, {Wischnewski}, {Wood},
  {Yamamoto}, {Yamazaki}, {Yanagita}, {Yang}, {Yoshida}, {Yoshiike},
  {Yoshikoshi}, {Zacharias}, {Zaharijas}, {Zampieri}, {Zandanel}, {Zanin},
  {Zavrtanik}, {Zavrtanik}, {Zdziarski}, {Zech}, {Zechlin}, {Zhdanov},
  {Ziegler}, and {Zorn}]{CTA2019book}
{Cherenkov Telescope Array Consortium}.; {Acharya}, B.S.; {Agudo}, I.; {Al
  Samarai}, I.; {Alfaro}, R.; {Alfaro}, J.; {Alispach}, C.; {Alves Batista},
  R.; {Amans}, J.P.; {Amato}, E.;  et~al.
\newblock {\em {Science with the Cherenkov Telescope Array}};  2019.
\newblock
  doi:{\changeurlcolor{black}\href{https://doi.org/10.1142/10986}{\detokenize{10.1142/10986}}}.

\bibitem[{Huentemeyer} \em{et~al.}(2019){Huentemeyer}, {BenZvi}, {Dingus},
  {Fleischhack}, {Schoorlemmer}, and {Weisgarber}]{Huentemeyer2019BAAS}
{Huentemeyer}, P.; {BenZvi}, S.; {Dingus}, B.; {Fleischhack}, H.;
  {Schoorlemmer}, H.; {Weisgarber}, T.
\newblock {The Southern Wide-Field Gamma-Ray Observatory (SWGO): A
  Next-Generation Ground-Based Survey Instrument}.
\newblock  Bulletin of the American Astronomical Society,  2019, Vol.~51, p.
  109,  \href{http://xxx.lanl.gov/abs/1907.07737}{{\normalfont
  [arXiv:astro-ph.IM/1907.07737]}}.
\newblock
  doi:{\changeurlcolor{black}\href{https://doi.org/10.48550/arXiv.1907.07737}{\detokenize{10.48550/arXiv.1907.07737}}}.

\bibitem[{Tomsick} \em{et~al.}(2019){Tomsick}, {Zoglauer}, {Sleator}, {Lazar},
  {Beechert}, {Boggs}, {Roberts}, {Siegert}, {Lowell}, {Wulf}, {Grove},
  {Phlips}, {Brandt}, {Smale}, {Kierans}, {Burns}, {Hartmann}, {Leising},
  {Ajello}, {Fryer}, {Amman}, {Chang}, {Jean}, and {von
  Ballmoos}]{Tomsick2019BAAS}
{Tomsick}, J.; {Zoglauer}, A.; {Sleator}, C.; {Lazar}, H.; {Beechert}, J.;
  {Boggs}, S.; {Roberts}, J.; {Siegert}, T.; {Lowell}, A.; {Wulf}, E.;  et~al.
\newblock {The Compton Spectrometer and Imager}.
\newblock  Bulletin of the American Astronomical Society,  2019, Vol.~51,
  p.~98,  \href{http://xxx.lanl.gov/abs/1908.04334}{{\normalfont
  [arXiv:astro-ph.IM/1908.04334]}}.
\newblock
  doi:{\changeurlcolor{black}\href{https://doi.org/10.48550/arXiv.1908.04334}{\detokenize{10.48550/arXiv.1908.04334}}}.

\bibitem[{Tomsick} and {COSI Collaboration}(2022)]{Tomsick2022icrc}
{Tomsick}, J.; {COSI Collaboration}.
\newblock {The Compton Spectrometer and Imager Project for MeV Astronomy}.
\newblock  37th International Cosmic Ray Conference,  2022, p. 652,
  \href{http://xxx.lanl.gov/abs/2109.10403}{{\normalfont
  [arXiv:astro-ph.IM/2109.10403]}}.
\newblock
  doi:{\changeurlcolor{black}\href{https://doi.org/10.22323/1.395.0652}{\detokenize{10.22323/1.395.0652}}}.

\bibitem[{Adri{\'a}n-Mart{\'\i}nez}
  \em{et~al.}(2016){Adri{\'a}n-Mart{\'\i}nez}, {Ageron}, {Aharonian}, {Aiello},
  {Albert}, {Ameli}, {Anassontzis}, {Andre}, {Androulakis}, {Anghinolfi},
  {Anton}, {Ardid}, {Avgitas}, {Barbarino}, {Barbarito}, {Baret},
  {Barrios-Mart{\'\i}}, {Belhorma}, {Belias}, {Berbee}, {van den Berg},
  {Bertin}, {Beurthey}, {van Beveren}, {Beverini}, {Biagi}, {Biagioni},
  {Billault}, {Bond{\`\i}}, {Bormuth}, {Bouhadef}, {Bourlis}, {Bourret},
  {Boutonnet}, {Bouwhuis}, {Bozza}, {Bruijn}, {Brunner}, {Buis}, {Busto},
  {Cacopardo}, {Caillat}, {Calamai}, {Calvo}, {Capone}, {Caramete}, {Cecchini},
  {Celli}, {Champion}, {Cherkaoui El Moursli}, {Cherubini}, {Chiarusi},
  {Circella}, {Classen}, {Cocimano}, {Coelho}, {Coleiro}, {Colonges},
  {Coniglione}, {Cordelli}, {Cosquer}, {Coyle}, {Creusot}, {Cuttone},
  {D'Amico}, {De Bonis}, {De Rosa}, {De Sio}, {Di Capua}, {Di Palma},
  {D{\'\i}az Garc{\'\i}a}, {Distefano}, {Donzaud}, {Dornic},
  {Dorosti-Hasankiadeh}, {Drakopoulou}, {Drouhin}, {Drury}, {Durocher},
  {Eberl}, {Eichie}, {van Eijk}, {El Bojaddaini}, {El Khayati}, {Elsaesser},
  {Enzenh{\"o}fer}, {Fassi}, {Favali}, {Fermani}, {Ferrara}, {Filippidis},
  {Frascadore}, {Fusco}, {Gal}, {Galat{\`a}}, {Garufi}, {Gay}, {Gebyehu},
  {Giordano}, {Gizani}, {Gracia}, {Graf}, {Gr{\'e}goire}, {Grella}, {Habel},
  {Hallmann}, {van Haren}, {Harissopulos}, {Heid}, {Heijboer}, {Heine},
  {Henry}, {Hern{\'a}ndez-Rey}, {Hevinga}, {Hofest{\"a}dt}, {Hugon},
  {Illuminati}, {James}, {Jansweijer}, {Jongen}, {de Jong}, {Kadler},
  {Kalekin}, {Kappes}, {Katz}, {Keller}, {Kieft}, {Kie{\ss}ling}, {Koffeman},
  {Kooijman}, {Kouchner}, {Kulikovskiy}, {Lahmann}, {Lamare}, {Leisos},
  {Leonora}, {Clark}, {Liolios}, {Llorens Alvarez}, {Lo Presti}, {L{\"o}hner},
  {Lonardo}, {Lotze}, {Loucatos}, {Maccioni}, {Mannheim}, {Margiotta},
  {Marinelli}, {Mari{\c{s}}}, {Markou}, {Mart{\'\i}nez-Mora}, {Martini},
  {Mele}, {Melis}, {Michael}, {Migliozzi}, {Migneco}, {Mijakowski}, {Miraglia},
  {Mollo}, {Mongelli}, {Morganti}, {Moussa}, {Musico}, {Musumeci}, {Navas},
  {Nicolau}, {Olcina}, {Olivetto}, {Orlando}, {Papaikonomou}, {Papaleo},
  {P{\u{a}}v{\u{a}}la{\c{s}}}, {Peek}, {Pellegrino}, {Perrina}, {Pfutzner},
  {Piattelli}, {Pikounis}, {Poma}, {Popa}, {Pradier}, {Pratolongo},
  {P{\"u}hlhofer}, {Pulvirenti}, {Quinn}, {Racca}, {Raffaelli}, {Randazzo},
  {Rapidis}, {Razis}, {Real}, {Resvanis}, {Reubelt}, {Riccobene}, {Rossi},
  {Rovelli}, {Salda{\~n}a}, {Salvadori}, {Samtleben}, {S{\'a}nchez
  Garc{\'\i}a}, {S{\'a}nchez Losa}, {Sanguineti}, {Santangelo}, {Santonocito},
  {Sapienza}, {Schimmel}, {Schmelling}, {Sciacca}, {Sedita}, {Seitz}, {Sgura},
  {Simeone}, {Siotis}, {Sipala}, {Spisso}, {Spurio}, {Stavropoulos},
  {Steijger}, {Stellacci}, {Stransky}, {Taiuti}, {Tayalati}, {T{\'e}zier},
  {Theraube}, {Thompson}, {Timmer}, {T{\"o}nnis}, {Trasatti}, {Trovato},
  {Tsirigotis}, {Tzamarias}, {Tzamariudaki}, {Vallage}, {Van Elewyck},
  {Vermeulen}, {Vicini}, {Viola}, {Vivolo}, {Volkert}, {Voulgaris}, {Wiggers},
  {Wilms}, {de Wolf}, {Zachariadou}, {Zornoza}, and
  {Z{\'u}{\~n}iga}]{Adrian-Martinez16JPG}
{Adri{\'a}n-Mart{\'\i}nez}, S.; {Ageron}, M.; {Aharonian}, F.; {Aiello}, S.;
  {Albert}, A.; {Ameli}, F.; {Anassontzis}, E.; {Andre}, M.; {Androulakis}, G.;
  {Anghinolfi}, M.;  et~al.
\newblock {Letter of intent for KM3NeT 2.0}.
\newblock {\em Journal of Physics G Nuclear Physics} {\bf 2016}, {\em
  43},~084001,  \href{http://xxx.lanl.gov/abs/1601.07459}{{\normalfont
  [arXiv:astro-ph.IM/1601.07459]}}.
\newblock
  doi:{\changeurlcolor{black}\href{https://doi.org/10.1088/0954-3899/43/8/084001}{\detokenize{10.1088/0954-3899/43/8/084001}}}.

\bibitem[{Agostini} \em{et~al.}(2020){Agostini}, {B{\"o}hmer}, {Bosma},
  {Clark}, {Danninger}, {Fruck}, {Gernh{\"a}user}, {G{\"a}rtner}, {Grant},
  {Henningsen}, {Holzapfel}, {Huber}, {Jenkyns}, {Krauss}, {Krings}, {Kopper},
  {Leism{\"u}ller}, {Leys}, {Macoun}, {Meighen-Berger}, {Michel}, {Moore},
  {Morley}, {Padovani}, {Papp}, {Pirenne}, {Qiu}, {Rea}, {Resconi}, {Round},
  {Ruskey}, {Spannfellner}, {Traxler}, {Turcati}, and
  {Yanez}]{Agostini20NatAst}
{Agostini}, M.; {B{\"o}hmer}, M.; {Bosma}, J.; {Clark}, K.; {Danninger}, M.;
  {Fruck}, C.; {Gernh{\"a}user}, R.; {G{\"a}rtner}, A.; {Grant}, D.;
  {Henningsen}, F.;  et~al.
\newblock {The Pacific Ocean Neutrino Experiment}.
\newblock {\em Nature Astronomy} {\bf 2020}, {\em 4},~913--915,
  \href{http://xxx.lanl.gov/abs/2005.09493}{{\normalfont
  [arXiv:astro-ph.HE/2005.09493]}}.
\newblock
  doi:{\changeurlcolor{black}\href{https://doi.org/10.1038/s41550-020-1182-4}{\detokenize{10.1038/s41550-020-1182-4}}}.

\bibitem[{Baikal-GVD Collaboration} \em{et~al.}(2019){Baikal-GVD
  Collaboration}, {:}, {Avrorin}, {Avrorin}, {Aynutdinov}, {Bannash},
  {Belolaptikov}, {Brudanin}, {Budnev}, {Domogatsky}, {Doroshenko},
  {Dvornicky}, {Dyachok}, {Dzhilkibaev}, {Fajth}, {Fialkovsky}, {Gafarov},
  {Golubkov}, {Gorshkov}, {Gress}, {Ivanov}, {Kebkal}, {Kebkal}, {Khramov},
  {Kolbin}, {Konischev}, {Korobchenko}, {Koshechkin}, {Kozhin}, {Kruglov},
  {Kryukov}, {Kulepov}, {Milenin}, {Mirgazov}, {Nazari}, {Panfilov},
  {Petukhov}, {Pliskovsky}, {Rozanov}, {Rjabov}, {Rushay}, {Safronov},
  {Shaybonov}, {Shelepov}, {Simkovic}, {Skurikhin}, {Solovjev}, {Sorokovikov},
  {Stekl}, {Suvorova}, {Sushenok}, {Tabolenko}, {Tarashansky}, and
  {Yakovlev}]{Baikal19arxiv}
{Baikal-GVD Collaboration}.; {:}.; {Avrorin}, A.D.; {Avrorin}, A.V.;
  {Aynutdinov}, V.M.; {Bannash}, R.; {Belolaptikov}, I.A.; {Brudanin}, V.B.;
  {Budnev}, N.M.; {Domogatsky}, G.V.;  et~al.
\newblock {Neutrino Telescope in Lake Baikal: Present and Future}.
\newblock {\em arXiv e-prints} {\bf 2019}, p. arXiv:1908.05427,
  \href{http://xxx.lanl.gov/abs/1908.05427}{{\normalfont
  [arXiv:astro-ph.HE/1908.05427]}}.

\bibitem[{Krumholz} \em{et~al.}(2022){Krumholz}, {Crocker}, and
  {Sampson}]{Krumholz2022MNRAS}
{Krumholz}, M.R.; {Crocker}, R.M.; {Sampson}, M.L.
\newblock {Cosmic ray interstellar propagation tool using It{\^o} Calculus
  (CRIPTIC): software for simultaneous calculation of cosmic ray transport and
  observational signatures}.
\newblock {\em \mnras} {\bf 2022}, {\em 517},~1355--1380,
  \href{http://xxx.lanl.gov/abs/2207.13838}{{\normalfont
  [arXiv:astro-ph.HE/2207.13838]}}.
\newblock
  doi:{\changeurlcolor{black}\href{https://doi.org/10.1093/mnras/stac2712}{\detokenize{10.1093/mnras/stac2712}}}.

\bibitem[{Hanasz} \em{et~al.}(2021){Hanasz}, {Strong}, and
  {Girichidis}]{Hanasz2021LRCA}
{Hanasz}, M.; {Strong}, A.W.; {Girichidis}, P.
\newblock {Simulations of cosmic ray propagation}.
\newblock {\em Living Reviews in Computational Astrophysics} {\bf 2021}, {\em
  7},~2,  \href{http://xxx.lanl.gov/abs/2106.08426}{{\normalfont
  [arXiv:astro-ph.HE/2106.08426]}}.
\newblock
  doi:{\changeurlcolor{black}\href{https://doi.org/10.1007/s41115-021-00011-1}{\detokenize{10.1007/s41115-021-00011-1}}}.

\bibitem[{Zweibel}(2020)]{Zweibel2020ApJ}
{Zweibel}, E.G.
\newblock {The Role of Pressure Anisotropy in Cosmic-Ray Hydrodynamics}.
\newblock {\em \apj} {\bf 2020}, {\em 890},~67,
  \href{http://xxx.lanl.gov/abs/1910.03052}{{\normalfont
  [arXiv:astro-ph.HE/1910.03052]}}.
\newblock
  doi:{\changeurlcolor{black}\href{https://doi.org/10.3847/1538-4357/ab67bf}{\detokenize{10.3847/1538-4357/ab67bf}}}.

\bibitem[{Ji} \em{et~al.}(2022){Ji}, {Squire}, and {Hopkins}]{Ji2022MNRAS}
{Ji}, S.; {Squire}, J.; {Hopkins}, P.F.
\newblock {Numerical study of cosmic ray confinement through dust resonant drag
  instabilities}.
\newblock {\em \mnras} {\bf 2022}, {\em 513},~282--295,
  \href{http://xxx.lanl.gov/abs/2112.00752}{{\normalfont
  [arXiv:astro-ph.HE/2112.00752]}}.
\newblock
  doi:{\changeurlcolor{black}\href{https://doi.org/10.1093/mnras/stac895}{\detokenize{10.1093/mnras/stac895}}}.

\end{thebibliography}

\end{adjustwidth}
\end{document}